\newcommand{\multidoc}{}
\newcommand{\ifstandalone}[1]{\ifdefined\multidoc\else #1 \fi}

\documentclass{report}
\usepackage{suthesis-2e}
\dept{Physics}
\usepackage{graphicx}
\usepackage{ctable}
\usepackage{subfigure}
\usepackage{hyperref}
\usepackage{amsmath}
\usepackage{mathtools}
\usepackage{amssymb}
\usepackage{amsfonts}
\usepackage{parskip}
\usepackage{empheq}
\usepackage{titling}
\usepackage{cite}
\usepackage{color}
\usepackage{tablefootnote}

\newcommand{\bra}[1]{\left\langle {#1} \right|}
\newcommand{\ket}[1]{\left| {#1} \right\rangle}
\newcommand{\ip}[2]{\left\langle {#1} | {#2} \right\rangle}
\newcommand{\avg}[1]{\left\langle {#1} \right\rangle}
\newcommand{\bracket}[3]{\left\langle {#1} | {#2} | {#3} \right\rangle}
\newcommand{\bbracket}[3]{\bigl\langle {#1} \bigl| {#2} \bigr| {#3} \bigr\rangle}
\newcommand{\sbracket}[3]{\langle {#1} | {#2} | {#3} \rangle}

\newcommand{\beq}{\begin{equation}}
\newcommand{\eeq}{\end{equation}}
\newcommand{\bea}{\begin{eqnarray}}
\newcommand{\eea}{\end{eqnarray}}

\newcommand{\dbl}[1]{\bar{#1}}
\renewcommand{\k}{\textbf{k}}
\renewcommand{\l}{\textbf{l}}
\renewcommand{\a}{\textbf{a}}
\renewcommand{\b}{\textbf{b}}
\renewcommand{\d}{{\rm d}}
\newcommand{\x}{\textbf{x}}

\newcommand{\sig}{{\rm s}}
\newcommand{\idl}{{\rm i}}
\newcounter{appsection}[chapter]
\begin{document}

\title{Coherent LQG Control, Free-Carrier Oscillations,\\
       Optical Ising Machines and Pulsed OPO Dynamics}
\author{Ryan Hamerly}
\principaladviser{Hideo Mabuchi}
\firstreader{Harold Hwang}
\secondreader{Surya Ganguli}
\thirdreader{Patrick Hayden}

\beforepreface
\prefacesection{Abstract}

Broadly speaking, this thesis is about {\it nonlinear optics}, {\it quantum mechanics}, and {\it computing}.  These fields have been around for quite a while, but only recently have scientists started to draw connections between them.  {\it Quantum optics} has been around since the laser, but it was advances in squeezing and single-atom cavity QED that caused the field to take off.  {\it Quantum computing} has grown from vague statements about simulation and factoring to a rigorous field of engineering, although large-scale quantum computers remain a distant goal.  {\it Optical computing}, by contrast, peaked in the 1970's and was eclipsed by electronics, but has made a comeback in the last decade as electronics run into physical limits in energy consumption.

In the next decade, nanophotonics will merge these fields.  By confining light in high-quality wavelength-scale resonators, optical nonlinearities can be enhanced by orders of magnitude.  This technology will first be used for low-power modulators and detectors in interconnects, and later for all-optical computing.  As fabrication improves and highly nonlinear materials become available, optical nonlinearities will reach the single-photon level, leading to quantum computing.  Who knows what will come next?

To reach this goal, we need a solid theoretical understanding of open quantum systems (in particular quantum-optical systems), quantum control and feedback networks.  This quantum ``circuit theory'' will resemble classical circuit theory, i.e.\ it will be modular and hierarchical, masking the underlying complexity of the components -- but should describe the full {\it quantum} dynamics of a circuit.  Although our work is aimed at quantum-optical systems, the theory will be applicable to any system that interacts through bosonic channels: optomechanics, superconducting circuits, etc.  Hand in hand with the theory, we aim to develop software to simulate quantum circuits: starting from a list of components and a network diagram, the computer automatically computes the correct quantum model and performs simulations, hiding most of the complexity from the user.

\subsection*{Organization of Thesis}

This thesis covers the four main projects I worked on as a Stanford PhD student: Coherent LQG Control, Free-Carrier Oscillations, Optical Ising Machines and Pulsed OPO Dynamics.  Tying them all together is a theory of open quantum systems called the {\it SLH model}, which I introduce in Chapter \ref{ch:01}.  The SLH model is a general framework for open quantum systems that interact through bosonic fields, and is the basis for the quantum circuit theory we develop.  It is modular in the sense that any circuit of SLH components has its own SLH model, derived through {\it Gough-James circuit algebra} rules.  Chapter \ref{ch:02} discusses SLH models for common quantum-optical components.  Both chapters are background material, but are a key prerequisite for what comes next.

My first project in the Mabuchi group was on Coherent LQG Control.  LQG stands for Linear Quadratic Gaussian: control of a linear system ({\it plant}) subject to Gaussian noise, where the cost function is quadratic.  This is a well-studied classical control problem, and the answer can be obtained by solving a Ricatti equation.  The optimal control involves a state estimator ({\it Kalman filter}) and a feedback element based on the estimated state of the plant.  Translating this to quantum systems, one can define an optimal {\it measurement-based} controller, where the outputs of the plant are sent into a homodyne detector and we perform LQG-optimal control on the measurement signal.

In two papers with Hideo Mabuchi, I showed that {\it coherent} LQG control, where a quantum system coherently processes the plant output rather than measuring it, does better than measurement-based control for two systems: an optical cavity and an optomechanical oscillator.  The intuition is that the coherent controller, being a quantum system, can process both quadratures simultaneously without adding extra noise, whereas the measurement-based controller must measure one quadrature and throw the other away (homodyne) or measure both with a noise penalty (heterodyne) \cite{Hamerly2012, Hamerly2013}.  Chapter \ref{ch:04b} discusses linear systems using the SLH model, and Chapter \ref{ch:11} presents our results.

After the LQG project, I worked on a software project with Gopal Sarma, Dmitri Pavlichin and Nikolas Tezak on a quantum circuit and computer-algebra software project.  We developed a set of circuit tools, based on term-rewriting in Mathematica, that could be used to model photonic networks, and Gopal applied this to his PhD work on error-correcting codes \cite{Sarma2013}.  It was a reduction of Chapters \ref{ch:01}-\ref{ch:02} to software.  After that, I spent a year working on useless stuff.

The next interesting project I joined was a collaboration with Charlie Santori at HP Labs.  Charlie showed that classical photonic networks, based on Kerr resonators with $\gtrsim 20$ photons per cavity, could be accurately simulated using the truncated Wigner method, a semiclassical approximation whose computation time scales linearly with circuit size.  We wrote code applying the Wigner method to arbitrary quantum networks, and simulated optical latches, flip-flops and digital counters \cite{Santori2014}.  With the Wigner method, one cannot model fully quantum behavior, but we could make strong statements about quantum limits to low-power classical photonic computing.  The Wigner method is discussed in Chapter \ref{ch:04}.

Inspired by the HP work, I extended the Wigner method to optical cavities with free-carrier nonlinearities, since in most materials, free-carrier dispersion is orders of magnitude stronger than the Kerr effect.  This was a challenge because the carriers are defined by fermionic operators, but the right bosonization did the trick.  In the end, I derived a set of stochastic differential equations that resembled the Kerr equations from the HP paper, but had additional noise terms due to free-carrier excitation and decay, which are incoherent processes \cite{Hamerly2015-1}; see Chapter \ref{ch:05b}.  These equations were used to study phase-sensitive amplifiers and latches (Ch.~\ref{ch:06b}) and limit-cycle behavior associated with the free-carrier Hopf bifurcation \cite{Hamerly2015-2} (Ch.~\ref{ch:07}).

After spending the summer of 2014 in Beijing studying Chinese, I visited Yoshi Yamamoto at National Institute for Informatics (NII) in Tokyo.  Through Yoshi I met Alireza Marandi, Peter McMahon, Shoko Utsunomiya (NII) and Hiroki Takesue (NTT), who were working on a ``coherent Ising machine''.  The Ising machine is a network of coupled optical parametric amplifiers (OPOs), driven slowly through threshold, which starts from squeezed vacuum, bifurcates and relaxes into a final state that solves for the ground state of the Ising problem.  This generated a lot of interest because general Ising problem is NP-hard, meaning that no one knows how to solve it efficiently on a computer.  Even quantum computers can't solve NP-hard problems.

Chapter \ref{ch:09} covers my work with Hiroki's group modeling their prototype 10000-bit Ising machine with 1D nearest-neighbor couplings.  Instead of relaxing into the ground state, they noticed the machine tended to form discrete ferromagnetic domains separated by defects, and that the defect density depended on the pump power \cite{Inagaki2016}.  I wrote code to simulate their system, based loosely on the truncated Wigner theory of Chapter \ref{ch:04}, correctly predicting the domain-wall density in 1D systems and making predictions for 2D and frustrated lattices.  In the process, I developed a theory on how Ising machines work: a {\it growth stage} where linear dynamics selects out the dominant eigenvectors of the coupling matrix, and a {\it saturation stage} where the system relaxes into a valid Ising state with amplitudes $\pm$1 \cite{Hamerly2016-2}.  Time will tell if this theory holds up to more complex experiments.

Discussions with Alireza and Yoshi led to a separate ``multimode'' theory of OPO Ising machines, which attracted the interest of Marty Fejer and his student Marc Jankowski, who study OPO pulse dynamics.  Together with Marc, Marty and Alireza, I wrote code to simulate their OPO system, but because of the separation of lengthscales in the problem, the code ran very slow.  So I learned CUDA, put it on a GPU and it ran 20 times faster, but that wasn't good enough, so in Chapter~\ref{ch:10}, I developed a number of {\it reduced models} for pulsed OPO dynamics, which accurately model the pulsed OPO in different regimes of operation \cite{Hamerly2016MM}.  This should be useful both for the Ising machine and also as a tool to aid the design and optimization of synchronously pumped OPOs.

During my final year, Hideo and I started a collaboration with Kambiz Jamshidi (TU-Dresden), an expert in slow light and silicon photonics who is interested in realizing the pulsed OPO (Ch.~\ref{ch:09}-\ref{ch:10}) and free-carrier (Ch.~\ref{ch:06b}-\ref{ch:07}) effects I studied in previous years.  As this is an ongoing project that I will continue post-Stanford, it is fitting to end this thesis with some of our early work: modeling optical waveguides in silicon (Chapter \ref{ch:12}).  While this is primarily a summary of existing literature with an eye towards the future, it is my hope that this chapter will lay the foundation for significant, impactful results as the collaboration begins to bear fruit.

\begin{figure}[tbp]
\begin{center}
\includegraphics[width=0.95\textwidth]{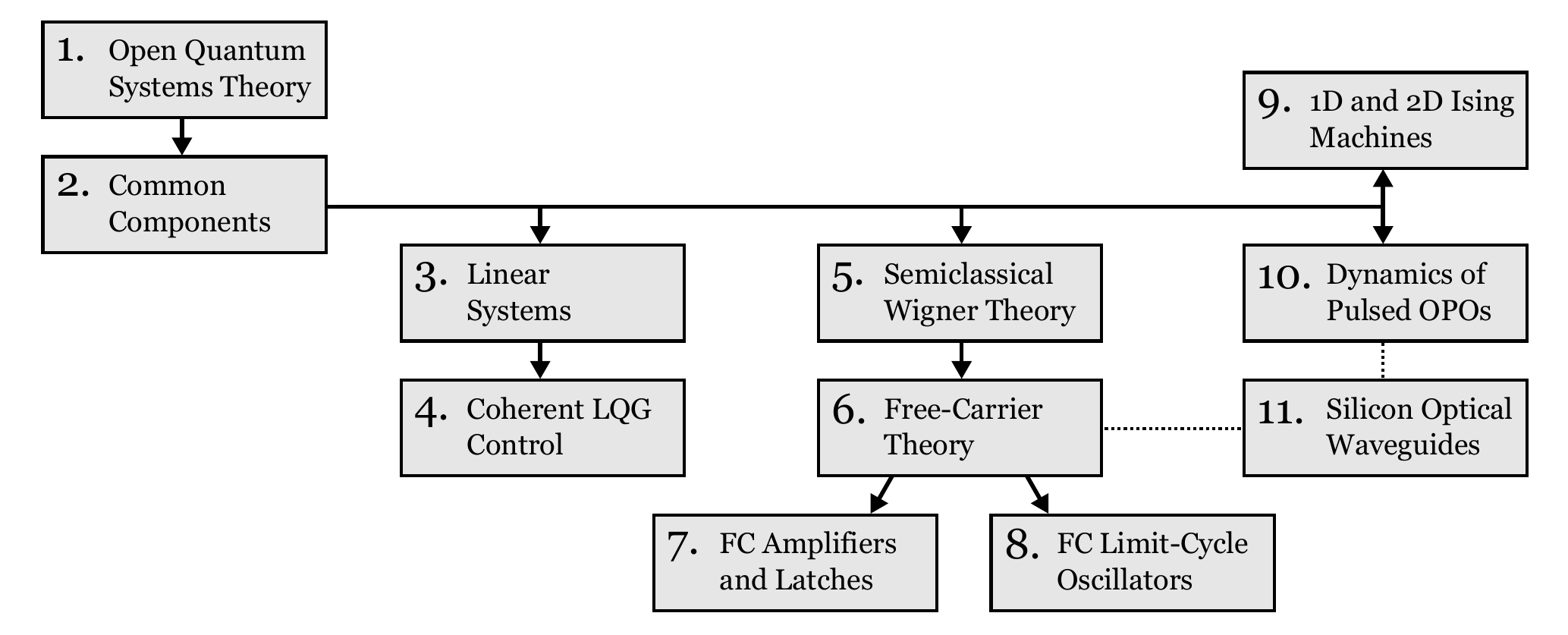}
\caption{Dependency tree for thesis chapters.}
\label{fig:00-f1}
\end{center}
\end{figure}

Experts may start at any chapter they want, but for the novice, I would highly recommend reading Ch.~\ref{ch:01}-\ref{ch:02} first.  For those who have trouble following Ch.~\ref{ch:01}-\ref{ch:02}, hopefully the references will be of use.  The rest of the thesis chapters group into projects, which are largely independent of each other.  Readers interested in the LQG work should familiarize themselves with Ch.~\ref{ch:04b} (Linear Systems) before proceeding to Ch.~\ref{ch:11} (Coherent LQG Control).  The free-carrier results in Ch.~\ref{ch:06b}-\ref{ch:07} depend on the equations derived in Ch.~\ref{ch:05b}, which in turn depends on the Wigner theory of Ch.~\ref{ch:04}.  The Ising machine chapters Ch.~\ref{ch:09}-\ref{ch:10} complement each other but are largely independent, although familiarity with the Wigner method (Ch.~\ref{ch:04}) might be helpful.  Chapter \ref{ch:12} is largely independent of the other chapters, but the material is most related to Ch.~\ref{ch:06b} and Ch.~\ref{ch:10}.
	
\subsection*{List of Publications}

\begin{itemize}
	\item \href{http://dx.doi.org/10.1103/PhysRevLett.109.173602}{Ryan Hamerly and Hideo Mabuchi, ``Advantages of Coherent Feedback for Cooling Quantum Oscillators.''  Phys.\ Rev.\ Lett.\ 109, 173602 (2012)}
	\item \href{http://dx.doi.org/10.1103/PhysRevA.87.013815}{Ryan Hamerly and Hideo Mabuchi, ``Coherent controllers for optical-feedback cooling of quantum oscillators.''  Phys.\ Rev.\ A 87, 013815 (2013)}
	\item \href{http://dx.doi.org/10.1109/JPHOT.2013.2243721}{Gopal Sarma, Ryan Hamerly, Nikolas Tezak, Dmitri S.\ Pavlichin, and Hideo Mabuchi, ``Transformation of Quantum Photonic Circuit Models by Term Rewriting.''  IEEE Photonics Journal 5:1, 7500111 (2013)}
	\item \href{http://dx.doi.org/10.1103/PhysRevApplied.1.054005}{Charles Santori, Jason S.\ Pelc, Raymond G.\ Beausoleil, Nikolas Tezak, Ryan Hamerly, and Hideo Mabuchi, ``Quantum Noise in Large-Scale Coherent Nonlinear Photonic Circuits.''  Phys.\ Rev.\ Applied 1, 054005 (2014)}
	\item \href{http://dx.doi.org/10.1364/CLEO_QELS.2014.FM2A.6}{Charles M.\ Santori, Jason S.\ Pelc, Raymond G.\ Beausoleil, Nikolas Tezak, Ryan Hamerly, and Hideo Mabuchi, ``Quantum Noise in Large-Scale Photonic Circuits.'' CLEO:2014, paper FM2A.6 (2014)}
	\item \href{http://dx.doi.org/10.1103/PhysRevA.92.023819}{Ryan Hamerly and Hideo Mabuchi, ``Quantum noise of free-carrier dispersion in semiconductor optical cavities.'' Phys.\ Rev.\ A 92, 023819 (2015)}
	\item \href{http://dx.doi.org/10.1103/PhysRevApplied.4.024016}{Ryan Hamerly and Hideo Mabuchi, ``Optical Devices Based on Limit Cycles and Amplification in Semiconductor Optical Cavities.'' Phys.\ Rev.\ Applied 4, 024016 (2015)}
	\item \href{http://dx.doi.org/10.1038/nphoton.2016.68}{Takahiro Inagaki, Kensuke Inaba, Ryan Hamerly, Kyo Inoue, Yoshihisa Yamamoto, and Hiroki Takesue, ``Large-scale Ising spin network based on degenerate optical parametric oscillators.'' {\it Nature Photonics} 10, 415Ð419 (2016)}
	\item \href{http://dx.doi.org/10.1117/12.2227308}{Ryan Hamerly, Kambiz Jamshidi, and Hideo Mabuchi, ``Quantum noise in energy-efficient slow light structures for optical computing: squeezed light from slow light'' Proc. SPIE 9900, Quantum Optics, 990012 (2016)}
	\item \href{http://dx.doi.org/10.1364/CLEO_AT.2016.JW2A.68}{Ryan Hamerly, Alireza Marandi, Marc Jankowski, Martin Fejer, Yoshihisa Yamamoto and Hideo Mabuchi, ``Reduced Models for Pulse Shaping and Nonlinear Dynamics in Optical Parametric Oscillators.'' CLEO:2016, paper JW2A.68 (2016)}
	\item \href{http://arxiv.org/abs/1605.03847}{Kenta Takata, Alireza Marandi, Ryan Hamerly, Yoshitaka Haribara, Daiki Maruo, Shuhei Tamate, Hiromasa Sakaguchi, Shoko Utsunomiya, Yoshihisa Yamamoto, ``A 16-bit Coherent Ising Machine for One-Dimensional Ring and Cubic Graph Problems.'' (to appear in {\it Scientific Reports})}
	\item \href{http://arxiv.org/abs/1605.03673}{Daniel B.\ S.\ Soh, Ryan Hamerly, Hideo Mabuchi, ``Comprehensive analysis of the optical Kerr coefficient of graphene.'' (to appear in Phys.\ Rev.\ A)}
	\item \href{http://arxiv.org/abs/1605.08121}{Ryan Hamerly, Kensuke Inaba, Takahiro Inagaki, Hiroki Takesue, Yoshihisa Yamamoto, and Hideo Mabuchi, ``Topological defect formation in 1D and 2D spin chains realized by network of optical parametric oscillators.'' (to appear in Intl.\ J.\ Mod.\ Phys.\ B)}
	\item \href{http://arxiv.org/abs/1608.02042}{Ryan Hamerly, Alireza Marandi, Marc Jankowski, Martin M.\ Fejer, Yoshihisa Yamamoto, and Hideo Mabuchi, ``Reduced models and design principles for half-harmonic generation in synchronously-pumped optical parametric oscillators'' (submitted to Phys.\ Rev.\ A)}
	\item Peter L. McMahon, Alireza Marandi, Yoshitaka Haribara, Ryan Hamerly, Carsten Langrock, Shuhei Tamate, Takahiro Inagaki, Hiroki Takesue, Shoko Utsunomiya, Kazuyuki Aihara, Robert L.\ Byer, M.\ M.\ Fejer, Hideo Mabuchi, Yoshihisa Yamamoto, ``A fully-programmable 100-spin coherent Ising machine with all-to-all connections.'' (submitted)
	\item Meysam Namdari, Mahmoud Jazayerifar, Ryan Hamerly, and Kambiz Jamshidi, ``CMOS Compatible Ring Resonators for Phase-Sensitive Optical Parametric Amplification.'' Photonics West OPTO 2017 (submitted)
	\item Ryan Hamerly, Levon Mirzoyam, Meysam Namdari, and Kambiz Jamshidi, ``Optical bistability, self-pulsing and soliton formation in silicon micro-rings with active carrier removal.'' Photonics West OPTO 2017 (submitted)
	\item Alireza Marandi, Marc Jankowski, Ryan Hamerly, Stephen J.\ Wolf, Evgeni Sorokin, Irina T.\ Sorokina, Martin M.\ Fejer, and Robert L.\ Byer, ``Efficient cascaded half-harmonic generation of mid-IR frequency combs.'' Photonics West LASE 2017 (submitted)
	\item Hiroki Takesue, Takahiro Inagaki, Kensuke Inaba, Ryan Hamerly, Kyo Inoue, and Yoshihisa Yamamoto, ``Large-scale artificial spin network based on time-multiplexed degenerate optical parametric oscillators for coherent Ising machine.'' Photonics West LASE 2017 (submitted)
\end{itemize}

I did a few interesting projects outside the Mabuchi Lab.  Before coming to Stanford, I worked on black hole dynamics with Yanbei Chen at Caltech \cite{Hamerly2011}.  During my winter-quarter rotation, I studied the effects of dark matter on helioseismology \cite{Hamerly2011b}.  The work was submitted to a conference in Hakone, Japan, but that winter Japan was struck by the terrible 2011 earthquake, so we could not go.  In 2013, I worked with a team of computer-science students on an app that visualizes the user's browsing history, \href{http://chrome.google.com/webstore/detail/webmapper/foachceonkmkeiigdbkjcihnaabppicf}{\it Webmapper}, which was later published on the Chrome Webstore.  It's free and you should check it out.

\begin{itemize}
	\item \href{http://arxiv.org/abs/1110.1169}{Ryan Hamerly and Alexander Kosovichev, ``Dark matter and its effects on helioseismology.'' in Proceedings of 61$^{\rm st}$ Fujihara Seminar: Progress in Solar / Stellar Physics with Helio- and Astroseismology, ed. Hiromoto Shibahashi, Masao Takata, Anthony Lynas-Gray (2011)}
	\item \href{http://dx.doi.org/10.1103/PhysRevD.84.124015}{Ryan Hamerly and Yanbei Chen, ``Event horizon deformations in extreme mass-ratio black hole mergers.'' Phys.\ Rev.\ D 84, 124015 (2011)}
	\item \href{http://chrome.google.com/webstore/detail/webmapper/foachceonkmkeiigdbkjcihnaabppicf}{Ryan Hamerly, Scott Chung, Sheta Chatterjee, and Milinda Lakkam, {\it Webmapper}, available at Chrome Webstore}, see also \url{http://github.com/rhamerly/webmapper/wiki} (2013)
	\item \href{http://allrecipes.com/recipe/236455/peanut-butter-rice-krispies-brownies/}{Ryan Hamerly, ``Peanut Butter Rice Krispies Brownies'', available on Allrecipes.com (2014)}
\end{itemize}

\prefacesection{Acknowledgments}

Lots of people contributed to the success of my PhD work.

On the LQG project, I benefited from discussions with Hideo Mabuchi, Nikolas Tezak (PhD 2016), Gopal Sarma (PhD 2013) and Orion Crisafulli (PhD 2012).  The work was based on some earlier results of Hendra Nurdin (UNSW), Matt James (ANU) and Ian Petersen (UNSW), and I would like to thank Hendra for useful discussions during his visits.

The circuit algebra work was done with Gopal, Nikolas, Dmitri, Hideo, and also Armand Niederberger.  Follow-up work has been done by Nik, Gil Tabak, Michael Celentano (MS 2015), and Michael Goerz (ARL)

For the free-carrier project, I would like to acknowledge Hideo and Nik, and on the HP side, Charles Santori, Jason Pelc, Ranojoy Bose and Ray Beausoleil.  For ongoing (as of this thesis) follow-up work I should acknowledge Dave Kielpinski, Thomas Van Vaerenbergh and Gabriel Mendoza, too.

My main collaborators for the Ising-machine projects were Alireza Marandi, Peter McMahon, Hiroki Takesue (NTT) and Yoshihisa Yamamoto (NII).  In addition, I should acknowledge Shoko Utsunomiya, Shuhei Tamate, Hiromasa Sakaguchi, Daiki Maruo (NII), Yoshitaka Haribara, Timothee Leleu (U.\ Tokyo), Kensuke Inaba, and Takahiro Inagaki (NTT).  For ongoing work at Stanford (as of this thesis), I should credit Tatsuhiro Onodera and Edwin Ng.

The pulsed OPO work was done with Alireza, Yoshi, Marc Jankowski and Marty Fejer.  I learned a lot from discussions with Marc and Marty.

The work on silicon waveguides was done with Kambiz Jamshidi, Meysam Namdari, and Levon Mirzoyam (Technische Universit\"{a}t Dresden), as well as Dodd Gray (Stanford).  Dodd is especially helpful and knowledgeable in areas such as electronics, nonlinear optics, and materials science.

I should also acknwledge Daniel Soh (Stanford / Sandia) for work on nonlinear properties of 2D materials.

\begin{figure}[tbp]
\begin{center}
\includegraphics[width=1.00\textwidth]{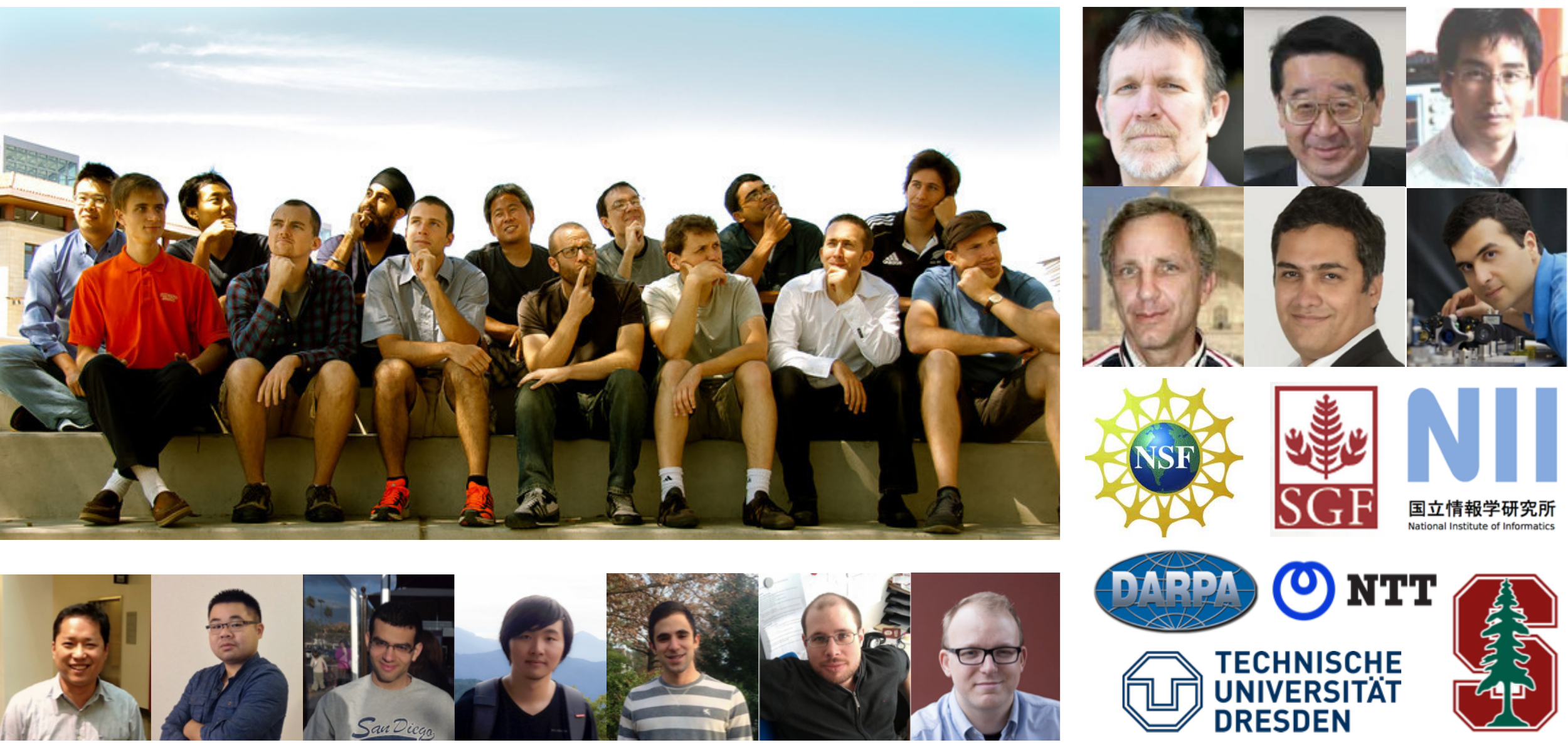}
\caption{Lab members and colleagues.  Top (left to right): Jie Wu, Ryan Hamerly, Yeong-Dae Kwon, Nate Bogdanowicz, Hardeep Sanghera, Dodd Gray, Hideo Mabuchi, Mike Armen, Orion Crisafulli, Dmitri Pavlichin, Gopal Sarma, Armand Niederberger, Charles Limouse, Nikolas Tezak.  Bottom (left to right): Daniel Soh, Edwin Ng, Gil Tabak, Tatsuhiro Onodera, Michael Celentano, Michael Goerz, Peter McMahon.  Right: Ray Beausoleil, Yoshi Yamamoto, Hiroki Takesue, Marty Fejer, Kambiz Jamshidi, Alireza Marandi.}
\label{default}
\end{center}
\end{figure}

Hardeep Sanghera is a real coffee guru.  I would like to thank him for maintaining the Mabuchilab Espresso Machine for the last three years, and for teaching me how to make cappuccinos and lattes.  Now I'm hooked, and I have a backup job in case science gets boring.

Additional lab members who contributed to the intellectual and social environment of the group include Jie Wu, Yeong-Dae Kwon, Nate Bogdanowicz, Eric Chatterjee, Mike Armen, Charles Limouse, Nina Amini, and Jeff Hill.

The organizers of the Stanford OSA chapter did a great job facilitating interaction among optics researchers.  The SUPR retreat was an excellent way to learn about the research of Stanford colleagues, learn how to make good posters and prepare for real conferences.  The PRACQSYS conferences, which I was fortunate to attend in 2012 (Tokyo), 2013 (Monterey), and 2015 (Sydney) were also highlights of my time at Stanford.

Our admins Suki Ungson (Mabuchi group), Yurika Peterman and Rieko Sasaki (Yamamoto group) did a lot of paperwork to keep the lab running, and were very helpful when I had questions.  I should also thank Maria Frank, Elva Carbajal and Paula Perron for their work in the physics / AP offices.

From my days as a teaching assistant, I am grateful to Rick Pam and Chaya Nanavati for teaching me to teach.  Learning from Chaya can be a bit rough at times, but it instills a habit of discipline and orderliness that all teachers need to be effective.  This is especially true for students in theoretical research like me, who tend to be dreamy and disorganized.

During my career, my work was supported by the NSF Graduate Research Fellowship Program (GRFP), a Stanford Graduate Fellowship (SGF), NSF grant PHY-1005386, AFOSR grant FA9550-11-1-0238, DARPA-MTO grant N66001-11-1-4106, a seed grant from the Precourt Institute for Energy at Stanford, and the Impulsing Paradigm Change through Disruptive Technologies (ImPACT) Program of the Cabinet Office of Japan.

I would like to thank Tim Zerlang and Richard Powers for teaching piano and dance classes, widening my interests beyond science, and also Michael and Kathryn Hamerly, as well as my friends and family at Stanford, Caltech and Colorado.

\includegraphics[width=1.00\textwidth]{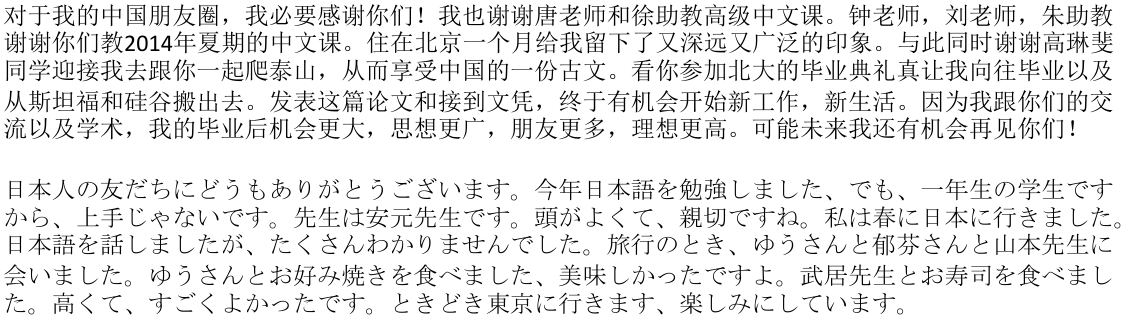}

\afterpreface

\ifdefined\multidoc\else\input{Header}\fi

\chapter{Open Quantum Systems Theory}
\label{ch:01}

Fake quantum systems are \textit{closed}.  They do not interact with the surroundings.  They are represented with a wavefunction $\ket{\psi}$ and evolve according to the Schrodinger Equation, which depends only on the Hamiltonian $H$.  Measurements are made by projecting $\ket{\psi}$ onto an operator eigenspace.  Most college-level quantum courses, including some courses on quantum information, deal only with closed quantum systems.

Real quantum systems are \textit{open}.  An open system interacts constantly with its surroundings (the \textit{bath}), and a full description must include the dynamics of both the system and the bath.  Rather than an isolated device, an open system is best described with both an internal Hamiltonian $H$ and couplings $L$ to \textit{input-output modes} of the bath.  By averaging over the bath degrees of freedom, the system can be represented with a restricted density matrix $\rho$, and in the absence of measurement, evolves according to the \textit{Master Equation}.  Measurements are performed by conditioning the master equation on the measured values of the output modes.

Closed quantum systems (without measurement) are Hamiltonian, deterministic, and conservative.  They conserve phase space and they conserve energy.  Two identical systems, initialized in orthogonal states, will remain forever orthogonal.  Not so with open systems.  They are dissipative and stochastic.  A very different set of tools must be developed to model open quantum systems.

This chapter introduces the basic theory of open quantum systems.  While these are all old results, it is helpful to restate them here to make the thesis self-contained.  The reader looking for more detail and background should consult {\it Quantum Optics} by Walls \& Milburn \cite{WallsMilburn}, {\it Quantum Noise} by Gardiner \& Zoller \cite{GardinerBook}, {\it An open systems approach to quantum optics} by Carmichael \cite{CarmichaelBook}, and Joe Kerckhoff's thesis \cite{KerckhoffThesis}.  Stochastic calculus theory is also helpful; Gardiner's {\it Handbook of Stochastic Methods} \cite{GardinerHandbook} is a good reference.

A basic understanding of quantum mechanics is assumed, but no knowledge formal open quantum systems theory is necessary.  These will be built up from the fundamentals.

\section{Opening Example: Optical Cavity}
\label{sec:01-cav}

\begin{figure}[t]
\begin{center}
\includegraphics[width=0.80\textwidth]{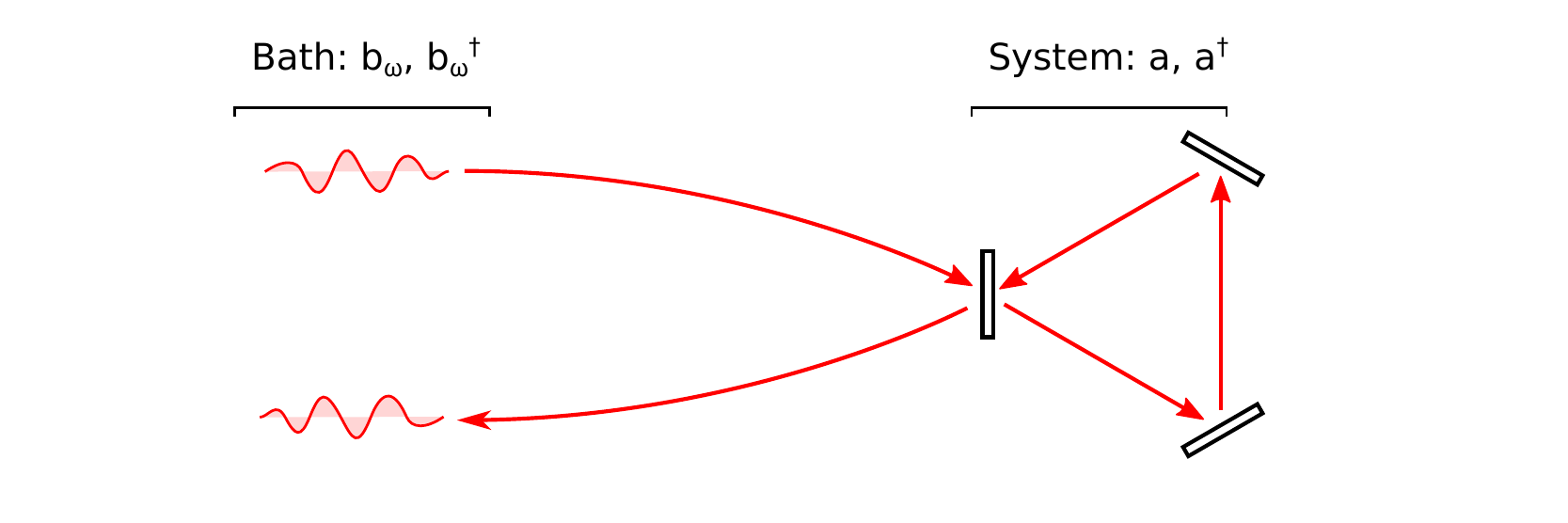}
\caption{Example system: optical cavity coupled to a single input-output field.}
\label{fig:01-f2}
\end{center}
\end{figure}

As a very simple example, consider an optical cavity with resonance $\omega_0$ and loss $\kappa$.  This is illustrated in Figure \ref{fig:01-f2}.  The cavity couples to an optical input and output, which has many modes $\omega$.  The cavity is the ``system''; these modes form the ``bath''.  The combined Hamiltonian is:
\beq
	H_{\rm full} = \omega_0 a^\dagger a + \int{\frac{\d\omega}{2\pi} \omega b_\omega^\dagger b_\omega} - i\sqrt{\kappa} \int{\frac{\d\omega}{2\pi}(a^\dagger b(\omega) - a b(\omega)^\dagger)} \label{eq:01-cav}
\eeq
where $(a, a^\dagger)$ are the cavity creation/annihilation operators $[a, a^\dagger] = 1$, and the $b(\omega)$ are operators for the continuum of modes that couple to the cavity: $[b(\omega), b(\omega')^\dagger] = 2\pi\delta(\omega - \omega')$.

\begin{figure}[t]
\begin{center}
\includegraphics[width=0.90\textwidth]{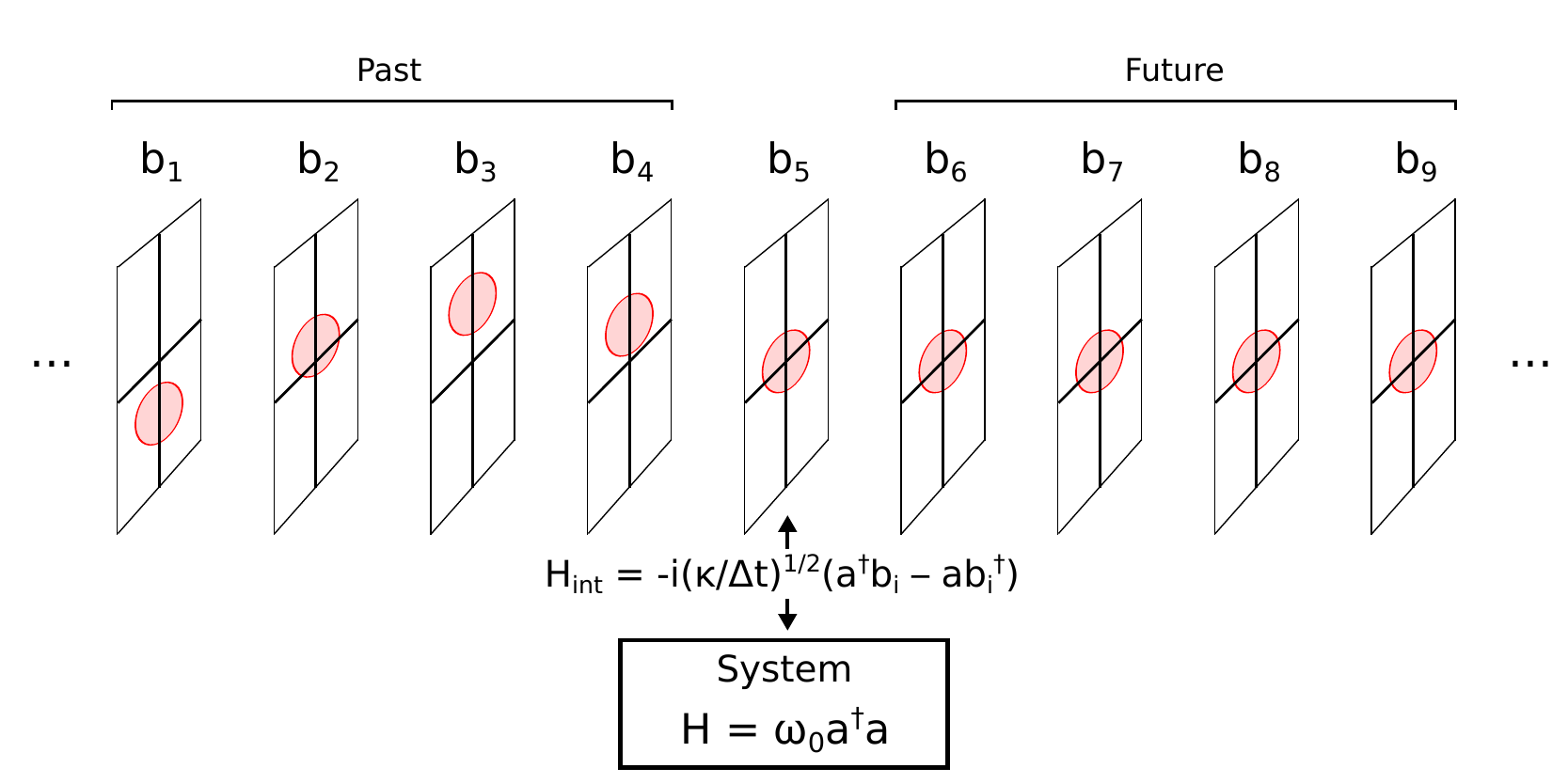}
\caption{Visualizing the optical cavity as an open quantum system, at time $t \in [t_5, t_5+\Delta t]$.}
\label{fig:01-f1}
\end{center}
\end{figure}

The Heisenberg equations become:
\bea
	\frac{\d a}{\d t} & = & -i\omega_0 a - \sqrt{\kappa} \int{\frac{\d\omega}{2\pi} b_\omega} \\
	\frac{\d b(\omega)}{\d t} & = & -i\omega b(\omega) + \sqrt{\kappa} a
\eea
Now we can perform a gauge transformation $\ket{\psi} \rightarrow \ket{\psi'}$ on the Hilbert space: in the Schrodinger picture, $\ket{\psi} = e^{-iH_0 t}\ket{\psi}$, where $H_0 = \int{\frac{\d\omega}{2\pi} \omega b_\omega^\dagger b_\omega}$.  In the Heisenberg picture, $A \rightarrow A'$, where $A = e^{iH_0 t} A' e^{-iH_0 t}$.  In this particular case:
\bea
	b(\omega) \rightarrow e^{-i\omega t} b(\omega) \label{eq:01-rwa}
\eea
The Heisenberg equations for the new operators become:
\bea
	\frac{\d a}{\d t} & = & -i\omega_0 a - \sqrt{\kappa} \int{\frac{\d\omega}{2\pi} e^{-i\omega t} b_\omega} \\
	\frac{\d b(\omega)}{\d t} & = & \sqrt{\kappa} e^{i\omega t} a
\eea
Now we Fourier-transform the input-output field $b_\omega$ into the spatial domain:
\beq
	b(\omega) = \int{e^{i\omega \tau} b(\tau)\d\tau},\ \ \ b(\tau) = \frac{1}{2\pi} \int{e^{-i\omega \tau} b(\omega)\d\omega} \label{eq:01-fourier}
\eeq
which satisfy $[b(\tau), b(\tau')] = \delta(\tau-\tau')$.  This gives equations of the form:
\bea
	\frac{\d a}{\d t} & = & -i\omega_0 a - \sqrt{\kappa}\,b(\tau=t) \\
	\frac{\d b(x)}{\d t} & = & \sqrt{\kappa}\,\delta(\tau-t)a
\eea
It is, however, difficult to work with a continuum of input and output modes.  So we discretize time into intervals of $\Delta t$, and define operators $b_i$:
\beq
	b_i = \frac{1}{\sqrt{\Delta t}} \int_{t_i}^{t_i+\Delta t}{b_i(\tau)\d\tau} \label{eq:01-disc}
\eeq
These have been normalized so that $[b_i, b_j^\dagger] = \delta_{ij}$.  Now notice that in the continuous-time equations, only one $b(\tau)$ interacts with the system at any time.  This carries over to the discrete picture.  In the interval $[t_i, t_i+\Delta t]$, only the mode $b_i$ interacts with the system, and the interaction is:
\bea
	\Delta a & = & -i\omega_0 a \Delta t - \sqrt{\kappa\,\Delta t}\,b_i \label{eq:01-qsde-proto} \\
	\Delta b_i & = & \sqrt{\kappa\,\Delta t}\,a \label{eq:01-qsde-proto-2}
\eea
Over time $\Delta t$, this corresponds to a Hamiltonian of the following form:
\beq
	H_{[t_i,t_i+\Delta t]} = \omega_0 a^\dagger a - i\sqrt{\kappa/\Delta t}(a^\dagger b_i - a b_i^\dagger) \label{eq:01-ham-proto}
\eeq
This is visualized in Figure \ref{fig:01-f1}.  This figure shows a system (optical cavity in this case) interacting with an input-output mode.  Following the derivation, this continuum of modes is discretized on a time scale $\Delta t$, and becomes an infinite train of harmonic oscillators.  The cavity interacts with each oscillator in succession, starting with the leftmost ($b_1$, at time $t_1$) and moving rightward.  Each harmonic oscillator $b_i$ is represented by a phase-space plot, though we caution that this is highly abstract, since all the modes can be entangled with the system and with each other.  All the ``future'' oscillators are in their ground state because they have not yet interacted with the system (and we are assuming vacuum inputs here).  The ``past'' oscillators, on the other hand, are not in their ground state, as these {\it have} interacted with the system.

We have succeeded in reducing an open quantum system with an infinite number of degrees of freedom, Eq.~(\ref{eq:01-cav}), into a (countably) infinite train of harmonic oscillators, each interacting with the system one after the other.  A further simplification can help here: over the time interval $[t_i, t_i + \Delta t]$, $U(\Delta t) \equiv e^{-iH_i \Delta t} = 1 + O(\sqrt{\Delta t})$.  Since each oscillator only interacts with the plant over a single interval, this means that for small $\Delta t$, all of the oscillators stay close to their ground states.

\section{Generalization}

Now let's repeat this for a general open quantum system.  Start by generalizing Eq.~(\ref{eq:01-cav}), as follows:
\bea
	H_{\rm full} & = & H_0 + \sum_m\int{\frac{\d\omega}{2\pi} \omega b_m(\omega)^\dagger b_m(\omega)} - i \sum_m\int{\frac{\d\omega}{2\pi}(M_m^\dagger b_m(\omega) - M_m b_m(\omega)^\dagger)} \nonumber \\
	& & +\ \sum_{mn} \int{\frac{\d\omega\,\d\omega'}{4\pi^2} N_{mn} b_m(\omega)^\dagger b_n(\omega')} \label{eq:01-slh-fuller}
\eea
This has a few differences from the original, but first notice what does \textit{not} change.  The bath still consists of a continuum of bosonic, harmonic-oscillator modes $b_m(\omega)$.  This bath is also Markovian -- since it respects the commutation relations $[b_m(\omega), b_n(\omega')] = 2\pi\delta_{mn}\delta(\omega - \omega')$, it can be Fourier-transformed into the time domain to give $[b_m(\tau), b_n(\tau')] = \delta(\tau-\tau')]$, which means that different parts of the input noise are independent of each other.  Additionally, the system-bath interaction takes the same form -- a simple product of system and bath operators.

Now consider the differences.  First, the system Hilbert space doesn't have to be that of a harmonic oscillator.  It could be a two-level atom, an oscillator connected to an atom -- anything, really.  And the harmonic-oscillator potential, $\omega_0 a^\dagger a$, is replaced by a more general $H$.  Note also that in the coupling part, $a$ is replaced by an operator $L_m$, which acts on the system Hilbert space.  In the previous equation, this coupling could be understood as a process that annihilated a photon from the cavity and created one in the bath, or vice versa.  Here, in a loose sense whenever a photon is ejected to the bath, the state of the system gets multiplied by $L$, likewise whenever it absorbs a photon from the bath, the system state is multiplied by $L^\dagger$.  So $L = \sqrt{\kappa} a$ makes sense for photon leakage, but other $L$ operators are possible.  For example, if the system were a two-level atom with ground state $\ket{0}$ and excited state $\ket{1}$, a good choice for $L$ would be the lowering operator $\sigma_- = \ket{0}\bra{1}$, since this would correspond to emitting a photon when the atom jumps from state $\ket{1}$ to $\ket{0}$, and absorbing a photon to jump back.

The final term -- $N_{mn}$ -- is new here, and facilitates inter-mode scattering.  It is an operator-valued matrix, the $N$ operators living on the system Hilbert space, and the matrix must be Hermitian.  This term is quadratic in the bath modes.  The double integral is necessary because any other quadratic form, like $N_{mn} b_m(\omega)^\dagger b_n(\omega)$, ends up being nonlocal in time and thus unphysical.  Other contributions, like $b_m b_n$ or $b_m^\dagger b_n^\dagger$, are also unphysical because they introduce infinite-bandwidth squeezing, which takes an infinite amount of energy.  The $N_{mn}$ term above is therefore the most general quadratic system-bath coupling for a open system coupled to a Markovian bath.

Higher powers in $b$ end up being negligible, since the continuum of bath states means that each oscillator $b_m$ always stays close to its ground state -- so (\ref{eq:01-slh-fuller}) is the most general Hamiltonian for an open Markovian quantum system.  Just as a closed system is fully described by a Hilbert space and a Hamiltonian $H$, an open quantum system is fully described by a Hilbert space, a Hamiltonian $H$, coupling terms $M_m$, and a scattering term $N_{mn}$.

We will use the Hamiltonian (\ref{eq:01-slh-fuller}) to derive Heisenberg equations of motion for system and bath operators.  These are best written out as It\^{o} SDEs, and are therefore called Quantum Stochastic Differential Equations (QSDEs).  From the QSDEs, it will be straightforward to derive a Schrodinger-like equation, the Master Equation, for the system density matrix, as well as conditional equations that depend on measurements of the output fields.

\section{Unitary Equation}

To derive the quantum stochastic differential equations, we follow the same procedure used in Sec.~\ref{sec:01-cav}.  Start by going into the rotating-wave frame $b_m(\omega) \rightarrow e^{-i\omega t} b_m(\omega)$.  This causes the quadratic part in $H_{\rm full}$ to drop out, leaving us with: 
\beq
	H_{\rm full} = H_0 - i \sum_m\int{\frac{\d\omega}{2\pi}(M_m^\dagger e^{-i\omega t} b_m(\omega) - M_m e^{i\omega t} b_m(\omega)^\dagger)} + \sum_{mn} \int{\frac{\d\omega\,\d\omega'}{4\pi^2} N_{mn} e^{i(\omega-\omega')t} b_m(\omega)^\dagger b_n(\omega')} \label{eq:01-slh-rwa}
\eeq
Now Fourier-transform the input-output fields
\beq
	b_m(\omega) = \int{e^{i\omega \tau} b_m(\tau)\d\tau},\ \ \ b_m(\tau) = \frac{1}{2\pi} \int{e^{-i\omega \tau} b_m(\omega)\d\omega}
\eeq
to get:
\beq
	H_{\rm full} = H_0 - i \sum_m(M_m^\dagger b_m(t) - M_m b_m(t)^\dagger) + \sum_{mn} N_{mn} b_m(t)^\dagger b_n(t) \label{eq:01-slh-rwa}
\eeq
(Now we see why the double-integral was needed for the $N_{mn}$ term -- when transformed into the time domain, it ensures that the coupling is local -- $b_m(t)$ is not coupling to fields before or after it.)

To construct the QSDEs, we first need an equation for the unitary $U_t$, defined as the solution to $\d U_t/\d t = -iH_{\rm full}(t) U_t$.  For small increments $\d t$ (using $\Delta t\rightarrow \d t$ in what follows), we can write:
\beq
	\d U_t \equiv U_{t+\d t}-U_t = -iH_{\rm full}\int_t^{t+\d t}{U_\tau \d\tau} \approx \left(-i\int_t^{t+\d t}{H_{\rm full}(\tau)\d\tau}\right)U_{t+\d t/2} \label{eq:01-du-sde}
\eeq
The extra $\d t/2$ in the final $U$ is critical.  It comes from making the \textit{midpoint} approximation $\int_t^{t+\d t}{U_\tau \d\tau} = U_{t+\d t/2}$ rather than the \textit{endpoint} approximation $\int_t^{t+\d t}{U_\tau \d\tau} = U_{t}$.  The midpoint approximation is more accurate than the endpoint approximation.  For smooth differential equations, both work fine, but for stochastic processes, the endpoint approximation is not good enough.  Open quantum systems follow stochastic equations, as we shall see, so the midpoint approximation must be used.

Going forward, one computes the $H_{\rm full}$ integral in (\ref{eq:01-du-sde}):
\bea
	-i\int_t^{t+\d t}{H_{\rm full}(\tau)\d\tau} & = & -i\sum_{mn} N_{mn} \int_t^{t+\d t}{b_m(\tau)^\dagger b_n(\tau)\d\tau} \nonumber \\
	& & \quad+\ \sum_m(M_m \int_t^{t+\d t}{b_m(\tau)^\dagger \d\tau} - M_m^\dagger \int_t^{t+\d t}{b_m(\tau)\d\tau}) - iH_0 \d t \nonumber \\
	& = & -i\sum_{mn} N_{mn}\d\Lambda_{mn} + \sum_m(M_m \d B_m^\dagger - M_m^\dagger \d B_m) - iH_0 \d t \label{eq:01-int-sde}
\eea
In (\ref{eq:01-int-sde}), we have defined the following stochastic increments:
\bea
	\d\Lambda_{mn} & = & \int_t^{t+\d t}{b_m(\tau)^\dagger b_n(\tau)\d\tau} \\
	\d B_m & = & \int_t^{t+\d t}{b_m(\tau)\d\tau} \\
	\d B_m^\dagger & = & \int_t^{t+\d t}{b_m(\tau)^\dagger \d\tau}
\eea
These are quantum, operator-valued white noise processes.  Like all white noise processes, they are not differentiable.  They satisfy the following It\^{o} relations:
\bea
	\d\Lambda_{mn}\d\Lambda_{pq} & = & \delta_{np} \d\Lambda_{mq} \\
	\d\Lambda_{mn}\d B_p^\dagger & = & \delta_{np} \d B_m^\dagger \\
	\d B_m \d\Lambda_{np} & = & \delta_{mn} \d B_p \\
	\d B_m \d B_n^\dagger & = & \delta_{mn} \d t
\eea
and all of the other products are zero.  These are very different from non-stochastic increments like $\d t$.  In calculus, we always assume that $\d t^2 = 0$.  Not so with stochastic increments.  This will become important in our derivation of $dU_t$, below.

Intuitively, $\d B_m$ and $\d B_m^\dagger$ act like non-normalized creation and annihilation operators for the field at $x \in [t, t+\d t]$.  In fact, they can be related to the discretized operators in Eq.~(\ref{eq:01-disc}) as follows: $\d B_{t_i} \leftrightarrow \sqrt{\d t}\,b_i$.  The $\d\Lambda_{mn}$ corresponds to a process that annihilates a photon in mode $n$ and creates one in $m$.  With only a single field, we can relate it to the discretized operators in (\ref{eq:01-disc}): $\d\Lambda_{t_i} \leftrightarrow b_i^\dagger b_i$.

Deriving $dU_t$ is straightforward, but a little tedious, starting with (\ref{eq:01-du-sde}) and (\ref{eq:01-int-sde}), we obtain:
\beq
	dU_t = \left[-i\sum_{mn} N_{mn}\d\Lambda_{mn} + \sum_m(M_m \d B_m^\dagger - M_m^\dagger \d B_m) - iH_0 \d t\right] U_{t+\d t/2} \label{eq:01-strat}
\eeq
This is a linear \textit{Stratonovich SDE} because the increment is expressed in terms of $U_{t+\d t/2}$ evaluated at the midpoint rather than the endpoint.  More common in quantum optics is the \textit{It\^{o} SDE}, which expresses the increment in terms of the endpoint $U_t$.  The It\^{o} SDE is easier to integrate, and in general more convenient to use, so the next step is to convert Eq.~(\ref{eq:01-strat}) into It\^{o} form.  This is done by approximating $U_{t+\d t/2} \approx U_t + dU_t/2$ to give
\beq
	dU_t = \left[-i\sum_{mn} N_{mn}\d\Lambda_{mn} + \sum_m(M_m \d B_m^\dagger - M_m^\dagger \d B_m) - iH_0 \d t\right] \left[U_t + \frac{1}{2}dU_t\right] \label{eq:01-strat-2}
\eeq
and evaluating (\ref{eq:01-strat-2}) recursively, by Picard's method.
\bea
	dU_t & = & \biggl[-i\sum_{mn} N_{mn}\d\Lambda_{mn} + \sum_m(M_m \d B_m^\dagger - M_m^\dagger \d B_m) - iH_0 \d t \nonumber \\
	& & +\ \frac{1}{2}\left(-\sum_{mn} (N^2)_{mn} \d\Lambda_{mn} + \sum_m(-i(NM)_m \d B_m^\dagger + i(M^\dagger N)_m \d B_m) - M^\dagger M \d t\right) \nonumber \\
	& & +\ \frac{1}{4}\left(i\sum_{mn} (N^3)_{mn} \d\Lambda_{mn} + \sum_m(-(N^2M)_m \d B_m^\dagger + (M^\dagger N^2)_m \d B_m) + iM^\dagger N M \d t\right) \nonumber \\
	& & +\ \frac{1}{8}\left(\sum_{mn} (N^4)_{mn} \d\Lambda_{mn} + \sum_m(i(N^3M)_m \d B_m^\dagger - i(M^\dagger N^3)_m \d B_m) + M^\dagger N^2 M \d t\right) \nonumber \\
	& & +\ (\cdots)\ \biggr]U_t \\
	& = & \left[\sum_{mn} \left(\frac{1-iN/2}{1+iN/2} - 1\right)_{mn} \d\Lambda_{mn} + \sum_m\left(\frac{1}{1+iN/2}M\right)_m \d B_m^\dagger\right. \nonumber \\
	& & \ \ \ - \sum_m\left(\frac{1}{1-iN/2}M\right)_m^\dagger \d B_m + \left.\left(-iH_0 - \frac{1}{2}M^\dagger \frac{1}{1+iN/2} M\right)\d t\right]U_t
\eea
Now define $S = (1+iN/2)/(1-iN/2)$ and $L = (1+iN/2)^{-1}M$.  The unitary equation becomes:
\bea
	\d U_t & = & \left[\sum_{mn} (S - 1)_{mn} \d\Lambda_{mn} + \sum_m L_m \d B_m^\dagger\right. \nonumber \\
	& & \ \ \ - \sum_m(L^\dagger S)_m \d B_m + \left.\left(-i\left(H_0 - \frac{1}{4}L^\dagger N L\right) - \frac{1}{2}L^\dagger L\right)\d t\right]U_t
\eea
One last thing -- the energy levels get renormalized by the interaction: $H = H_0 - L^\dagger N L/4$.  The unitary evolves as:
\beq
\boxed{
	\d U_t = \left[\sum_{mn} (S - 1)_{mn} \d\Lambda_{mn} + \sum_m L_m \d B_m^\dagger - \sum_m(L^\dagger S)_m \d B_m + \left(-iH - \frac{1}{2}L^\dagger L\right)\d t\right]U_t
	  } \label{eq:01-unitary}
\eeq

\section{Operator QSDEs}

Equation (\ref{eq:01-unitary}) is the root from which everything else is derived.  Start with the operator QSDEs, the Heisenberg equations for open quantum systems.  There will be three QSDEs -- one for system operators, one for the external fields $\d B, \d B^\dagger$, and one for the $\d\Lambda$.  Start by writing the operator in Heisenberg form:
\beq
	A(t) = U_t^\dagger A\,U_t \Rightarrow dA(t) = dU_t^\dagger A\,U_t + U_t^\dagger A\,dU_t + dU_t^\dagger A\,dU_t \label{eq:01-heis}
\eeq
Because $dU_t$ is a stochastic process, the final term in (\ref{eq:01-heis}) is not negligible.  For system operators $X$, this gives:
\bea
	\d X & = & \left[-i[X, H] + \frac{1}{2}\left(L_m^\dagger[X, L_m] + [L_m^\dagger, X]L_m\right)\right]\d t \nonumber \\
	& & +\ \d B_m^\dagger S_{mn}^\dagger[X, L_n] + [L_n^\dagger, X]S_{nm}\d B_m + \left(S_{mp}^\dagger X S_{pn} - X\delta_{mn}\right)\d\Lambda_{mn} \label{eq:01-qsde-1}
\eea
The input-output operators only interact with the system momentarily -- at this moment, Eq.~(\ref{eq:01-heis}) causes their values to change instantaneously.  One can therefore view these QSDEs as input-output relations:
\bea
	\d\tilde{B}_m & = & S_{mn} \d B_n + L_m \d t  \label{eq:01-qsde-2} \\
	\d\tilde{\Lambda}_{mn} & = & S_{mp}^*\d\Lambda_{pq}S_{qn}^{\rm T} + S_{mp}\d B_p^\dagger L_n + L_m \d B_p S_{pn}^{\rm T} + L_m L_n \d t \label{eq:01-qsde-3}
\eea
where $\tilde{B}, \tilde{\Lambda}$ are the outputs and $B, \Lambda$ are the inputs.  Here, $S_{mn}^*$ is the adjoint of $S_{mn}$ (not transposed in the indices), while $S_{mn}^\dagger = S_{nm}^*$ is the transpose adjoint.

Stochastic processes are intimately connected to filtrations of $\sigma$-algebras in probability theory \cite{ShreveBook}.  A similar theory can be derived relating quantum stochastic processes to a {\it non}-commutative probability space \cite{Hudson1984, ParthasarathyBook, AP225Notes}.  It is related to the interpretation of quantum mechanics as a noncommutative extension to probability theory.  While elegant and mathematically rigorous, this approach requires a very deep understanding of probability theory and would confuse most readers; therefore, in this thesis I have chosen to shut up and calculate, and model open quantum systems using the standard (Copenhagen) interpretation.

\section{Master Equation}

Just as the QSDEs are open-system analogues for the Heisenberg equations, the Master Equation is the counterpart to the Schrodinger Equation.  The QSDEs are good for proving formal results and for deriving reduced equations in particular cases (like linear systems), but are not very useful for numerical simulations.  The master equation, on the other hand, is well suited for numerical studies, and can be rewritten as trajectory equations and semiclassical SDEs, useful for simulation in their own right.

\begin{figure}[htbp]
\begin{center}
\includegraphics[width=0.90\textwidth]{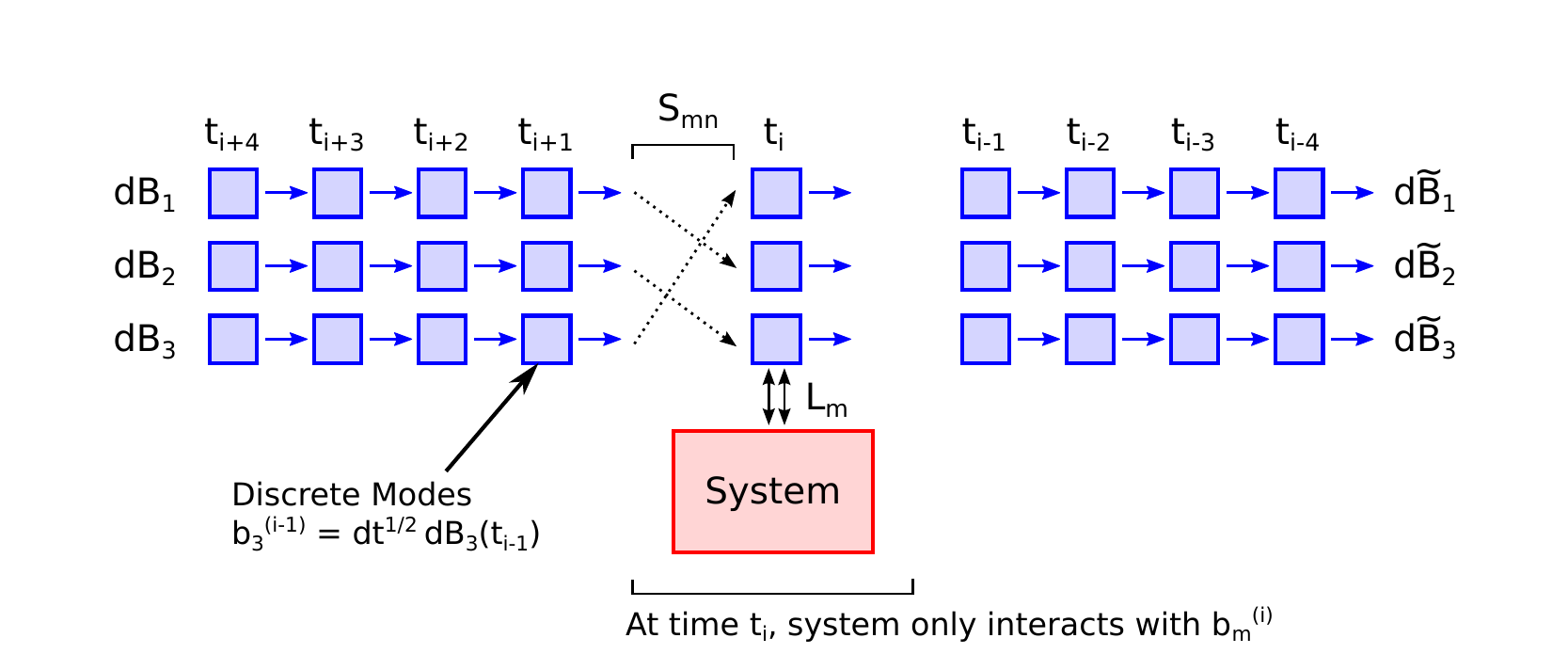}
\caption{Open system with 3 input-output modes.}
\label{fig:01-f3}
\end{center}
\end{figure}

The easiest way to obtain the input-output relations is to discretize the input-output fields in time, as was done in Section (\ref{sec:01-cav}).  The input-output modes become a train of independent harmonic oscillators with creation / annihilation operators $b_m^{(i)}, (b_m^{(i)})^\dagger$, with:
\beq
	b_m^{(i)} = \frac{1}{\sqrt{\d t}} \d B_m(t_i),\ \ \ (b_m^{(i)})^\dagger = \frac{1}{\sqrt{\d t}} \d B_m(t_i)^\dagger,\ \ \ (b_m^{(i)})^\dagger b_n^{(i)} = \d\Lambda_{mn}
\eeq
This is illustrated in Figure \ref{fig:01-f3}.  Now consider what happens on the time interval $[t_i, t_i+\d t]$.  On this interval, the system interacts with the current $b_m^{(i)}$ modes, but not with any of the past modes ($b_m^{(i-1)}$, $b_m^{(i-2)}$, etc.) or future modes ($b_m^{(i+1)}$, $b_m^{(i+2)}$, etc).  Assuming vacuum inputs, each $b_m^{(i)}$ is in the ground state at the beginning of the interval, $t = t_i$, and the density matrix, restricted to the system plus $b_m^{(i)}$ modes, is:
\beq
	\rho(t_i) = \rho_s(t_i) \otimes \ket{0}\bra{0}
\eeq
where $\rho_s$ is the density matrix of the system (tracing over all outputs) and $\ket{0}$ is the ground state of the $b_m^{(i)}$ modes.

We want an equation for $\rho_s(t)$.  By applying the unitary in (\ref{eq:01-unitary}), we can obtain the density matrix, for system plus $b_m^{(i)}$ modes, at time $t_i + \d t$.  It is like the Schrodinger equation, but a second-order term must be kept because $dU_t$ is a stochastic process:
\bea
	\rho(t_i+\d t) & = & \rho(t_i) + \d U_t \rho(t_i) + \rho(t_i) \d U_t^\dagger + \d U_t \rho(t_i) \d U_t^\dagger \\
	& = & \left[\rho_s + \left(-i[H,\rho_s] - \frac{1}{2} \sum_m\{L_m^\dagger L_m, \rho_s\}\right)\d t\right] \otimes \ket{0}\bra{0} \nonumber \\
	& & +\ \sqrt{\d t} \sum_m{\left(L_m \rho_s \otimes \ket{m}\bra{0} + \rho_s L_m^\dagger \otimes \ket{0}\bra{m} \right)} \nonumber \\ 
	& & +\ \d t \sum_{mn}{L_m\rho_s L_n^\dagger \otimes \ket{m}\bra{n}} \label{eq:01-rhome}
\eea
where $\ket{m} = (b_m^{(i)})^\dagger \ket{0}$ is the state with a single photon in mode $m$.

We want an unconditional equation of motion.  This means that $\d\rho/\d t$ is not conditioned on the measured value of the output field.  Mathematically, this means that we trace over the output degrees of freedom (here, the $b_m^{(i)}$), giving the following equation for $\rho_s$:
\beq
\boxed{
	\frac{\d\rho_s}{\d   t} = -i[H, \rho_s] + \frac{1}{2}\sum_m(2L_m \rho_s L_m^\dagger - L_m^\dagger L_m \rho_s - \rho_s L_m^\dagger L_m)
	  } \label{eq:01-master}
\eeq
This is the {\it Master Equation}.  Elsewhere in the literature it is called the {\it Kossakowski-Lindblad equation} or simply {\it Lindblad equation}.  It is the most general quantum equation of motion for systems with the Markov property \cite{Lindblad1976}, a fact that arises from the dynamical semigroup nature of time evolution \cite{Kossakowski1972}.

It is always straightforward, often easy, to solve (\ref{eq:01-master}) numerically, and is the workhorse of numerical quantum optics.  The Quantum Optics Toolbox (Matlab) \cite{Tan1999} and QuTIP (Python) \cite{Johansson2012} both have numerical master equation solvers.

\section{Conditional Master Equations and Trajectories}

\begin{figure}[b!]
\begin{center}
\includegraphics[width=0.90\textwidth]{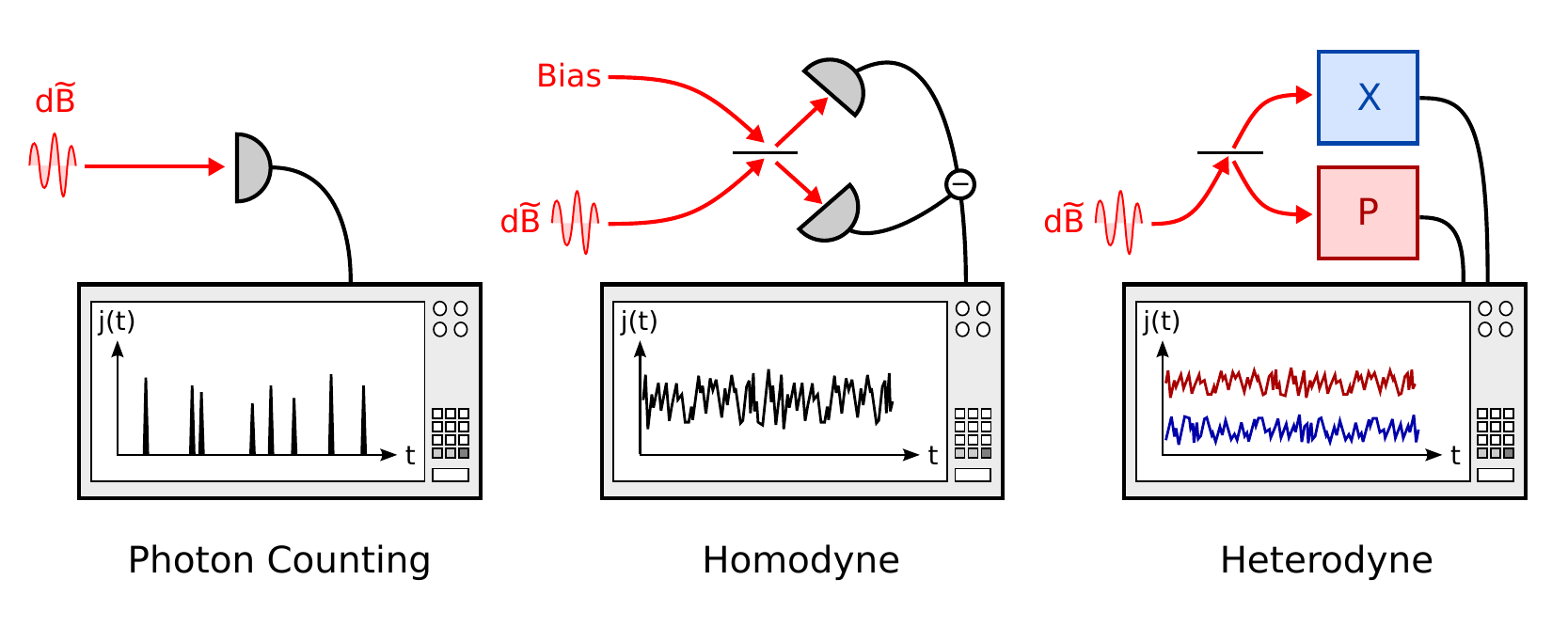}
\caption{Three common optical detection schemes.  The \texttt{X} and \texttt{P} boxes in the heterodyne setup are homodyne detectors.}
\label{fig:01-f4}
\end{center}
\end{figure}

Often we want to measure the output fields.  Unless we actively insert a probe into the system, measuring the outputs is the only way we can learn information about the system's state.  In many devices, it is the only practical way to do so.

There are three ways an optical field can be measured (Fig.~\ref{fig:01-f4}).  In \textit{photon counting}, we count the number of photons received -- this becomes a measurement of $\d\Lambda_{ii}$, the photon-number process.  In \textit{homodyne detection}, the output is sent through a beamsplitter with a strong bias field injected in the dark port.  Photons are counted at each output, and the difference is a measurement of $\d\tilde{B}$.  Depending on the phase of the bias field, the detector may measure $X = \d\tilde{B}+\d\tilde{B}^\dagger$, $P = (\d\tilde{B}-\d\tilde{B}^\dagger)/i$, or any combination of them.  \textit{Heterodyne detection} is simultaneous homodyne detection on two quadratures.  One way to realize this is to split the beam first, then the outputs into homodyne detectors to measure $X$ and $P$.  This is a simultaneous measurement of both quadratures of the field, but is not prohibited by the uncertainty principle because of the additional quantum noise injected at the dark port.  Another way to do a heterodyne measurement is to slowly sweep the bias-field phase in the homodyne setup.  If the output is slowly-varying (compared to the optical period), these two methods are equivalent.

So in short, we can use photodetectors to measure either the photon number, the real part of the output field, the imaginary part, or (with extra noise) both real and imaginary parts.  These measurements provide information about the system state; the \textit{conditional master equations} tell us how.

These equations will be derived from standard quantum measurement theory.  From quantum mechanics, we know that when operator $A$ is measured, output $a$ is obtained with probability $P(a) = \mbox{Tr}[P_a\rho]$, where $P_a$ is the projector onto the $A = a$ eigenspace -- and the post-measurement state is $\rho(a) = P_a \rho P_a/\mbox{Tr}[P_a\rho]$.

In the current setup, the Hilbert space can be decomposed into two parts -- system and input-output field -- and all the measurements are made on the latter.  Measuring the output field in state $\ket{i}$ will take place with probability
\beq
	P(i) = \mbox{Tr}_{\rm sys} \bigl[\bra{i}\rho\ket{i}\bigr]
\eeq
and will result in the following density matrix:
\beq
	\rho(i) = \frac{\bra{i}\rho\ket{i}}{\mbox{Tr}_{\rm sys} \bigl[\bra{i}\rho\ket{i}\bigr]} \otimes \ket{i}\bra{i}
\eeq
(Note that $\bra{i}\rho\ket{i}$ is \textit{not} a scalar -- it is a density matrix on the \textit{system} Hilbert space.  To obtain a scalar, we need to take the trace).  As with the QSDEs, conditional master equations can be understood in the probability interpretation through a theory called {\it quantum filtering} \cite{Bouten2007, VanHandelThesis}, but for the convenience of the reader I will stick to the Copenhagen interpretation here.

\subsection{Photon Counting}

In photon counting, we measure whether there is a photon in any of the output modes.  This collapses the output onto $\ket{0}$ if no photon is detected, and $\ket{m}$ is a photon is detected in mode $m$.

Applying Eq.~(\ref{eq:01-rhome}), in the absence of a detection, the state becomes:
\beq
	\rho_s(t+\d t) = N \bra{0} \rho(t+\d t) \ket{0} = N \left[\rho_s(t) + \left(-i[H,\rho_s] - \frac{1}{2} \sum_m\{L_m^\dagger L_m, \rho_s\}\right)\d t\right]
\eeq
where $N$ is the normalization term.  This can be recast as a differential equation:
\beq
\boxed{	\frac{\d\rho_s}{\d t} = -i[H, \rho_s] - \frac{1}{2} \sum_m\{L_m^\dagger L_m, \rho_s\} + \underbrace{\frac{1}{2} \rho_s \sum_m\mbox{Tr}\left[\{L_m^\dagger L_m, \rho_s\}\right]}_{\rm normalization}  
	  } \label{eq:01-pc-smooth}
\eeq
The first two terms describe Hamiltonian evolution and dissipation.  The last term is in there to keep $\rho$ normalized.  This is often unnecessary, and since it makes the equation nonlinear, is generally excluded.

If a photon is detected in mode $m$, the state becomes:
\beq
\boxed{
	\rho_s \rightarrow \frac{\bra{m} \rho(t+\d t) \ket{m}}{\mbox{Tr}_{\rm sys}\bigl[\bra{m} \rho(t+\d t) \ket{m}\bigr]} = \frac{L_m \rho_s L_m^\dagger}{\mbox{Tr}[L_m \rho_s L_m^\dagger]} \label{eq:01-pc-jump}
	  }
\eeq
This happens with probability $P(m) = \mbox{Tr}[L_m \rho_s L_m^\dagger]\d t$.

These are the {\it conditional master equations for photon counting}.  They correspond to a motion under (\ref{eq:01-pc-smooth}) for most of the time, punctuated by discrete jumps under (\ref{eq:01-pc-jump}), whose probabilities were given above.

If the system starts in a pure state $\rho_s = \ket{\psi}\bra{\psi}$, it will remain in a pure state forever because  perfect photon counting recovers all the information ``lost'' to the output fields.  Equations (\ref{eq:01-pc-smooth}) and (\ref{eq:01-pc-jump}) can be rewritten for $\psi$.
\bea
	\frac{\d}{\d t}\ket{\psi} & = & \left(-iH - \sum_m\frac{1}{2}L_m^\dagger L_m + \frac{1}{2}\bra{\psi}L_m^\dagger L_m\ket{\psi}\right)\ket{\psi} \nonumber \\
	& & \mbox{(smooth motion, no detections)} \\
	\ket{\psi} & \rightarrow & L_m \ket{\psi} \nonumber \\
	& & \mbox{(detection at field $m$, probability $P(m) = \bra{\psi}L_m^\dagger L_m \ket{\psi}\d t$)}
\eea
These are the photon-counting \textit{jump trajectory equations} \cite{CarmichaelBook, GardinerBook, Gardiner1992}.  They are especially useful for large numerical studies because for a large Hilbert space of dimension $N$, the density matrix has $n^2$ entries, while the wavevector only has $n$.  Averaging over many trajectories is equivalent to sampling from the master equation, and often encodes additional information about jumping and dynamical processes that master-equation solutions can miss.

\subsection{Homodyne Detection}

In homodyne detection, one measures the real part $X \d t = \d B_m + \d B_m^\dagger$ or imaginary part $P \d t = (\d B_m - \d B_m^\dagger)/i$ of the output field.  Consider the real part for now.  Measuring $X$ projects the state on to an eigenstate $\ket{x}$, where $X\ket{x} = x\ket{x}$.  The states $\ket{x}$ are non-normalizable, so assume it is scaled so that $\ip{0}{x} = 1$.  Then $\ip{1}{x} = \d t^{-1/2}\bra{0}X\,\d t\ket{x} = \sqrt{\d t}\,x$, and $\ket{x}$ has the following expansion:
\beq
	\ket{x} = \ket{0} + \sqrt{\d t}\,x \ket{1} + \ldots \label{eq:01-xstate}
\eeq
For now, consider homodyne detection on only a single field (the many-field case is a straightforward extension -- it just simplifies the math).  Applying (\ref{eq:01-rhome}), we find that a measurement of $x$ changes the density matrix by:
\bea
	\rho_s(t+\d t) - \rho_s(t) & = & N\bra{x}\rho(t+\d t)\ket{x} - \rho_s(t) \nonumber \\
	& = & \left(-i[H,\rho_s] - \frac{1}{2}\{L^\dagger L, \rho_s\}\right)\d t + x(L \rho_s + \rho_s L^\dagger)\d t + (x\,\d t)^2L^\dagger\rho_s L + N' \rho_s\nonumber \\
		& & 
\eea
where $N, N'$ are normalization constants.  Now $x$ is a Gaussian random variable with mean $\avg{x} = \avg{L + L^\dagger}$ and variance $\sigma_x = \avg{(\d B+\d B^\dagger)^2/\d t} = 1/\sqrt{\d t}$.  We can represent $x\,\d t = \d M_x(t)$, where $\d M_x(t) = \avg{L + L^\dagger}_\rho \d t + \d w$ is the measurement process, and is the sum of a continuous part, and a noisy Wiener process: $\d w\,\d w = \d t$.  We can also replace $(x\,\d t)^2 \rightarrow \d M_x^2 \rightarrow \d t$.  The stochastic master equation may be written as an SDE:
\begin{align}
	\mathbf{(X)}\quad\quad & \d\rho_s = \left[-i[H, \rho_s] + \frac{1}{2}\left(2L\rho_s L^\dagger - L^\dagger L \rho_s - \rho_s L^\dagger L\right)\right]\d t \nonumber \\ 
	& \qquad+ (L\rho_s + \rho_s L^\dagger)\d M_x(t) - \underbrace{\avg{L+L^\dagger}_\rho \rho_s \d M_x(t)}_{\rm normalization} \label{eq:01-sme}
\end{align}
As in photon counting, if the system starts in a pure state, it remains so indefinitely.  Thus, Eq.~(\ref{eq:01-sme}) can be used to derive a {\it stochastic Schrodinger Equation} \cite{CarmichaelBook, GardinerBook, WallsMilburn, KerckhoffThesis}:
\beq
	\mathbf{(X)}\quad\quad \boxed{\d\ket{\psi} = \left[-iH - \frac{1}{2}L^\dagger L \right] \ket{\psi}\d t + L \ket{\psi} \d M_x(t) + \underbrace{\avg{L}_\psi^*\left[L - \frac{1}{2}\avg{L}_\psi\right] \ket{\psi} \d t - \avg{L}_\psi \ket{\psi} \d M_x(t)}_{\rm normalization}}
\eeq
A similar set of equations can be derived for $P$-quadrature detection.  In this case, the state is projected down onto a state $\ket{p}$, where $P\ket{p} = p\ket{p}$, and $\ket{p}$ has the expansion:
\beq
	\ket{p} = \ket{0} + \sqrt{\d t}\,ip\ket{1} + \ldots
\eeq
Following the same derivation as before, we find:
\bea
	\mathbf{(P)}\quad\quad \d\rho_s & = & \left[-i[H, \rho_s] + \frac{1}{2}\left(2L\rho_s L^\dagger - L^\dagger L \rho_s - \rho_s L^\dagger L\right)\right]\d t \nonumber \\
	& & + \left(\frac{L\rho_s - \rho_s L^\dagger}{i}\right)\d M_p(t) - \underbrace{\avg{\frac{L-L^\dagger}{i}}_\rho \rho_s \d M_p(t)}_{\rm normalization} \\
	\mathbf{(P)}\quad\quad \d\ket{\psi} & = & \left[-iH - \frac{1}{2}L^\dagger L \right] \ket{\psi}\d t - iL \ket{\psi} \d M_p(t) \nonumber \\
	& & + \underbrace{\avg{L}_\psi^*\left[L - \frac{1}{2}\avg{L}_\psi\right] \ket{\psi} \d t + i\avg{L}_\psi \ket{\psi} \d M_p(t)}_{\rm normalization}
\eea
And the case of many measurements is just the obvious generalization of this.

\subsection{Heterodyne Detection}

In heterodyne detection, we measure both X and P quadratures of $\d B$ simultaneously.  This is not possible in the strict sense of quantum measurement because they do not commute, but we can make a quantum-limited noisy measurement of both.  If the noise in X equals the noise in P, then this amounts to projecting the output onto a coherent state $\ket{\alpha}$, where $\d B \ket{\alpha} = \alpha\,\d t\ket{\alpha}$.  As a coherent state, $\ket{\alpha}$ has the expansion:
\beq
	\ket{\alpha} = e^{-\alpha^*\alpha\,\d t/2} \left[\ket{0} + \sqrt{\d t}\,\alpha\ket{1} + \ldots\right]
\eeq
Applying Eq.~(\ref{eq:01-rhome}) to this problem, we obtain:
\beq
	\rho(t+\d t) = N e^{-\alpha^*\alpha \d t} \left[\rho_s + \left(-i[H, \rho_s] - \frac{1}{2} \{L^\dagger L, \rho_s\}\right)\d t + \left(L\rho_s \alpha^* + \rho_s L^\dagger \alpha\right)\d t + L^\dagger\rho_s L(\alpha^*\alpha \d t^2)\right]
\eeq
for some normalization constant $N$.  From this we can deduce that the probability distribution for $\alpha$ is:
\beq
	P(\alpha) \sim \mbox{Tr}\left[1 + O(\d t) + \left(\avg{L}_\rho\alpha^* + \avg{L}_\rho^*\alpha\right)\d t\right] \approx e^{-|\alpha-\avg{L}_\rho|^2 \d t}
\eeq
So $\alpha$ is a complex Gaussian random variable with mean $\avg{L}$ and standard deviation $1/\sqrt{2\d t}$ (for both real and imaginary parts).  We can relate $\alpha$ to a stochastic measurement process $M$ by $\d M = \alpha\,\d t$, where
\beq
	\d M = \avg{L}_\rho \d t + \frac{\d w_1(t) + i\,\d w_2(t)}{\sqrt{2}}
\eeq
In terms of this process, the stochastic master equation for heterodyne detection is:
\bea
	\d\rho_s & = & \left[-i[H, \rho_s] + \frac{1}{2}\left(2L\rho_s L^\dagger - L^\dagger L \rho_s - \rho_s L^\dagger L\right)\right]\d t \nonumber \\
	& & + L\rho_s\,\d M^* + \rho_s L^\dagger \d M - \underbrace{\left[\avg{L}_\rho \d M^* + \avg{L}_\rho^* \d M\right]\rho_s}_{\rm normalization}
\eea
As in the homodyne case, if the system starts in a pure state it remains in a pure state, and the pure state evolves as:
\beq
	\boxed{\d\ket{\psi} = \left[-iH - \frac{1}{2}L^\dagger L\right]\ket{\psi}\d t + L\ket{\psi}\d M^*(t) - \underbrace{\left[\frac{1}{2}\avg{L}_\psi^*\avg{L}_\psi \d t + \avg{L}_\psi^* \d M(t)\right]\ket{\psi}}_{\rm normalization}}
\eeq

\section{SLH Circuit Algebra}
\label{sec:01-circalg}

The SLH models are useful because they are \textit{cascadeable} -- multiple elements can be linked together to form a circuit.  Any photonic circuit can be represented by a directed graph, where the nodes represent components and the edges represent propagating fields.  However, it is difficult to determine the quantum model for a circuit directly from its graph.  Rather, we proceed by writing the circuit in the language of the \textit{Gough-James circuit algebra} \cite{Gough2009a, Gough2009b}, from which there is a straightforward, algorithmic way to compute the quantum model.

\begin{figure}[tb]
\begin{center}
\includegraphics[width=0.60\textwidth]{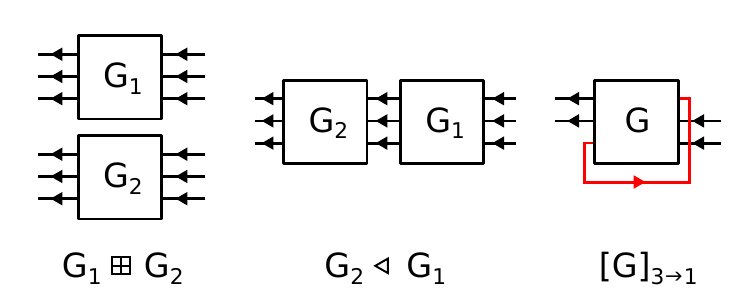}
\caption{Concatenation, series and feedback products}
\label{fig:01-f5}
\end{center}
\end{figure}

The Gough-James circuit algebra is an algebra of components.  To start, we define a basis set of simple components like beamsplitters, phase shifters, lasers and optical cavities.  More complex components are built from these using the \textit{concatenation product} $G_1 \boxplus G_2$, the \textit{series product} $G_2 \triangleleft G_1$, and the \textit{feedback operator} $[G]_{k\rightarrow l}$ (Fig.~\ref{fig:01-f5}).

The \textit{concatenation product} builds a larger open system by placing two smaller systems together, without interaction.  This is the open-systems analogue to a tensor product of Hilbert spaces.  Concatenating two closed quantum systems involves tensoring their Hilbert spaces $\mathcal{H} = \mathcal{H}_1 \otimes \mathcal{H}_2$ and adding their Hamiltonians $H = H_1 \otimes 1 + 1 \otimes H_2$.  For open systems, the SLH model for $G_1 \boxplus G_2$ is:
\beq
	\left(\begin{bmatrix} S_1 & 0 \\ 0 & S_2 \end{bmatrix},\ \ \begin{bmatrix} L_1 \\ L_2 \end{bmatrix},\ \ H_1 + H_2 \right) \label{eq:01-slh-concat}
\eeq
The \textit{series product} $G_2 \triangleleft G_1$ takes the output of the first device and feeds it into the input of the second device.  For this to work, both devices must have the same number of inputs.  The cascading operation means that the unitary evolution operators (Eq.~(\ref{eq:01-unitary})) are themselves cascaded -- first $dU_1$ is applied to the system, which alters the state of the inputs, and then $dU_2$ is applied to the result.  This has the effect of feeding $G_1$'s output into $G_2$.  One can show that the series product $G_2 \triangleleft G_1$ has an SLH model:
\beq
	\left(S_2 S_1,\ \ L_2 + S_2 L_1,\ \ H_1 + H_2 + \frac{L_2^\dagger S_2 L_1 - L_1^\dagger S_2^\dagger L_2}{2i}\right) \label{eq:01-slh-series}
\eeq
One can prove (\ref{eq:01-slh-concat}-\ref{eq:01-slh-series}) from the QSDEs.  Define the {\it infinitesimal} of an SLH model $\d G(t)$ by $U(t+\d t) = \d G(t) U(t)$.  One can show that $\d G(t) = \d G_1(t) \otimes \d G_2(t)$ for the concatenation product and $\d G(t) = \d G_1(t) + \d G_2(t) + \d G_2(t)\d G_1(t)$ for the series product, and applying Eq.~(\ref{eq:01-unitary}), derive the SLH model for the full system \cite{Gough2009b}.

The \textit{feedback operator} $[G]_{k \rightarrow l}$ takes output $k$ and sends it into input $l$.  This reduces the number of external ports by one.  The resulting model is more complicated because of the loop; it is:
\bea
	S & \rightarrow & S_{!k,!l} + S_{!k,l} (1 - S_{kl})^{-1} S_{k,!l} \label{eq:01-slh-f1} \\
	L & \rightarrow & L_{!k} + S_{!k,l} (1 - S_{kl})^{-1} L_k \label{eq:01-slh-f2} \\
	H & \rightarrow & H + \mbox{Im} \left[\sum_j L_j^\dagger S_{jl} (1 - S_{kl})^{-1} L_k\right] \label{eq:01-slh-f3}
\eea
where $S_{!k,!l}$ is a matrix with the $k^{\rm th}$ row and $l^{\rm th}$ column removed, $L_{!k}$ is the vector $L$ with the $k^{\rm th}$ entry removed, and so on.

In Eqs.~(\ref{eq:01-slh-series}-\ref{eq:01-slh-f3}), we assume instantaneous propagation of fields.  This is equivalent to saying that the time delays for the feedback / feedforward paths are negligible compared to the timescales of the system, so that there is no dynamics in the connections.  If the connections get long enough, they start introducing time delays and (\ref{eq:01-slh-series}-\ref{eq:01-slh-f3}) break down.  The problem can then be treated using matrix product states for long delays \cite{Grimsmo2015, Pichler2015} or trapped modes for small delays \cite{Tabak2015}, but this adds considerable complexity to the model, so when possible we will assume connection delays are negligible so that the Gough-James expressions can be used.

\begin{figure}[b!]
\begin{center}
\includegraphics[width=1.00\textwidth]{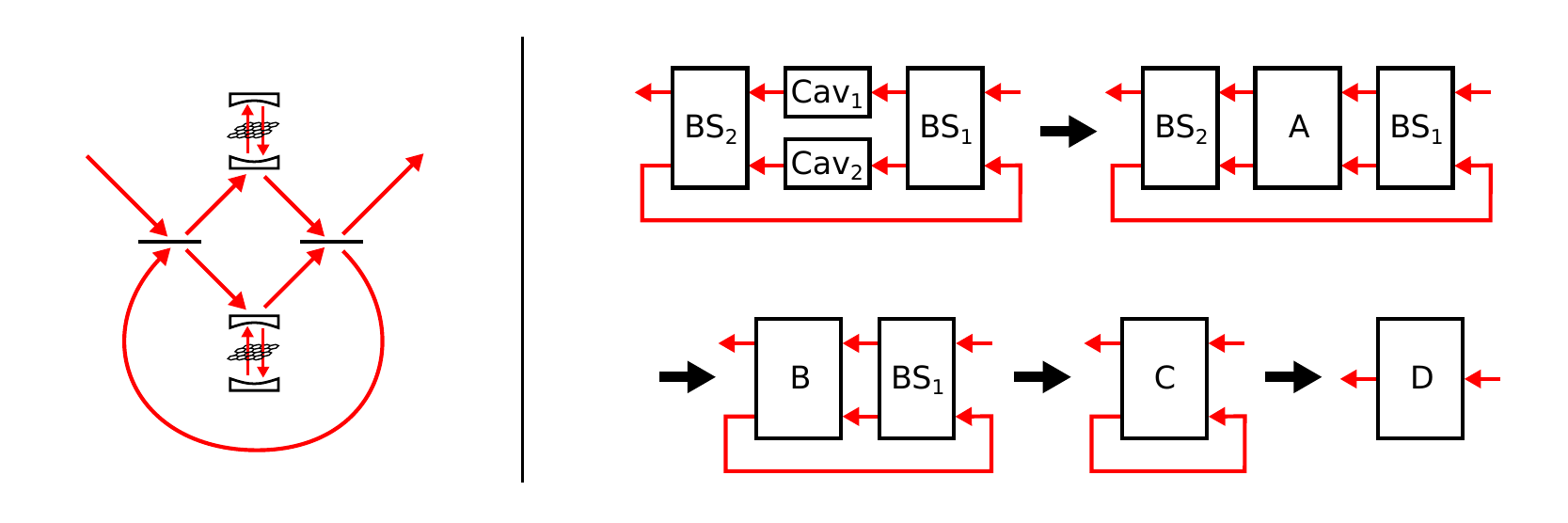}
\caption{Example of a circuit built up from the SLH algebra.  Left: Circuit shown as a netlist.  Right: Algorithmic reduction using the circuit algebra.}
\label{fig:01-f6}
\end{center}
\end{figure}

As mentioned, any circuit can be constructed from basic components using the circuit algebra.  As an example, consider the circuit in Figure~\ref{fig:01-f6}.  This has two cavities with nonlinear materials (graphene in this case) sandwiched in an interferometer with feedback.  An algorithm can reduce this graph to a Gough-James expression by applying standard rules: elements in parallel are replaced with concatenation products, cascaded elements are replaced with series products, and elements with feedback loops are replaced with the feedback operator.  The graph in the figure is sequentially simplified with the following replacements:
\bea
	\mathbf{A} & = & \mathbf{Cav}_1 \boxplus \mathbf{Cav}_2 \nonumber \\
	\mathbf{B} & = & \mathbf{BS}_2 \triangleleft \mathbf{A} \nonumber \\
	\mathbf{C} & = & \mathbf{B} \triangleleft \mathbf{BS}_1 \nonumber \\
	\mathbf{D} & = & [\mathbf{C}]_{2\rightarrow 2}
\eea
The final Gough-James expression is:
\beq
	[\mathbf{BS}_1 \triangleleft (\mathbf{Cav}_1 \boxplus \mathbf{Cav}_2) \triangleleft \mathbf{BS}_2]_{2\rightarrow 2}
\eeq
If the SLH models for the components are known, one can algorithmically apply the rules (\ref{eq:01-slh-concat}--\ref{eq:01-slh-f3}) to obtain the quantum model for the full circuit.  This can then be analyzed or simulated using the master equation and the QSDEs.

For small circuits, the Gough-James models can be computed by hand, but this becomes increasingly tedious as the system size grows, especially for networks with feedback.  For large networks, one can use the QHDL language to specify the circuit elements and their connections, which can then be parsed by a computer algebra system to obtain the SLH model \cite{Tezak2012, Sarma2013}.  At the time of the writing of this thesis, the mode complete set of tools for this job is the QNET package managed by Nikolas Tezak and Michael Goerz \cite{QNET}.

\section{Adiabatic Elimination}
\label{sec:01-adiabatic}

Many classical dynamical systems naturally have widely separated timescales, and it is possible to divide the system state into ``slow'' degrees of freedom, which evolve on slowly, and ``fast'' degrees of freedom that evolve very quickly.  This also happens in quantum optical systems -- for example, in a cavity with trapped atoms, the atomic decay time is often much longer, or much shorter, than the photon lifetime.  In this case, a simpler model can be derived by \textit{adiabatic elimination} of the fast variables, leaving a reduced system that only depends on the slow variables \cite{StocktonThesis, GardinerBook, GardinerHandbook}.  This is very useful because, in many cases, we do not particularly care how the fast variables change; all that matters is how they affect the slow variables.

The most rigorous way to do this is via the QSDE limit theorem \cite{Bouten2008a, Bouten2008b}, which proves convergence on the propagator $U(t)$.  To apply this theorem, we find some parameter $k$ in our SLH model that we wish to take to infinity, and write the SLH model in the following form:
\bea
	-iH - \frac{1}{2}\sum_m L_m^\dagger L_m & = & k^2 Y^\dagger + k A^\dagger + B^\dagger \\
	L_m & = & k F_m^\dagger + G_m^\dagger \\
	S_{ij} & = & (W^\dagger)_{ij}
\eea
The preponderance of daggers is due to a peculiar convention used by Bouten et al.\ in the derivation of the theorem.  Let $P_0$ and $P_1$ be projectors onto the ``slow'' and ``fast'' spaces, respectively.  If the following conditions hold:

\begin{enumerate}
	\item $P_0 Y^\dagger = 0$
	\item There exists a $\tilde{Y}$ for which $Y\tilde{Y} = \tilde{Y}Y = P_1$
	\item $F_m^\dagger P_0 = 0$
	\item $P_0 A^\dagger P_0 = 0$
\end{enumerate}

then the QSDE for this model approaches the QSDE for the following reduced model \cite{Bouten2008b, KerckhoffThesis}:
\bea
	-iH - \frac{1}{2}\sum_m L_m^\dagger L_m & \rightarrow & P_0(B^\dagger - A^\dagger \tilde{Y}^\dagger A^\dagger)P_0 \label{eq:01-adel1} \\
	L_m & \rightarrow & P_0(G_m^\dagger - F_m^\dagger \tilde{Y}^\dagger A^\dagger)P_0 \label{eq:01-adel2} \\
	S_{ij} & \rightarrow & P_0(F_i^\dagger \tilde{Y}^\dagger F_k + \delta_{ik})(W^\dagger)_{kj} P_0 \label{eq:01-adel3}
\eea
The intuition behind the theorem is that the adiabatic elimination works when the system quickly relaxes back to a ``slow'' subspace, on timescales $O(k^{-2})$ with $k \rightarrow \infty$.  If everything else were finite, we could simply ignore the fast modes and treat the slow ones without modification (the special case $A = F_m = 0$ does exactly that).  But in the case when coupling between fast and slow modes is $O(k)$, then the fast degrees of freedom will always deviate $O(1/k)$ from the slow subspace.  These deviations couple back to the slow degrees of freedom, again with a coupling constant $O(k)$, to give a finite, nonnegligible effect.

In many cases, especially with linear systems, the adiabatic elimination is trivial enough to guess by inspection, but for complex nonlinear systems, this theorem is useful because it is rigorous and does not rest on any hidden assumptions.

Many cavity QED systems can be simplified using this theorem if one invokes the ``bad-cavity limit'' in which the photon loss rate is much faster than other rates in the system.  In this limit, a two-mode cavity can be reduced to an optical ``relay'' \cite{Mabuchi2009}, which can route signals and form the backbone for autonomous error-correction schemes \cite{Mabuchi2009b, Kerckhoff2010, Kerckhoff2011, Sarma2013b}.  In the opposite limit of long photon lifetime, one can show that a Kerr cavity reduces to a qubit \cite{Mabuchi2012}.

\ifstandalone{\bibliography{NoteRefs}{}
\bibliographystyle{alpha}
}
\ifdefined\multidoc\else\input{Header}\fi

\ifstandalone{\setcounter{chapter}{1}}

\chapter{Common Components}
\label{ch:02}

The previous section discussed the general theory of quantum circuits.  Given a basis set of simple components, one can apply those results to construct and simulate large circuits.  But to understand the whole, we must understand its parts.  This section is about the parts.

The components discussed in this chapter fall into three categories: (1) \textit{scattering components}, which scatter input-output fields and do not have any memory or dynamics of their own, (2) \textit{linear components}, which have (linear) internal dynamics, and (3) \textit{nonlinear components} -- everything else.  Here is a list of the components covered in this chapter:

\begin{enumerate}
\item Scattering Components: Beamsplitter, Phase shifter, Displacement, Permutation, Identity
\item Linear Components: Cavity, General Passive Linear, Linearized OPO
\item Nonlinear Components: Atom Cavity, Kerr Cavity, OPO, Optomechanical Cavity
\end{enumerate}

Of course, the third category is the one with all the interesting stuff, but it is also the most difficult, so we start with the first two which, although boring, are conceptually simple.  Having built up an intuition studying the easy things, the reader will be well prepared for the more interesting things that make up the bulk of this chapter.

Like the last chapter, this is a compilation of existing results, and none of the the work is my own.  I include it here because the systems in this chapter span a range of fields, and there is no good reference that covers them all.  What unites them is the Markov property of the bath, which allows the open quantum systems theory of Chapter~\ref{ch:01} to be applied.  All of these components will be used, in one form or another, in my own work in later chapters.

\section{Scattering Components}

\begin{figure}[tbp]
\begin{center}
\includegraphics[width=1.00\textwidth]{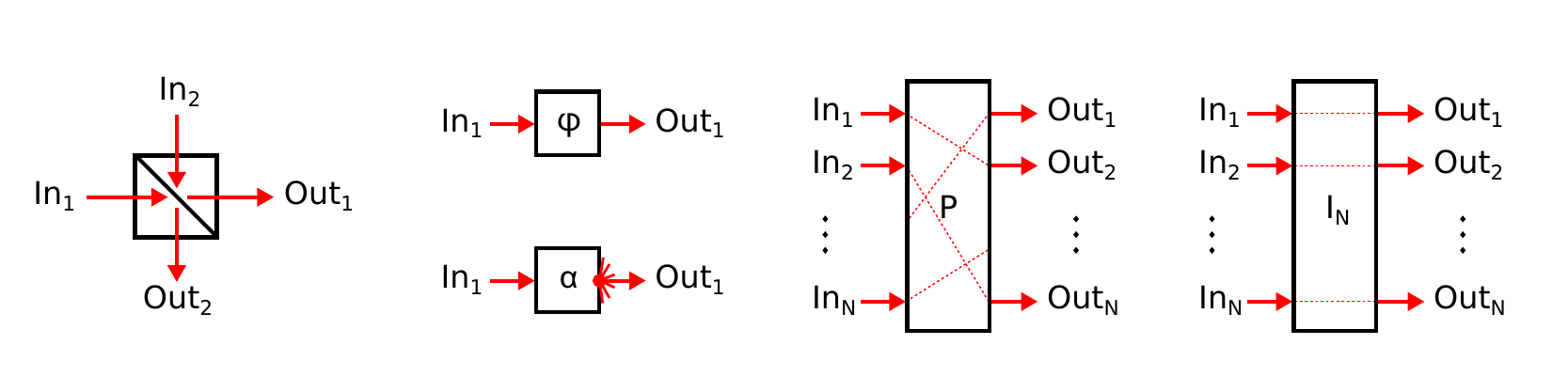}
\caption{Scattering components: Beamsplitter, Phase Shifter, Displacement, Permutation, Identity}
\label{fig:02-f1}
\end{center}
\end{figure}

There are five important scattering components: the beamsplitter, phase shifter, displacement, permutation, and identity (Fig.~\ref{fig:02-f1}).  They all have SLH models that depend only on S and L -- since none have internal degrees of freedom, so $H$ is undefined, and S and L are scalar, not operator, valued.

\subsubsection{Beamsplitter}

The beamsplitter has two input and output ports: \texttt{(In1, In2), (Out1, Out2)}.  It depends only on the scattering angle $\theta$ ($\theta = 0$ for perfect transmission, $\theta = 1$ for perfect reflection).  There are two sign conventions.  The \textit{symmetric} sign convention has the following SLH model \cite{Gough2009b}:
\beq
	G = \left(\begin{bmatrix} \cos(\theta) & i\sin(\theta) \\ i\sin(\theta) & \cos(\theta) \end{bmatrix}, \_, \_ \right)
\eeq
Some writers do not like this model because the $S$ matrix is not real, but for a symmetric $S$ matrix, this is necessary for unitarity to hold.  It is possible to construct a beamsplitter model with real off-diagonal coefficients, but the $S$ matrix is no longer symmetric:
\beq
	  G = \left(\begin{bmatrix} \cos(\theta) & -\sin(\theta) \\ \sin(\theta) & \cos(\theta) \end{bmatrix}, \_, \_ \right)
\eeq
This is just the symmetric beam splitter with a $-i$ phase shift on the \texttt{Out2} end and an $i$ phase shift on the \texttt{In2} end.

\subsubsection{Phase Shifter}

This has a $1\times 1$ scattering matrix $e^{i\phi}$ \cite{Tezak2012}:
\beq
	G = \left(\begin{bmatrix} e^{i\phi} \end{bmatrix}, \_, \_ \right)
\eeq

\subsubsection{Displacement}

Displaces the field by an amplitude $\alpha$ \cite{Tezak2012}:
\beq
	G = (1, \alpha, \_)
\eeq
Equivalent to passing the light through a beamsplitter with a very high transmittance, with a strong laser beam incident from the opposite side.  We are assuming, of course, an ideal, coherent laser beam without any extra noise.

Input-output relation:
\beq
	\d\tilde{B} = \d B + \alpha\,\d t
\eeq

\subsubsection{Permutation}

Permutes the fields $\d B_1, \ldots, \d B_N$ \cite{Tezak2012}:
\beq
	G = \left(P_\sigma, \_, \_\right)
\eeq
The input-output relation sends input $i$ to output $\sigma_i$:
\beq
	\d\tilde{B}_{\sigma_i} = \d B_i
\eeq

\subsubsection{Identity}

Does nothing, sends everything straight through \cite{Tezak2012}:
\beq
	G = (1_{N\times N}, \_, \_)
\eeq
The displacement, permutation and identity are components only in the formal sense -- they aren't fabricated on chip or inserted into the circuit diagram.  However, when converting a circuit into a Gough-James expression, they are essential.  For example, if components $\mathbf{A}$ and $\mathbf{B}$ both have two ports and outputs 1 and 2 of $\mathbf{B}$ are fed into inputs 2 and 1 of $\mathbf{A}$, respectively, the resulting circuit is \textit{not} the series product $\mathbf{A} \triangleleft \mathbf{B}$.  A permutation must be inserted between them: $\mathbf{B} \triangleleft P_{(2,1)} \triangleleft \mathbf{A}$.

\section{Linear Components}

Most input-output systems are linear.  This is especially true for optics, since the light-matter interaction is weak and great effort must be put into making devices nonlinear at reasonable energy scales.  In this section, the single-mode optical cavity is introduced as the classic example of a linear device.  Cavities are an example of {\it passive linear systems} which conserve photon number.  Non-passive systems (degenerate, non-degenerate OPO) generally involve some type of nonlinearity, but are linearized with a strong classical pump.  This section just discusses the components.  For a more detailed treatment, see Chapters \ref{ch:04b}-\ref{ch:11}.

\subsection{Optical Cavity}
\label{sec:02-cavity}

A simple passive optical cavity has a single internal field $a$, satisfying $[a, a^\dagger] = 1$, and an arbitrary number of input-output ports.  The SLH model is \cite{Gough2009b}:
\beq
	\left(1_{N\times N},\ \ \left[\sqrt{\kappa_1}e^{i\psi_1}, \ldots, \sqrt{\kappa_N}e^{i\psi_N}\right],\ \ 
      \Delta a^\dagger a\right)
\eeq
Here $\Delta$ is the cavity detuning (resonance frequency minus reference frequency), $\kappa_i$ is the loss from mirror $i$, and $E$ is the external drive, if any.

Defining $\kappa = \sum_i \kappa_i$, the QSDEs are:
\bea
	da & = & \left[(-i\Delta - \kappa/2)a\right]\,\d t - \sum_i \sqrt{\kappa_i}e^{-i\psi_i} \d B_i \\
	\d\tilde{B}_i & = & \d B_i + \sqrt{\kappa_i}e^{i\psi_i} a\,\d t
\eea
The master equation is:
\beq
	\frac{\\d\rho}{\d t} = -i[H,\rho] + \frac{1}{2}\kappa \left(2a\rho a^\dagger - a^\dagger a\rho - \rho a^\dagger a\right)
\eeq

\subsection{General Passive Linear Component}
\label{sec:02-genlim}

The most general linear component will have an SLH model that is quadratic in $H$, linear in $L$, and constant in $S$.  Anything else will give rise to nonlinear QSDEs.  The Hilbert space is consists of $N$ harmonic oscillators, with modes $x = (a_1, \ldots, a_N)$.  Valid terms for the SLH model are thus:
\beq
	S \sim \mbox{(const)},\ \ \ L \sim a_i\ \&\ a_i^\dagger\ \&\ \mbox{(const)},\ \ \ 
	H \sim a_i^\dagger a_j\ \&\ a_i^\dagger a_j^\dagger\ \&\ a_i a_j\ \&\ a_i^\dagger \ \&\ a_i \ \&\ \mbox{(const)}
\eeq
A coherent displacement of the fields \cite{WallsMilburn} can remove the constant and linear terms in $H$, and the constant terms in $L$.  Assuming that the system is {\it passive} (and thus conserves photon number), the $a^\dagger$ terms are inadmissible in $L$ and $a^2, (a^\dagger)^2$ terms are inadmissible in $H$ (these terms create photon pairs, violating conservation).  Given matrices $S, R, \Lambda$ and vectors $r, \lambda$, the SLH model takes the form:
\beq
	G = \left(S,\ \ \Lambda a + \lambda,\ \ a^\dagger R a + r^\dagger a + a^\dagger r\right) \label{eq:02-genlin}
\eeq
The QSDEs take a form familiar to those who have worked with linear dynamical systems:
\bea
	\d x & = & (Ax+a)\,\d t + B\,\d B \nonumber \\
	\d\tilde{B} & = & (Cx+c)\,\d t + D\,\d B \label{eq:02-qsde-abcd}
\eea
The matrices $A, B, C, D$ and vectors $a, c$ are:
\bea
	A & = & -iR - \frac{1}{2}\Lambda^\dagger \Lambda \nonumber\\
	B & = & -\Lambda^\dagger S \nonumber\\
	C & = & \Lambda \nonumber\\
	D & = & S \nonumber\\
	a & = & -ir - \frac{1}{2}\Lambda^\dagger \lambda \nonumber\\
	c & = & \lambda \label{eq:02-abcd}
\eea
This is the \textit{ABCD representation} of the linear SLH model.  Any passive linear component has a valid ABCD model and follows Eqs.~(\ref{eq:02-qsde-abcd}, \ref{eq:02-abcd}).  But not every ABCD model corresponds to a valid component.  We say an ABCD model is \textit{physically realizable} if it can be realized by a valid linear SLH model.  The same is true for non-passive components.  The general theory is discussed in Chapter \ref{ch:04b}.

\subsection{Linearized OPO}
\label{sec:02-opolin}

The OPO is technically a nonlinear device (Sec.~\ref{sec:02-opo}), but in the adiabatic limit below threshold, the nonlinear terms drop out.  The resulting system resembles a linear optical cavity, but with an extra term in the Hamiltonian \cite{WallsMilburn}.

\subsubsection{Degenerate}

For the degenerate OPO with a single input-output port, the SLH model is:
\beq
	\left(1,\ \ \sqrt{\kappa}\,e^{i\psi} a,\ \ \Delta a^\dagger a + \frac{\epsilon^* a^2 - \epsilon (a^\dagger)^2}{2i}\right) \label{eq:02-ldopo-slh}
\eeq
The new parameter $\epsilon$ is proportional to the strength of the OPO pump (at $2\omega$, not shown here).  The QSDEs for the OPO are:
\bea
	\d a & = & \left[(-i\Delta - \kappa/2)a + \epsilon\,a^\dagger\right]\,\d t - \sqrt{\kappa}\,e^{-i\psi} \d B \\
	\d\tilde{B} & = & \d B + \sqrt{\kappa}\,e^{i\psi} a\,\d t
\eea
These are linear and can be solved exactly.  Two QSDE eigenvectors -- linear combinations of $(a, a^\dagger)$ -- can be constructed.  Their eigenvalues are:
\beq
	\lambda_\pm = -\frac{\kappa}{2} \pm \sqrt{\epsilon^*\epsilon - \Delta^2}
\eeq
Instability will set in when $\epsilon \geq \sqrt{\Delta^2 + \kappa^2/4}$.  This is the threshold condition for an OPO -- above threshold, it turns into a laser.

The steady state of an OPO is a squeezed state, so one quadrature has less noise than the vacuum.  This is discussed in more detail in Sec.~\ref{sec:04b-intstate}.

\subsubsection{Nondegenerate}

For the degenerate OPO there are two fields $a$, $b$ with the SLH model:
\beq
	\left(1,\ \ \begin{bmatrix}\sqrt{\kappa_a}\,e^{i\psi_a} a \\ \sqrt{\kappa_b}\,e^{i\psi_b} b\end{bmatrix},\ \ \Delta a^\dagger a + \Delta_b b^\dagger b + \frac{\epsilon^* a b - \epsilon a^\dagger b^\dagger}{i}\right) \label{eq:02-lndopo-slh}
\eeq
which gives the QSDEs:
\bea
	\d a & = & \left[(-i\Delta_a - \kappa_a/2)a + \epsilon\,b^\dagger\right]\,\d t - \sqrt{\kappa_a}\,e^{-i\psi_a} \d B_a \\
	\d b & = & \left[(-i\Delta_b - \kappa_b/2)b + \epsilon\,a^\dagger\right]\,\d t - \sqrt{\kappa_b}\,e^{-i\psi_b} \d B_b \\
	\d\tilde{B}_a & = & \d B_a + \sqrt{\kappa_a}e^{i\psi_a} a\,\d t \\
	\d\tilde{B}_b & = & \d B_b + \sqrt{\kappa_b}e^{i\psi_b} b\,\d t
\eea
As before, these are linear and can be solved exactly.  If $\kappa_a = \kappa_b$ then the instability condition is $\epsilon^*\epsilon \geq \kappa^2/4 + \tfrac{1}{4}(\Delta_a+\Delta_b)^2$.

\section{Kerr Cavity}
\label{sec:02-kerr}

In many optical cavities, light lives inside a material and the material has some nonlinearity.  The most common such nonlinearity is the Kerr ($\chi^{(3)}$) effect, which is present in all materials, though very small in most.  The effect is manifested in a nonlinear polarization \cite{BoydBook}:
\beq
	P_i = \chi^{(1)}_{ij} E_j + \chi^{(2)}_{ijk} E_j E_k + \chi^{(3)}_{ijkl} E_j E_k E_l + \ldots
\eeq
The electric Hamiltonian $H = \frac{1}{2}\int{E\cdot D}$ will take the following form:
\beq
	H = \frac{1}{2}\epsilon_0 (\delta_{ij} + \chi^{(1)}_{ij}) \int{E_i E_j} + \frac{1}{3}\epsilon_0 \chi^{(2)}_{ijk}\int{E_i E_j E_k} + \frac{1}{4}\epsilon_0 \chi^{(3)}_{ijkl} \int{E_i E_j E_k E_l} + \ldots
\eeq
For most (centrosymmetric) materials, $\chi^{(2)}$ vanishes.  The lowest-order nonlinearity is $\chi^{(3)}$, which is quartic in the field strength.  Thus we expect a contribution to the cavity Hamiltonian that goes as $a^\dagger a^\dagger a a$.

The actual SLH model for a simple Kerr system is:
\beq
	G = \left(1,\ \ \begin{bmatrix} \sqrt{\kappa}a \\ \sqrt{\beta} a^2 \end{bmatrix},\ \ 
          \Delta a^\dagger a + \frac{1}{2}\chi a^\dagger a^\dagger a a \right)
\eeq
Two nonlinear effects show up -- dispersion ($\chi$) and absorption ($\beta$).  It is a general fact that these two are related and you cannot have one without the other.  They are two-photon processes, resulting from two-photon coupling between filled valence-band states and empty conduction-band states.

\subsection{Derivation of the Kerr Nonlinearity}

Both the dispersive and absorptive Kerr nonlinearities come from two-photon processes.    The simplest description is a single two-level atom coupled to an optical cavity.  The atom's levels are separated by around $2\omega$, and due to spontaneous decay, the excited state's lifetime is very short.  In the dispersive case, the atom absorbs two photons and re-emits them coherently.  In the absorptive case, the atom absorbs the photons but decays to the ground state through some non-radiative process.  Together, these processes comprise the Kerr nonlinearity.

\begin{figure}[tbp]
\begin{center}
\includegraphics[width=0.40\textwidth]{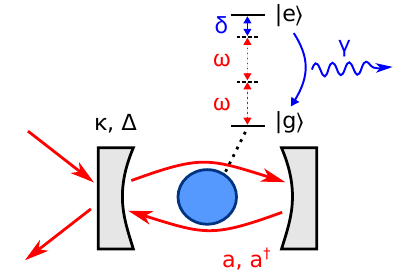}
\caption{Single-atom toy model for deriving the Kerr nonlinearity in a cavity.}
\label{default}
\end{center}
\end{figure}

The Hilbert space is a product of the optical ladder of states and the the atom, which has two states $\ket{g}, \ket{e}$.  The optical creation / annihilation operators are $a, a^\dagger$, and $\sigma_+ = \ket{e}\bra{g}, \sigma_- = \ket{g}\bra{e}$ are the atom raising-lowering operators.  This system has a Hilbert space simple SLH model:
\beq
	G = \left(1,\ \ \ \begin{bmatrix} \sqrt{\kappa}a \\ \sqrt{\gamma}\sigma_- \end{bmatrix},\ \ \  \Delta a^\dagger a + \delta \sigma_+\sigma_- + \frac{\eta\bigl[(a^\dagger)^2\sigma_- + a^2\sigma_+\bigr]}{\sqrt{2}}\right) \label{eq:02-kerr-slh}
\eeq
The $\Delta$ and $\kappa$ terms are the detuning and loss for a linear cavity.  All of the other terms involve for the two-level atom -- $\gamma$ is the spontaneous decay rate, $\delta = E_1 - E_0 - 2\omega$ is the atomic detuning, and $\eta$ couples the atom to the cavity field.

In solid-state systems, these ``atoms'' are electron states in a band structure, and tend to have very short excited-state lifetimes, so the atomic degrees of freedom can be adiabatically eliminated.  Adiabatic elimination is carried out via the QSDE limit theorem (Sec.~\ref{sec:01-adiabatic}).  Setting $\gamma, \delta \sim k^2$, $\eta \sim k$ and taking $k \rightarrow \infty$ projects the dynamics onto the ``slow'' subspace $P_0 \ket{\psi} = 0$, where $P_0 = \sigma_+\sigma_-$ is the projector onto the atom ground state.  A brief calculation gives the reduced SLH model:
\bea
	H & = & \Delta a^\dagger a + \frac{1}{2}\frac{\eta^2\delta}{\delta^2 + (\gamma/2)^2}a^\dagger a^\dagger a a \label{eq:02-kerr-h01}\\
	L_1 & = & \sqrt{\kappa}a \label{eq:02-kerr-l01} \\
	L_2 & = & \frac{\eta\sqrt{\gamma/2}}{-i\delta - \gamma/2} a^2 \label{eq:02-kerr-l02}
\eea
Changing the output $\d\tilde{B}_2$ field by a phase (which does not affect the internal dynamics) and defining $\chi, \beta$ by
\beq
	i\chi + \beta = \frac{\eta^2}{\gamma/2 + i\delta}
\eeq
converts (\ref{eq:02-kerr-h01}-\ref{eq:02-kerr-l02}) to the well-known form (\ref{eq:02-kerr-slh}).  The parameters $\chi$ and $\beta$ are the cavity's dispersive $\chi^{(3)}$ (self-phase modulation) and two-photon absorption.  The dispersive and absorptive parts are, in a sense, related by Kramers-Kronig \cite{SheikBahae1991}; therefore it is not possible to have one without the other.  One can turn off the dispersive effect by working at resonance: $\delta = 0$.  By working very far from the resonance, where $\chi \sim O(\delta^{-1})$ and $\beta \sim O(\delta^{-2})$ and thus the dispersive term dominates, but at the cost of a smaller $\chi$.  If we want to make $\chi$ as large as possible, there will inevitably be some two-photon absorption.

\subsection{Many Atoms, Multiple Fields}
\label{sec:02-multikerr}

Suppose that we have several resonant modes, not just one.  The most general case of varying polarization seems very complicated.  Instead, let's consider the case where all of the fields have the same polarization.  Then in the original $\eta$ term in the Hamiltonian, one should substitute $a \rightarrow \sum_i\psi_i(x)a_i$, where $\psi_i$ is the spatial profile of the field, which may be complex for traveling modes.  The effective Hamiltonian $H_{\rm eff} \equiv H - (i/2)L_m^\dagger L_m$ (for the Kerr process) becomes:
\beq
	H_{\rm eff} = \frac{1}{2}\eta^2(\chi-i\beta) a^\dagger a^\dagger a a \longrightarrow
\frac{1}{2}\eta^2(\chi-i\beta) \sum_{ijkl} \psi_i^\ast(x)\psi_j^\ast(x)\psi_k(x)\psi_l(x) a_i^\dagger a_j^\dagger a_k a_l
\eeq
But there are also multiple atoms.  Each atom has its own position $x$, and the total $H_{\rm eff}$ is the sum of each atom's contribution.  If there are enough atoms, this sum can be replaced by an integral weighted by the atom density $\rho(x)$.  Assuming that they all have the same $\chi$ and $\beta$ (this doesn't actually matter in most cases; it just makes the derivation simpler), the total $H_{\rm eff}$ is:
\beq
	H_{\rm eff} = \frac{1}{2}(\chi-i\beta) \sum_{ijkl} \left(\int{\rho(x) \psi_i^\ast(x)\psi_j^\ast(x)\psi_k(x)\psi_l(x)dx}\right) a_i^\dagger a_j^\dagger a_k a_l
\eeq
Making the substitution
\beq
	\Psi_{ijkl} = \int{\rho(x) \psi_i^\ast(x)\psi_j^\ast(x)\psi_k(x)\psi_l(x)dx}
\eeq
so the Hamiltonian and Lindblad terms become
\bea
	H & = & \frac{1}{2}\chi \sum_{ijkl} \Psi_{ijkl}a_i^\dagger a_j^\dagger a_k a_l \label{eq:02-kerr-h02} \\
	\sum_m{L_m^\dagger L_m} & = & \beta \sum_{ijkl} \Psi_{ijkl}a_i^\dagger a_j^\dagger a_k a_l \label{eq:02-kerr-ll02}
\eea
We can work back from (\ref{eq:02-kerr-h02}-\ref{eq:02-kerr-ll02}) to obtain the SLH model.  It will take the form:
\beq
	L_m = \sqrt{\beta} \sum_{ij} \Lambda_{m,ij} a_i a_j
\eeq
where the $\Lambda$'s must satisfy:
\beq
	\sum_m\Lambda^\ast_{m,ij}\Lambda_{m,kl} = \Psi_{ijkl} \label{eq:02-lambdadec}
\eeq
Grouping $I = (ij), J = (kl)$, $\Lambda_{m,K}$ and $\Psi_{K,L}$ become $n^2 \times n^2$ matrices, where $n$ is the number of fields, and $\Lambda$ satisfies: $\Lambda^\dagger\Lambda = \Psi$.  One valid possibility for $\Lambda$ is the Cholesky decomposition.  (The particular decomposition of $\Psi_{ijkl}$ does not matter; given any two $\Lambda_{m,ij}$ that satisfy (\ref{eq:02-lambdadec}), the SLH models will be equivalent up to a permutation of the outputs, and the master equations will also be the same.)

Putting this all together, the SLH model for the multi-field cavity is:
\beq
	\boxed{G = \left(1,\ \ \sqrt{\beta} \sum_{ij} \Lambda_{m,ij} a_i a_j,\ \ \frac{1}{2}\chi \sum_{ijkl} \Psi_{ijkl}a_i^\dagger a_j^\dagger a_k a_l\right)} \label{eq:02-slh-kerr-general}
\eeq
Adding detuning terms and external couplings gives a model like (\ref{eq:02-kerr-slh}), but with multiple fields and atoms.  This is the most general SLH model cavity systems with a Kerr nonlinearity.

\subsubsection{Example: Two-Mode Ring Cavity}

Consider a two-mode ring cavity.  A perfect ring cavity will support two degenerate optical fields -- the left and right traveling modes (Fig.~\ref{fig:02-f3}).  Imperfections will lift the degeneracy between these modes, but let's assume that they are negligible here (negligible means that the energy splitting is much less than $\kappa$).  Let $a_+$ and $a_-$, be the annihilation operators for these modes.  In cylindrical coordinates, the fields have the form:
\beq
	\psi_+ = f_{\perp} (r, z) e^{i\phi},\ \ \ \psi_- = f_{\perp} (r, z) e^{-i\phi}
\eeq
where the cross section $f_{\perp}$ is the same for both modes.  It is not hard to see that
\beq
	\Psi_{ijkl} = \delta_{i+j,k+l}
\eeq
Or, if we group the indices $I, J \in (++,+-,-+,--)$, then $\Psi$ becomes a matrix:
\beq
	\Psi_{IJ} = \begin{bmatrix} 1 & 0 & 0 & 0 \\ 0 & 1 & 1 & 0 \\ 0 & 1 & 1 & 0 \\ 0 & 0 & 0 & 1 \end{bmatrix}
\eeq
\begin{figure}[tbp]
\begin{center}
\includegraphics[width=1.00\textwidth]{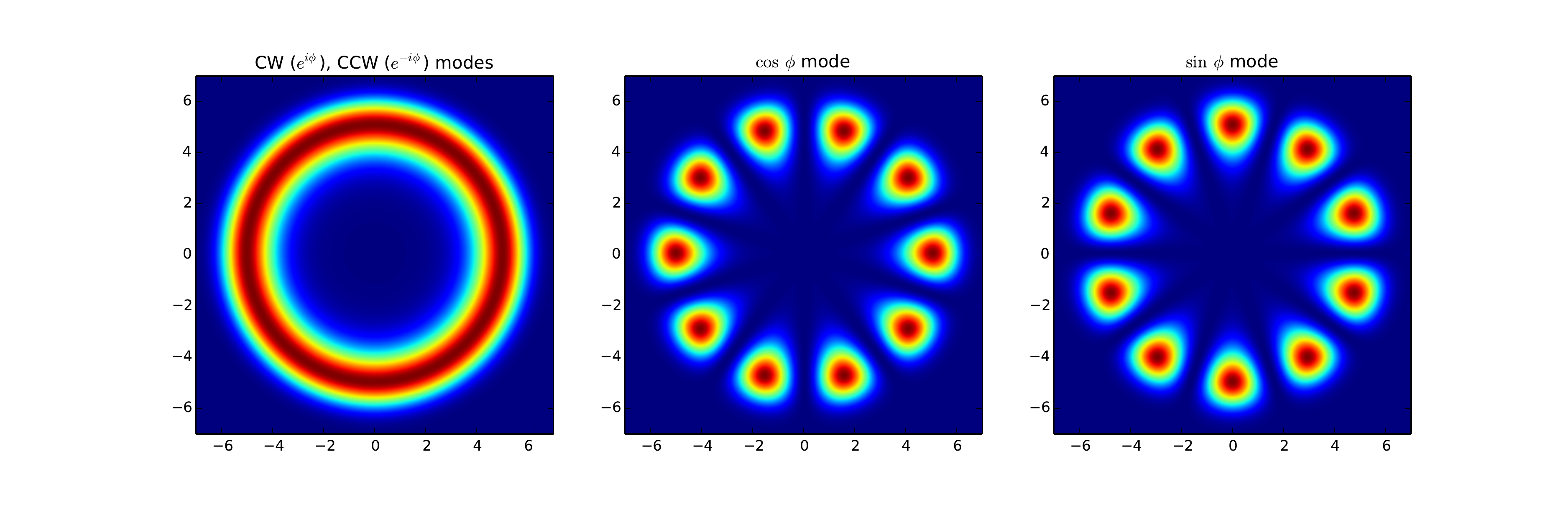}
\caption{Left: Left- and right-moving modes in a perfect ring cavity are degenerate and have a uniform energy distribution.  Center and right: Imperfections lift the mode degeneracy, giving rise to standing-wave eigenmodes.}
\label{fig:02-f3}
\end{center}
\end{figure}
A valid choice of $\Lambda$ satisfying $\Lambda^\dagger \Lambda = \Psi$ is:
\beq
	\Lambda_{mI} = \begin{bmatrix} 1 & 0 & 0 & 0 \\ 0 & 1 & 1 & 0 \\ 0 & 0 & 0 & 1 \end{bmatrix}
\eeq
Now that $\Psi$ and $\Lambda$ have been computed, the SLH model follows from Eq.~(\ref{eq:02-slh-kerr-general}):
\beq
	G = \left(1_{3\times 3},\ \ \sqrt{\beta}\begin{bmatrix} a_+^2 \\ 2a_+a_- \\ a_-^2 \end{bmatrix},\ \ \frac{1}{2}\chi\left(a_+^\dagger a_+^\dagger a_+ a_+ + 4a_+^\dagger a_+ a_-^\dagger a_- + a_-^\dagger a_-^\dagger a_- a_- \right)\right)
\eeq

\subsubsection{Example: Ring Cavity, Standing-Wave Modes}

Now consider the case of a ring cavity with standing-wave modes $\psi_1, \psi_2$.  For convenience, we take these to be sine and cosine modes, but offsetting them by a phase does not affect any of our results.  The field amplitudes may be written as:
\beq
	\psi_1 = f_{\perp} (r, z) \sin(\phi),\ \ \ \psi_2 = f_{\perp} (r, z) \cos(\phi)
\eeq
Up to a constant, $\Psi$ takes the form:
\beq
	\Psi_{IJ} = \begin{bmatrix} 1 & 0 & 0 & \frac{1}{3} \\ 0 & \frac{1}{3} & \frac{1}{3} & 0 \\
		0 & \frac{1}{3} & \frac{1}{3} & 0 \\
		\frac{1}{3} & 0 & 0 & 1 \end{bmatrix}
\eeq
which has the following $\Psi = \Lambda^\dagger \Lambda$ decomposition:
\beq
	\Lambda_{mI} = \begin{bmatrix} 1 & 0 & 0 & \frac{1}{3} \\
		0 & \frac{1}{\sqrt{3}} & \frac{1}{\sqrt{3}} & 0 \\
		0 & 0 & 0 & \frac{2\sqrt{2}}{3} \end{bmatrix}
\eeq
and the following SLH model:
\bea
	S & = & 1_{3\times 3} \\
	L & = & \sqrt{\beta}\begin{bmatrix} a_1^2 + \frac{1}{3} a_2^2 \\ \frac{2}{\sqrt{3}} a_1 a_2 \\ \frac{2\sqrt{2}}{3} a_2^2 \end{bmatrix} \\
	H & = & \frac{1}{2}\chi \left[a_1^\dagger a_1^\dagger a_1 a_1 + a_2^\dagger a_2^\dagger a_2 a_2 
		+ \frac{1}{3}\left(a_1^\dagger a_1^\dagger a_2 a_2 + a_2^\dagger a_2^\dagger a_1 a_1\right) 
		+ \frac{4}{3} a_1^\dagger a_1 a_2^\dagger a_2 \right]
\eea

\section{OPO}
\label{sec:02-opo}

\subsection{Degenerate}

An optical parametric oscillator (OPO) exploits the $\chi^{(2)}$ nonlinearity that some materials, like LiNbO$_3$, exhibit \cite{BoydBook}.  The $\chi^{(2)}$ effect gives rise to sum- and difference-frequency generation, or in the degenerate case, frequency doubling.  A degenerate OPO has two modes: a pump $b$ and signal $a$, that satisfy the condition $2\omega_a = \omega_b$.  The $\chi^{(2)}$ process converts pump photons into pairs of signal photons.  The SLH model is:
\beq
	G = \left(1,\ \ \begin{bmatrix} \sqrt{\kappa_a} a \\ \sqrt{\kappa_b} b \end{bmatrix},\ \ 
          \Delta_a a^\dagger a + \Delta_b b^\dagger b
          + \frac{\epsilon^* a^2 b^\dagger - \epsilon (a^\dagger)^2 b}{2i} \right)
\eeq
The QSDEs are:
\bea
	\d a & = & \left[(-\kappa_a/2 - i\Delta_a)a + \epsilon\,a^\dagger c\right]\d t - \sqrt{\kappa_a} \d B_a \\
	\d b & = & \left[(-\kappa_b/2 - i\Delta_b)b - \frac{1}{2}\epsilon^\ast\,a^2\right]\d t - \sqrt{\kappa_b} \d B_b \\
	\d\tilde{B}_a & = & \d B_a + \sqrt{\kappa_a} a\,\d t \\
	\d\tilde{B}_b & = & \d B_b + \sqrt{\kappa_b} b\,\d t
\eea

\subsubsection{Adiabatic Elimination of Pump}

Having a resonant pump and signal gives a {\it doubly-resonant} OPO.  In most OPOs, the pump is not resonant, giving a {\it singly-resonant} oscillator.  We can obtain the SLH model for the singly-resonant case by adiabatically eliminating $b$: scaling the pump loss and coupling to infinity $\kappa_c \sim O(k^2), \epsilon \sim O(k)$, $k \rightarrow \infty$ and following the adiabatic limit theorem (Sec.~\ref{sec:01-adiabatic}), we find that the ``slow'' space is the kernel of $Y^\dagger = (-\kappa_b/2-i\Delta_b) b^\dagger b$, i.e.\ all states without photons in the $b$ mode.  The projector is obviously $P_0 = I_a \otimes \left|0\rangle \langle 0\right|_b$, and $P_1 = I - P_0$.

Next, we construct the pseudo-inverse of $Y$, $\tilde{Y}$ such that $\tilde{Y} Y = Y\tilde{Y} = P_1$.  One can follow the procedure in Eqs.~(\ref{eq:01-adel1}-\ref{eq:01-adel3}) to obtain $-iH - \tfrac{1}{2}L_m^\dagger L_m$, $L$ and $S$.  By tweaking the input and output phases and defining a two-photon absorption and cross-Kerr coefficient 
\beq
    \beta + i\chi = \frac{\epsilon^*\epsilon}{2} \frac{1}{\kappa_b/2 + i\Delta_b}
\eeq
we convert this to the following SLH model:
\beq
    \left(1_{2\times 2},\ \ \ \begin{bmatrix} \sqrt{\kappa_a} a \\ \sqrt{\beta}\,a^2 \end{bmatrix},\ \ \ \Delta_a a^\dagger a + \frac{1}{2}\chi a^\dagger a^\dagger a a \right)
\eeq
This is the correct model in the absence of a pump field.  Generally OPOs are pumped, so one models the pump as a coherent displacement $\bar{\epsilon}/2\sqrt{\beta}$ in front of the input, where $\bar{\epsilon}$ is the normalized pump amplitude.  Without changing the internal dynamics, one can put a reverse displacement on the output.  The resulting Gough-James expression
\beq
    L(\bar{\epsilon}/2\sqrt{\beta}) \triangleleft G \triangleleft L(-\bar{\epsilon}/2\sqrt{\beta})
\eeq
has the SLH model:
\beq \label{eq:02-dopo-ad}
    \left(1_{3\times 3},\ \ \ \begin{bmatrix} \sqrt{\kappa_a} a \\ \sqrt{\beta}\,a^2 \end{bmatrix},\ \ \ \Delta_a a^\dagger a + \frac{\bar{\epsilon}^* a^2 - \bar{\epsilon} (a^\dagger)^2}{2i} + \frac{1}{2}\chi a^\dagger a^\dagger a a  \right)
\eeq
Typically we assume that $\Delta_b = 0$, so that $\chi = 0$ and $\beta = |\epsilon|^2/\kappa$.  The QSDEs for the degenerate OPO are:
\beq
\d a = \left[\left(-i\Delta_a - \frac{\kappa_a + (\beta+i\chi)\,a^\dagger a}{2}\right) a + \bar{\epsilon}\, a^\dagger\right]\,\d t 
- \sqrt{\kappa_a} \d B_a - \sqrt{\beta}\,a^\dagger \d B_b \label{eq:02-dopo-adsde} \\
\eeq
Taking the limit $\beta, \chi \rightarrow 0$ removes the nonlinear terms from (\ref{eq:02-ndopo-adsde1}-\ref{eq:02-ndopo-adsde2}).  This is the undepleted pump limit, which reduces to the degenerate OPO in Eq.~(\ref{eq:02-ldopo-slh}) if we have $N$ signal ports rather than one.

\subsection{Nondegenerate}
\label{sec:02-opo-nd}

In a nondegenerate OPO, there are three modes $a, b, c$ that satisfy the sum-frequency condition $\omega_a + \omega_b = \omega_c$, and the $\chi^{(2)}$ process is phase-matched to the process $\omega_a + \omega_b \leftrightarrow \omega_c$.  Photons at the pump frequency ($\omega_c$) are down-converted to pairs of photons at $\omega_a$ (signal) and $\omega_b$ (idler).  The SLH model is:
\beq
	G = \left(1,\ \ \begin{bmatrix} \sqrt{\kappa_a} a \\ \sqrt{\kappa_b} b \\ \sqrt{\kappa_c} c \end{bmatrix},\ \ 
          \Delta_a a^\dagger a + \Delta_b b^\dagger b + \Delta_c c^\dagger c
          + \frac{\epsilon^* a b c^\dagger - \epsilon\,a^\dagger b^\dagger c}{2i} \right)
\eeq
where $a$, $b$, and $c$ are the modes, with frequencies that add: $\omega_a + \omega_b = \omega_c$, $\epsilon$ is the strength of the nonlinearity, and $\Delta_a$, $\Delta_b$, $\Delta_c$ are the cavity detunings for each mode.

The QSDEs are:
\bea
	\d a & = & \left[(-\kappa_a/2 - i\Delta_a)a + \frac{1}{2}\epsilon\,b^\dagger c\right]\d t - \sqrt{\kappa_a} \d B_a \\
	\d b & = & \left[(-\kappa_b/2 - i\Delta_b)b + \frac{1}{2}\epsilon\,a^\dagger c\right]\d t - \sqrt{\kappa_b} \d B_b \\
	\d c & = & \left[(-\kappa_c/2 - i\Delta_c)c - \frac{1}{2}\epsilon^\ast\,a b\right]\d t - \sqrt{\kappa_c} \d B_c \\
	\d\tilde{B}_a & = & \d B_a + \sqrt{\kappa_a} a\,\d t \\
	\d\tilde{B}_b & = & \d B_b + \sqrt{\kappa_b} b\,\d t \\
	\d\tilde{B}_c & = & \d B_c + \sqrt{\kappa_c} c\,\d t
\eea
As in the previous section, we adiabatically eliminate the pump by letting he pump loss and coupling scale a $\kappa_c \sim O(k^2), \epsilon \sim O(k)$ with $k \rightarrow \infty$.  The QSDE limit theorem projects the dynamics down to the vacuum in $c$ ($P_0 = I_a \otimes I_b \otimes \left|0\rangle \langle 0\right|_c$).  Defining a two-photon absorption and cross-Kerr coefficient 
\beq
    \beta + i\chi = \frac{\epsilon^*\epsilon}{2} \frac{1}{\kappa_c/2 + i\Delta_c}
\eeq
we obtain the adiabatically eliminated SLH model:
\beq
    \left(1_{3\times 3},\ \ \ \begin{bmatrix} \sqrt{\kappa_a} a \\ \sqrt{\kappa_b} b \\ \sqrt{\beta}\,ab \end{bmatrix},\ \ \ \Delta_a a^\dagger a + \Delta_b b^\dagger b + \frac{1}{2}\chi a^\dagger a b^\dagger b \right)
\eeq
Next, we add a pump field $\bar{\epsilon}/\sqrt{\beta}$; the resulting cascade $L(\bar{\epsilon}/\sqrt{\beta}) \triangleleft G \triangleleft L(-\bar{\epsilon}/\sqrt{\beta})$ has the SLH model:
\beq \label{eq:02-ndopo-ad}
    \left(1_{3\times 3},\ \ \ \begin{bmatrix} \sqrt{\kappa_a} a \\ \sqrt{\kappa_b} b \\ \sqrt{\beta}\,ab \end{bmatrix},\ \ \ \Delta_a a^\dagger a + \Delta_b b^\dagger b + \frac{\bar{\epsilon}^* a b - \bar{\epsilon}\,a^\dagger b^\dagger}{i} + \frac{1}{2}\chi a^\dagger a b^\dagger b  \right)
\eeq
The QSDEs are:
\begin{eqnarray}
\d a & = & \left[\left(-i\Delta_a - \frac{\kappa_a + (\beta+i\chi)\,b^\dagger b}{2}\right) a + \bar{\epsilon}\, b^\dagger\right]\,\d t 
- \sqrt{\kappa_a} \d B_a - \sqrt{\beta}\,b^\dagger \d B_c \label{eq:02-ndopo-adsde1} \\
\d b & = & \left[\left(-i\Delta_b - \frac{\kappa_b + (\beta+i\chi)\,a^\dagger a}{2}\right) b + \bar{\epsilon}\, a^\dagger\right]\,\d t
- \sqrt{\kappa_b} \d B_b - \sqrt{\beta}\,a^\dagger \d B_c \label{eq:02-ndopo-adsde2}
\end{eqnarray}
The nonlinear terms go away when $\beta, \chi \rightarrow 0$, reducing this to the linearized model (\ref{eq:02-lndopo-slh}).

\begin{figure}[b!]
\begin{center}
\includegraphics[width=0.66\textwidth]{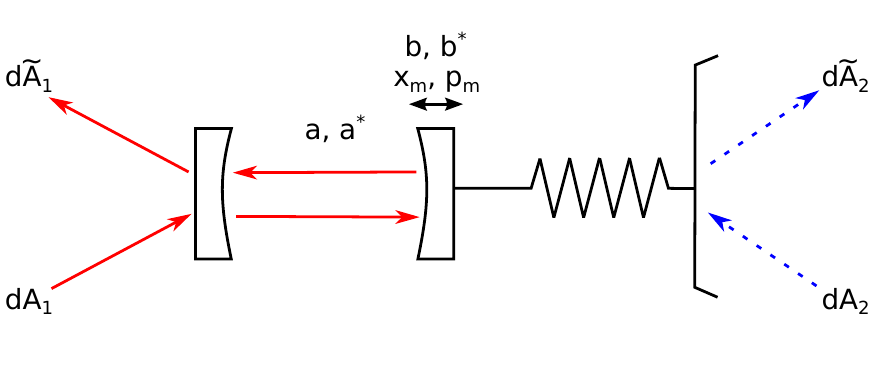}
\caption{Diagram of a mechanical oscillator and its fields.}
\label{fig:02-f4}
\end{center}
\end{figure}

\section{Optomechanical Cavity}

An optomechanical cavity is an optical cavity connected to a mechanical mode.  Sometimes this coupling is obvious, as in a mirror on a spring.  Sometimes not.  In any case, we can model it as a system with two bosonic modes -- $a$ for the cavity (photons) and $b$ for the mechanical oscillator (phonons).  The mechanical displacement is encoded in the $X$ quadrature of the mechanical mode: $x_m = b + b^\dagger$.  In the absence of any external forces, this mode exhibits damped oscillations governed by the equation $\ddot{x}_m = -\Omega_m^2 x_m - \gamma \dot{x}_m$ \cite{Aspelmeyer2014}.  This equation is realized by the SLH model
\beq
	\left(1,\ \ \sqrt{\gamma}\,b,\ \ \Omega_m b^\dagger b + \frac{\gamma}{2}\frac{b^2 - (b^\dagger)^2}{2i}\right) \label{eq:02-mechslh1}
\eeq
Note the resemblance between this model and the degenerate OPO (Eq.~\ref{eq:02-ldopo-slh}), where $\epsilon \rightarrow \gamma/2$ is the OPO pump.  However, optomechanical cavities are engineered to have high $Q$, so $\gamma \ll \Omega_m$ and the $O(\gamma)$ term in $H$ can be dropped.  The resulting SLH model matches that of an optical cavity.

Now add the optomechanical coupling.  Because this coupling must conserve energy and the the photon energy is much greater than that of the the phonons, and because the mechanical motion is nonrelativistic, it must take the form $a^\dagger a f(x_m)$ for some function $f$.  The strongest couplings are linear in $x_m$ (although $x_m^2$ couplings have been considered as well \cite{Hill2011}); this section assumes a coupling $\eta\,x_m a^\dagger a$, which is common in the literature.  One obtains the SLH model \cite{Hamerly2012}:
\beq
	G = \left(1_{4\times 4},\ \ \begin{bmatrix} {\sqrt{\kappa} a} \\ \sqrt{k_m} b \\ \sqrt{k_N} b \\ \sqrt{k_N} b^\dagger \end{bmatrix},\ \ 
	\Delta_{a} a^\dagger a + \eta\,x_m a^\dagger a + \omega_m b^\dagger b + i(E a^\dagger - E^* a)\right)
\eeq
A few things are worth noting in this model.  First, there extra couplings $k_N$ are auxiliary noise terms -- they add thermal noise to the cavity but do not change the dissipation rate.  This is necessary for mechanical modes, which tend to be thermally excited even at very low temperatures.

The QSDEs for this system are:
\bea
	\d a & = & \left[\left(-i\Delta_{a} - \frac{1}{2}\kappa\right)a - i\eta x_m a + E\right]\d t - \sqrt{\kappa_a}\d B_1\\
	\d b & = & \left[\left(-i\omega_m - \frac{1}{2}k_m\right)b - i\eta a^\dagger a\right]\d t - \sqrt{\kappa_m}\d B_2 - \sqrt{\kappa_N}\d B_3 + \sqrt{\kappa_N}\d B_4^\dagger\\
	\d\tilde{B}_1 & = & \d B_1 + \sqrt{\kappa_a} a\,\d t \\
	\d\tilde{B}_2 & = & \d B_2 + \sqrt{\kappa_m} b\,\d t \\
	\d\tilde{B}_3 & = & \d B_3 + \sqrt{\kappa_N} b\,\d t \\
	\d\tilde{B}_4 & = & \d B_4 + \sqrt{\kappa_N} b^\dagger \d t
\eea
Suppose that the field $E$ is considerable.  Then the steady-state value of the QSDEs (ignoring the stochastic part) will be some large photon field $a_0 \gg 1$ with some large steady-state displacement $x_m \gg 1$.  The QSDEs can be linearized about this fixed point by making the replacements $a \rightarrow \alpha + a$, $b \rightarrow \beta + b$, and assuming $a \ll a_0$, $b \ll b_0$:
\bea
	\d a & = & \left[\left(-i\Delta_{a} - \frac{1}{2}\kappa\right)a - i\eta (x_m \alpha + x_0 a)\right]\d t - \sqrt{\kappa_a}\d B_1\\
	\d b & = & \left[\left(-i\omega_m - \frac{1}{2}k_m\right)b - i\eta (\alpha a^\dagger + \alpha^* a)\right]\d t - \sqrt{\kappa_m}\d B_2 - \sqrt{\kappa_N}\d B_3 + \sqrt{\kappa_N}\d B_4^\dagger 
\eea
This corresponds to the following (linear) SLH model:
\beq
	G = \left(1_{4\times 4},\ \ \begin{bmatrix} {\sqrt{\kappa} a} \\ \sqrt{k_m} b \\ \sqrt{k_N} b \\ \sqrt{k_N} b^\dagger \end{bmatrix},\ \ 
	(\Delta_{a}+\eta (\beta+\beta^*)) a^\dagger a + \eta(\alpha^*a+\alpha a^\dagger)(b+b^\dagger) + \omega_m b^\dagger b \right)
\eeq
This is the SLH model of two harmonic oscillators with a linear coupling term.  It can be solved analytically.

\begin{figure}[b!]
\begin{center}
\includegraphics[width=0.35\textwidth]{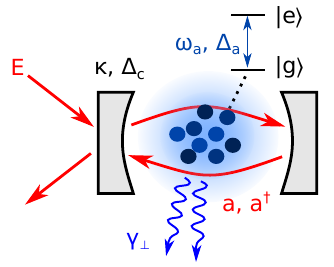}
\caption{Atom cloud in an optical cavity}
\label{fig:02-f5}
\end{center}
\end{figure}

\section{Atom Cavity}
\label{sec:02-atom}

Cavity QED is the study of atoms that couple to resonant modes in cavities \cite{Mabuchi2002, Doherty2004}.  The most common system is the {\it Jaynes-Cummings model}, a two-level atom coupled to a cavity with a single mode \cite{CarmichaelBook}.  This is a useful quantum system to study because it is possible to create very strong nonlinearities using experimental techniques that have been around for several decades.  More recently, this framework has been extended to other two-level systems, including superconducting qubits \cite{Girvin2011, Blais2004} and quantum dots \cite{Reithmaier2004, Hennessy2007, Englund2007, Hu2008}.

Start with the full quantum model of an atom-cavity system.  The Hilbert space is $\mathcal{H} = \mathcal{H}_{\rm cav} \otimes (\mathcal{H}_{\rm at})^N$ -- a single cavity mode $\mathcal{H}_{\rm cav}$ coupled to $N$ identical two-level atoms $\mathcal{H}_{\rm at}$.  Let $(a^\dagger, a)$ be the creation and annihilation operators on $\mathcal{H}_{\rm cav}$.  For each atom, define raising and lowering operators $\sigma_{+,i} = |e\rangle\langle g|$ and $\sigma_{-,i} = |g\rangle\langle e|$, and an energy operator $\sigma_{z,i} = |e\rangle\langle e| - |g\rangle\langle g|$.  The operator algebra for this Hilbert space is:
\begin{eqnarray}
	[a, a^\dagger] & = & 1 \\
	{[}\sigma_{+,i}, \sigma_{-,i}] & = & \sigma_{z,i} \\
	{[}\sigma_{+,i}, \sigma_{z,i}] & = & -2\sigma_{+,i} \\
	{[}\sigma_{-,i}, \sigma_{z,i}] & = & 2\sigma_{-,i}
\end{eqnarray}
The SLH model is \cite{Armen2006}:
\begin{eqnarray}
	S & = & 1 \\
	L & = & \begin{bmatrix} \sqrt{\kappa} a \\ \sqrt{\gamma_{||}}\;\sigma_{-,i} \\ \sqrt{\gamma_{nr}/2}\;\sigma_{z,i} \end{bmatrix} \\
	H & = & \Delta_c a^\dagger a + \frac{1}{2}\Delta_a \sum_k\sigma_{z,k} + i g_0 \sum_k (a^\dagger \sigma_{-,k} - a \sigma_{+,k}) \\
	& = & \Delta_c a^\dagger a + \frac{1}{2}\Delta_a \sigma_{z} + i g_0 (a^\dagger \sigma_{-} - a \sigma_{+})
\end{eqnarray}
where in the last line, the total spin $\sigma = \sum_k\sigma_k$ has been substituted.

As far as an observer is concerned, all of the atoms in the cavity are identical.  An observer can only measure things related to spin sums, $\sigma_z = \sum_k\sigma_{z,k}$, $\sigma_\pm = \sum_k\sigma_{\pm,k}$, because an observer can only measure the optical field, and the optical field couples to spin sums.  Therefore, as much as possible, we would like to only keep track of the optical field and spin sums, $(a, a^\dagger, \sigma_-, \sigma_+, \sigma_z)$, when we model the system.

An important semiclassical way to model the system is through the Maxwell-Bloch equations \cite{Mandel2005}.  These are derived from the QSDEs for the atom cavity, which take the following form:
\bea
	\d a & = & \left[\left(-i\Delta_c - \frac{1}{2}\kappa\right)a + g_0 \sigma_-\right]\d t - \sqrt{\kappa}\d B_a \\
	\d\sigma_- & = & \left[\left(-i\Delta_a - \gamma_\perp\right)\sigma_- + g_0a\sigma_z\right]\d t + \sum_i \left[\sqrt{\gamma_{||}}\sigma_{z,i}\d B_{||,i} + \sqrt{2\gamma_{nr}}\sigma_{-,i}(\d B_{z,i}-\d B_{z,i}^\dagger)\right]  \\
	\d\sigma_z & = & \left[-\gamma_\perp(\sigma_z + N) - 2g_0 (a\sigma_+ + a^\dagger\sigma_-)\right]\d t - \sum_i 2\sqrt{\gamma_{||}} (\sigma_{+,i}\d B_{nr,i} + \sigma_{-,i}\d B_{nr,i}^\dagger)
\eea
where $\gamma_{\perp} = \gamma_{nr} + \gamma_{||}/2$.
	
If we throw out the stochastic terms -- as we would do if we were interested in expected values, the equations only depend on spin-sum quantities, and take the following form:
\bea
	\frac{\d}{\d t}\avg{a} & = & \left(-i\Delta_c - \frac{1}{2}\kappa\right)\avg{a} + g_0 \avg{\sigma_-} \\
	\frac{\d}{\d t}\avg{\sigma_-} & = & \left(-i\Delta_a - \gamma_\perp\right)\avg{\sigma_-} + g_0\avg{a\sigma_z} \\
	\frac{\d}{\d t}\avg{\sigma_z} & = & -\gamma_\perp\bigl(\avg{\sigma_z} + N\bigr) - 2g_0 (\avg{a\sigma_+} + \avg{a^\dagger\sigma_-})
\eea
These are not closed, since the derivatives for $\avg{\sigma_-}$ and $\avg{\sigma_z}$ depend on operator products.  A semiclassical way to get around this is to factorize the operator products, as follows:
\bea
	\frac{\d}{\d t}\avg{a} & = & \left(-i\Delta_c - \frac{1}{2}\kappa\right)\avg{a} + g_0 \avg{\sigma_-} \\
	\frac{\d}{\d t}\avg{\sigma_-} & = & \left(-i\Delta_a - \gamma_\perp\right)\avg{\sigma_-} + g_0\avg{a}\avg{\sigma_z} \\
	\frac{\d}{\d t}\avg{\sigma_z} & = & -\gamma_\perp\bigl(\avg{\sigma_z} + N\bigr) - 2g_0 (\avg{a}\avg{\sigma_-}^* + \avg{a}^*\avg{\sigma_-})
\eea
The Maxwell-Bloch equations are accurate for classical systems with many atoms and photons, but even in the single-atom case, they are still approximately correct provided that the photon number is sufficiently large \cite{Armen2006, Mandel2005}.  Maxwell-Bloch-like equations can also be derived from phase-space methods \cite{Lugiato1978} and manifold projection \cite{Mabuchi2008b}; both methods are semiclassical in nature and do not apply to the quantum case of strong coupling ($g_0 \gtrsim \kappa, \gamma$).

\section{Reciprocity Rules}

\begin{figure}[!bp]
\begin{center}
\includegraphics[width=1.00\textwidth]{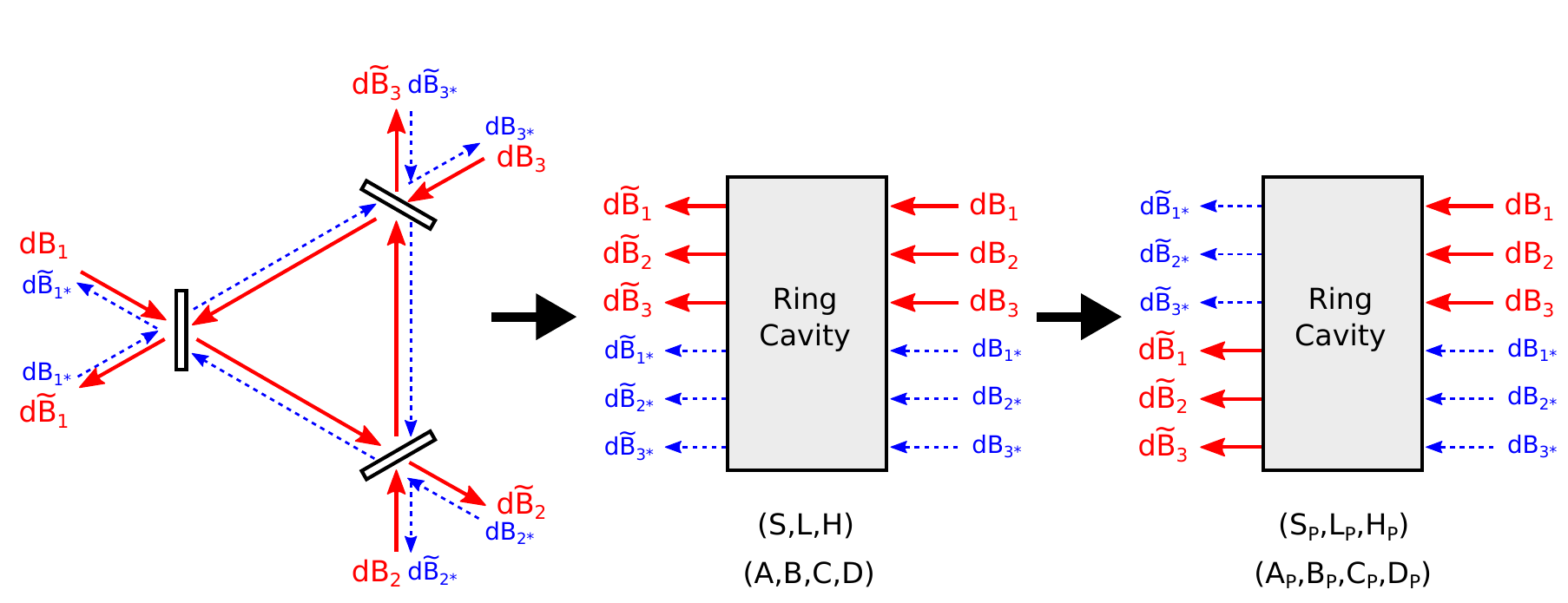}
\caption{Left: Ring cavity with forward- and reverse-propagating modes.  Center: QHDL diagram of ring cavity.  Right: Ring cavity with canonical port ordering.}
\label{fig:02-f6}
\end{center}
\end{figure}

Maxwell's Equations place some constraints on the models that can be realized in photonic circuits.  One important aspect of Maxwell's equations is \textit{time-inversion symmetry} -- if a forward-propagating solution $\mbox{Re}[E(x)e^{i\omega t}]$ is allowed, so is the time-reversed solution $\mbox{Re}[E(x)^*e^{i\omega t}]$.  

Time-reversal symmetry imposes several \textit{reciprocity relations} on the SLH model (or ABCD matrices in Eq.~\ref{eq:02-abcd}, in the linear case) of a quantum system.  External fields, such as applied magnetic fields or laser beams, can break this symmetry -- but in the absence of such fields, the reciprocity relations always hold.

As Figure \ref{fig:02-f6} shows, input-output ports come in reciprocal pairs.  Each of the three red input fields shown in the figure, $\d B_1, \d B_2, \d B_3$, has a time-reversed output $\d\tilde{B}_{1*}, \d\tilde{B}_{2*}, \d\tilde{B}_{3*}$.  These are \textit{not}, in general, the same as the outputs $\d\tilde{B}_1, \d\tilde{B}_2, \d\tilde{B}_3$.

In essence, reciprocity states that if we send a field $f(t)$ into input $\d B_i$ and measure $g(t)$ from output $\d\tilde{B}_j$, then the reciprocal should also be true -- sending $f(t)^*$ into $\d B_{j*}$ should result in an output $g(t)^*$ at $\d\tilde{B}_{i*}$.  This can be reduced to a set of constraints on S, L, and H.  However, these relations make the most sense if the input and output ports are \textit{canonically ordered} -- if $\d B_i$ is the $n^{\rm th}$ input port in the device, $\d\tilde{B}_{i*}$ must be $n^{\rm th}$ output (right panel in Fig.~\ref{fig:02-f6}).  Define $(S_P, L_P, H)$ as the canonically ordered SLH model.  If it is a linear system, we can also define canonically ordered ABCD matrices $(A_P, B_P, C_P, D_P)$.

\subsection{Static Case}

The reciprocity relations are easy to establish for a nondynamical system with scattering matrix $S$.  First, as noted above, arrange the rows and columns so that row $n$ and column $n$ correspond to the same channel but different propagation directions.  For example, for a two-way beamsplitter, the typical S matrix is:
\beq
	S = \begin{bmatrix} t & -r &  u & -v \\
	                    r &  t & -v & -u \\
	                    u &  v &  t &  r \\
	                    v & -u & -r &  t \end{bmatrix}
\eeq
in the typical ordering \texttt{(In1, In2, In1*, In2*)} for inputs and \texttt{(Out1, Out2, Out1*, Out2*)}.  To bring this to canonical ordering, we would permute the outputs to \texttt{(Out1*, Out2*, Out1, Out2)} and construct the permuted scattering matrix:
\beq
	S_P = \begin{bmatrix} u &  v &  t &  r \\
	                      v & -u & -r &  t \\
	                      t & -r &  u & -v \\
	                      r &  t & -v & -u \end{bmatrix}
\eeq
For any input $\d B$, the output field will be $\d\tilde{B} = S_P \d B$.  Reciprocity tells us that all of the fields can be reversed, so $\d\tilde{b}^*$ can be fed into the device to produce $\d B^*$ as output: $\d B^* = S_P \d\tilde{B}^*$.  Taking the complex conjugate, $\d B = S_P^* \d\tilde{B}$.  This implies that $S_P^* = S_P^{-1}$, By unitarity, $S_P^{-1} = S_P^\dagger$, so $S_P^* = S_P^\dagger$.  Taking the conjugate of this gives the reciprocity relation:
\beq
	\boxed{S_P = S_P^T}
\eeq
This relation only holds once we have permuted the rows and columns to identify same-channel modes.  If this is not done, the solution $\d\tilde{B}^*$ will \textit{not} represent the time-reversed version of the inputs $\d B$, and the relation will not hold.

\subsection{Linear, Dynamic Case}

Following Eqs.~(\ref{eq:02-qsde-abcd}, \ref{eq:02-abcd}), any linear dynamical system can be represented by its $A, B, C, D$ matrices ($a$ and $c$ are not present since these are due to external driving which always breaks reciprocity).  The system, if it takes $\d B$ as an input, outputs $\d\tilde{B}$ by way of an internal state $x$.  If reciprocity holds, then $\d\tilde{B}^*$ may be taken as the input, leading to $\d B^*$ as the output, with some $\bar{x}^*$ as the internal state.  The original dynamics (in the frequency domain) are given by:
\bea
	-i\omega x(\omega) & = & A x(\omega) + B\,b(\omega) \label{eq:02-ld1} \\
	\tilde{b}(\omega) & = & C x(\omega) + D\,b(\omega) \label{eq:02-ld2} 
\eea
The time-reversed dynamics are given by:
\bea
	-i\omega \bar{x}(\omega)^* & = & A \bar{x}(\omega)^* + B\,\tilde{b}(\omega)^* \label{eq:02-ld3} \\
	b(\omega)^* & = & C \bar{x}(\omega)^* + D\,\tilde{b}(\omega)^* \label{eq:02-ld4} 
\eea
which may be rewritten as:
\bea
	-i\omega \bar{x}(\omega) & = & \left(-A^* + B^* D^T C^*\right) \bar{x}(\omega) - B^* D^T b(\omega) \label{eq:02-ld5} \\
	\d\tilde{b}(\omega) & = & -D^T C^* \bar{x}(\omega) + D^T b(\omega) \label{eq:02-ld6}
\eea
Comparing these to the original input-output equations, we see immediately that $D = D^T$.  Making the SLH model substitutions (\ref{eq:02-abcd}) makes things clearer:
\beq
	A = -i R - \frac{1}{2} \Lambda^\dagger \Lambda,\ \ \ 
	B = -\Lambda^\dagger S,\ \ \ 
	C = \Lambda,\ \ \ 
	D = S
\eeq
This gives the original input-output equations (\ref{eq:02-ld1}-\ref{eq:02-ld2}):
\bea
	-i\omega x(\omega) & = & \left(-i R - \frac{1}{2}\Lambda^\dagger\Lambda\right) x(\omega) - \Lambda^\dagger S\,b(\omega) \\
	\tilde{b}(\omega) & = & \Lambda x(\omega) + S\,b(\omega)
\eea
and the time-reversed equations (\ref{eq:02-ld5}-\ref{eq:02-ld6}):
\bea
	-i\omega \bar{x}(\omega) & = & \left(-i R^* - \frac{1}{2}\Lambda^T\Lambda^*\right) \bar{x}(\omega) + \Lambda^T b(\omega) \label{eq:02-ld7} \\
	\tilde{b}(\omega) & = & -S \Lambda^* \bar{x}(\omega) + S\,b(\omega) \label{eq:02-ld8}
\eea
These equations must be consistent.  Since both are linear equations, it is clear that there must be a linear, one-to-one relationship between $x$ and $\bar{x}$, namely $\bar{x} = U x$.  Then the time-reversed equations (\ref{eq:02-ld7}-\ref{eq:02-ld8}) may be written as:
\bea
	-i\omega x(\omega) & = & \left(-i U^{-1} R^* U - \frac{1}{2}U^{-1}\Lambda^T \Lambda^* U\right) x(\omega) + U^{-1}\Lambda^T b(\omega) \\
	\tilde{b}(\omega) & = & -S \Lambda^* U x(\omega) + S\,b(\omega)
\eea
Comparing this to the original set of equations, one finds four (partly redundant) constraints on $U$:
\bea
	U^{-1} R^* U & = & R \\
	U^{-1} \Lambda^T \Lambda^* U & = & \Lambda^\dagger \Lambda \\
	U^{-1}\Lambda^T & = & -\Lambda^\dagger S \\
	S \Lambda^* U & = & -\Lambda
\eea
The third and fourth equations, taken together, require $U$ to be unitary and symmetric, $U^T = U$ and $U^\dagger U = 1$.  The second is redundant on these.  The four constraints can then be simplified to:
\begin{empheq}[box=\fbox]{align}
	U R U^\dagger & = R^* \\
	S^\dagger \Lambda U^\dagger & = -\Lambda^*
\end{empheq}
These are the reciprocity relations for linear systems.

\subsection{Components with Back-Reflection}

Reciprocity can usually be ignored as long as all of the components have negligible back-reflection.  The component can be divided into two identical sub-components -- a ``forward'' one and a ``reverse'' one -- with no coupling between them and no other constraints on the sub-components.  When back-reflection becomes important, reciprocity places important restrictions on how these sub-components couple.  The most common components are given as examples here.

\subsubsection{Beamsplitter}

\begin{figure}[tbp]
\begin{center}
\includegraphics[width=0.3\textwidth]{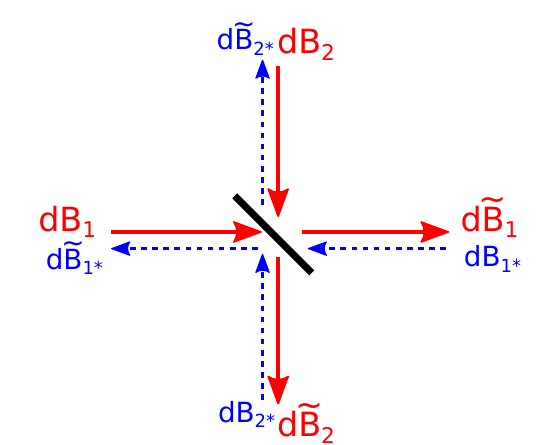}
\caption{Input-output ports of a beamsplitter with back-reflection.}
\label{fig:02-f7}
\end{center}
\end{figure}

Consider a beamsplitter with inputs \texttt{In1, In1*, In2, In2*}, and corresponding outputs \texttt{Out1, Out1*, Out2, Out2*} (Fig.~\ref{fig:02-f7}).  The scattering matrix is unitary; the most general unitary matrix can be written as the exponential of a Hermitian:
\beq
	S = e^{iG},\ \ \ G = G^\dagger = \begin{bmatrix} a & b & c & d \\ b^\ast & e & f & g \\ 
	    c^\ast & f^\ast & h & k \\ d^\ast & g^\ast & k^\ast & l \end{bmatrix}
\eeq
where $a, e, h, j$ are real and the rest are complex.  When the inputs and outputs are canonically ordered: \texttt{(In1, Out1*), (In1*, Out1), (In2, Out2*), (In2*, Out2)}, reciprocity imposes the constraint $S_P = S_P^T$.  Canonical ordering is obtained by permuting the outputs:
\beq
	S_P = P_P S \equiv \begin{bmatrix} 0&1&0&0\\1&0&0&0\\0&0&0&1\\0&0&1&0 \end{bmatrix} S
\eeq
This gives the constraint $P e^{iG} = (P e^{iG})^T$, which can be simplified to $P G P = G^T$.  This imposes the constraints:
\beq
	e = a,\ \ \ g = c^*,\ \ \ f = d^*,\ \ \ l = h  
\eeq
giving the matrix
\beq
	G = \begin{bmatrix} a & b & c & d \\ 
	                    b^\ast & a & d^\ast & c^\ast \\ 
	                    c^\ast & d & h & k \\
	                    d^\ast & c & k^\ast & h \end{bmatrix}
\eeq
This is the most general physically realizable two-way beamsplitter.  It has 10 degrees of freedom (2 real plus 4 complex).  This is generally too complicated, so additional symmetry assumptions are often made -- $P_{\rm inv} G P_{\rm inv} = G$, $P_{\rm flip} G P_{\rm flip} = G$, where:
\beq
	P_{\rm inv} = \begin{bmatrix} 0&1&0&0\\1&0&0&0\\0&0&0&1\\0&0&1&0 \end{bmatrix},\ \ \ 
	P_{\rm flip} = \begin{bmatrix} 0&0&1&0\\0&0&0&1\\1&0&0&0\\0&1&0&0 \end{bmatrix}
\eeq
These give the following constraints:
\begin{itemize}
	\item Inversion symmetry constraints: $b, c, d, k \in \mathbb{R}$
	\item Flip symmetry constraints: $c \in \mathbb{R}, a = h, b = k$
\end{itemize}
Satisfying both symmetries gives a device with 3 real degrees of freedom plus a phase:
\beq
	G = \phi \begin{bmatrix} 1&0&0&0\\0&1&0&0\\0&0&1&0\\0&0&0&1 \end{bmatrix} + 
	    \alpha \begin{bmatrix} 0&0&1&0\\0&0&0&1\\1&0&0&0\\0&1&0&0 \end{bmatrix} + 
	    \beta \begin{bmatrix} 0&1&0&0\\1&0&0&0\\0&0&0&1\\0&0&1&0 \end{bmatrix} + 
	    \gamma \begin{bmatrix} 0&0&0&1\\0&0&1&0\\0&1&0&0\\1&0&0&0 \end{bmatrix}
\eeq
This gives the beamsplitter in ``symmetric'' form.  One alternative to symmetric form is ``real form'' (which, it turns out, is only real when there are no back-reflections).  In this form, one places an $i$ phase shift in front of the \texttt{In2} port and a $-i$ phase shift in front of the \texttt{Out2} port, as is done for the one-way beamsplitter.  This alters the scattering matrix as follows:
\beq
	S_r \rightarrow U S U^\dagger,\ \ \ U = \begin{bmatrix} 1&0&0&0\\0&1&0&0\\0&0&-i&0\\0&0&0&i \end{bmatrix}
\eeq
This transforms the generator matrix as follows:
\beq
	G \rightarrow U G U^\dagger = 
	    \phi\begin{bmatrix} 1&0&0&0\\0&1&0&0\\0&0&1&0\\0&0&0&1 \end{bmatrix} + 
	    \alpha \begin{bmatrix} 0&0&i&0\\0&0&0&-i\\-i&0&0&0\\0&i&0&0 \end{bmatrix} + 
	    \beta \begin{bmatrix} 0&1&0&0\\1&0&0&0\\0&0&0&-1\\0&0&-1&0 \end{bmatrix} + 
	    \gamma \begin{bmatrix} 0&0&0&-i\\0&0&i&0\\0&-i&0&0\\i&0&0&0 \end{bmatrix}
\eeq
The simplest case is the one with no back-reflection ($\phi = \beta = \gamma = 0$).  The scattering matrix, in symmetric form, is:
\beq
	S = \begin{bmatrix} \cos\alpha & 0 & i\sin\alpha & 0 \\
	                    0 & \cos\alpha & 0 & i\sin\alpha \\
	                    i\sin\alpha & 0 & \cos\alpha & 0 \\
	                    0 & i\sin\alpha & 0 & \cos\alpha \end{bmatrix} =
	    \begin{bmatrix} \cos\alpha & i\sin\alpha \\ i\sin\alpha & \cos\alpha \end{bmatrix} \otimes
	    \begin{bmatrix} 1 & 0 \\ 0 & 1 \end{bmatrix}
\eeq
The system is decomposable into two subblocks -- a forward-propagating system and a reverse-propagating system.  Both subblocks act as symmetric beamsplitters:
\beq
{\rm BS}_{\rm sym}(r)_F \boxplus {\rm BS}_{\rm sym}(r)_R
\eeq
In real form, the scattering matrix is:
\beq
	S = \begin{bmatrix} \cos\alpha & 0 & -\sin\alpha & 0 \\
	                    0 & \cos\alpha & 0 & \sin\alpha \\
	                    \sin\alpha & 0 & \cos\alpha & 0 \\
	                    0 & -\sin\alpha & 0 & \cos\alpha \end{bmatrix} =
	    \begin{bmatrix} \cos\alpha & -\sin\alpha \\ \sin\alpha & \cos\alpha \end{bmatrix} \otimes 
	    \begin{bmatrix} 1 & 0 \\ 0 & 0 \end{bmatrix} + 
	    \begin{bmatrix} \cos\alpha & \sin\alpha \\ -\sin\alpha & \cos\alpha \end{bmatrix} \otimes 
	    \begin{bmatrix} 0 & 0 \\ 0 & 1 \end{bmatrix}
\eeq
This is also decomposable into subblocks, but the blocks are not equivalent.  The reverse-propagating block has an opposite reflection coefficient:
\beq
	BS_{\rm re}(r)_F \boxplus BS_{\rm re}(-r)_R
\eeq

\subsubsection{Ring Cavity}

\begin{figure}[bp]
\begin{center}
\includegraphics[width=0.3\textwidth]{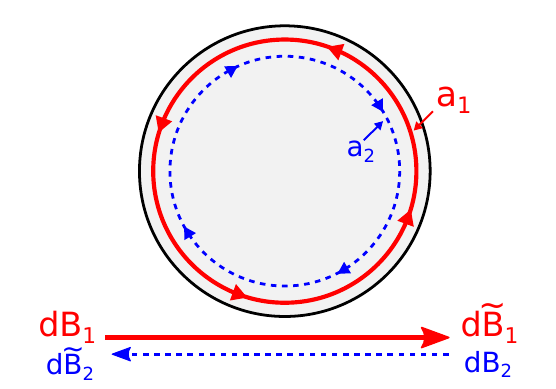}
\caption{Optical ring cavity with back-reflection.}
\label{fig:02-f8}
\end{center}
\end{figure}

Consider an oscillator with two modes, $a = (a_1, a_2)$ (Fig.~\ref{fig:02-f8}).  Reciprocity tells us that there exists a unitary $U$ such that, switching the directions of all the fields, the dynamics remain the same if we send $a \rightarrow U a$ (see above).  If the oscillator is symmetric, there exists a basis in which this $U$ just permutes the two modes, which are identified as the clockwise and counterclockwise modes of the oscillator:
\beq
U = \begin{bmatrix} 0 & -1 \\ -1 & 0 \end{bmatrix}
\eeq
The most general SLH model (up to a phase) that satisfies the reciprocity relations has the following canonical form:
\beq
	G = \left(S, \Lambda a, a^\dagger R a \right)
\eeq
with
\bea
	S & = & \begin{bmatrix} 0 & 1 \\ 1 & 0 \end{bmatrix} R(\theta) \\
	R & = & \begin{bmatrix} \Delta & \Delta_{XC}^\ast \\ \Delta_{XC} & \Delta \end{bmatrix} \\
	\Lambda & = & R(\theta/2)\begin{bmatrix} \alpha & \beta \\ \beta^* & \alpha^* \end{bmatrix}
\eea
where $R(\xi)$ is the symmetrized beamsplitter matrix:
\beq
	R(\xi) = \begin{bmatrix} \cos(\xi) & i\sin(\xi) \\ i\sin(\xi) & \cos(\xi) \end{bmatrix}
\eeq
This is ``canonical form'', where \texttt{In1} is identified with \texttt{Out1}, etc.  In standard form, \texttt{In1} is identified with \texttt{Out1}, etc.  This just involves permuting the input ports: $G' = G \triangleleft P_{\rm in}$, which alters the $S$ matrix and nothing else:
\beq
	S = R(\theta)
\eeq
The ABCD matrices are now easily computed:
\bea
	A & = & -i R - \frac{1}{2} \Lambda^\dagger \Lambda \nonumber \\
	& = & \begin{bmatrix} -i\Delta - \frac{1}{2}(\alpha^\ast\alpha + \beta\beta^\ast) & -i\Delta_{XC} - \alpha^\ast\beta \\ -i\Delta_{XC} - \alpha\beta^* & -i\Delta - \frac{1}{2}(\alpha^\ast\alpha + \beta\beta^\ast) \end{bmatrix} \\
	B & = & -\Lambda^\dagger S = -\begin{bmatrix} \alpha^* & \beta \\ \beta^* & \alpha \end{bmatrix} R(\theta/2) \\
	C & = & \Lambda = R(\theta/2) \begin{bmatrix} \alpha & \beta \\ \beta^* & \alpha^* \end{bmatrix} \\
	D & = & S = R(\theta)
\eea

\ifstandalone{\bibliography{NoteRefs}{}
\bibliographystyle{alpha}
}
\ifdefined\multidoc\else\input{Header}\fi

\ifstandalone{\setcounter{chapter}{2}}

\chapter{Linear and Linearized Systems}
\label{ch:04b}

Linear dynamical systems are the simplest, most common, and arguably most important systems in physics and engineering.  Entire branches of physics, from cosmology to the Standard Model, are based on linear or nearly-linear systems.  The ubiquity of linear systems results from the fact that most things in physics are only {\it weakly-coupled} to each other.  In this weak-coupling regime, we can distinguish the principal from the peripheral, the dominant from the perturbation, the fundamental from the emergent; and reductionism -- that is, science -- becomes possible.

This chapter discusses linear and linearized models in the open quantum systems framework.  First, I define the doubled-up ABCD model, a useful way to express linear open quantum systems.  The QSDEs are then calculated.  Because the QSDEs are linear, the system state is always Gaussian, and can be obtained analytically.

First, I obtain equations of motion for the moments of the Gaussian state in a linear quantum system.  The centroid of the Gaussian obeys a mean-field equation, while the covariance matrix satisfies a Lyapunov equation.  These can be used to obtain the internal field at steady state.

Next, I compute the quantum transfer function between input and output fields, and relate it to the input-output squeezing spectrum.  Optical squeezing is a particularly important field, given its potential for sensing and quantum information applications, I derive formulas the free-field squeezing from an arbitrary linear model.

Several systems are considered: the linear cavity, the degenerate OPO, and the linearized Kerr cavity.  With the methods developed in this chapter, we can study these systems in more detail than in Chapter~\ref{ch:02}.

The content of this chapter is an amalgam of old results, results with updated notation, derivations, and a few of my own ideas.  The new results are in the next chapter, which relies on the theory developed here.

\section{Basic Theory}

A bosonic linear system is defined as any open quantum system with bosonic fields, a quadratic Hamiltonian $H$, linear couplings $L$, and a constant scattering matrix $S$.  Recalling Eq.~\ref{eq:01-slh-fuller}, this implies that the full Hamiltonian, system and bath combined, is quadratic.  The fields must be bosonic so that the operator QSDEs Eqs.~(\ref{eq:01-qsde-1}-\ref{eq:01-qsde-3}) are linear (as a contrapositive consider the atom-cavity in Sec.~(\ref{sec:02-atom}); the Hamiltonian is quadratic but the equations of motion are not linear).  Note that bosonic does not necessarily mean optical.  Mechanical excitations \cite{Aspelmeyer2014}, exciton-polaritons \cite{Deng2010} and plasmons \cite{HaugBook} are also linear (in some limit) and bosonic.

Non-bosonic systems will not be treated in this chapter.  However, it is possible to analyze them with phase-space methods using a Wigner or positive-P distribution, and the dynamical equations can then be linearized.  This will be covered in Chapter \ref{ch:04}.  The resulting equations are very similar to those derived here.

\subsection{SLH Model}
\label{sec:04b-linslh}

Let $x = (a_1, \ldots, a_n)$ be the state vector for the system, where $a_i$ is the annihilation operator for mode $i$.  The most general SLH model for a linear system takes the form:
\beq
	\left(S,\ \ \ \Lambda_- x + \Lambda_+ x^\dagger + \lambda,\ \ \ x^\dagger R_- x + \frac{x^T R_+^* x + x^\dagger R_+ x^*}{2} + r^T x + r^\dagger \bar{x}
	\right)
\eeq
While the $\Lambda_\pm$ matrices can be arbitrary, $R_-$ must be Hermitian and $R_+$ must be symmetric.  It is most convenient to express this model in \textit{doubled-up notation} \cite{Gough2009c, Gough2010b}.  This notation combines the $x$ and $x^\dagger$ into a single vector.  Similarly, $\d B$ and $\d B^\dagger$ are doubled-up:
\beq
	\dbl{x} = \left(a_1,\ \ldots,\ a_n,\ a_1^\dagger,\ \ldots,\ a_n^\dagger\right)^T,\ \ \ 
	\d\dbl{B} = \left(\d B_1,\ \ldots,\ \d B_m,\ \d B_1^\dagger,\ \ldots,\ \d B_m^\dagger\right)^T \label{eq:04b-xb}
\eeq
The doubled-up state vector satisfies the commutation relations $[\dbl{x}_i, \dbl{x}_j] = (\Theta_{a})_{ij}$; likewise, the external modes satisfy $[\d\dbl{B}_i, \d\dbl{B}_j] = (\Theta_{\d B})_{ij}$.  Here, $\Theta_a = i\sigma_y\otimes I_n$ and $\Theta_{\d B} = i\sigma_y \otimes I_m$ have the block form:
\beq
	\Theta_a = \begin{bmatrix} 0 & I_{n\times n} \\ -I_{n\times n} & 0 \end{bmatrix},\ \ \ 
	\Theta_{\d B} = \begin{bmatrix} 0 & I_{m\times m} \\ -I_{m\times m} & 0 \end{bmatrix}
\eeq
where $m$ and $n$ are the number of ports and input-output modes, respectively.  Also important are the conjugation matrices $J$: $\dbl{a}^\dagger = J_a \dbl{a}$, $\d\dbl{B}^\dagger = J_{\d B} \d\dbl{B}$; these take the form
\beq
	J_a = \begin{bmatrix} 0 & I_{n\times n} \\ I_{n\times n} & 0 \end{bmatrix},\ \ \ 
	J_{\d B} = \begin{bmatrix} 0 & I_{m\times m} \\ I_{m\times m} & 0 \end{bmatrix}
\eeq
From now on, we suppress the subscripts on $J$ and $\Theta$, since they have the same form and picking the right one is obvious because of their different sizes.  For vectors and matrices, let $(A^*)_{ij} = (A_{ij}^\dagger)$ represent element-wise conjugation while $(A^\dagger)_{ij} = (A_{ji}^*)$ represents the conjugate transpose.  It is important to distinguish these because they will both play a role in what follows.

With the doubled-up state vector $\dbl{x}$, we rewrite the SLH model in terms of $\dbl{\Lambda}$ and $\dbl{R}$ as follows:
\beq
	\biggl(S,\ \ \ \underbrace{\begin{bmatrix} \Lambda_- & \Lambda_+ \end{bmatrix}}_{\dbl{\Lambda}} \dbl{x} + \lambda,\ \ \ \frac{1}{2}\dbl{x}^T \underbrace{\begin{bmatrix} R_+^* & R_-^* \\ R_- & R_+ \end{bmatrix}}_{\dbl{R}} \dbl{x} + {\underbrace{\begin{bmatrix} r \\ r^* \end{bmatrix}}_{\dbl{r}}}^T \dbl{x}
	\biggr) \label{eq:04b-slh}
\eeq

\subsection{ABCD Notation}

The dynamics of the quantum model (\ref{eq:01-qsde-1}) come from its operator QSDEs.  For a linear system, the QSDEs are:
\bea
	\d\dbl{x}_i & = & \left[-i[\dbl{x}_i, H] + \frac{1}{2}\left(L_m^\dagger[\dbl{x}_i, L_m] + [L_m^\dagger, \dbl{x}_i]L_m\right)\right]\d t \nonumber \\
	& & +\ \d B_m^\dagger S_{mn}^\dagger[\dbl{x}_i, L_n] + [L_n^\dagger, \dbl{x}_i]S_{nm}\d B_m + \left(S_{mp}^\dagger \dbl{x}_i S_{pn} - \dbl{x}_i\delta_{mn}\right)\d\Lambda_{mn} \nonumber \\
	& = & \left[\Theta \left(-i\dbl{R} + \frac{\dbl{\Lambda}^T\dbl{\Lambda}^* J - J \dbl{\Lambda}^\dagger\dbl{\Lambda}}{2}\right) \dbl{x} + \Theta\left(-i\dbl{r} + \frac{\dbl{\Lambda}^T\lambda^* - J\dbl{\Lambda}^\dagger \lambda}{2}\right) \right]_i \d t \nonumber \\
	& & + \left[\Theta \begin{bmatrix} -J\dbl{\Lambda}^\dagger S & \dbl{\Lambda}^T S^* \end{bmatrix} \d\dbl{B}_{\rm in}\right]_i \\
	\d B_{{\rm out}, m} & = & L_m\,\d t + S_{mn} \d B_{{\rm in}, n} \nonumber \\
	& = & \left(\dbl{\Lambda} \dbl{x} + \lambda\right)_m \d t + (S\,\d B_{\rm in})_m
\eea
These equations let us recast the model (\ref{eq:04b-slh}) as a linear dynamical system, albeit with a doubled-up, operator-valued state.
\bea
	\d\bar{x} & = & \biggl[\underbrace{\Theta \left(-i\dbl{R} + \frac{\dbl{\Lambda}^T\dbl{\Lambda}^* J - J \dbl{\Lambda}^\dagger\dbl{\Lambda}}{2}\right)}_{\dbl{A}} \dbl{x}
	+ \underbrace{\Theta \left(-i\dbl{r} + \frac{\dbl{\Lambda}^T\lambda^* - J\dbl{\Lambda}^\dagger \lambda}{2}\right)}_{\dbl{a}}
	\biggr]\d t \nonumber \\
	& & + \underbrace{\Theta \begin{bmatrix} -J\dbl{\Lambda}^\dagger S & \dbl{\Lambda}^T S^* \end{bmatrix}}_{\dbl{B}} \d\dbl{B}_{\rm in} \label{eq:04b-abcd-1} \\
	\d\dbl{B}_{\rm out} & = & \biggl(\underbrace{\begin{bmatrix} \dbl{\Lambda} \\ \dbl{\Lambda}^*J \end{bmatrix}}_{\dbl{C}}\dbl{x}
	+ \underbrace{\begin{bmatrix}\lambda \\ \lambda^*\end{bmatrix}}_{\dbl{c}}
	\biggr)\d t + \underbrace{\begin{bmatrix} S & 0 \\ 0 & S^* \end{bmatrix}}_{\dbl{D}} \d\dbl{B}_{\rm in} \label{eq:04b-abcd-2}
\eea
This is the {\bf ABCD model} of a linear quantum system.  In terms of the SLH components, the ABCD matrices are:
\begin{align}
	\dbl{A} & = \begin{bmatrix} -iR_- & -iR_+ \\ iR_+^* & iR_-^* \end{bmatrix} + 
		\frac{1}{2}\begin{bmatrix} -\Lambda_-^\dagger\Lambda_- + \Lambda_+^T\Lambda_+^* &
		-\Lambda_-^\dagger\Lambda_+ + \Lambda_+^T\Lambda_-^* \\
		(-\Lambda_-^\dagger\Lambda_+ + \Lambda_+^T\Lambda_-^*)^* & 
		(-\Lambda_-^\dagger\Lambda_- + \Lambda_+^T\Lambda_+^*)^* \end{bmatrix} \nonumber \\
	\dbl{B} & = \begin{bmatrix} -\Lambda_-^\dagger S & \Lambda_+^T S^* \\
		\Lambda_+^\dagger S & -\Lambda_-^T S^* \end{bmatrix},\ \ \ 
	\dbl{C} = \begin{bmatrix} \Lambda_- & \Lambda_+ \\ \Lambda_+^* & \Lambda_-^* \end{bmatrix},\ \ \ 
	\dbl{D} = \begin{bmatrix} S & 0 \\ 0 & S^* \end{bmatrix} \nonumber \\
	\dbl{a} & = \begin{bmatrix} -i r^* + \tfrac{1}{2}(-\Lambda_-^\dagger \lambda + \Lambda_+^T \lambda^*) \\
		(-i r^* + \tfrac{1}{2}(-\Lambda_-^\dagger \lambda + \Lambda_+^T \lambda^*))^* \end{bmatrix},\ \ \ 
	\dbl{c} = \begin{bmatrix} \lambda \\ \lambda^* \end{bmatrix} \label{eq:04-abcd}
\end{align}
Each of these matrices takes the following, doubled-up form:
\beq
	M = \begin{bmatrix} M_{-} & M_{+} \\ M_{+}^* & M_{-}^* \end{bmatrix} \label{eq:04b-abcd-conj}
\eeq
This is necessary because $\dbl{x}$ and $\d\dbl{B}$ consist of conjugate components.  If the top of the vector evolves one way, the lower part must evolve in the conjugate manner.  This forces all ABCD matrices to have the form (\ref{eq:04b-abcd-conj}).

To verify that this is a valid quantum model, one can check that it satisfies the {\it physical realizability conditions}.  These conditions arise from the fact that the commutators $[x_i, x_j^\dagger] = \delta_{ij}$ do not change in time.  However, the operators themselves evolve according to (\ref{eq:04-abcd}), maintaining the commutation relations places some constraints on $A, B, C, D$ \cite{James2008, Nurdin2009}:
\bea
	\d[\dbl{x}_i,\dbl{x}_j] = 0 & \Rightarrow & \dbl{A}\Theta + \Theta \dbl{A}^T + \dbl{B}\Theta \dbl{B}^T = 0 \label{eq:04b-re1} \\
	\d[\dbl{x}_i, \d\dbl{B}_m] = 0 & \Rightarrow & \Theta \dbl{C}^T = -\dbl{B} \Theta \dbl{D}^T \label{eq:04b-re2} \\
	\d[\d\dbl{B}_m, \d\dbl{B}_n] = 0 & \Rightarrow & \dbl{D}\Theta \dbl{D}^T = \Theta \label{eq:04b-re3}
\eea
Condition (\ref{eq:04b-re3}) is similar to the unitarity condition for scattering matrices.  It is a little more general, though, since it allows arbitrary Bogoliubov transformations, e.g.\ $\d B_{\rm out} = \d B_{\rm in}\cosh(\eta) + \d B_{\rm in}^\dagger\sinh(\eta)$.  Condition (\ref{eq:04b-re2}) relates the input matrix $B$ to the output matrix $C$; in a sense it says that the amount of information entering the system is the same as the amount leaving.  Condition (\ref{eq:04b-re1}) is the quantum analog of the fluctuation-dissipation theorem, relating the system loss to its coupling to the environment.  This arises because, in the absence of extra vacuum fluctuations, in a lossy system the commutator $[x_i, x_j]$ would decay to zero.  There can be no dissipation without a coupling to a bath.

Any open quantum oscillator satisfies the physical realizability conditions (\ref{eq:04b-re1}-\ref{eq:04b-re3}).  However, some solutions to (\ref{eq:04b-re1}-\ref{eq:04b-re3}) give infinite-bandwidth squeezing (Bogoliubov components) and thus do not admit an SLH representation.  In addition to (\ref{eq:04b-re1}-\ref{eq:04b-re3}), a linear input-output system must have block-diagonal $\dbl{D}$, with $S^\dagger S = 1$ in order to be realizable as an open quantum oscillator \cite{Nurdin2009b}.

\subsection{Gaussian Moment Equations}

Linear systems preserve the Gaussianity of states: if the internal state starts in a Gaussian, it remains Gaussian for all time.  A dissipative linear system will always tend to a Gaussian steady state.  A Gaussian state can be represented by its moments:
\beq
	\dbl{\mu}_i = \langle \dbl{x}_i \rangle,\ \ \ 
	\dbl{\sigma}_{ij} = \frac{1}{2}\langle{\dbl{x}_i \dbl{x}_j^\dagger + \dbl{x}_j^\dagger \dbl{x}_i \rangle}
\eeq
Applying the QSDE (\ref{eq:04b-abcd-1}) and the It\^{o} rule $\d B_i \d B_j^\dagger = \delta_{ij} \d t$ (from which $\tfrac{1}{2}(\d\dbl{B}_i \d\dbl{B}_j + \d\dbl{B}_j \d\dbl{B}_i) = \tfrac{1}{2} J_{ij} \d t$), we obtain equations of motion for the moments \cite{Nurdin2009}:
\bea
	\frac{\d\dbl{\mu}}{\d t} & = & \dbl{A}\dbl{\mu} \label{eq:04b-mom1} \\
	\frac{\d\dbl{\sigma}}{\d t} & = & \dbl{A}\dbl{\sigma} + \dbl{\sigma}\dbl{A}^\dagger  + \frac{1}{2}\dbl{B} \dbl{B}^\dagger \label{eq:04b-mom2}
\eea

\subsection{Circuit Algebra}
\label{sec:04b-algebra}

The circuit algebra discussed in Sec.~\ref{sec:01-circalg} can also be used to represent networks of linear quantum systems.  The rules derived in that section carry over to the linear case, but often it is more useful to have a set of rules that act directly on the ABCD matrices.  

\subsubsection{Concatenation}

The concatenation product $G_1 \boxplus G_2$ is the easiest.  Since there is no coupling between the components, one might expect the matrices to stack in a block-diagonal form:
\beq 
	\dbl{a}\ \mbox{``}\!=\!\mbox{''} \begin{bmatrix} \dbl{a}^{(1)} \\ \dbl{a}^{(2)} \end{bmatrix},\ \ \ 
	\d\dbl{B}\ \mbox{``}\!=\!\mbox{''} \begin{bmatrix} \d\dbl{B}^{(1)} \\ \d\dbl{B}^{(2)} \end{bmatrix},\ \ \
	\dbl{A}\ \mbox{``}\!=\!\mbox{''} \begin{bmatrix} \dbl{A}^{(1)} & 0 \\ 0 & \dbl{A}^{(2)} \end{bmatrix},\ \ \ 
	\dbl{B}, \dbl{C}, \dbl{D}
	 = \mbox{(likewise)} \label{eq:04b-concatwrong}
\eeq
I have put quotes around the equals signs because Eq.~(\ref{eq:04b-concatwrong}) is wrong.  The state vector $\dbl{a} \equiv (a^{(1)},\ (a^{(1)})^*,\ a^{(2)},\ (a^{(2)})^*)$ is not in doubled-up form with $a$ terms in the upper half and $a^*$ terms in the lower half.  The same is true for $\d\dbl{B}$.  To correct, this, we need to permute the rows of $\dbl{a}$ and $\d\dbl{B}$, and correspondingly permute the rows and columns of the ABCD matrices.  Define permutation matrices $P_a$, $P_{\d B}$ to perform this transformation.  The correct ABCD model for $G_1 \boxplus G_2$ is:
\begin{align} 
	\dbl{a} &= P_a\begin{bmatrix} \dbl{a}^{(1)} \\ \dbl{a}^{(2)} \end{bmatrix}, 
	&\dbl{A} &= P_a\begin{bmatrix} \dbl{A}^{(1)} & 0 \\ 0 & \dbl{A}^{(2)} \end{bmatrix}P_a^{-1},
	&\dbl{B} &= P_a\begin{bmatrix} \dbl{B}^{(1)} & 0 \\ 0 & \dbl{B}^{(2)} \end{bmatrix}P_{\d B}^{-1}, \nonumber \\ 
	\d\dbl{B} &= P_{\d B}\begin{bmatrix} \d\dbl{B}^{(1)} \\ \d\dbl{B}^{(2)} \end{bmatrix},
	&\dbl{C} &= P_{\d B}\begin{bmatrix} \dbl{C}^{(1)} & 0 \\ 0 & \dbl{C}^{(2)} \end{bmatrix}P_a^{-1},
	&\dbl{D} &= P_{\d B}\begin{bmatrix} \dbl{D}^{(1)} & 0 \\ 0 & \dbl{D}^{(2)} \end{bmatrix}P_{\d B}^{-1} \label{eq:04b-concat} 
\end{align}

\subsubsection{Series}

The series product $G_2 \triangleleft G_1$ is computed by taking the output fields of the first component, $G_1$, and applying them as inputs to the second one.  This is a straightforward cascaded linear dynamical system, which has the following ABCD model:
\begin{align} 
	\dbl{A} &= P_a\begin{bmatrix} \dbl{A}_1 & 0 \\ \dbl{B}_2 \dbl{C}_1 & \dbl{A}_2 \end{bmatrix}P_a,
	&\dbl{B} &= P_a\begin{bmatrix} \dbl{B}_1 & \dbl{B}_2 \dbl{D}_1 \end{bmatrix},\nonumber \\
	\dbl{C} &= \begin{bmatrix} \dbl{D}_2 \dbl{C}_1 \\ \dbl{C}_2 \end{bmatrix}P_{a},
    &\dbl{D} &= \dbl{D}_2 \dbl{D}_1 \label{eq:04b-series} 
\end{align}
This is very similar to the case for purely passive systems [cite some of the Gough, James, Nurdin, etc. papers on ABCD models].  The only difference lies in the extra permutations $P_a$, $P_{\d B}$ that put the internal and input-output fields into doubled-up form.

\subsubsection{Feedback}

The feedback operator $[G]_{k\rightarrow l}$ sends output $k$ into input $l$.  This is realized by:
\begin{align} 
	\dbl{A} &= \dbl{A} + \dbl{B}_{:,l}(1 - \dbl{D}_{kl})^{-1}\dbl{C}_{k,:} 
	&\dbl{B} &= \dbl{B}_{:,!l} + \dbl{B}_{:,l}(1 - \dbl{D}_{kl})^{-1}\dbl{D}_{k,!l} \nonumber \\
	\dbl{C} &= \dbl{C}_{!k,:} + \dbl{D}_{!k,l}(1 - \dbl{D}_{kl})^{-1}\dbl{C}_{k,:} 
	&\dbl{D} &= \dbl{D}_{!k,!l} + \dbl{D}_{!k,l}(1 - \dbl{D}_{kl})^{-1}\dbl{D}_{k,!l} \label{eq:04b-feedback}
\end{align}
Here, $:$ refers to all indices, $k$ refers to a pair of indices (the $a_k$ one as well as the $a_k^\dagger$ one), and $!k$ refers to all indices except that pair; and likewise for $l$, $!l$.  Thus, unlike in Sec.~\ref{sec:01-circalg}, the matrix we have to invert, $(1 - \dbl{D}_{kl})$, is 2-by-2, not 1-by-1.  

\subsubsection{Adiabatic Elimination}

Although it is not part of the Gough-James algebra, adiabatic elimination is very important in linear systems, and relatively straightforward to compute.  The simplest case involves eliminating {\it all} degrees of freedom from the system, turning a dynamic SLH model into a static scattering element.  This limit is valid for a very quickly-evolving system that is strongly coupled to the environment:
\beq 
	\dbl{A} \sim k^2,\ \ \ \dbl{a}, \dbl{B}, \dbl{C}
	 \sim k,\ \ \ \dbl{c}, \dbl{D} \sim 1,\ \ \ k \rightarrow \infty
\eeq
In this limit, the state tracks the input, and we may approximate: $\dbl{a} \rightarrow \dbl{A}^{-1}(\dbl{B}\,\d\dbl{B}/\d t + \dbl{a}
)$.  Substituting this and solving for $\d\dbl{B}_{\rm out}$, we find a static input-output relation:
\beq 
	\d\dbl{B}_{\rm out} = \left(\dbl{D} - \dbl{C} \dbl{A}^{-1} \dbl{B}\right) \d\dbl{B}_{\rm in} 
	+ (\dbl{c} - \dbl{C}\dbl{A}^{-1}\dbl{a}) \d t
	\label{eq:04b-ad-el}
\eeq
This will be useful because many of the linear components we want are static components -- squeezing and linear amplification being the most common.  The adiabatic elimination result (\ref{eq:04b-ad-el}) tells us the whether this is possible with a given model.

\subsection{Quadrature Notation}
\label{sec:04b-quad}

An alternative approach is to write $x$ and $\d B$ in quadrature notation (compare Eq.~(\ref{eq:04b-xb})):
\beq
	x \equiv \left(X_1, P_1,\ \ldots,\ X_n, P_n\right)^T,\ \ \ 
	\d a \equiv \left(\d B_{1,x}, \d B_{1,p}\ \ldots,\ \d B_{m,x}, \d B_{m,p}\right)^T
\eeq
where $X_i = (a_i + a_i^\dagger)$, $P_i = (a_i - a_i^\dagger)/i$ and likewise for the $\d a$.  The commutators are complex, so we write $[\dbl{x}_i, \dbl{x}_j] = 2i(\Theta_{x})_{ij}$; likewise, the external modes satisfy $[\d a_i, \d a_j] = 2i(\Theta_{\d a})_{ij}$, with $\Theta_x = i\sigma_y\otimes I_n$ and $\Theta_{\d a} = i\sigma_y \otimes I_m$ (compare Sec.~\ref{sec:04b-linslh}).  Since these have the same form, we suppress the subscripts in what follows.

The most general SLH model takes the form:
\beq
	\left(S,\ \ \Lambda x + \lambda,\ \ \frac{1}{2} x^T R x + r^T x\right) \label{eq:04b-slhquad}
\eeq
In this case, $S$ is unitary, $R$ is real symmetric, and $\Lambda$ is arbitrary.  Because of the complex commutators, the ABCD matrices take a slightly modified form:
\beq
\begin{array}{rclrcl}
	A & = & 2\Theta\left(R + \frac{1}{4}\tilde{\Lambda}^{\rm T} \Theta \tilde{\Lambda}\right),\ \ \  &
	B & = & \Theta \tilde{\Lambda}^{\rm T} \Theta \tilde{S}, \ \ \ \\
	C & = & \tilde{\Lambda}, &
	D & = & \tilde{S}, \\
	a & = & 2\Theta\left(r + \frac{1}{4}\tilde{\Lambda}^{\rm T} \Theta \tilde{\lambda}\right), &
	c & = & \tilde{\lambda} \label{eq:04b-slh-abcd}
\end{array}
\eeq
We form matrices $\tilde{S}$ and $\tilde{\Lambda}$, and vector $\tilde{\lambda}$ by stacking $S$, $\Lambda$, and $\lambda$, as follows:
\begin{eqnarray}
	\tilde{S}_{ab} & = & 2M^\dagger \left[\begin{array}{cc} S_{ab} & 0 \\ 0 & S_{ab}^* \end{array}\right] M \nonumber \\
	\tilde{\Lambda}_a & = & 2M^\dagger \left[\begin{array}{c} \Lambda_{a} \\ \Lambda_{a}^* \end{array}\right],\ \ \ \tilde{\lambda}_a = 2M^\dagger \left[\begin{array}{c} \lambda_{a} \\ \lambda_{a}^* \end{array}\right]
\end{eqnarray}
where $J_{2n\times 2n}$ is the canonical antisymmetric matrix of dimension $2n$ (written above as $J$, where the dimension is inferred), and $M_{2n\times 2n}$ is used to convert between standard $dA_i, dA_i^\dagger$ and Hermitian $da_x, da_p$ input-output fields:
\beq
	J_{2n\times 2n} = I_n \otimes \left[\begin{array}{cc} 0 & 1 \\ -1 & 0 \end{array}\right],\ \ \ 
	M_{2n\times 2n} = I_n \otimes \frac{1}{2}\left[\begin{array}{cc} 1 & i \\ 1 & -i\end{array}\right]
\eeq
The matrix $\tilde{S}$ is made from the blocks $\tilde{S}_{ab}$ above, and likewise for $\tilde{\Lambda}$ and $\tilde{\lambda}$.

These match the formulas used in \cite{James2008}, the difference being that we have defined the ABCD matrices in terms of the real stacked matrices $\tilde{S}, \tilde{\Lambda}$, rather than in terms of $S$, $L$, and $H$ directly.

The moment equations (\ref{eq:04b-mom1}-\ref{eq:04b-mom2}) change to:

\beq
	\frac{\d\dbl{\mu}}{\d t} = \dbl{A}\dbl{\mu},\ \ \ 
	\frac{\d\dbl{\sigma}}{\d t} = \dbl{A}\dbl{\sigma} + \dbl{\sigma}\dbl{A}^\dagger  + \dbl{B} \dbl{B}^\dagger \label{eq:04b-mom3}
\eeq

Since all the operators are Hermitian, the ABCD matrices are all real (and no longer have the doubled-up form).  Many numerical methods for matrix optimization only apply to real matrices, and in this context, the quadrature notation is the more convenient one to use.

\subsubsection{Circuit Algebra, SLH form}

The circuit algebra relations are a little different in quadrature notation.  Most of it is just index permutations, arising from the non-doubled-up nature of quadrature notation.  To form the concatenated system $G_1 \boxplus G_2$, we define a new state variable $x = [\begin{array}{cc}x_1 & x_2\end{array}]$ that includes both the state of $G_1$ and the state of $G_2$.  Applying the concatenation rule,
\begin{eqnarray}
	& & S = \left[\begin{array}{cc} S_1 & 0 \\ 0 & S_2 \end{array}\right],\ \ \ 
	\Lambda = \left[\begin{array}{cc} \Lambda_1 & 0 \\ 0 & \Lambda_2 \end{array}\right], \nonumber \\
	& & R = \left[\begin{array}{cc} R_1 \\ R_2 \end{array}\right],\ \ \ 
	\lambda = \left[\begin{array}{c} \lambda_1 \\ \lambda_2 \end{array}\right],\ \ \ 
	r = \left[\begin{array}{c} r_1 \\ r_2 \end{array}\right]
\end{eqnarray}
Likewise, applying the series product rule, one finds the parameters for the system $G_2 \triangleleft G_1$:
\begin{eqnarray}
	& & S = S_2 S_1,\ \ \ 
	\Lambda = \left[\begin{array}{cc} \Lambda_1 & \Lambda_2 \end{array}\right],\ \ \ 
	\lambda = \lambda_1 + \lambda_2,\nonumber \\
	& & R = \left[\begin{array}{cc} R_1 & \mbox{Im}(\Lambda_2^\dagger S_2 \Lambda_1) \\
		\mbox{Im}(\Lambda_2^\dagger S_2 \Lambda_1) & R_2 \end{array}\right], \nonumber \\
	& & r = \left[\begin{array}{cc} r_1 + \mbox{Im}(\lambda_2^\dagger S_2 \Lambda_1)^{\rm T} \\
		r_2 - \mbox{Im}(\lambda_2^\dagger S_2 \Lambda_1)^{\rm T}\end{array}\right]
\end{eqnarray}
Similarly, applying the feedback equations can give us the parameters for the system $G' = [G]_{i\rightarrow j}$:
\begin{eqnarray}
	S' & = & \left[S + S_{\ast j}(1-S_{ij}) S_{i\ast}\right]_{!i,!j} \nonumber \\
	\Lambda' & = & \left[\Lambda + S_{\ast j}(1 - S_{ij})^{-1} \Lambda_{i\ast}\right]_{!i,\ast} \nonumber \\
	R' & = & R + \mbox{Im}\left[\Lambda^\dagger S_{\ast j}(1 - S_{ij}) \Lambda - h.c.\right] \nonumber \\
	r' & = & r + \mbox{Im}\left[\Lambda^\dagger S_{\ast j}(1 - S_{ij}) \lambda + \Lambda^{\rm T} (1 - S_{ij}) S_{j\ast}^{\rm T} \lambda^*\right] \nonumber \\
	\lambda' & = & \left[\lambda + S_{\ast j}(1 - S_{ij})^{-1} \lambda_{i\ast}\right]_{!i,\ast}
\end{eqnarray}
where the notation $\ast$ means ``take all rows (columns) of the given matrix'', while $!j$ means ``take all rows (columns) except $j$''.  For example, $S_{\ast, j}$ would be the $j^{\rm th}$ column of $S$, while $M_{!i,!j}$ would be obtained by removing row $i$ and column $j$ from the matrix $M$.

\subsubsection{Circuit Algebra, ABCD form}

One can also use the ABCD form in quadrature notation.  Concatenating two models in ABCD form to create $G_1 \boxplus G_2$ is straightforward:
\begin{eqnarray}
	& & A = \left[\begin{array}{cc} A_1 & 0 \\ 0 & A_2 \end{array}\right],\ \ \ 
	B = \left[\begin{array}{cc} B_1 & 0 \\ 0 & B_2 \end{array}\right],\ \ \ \nonumber \\ 
	& & C = \left[\begin{array}{cc} C_1 & 0 \\ 0 & C_2 \end{array}\right],\ \ \ 
	D = \left[\begin{array}{cc} D_1 & 0 \\ 0 & D_2 \end{array}\right],\ \ \ \nonumber \\
	& & a = \left[\begin{array}{c} a_1 \\ a_2 \end{array}\right],\ \ \ 
	c = \left[\begin{array}{c} c_1 \\ c_2 \end{array}\right],\ \ \ 
\end{eqnarray}
One can arrive at the series product in ABCD form by first taking the series product in SLH form and then converting to the ABCD matrices.  The series product $G = G_2 \triangleleft G_1$ is:
\begin{eqnarray}
	& & A = \left[\begin{array}{cc} A_1 & B_2C_1 \\ 0 & A_2 \end{array}\right],\ \ \ 
	B = \left[\begin{array}{c} B_1 \\ B_2D_1 \end{array}\right], \nonumber \\
	& & C = \left[\begin{array}{cc} D_2C_1 & C_2 \end{array}\right],\ \ \ 
	D = D_2 D_1, \nonumber \\
	& & a = \left[\begin{array}{c} a_1 \\ a_2 + B_2 c_1 \end{array}\right],\ \ \ 
	c = c_2 + D_2 c_1
\end{eqnarray}
Likewise, the internal feedback $G' = [G]_{i\rightarrow j}$ is given by:
\begin{eqnarray}
	A' & = & A + B_{\ast j}(1-D_{ij})^{-1} C_{i\ast} \nonumber \\
	B' & = & \left[B + B_{\ast j}(1-D_{ij})^{-1} D_{i\ast}\right]_{\ast,!j} \nonumber \\
	C' & = & \left[C + D_{\ast j}(1-D_{ij})^{-1} C_{i\ast}\right]_{!i,\ast} \nonumber \\
	D' & = & \left[D + D_{\ast j}(1-D_{ij})^{-1} D_{i\ast}\right]_{!i,!j} \nonumber \\
	a' & = & a + B_{\ast j}(1-D_{ij})^{-1} c_{i} \nonumber \\
	c' & = & \left[c + D_{\ast j}(1-D_{ij})^{-1} c_{i}\right]_{!i,\ast}
\end{eqnarray}
Note that there are no permutation matrices here, in contrast to Sec.~\ref{sec:04b-algebra}, since the matrices are not in doubled-up form.

\section{Examples}

Passive systems (empty cavities), OPOs, and some optomechanical devices are genuine, bosonic linear quantum systems, in the sense that the degrees of freedom are all bosonic and the Hamiltonian (at least in the appropriate limit) is quadratic.

\subsection{Passive Systems}
\label{sec:04b-passive}

A {\it passive linear system} is defined as any system that does not create or destroy photons, but merely pushes them around \cite{Nurdin2010, Nurdin2010b}.  Any network of empty cavities qualifies.  Since photons within the system are conserved, the Hamiltonian must take the form $R_{ij}a_i^\dagger a_j$, and since photons are conserved between the cavity and the bath, the $L$ operators must look like $\Lambda_{mj} a_j$.  One finds the same model introduced in Sec.~\ref{sec:02-genlim} (ignoring the inhomogeneous terms $r, \lambda$, which can be eliminated by a change of variables).

Given these constraints, the SLH model for this system can be expressed in doubled-up notation, as follows:
\beq
	\biggl(S,\ \ \ \underbrace{\begin{bmatrix} \Lambda & 0 \end{bmatrix}}_{\dbl{\Lambda}} \dbl{a},\ \ \ \frac{1}{2}\dbl{a}^T \underbrace{\begin{bmatrix} 0 & R^* \\ R & 0 \end{bmatrix}}_{\dbl{R}} \dbl{a} \biggr)
\eeq
This gives the following ABCD matrices:
\begin{align}
	\dbl{A} & = \begin{bmatrix} -iR - \frac{1}{2}\Lambda^\dagger\Lambda & 0 \\ 0 & -iR^* - \frac{1}{2}\Lambda^T\Lambda^* \end{bmatrix} & 
	\dbl{B} & = \begin{bmatrix} -\Lambda^\dagger S & 0 \\ 0 & -\Lambda^T S^* \end{bmatrix} \nonumber \\
	\dbl{C} & = \begin{bmatrix} \Lambda & 0 \\ 0 & \Lambda^* \end{bmatrix} & 
	\dbl{D} & = \begin{bmatrix} S & 0 \\ 0 & S^* \end{bmatrix} \label{eq:04b-passive-abcd}
\end{align}
Because the system is passive, the ABCD matrices factor.  In this case, the doubled-up notation was superfluous; instead of evolving $a, a^\dagger$ together, we only need to keep track of $a$ because the two do not mix.  One finds the following QSDEs:
\bea
	\d a & = & \left(-iR - \frac{1}{2}\Lambda^\dagger\Lambda\right)a\,dt + \left(-\Lambda^\dagger S\right)\d B_{\rm in} \\
	\d B_{\rm out} & = & \left(\Lambda\right)a\,dt + \left(S\right)\d B_{\rm in}
\eea
These are equivalent to the QSDEs derived in Sec.~\ref{sec:02-genlim}.  While these are obviously much simpler than the general model (\ref{eq:04-abcd}), they only work when the system is passive.  Passive systems are generally limited in their usefulness.  While they can perform some useful operations, like filtering, they do not exhibit gain and therefore cannot amplify signals.  They are also very classical in nature.  For a given a coherent input, a passive system always outputs light in a coherent state, and the internal state is always coherent.  To create interesting quantum states of light, one needs to go to active systems.

\subsection{Non-degenerate OPO}

The simplest active linear system is the optical parametric oscillator, discussed in Sec.~\ref{sec:02-opo}.  In full generality, this is a nonlinear system, but if one assumes a strong pump field, it linearizes and reduces to a cavity with an extra pump term.  Depending on the phase matching, the pump can split into a two photons of equal (degenerate) or unequal (non-degenerate) frequency.
\begin{figure}[tb]
	\centering
	\includegraphics[width=0.50\textwidth]{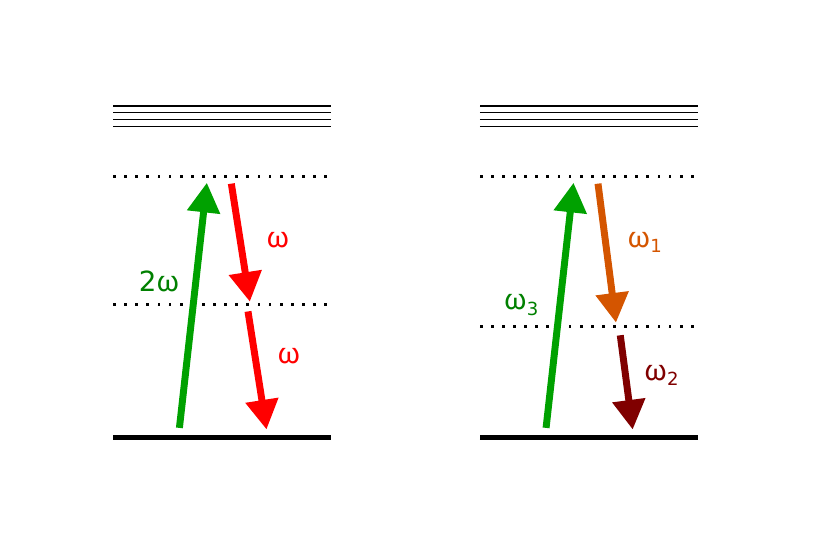}
	\caption{Left: Energy conservation for degenerate OPO pump process.  Right: Non-degenerate pump process.}
	\label{fig:04b-f1}
\end{figure}
Taking the non-degenerate SLH model from Sec.~\ref{sec:02-opo} and replacing the pump field $c$ with a constant, we arrive at the linearized OPO:
\beq
	G = \left(1,\ \ \ \begin{bmatrix} \sqrt{\kappa_a} a \\ \sqrt{\kappa_b} b \end{bmatrix},\ \ \ 
	\Delta_a a^\dagger a + \Delta_b b^\dagger b + \frac{\epsilon^* a b - \epsilon a^\dagger b^\dagger}{2i}\right)
\eeq
Define the doubled-up state vector $\dbl{a} = (a, b, a^\dagger, b^\dagger)$.  In terms of it, the doubled-up $\Lambda$ and $R$ are:
\beq
	\dbl{\Lambda} = \begin{bmatrix} \sqrt{\kappa_a} & 0 & 0 & 0 \\ 0 & \sqrt{\kappa_b} & 0 & 0 \end{bmatrix},\ \ \ 
	\dbl{R} = \begin{bmatrix} 0 & \frac{1}{2i}\epsilon^* & \Delta_a & 0 \\ \frac{1}{2i}\epsilon^* & 0 & 0 & \Delta_b \\
		\Delta_a & 0 & 0 & -\frac{1}{2i}\epsilon \\ 0 & \Delta_b & -\frac{1}{2i}\epsilon & 0 \end{bmatrix}
\eeq
From equations (\ref{eq:04b-abcd-1}-\ref{eq:04b-abcd-2}) we can compute the ABCD model:
\begin{align}
	\d\dbl{a} & = \begin{bmatrix} -i\Delta_a - \frac{1}{2}\kappa_a\!\! & 0 & 0 & \frac{1}{2}\epsilon \\
	0 & \!\!-i\Delta_b - \frac{1}{2}\kappa_b\!\! & \frac{1}{2}\epsilon & 0 \\
	0 & \frac{1}{2}\epsilon^* & \!\!i\Delta_a - \frac{1}{2}\kappa_a\!\! & 0 \\
	\frac{1}{2}\epsilon^* & 0 & 0 & \!\!i\Delta_b - \frac{1}{2}\kappa_b \end{bmatrix}\bar{a}\,dt - 
	\begin{bmatrix} \sqrt{\kappa_a}\!\! & 0 & 0 & 0 \\ 0 & \!\!\sqrt{\kappa_b}\!\! & 0 & 0 \\ 0 & 0 & \!\!\sqrt{\kappa_a}\!\! & 0 \\ 0 & 0 & 0 & \!\!\sqrt{\kappa_b} \end{bmatrix} \d\dbl{B}_{\rm in} \\
	\d\dbl{B}_{\rm out} & = \begin{bmatrix} \sqrt{\kappa_a}\!\! & 0 & 0 & 0 \\ 0 & \!\!\sqrt{\kappa_b}\!\! & 0 & 0 \\ 0 & 0 & \!\!\sqrt{\kappa_a}\!\! & 0 \\ 0 & 0 & 0 & \!\!\sqrt{\kappa_b} \end{bmatrix} \dbl{a}\,dt + 
		\begin{bmatrix} 1&0&0&0\\0&1&0&0\\0&0&1&0\\0&0&0&1 \end{bmatrix} \d\dbl{B}_{\rm in}
\end{align}
In the broad-band limit, where $\kappa_a, \kappa_b \rightarrow \infty$, we can apply the adiabatic elimination result (\ref{eq:04b-ad-el}) to get a static input-output model:
\beq
	\d\dbl{B}_{\rm out} = -\begin{bmatrix} \cosh\eta & 0 & 0 & e^{i\phi} \sinh\eta \\ 0 & \cosh\eta & e^{i\phi}\sinh\eta & 0 \\ 0 & e^{-i\phi}\sinh\eta & \cosh\eta & 0 \\ e^{-i\phi}\sinh\eta & 0 & 0 & \cosh\eta \end{bmatrix} \d\dbl{B}_{\rm in}
\eeq
This is a perfect two-mode squeezer:
\begin{align}
	-\d B_{{\rm out},1} &= (\cosh\eta)\d B_{{\rm in},1} + (e^{i\phi}\sinh\eta)\d B_{{\rm in},2}^*,
	&-\d B_{{\rm out},2} &= (\cosh\eta)\d B_{{\rm in},2} + (e^{i\phi}\sinh\eta)\d B_{{\rm in},1}^*
\end{align}
with is the squeezing parameter $\eta$ that diverges when the device is pumped to threshold, $|\epsilon|^2 \rightarrow \kappa_a\kappa_b$:
\beq
	\eta = \log\left[\frac{\sqrt{\kappa_a\kappa_b} + |\epsilon|}{\sqrt{\kappa_a\kappa_b} - |\epsilon|}\right],\ \ \ 
	\phi = \mbox{arg}(\epsilon)
\eeq

\subsection{Degenerate OPO}
\label{sec:04b-dopo}

The degenerate OPO only has a single mode if we consider the pump as a classical field.  Its SLH model is given by:

\beq
	G = \left(1,\ \ \ \begin{bmatrix} \sqrt{\kappa}a \end{bmatrix},\ \ \ \Delta a^\dagger a + \frac{\epsilon^* a^2 + \epsilon(a^\dagger)^2}{2i}\right)
\eeq

In terms of the doubled-up state vector $\dbl{a} = (a, a^\dagger)$, $\Lambda$ and $R$ are:

\beq
	\dbl{\Lambda} = \begin{bmatrix} \sqrt{\kappa} & 0 \end{bmatrix},\ \ \ 
	\dbl{R} = \begin{bmatrix} -i\epsilon^* & \Delta \\ \Delta & i\epsilon \end{bmatrix}
\eeq

This gives the ABCD model:

\bea
	\d\dbl{a} & = & \begin{bmatrix} -i\Delta - \frac{1}{2}\kappa & \epsilon \\ \epsilon^* & i\Delta - \frac{1}{2}\kappa \end{bmatrix} \dbl{a}\,dt - \begin{bmatrix} \sqrt{\kappa} & 0 \\ 0 & \sqrt{\kappa} \end{bmatrix}\d\dbl{B}_{\rm in} \\
	\d\dbl{B}_{\rm out} & = & \begin{bmatrix} \sqrt{\kappa} & 0 \\ 0 & \sqrt{\kappa}\end{bmatrix} \dbl{a}\,dt + \begin{bmatrix} 1 & 0 \\ 0 & 1 \end{bmatrix} \d\dbl{B}_{\rm in}
\eea

which is equivalent to the familiar OPO equation of motion

\beq
	da = \left[\left(-i\Delta - \frac{1}{2}\kappa\right)a + \epsilon\,a^\dagger\right]dt - \sqrt{\kappa}\,\d B_{\rm in}
\eeq

As with the non-degenerate OPO, the adiabatic limit is very important in the degenerate case.  Applying Eq.~(\ref{eq:04b-ad-el}) we find that:

\beq
	\d\dbl{B}_{\rm out} = -\begin{bmatrix} \cosh\eta & e^{i\phi}\sinh\eta \\ e^{-i\phi}\sinh\eta & \cosh\eta \end{bmatrix} d\dbl{B}_{\rm in}
\eeq

which is a perfect single-mode squeezer:

\beq
	-\d B_{\rm out} = (\cosh\eta)\d B_{\rm in} + (e^{i\phi}\sinh\eta)\d B_{\rm in}^*,\ \ \ \eta = \log\left[\frac{\kappa + 2|\epsilon|}{\kappa - 2|\epsilon|}\right],\ \ \ \phi = \mbox{arg}(\epsilon)
\eeq

Both degenerate and non-degenerate OPOs produce ideally squeezed light, but in the degenerate case the mode is squeezed with itself, rather than being squeezed with another mode.  This single-mode squeezing gives an output that is less noisy than the vacuum along certain quadratures.  This is very useful for sensing applications where the accuracy of a measurement is limited by photon shot noise.

\section{The Internal State}
\label{sec:04b-intstate}

Any stable linear system driven by Gaussian noise will tend to a Gaussian equilibrium state.  Because quantum linear systems can be described by Gaussian processes, they are no exception.  Recall that the doubled-up covariance matrix is defined by $\dbl{\sigma}_{ij} = \tfrac{1}{2}\langle \dbl{a}_i \dbl{a}_j^\dagger + \dbl{a}_j^\dagger \dbl{a}_i \rangle$.  As we saw in Eq.~(\ref{eq:04b-mom2}), this evolves as $d\dbl{\sigma}/dt = \dbl{A}\dbl{\sigma} + \dbl{\sigma}\dbl{A}^\dagger + \dbl{B}\dbl{B}^\dagger$.  Over time, this tends to a steady state with $\dbl{\sigma}$ given by the {\it Lyapunov Equation}:
\beq 
	\boxed{\dbl{A}\dbl{\sigma} + \dbl{\sigma}\dbl{A}^\dagger + \frac{1}{2}\dbl{B}\dbl{B}^\dagger = 0 
	}\label{eq:04b-lyapunov}
\eeq
For a single internal field, $\dbl{\sigma}$ takes the following form:
\beq
	\sigma = \begin{bmatrix} \langle a^\dagger a \rangle + \frac{1}{2} & \langle a^2 \rangle \\ \langle a^2\rangle^* & \langle a^\dagger a \rangle + \frac{1}{2} \end{bmatrix}
\eeq
For a coherent state, this is obviously half the identity: $\dbl{\sigma} = \frac{1}{2}I$.  Squeezed states will have off-diagonal terms in $\dbl{\sigma}$.  The Heisenberg uncertainty relation is a condition on the determinant: $\det(\dbl{\sigma}) \geq \frac{1}{4}$, equality holding for the pure states.  (For multiple fields, this determinant condition is necessary but not sufficient).

\subsubsection{Passive Systems}

In a passive system, the field is always in a coherent state.  This can be seen from the Lyapunov equation.  The ABCD model derived in Sec.~(\ref{sec:04b-passive}) is block-diagonal; from this we can infer that $\dbl{\sigma}$ must take the block-diagonal form:
\beq
	\dbl{\sigma} = \begin{bmatrix} \sigma & 0 \\ 0 & \sigma^* \end{bmatrix}
\eeq
The Lyapunov equation, applying the substitutions from the ABCD model (\ref{eq:04b-passive-abcd}), gives:
\beq
	\left(-iR - \frac{1}{2}\Lambda^\dagger\Lambda\right)\sigma + \sigma\left(-iR - \frac{1}{2}\Lambda^\dagger\Lambda\right) + \frac{1}{2}\Lambda^\dagger\Lambda = 0
\eeq
It is obvious from inspection that $\sigma = \frac{1}{2}I$ is the correct solution.  This is the covariance matrix of a vacuum state, proving that all passive linear systems driven by coherent fields always remain in a coherent state.

\subsubsection{Degenerate OPO}

Applying the Lyapunov equation to the degenerate OPO discussed in (\ref{sec:04b-dopo}), we arrive at: 
\beq
	\dbl{\sigma} = \frac{1}{2}\begin{bmatrix}
		\frac{\left|-i\Delta+\kappa/2\right|^2}{\left|-i\Delta+\kappa/2\right|^2 - \left|\epsilon\right|^2} & 
		\frac{\epsilon(-i\Delta+\kappa/2)}{\left|-i\Delta+\kappa/2\right|^2 - \left|\epsilon\right|^2} \\
		\frac{\epsilon^*(i\Delta+\kappa/2)}{\left|-i\Delta+\kappa/2\right|^2 - \left|\epsilon\right|^2} & 
		\frac{\left|-i\Delta+\kappa/2\right|^2}{\left|-i\Delta+\kappa/2\right|^2 - \left|\epsilon\right|^2}
	\end{bmatrix} \label{eq:04b-oposig}
\eeq
The maximum and minimum covariance happen at the angles:
\beq
	\sigma_\pm = \frac{1}{2}\frac{1}{1 \mp \left|\epsilon/(-i\Delta + \kappa/2)\right|},\ \ \ 
	\phi_+ = \mbox{arg}\bigl(\epsilon(-i\Delta + \kappa/2)\bigr),\ \ \ 
	\phi_- = \mbox{arg}\bigl(\epsilon(-i\Delta + \kappa/2)\bigr) + \frac{\pi}{2}
\eeq
\begin{figure}[bt]
	\centering
	\includegraphics[width=1.00\textwidth]{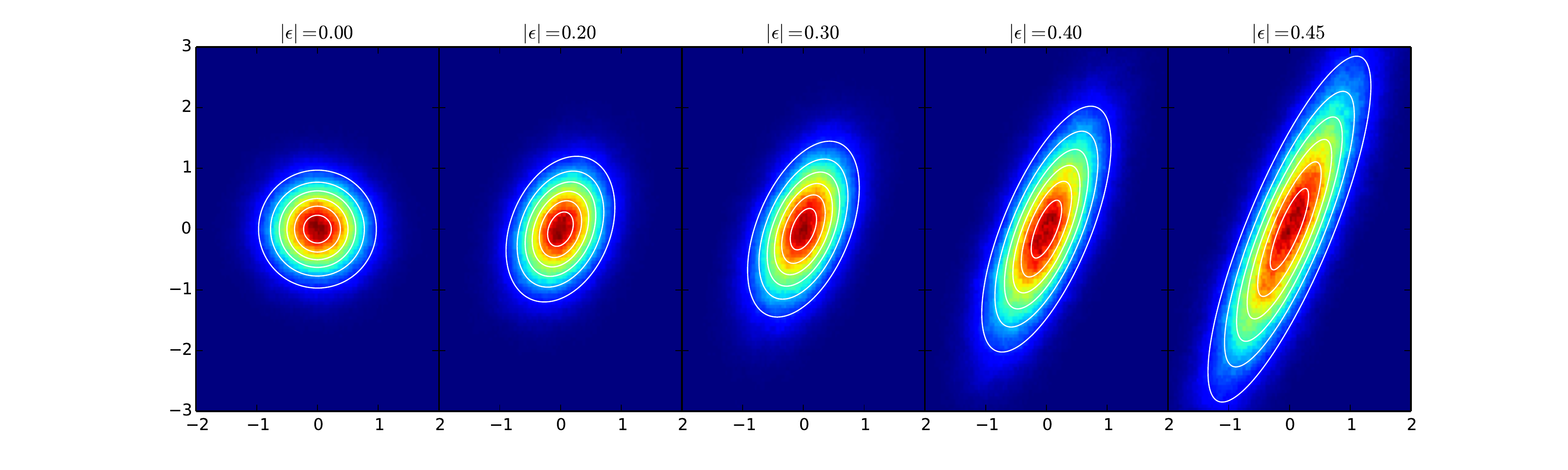}
	\caption{Wigner function for the OPO internal state, computed from numerical simulation (density plot) and from 	\label{fig:04b-f2}
the analytic formula in Eq.~(\ref{eq:04b-oposig}) (white contours)}
\end{figure}
As Figure \ref{fig:04b-f2} shows, the internal state of the OPO is squeezed, but it is not a pure state.  In the limit $|\epsilon| \rightarrow |-i\Delta + \kappa/2|$, the noise of the larger quadrature goes to infinity while the smaller one goes to $1/4$, half the value for a coherent state, so the product $\sigma_+\sigma_-$ is much larger than the Heisenberg limit.  For a simple OPO with coherent inputs and no feedback, it is not possible to squeeze the {\it internal} field by more than a factor of two (``3 \d B of squeezing'').

\section{Input-Output Relations}
\label{sec:04b-inout}

Linear systems are useful in engineering because they transform input signals into output signals.  Thus, to an engineer, what really matters is not the internal dynamics but rather the input-output relations of the device.  For a linear system without explicit time dependence, the input-output relations are fully determined by the doubled-up {\it transfer function} $\dbl{T}(\omega)$. 

The transfer function 
connects inputs and outputs in the frequency domain.  It is straightforward to convert doubled-up ABCD models to the frequency domain, but there are a few subtleties.  First, we define frequency-domain doubled-up vectors by the Fourier transform:
\beq
	\dbl{a}_\omega = \frac{1}{2\pi}\int{e^{i\omega t} \dbl{a}(t)dt} = \frac{1}{2\pi}\int{e^{i\omega t} \begin{bmatrix} e^{i\omega t} a(t) \\ e^{i\omega t} a^*(t) \end{bmatrix} dt} = \begin{bmatrix} a_\omega \\ a_{-\omega}^* \end{bmatrix} \label{eq:04b-adbl}
\eeq
Note the $-\omega$ subscript in the conjugated term.  In other words, $\dbl{a}_\omega \neq (a_\omega,\ a_\omega^*)$.  Thus, the component can convert a signal at $e^{-i\omega t}$ to one at $e^{+i\omega t}$, since $\d a/\d t$ depends not only on $a$ but also $a^\dagger$.

Similarly, we define $\dbl{b}_\omega = (b_\omega,\ b_{-\omega}^*)$ in terms of the input-output field $b(t) = \d B(t)/\d t$.  
The ABCD equations, in the frequency domain, become:
\bea
	-i\omega \dbl{a}_\omega & = & \dbl{A}\dbl{a}_\omega + \dbl{B}\dbl{b}_{{\rm in},\omega} \label{eq:04b-abcdf1} \\ 
	\dbl{b}_{{\rm out},\omega} & = & \dbl{C}\dbl{a}_\omega + \dbl{D}\dbl{b}_{{\rm in},\omega} \label{eq:04b-abcdf2}
\eea
The input and output are related by a matrix and some noise:
\beq
	\boxed{\dbl{b}_{{\rm out},\omega} = \underbrace{\left[\dbl{D} + \dbl{C} \frac{1}{-i\omega - \dbl{A}}\dbl{B}\right]}_{\dbl{T}(\omega)}\dbl{b}_{{\rm in},\omega}} \label{eq:04b-tf}
\eeq
This defines the transfer function $\dbl{T}(\omega)$ 
 for any system.  Note that is matrix has a doubled-up structure similar to $\dbl{A}$, $\dbl{B}$, $\dbl{C}$, $\dbl{D}$, but with $\omega$-dependence:
\beq
    \dbl{T}(\omega) = \begin{bmatrix} T_{-}(\omega) & T_{+}(\omega) \\ 
        T_{+}(-\omega)^* & T_{-}(-\omega)^* \end{bmatrix} \label{eq:04b-tstruc} 
\eeq

\subsection{Gain}
\label{sec:04b-gain}

An input signal $\dbl{b}_{{\rm in},\omega}$ becomes $\dbl{T}(\omega)\dbl{b}_{{\rm in},\omega}$ on output, plus some noise.  The amplitude of this output is: $|\dbl{b}_{{\rm out},\omega}|^2 = |\dbl{T}(\omega)\dbl{b}_{{\rm in},\omega}|^2$.  The {\it amplitude gain} is:

\beq
	G^2 = \frac{\dbl{b}_{{\rm in},\omega}^\dagger \dbl{T}(\omega)^\dagger \dbl{T}(\omega) \dbl{b}_{{\rm in},\omega}}{\dbl{b}_{{\rm in},\omega}^\dagger\dbl{b}_{{\rm in},\omega}} \label{eq:04b-gain}
\eeq

This is maximized for the largest eigenvalue of $\dbl{T}(\omega)^\dagger \dbl{T}(\omega)$ and minimized for its smallest eigenvalue.  For a single-input single-output system, there are only two eigenvalues, corresponding to the different quadratures of the device.  The device may amplify one quadrature more than another (phase-sensitive amplification) or amplify them both equally (phase-insensitive amplification).

Amplification at DC for a 2-by-2 matrix is especially simple.  The transfer-function matrix takes the following form:

\beq
	\dbl{T} = \begin{bmatrix} T_{-} & T_{+} \\ T_{+}^* & T_{-}^* \end{bmatrix}
\eeq

since $\omega = -\omega$ at DC.  The gain is given by the singular values of the matrix, given by the following SVD:

\beq
	\dbl{T} = \begin{bmatrix} \eta' & i\eta' \\ (\eta')^* & (i\eta')^* \end{bmatrix}
		\begin{bmatrix} |T_{-}| + |T_{+}| & 0 \\ 0 & |T_{-}| - |T_{+}| \end{bmatrix}
		\begin{bmatrix} \eta & i\eta \\ \eta^* & (i\eta)^* \end{bmatrix}^{-1}
\eeq

where $\eta$ is the maximally amplified input quadrature, and $\eta'$ is its respective image, given by:

\beq
	\eta = \sqrt{\frac{T_{-}^* T_{+}}{|T_{-}^* T_{+}|}},\ \ \ 
	\eta' = \sqrt{\frac{T_{-} T_{+}}{|T_{-}^* T_{+}|}}
\eeq

The minimally amplified quadrature is $i\eta$, which maps to $i\eta'$ in the output.

\subsection{Noise}
\label{sec:04b-noise}

For a single mode, any state can be fully described by its photon number and squeezing (at least if it's a Gaussian state, and in this chapter we only deal with Gaussian states).  Likewise, the output field $\dbl{b}_{\rm out}$ can be described in terms of its photon number and squeezing.  But since the output channel has an infinite number of degrees of freedom, we replace photon number and squeezing with a {\it power spectrum} $\mathcal{N}(\omega)$ and a {\it squeezing spectrum} $\mathcal{M}(\omega)$.  Any output from a linear system can be fully specified in terms of $\mathcal{M}$ and $\mathcal{N}$.

\begin{figure}[tb]
	\centering
	\includegraphics[width=0.66\textwidth]{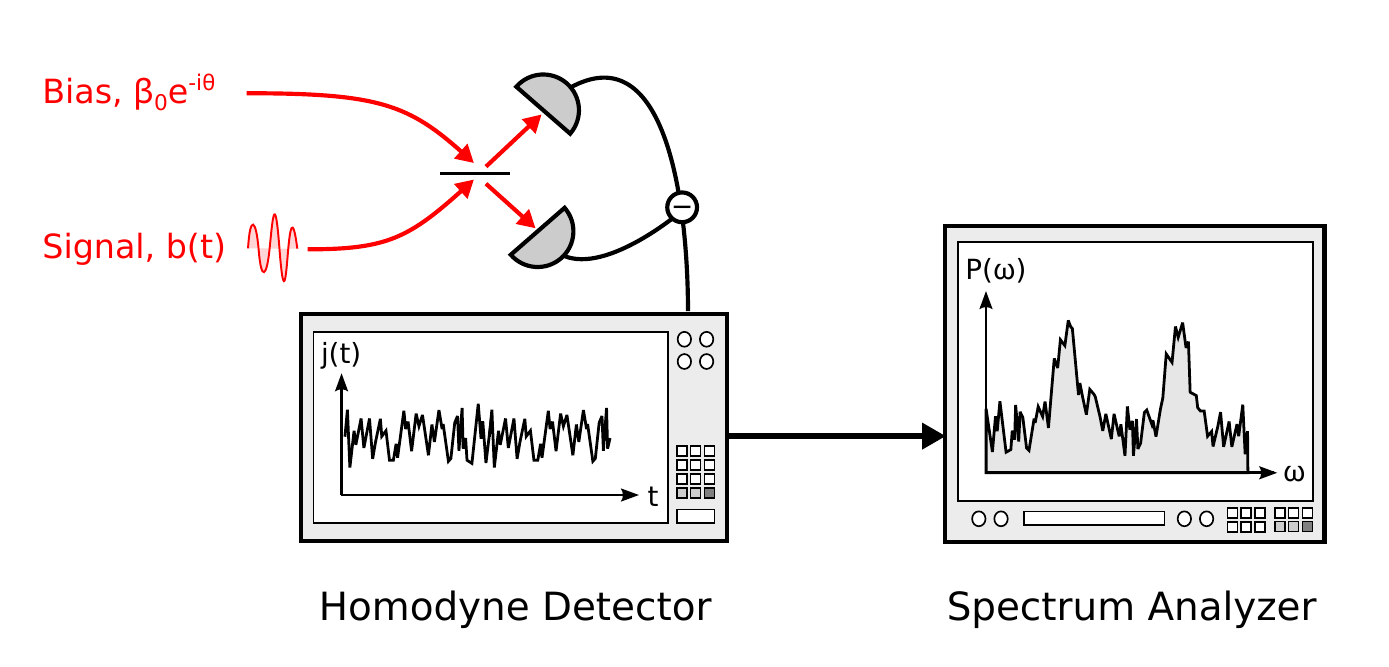}
	\caption{Measurement of the noise spectrum.}
	\label{fig:04b-f4}
\end{figure}

The power and squeezing spectra are defined by way of a homodyne measurement, illustrated in Figure~\ref{fig:04b-f4}.  Suppose that an output $b(t)$ is passed through a 50-50 beamsplitter with a coherent field $\beta e^{-i\theta} + b_{\rm vac}$ coming in from the dark port.  Both beamsplitter outputs are measured, denoted $b_+$ and $b_-$ here:
\beq
	b_\pm(t) = \sqrt{2}\bigl(b(t) \pm (\beta_0 e^{-i\theta} + b_{\rm vac}(t))\bigr)
\eeq
The homodyne output is the difference between the $b_+$ photocurrent and the $b_-$ photocurrent.  In the limit of large $\beta$, this becomes:
\beq
	j_\theta(t) = b_+(t)^\dagger b_+(t) - b_-(t)^\dagger b_-(t) \rightarrow \beta_0 (e^{-i\theta} b + e^{i\theta}b^\dagger)
\eeq
That is, it measures a quadrature of the field $b(t)$.  We can Fourier transform $j_\theta(t)$ to obtain the frequency-domain homodyne signal which, up to the factor of $\beta$, is $j_\theta(\omega) = e^{-i\theta} b_\omega + e^{i\theta}b_{-\omega}^\dagger$.  Squeezing is defined in terms of the {\it power spectral density} $P_\theta(\omega)$ of $j_\theta$, namely:
\begin{eqnarray}
S_\theta(\omega) & = & \sqrt{2P_\theta(\omega)} \\
2\pi\delta(\omega-\omega')P_\theta(\omega) & = & \left\langle j_\theta(\omega)^\dagger j_\theta(\omega') \right\rangle \nonumber \\
& = & \left\langle (e^{i\theta}b_\omega^\dagger + e^{-i\theta}b_{-\omega})
    (e^{-i\theta}b_{\omega'} + e^{i\theta}b_{-\omega'}^\dagger) \right\rangle \nonumber \\
& = & \left\langle b_\omega^\dagger b_{\omega'} + b_{-\omega} b_{-\omega'}^\dagger \right\rangle 
    + e^{2i\theta} \left\langle b_\omega^\dagger b_{-\omega'}^\dagger \right\rangle
    + e^{-2i\theta} \left\langle b_{\omega} b_{-\omega'} \right\rangle \nonumber \\
& = & 2\pi\delta(\omega - \omega')\left[\left(2\mathcal{N} + 1\right) + e^{2i\theta} \mathcal{M}^* + e^{-2i\theta} \mathcal{M}\right]
\end{eqnarray}
$\mathcal{N}$ is a measure of power at detuning $\omega$, while $\mathcal{M}$ is a measure of squeezing \cite{Gough2009c}.  From the equation above, they are given by:
\begin{eqnarray}
2\pi\delta(\omega-\omega')\bigl(\mathcal{N}(\omega) + 1/2\bigr) & = & \frac{1}{2}\left\langle b_\omega^\dagger b_{\omega'} + b_{-\omega} b_{-\omega'}^\dagger \right\rangle \label{eq:04b-sqspec-a1} \\
2\pi\delta(\omega-\omega')\mathcal{M}(\omega) & = & \left\langle b_{\omega} b_{-\omega'} \right\rangle \label{eq:04b-sqspec-a2}
\end{eqnarray}
From the uncertainty principle and the commutator $[j_\theta(t),j_{\theta+\pi/2}(t')] = 2i\,\delta(t-t')$, we can show that $\mathcal{M}$ and $\mathcal{N}$ satisfy the following inequality:
\beq
	\bigl(\mathcal{N} + 1/2\bigr)^2 - \mathcal{M}^2 \geq \frac{1}{4}
\eeq
We want to find a formula for $\mathcal{M}$ and $\mathcal{N}$ in terms of the transfer function.  To start, Eq.~(\ref{eq:04b-tf}), together with definition of doubled-up matrices, gives the following input-output relations:
\begin{eqnarray}
b_{{\rm out},\omega} & = & T_{-}(\omega)b_{{\rm in},\omega} + T_{+}(\omega)b_{{\rm in},-\omega}^* \\
b_{{\rm out},-\omega} & = & T_{+}(-\omega)b_{{\rm in},\omega}^* + T_{-}(-\omega)b_{{\rm in},-\omega} \\ 
\end{eqnarray}
Assuming vacuum inputs, the outputs have the following statistics:
\begin{eqnarray}
\tfrac{1}{2}\langle b_{{\rm out},\omega}^\dagger b_{{\rm out},\omega'} + b_{{\rm out},\omega'} b_{{\rm out},\omega}^\dagger \rangle & = & 2\pi\delta(\omega - \omega')\left[\frac{|T_{-}(\omega)|^2 + |T_{+}(\omega)|^2}{2} 
\right] \\
\tfrac{1}{2}\langle b_{{\rm out},-\omega}^\dagger b_{{\rm out},-\omega'} + b_{{\rm out},-\omega'} b_{{\rm out},-\omega}^\dagger \rangle & = & 2\pi\delta(\omega - \omega')\left[\frac{|T_{+}(-\omega)|^2 + |T_{-}(-\omega)|^2}{2} 
\right] \\
\langle \beta_{{\rm out},\omega}\beta_{{\rm out},-\omega'} \rangle & = & 2\pi\delta(\omega - \omega')\left[\frac{T_{-}(\omega)T_{+}(-\omega) + T_{+}(\omega)T_{-}(-\omega)}{2} 
\right]
\end{eqnarray}
which leads to
\begin{eqnarray}
    \mathcal{N} + \frac{1}{2} & = & \frac{1}{2}\left[\frac{|T_{-}(\omega)|^2 + |T_{+}(\omega)^2| + |T_{+}(-\omega)^2| + |T_{-}(-\omega)^2|}{2} 
    \right] \\
    \mathcal{M} & = & \left[\frac{T_{-}(\omega)T_{+}(-\omega) + T_{+}(\omega)T_{-}(-\omega)}{2} 
    \right]
\end{eqnarray}
These can be expressed in matrix notation as
\begin{empheq}[box=\fbox]{align}
	2(\mathcal{N}+1/2) &= \mathcal{N}_1 + \mathcal{N}_2, 
	&\begin{bmatrix} \mathcal{N}_1 & \mathcal{M} \\ \mathcal{M}^* & \mathcal{N}_2 \end{bmatrix} & = 
	\frac{1}{2} \dbl{T}(\omega)\dbl{T}(\omega)^\dagger 
	\label{eq:04b-mn}
\end{empheq}
The maximum and minimum values of the amplitude $S_\theta(\omega)$ occur when $e^{2i\theta} \mathcal{M}^* \in \Re$.  These values are:
\beq
	S_\pm(\omega) = \sqrt{2\left((\mathcal{N}(\omega)+1/2) \pm |\mathcal{M}(\omega)|\right)} \label{eq:04b-spm}
\eeq
Vacuum noise has $S_+(\omega) = S_-(\omega) = 1$.  A squeezed vacuum has $S_+(\omega) = r(\omega), S_-(\omega) = 1/r(\omega)$.  The Heisenberg uncertainty principle becomes a condition on the product of the extrema: $S_+ S_- \geq 1$.
\begin{figure}[tb]
	\centering
	\includegraphics[width=0.90\textwidth]{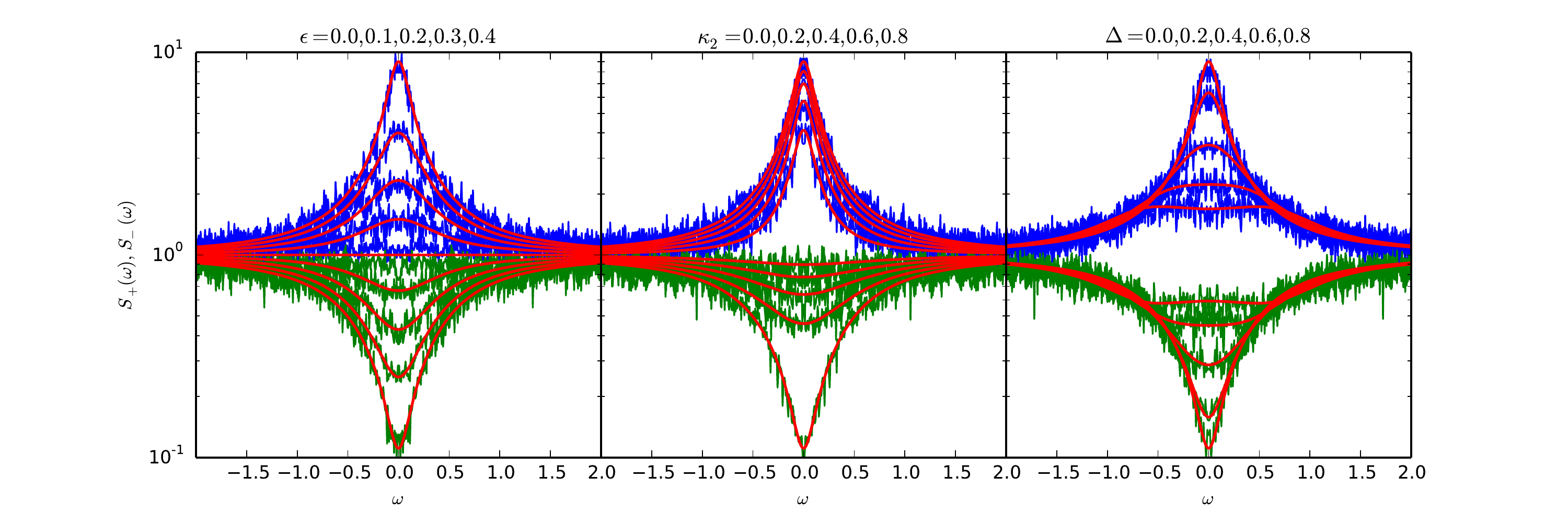}
	\caption{Quadrature noises $S_+(\omega)$, $S_-(\omega)$ for a degenerate OPO.  Effects of varying pump $\epsilon$ (left), external loss $\kappa_2$ (center), and detuning $\Delta$ (right) on the spectrum.}
	\label{fig:04b-f5}
\end{figure}

\subsubsection{Example: Degenerate OPO}

Since the degenerate OPO has a linear SLH model, the predicted gain (\ref{eq:04b-gain}) and squeezing spectrum (\ref{eq:04b-spm}) should exactly match simulations.  Even though the OPO model is quite simple, the analytic form for these quantities is rather cumbersome.  However, in the resonant case $\Delta = 0$, without any additional losses, it is:

\beq
S_+(\omega) = \left|\frac{i\omega + \frac{1}{2}\kappa + |\epsilon|}{i\omega + \frac{1}{2}\kappa - |\epsilon|}\right|,\ \ \ S_-(\omega) = \left|\frac{i\omega + \frac{1}{2}\kappa - |\epsilon|}{i\omega + \frac{1}{2}\kappa + |\epsilon|}\right|
\eeq

The output satisfies $S_+ S_- = 1$, so this is perfect squeezing.  The larger the pump $\epsilon$, the larger the squeezing, a fact confirmed in the left panel of Fig.~\ref{fig:04b-f5}.  Adding an additional loss channel will decrease both the squeezing and the anti-squeezing,, but it primarily affects the squeezing (center plot).  Rather than shift the spectrum, changing the detuning broadens it, since the squeezing spectrum depends on both the $\omega$ and $-\omega$ modes.

\ifstandalone{\bibliography{NoteRefs}{}
\bibliographystyle{alpha}
}
\ifdefined\multidoc\else\input{Header}\fi

\ifstandalone{\setcounter{chapter}{3}}
\chapter{Coherent Quantum LQG Control}
\label{ch:11}

This chapter is based on the following papers:

\begin{itemize}
\item \href{http://dx.doi.org/10.1103/PhysRevLett.109.173602}{Ryan Hamerly and Hideo Mabuchi, ``Advantages of coherent feedback for cooling quantum oscillators.'' Physical Review Letters 109, 173602 (2012)}
\item \href{http://dx.doi.org/10.1103/PhysRevA.87.013815}{Ryan Hamerly and Hideo Mabuchi, ``Coherent controllers for optical-feedback cooling of quantum oscillators.'' Physical Review A 87, 013815 (2013)}
\end{itemize}

As present-day engineering relies broadly and implicitly on real-time feedback control methodology~\cite{AstromBook}, it is difficult to imagine our nascent explorations of quantum engineering advancing to technological relevance without rigorous extensions of core control theory to incorporate novel features of quantum dynamics, stochastics and measurement. While significant progress has been made recently in terms of analyzing quantum feedback systems~\cite{Belavkin1983,Wiseman1993,Doherty2000,Mabuchi2005,Dong2010,Brif2010} and in experimental demonstrations of quantum feedback control~\cite{Smith2002,Armen2002,Bushev2006,Gigan2006,Arcizet2006b,Kleckner2006,Mabuchi2008,Gillett2010,Sayrin2011,Iida2012}, we still have a relatively limited understanding of systematic approaches to quantum control design and of the qualitative role of quantum coherence and entanglement between the plant and controller in a feedback loop.

Within the elementary context of linear open quantum systems, James, Nurdin and Petersen~\cite{James2008,Nurdin2009} have utilized interconnection models based quantum stochastic differential equations (QSDEs)~\cite{Hudson1984,Carmichael1993,Gardiner1993,Barchielli2006} to develop quantum generalizations of the traditional paradigms of ${\cal H}^\infty$ and Linear Quadratic Gaussian (LQG) optimal control. While some of the most exciting potential applications of quantum feedback control involve nonlinear dynamics and/or non-Gaussian noises~\cite{Kerckhoff2010,Kerckhoff2011,Mabuchi2011a,Gough2012}, the linear setting is an essential starting point for rigorous study and presents crucial advantages in terms of analytic and computational tractability.

Here we focus on a theoretical investigation of steady-state cooling of open quantum oscillators such as optical and optomechanical resonators subject to stationary heating, damping, and optical probing and feedback. We work within an LQG framework as in the recent paper of Nurdin, James and Petersen~\cite{Nurdin2009} and utilize numerical optimization together with fundamental analytic results~\cite{AstromBook} bounding the best possible LQG performance of measurement-based feedback control schemes to establish and to interpret quantitative advantages of coherent feedback for cooling-type performance metrics in certain parameter regimes.

Following recent convention, as in~\cite{James2008,Nurdin2009,Mabuchi2008}, we will here refer to measurement-based controllers as ``classical'' controllers and to coherent feedback controllers as ``quantum'' controllers. This terminology reflects the general distinction that the signal processing required to determine LQG-optimal control actions from a real-time measurement signal can be implemented by a classical electric circuit, while all of the hardware in a coherent feedback loop must be physically describable using quantum mechanics (typically with weak damping).

\section{Linear Systems}

Quantum harmonic oscillators can be modeled as cascadable open quantum systems using the SLH framework \cite{Gough2009a,Gough2009b} and the associated QSDEs. In the SLH framework, any open quantum system may be described as a triple:
\beq
	G = (S, L, H)
\eeq
where $S$ is a scattering matrix, $L$ is a coupling vector and $H$ is the Hamiltonian operator for the system's internal degrees of freedom. For a {\em linear} system with an internal state $x$, $S_{ij}$ is independent of the internal state, $L_i = \Lambda_i x + \lambda_i$ is at most linear, and $H = \frac{1}{2} x^{\rm T} R x + r^{\rm T} x$ is at most quadratic.

Armed with an SLH representation the most efficient way to simulate a {\it linear} quantum system is to solve the QSDEs, which represent coupled Heisenberg equations of motion for system operators and input-output quantum stochastic processes. Following the work of James, Nurdin and Petersen~\cite{James2008} we write the QSDEs for a linear system in the state-space form,
\begin{eqnarray}
	\d x(t) & = & \left[A\,x(t) + a\right] \d t + B\,\d a(t) \nonumber \\
	\d\tilde{a}(t) & = & \left[C\,x(t) + c\right] \d t + D\,\d a(t) \label{eq:11-abcd}
\end{eqnarray}
Here $x(t)$ gives the plant's internal variables; this is a Hermitian, operator-valued vector. $A$, $B$, $C$ and $D$ are real matrices; $a$ and $c$ are real vectors.  The processes $\d a(t)$ and $\d\tilde{a}(t)$ are quantum stochastic processes for the inputs and outputs, respectively. For convenience, we make them Hermitian as well; for a given port, one has $\d a_i = \bigl(\d B_i + \d B_i^\dagger, (\d B_i - \d B_i^\dagger)/i\bigr)$, where $\d B(t)$ is the quantum Wiener process~\cite{GardinerBook,Bouten2007} following the It\^{o} rule $\d B_i\,\d B_j^\dagger = \delta_{ij} \d t$ (Sec.~\ref{sec:04b-quad}).

Defining $(\Theta_x)_{ij} = [x_i, x_j]/2i$ as the commutator matrix, the ABCD parameters of (\ref{eq:11-abcd}) can be related to the SLH parameters as follows:
\beq
\begin{array}{rclrcl}
	A & = & 2\Theta\left(R + \frac{1}{4}\tilde{\Lambda}^{\rm T} J \tilde{\Lambda}\right),\ \ \  &
	B & = & \Theta \tilde{\Lambda}^{\rm T} J \tilde{S}, \\
	C & = & \tilde{\Lambda}, &
	D & = & \tilde{S}, \\
	a & = & 2\Theta\left(r + \frac{1}{4}\tilde{\Lambda}^{\rm T} J \tilde{\lambda}\right), &
	c & = & \tilde{\lambda}
\end{array}
\eeq
(Here $\tilde{S}$, $\tilde{\Lambda}$, and $\tilde{\lambda}$ are real matrices which can be easily constructed from $S$, $\Lambda$ and $\lambda$, which are in general complex.  $J$ is a canonical antisymmetric matrix of the appropriate size.  See Sec.~\ref{sec:04b-quad}.)

To measure the performance of a given controller we need to define a cost function.  For example, to minimize the plant's response to a noisy input one could minimize the steady-state expectation value of the excitation number $\avg{a^\dagger a}$. With (classical) state feedback and in the absence of exogenous noise such a quadratic cost function would result in a Linear Quadratic Regulator (LQR) optimal control problem~\cite{AstromBook}, but in our optical feedback scenario with Gaussian input fields (vacuum or thermal noise) this becomes a quantum LQG problem~\cite{AstromBook,Nurdin2009}.

It is straightforward to concatenate and cascade linear systems once we have the ABCD models. We have written software in {\it Mathematica} to compute the ABCD matrices for an arbitrary linear quantum system.  This borrows many elements from the Modelica quantum circuit toolkit of Sarma et al.\ \cite{Sarma2013}, and is similar to the QHDL framework of Tezak et al.\ \cite{Tezak2012}.  The code computes the LQR cost function as a function of the plant and controller properties, and it would not be difficult to extend it to more general cost functions.  Thanks to the linearity of our system, simulation is very fast: the complexity is polynomial in the size of the circuit, not exponential as is usually the case for quantum simulations, and for a simple system, it computes the LQR in well under 50 microseconds.

Given a particular plant, the code is fast enough to perform a multivariate Newton-Raphson optimization scheme to find the (locally) optimal controller parameters.  This is possible regardless of whether the controller has any particular structure -- if the controller's structure is left arbitrary, the code can simply optimize with respect to the controller's ABCD matrices, subject to the physical realizability conditions
\begin{eqnarray}
	A \Theta + \Theta A^{\rm T} + B J B^{\rm T} & = & 0 \nonumber \\
	\Theta C^{\rm T} + B J D^{\rm T} & = & 0 \nonumber \\
	D J D^{\rm T} & = & J \label{eq:11-prc}
\end{eqnarray}
that arise from the fact that time evolution should preserve the commutation relations between system and input/output fields \cite{James2008, Nurdin2009}.  A variant of the algorithm was published independently \cite{Sichani2015}.  Optimizing with respect to an ``arbitrary'' controller takes longer because there are more free parameters, but the code is fast enough for each Newton step to take no more than 1.5 milliseconds on a standard laptop.  

We note that the classical steady-state LQG problem is a convex problem, and the optimal steady-state controller parameters can be derived via solution of algebraic Riccati equations~\cite{AstromBook}. In the quantum case, no such closed-from solutions are known and the realizability constraints (\ref{eq:11-prc}) make the landscape for numerical optimization non-convex~\cite{Nurdin2009}. Hence while we can be sure about the classical optimality of measurement-based controllers for the oscillator cooling scenarios we consider, the coherent controllers we find via numerical optimization are merely {\it local} minima and can only be considered as candidates for quantum optimality.

\section{Control of an Optical Cavity}

As a simple example of a quantum ``plant'' system, consider an optical cavity with a noisy input, Fig.\ \ref{fig:11-f1}.  In the controller's absence, the cavity is driven by two vacuum inputs (mirrors $k_1$ and $k_2$, and one thermal input (mirror $k_3$).  Any noise process that is much broader spectrally than the cavity linewidth can be approximated as a ``white noise'' thermal input.  Without such noise, the cavity's internal mode decays quickly to the ground state.  The objective in this control problem is to minimize the effect of the noise on the cavity's internal state -- in other words, to minimize the photon number $\avg{a^\dagger a}$ of the cavity.  We accomplish this by sending output $1$ through a control circuit and feeding the result back into input $2$.  This is an LQG feedback control problem.

\begin{figure}[t]
	\centering
	\includegraphics[width=0.60\textwidth]{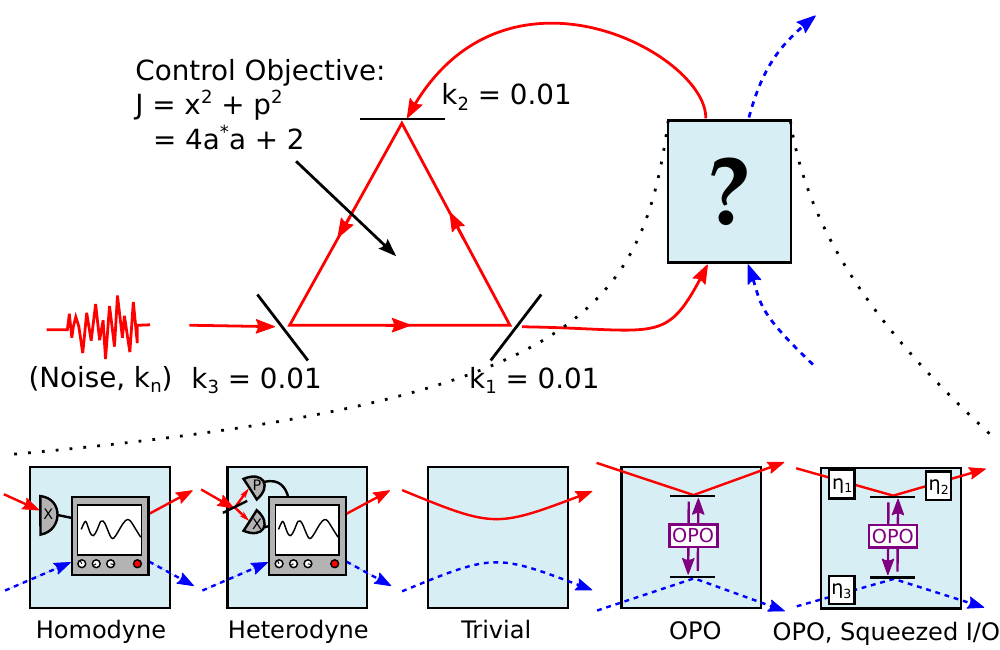}
	\caption{Optical cavity plant system with five possible classical and coherent feedback controllers}
	\label{fig:11-f1}
\end{figure}

Five possible controllers are shown in Figure \ref{fig:11-f1}.  The classical controllers work by measuring a quadrature from the cavity's output (or in the heterodyne case, splitting the beam and measuring two different quadratures), and applying a feedback signal based on this measurement and the controller's internal state.  The ``trivial controller'' works by feeding the output directly back into mirror 2 of the plant, perhaps with a phase shift.  If the light reflecting off of mirror 2 is in phase with the light leaking out of the mirror, the light lost through both mirrors interferes constructively, reducing the control objective $\avg{a^\dagger a}$ (see also~\cite{Mabuchi2011a}).

The remaining two controllers shown in the figure are coherent controllers with memory.  Unlike the trivial controller, the control signal is a function not only of the input field, but also the input's history.  But unlike the classical controllers, the input field is not measured; instead, it is coherently processed and the result is fed back into the plant cavity.  These designs use an optical parametric oscillator (OPO, as in Fig.~\ref{fig:11-f2}) to squeeze the optical field.

\begin{figure}[t]
	\centering
	\includegraphics[width=0.65\textwidth]{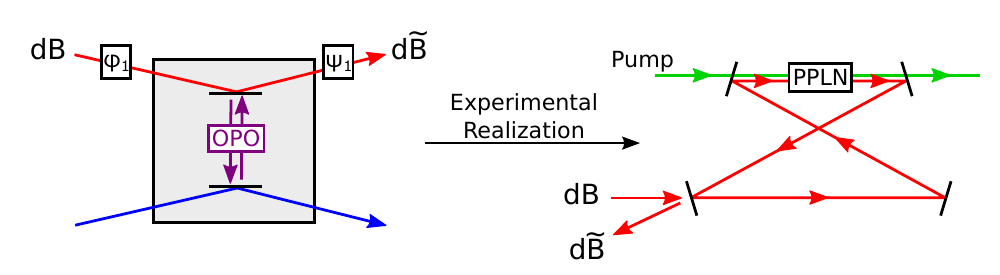}
	\caption{Experimental realization of an OPO with a cavity and a nonlinear crystal.}
	\label{fig:11-f2}
\end{figure}

The OPO will have the following SLH model:
\begin{eqnarray}
	& & S = 1_{2\times 2},\ \ L = \left[\sqrt{\kappa_1} a,\ \sqrt{\kappa_2} a\right], \nonumber \\
	& & H = \frac{1}{4} x^{\rm T} \left[\begin{array}{cc} \Delta - \mbox{Im}(\epsilon) & \mbox{Re}(\epsilon) \\ \mbox{Re}(\epsilon) & \Delta + \mbox{Im}(\epsilon) \end{array}\right]x
	\label{eq:11-opo-slh}
\end{eqnarray}
Here, $\kappa_1$ is related to the input/output mirror reflectance, $\kappa_2$ to other losses; $\Delta$ is the cavity detuning, and $\epsilon$ is a complex number, whose amplitude encodes the strength of the pump field and the nonlinear medium's $\chi^{(2)}$, and whose phase encodes the pump field's phase \cite{Nurdin2009b}.

The plant system, an optical cavity with a noisy input, can be modeled as an open quantum system with three couplings, one for each mirror.  The SLH model for this system is:
\beq
	S = 1_{3\times 3},\ \ L = \left[\sqrt{k_1} a,\ \sqrt{k_2} a,\ \sqrt{k_3} a\right],\ \
	 H = \Delta a^\dagger a
\eeq
We also need to find the covariance matrix $F_{ij}$ for the noisy inputs $\d a_i$, defined  by $\frac{1}{2}\avg{\d a_i \d a_j + \d a_j \d a_i} = F_{ij}\d t$.  Recall that, for {\it vacuum} inputs, the fields $\d B$ and $\d B^\dagger$ satisfy the It\^{o} relations $\d B\,\d B = \d B^\dagger \d B^\dagger = \d B^\dagger \d B = 0$, $\d B\,\d B^\dagger = \d t$ \cite{Hudson1984,Bouten2007}, leading to the It\^{o} tables:

\begin{table}[h]
	\centering
	\begin{tabular}{ccc}
	
	\begin{tabular}{c|cc}
		$\d X$/$\d Y$ & $\d B$ & $\d B^\dagger$ \\ \hline
		$\d B$ & $0$ & $\d t$ \\
		$\d B^\dagger$ & $0$ & $0$
	\end{tabular}
	&\ \ \ $\leftrightarrow$\ \ \ &
	\begin{tabular}{c|cc}
		$\d x$/$\d y$ & $\d a_x$ & $\d a_p$ \\ \hline
		$\d a_x$ & $\d t$ & $i\,\d t$ \\
		$\d a_p$ & $-i\,\d t$ & $\d t$
	\end{tabular}

	\end{tabular}	
	\caption{Left: It\^{o} table for the second-order increments $\d X\,\d Y$ (in terms of $\d B$, $\d B^\dagger$.  Right: It\^{o} table for the increments $dx\,dy$, in terms of $\d a_x = \d B + \d B^\dagger$, $\d a_p = (\d B - \d B^\dagger)i$.}
\end{table}

For a {\it non-vacuum}, thermal input, the field $\d B$ has additional (unsqueezed) noise, so $\d B^\dagger \d B = k_n \d t$ for some noise strength $k_n > 0$, and the rest of the relations are adjusted accordingly, leading to the following It\^{o} tables:

\begin{table}[h]
	\centering
	\begin{tabular}{ccc}
	
	\begin{tabular}{c|cc}
		$\d X$/$\d Y$ & $\d B$ & $\d B^\dagger$ \\ \hline
		$\d B$ & $0$ & $\!\!(1 + k_n)\d t$ \\
		$\d B^\dagger\!\!$ & $k_n \d t$ & $0$
	\end{tabular}
	&\ \ $\leftrightarrow$\ \ &
	\begin{tabular}{c|cc}
		$dx$/$dy$ & $\d a_x$ & $\d a_p$ \\ \hline
		$\d a_x$ & $(1+2k_n)\d t\!\!\!$ & $i\,\d t$ \\
		$\d a_p$ & $-i\,\d t$ & $\!\!\!(1+2k_n)\d t$
	\end{tabular}

	\end{tabular}	
	\caption{It\^{o} table for the increments $\d X,\d Y$ and $\d x,\d y$, assuming non-vacuum, thermal input.}
\end{table}

\begin{figure}[b]
	\centering
	\includegraphics[width=0.65\textwidth]{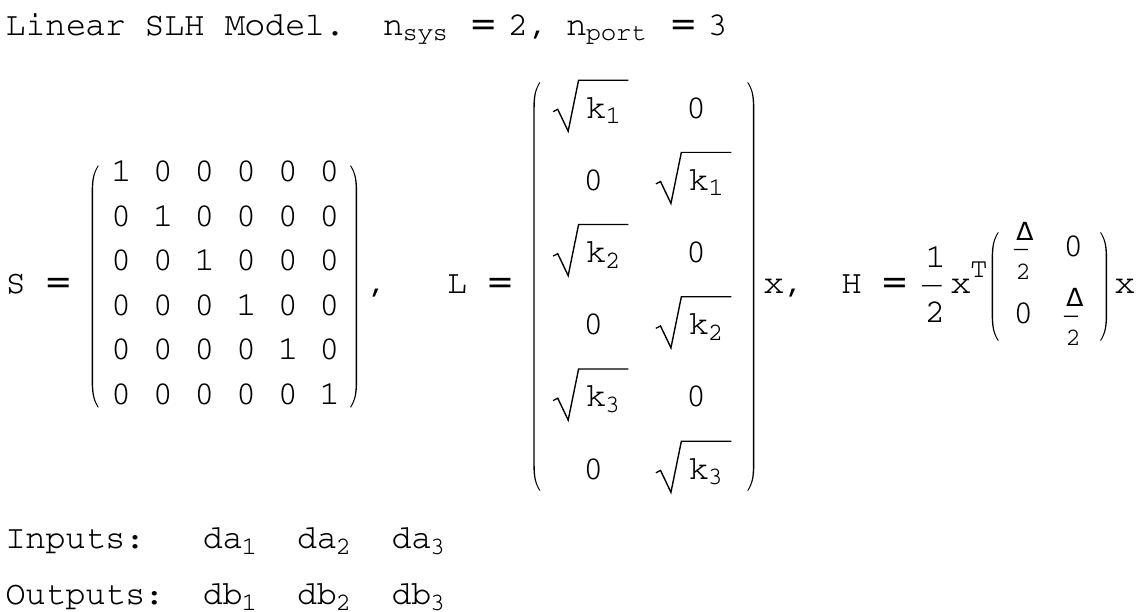}
	\caption{Output from our {\it Mathematica} package, describing the plant system.}
	\label{fig:11-f3}
\end{figure}

In the present system, inputs $1$ and $2$ are vacuum, and $3$ is thermal noise.  This gives the following covariance matrix:
\beq
	F = \left[\begin{array}{cccccc}
		1 & 0 & 0 & 0 & 0 & 0 \\
		0 & 1 & 0 & 0 & 0 & 0 \\
		0 & 0 & 1 & 0 & 0 & 0 \\
		0 & 0 & 0 & 1 & 0 & 0 \\
		0 & 0 & 0 & 0 & 1+2k_n & 0 \\
		0 & 0 & 0 & 0 & 0 & 1+2k_n
		\end{array}\right]
\eeq
The plant system is easy to set up in our {\it Mathematica} package; a sample output is shown in Figure \ref{fig:11-f3}.  The package, based on the circuit modeling and analysis framework of Sarma et al.\ \cite{Sarma2013}, allows one to arbitrarily concatenate and link smaller elements to form larger quantum circuits, as long as all of the components are linear.  The feedback control circuit is one example system the package can be used to simulate.

Once the combined plant / controller system is set up, with its associated $A$, $B$, $C$ and $D$ matrices, the covariance matrix $\sigma_{ij} = \frac{1}{2} \avg{x_i x_j + x_j x_i}$ can be computed with the Lyapunov equation
\beq
	A \sigma + \sigma A^{\rm T} + B F B^{\rm T} = 0
\eeq

\begin{figure}[tbp]
	\centering
	\includegraphics[width=0.56\textwidth]{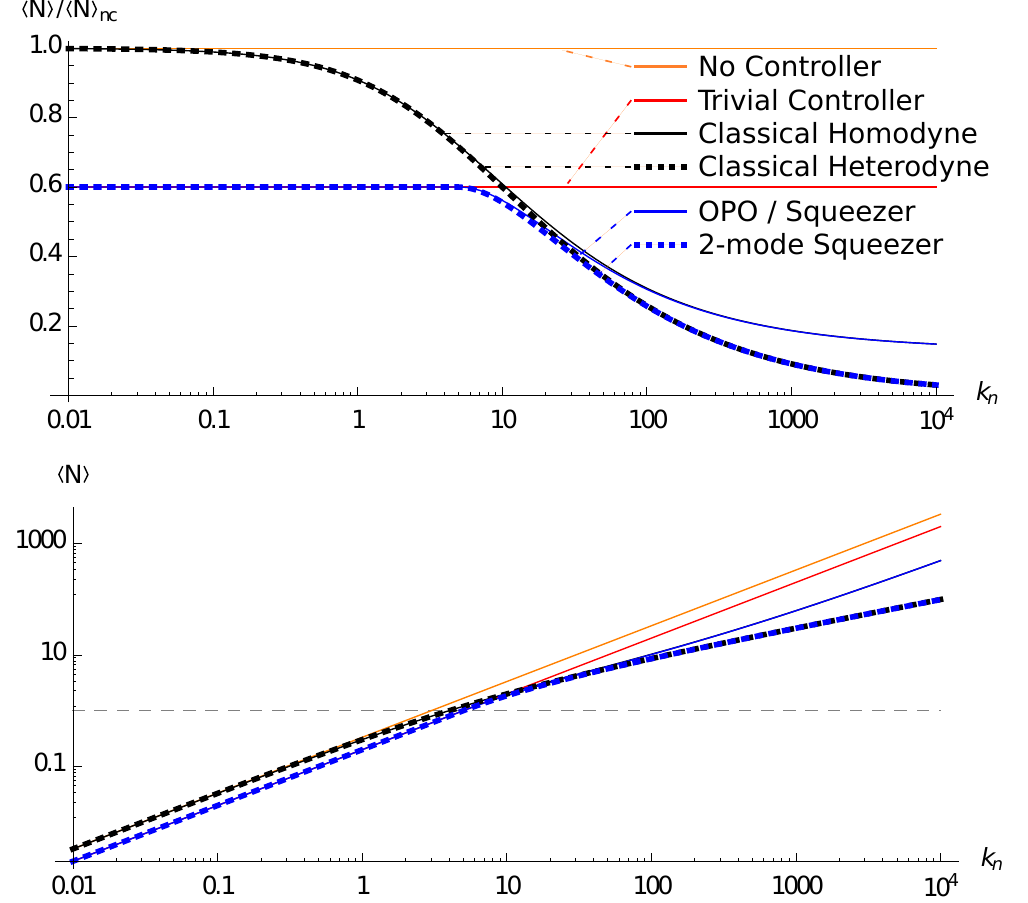}
	\caption{Bottom: Plant Ccavity photon number, as a function of noise strength $k_N$.  The uncontrolled case is shown, as well as the photon number for various control schemes.  Top: Photon number relative to the no-control case.  Smaller is better.}
	\label{fig:11-f4}
\end{figure}

For a model system with the parameters
\beq
	k_1 = k_2 = k_3 = 0.01, \Delta = 0.1
\eeq
we plot the cost function $\avg{a^\dagger a}$ as a function of noise $k_n$ for the various controller types in Figure \ref{fig:11-f4}.  The orange line gives the performance of the plant without a controller.  As expected, the photon number rises linearly with the noise power.  It is not hard to show that this matches the analytic result
\beq
	\avg{a^\dagger a}_{\rm nc} = \frac{k_3}{k_1 + k_2 + k_3} k_n \label{eq:11-cav-nc}
\eeq
that one can derive from the QSDEs.

The trivial controller is simple enough that it also has an analytic solution.  The two mirrors, rather than leaking photons separately, do so constructively so that the leakage {\it amplitudes} (rather than their {\it powers}) add up.  This requires the replacement $k_1+k_2 \rightarrow (\sqrt{k_1} + \sqrt{k_2})^2$ in (\ref{eq:11-cav-nc}), leading to the following result
\beq
	\avg{a^\dagger a}_{\rm tr} = \frac{k_3}{k_1 + k_2 + k_3 + 2\sqrt{k_1 k_2}} k_n \label{eq:11-cav-tr}
\eeq
which agrees with the numerical data plotted in Fig. \ref{fig:11-f4}.

\subsection{Classical Controllers}

More sophisticated are the classical measurement controllers.  The first simply makes a homodyne measurement of the $\d\tilde{a}_{1x}$ field.  This signal is fed through a classical circuit which generates an output.  The heterodyne controller is slightly more complicated, and can be modeled as a two-input homodyne measurement controller in the following circuit (using the notation of~\cite{Gough2009b,Tezak2012}; see Sec.~\ref{sec:04b-algebra}):
\beq
	(Hom)_{2-\rm in} \triangleleft (I_1 \boxplus e^{i\pi/2}) \triangleleft BS(\alpha) \label{eq:11-gj-het}
\eeq
In addition to the homodyne controller's parameters, we can also vary the beamsplitter transmittance.  (Setting the beam-splitter transmission coefficient $\alpha \rightarrow 1$ would send all the light entering controller input $1$ into the $x$-quadrature homodyne detector, so the classical homodyne controller is really a special case of the classical heterodyne controller.)

This example, in particular, illustrates the power of the Gough-James circuit algebra in treating control problems when the controller has a more complex, ``circuit-like'' structure.  Having written code to output the $ABCD$ model for a general $n$-input homodyne controller, it would have been straightforward, albeit tedious, to write additional code for the $n$-input heterodyne controller.  But using the Gough-James circuit algebra allows us to write the $n$-input heterodyne system in terms of a $2n$-input homodyne system, plus some beamsplitters and phase shifters, so we get the heterodyne controller for free.  By breaking the system into smaller components, we can reduce the total amount of work we need to do in quantum control and simulation problems.

There also exist ``analytic'' formulas for LQG-optimal classical controllers in the classical case.  It is not difficult to rewrite Eq.\ (\ref{eq:11-abcd}) in the standard form for an LQG problem:
\begin{eqnarray}
	\d x & = & A_p x\,\d t + B_p\,\d u + \d w \nonumber \\
	\d y & = & C_p x\,\d t + \d v \label{eq:11-class-abcd}
\end{eqnarray}
Here $\d y$ is the measurement signal, $du$ is the controller output, and $\d w$ and $\d v$ are the plant and controller noises, $\d w \sim N(0, F_w \d t)$, $dv \sim N(0, F_v \d t)$.  Unfortunately, in this system the noises are correlated; the vacuum noise $\d a_1$ acts on both the plant and, after reflection off mirror $k_1$, the controller.  One can define a covariance matrix $M_{ik} = \avg{\d w_i \d v_k}$ to account for this correlation.

A common trick is to remove the noise correlations by performing a change of variables \cite{Simon2006}.  Since $\d y - C_p x\,\d t - \d v = 0$, we can subtract this quantity from the first line of (\ref{eq:11-class-abcd}) to find an equivalent equation of motion:
\begin{eqnarray}
	\d x & = & A_p x\,\d t + B_p\,\d u + \d w + M F_v^{-1}(\d y - C_p x\,\d t - \d v) \nonumber \\
	& = & (A_p - M F_v C_p) x + B_p (\d u + B_p^{-1}M F_v^{-1}\d y) \nonumber \\
	& & + (\ dw - M F_v^{-1}\d y) \nonumber \\
	& = & \tilde{A} x + B_p \d\tilde{u} + \d\tilde{w}
\end{eqnarray}
Here, the noises $\d v$ and $\d\tilde{w}$ are uncorrelated.  The controller for this plant will consist of a Kalman filter and a feedback:
\begin{eqnarray}
	\d\hat{x} & = & (\tilde{A} - B_p L - K C_p) \hat{x}\,\d t + K\,\d y \nonumber \\
	\d\tilde{u} & = & -L \hat{x}\,\d t
\end{eqnarray}
The Kalman gain and feedback matrices can be obtained by solving the Riccati Equations:
\begin{eqnarray}
	K = \sigma C^{\rm T} F_v^{-1} & & (A\sigma + \sigma A^{\rm T} - \sigma C^{\rm T} F_v^{-1} C \sigma = 0) \nonumber \\
	L = R^{-1} B^{\rm T} \lambda  & & (A^{\rm T}\lambda + \lambda A + Q - \lambda B R^{-1} B \lambda = 0) \nonumber \\
\end{eqnarray}
(Here $Q$ and $R$ are LQR optimization weights for the plant and controller states; we assume $Q\gg R$). For this particular case we optimized the classical controllers numerically, but the results agree with the analytical expression.  When optimizing the measurement controllers, we found that the best controllers always had dynamics that were much faster than the plant timescales. When this happens, the controller's internal dynamics can be {\it adiabatically eliminated} and the controller can be replaced by a simplified ``limit model'' of the original component~\cite{Bouten2008a,Bouten2008b,Gough2010}; see Eq.~(\ref{eq:04b-ad-el}) in Sec.~\ref{sec:04b-algebra}.  When a linear component is adiabatically eliminated, its internal variables are removed and its ABCD model is replaced by the input-output relations:
\beq
	\d\tilde{a} = (D - C A^{-1} B) \d a \label{eq:11-ad-el-eq}
\eeq
The homodyne controller, adiabatically eliminated, becomes:
\begin{eqnarray}
	\d\tilde{a}_x & = & \xi_1 \d a_x + \d a_{k1,x} \nonumber \\
	\d\tilde{a}_p & = & \xi_2 \d a_x + \d a_{k1,p}
\end{eqnarray}
In this device, the signal $\d a_x$ is measured, amplified by factors $c_1$ and $c_2$, and imprinted onto the output field.  The downside of this measurement is the additional noise $\d a_{k1}$ that the output accrues.

\begin{figure}[t]
	\centering
	\includegraphics[width=0.7\textwidth]{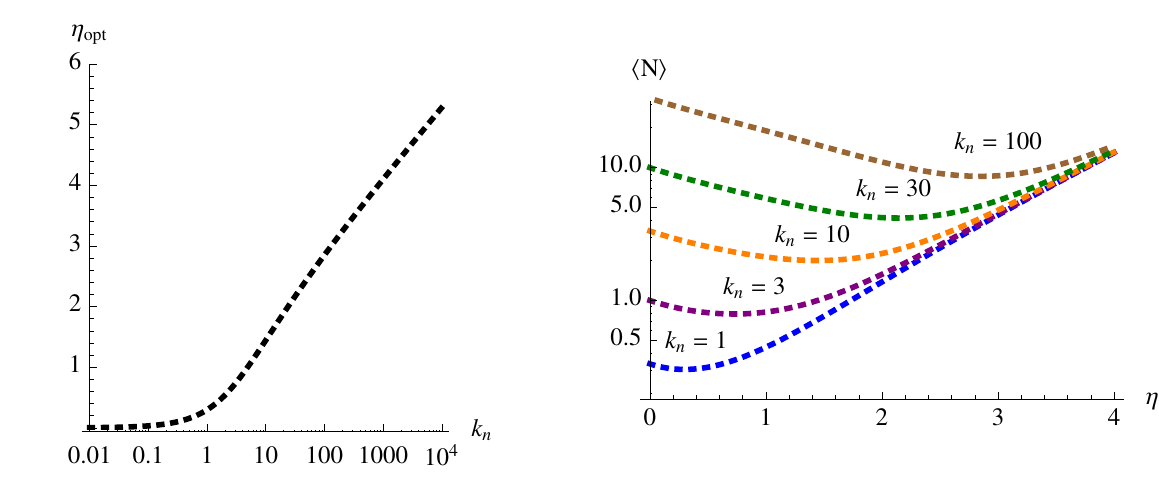}
	\caption{Left: Optimal heterodyne amplification $\eta$ as a function of plant noise.  Right: LQR as a function of controller amplification, for five different noise values.}
	\label{fig:11-f6b}
\end{figure}

The optimal heterodyne controller uses a 50-50 beamsplitter so we set $\alpha = 1/\sqrt{2}$ in (\ref{eq:11-gj-het}).  It too has very fast dynamics that can be adiabatically eliminated to give:
\begin{eqnarray}
	\d\tilde{a}_x & = & \xi (\d a_x + \d a_{k1,x}) + \d a_{k2,p} \nonumber \\
	\d\tilde{a}_p & = & \xi (\d a_p - \d a_{k1,p}) + \d a_{k2,p}
\end{eqnarray}
Or equivalently:
\beq
	\d\tilde{B} = \xi (\d B + \d B_{k1}^\dagger) + \d B_{k2}
\eeq
The heterodyne controller amplifies both quadratures, but there is an additional noise due to splitting the beam before measurement, $\d B_{k1}$, as well as the measurement noise itself.  The LQR can be computed analytically, and the analytic result agrees with the numerical optimizer.  Setting $\xi = \sinh(\eta)$, we have:
\beq
	\avg{a^\dagger a}_{\rm cl} = \frac{k_2 \sinh^2\eta + k_3 k_n}{k_1 + k_2 + k_3 + 2\sqrt{k_1 k_2} \sinh\eta} \label{eq:11-lqr-cav-cl}
\eeq
This is plotted in Fig. \ref{fig:11-f6b}.  As the plant noise increases, so does the controller's optimal amplification.  It does not do well to increase the amplification indefinitely, however, since this also adds noise into the system.  From Fig. \ref{fig:11-f4}, one can also see that measurement control does well at reducing the photon number for large $k_n$, but in the quantum regime, $k_n \lesssim 1$, it has hardly any effect at all.

\subsection{Coherent Control}

The three coherent controllers of interest are the cavity controller and the two OPO setups, as shown in Figure \ref{fig:11-f1}.  The optimizer consistently showed that the best cavity controller is in fact the trivial controller (which is the special case of a cavity with mirror transmittivity set to zero).  Because of this, we do not consider empty cavity controllers in this section.  The OPO controllers, on the other hand, have more interesting behavior.

\begin{figure}[t]
	\centering
	\includegraphics[width=0.60\textwidth]{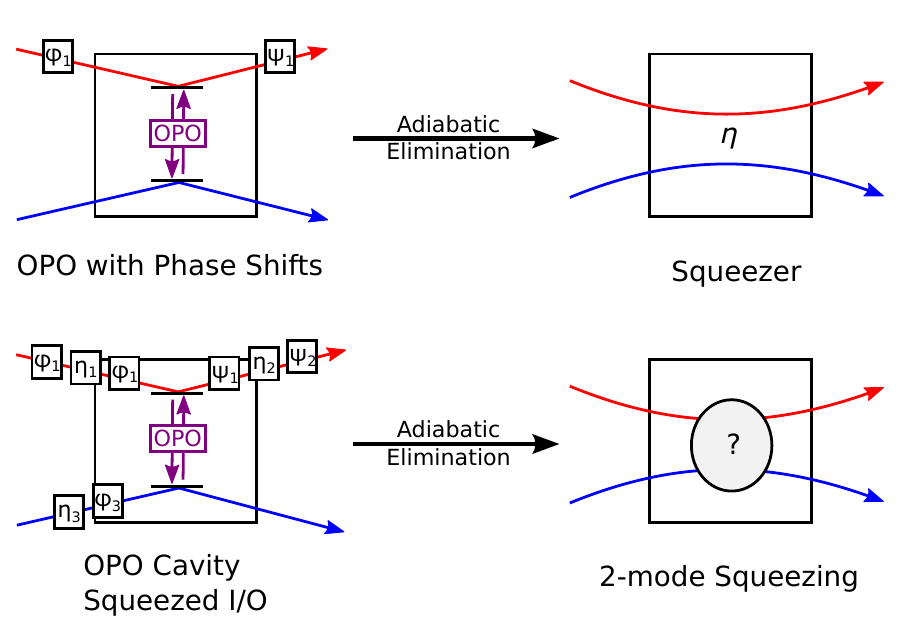}
	\caption{Optical parametric oscillators adiabatically eliminate into ideal squeezers.}
	\label{fig:11-f5}
\end{figure}

As in the classical case, it was discovered that the best coherent controllers always had dynamics that were much faster than the plant timescales and could be adiabatically eliminated.  A single OPO will adiabatically eliminate to a squeezer with the following input-output relations:
\beq
	\d\tilde{a}_x = e^{\eta} \d a_x,\ \ \ \d\tilde{a}_p = e^{-\eta} \d a_p \label{eq:11-hom-squ}
\eeq
(up to input and output phase shifts).  An OPO system with squeezed inputs and outputs, which can in principle replicate any 2-port linear quantum system with a single internal degree of freedom \cite{Nurdin2009b}, will adiabatically eliminate to arbitrary two-mode squeezer (Fig.\ \ref{fig:11-f5}).  As far as this control problem is concerned, the {\it best} two-mode squeezer is the linear amplifier, given by the input-output relations:
\begin{eqnarray}
	\d\tilde{a}_{1} & = & \cosh(\eta) \d a_1 + \sinh(\eta) \d a_2 \nonumber \\
	\d\tilde{a}_{2} & = & \sinh(\eta) \d a_2 + \cosh(\eta) \d a_2
\end{eqnarray}

Analytic formulas can be derived straightforwardly from the quantum stochastic differential equations.  For the squeezer:
\beq
\avg{a^\dagger a}_{\rm sq} = \frac{\mbox{Re}\left[\left(k_2\sinh^2\eta  + 2 k_n\right) - 2 \frac{k_2\sqrt{k_1k_2}\cosh\eta\sinh^2\eta}{G + 2i \Delta} e^{i \phi}\right]}{\mbox{Re}\left[G - 4k_1k_2\sinh^2\eta/(G + 2i\Delta)\right]} \label{eq:11-sq-1md}
\eeq
where
\beq
	G \equiv k_1 + k_2 + k_3 + 2\sqrt{k_1 k_2}\cosh(\eta) e^{i\phi}
\eeq
For the linear amplifier:
\beq
	\avg{a^\dagger a}_{\rm 2-sq} = \frac{k_2 \sinh^2\eta + k_3 k_n}{k_1 + k_2 + k_3 + 2\sqrt{k_1 k_2}\cosh\eta} \label{eq:11-sq-2md}
\eeq
Qualitatively, the results for the heterodyne controller, Eq.\ (\ref{eq:11-lqr-cav-cl}) and the linear amplifier, Eq.\ (\ref{eq:11-sq-2md}) look very similar.  Both the heterodyne controller and the linear amplifier reduce the cavity's photon number by amplifying the feedback signal, but also add noise to the system.  For equivalent levels of amplification (compare (\ref{eq:11-lqr-cav-cl}), substituting $\sinh\eta \rightarrow \cosh\eta$, to (\ref{eq:11-sq-2md})) the classical controller adds extra noise into the system from the measurement process.  When $k_n$ and $\eta$ are large, this extra noise is negligible, but in the quantum regime where $k_n$ and $\eta$ are $\lesssim 1$, this noise can play a major role in making the linear amplifier outperform the heterodyne controller.

As far as optimization is concerned, Equations (\ref{eq:11-sq-1md}--\ref{eq:11-sq-2md}) are simple enough to apply.  Finding the best controller just involves minimizing these functions with respect to $\eta$.  But remember that it was not at all obvious that the best quantum controller should be an adiabatically eliminated squeezer.  This had to be demonstrated by optimizing the general OPO controller, which has many more parameters, and comparing the result to that of the squeezer.  This required a {\it Mathematica} package to quickly convert circuit diagrams to ABCD models, and an efficient optimizer to find the best controller parameters.

Notice that, for large $k_n$, the performance of the two quantum controllers follows the classical performance.  In the classical limit, the OPO / squeezer is amplifying a single quadrature and feeding this back into the plant (with the proper phase shift).  Likewise, the classical controller measures a single quadrature, amplfies that signal and sends this back into the plant.  Thus, the OPO / squeezer is a ``homodyne-like'' controller in the classical limit.  By contrast, the linear amplifier amplifies both modes equally and feeds back the result, making it a ``heterodyne-like'' controller which tracks the performance of the heterodyne controller in the classical limit.

\begin{figure}[t]
	\centering
	\includegraphics[width=0.70\textwidth]{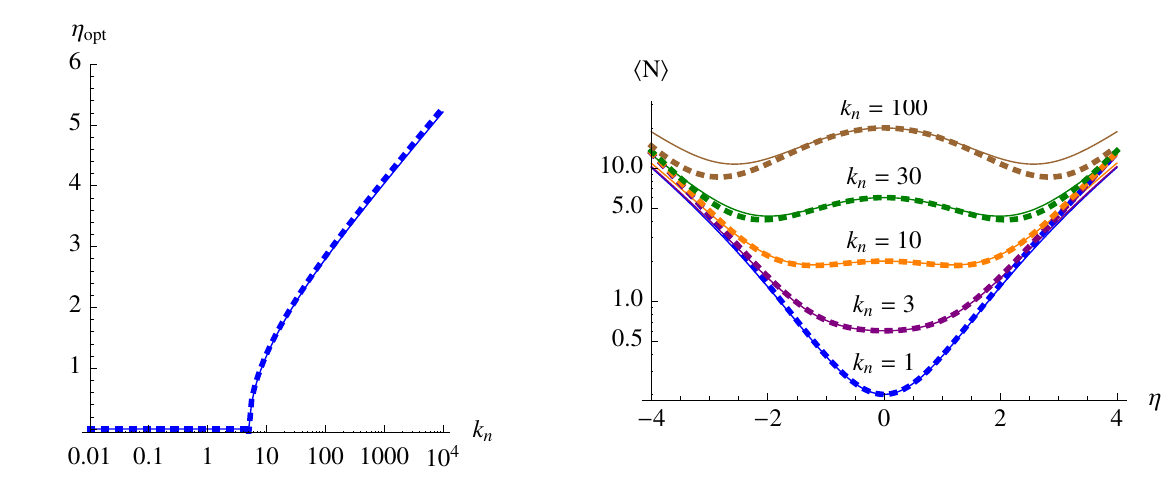}
	\caption{Left: Optimal squeezing for the squeezer (solid) and two-mode squeezer (dotted) controllers.  Right: Performance as a function of squeezing for multiple noise levels.}
	\label{fig:11-f6}
\end{figure}

However, in the quantum regime, this distinction is lost and both of the quantum controllers track the performance of the trivial controller.  Below a threshold value of
\beq
	k_{n,\rm min} = \frac{k_1(k_1 + k_2 + k_3 + 2\sqrt{k_1 k_2})}{\sqrt{k_1 k_2}k_3}
\eeq
(for this system, $k_{n,\rm min} = 5$), any squeezing will {\it increase} the noise in the cavity, so the optimal value of $\eta$ is zero -- in other words, for $k_n \leq 5$, the best controller is the trivial controller.

As Fig.\ \ref{fig:11-f6} illustrates, when $k_n > 5$, the best controller has a nonzero amount of squeezing.  We plot the controller performance as a function of squeezing for five different noise levels on the right pane of the figure.  Intuitively, this is a battle between the noise {\it introduced} by squeezing and the noise {\it removed} by constructive interference with the light leaking out mirror $2$.  When $\eta$ is low, the latter dominates.  By increasing the squeezing, we effectively increase the amplitude of the field impinging upon mirror $2$.  Recall that the trivial controller worked by constructive interference between this field and the light leaking out of mirror $2$.  By increasing this field's amplitude, we magnify the effect of this interference; this reduces the overall cavity photon number.  This explains the $\cosh\eta$ term in the denominator of (\ref{eq:11-sq-2md}).  But a squeezed vacuum carries photons of its own, and some of these photons leak back into the cavity.  If the squeezing is too high, this winds up increasing the photon number, giving rise to the $\sinh^2\eta$ term in (\ref{eq:11-sq-2md}).  Above the threshold temperature $k_{n,\rm min}$, the ideal $\eta$ lies somewhere between these extremes.

Below the threshold temperature, the cavity photon number is so low that the interference effect never wins out -- squeezing the control field always introduces more photons in the cavity, and the best controller involves looping the output from mirror $1$ into mirror $2$ without squeezing -- the trivial controller.

The optimal coherent controller is, in principle, an ideal squeezer -- i.e.\ a squeezer of infinite bandwidth.  In a realistic device, the controller will have a finite bandwidth that is limited by design constraints and may not be much larger than the bandwidth of the plant.  Since this results in a control output that is not equally squeezed at all wavelengths, the performance of the real squeezer will be worse than that of the ideal squeezer.

\begin{figure}[t]
	\centering
	\includegraphics[width=0.56\textwidth]{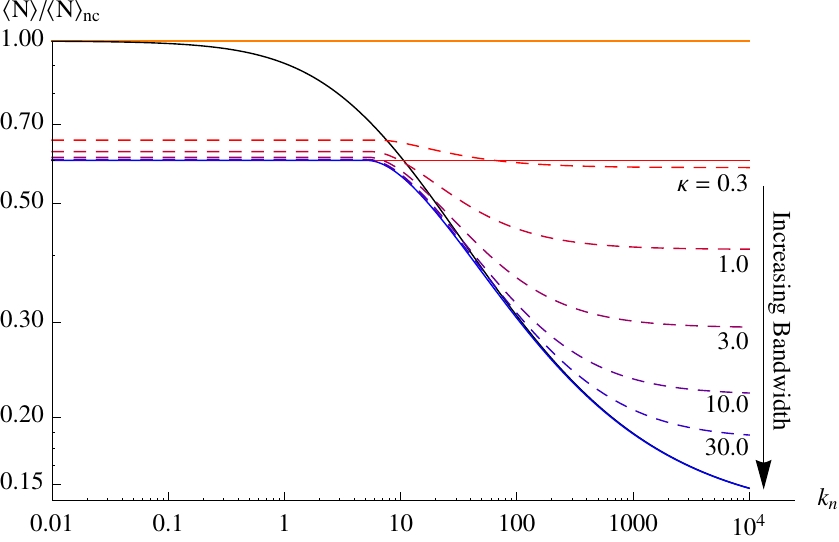}
	\caption{Performance plots for the trivial controller, homodyne measurement, and the ideal squeezer (solid), and OPO controllers of bandwidth $\kappa = 0.3, 1.0, 3.0, 10.0$, and $30.0$ (dashed).}
	\label{fig:11-f20}
\end{figure}

As an example, we compare the ideal ``homodyne-like'' controller, the squeezer in (\ref{eq:11-hom-squ}), to the OPO in (\ref{eq:11-opo-slh}).  As design constraints, we require that the OPO be driven on resonance (no detuning, $\Delta = 0$), with a fixed cavity bandwidth ($\kappa_1 \equiv \kappa$ fixed and nonzero, $\kappa_2 = 0$).  Figure \ref{fig:11-f20} plots the OPO performance for five different bandwidths $\kappa$.  As expected, the best OPO is comparable to the optimal squeezer when the bandwidth is much larger than the plant's ($\kappa \gg k_1 + k_2 + k_3$), but when the two are of the same order of magnitude, the OPO hardly performs any better than the trivial controller.

The reduced performance of the OPO can be understood in the context of the interference arguments made previously.  The controller minimizes the number of photons in the plant by maximizing the amount of light that leaves the cavity through mirror 2 without injecting too much additional noise.  The controller output must have a large amplitude, and must be in phase with the light leaking out of mirror 2 from the plant.  The squeezer amplifies all frequencies without altering their phase.  Signals passing through the OPO, on the other hand, pick up a phase shift depending on whether they are above or below the OPO's resonance, and only get amplified if they are on resonance.  For a narrow-bandwidth OPO, only a small fraction of the input signal passes through the OPO amplified and have the phase shift needed to produce the desired interference, reducing the performance gain of the controller.

\section{Optical Feedback Control of a Mechanical Oscillator}

Optomechanical oscillators -- mechanical springs that couple to an optical field via a cavity -- have been a topic of tremendous recent interest in the physics community~\cite{Marquardt2009}. A central goal has been to find ways to exploit optomechanical coupling to cool the mechanical oscillator from ambient temperature to its ground state, using optical feedback.

\begin{figure}[t]
	\centering
	\includegraphics[width=0.58\textwidth]{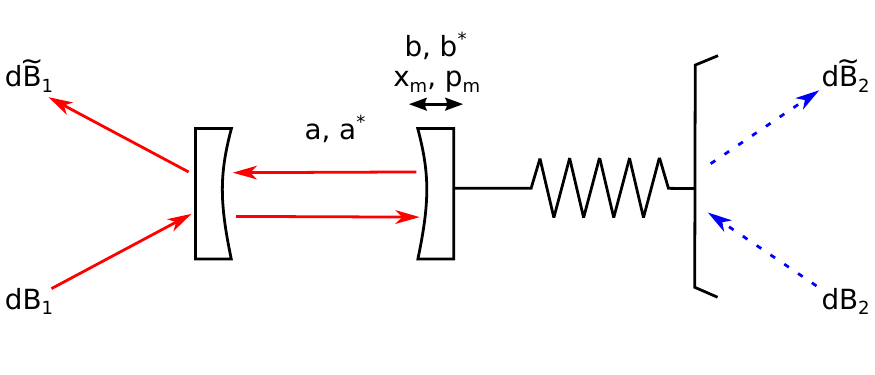}
	\caption{Single cavity with modes $a, a^\dagger$, coupled to a mechanical oscillator with modes $b, b^\dagger$.}
	\label{fig:11-f7}
\end{figure}

In this section we analyze the optomehcanical oscillator as a coherent control system, with the spring comprising the plant, and with optical probing and feedback.
We optimally cool the oscillator by solving the LQG control problem for the cost function $\avg{b^\dagger b}$, where $b$ is the spring's annihilation operator. While the control setups we consider may appear impractical from an experimental perspective, we will discuss how they can be related to systems that are more realistic to implement.

At the heart of this control problem is the ``adiabatically eliminated cavity,'' depicted in Figure \ref{fig:11-f7}.  If we go into the rotating frame for the light, this has the SLH model
\beq
	S = 1_{2\times 2},\ \ \ L = \left[\sqrt{\kappa}a, \sqrt{\Omega/Q}b\right],\ \ \
	H = \hbar\Omega b^\dagger b + \eta a^\dagger a x_m \label{eq:11-slh-mems}
\eeq
where $\Omega$ is the natural spring frequency, $Q$ is the Q-factor, $\kappa$ is the cavity decay parameter, and $m$ is the mirror mass.  See Table \ref{tbl:11-t1}.

\begin{table}[b]
	\begin{center}
	\begin{tabular}{r|cl}
		Qty & & Value \\ \hline
		$K_i$ & = & $4\eta r_i/\kappa_i$ \\
		$r_i$ & = & $\sqrt{P_i/\hbar\omega}$ \\
		$\kappa_i$ & = & $t_i c/2l_i$ \\
		$\eta$ & = & $(\omega/l_i)\sqrt{\hbar/2m\Omega}$ \\
		$k_n$ & = & $(1 - e^{-\hbar\Omega/kT})^{-1}$ \\
		$k_m$ & = & $\Omega/Q$ \\
		& &
	\end{tabular}
	\begin{tabular}{p{0.05\textwidth}|p{0.3\textwidth}|p{0.15\textwidth}}
		Qty & Description & Typical Values \\ \hline
		$P_i$ & Laser power in coherent displacement $r_i$, $i = 1, 2$ & $1\,\mu$W--$1\,$mW \\
		$t_i$ & Power transmittance for cavity mirror $i$.  Inversely proportional to finesse. & $10^{-6}$--$10^{-3}$ \\
		$l_i$ & Length of cavity $i$ & $10^{-6}$--$10^{-1}$m \\
		$m$ & Mass of spring-mounted mirror & $10^{-15}$--$10^{-10}$kg \\
		$\Omega$ & Spring oscillation frequency & kHz--GHz \\
		$Q$ & Spring quality factor & $10^{3}$--$10^7$ \\
		$\omega$ & Laser frequency & $2$--$4\times10^{15}$/s
	\end{tabular}
	\caption{Parameters for the optical cavity controller problem.  See, e.g.\ \cite{Marquardt2009}}
	\end{center}
	\label{tbl:11-t1}
\end{table}

System (\ref{eq:11-slh-mems}) is nonlinear by virtue of the interaction term $\eta a^\dagger a X$.  This term is due to the photon pressure of the field in a cavity, which exerts a physical force on the mirror.  In the limit that the light mode $a$ evolves much faster than the mechanical mode $b$, we can adiabatically eliminate the former to give an SLH system of the form:
\beq
	S = \left[\begin{array}{cc} e^{i\phi(x_m - x_{m0};\eta/\kappa)} & 0 \\ 0 & 1\end{array}\right],
	\ \ \ L = \left[0,\sqrt{\Omega/Q}b\right],\ \ \ H = \Omega b^\dagger b \label{eq:11-el-cav}
\eeq
where
\beq
	\phi(z; \eta/\kappa) = 2\tan^{-1}(2\eta z/\kappa)
\eeq
is the phase shift of the cavity reflected light, as a function of the mirror position (we have absorbed a factor $-1$ in $S$ for convenience).  This is still a highly nonlinear system.  A real optomechanical oscillator is usually driven by a coherent field, and the output that is measured is generally interfered with an equal and opposite field, so as to discern the phase fluctuations on a homodyne detector.  Thus, the real plant system we are interested in is the adiabatically eliminated cavity sandwiched between two coherent displacements.  For a cavity subject to a coherent input of amplitude $r_1$, we write this as:
\beq
	(\mbox{Cav}_1) = L(-r) \triangleleft (\mbox{Cav}) \triangleleft L(r)
\eeq
This has the simple, linear SLH model:
\beq
	S_1 = 1_{2\times 2},\ \ \ L_1 = \left[K_1 x_m, \sqrt{\Omega/Q}b\right],\ \ \
	H_1 = \Omega b^\dagger b \label{eq:11-sand-cav}
\eeq
with $K_1 = 4\eta r_1/\kappa_1$ is the effective coupling between the spring and the field, which need not be positive or even real.  The $x_m$-coupling to the field $\d a_1$ gives rise to the following input-output relations:
\begin{eqnarray}
	& & \left\{\begin{array}{rcl}
		(\d x_m)_1 & = & 0 \\
		(\d p_m)_1 & = & -2K_1 \d a_{1p}
		\end{array}\right. \nonumber \\
	& & \left\{\begin{array}{rcl}
		\d\tilde{a}_{1x} & = & \d a_{1x} + 2K_1 x_m \d t \\
		\d\tilde{a}_{1p} & = & \d a_{1p}
		\end{array}\right.
\end{eqnarray}
The state variable $x_m$ is imprinted on the output $\d\tilde{a}_{1x}$, so by measuring the $x$-quadrature of the output field, we can deduce the value of $x_m$; this allows us to use the mirror as a ``measurement'' device, learning information from the output field.  Note that this only works for $\d\tilde{a}_{1x}$; no information is imprinted onto the $p$-quadrature of the output.  Conversely, by sending in a particular input $\d a_{1p}$, we can alter the state of the system; this allows us to use the mirror as a ``feedback'' device.  Note likewise that feedback is not possible via the $\d a_{1x}$ channel, which does not affect the system.

\begin{figure}[t]
	\centering
	\includegraphics[width=0.57\textwidth]{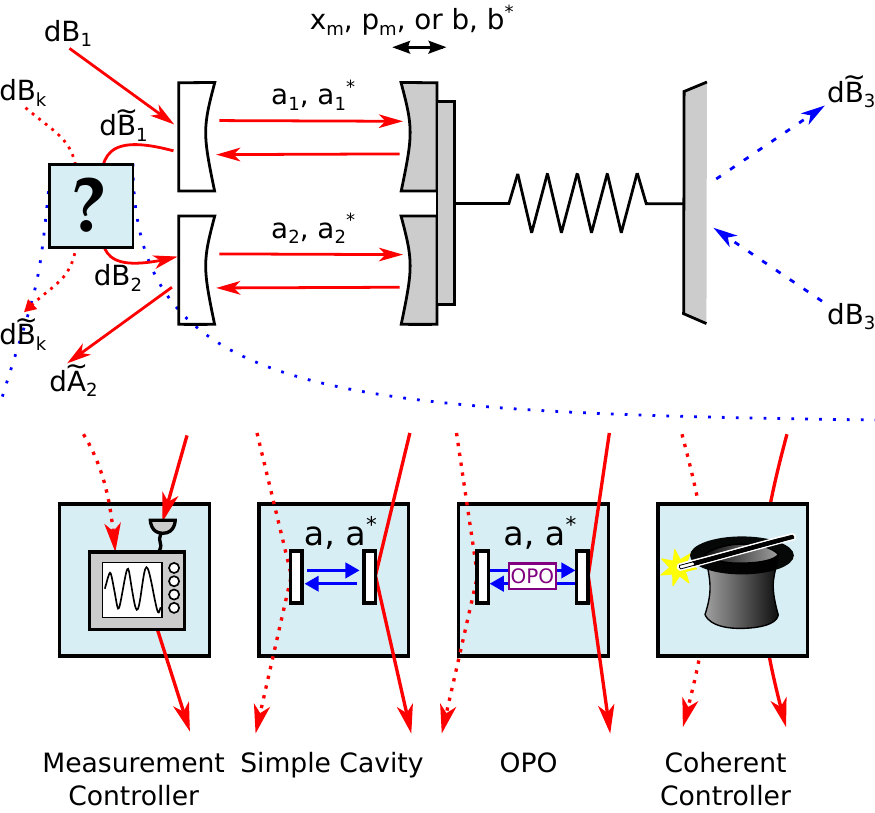}
	\caption{Control-system setup for the mechanical oscillator cooling problem.  Four potential controller designs.}
	\label{fig:11-f8}
\end{figure}

\subsection{Plant System}

The plant-controller setup is shown in Figure \ref{fig:11-f8}.  The plant system consists of two (adiabatically eliminated) cavities coupled to the same mirror.  The output from the first cavity, $\d\tilde{B}_1$, goes into the controller, and the controller output is fed back into the second cavity input $\d B_2$.  Not shown are the two coherent displacements (lasers) putting fields $\d B_1$ and $\d B_k$ into nonvacuum coherent states.  These coherent fields allow us to replace the cavity with model (\ref{eq:11-el-cav}) with the linearized model (\ref{eq:11-sand-cav}).  Since the system is now linear, this becomes an LQG control problem.  The combined plant-controller system can be viewed as a feedback loop from output $\d\tilde{B}_1$ to input $\d B_2$, or conversely, we can write it as a series product
\beq
	(\mbox{Sys}) = \left[(\mbox{Cav}_2 \boxplus I_1) \triangleleft \mathcal{K} \triangleleft
		(\mbox{Cav}_1 \boxplus I_1)\right] \boxplus (\mbox{Spr}) \label{eq:11-alg-spring}
\eeq
where (Sys) is the combined system, $\mbox{Cav}_i$ is the $i^{th}$ cavity, with SLH model $(1, \sqrt{k_i}x_m, \_)$, (Spr) gives the spring and phonon couplings, SLH model $(1, \sqrt{k_m} x_m, \Omega b^\dagger b)$, and $\mathcal{K}$ is the controller.  See Figure \ref{fig:11-f9}.

\begin{figure}[t]
	\centering
	\includegraphics[width=0.4\textwidth]{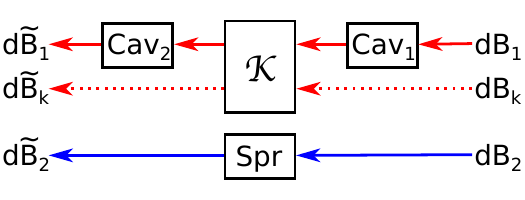}
	\caption{Equivalent view of the plant-controller setup shown in Figure \ref{fig:11-f8}.  See Eq.\ (\ref{eq:11-alg-spring}).}
	\label{fig:11-f9}
\end{figure}

The controllers we consider here are not unlike those for the simple cavity.  It is not difficult to show using Eq.\ (\ref{eq:11-alg-spring}) that the trivial controller amounts to no control at all at best, and additional noise at worst.  The classical controller measures the output from mirror cavity 1 and sends an input in to cavity 2, as a function of the controller's internal state.  (Note that we only need to consider a classical controller that measures the $x$ quadrature $\d\tilde{a}_{1x}$; $\d\tilde{a}_{1p}$ contains no information about the plant's state.)  The simple cavity and OPO cavity coherently process the signal rather than destroying it in a measurement.  Finally, we considered the most general coherent controller, an open quantum system specified by arbitrary $A, B, C, D$ matrices satisfying the realizability relations.  For the LQG problem of minimizing $\avg{b^\dagger b}$, we found optimal controllers in each class for the following plant system:
\begin{eqnarray}
	\Omega & = & 100\ \mbox{(arbitrary units)} \nonumber \\
	k_m & = & 0.01 \nonumber \\
	Q & = & 10000 \nonumber \\
	k_n & = & 10^{-9}\mbox{--}10^9
\end{eqnarray}
In the optimization, we are allowed to vary both the controller parameters {\it and} the couplings $K_1$, $K_2$ to the cavities in (\ref{eq:11-sand-cav}).  This is because the couplings depend on the input laser powers $P_i$ (in addition to the mirror transmittances $t_i$), which are external quantities (see Table \ref{tbl:11-t1}) rather than fixed properties of the plant itself.  Here we will operate primarily under the assumption $K_1 = -K_2 \equiv K$; this is a reasonable assumption that avoids classical solutions with divergent controller gain, but we also show that the coherent controllers discussed here outperform the best classical controllers even when this assumption is relaxed.

\begin{figure}[t]
	\centering
	\includegraphics[width=0.56\textwidth]{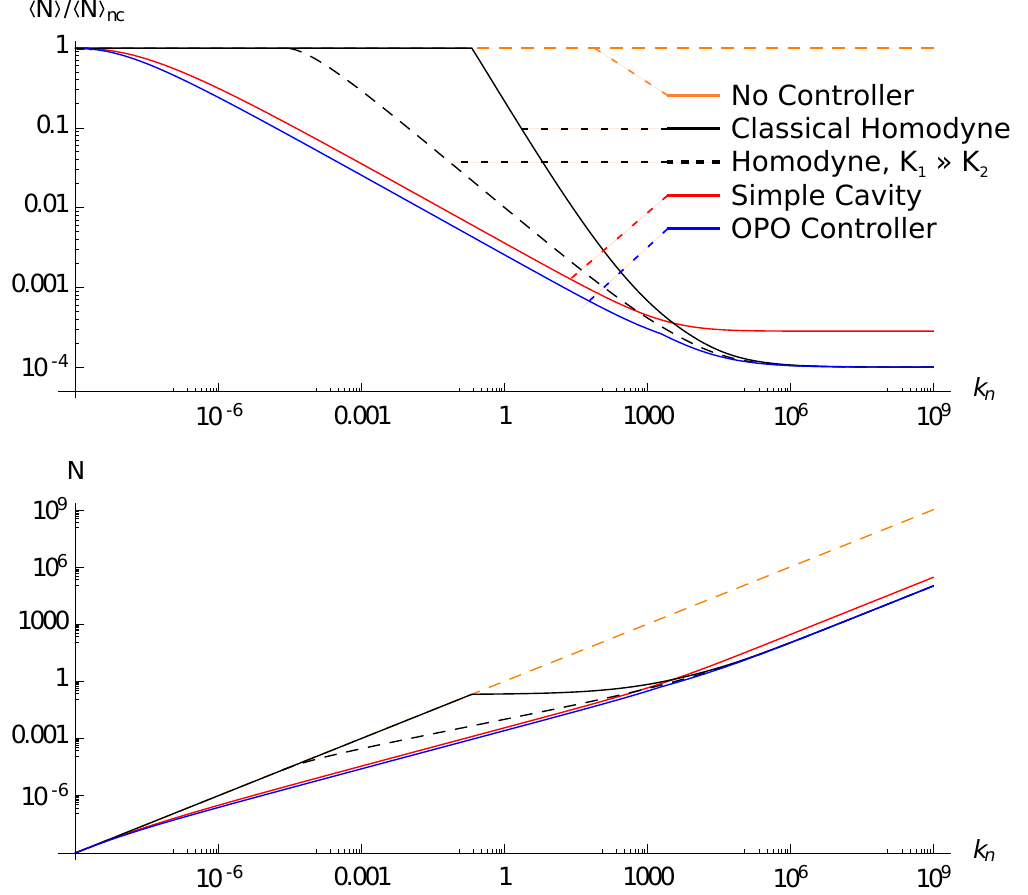}
	\caption{Bottom: Plot of the average phonon number $\avg{N} = \avg{b^\dagger b}$ of the mechanical oscillator for three different control schemes.  Top: Phonon number reduction, relative to no-control case.  The general coherent controller result is not shown, since it overlaps the OPO line, the optimal coherent controller being an OPO cavity.}
	\label{fig:11-f10}
\end{figure}

Figure \ref{fig:11-f10} plots the performance of the measurement, simple cavity, and OPO controllers.  For very low ambient temperatures where the noise is weak, the plant is nearly in its ground state to begin with, and none of the controllers can reduce its value.  This differs from the optical cavity.  In the cavity, we used a ``trivial controller'' to cause the light leaking out of mirror $1$ to interfere constructively with the light leaking out mirror $2$, increasing the net dissipation from $k_1+k_2$ to $(\sqrt{k_1} + \sqrt{k_2})^2$.  No such scheme exists in the oscillator because phonons do not ``leak out'' of the system the same way photons leak out of an optical cavity.

At high temperatures, the best classical controller and the OPO controller do equally well, each reducing the phonon number by a factor of exactly $Q = 10000$.  The cavity controller does reasonably well, reducing the phonon number by a factor of about $0.354Q = 3540$.  These results are not very surprising.  The high-temperature limit takes our oscillator into the classical regime, where vacuum noise is negligible and no coherent controller can hope to outperform the best classical controller.

The interesting region lies between these two limits.  Here, there is a sharp cutoff, near $k_n \approx 0.2$, below which the classical measurement controller becomes useless.  As explained below, the classical controller must add noise to the system to make a measurement; below a certain threshold, the gains from control are offset by the noise from measurement.  In this region, the cavity and OPO controllers do significantly better than the classical controller, in some places by a factor of 100--200.

\subsection{The Classical Controller}

\begin{figure}[t]
	\centering
	\includegraphics[width=0.66\textwidth]{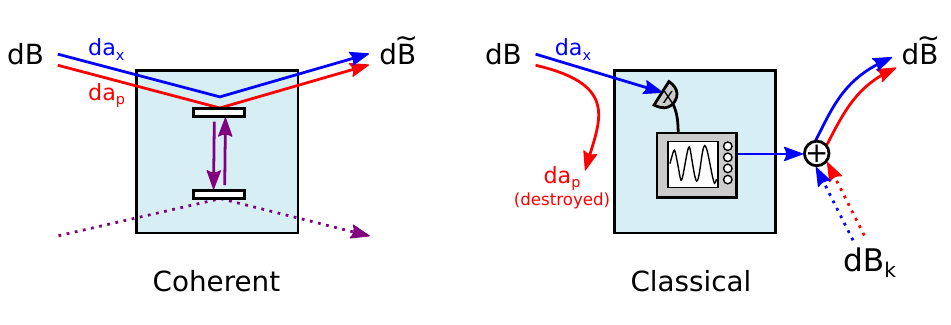}
	\caption{Flow of $x$- and $p$-quadrature signals (blue and red, respectively) in the classical and coherent controllers.}
	\label{fig:11-f16}
\end{figure}

The classical controller works by measuring the plant output field ($\d B$ in Figure \ref{fig:11-f16}) and inferring the plant's state from this measurement.  From the inferred plant state, the controller applies a feedback signal, which is added to an auxiliary vacuum input $\d B_k$ and sent back to the plant.

The plant output contains two quadratures, but only one of them contains information about the system.  Thus, in our classical controller we choose to measure the $x$-quadrature of the output, and necessarily discard the $\d B_p$.  This is the optimal control strategy in the classical case because $\d B_p$ does not contain any information about the system.  Like any LQG-optimal controller, the classical controller consists of a Kalman filter, which estimates the plant state, plus a feedback element.

The classical controller adds two sources of noise to the plant.  First, by sending a laser through the measurement cavity $(\mbox{Cav}_1)$, it adds {\it measurement noise}, with an amplitude that scales as $O(K)$.  Second, the feedback field $\d\tilde{B}$ (with a vacuum noise component due to the auxiliary field $\d B_k$) is sent through the controller, adding a {\it feedback noise} of equal magnitude, also $O(K)$.  Both of these factors increase the cavity phonon number by $O(K^2)$, independent of the noise $k_n$.  The control loop will decrease the cavity phonon number by an amount proportional to the present phonon number, which increases with $k_n$.  In the high-$k_n$ limit, the ``control'' term dominates and the coupling $K$ is large.  By contrast, in the low-$k_n$ limit, the ``noise'' term is dominant, and the optimal value of $K$ is small or zero -- no measurement controller can effectively reduce the phonon number, since the noise incurred will more than offset any gains from control.

An important thing to note is the role the $p$-quadrature field $\d a_p$ plays in this noise budget.  It is true that $\d a_p$ does not contain any information about the plant state.  But this quadrature still plays an important part, since $\d a_p$ gives rise to the noise in the measurement cavity, and $\d\tilde{a}_p$ gives rise to the noise in the feedback cavity.  Because $\d a_p$ and $\d\tilde{a}_p$ are independent (the former being destroyed in the $\d a_x$ measurement), their noises add up.  The beauty of coherent control is that we can process the $x$-field without destroying $\d a_p$ and the measurement and feedback noises become correlated.  If done right, they cancel each other out.

\begin{figure}[b]
	\centering
	\includegraphics[width=0.5\textwidth]{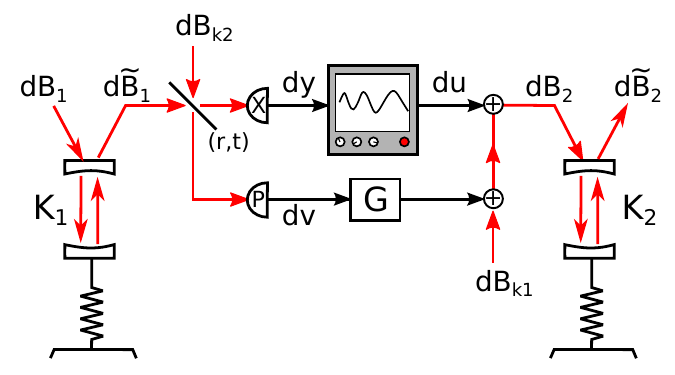}
	\caption{Heterodyne-based measurement controllers, which measure both quadratures of the beam by splitting it, do not not outperform the best homodyne controller for this system.}
	\label{fig:11-f17}
\end{figure}

If we are free to relax the $K_1 = -K_2$ assumption, then the classical controller does somewhat better (dashed line in Figure \ref{fig:11-f10}), but still underperforms the coherent schemes discussed below.  When $K_1 \neq K_2$, the optimal classical controller tends to have $K_2 \ll K_1$, which greatly suppresses the measurement noise.  To compensate for this disparity, the controller must have a large classical gain.

It might be thought that a heterodyne-based control scheme, like that in Figure \ref{fig:11-f17} could perform better than the best homodyne controller.  After all, the homodyne controller is just a special case of the heterodyne controller, where the beamsplitter has a transmissivity of 100\%.  Moreover, one might imagine using a heterodyne scheme to cycle part of the $\d\tilde{a}_p$ quadrature back into the plant, canceling out part of the measurement noise with the feedback noise.  However, we find numerically that the most general heterodyne controller does not perform any better -- either with $K_1 = K_2$ or not.  The extra noise added from splitting the beam outweighs any of the benefits of the control scheme.

\subsection{Simple Cavity Controller}

An empty optical cavity with two input / output ports has the following SLH model:
\beq
	S = 1_{2\times 2},\ \ \ L = \left[\sqrt{\kappa_1}a, \sqrt{\kappa_2}a\right],\ \ \
	H = \Delta a^\dagger a
\eeq
Here the $\kappa$'s are mirror decay parameters and $\Delta$ is the detuning of the cavity.  The QSDEs for the cavity are easy to derive:
\begin{eqnarray}
	\d a & = & (-i\Delta - \kappa/2)a\,\d t + \sqrt{\kappa_1} \d\tilde{B}_1 + \sqrt{\kappa_2} \d\tilde{B}_2 \nonumber \\
	\d\tilde{B}_i & = & \d B_i + \sqrt{\kappa_1}a\,\d t
\end{eqnarray}
Remember that, in addition to the controller parameters, we can vary the input coherent fields, which allows us to vary the plant's  $x_m$-coupling $K$.  The laser field impinging on cavity 1 adds shot noise to the mirror; in this setup, since $K_1 = -K_2 \equiv K$, the shot noise from cavity 1 will exactly cancel the shot noise from cavity 2 (if we let $K_1$ and $K_2$ vary freely, we find that the optimal controller has $K_1 = -K_2$).  As a consequence, the cavity controller has {\it neither} measurement nor feedback noise.

\begin{figure}[b!]
	\centering
	\includegraphics[width=0.6\textwidth]{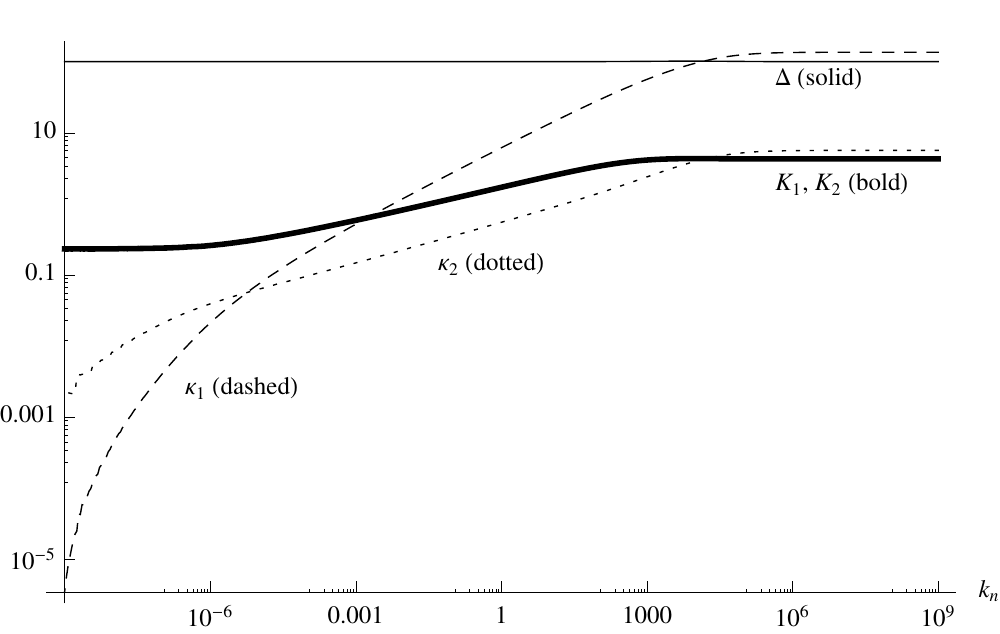}
	\caption{Parameters of the optimal simple cavity controller, as a function of noise strength.}
	\label{fig:11-f11}
\end{figure}

The optimal detuning and couplings are plotted in Figure \ref{fig:11-f11}.  Not surprisingly, as the noise on the mirror is increased, the couplings $K_{1,2}$ and $\kappa_{1,2}$ increase as well.  The detuning $\Delta$, which shows no dependence on the noise power, always remaining at a constant value $\Delta \approx \Omega = 100$ for this system, making the cavity controller setup analogous to two coupled harmonic oscillators, one mechanical and the other optical \cite{Botter2012}.  Absent the couplings, the quadratures $x = a + a^\dagger, p = (a - a^\dagger)/i$ would evolve just as the mirror variables $x_m, p_m$.

This can also be interpreted as a form of sideband cooling.  The detuning $\Delta \approx \Omega$ indicates that our control system is being driven by laser light at a frequency $\omega_{\rm cav} - \Omega$, where $\omega_{\rm cav}$ is the cavity resonance frequency.  The plant-controller coupling serves to convert photons of frequency $\omega_{\rm cav} - \Omega$ to photons of frequency $\omega_{\rm cav}$, cooling the oscillator.  At high temperatures, we need a large cooling rate to counter the noise; this is achieved by using a cavity with a broad bandwidth $\kappa$, so that both $\omega_{\rm cav} - \Omega$ and $\omega_{\rm cav}$ photons are interact effectively with the cavity.  Conversely, at low temperatures, we need to work in the resolved sideband limit $\kappa \ll \Omega$ to suppress quantum fluctuations of the radiation-pressure force \cite{Miao2010, Marquardt2007, WilsonRae2007}.

The effects of this cooling are made manifest on the output power spectrum of the photon channel $\tilde{P}_1(\omega) = \tilde{A}_1(\omega)^\dagger \tilde{A}_1(\omega)$, where $\tilde{A}_1(\omega)$ is the Fourier transform of the stochastic process $\d\tilde{B}_1(t)$.  In the frequency domain, the relevant QSDEs for the combined plant-cavity system are
\begin{eqnarray}
	-i\omega a & = & \Bigl[\bigl(-i\Delta - (\kappa_1+\kappa_2)/2\bigr)a + \sqrt{\kappa_1} K (b + b^\dagger)\Bigr] \nonumber \\ 
	& & + i\omega \sqrt{\kappa_1} A_1 + i\omega \sqrt{\kappa_2} A_2 \nonumber \\
	-i\omega b & = & \left[\left(-i\Omega - \Omega/2Q\right)b - \sqrt{\kappa_1} K (a - a^\dagger)\right] + i\omega \sqrt{\Omega/Q} A_3 \nonumber \\
	-i\omega \tilde{A}_1 & = & -i\omega A_1 + \sqrt{\kappa_1}a
\end{eqnarray}
This power spectrum is plotted in Figure \ref{fig:11-f19}.  As the exiting light is blue-detuned, it reduces the phonon number in the oscillator, driving it towards the ground state.  For small $k_n$, when the plant and controller are weakly coupled, there is a single sideband corresponding to the plant's oscillation frequency $\Omega$.  When $k_n$ is large, the plant and controller become strongly coupled and the combined system resonates at two different frequencies, one larger than $\Omega$ and one smaller.  This is the origin of the sideband splitting in the figure.

\begin{figure}[b!]
	\centering
	\includegraphics[width=0.60\textwidth]{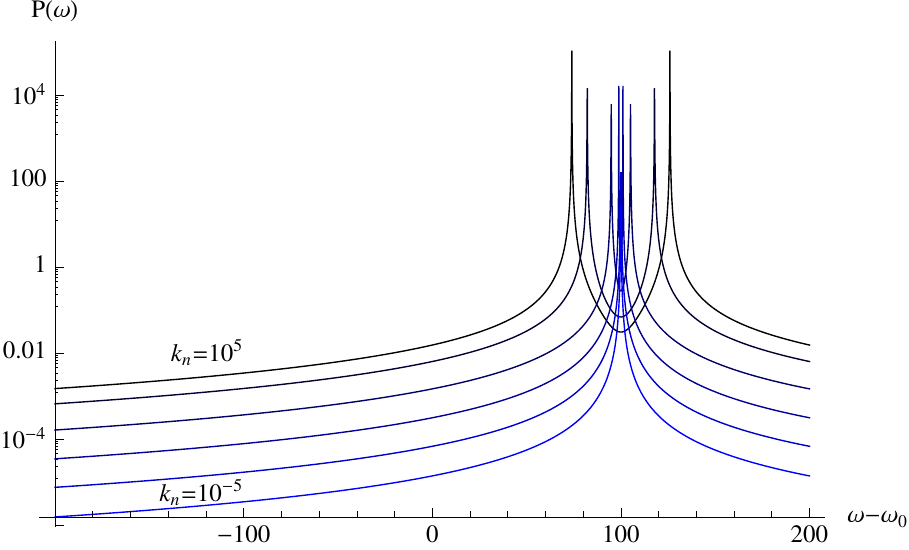}
	\caption{Calculated output spectrum of light exiting the optimal simple cavity controller.  Six values of $k_n$ are plotted, $10^5$ (darkest), $10^3$, $10^1$, $10^{-1}$, $10^{-3}$, and $10^{-5}$ (lightest).}
	\label{fig:11-f19}
\end{figure}

The system can also be understood as a form of coherent Kalman filtering.  Recall that the optimal classical controller works as a Kalman filter, reproducing the state of the plant by measuring one of its outputs.  The cost we paid for the Kalman filtering was additional noise added to the system.  The cavity controller can also be thought of as a Kalman filter, but one that preserves the coherence of the input signal $\d B$.  From a quantum mechanical standpoint, in the classical controller, the $p$-quadrature $\d a_p$ is essentially discarded after the measurement.  In the cavity controller, the field retains its coherent properties and the $\d a_p$ coming out is the same as $\d a_p$ going in.  This makes the noises in the measurement and feedback cavities correlated.  In the present setup, they exactly cancel out.  This cancellation of the measurement noise is what gives the coherent cavity controller its superior performance, particularly in the low phonon-number regime.

Measurement sensing experiments \cite{Tsang2010}, particularly in the context of LIGO \cite{Arcizet2006b}, show similar improvements, but for a different performance metric.  This suggests that LQG control is far from the only problem to benefit from this noise cancellation and coherent feedback; similar gains should be expected in all types of control problems when the plant operates in the quantum regime.

\subsection{OPO Cavity Controller}

\begin{figure}[b!]
	\centering
	\includegraphics[width=0.60\textwidth]{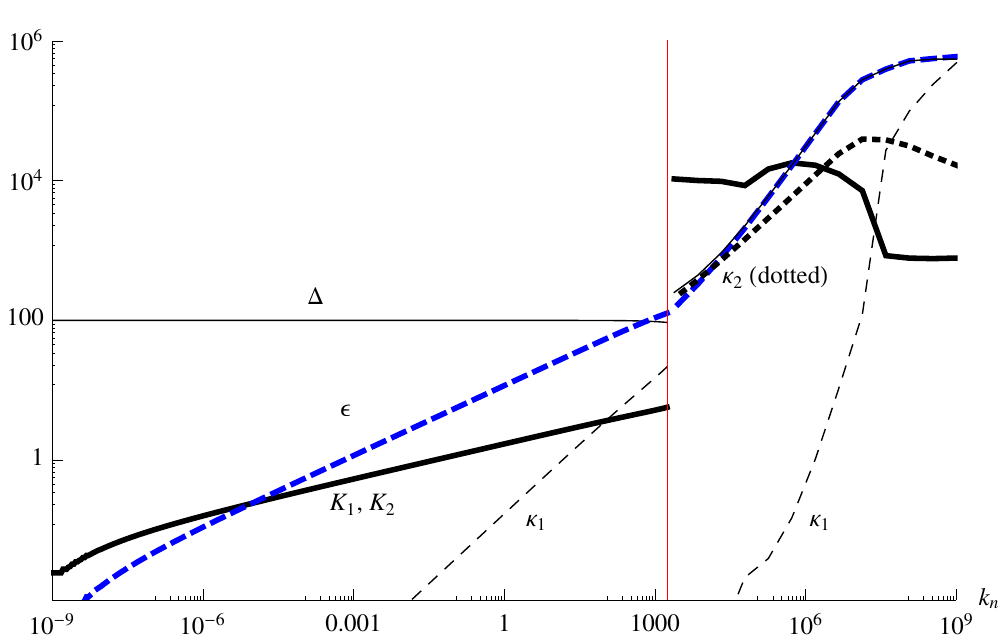}
	\caption{Parameters of the optimal OPO cavity controller, as a function of noise strength.}
	\label{fig:11-f12}
\end{figure}

Recall from Eq.\ (\ref{eq:11-opo-slh}) that the OPO has the following SLH model:
\begin{eqnarray}
	& & S = 1_{2\times 2},\ \ L = \left[\sqrt{\kappa_1} a,\ \sqrt{\kappa_2} a\right], \nonumber \\
	& & H = \frac{1}{4} x^{\rm T} \left[\begin{array}{cc} \Delta - \mbox{Im}(\epsilon) & \mbox{Re}(\epsilon) \\ \mbox{Re}(\epsilon) & \Delta + \mbox{Im}(\epsilon) \end{array}\right]x \nonumber \\
	& & \ \ = \Delta a^\dagger a + \frac{\epsilon^* a^2 - \epsilon (a^\dagger)^2}{2i} \label{eq:11-opo-slh2}
\end{eqnarray}
For fullest generality, the OPO controller is placed between two phase shifters, so the actual controller is $e^{i\phi_1} \triangleleft (\mbox{OPO}) \triangleleft e^{i\phi_2}$.  Between the controller, the phase shifters and the couplings $K_{1,2}$, there are nine free parameters in this LQG problem.  The best OPO controller parameters, found using the optimization code, are plotted in Figure \ref{fig:11-f12}.  As with the cavity controller, the best OPO controller has $K_1 = -K_2$.

For $k_n \lesssim 1800$, the OPO behaves much like the simple cavity.  Its detuning is close to $\Omega$, the coupling $K_1 = -K_2$ increases with $k_n$, and the mirror losses $\kappa_1, \kappa_2$, while small, increase with increasing noise ($\kappa_2$ is too small to be seen on this plot).  For the most part, $\epsilon \ll \Delta$ and the OPO squeezing is only a perturbation on the dynamics of an empty cavity.

At $k_n \approx 1800$, this changes suddenly.  This happens because the OPO controller has two local minima.  Below $k_n \approx 1800$, the empty cavity-like local minimum is smaller, but above this threshold, a new minimum dominates.  In this regime, the coupling $K$ is much stronger than before and the mirrors $\kappa_1, \kappa_2$ are much more lossy.

The OPO controller appears to be the best coherent controller one can make for this system.  We ran the optimizer for general coherent controller, subject to no constraints other than the realizability conditions (\ref{eq:11-prc}).  At no point did we find a coherent controller that outperformed the OPO for this system.  This in mind, ths discontinuity at $k_n$ can be better understood.  As the best relizable controller, the OPO must do at least as well as both the simple cavity and the classical controller.  For weak noise, the simple cavity outperforms the classical controller, so we expect the OPO to look more like a simple cavity.  For strong noise, the classical controller does better, so we expect the OPO to look more like a classical controller, inasmuch as this is possible.  There is no reason to assume that the transition between the two {\it must} be smooth.  It may be marked with bifurcation points, as in Figure \ref{fig:11-f6} for the cavity control problem, or it may occur with a discontinuity in the parameters.  What happens for a general plant / controller system will depend on the landscape of the cost function, and in particular, the behavior of local minima.

\subsection{More Realistic Control Systems}

\begin{figure}[t]
	\centering
	\includegraphics[width=0.60\textwidth]{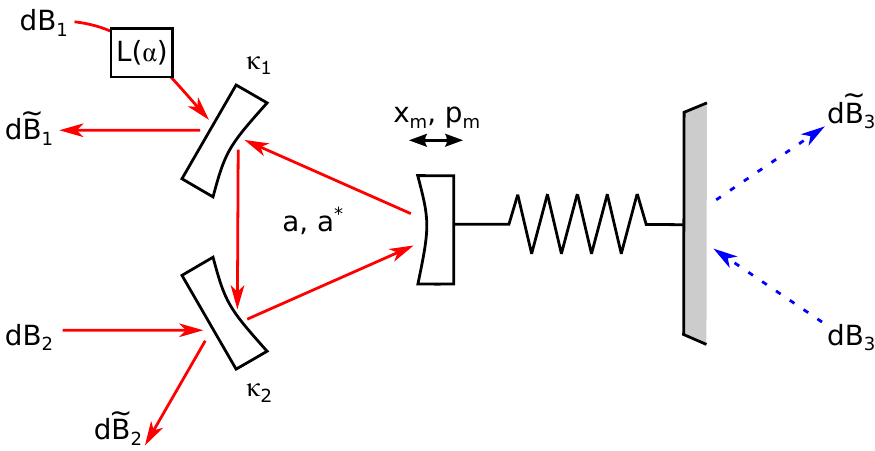}
	\caption{Model for a {\it non}-adiabatically eliminated cavity.}
	\label{fig:11-f13}
\end{figure}

The control systems discussed above can be implemented in principle, but they require two separate mirrors and two separate cavities to be coupled to the same mechanical oscillator, which may prove difficult to build in a laboratory.  Fortunately, one can show that for the cavity controller and the OPO controller, equivalent systems can be realized using a non-adiabatically eliminated cavity with one of its mirrors on a spring.

First, the simple cavity controller.  Recall from (\ref{eq:11-alg-spring}) that the cavity controller system can be modeled as
\beq
	\left[(\mbox{Cav}_2 \boxplus I_1) \triangleleft (\mbox{Cav}) \triangleleft (\mbox{Cav}_1 \boxplus I_1)\right] \boxplus (\mbox{Spr})
\eeq
which has the SLH model
\begin{eqnarray}
	& & S = 1_{2\times 2},\ \ \ L = \left[\sqrt{\kappa_1}a, \sqrt{\kappa_2}a, \sqrt{\kappa_m}b\right] \nonumber \\
	& & H = \omega_c a^\dagger a + \Omega b^\dagger b + \sqrt{\kappa_1}K_1 x_m p_c \label{eq:11-slh-spr1}
\end{eqnarray}

\begin{figure}[t]
	\centering
	\includegraphics[width=0.62\textwidth]{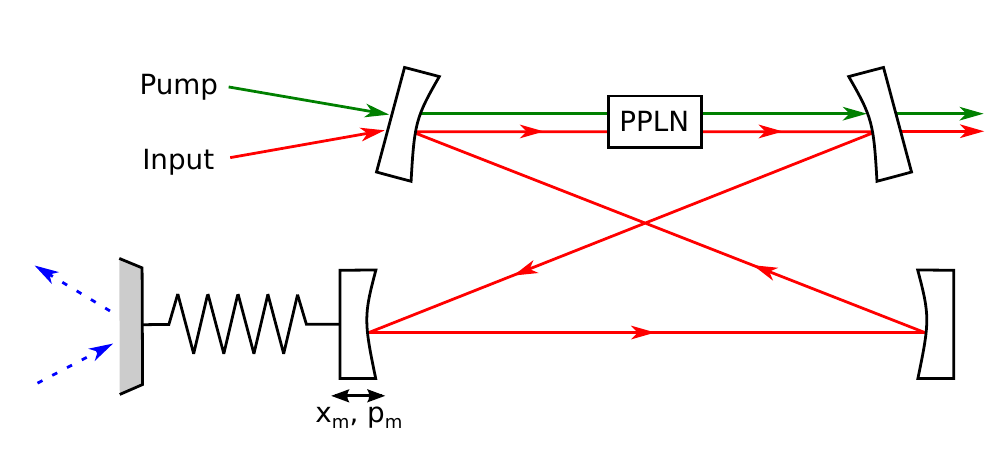}
	\caption{Model for a {\it non}-adiabatically eliminated OPO cavity with a spring mirror.}
	\label{fig:11-f14}
\end{figure}

Now consider a system, depicted in Figure \ref{fig:11-f13}, consisting of a {\it non}-adiabatically eliminated cavity with one of its mirrors attached to a spring.  This has the SLH model:
\begin{eqnarray}
	& & S = 1_{3\times 3},\ \ \ L = \left[\sqrt{\kappa_1}a, \sqrt{\kappa_2}a, \sqrt{k_m} b\right] \nonumber \\
	& & H = \Delta_0 a^\dagger a + \Omega b^\dagger b + \eta a^\dagger a x_m
\end{eqnarray}
A laser $L(\alpha)$ sends a coherent input into mirror $1$, giving the system $\mbox{Cav} \triangleleft (L(\alpha) \boxplus I_2)$.  Of course, the internal dynamics do not depend on anything downstream of the system, so we can just as well use $(L(\alpha') \boxplus I_2) \triangleleft \mbox{Cav} \triangleleft (L(\alpha) \boxplus I_2)$, for any $\alpha'$.  Making substitutions $a \rightarrow a - a_0, b \rightarrow b - b_0$ to center around the equilibrium point, the SLH model becomes:
\begin{eqnarray}
	& & S = 1_{3\times 3},\ \ \ L = \left[\sqrt{\kappa_1}a, \sqrt{\kappa_2}a, \sqrt{k_m} b\right] \nonumber \\
	& & H = \Delta a^\dagger a + \Omega b^\dagger b + \frac{\eta|\alpha|\sqrt{\kappa_1}}{\Delta^2 + (\kappa/2)^2} x_m x_c + \eta a^\dagger a x_m \nonumber \\
	& & \label{eq:11-slh-spr2}
\end{eqnarray}
Ignoring the nonlinear term, this is almost identical to (\ref{eq:11-slh-spr1}).  One can convert the $x_m x_c$ term to an $x_m p_c$ term with a canonical transformation, and the coefficients can be matched by varying $\alpha$.  Thus the systems in (\ref{eq:11-slh-spr1}) and (\ref{eq:11-slh-spr2}) are equivalent, and the ``simple cavity controller'' can be realized in the lab using a single cavity with a mirror attached to a spring.  Cooling an oscillator in this setup has been realized experimentally, though it was not interpreted as a control system \cite{Gigan2006,Arcizet2006b,Kleckner2006}.

The OPO controller is just like the cavity controller, but the Hamiltonian has an additional squeezing term; see (\ref{eq:11-opo-slh}).  The same procedure can be applied to show that the OPO plant-controller system is equivalent to a (non-adiabatically eliminated) OPO cavity with a spring mirror, as shown in Figure \ref{fig:11-f14}.

\subsection{Quantum Refrigerator Analogy}

\begin{figure}[t]
	\centering
	\includegraphics[width=0.57\textwidth]{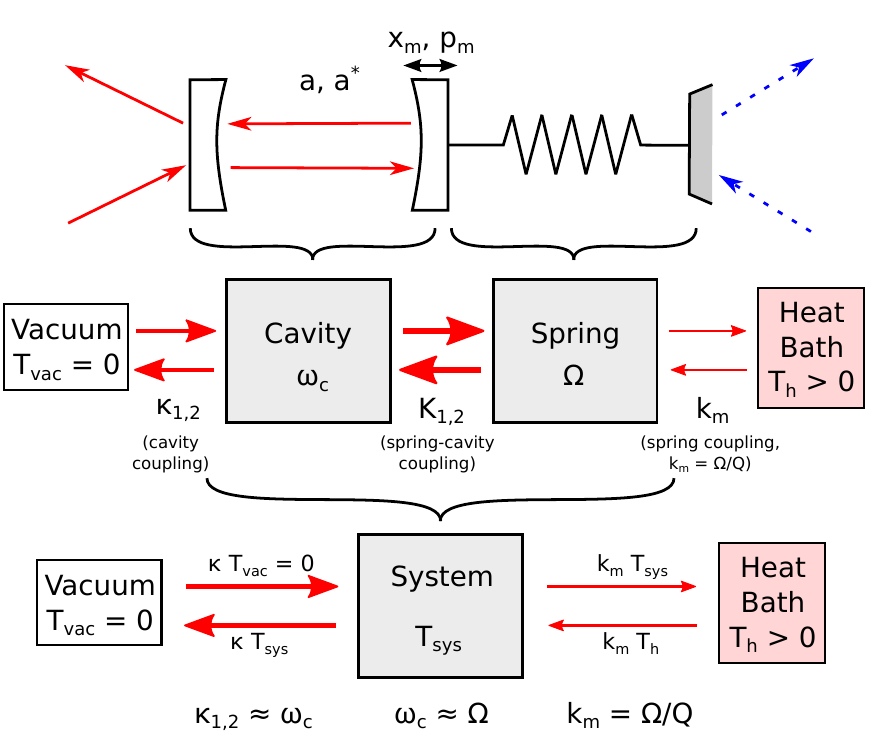}
	\caption{Coherent control problem represented as two coupled thermodynamic systems.}
	\label{fig:11-f15}
\end{figure}

One thing we notice from the optimal controller performance is that, in the strong-noise limit, the optimal controllers -- classical, OPO, cavity -- all reduce the spring phonon number by a factor of about $Q = \Omega / k_m$.  The classical and OPO controllers reduce it by exactly $Q$, while the cavity controller only reduces it by a factor $Q / 2.83$.  This factor-of-$Q$ reduction can be understood by viewing the plant and the controller as thermodynamic systems.

Figure \ref{fig:11-f15} illustrates our point.  Starting with a cavity with a spring mirror, we separate the system into the cavity, which oscillates at a frequency $\omega_c$, and the spring, which oscillates at a frequency $\Omega$.  Each system has its own coupling to the environment.  The cavity couples to a vacuum-state environment ($T = 0$) with coupling strengths $\kappa_1, \kappa_2$, the spring, couples to a heat bath with $T_h > 0$ with strength $k_m$, and a spring-cavity coupling $K_1 = -K_2$ couples the two modes.

If the spring and cavity oscillate at about the same frequency and the spring-cavity coupling is strong compared to the other two, then the ``temperature'' of the spring will be roughly equal to the ``temperature'' of the cavity.  We denote this temperature $T_{\rm sys}$.  One expects the combined system to be in thermal steady-state with both the heat bath and vacuum inputs and outputs; this gives us the energy balance equation:
\beq
	k_m T_h = k_m T_{\rm sys} + \kappa T_{\rm sys}
\eeq
where $\kappa = \kappa_1 + \kappa_2 \sim \omega_c \sim \Omega$, and $k_m = \Omega / Q$.  Solving for the system's steady-state temperature,
\beq
	T_{\rm sys} = \frac{k_m T_h}{\kappa + k_m} \sim \frac{T_h}{Q}
\eeq
From general arguments, we can therefore expect that most good controllers will reduce the spring phonon number by a factor of about $Q$, but that no controller will do significantly better.  Note that, since this argument is based on thermodynamic assumptions that are only approximately valid here, the factor-of-$Q$ reduction is only approximate, and only holds in ths classical limit.  These classical results, unsurprisingly, break down in the quantum regime because, among other things, the effects of vacuum noise inputs become important.

\section{Conclusions}

In this chapter, we have studied the coherent-feedback cooling of linear quantum systems from an LQG control perspective.  The systems were modeled using the SLH framework and the Gough-James circuit algebra, which allow arbitrarily large circuits be constructed in a straightforward and systematic manner.  The evolution of the system was studied using QSDEs, the open-system analogue to the Heisenberg Equations. We wrote {\it Mathematica} scripts based on the QHDL/M framework to model quantum LQG control systems, and designed algorithms to optimize a controller's parameters for a given setup.

For any LQG control problem, there is always a quantum controller that does at least as well as the optimal classical controller. In the quantum regime, when excitation number in the plant is of order unity, we have shown that the best quantum controller can do better -- in some cases, significantly so.  Two systems -- the optical cavity and the optomechanical oscillator -- were studied in detail.  For the former, modest gains were found using coherent control in the low-photon-number regime.  For the latter, the gains were much larger.

One could imagine extending these results to look at non-quadratic cost functions in linear control systems. Indeed, some work has already been done on this matter, focusing on using coherent feedback to maximize the squeezing in a cavity mode \cite{Iida2012}. Taking a control theory perspective may also provide insight into minimizing the noise in optomechanical sensors. In addition, the understanding the superior performance of coherent feedback in linear systems may provide important clues for the design of quantum controllers for nonlinear systems such as optical switches or error correcting codes.

\section*{Appendix}

\renewcommand\thesection{\arabic{chapter}.\Alph{appsection}}
\renewcommand\thesubsection{\arabic{chapter}.\Alph{appsubsection}}

\setcounter{appsection}{1}
\section{SLH and ABCD Models for Systems in this Chapter}
\label{sec:11-app-models}

This appendix introduces the three components mentioned in the chapter -- the empty cavity, the optomechanical oscillator, and the OPO cavity.  We follow the ``quadrature notation'' of Sec.~\ref{sec:04b-quad}.  Start with the empty cavity.  This has the SLH model
\beq
	S = I,\ \ L = \left[\sqrt{k_1} a,\ \sqrt{k_2} a,\ \sqrt{k_3} a\right],\ \
	 H = \Delta a^\dagger a
\eeq
Let $x = a + a^\dagger, p = (a - a^\dagger)/i$ be the Hermitian state variables for this system.  Then we can write out $(S, L, H)$ in the form of Eq.\ (\ref{eq:04b-slhquad}) with the linear system parameters
\begin{eqnarray}
	& & S = I_{3\times 3},\ \ \ 
	\Lambda = \frac{1}{2} \left[\begin{array}{cc}
		\sqrt{k_1} & i\sqrt{k_1} \\ \sqrt{k_2} & i\sqrt{k_2} \\ \sqrt{k_3} & i\sqrt{k_3}
		\end{array}\right],\nonumber \\
	& & R = \frac{1}{2} \Delta I_{2\times 2},\ \ \ 
	r = \lambda = 0,\ \ \ 
	\Theta = J_{2\times 2}
\end{eqnarray}
Using Eqs.\ (\ref{eq:04b-slh-abcd}), the ABCD model is:
\begin{eqnarray}
	& & A = \left[\begin{array}{cc} -\frac{k_1+k_2+k_3}{2} & \Delta \\ 
		-\Delta & -\frac{k_1+k_2+k_3}{2} \end{array}\right],\nonumber \\
	& & -B = C^{\rm T} = \left[\begin{array}{cccccc} \sqrt{k_1} & 0 & \sqrt{k_2} & 0 & \sqrt{k_3} & 0 \\ 0 & \sqrt{k_1} & 0 & \sqrt{k_2} & 0 & \sqrt{k_3} \end{array}\right] \nonumber \\
	& & D = 1_{6\times 6},\ \ \ a = c = 0
\end{eqnarray}
By inspection, one can see that this is equivelent to the well-known input-output equations for an empty cavity:
\begin{eqnarray}
	\d a & = & \left(-i\Delta - \frac{1}{2}\sum_i k_i\right) a\,\d t - \sum_i{\sqrt{k_i} \d B_i} \nonumber \\
	\d\tilde{B}_i & = & \sqrt{k_i} a\,\d t + \d B_i
\end{eqnarray}
In the chapter, we also study the control of an optomechanical oscillator.  Here, we posited an oscillator with one degree of freedom (the mechanical degree of freedom)
\beq
	S = 1_{2\times 2},\ \ \ L = \left[K x_m, \sqrt{\Omega/Q}b\right],\ \ \	H = \Omega b^\dagger b \label{eq:11-sand-cav}
\eeq
where $(x_m, p_m)$ are the Hermitian state variables and $b = (x_m + i p_m)/2$ is the phonon annihilation operator.  Again referring to Eq.\ (\ref{eq:04b-slhquad}) the linear system parameters are:
\begin{eqnarray}
	& & S = I_{2\times 2},\ \ \ 
	\Lambda = \left[\begin{array}{cc} K & 0 \\ \frac{1}{2}\sqrt{\Omega/Q} & \frac{i}{2}\sqrt{\Omega/Q} \end{array}\right], \nonumber \\
	& & R = \frac{\Omega}{2} I_{2\times 2},\ \ \ r = \lambda = 0,\ \ \ \Theta = J_{2\times 2}
\end{eqnarray}
Again, following the standard procedure, we derive ABCD matrices for the model:
\begin{eqnarray}
	& & A = \Omega \left[\begin{array}{cc} -1/2Q & 1 \\ -1 & -1/2Q \end{array}\right], \nonumber \\
	& & B = \left[\begin{array}{cccc} 0 & 0 & -\sqrt{\Omega/Q} & 0 \\ 
		0 & -2K & 0 & -\sqrt{\Omega/Q} \end{array}\right] \nonumber \\
	& & C = \left[\begin{array}{cc} 2K & 0 \\ 0 & 0 \\ \sqrt{\Omega/Q} & 0 \\ 
		0 & \sqrt{\Omega/Q}\end{array}\right],\nonumber \\
	& & D = 1_{4\times 4},\ \ \ a = c = 0
\end{eqnarray}
This model is consistent with the equations of motion
\begin{eqnarray}
	& & \left\{\begin{array}{rcl}
		\d x_m & = & (\Omega p_m - \Omega/2Q x_m)\d t - \sqrt{\Omega/Q} \d a_{2x}\\
		\d p_m & = & (-\Omega x_m - \Omega/2Q p_m)\d t - 2K \d a_{1p} - \sqrt{\Omega/Q} \d a_{2p}
		\end{array}\right. \nonumber \\
	& & \left\{\begin{array}{rcl}
		\d\tilde{a}_{1x} & = & \d a_{1x} + 2K x_m \d t \\
		\d\tilde{a}_{1p} & = & \d a_{1p} \\
		\d\tilde{a}_{2x} & = & \d a_{2x} + \sqrt{\Omega/Q} x_m \d t \\
		\d\tilde{a}_{2p} & = & \d a_{2p} + \sqrt{\Omega/Q} p_m \d t
		\end{array}\right.
\end{eqnarray}
which were stated without proof previously.  (For clarity, the phonon mode $\d a_{2x}, \d a_{2p}$ was omitted above).

Finally, we consider the OPO cavity.  Though not studied as a plant, the OPO has interesting properties as a controller for the mechanical oscillator system.  The OPO has the following SLH model:
\begin{eqnarray}
	& & S = 1_{2\times 2},\ \ \ L = \left[\begin{array}{cc} \sqrt{\kappa_1} a, & \sqrt{\kappa_2} a \end{array}\right],\nonumber \\
	& & H = \Delta a^\dagger a + \frac{\epsilon^* a^2 - \epsilon (a^\dagger)^2}{2i}
\end{eqnarray}
Once more referring to Eq.\ (\ref{eq:04b-slhquad}) and turning the crank, the linear system parameters are
\begin{eqnarray}
	& & S = 1_{2\times 2},\ \ \ \Lambda = \frac{1}{2}\left[\begin{array}{cc} \sqrt{\kappa_1} & i\sqrt{\kappa_1} \\ \sqrt{\kappa_2} & i\sqrt{\kappa_2} \end{array}\right],\nonumber \\
	& & R = \frac{1}{2} \left[\begin{array}{cc} \Delta - \mbox{Im}(\epsilon) & \mbox{Re}(\epsilon) \\ \mbox{Re}(\epsilon) & \Delta + \mbox{Im}(\epsilon) \end{array}\right],\nonumber \\
	& & r = \lambda = 0,\ \ \ \Theta = J_{2\times 2}
\end{eqnarray}
and the ABCD matrices are
\begin{eqnarray}
	& & A = \left[\begin{array}{cc} \mbox{Re}(\epsilon) - \frac{\kappa_1 + \kappa_2}{2} & \Delta + \mbox{Im}(\epsilon) \\ -\Delta + \mbox{Im}(\epsilon) & -\mbox{Re}(\epsilon) - \frac{\kappa_1 + \kappa_2}{2} \end{array}\right],\ \ \ \nonumber \\
	& & -B = C^{\rm T} = \left[\begin{array}{cccc} \sqrt{\kappa_1} & 0 & \sqrt{\kappa_2} & 0 \\ 0 & \sqrt{\kappa_1} & 0 & \sqrt{\kappa_2} \end{array}\right] \nonumber \\
	& & D = 1_{4\times 4},\ \ \ a = c = 0
\end{eqnarray}

\renewcommand\thesection{\arabic{chapter}.\arabic{section}}
\renewcommand\thesubsection{\arabic{chapter}.\arabic{section}.\arabic{subsection}}

\ifstandalone{\bibliography{NoteRefs}{}
\bibliographystyle{alpha}
}
\ifdefined\multidoc\else\input{Header}\fi

\ifstandalone{\setcounter{chapter}{4}}

\chapter{Semiclassical Wigner Theory}
\label{ch:04}

A key motivation driving photonics research is the ability to do nontrivial computations with complex low-power or quantum circuits.  Realizing this goal will require major advances in fabrication, e.g.\ creating reproducible, high-quality nonlinear devices -- and theory, e.g.\ building simulation tools and techniques that guide the design of quantum circuits \cite{Tezak2012, Sarma2013}.

In Chapter \ref{ch:01}, I introduced the basic theory of open quantum systems -- both in isolation and within circuits.  Chapter \ref{ch:02} derived quantum models for the basic components.  In principle, given these tools and enough simulation time, an arbitrarily complex quantum circuit can be modeled and simulated.

Unfortunately, large quantum simulations of the types described in Chapter \ref{ch:01} are not feasible because the dimension $\mbox{dim}(\mathcal{H})$ of the Hilbert space for a quantum circuit, being the tensor product of the Hilbert spaces of its elements, grows exponentially with the circuit size.  Since the wavevector $\ket{\psi}$ and density matrix $\rho$ have $\mbox{dim}(\mathcal{H})$ and $\mbox{dim}(\mathcal{H})^2/2$ independent components, both the memory use and the computation time scale exponentially with the size of the circuit.  In practice, master-equation simulations are only practical on large clusters for circuits of $\lesssim 2$ nonlinear cavities \cite{Mabuchi2011b, Santori2014} or $\lesssim 10$ qubits \cite{Kerckhoff2011b, Sarma2013b}, and trajectory simulations are practical for $\lesssim 4$ cavities or $\lesssim 20$ qubits.  Going beyond these limits will require some type of approximation.

In this chapter, I introduce a semiclassical approximation that allows one to sample from the density matrix $\rho(t)$ by solving a set of stochastic differential equations.  The procedure works by defining a generalized Wigner function $W_\rho(\alpha)$, a quasi-probability function that represents the quantum state, and converting the master equation into a linear PDE for $W$.  In the limit of $\gtrsim 20$ photons per cavity, where quantum noise plays a relevant but not dominant role, all derivatives higher than second-order in this PDE can be ignored, and the PDE becomes a Fokker-Planck Equation (FPE).  The solution to this FPE is a probability distribution which can be sampled from by solving an associated stochastic differential equation (SDE).  Thus, the solution to the quantum master-equation -- a problem that scales exponentially with circuit size -- can be approximated by solutions to SDEs -- a problem with linear scaling.  This allows very large circuits to be simulated, well beyond what was possible using the quantum approach.

Approximating quantum dynamics with classical noise is not new.  Early work focused on understanding the amplitude and phase fluctuations in masers \cite{Gordon1955, Schawlow1958}, culminating in semiclassical {\it laser rate equations} \cite{Haken1966, Haken1966b, Lax1967}.  This approach was subsequently extended to semiconductor lasers \cite{Haug1967, Haug1969, AgrawalDutta} and was used to predict squeezing in laser light \cite{Benkert1990, Yamamoto1986, Machida1987, Yamamoto1987, Marte1988}.  Later, this was extended to nonlinear-optical systems and put on more rigorous footing using quantum-optical {\it phase-space methods} \cite{Drummond1980, Carmichael1999}.  Fokker-Planck equations based on Wigner and positive-P functions were used to study spontaneous switching in a degenerate OPO \cite{Graham1973, Kinsler1991}, quantum fluctuations in nonlinear fibers \cite{Drummond1993, Carter1995}, and optical bistability in cavity QED \cite{Lugiato1978, Lugiato1982, Drummond1981}.  

Put in this perspective, the content of this chapter is not new.  Rather, it serves to recapitulate old results, using a notation consistent with open quantum systems theory.  The resulting {\it truncated Wigner} theory becomes a useful semiclassical approximation to quantum mechanics, which may be checked against quantum simulations for small systems.

\section{Wigner Function}

The state of an optical field is defined by the density matrix $\rho$.  For each state, the Wigner function can be defined as follows:
\beq
	W(\alpha) = \frac{1}{\pi^2} \int{\d^2\beta\,e^{-i(\alpha^*\beta^*+\alpha\beta)}\chi(\beta)},\ \ \ \chi(\beta) = \mbox{Tr}\left[e^{i(\beta^*a^\dagger + \beta a)}\rho\right] \label{eq:04-wigner}
\eeq
The intermediate function $\chi$ is the \textit{characteristic function}.  Different textbooks define $\chi$ differently, but up to a normalization, $W$ is always the same.  I use the form above because it is symmetric and easily generalizes to non-optical systems.

Coherent states have Gaussian Wigner functions:
\beq
	W_{\ket{\alpha_0}} = \frac{2}{\pi} e^{-2|\alpha-\alpha_0|^2}
\eeq
For a squeezed vacuum state, that is, the ground state of $\bar{\alpha}^\dagger\bar{\alpha}$, where $a = \bar{a}\cosh\eta + \bar{a}^\dagger e^{i\phi}\sinh\eta$, it is:
\beq
	W_{\rm sq} = \frac{2}{\pi} e^{-2|\bar{\alpha}|^2}
\eeq
where $\alpha = \bar{\alpha}\cosh\eta + \bar{\alpha}^*e^{i\phi}\sinh\eta$.  Wigner functions for number states are polynomials multiplied by a Gaussian:
\bea
	W_{\ket{0}} & = & \frac{2}{\pi} e^{-2\alpha^*\alpha} \\
	W_{\ket{1}} & = & \frac{2}{\pi} (4\alpha^*\alpha - 1)e^{-2\alpha^*\alpha} \\
	W_{\ket{2}} & = & \frac{4}{\pi} (8(\alpha^*\alpha)^2 - 8\alpha^*\alpha + 1) e^{-2\alpha^*\alpha}
\eea
These are plotted in Figure \ref{fig:04-f1}.  

\begin{figure}[tbp]
\begin{center}
\includegraphics[width=1.00\textwidth]{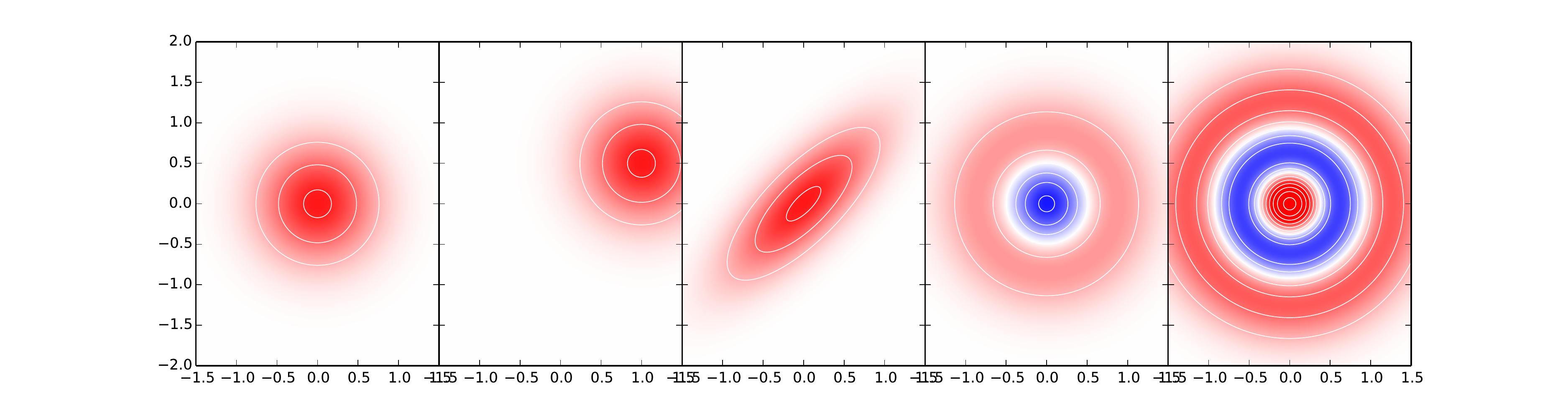}
\caption{Wigner functions.  Left to right: Vacuum state, coherent state, squeezed state, one-photon state, two-photon state.  Red is positive $W(\alpha)$, blue is negative.}
\label{fig:04-f1}
\end{center}
\end{figure}

The Wigner function resembles a probability density in many respects.  The probability distribution for any quadrature measurement ($X = a+a^\dagger$, $P = (a-a^\dagger)/i$, or any linear combination of the two) is obtained by \textit{marginalizing} the Wigner function -- integrating over the other quadrature, for example, for $X$, it would be $P(x) = \int_{-\infty}^\infty{W(\frac{x+ip}{2})dp}$.

Like a probability distribution, moments of the Wigner function map to operator moments of the quantum state.  Because $a$ and $a^\dagger$ do not commute, the \textit{operator ordering} of the moments is quite important.  Moments of the Wigner function correspond to \textit{symmetrically-ordered} operator moments in the state.
\beq
	\boxed{\avg{(\alpha^*)^m \alpha^n}_W \equiv \int{(\alpha^*)^m \alpha^n W(\alpha)\,\d^2\alpha} = \avg{(a^\dagger)^m a^n}_{\rm sym}} \label{eq:04-moments}
\eeq
For example, $\avg{\alpha}_W = \avg{a}$, $\avg{\alpha^*\alpha}_W = \frac{1}{2}\avg{a^\dagger a+a a^\dagger}$, $\avg{\alpha^*\alpha^2}_W = \frac{1}{3}\avg{a^\dagger a a + a a^\dagger a + a a a^\dagger}$.

However, the probability analogy only goes so far.  As we can see from Figure \ref{fig:04-f1} that some states have a negative Wigner function, so it is not strictly a probability density.  However, all classical states, and some quantum states like squeezed states, have positive Wigner functions.  Only highly nonclassical states, like number states and cat states, have a negative Wigner function -- indeed, some have posited that this is what \textit{defines} a nonclassical state \cite{Kenfack2004, Mari2012, Delfosse2015}.

The Wigner function can be generalized to a multiple fields: $W(\alpha) \rightarrow W(\alpha_1,\ldots,\alpha_n)$.  Tensor-product states also factorize in the Wigner function: $\rho = \rho_1\otimes\rho_2 \Rightarrow W(\alpha_1,\alpha_2) = W_1(\alpha_1)W_2(\alpha_2)$.  One also has \cite{WallsMilburn}:
\bea
	W[a\rho] & = & \left(\alpha + \frac{1}{2}\frac{\partial}{\partial\alpha^*}\right)W[\rho] \label{eq:04-wig-a1} \nonumber \\
	W[a^\dagger\rho] & = & \left(\alpha^* - \frac{1}{2}\frac{\partial}{\partial\alpha}\right)W[\rho] \nonumber \\
	W[\rho a] & = & \left(\alpha - \frac{1}{2}\frac{\partial}{\partial\alpha^*}\right)W[\rho] \nonumber \\
	W[\rho a^\dagger] & = & \left(\alpha^* + \frac{1}{2}\frac{\partial}{\partial\alpha}\right)W[\rho] \label{eq:04-wig-a4} 
\eea
The inner product between two operators is related to the overlap integral of their Wigner functions:
\beq
	\mbox{Tr}\left[\rho_1 \rho_2\right] = \pi \int{W_1(\alpha)W_2(\alpha)\d^2\alpha} \label{eq:04-wig-op}
\eeq
Non-positive and non-Hermitian operators can have Wigner functions, too.  For example, for the identity, $W_I = 1/\pi$.  Using Eqs.~(\ref{eq:04-wig-a1}--\ref{eq:04-wig-a4}), one can build up Wigner functions for operator products:
\beq
	W_a = \frac{\alpha}{\pi},\ \ \ W_{a^\dagger} = \frac{\alpha^*}{\pi},\ \ \ 
	W_{a^\dagger a} = \frac{\alpha^*\alpha - \frac{1}{2}}{\pi}, \ \ \ldots
\eeq

\section{Generalized Wigner Function}

In general, we will be interested in more than just optical fields -- for example, a system may contain a mechanical spring, an ensemble of atoms, or a sea of electrons and holes.  As long as the operators are \textit{bosonic} -- that is, they satisfy commutation rather than anticommutation relations -- this is no impediment to the Wigner approach.

The first step is to find a closed operator algebra for the system.  This algebra will have a basis $X \equiv (X_1, \ldots, X_N)$ and should contain all (relevant) system observables.  For an all-optical system, $X = (a, a^\dagger)$ is sufficient, since any operator on the field can be built from products of these two.  For an ensemble of identical atoms, we will show later that $X = (\sigma_x, \sigma_y, \sigma_z)$ -- the net spin of the ensemble -- is the algebra.

It is very important that the operator algebra be \textit{closed}.  Mathematically, this means that if we take the adjoint master equation for operators (equivalent to QSDEs without noise terms)
\beq
	\frac{\d A}{\d t} = -i[A, H] + \frac{1}{2}\left(2L^\dagger A L - L^\dagger L A - A L^\dagger L\right)
\eeq
that the time-derivatives $\dot{X}_1, \ldots, \dot{X}_N$ \textit{must be expressable} in terms of the $\dot{X}_i$'s.  If this is true, than the algebra spans all dynamically relevant quantities.  If it is \textit{not} true, then there are extra degrees of freedom, outside the algebra, that can ``sneak in'' to the algebra over time.  Closure is necessary for the semiclassical Wigner method to work (see Appendix~\ref{sec:05b-closedness} for more detail).

The generalized Wigner function is defined, up to normalization, as:
\beq
	\boxed{W(x; t) = \int{\d^Ny\,e^{-i \sum_k x_k y_k} \chi(y;t)},\ \ \ \chi(y;t) = \mbox{Tr}\left[e^{i \sum_k X_k y_k}\rho(t)\right]}
\eeq
This behaves a lot like the all-optical Wigner function.  For example, moments of the generalized Wigner function correspond to symmetrically ordered operator products.  This can be proved by relating moments of $W$ to derivatives of the characteristic function:
\bea
	\avg{x_a\ldots x_z}_W & \equiv & \int{x_a\ldots x_z W(x)\d^Nx} \nonumber \\
	& = & \int{\d^Nx\,\d^Ny\,\chi(y;t) \left(i\frac{\partial}{\partial y_a}\right)\ldots \left(i\frac{\partial}{\partial y_z}\right) e^{-i \sum_k x_k y_k}} \nonumber \\
	& = & \int{\d^Nx\,\d^Ny\,e^{-i \sum_k x_k y_k} \left(-i\frac{\partial}{\partial y_a}\right)\ldots \left(-i\frac{\partial}{\partial y_z}\right) \chi(y;t)} \nonumber \\
	& = & \left.\left(-i\frac{\partial}{\partial y_a}\right)\ldots \left(-i\frac{\partial}{\partial y_z}\right) \chi(y;t)\right|_{y=0} \label{eq:04-wig-char1}
\eea
Now the characteristic function is trace of the exponential of a sum, times $\rho$.  The exponential may be Taylor expanded, and like terms may be collected.  Because the exponential is symmetric, all operator orderings must contribute -- and the Taylor series must have symmetric coefficients:
\bea
	\!\!\!\!\!\!\!\chi(y;t) & \!\!\!\equiv\!\!\! & \avg{e^{i \sum_k X_k y_k}} = \sum_n{\frac{1}{n!}\avg{\left(i \sum_k X_k y_k\right)^n}} \nonumber \\
	& \!\!\!=\!\!\! & 1 + i\sum_k \avg{X_k}y_k - \frac{1}{2}\sum_{kl} \frac{\avg{X_k X_l + X_l X_k}}{2} y_k y_l + \ldots \nonumber \\
	& \!\!\!=\!\!\! & 1 + i\sum_k \avg{X_k}_{\rm sym}y_k - \frac{1}{2}\sum_{kl} \avg{X_k X_l}_{\rm sym} y_k y_l - \frac{i}{3!} \sum_{klm} \avg{X_k X_l X_m}_{\rm sym} y_k y_l y_m + \ldots
\eea
Taking derivatives width respect to the $y_k$ exposes symmetrically ordered moments.  Applying (\ref{eq:04-wig-char1}), it is clear that Wigner function moments correspond to symmetric operator moments, i.e.
\beq
	\boxed{\avg{x_a\ldots x_z}_W = \avg{X_a\ldots X_z}_{\rm sym}}
\eeq
This is a general fact that is independent of the commutation relations.  It will be very useful in converting the master equation for $\rho$ into a Fokker-Planck equation for $W$, below.

\section{Fokker-Planck Procedure}
\label{sec:04-procedure}

In this section, I will derive a Fokker-Planck equation for the generalized Wigner function.  This derivation will come from the master equation, by way of operator moments.  The Fokker-Planck equation is then converted into a stochastic differential equation.

To start, in this system, recall that $X = (X_1, \ldots X_n)$ is a basis for the operator algebra, and $x = (x_1, \ldots, x_n)$ is an n-dimensional vector.  As explained above, you get Weyl-ordered products form the moments of the Wigner function, as follows:
\beq
	\langle X_{a}\ldots X_{z} \rangle_{\rm sym} = \int{x_{a}\ldots x_{n} W(x, t)\d^nx}
\eeq
Now $W(x, t)$ contains a full description of the quantum state.  Since it is a function of position as well as time, it should satisfy a \textbf{partial} differential equation.  The general form of this equation is unknown, but most PDEs in math and science are fairly low-order, so it's likely that the crucial behavior is captured in the lowest-order derivatives (first and second derivatives, ideally).  It also has to be a linear PDE, since quantum mechanics is linear, and it must be first-order in time to match the Schr\"{o}dinger Equation.  So we write out the general form of a linear PDE, which looks something like this:
\beq
	\frac{\partial W}{\partial t} = -\frac{\partial}{\partial x_i} C^{(1)}_i(x) W(x) + 
	\frac{1}{2}\frac{\partial^2}{\partial x_i \partial x_j} C^{(2)}_{ij}(x) W(x) - 
	\frac{1}{6}\frac{\partial^3}{\partial x_i \partial x_j \partial x_k} C^{(3)}_{ijk}(x) W(x) + \ldots
\eeq
Before we solve this equation, we need to find the values of the $C^{(i)}$, since these functions dictate the equation's behavior.  Different systems will, of course, have different $C^{(i)}$.  Here we are only interested in $C^{(1)}$ and $C^{(2)}$.  Ignoring all the higher-order terms, the PDE for $W$ becomes a Fokker-Planck equation which can be solved using standard SDE methods.

The $C^{(i)}$ are intimately related to the equations of motion for the Weyl-ordered moments $\langle X_{i_1}\ldots X_{i_n} \rangle_S$ of the state.  Starting with the first moment:
\begin{eqnarray}
	\frac{\d}{\d t}\langle X_m\rangle_{\rm sym}
	& = & \int{x_m \frac{\partial W}{\partial t}\d^nx} 
	= \int{x_m \left[-\frac{\partial}{\partial x_i} C^{(1)}_i(x) W(x) + 
	\frac{1}{2}\frac{\partial^2}{\partial x_i \partial x_j} C^{(2)}_{ij}(x) W(x) + \ldots\right] \d^7x} \nonumber \\
	& = & \int{C^{(1)}_m(x) W(x)\d^7x} = \langle C^{(1)}_m(x) \rangle_W \label{eq:04-lin-p1}
\end{eqnarray}
For every operator $A$ in the algebra, there exists a polynomial representation $A_p$ for it.  $A_p$ is not an operator; it is a polynomial in $(x_1, \ldots, x_n)$, defined so that:
\beq
	A = \bigl(A_p(X_1, \ldots, X_n)\bigr)_{\rm sym}
\eeq
This is defined so that:
\beq
	\langle A_p(x) \rangle_W = \mbox{Tr}\bigl[A\,\rho\bigr]
\eeq
For example, in the optical algebra $(a, a^\dagger)$,
\beq
	(a)_p = \alpha,\ \ \ (a^\dagger)_p = \alpha^*,\ \ \ (a^\dagger a)_p = (\alpha^*\alpha - \tfrac{1}{2}),\ \ \ldots
\eeq
because $a = (a)_{\rm sym}, a^\dagger = (a^\dagger)_{\rm sym}, a^\dagger a = (a^\dagger a - \frac{1}{2})_{\rm sym}$.  Using property (\ref{eq:04-wig-op}) of the Wigner function, we can relate $(A)_p$ to its Wigner representation:
\beq
	\avg{A_p(x)}_W \stackrel{(\ref{eq:04-moments})}{=} \mbox{Tr}\left[\bigl(A_p(X)\bigr)_{\rm sym} \rho\right] \stackrel{(\ref{eq:04-wig-op})}{=} \pi\int{W_A(x)W(x)\d^nx} = \pi \avg{W_A(x)}_W
\eeq
which implies that
\beq
	\boxed{A_p(x) = \pi W_A(x)}
\eeq
In other words, the Wigner representation and the polynomial representation are the same up to a factor of $\pi$.  The polynomial representation is important because it is how one expresses the $C^{(k)}$.  Consider first the drift term, calculated above.  Another way to calculate the drift is to use the adjoint master equation:
\beq
	\dot{X}_m\bigr|_{\rm adj} = -i[X_m, H] + \frac{1}{2}\sum_k(2L_k^\dagger X_m L_k - L_k^\dagger L_k X_m - X_m L_k^\dagger L_k)
\eeq
This is just the QSDE without the stochastic terms.  If we are only interested in ensemble-averaged quantities, it is easy to show that this equation is equivalent to the master equation, i.e.
\beq
	\mbox{Tr}\left[\dot{X}_m\bigr|_{\rm adj} \rho\right] = \mbox{Tr}\left[X_m \dot{\rho}\bigr|_{\rm me}\right] = \frac{\d}{\d t}\langle X_m\rangle_\rho
\eeq
From this we have
\beq
	\frac{\d}{\d t} \langle X_m\rangle_{\rm sym} = \avg{(\dot{X}_m)_p}_W \label{eq:04-lin-p2}
\eeq
Combining Eqs.~(\ref{eq:04-lin-p1}) and (\ref{eq:04-lin-p2}), one can solve for the drift term $C^{(1)}$.  It is the polynomial representation of the $\dot{X}_m$ computed with the adjoint equation.
\beq
	C^{(1)} = \left(\dot{X}_m\right)_p
\eeq
So the first moment's time derivative encodes the drift term.  Let's look at the second moment:
\begin{eqnarray}
\frac{\d}{\d t}\langle X_m X_n\rangle_{\rm sym}
	& = & \int{x_m x_n \frac{\partial W}{\partial t}\d^nx} 
	= \int{x_m x_n \left[-\frac{\partial}{\partial x_i} C^{(1)}_i(x) W(x) + 
	\frac{1}{2}\frac{\partial^2}{\partial x_i \partial x_j} C^{(2)}_{ij}(x) W(x) + \ldots\right] \d^nx} \nonumber \\
	& = & \int{\left[x_m C^{(1)}_n(x) + x_n C^{(1)}_m(x) + C_{mn}^{(2)}(x)\right]W(x) \d^nx} \nonumber \\
	& = & \left\langle C_{mn}^{(2)}(x) + x_m C^{(1)}_n(x) + x_n C^{(1)}_m(x)\right\rangle_W
\end{eqnarray}
This term is equal to $\left(\frac{\d}{\d t}(X_mX_n)_{\rm sym}\right)_p$, the polynomial representation of the time derivative of the symmetric product $(X_mX_n)_{\rm sym}$.

This is an expression for the change in a \textit{moment} ($\d\langle X_m X_n\rangle_{\rm sym}/\d t$) in terms of its \textit{cumulants} ($C_m^{(1)}(x), C_{mn}^{(2)}(x)$).  In a sense, the coefficients in the Fokker-Planck equation are related to how quickly operator-product \textit{cumulants} change with time.  This is not unlike the cluster expansion \cite{KiraKoch2012}.  The first two terms are:
\begin{empheq}[box=\fbox]{align}
	C^{(1)}_m &\ =\  \left(\frac{\d}{\d t}X_m\right)_p \label{eq:04-c1-corr} \\
	C^{(2)}_{mn} &\ =\ \left(\frac{\d}{\d t}(X_m X_n)_{\rm sym}\right)_p - x_m\left(\frac{\d}{\d t} X_n\right)_p - x_n \left(\frac{\d}{\d t} X_m\right)_p \label{eq:04-c2-corr}
\end{empheq}
There are two contributions to these terms: one from the Hamiltonian $H$, and one from the coupling $L$.

\subsection{Hamiltonian Part}
\label{sec:04-ham}

For most ``reasonable'' Hamiltonians, the following identity is satisfied:
\beq
	\left([(X_mX_n)_{\rm sym}, H]\right)_p = \left(\left(X_m\right)_p\left([X_n, H]\right)_p + \left([X_m, H]\right)_p\left(X_n\right)_p\right)_{\rm sym} \label{eq:04-identity-h}
\eeq
This means that $C^{(2)}_{mn} = 0$ under pure Hamiltonian evolution.  This says that ``reasonable'' Hamiltonians do not add extra diffusion into the Wigner function -- they conserve phase space and satisfy Liouville's Theorem.  Thus, for Hamiltonian evolution, with ``reasonable'' Hamiltonians, $C^{(1)}$ and $C^{(2)}$ are:
\begin{eqnarray}
	C^{(1)}_m\Bigr|_H & = & -i\left([X_m, H]\right)_p \\
	C^{(2)}_{mn}\Bigr|_H & = & 0
\end{eqnarray}

\subsection{Coupling Part}

The same is not true when there are couplings to the environment.  Regardless of whether an identity like (\ref{eq:04-identity-h}) is satisfied, $C^{(2)}_{mn}$ will always be zero because, for adjoint time-derivatives:
\beq
	\left.\frac{\d (X_mX_n)}{\d t}\right|_{\rm ad} \neq \left.X_m \frac{\d X_n}{\d t} + \frac{\d X_m}{\d t} X_n\right|_{\rm ad} \label{eq:04-prodvio}
\eeq
Since the adjoint equation does not strictly define a time derivative, it does not satisfy the product rule.  This can be understood in light of the fact that the \textit{actual} Heisenberg equations are stochastic (QSDEs) -- these stochastic terms naturally give rise to diffusion in the Wigner function.  It is this effect, manifested in the product-rule violation (\ref{eq:04-prodvio}).  This is what makes the cumulant $C^{(2)}_{mn}$ nonzero.

Likewise, using the following property of the Lindbladian
\beq
	\mathcal{L}_L\bigl[X_mX_n\bigr] = \left(X_m \mathcal{L}_L(X_n) + \mathcal{L}_L(X_m) X_n\right) + [L^\dagger, X_m][X_n, L]
\eeq
we get
\beq
	\mathcal{L}_L\bigl[(X_mX_n)_{\rm sym}\bigr]_p = \left(\left(X_m \mathcal{L}_L(X_n)\right)_p + \left(\mathcal{L}_L(X_m) X_n\right)_p\right)_{\rm sym} + \left(([L^\dagger, X_m][X_n, L])_p\right)_{\rm sym}
\eeq
For most ``reasonable'' L terms, one has separability, and finds the following:
\beq
	\left(\mathcal{L}_L(X_mX_n)_{\rm sym}\right)_p = \left(\mathcal{L}_L(X_m)\right)_p(X_n)_p + (X_m)_p\left(\mathcal{L}_L(X_n)\right)_p + \left([L^\dagger, X_m]\right)_p \left([L^\dagger, X_n]\right)_p \label{eq:04-sep-l}
\eeq
This gives the following $C^{(k)}$ terms:
\bea
	C^{(1)}_m\Bigr|_L & = & \left(\mathcal{L}_L(X_m)\right)_p \\
	C^{(2)}_{mn}\Bigr|_L & = & \left([L^\dagger, X_m]\right)_p \left([X_n, L]\right)_p
\eea
Put together, one finds:
\begin{empheq}[box=\fbox]{align}
	C^{(1)}_m &\ =\ \left(-i[X_m, H] + \frac{1}{2}(2L^\dagger X_m L - L^\dagger L X_m - X_m L^\dagger L)\right)_p \\
	C^{(2)}_{mn} &\ =\ \left([L^\dagger, X_m]\right)_p \left([X_n, L]\right)_p
\end{empheq}
Note that these equations are only valid when separability conditions (\ref{eq:04-identity-h}, \ref{eq:04-sep-l}) on $H$ and $L$ are satisfied!  For simple models, like all-optical circuits with Kerr nonlinearities, this is true.  But for some more complex ones, like many-atom cavities or free-carrier devices, the separability conditions do not hold and one must resort to the less intuitive forms (\ref{eq:04-c1-corr}--\ref{eq:04-c2-corr}), which are always correct.

\section{SDEs}

A Fokker-Planck equation with a positive-definite diffusion matrix can be recast as an SDE.  The Wigner function becomes the probability distribution of a c-number stochastic process $x(t)$.  To see how this works, consider a stochastic process with a drift term and a diffusion term:
\beq
	\d x_i = \mu_i(x)\d t + R_{ij}(x) \d w_j(t)
\eeq
The moments of the distribution evolve as follows:
\begin{eqnarray}
	\frac{\d}{\d t}\langle x_a \rangle & = & \langle \mu_a(x) \rangle \\
	\frac{\d}{\d t}\langle x_a x_b \rangle & = & \langle x_a\mu_b(x) + x_b\mu_a(x) + R_{ac}(x)R_{bc}(x) \rangle
\end{eqnarray}
Now compare this to the Fokker-Planck equation.  Let $W(x,t)$ be the probability distribution of $x_i(t)$, and let $W(x,t)$ satisfy the Fokker-Planck equation:
\beq
	\frac{\partial W(x,t)}{\partial t} = -\frac{\partial}{\partial x_i} C^{(1)}_i(x) W(x,t) + 
	\frac{1}{2}\frac{\partial^2}{\partial x_i \partial x_j} C^{(2)}_{ij}(x) W(x,t)
\eeq
Now the distribution moments evolve as:
\begin{eqnarray}
	\frac{\d}{\d t}\langle x_a \rangle & = & \int{x_a \left[-\frac{\partial}{\partial x_i} C^{(1)}_i(x) W + 
	\frac{1}{2}\frac{\partial^2}{\partial x_i \partial x_j} C^{(2)}_{ij}(x) W\right]} = \langle C^{(1)}_a(x)\rangle \\
	\frac{\d}{\d t}\langle x_a x_b \rangle & = & \int{x_a x_b \left[-\frac{\partial}{\partial x_i} C^{(1)}_i(x) W + 
	\frac{1}{2}\frac{\partial^2}{\partial x_i \partial x_j} C^{(2)}_{ij}(x) W\right]} = \langle x_a C^{(1)}_b(x) + x_b C^{(1)}_a(x) + C^{(2)}_{ab}(x)\rangle \nonumber \\
\end{eqnarray}
The following identifications can be made:
\beq
	\boxed{\mu = C^{(1)},\ \ \ RR^T = C^{(2)}}
\eeq
To obtain an SDE that lets us sample from the Wigner distribution, we follow a simple two-step process:  First compute $C^{(1)}$ and $C^{(2)}$ from Heisenberg equations for operator moments, in the previous section.  Then, obtain the drift and diffusion terms, above.  Then we are done.

\section{Input-Output Relations}
\label{sec:04-inout}

Oftentimes we will be interested in the output fields as well as the internal dynamics.  Over a time interval $\d t$, we can treat the bath as a single oscillator mode $\d B = b\sqrt{\d t}$, where $[\d B, \d B^\dagger] = \d t$ and therefore $[b, b^\dagger] = 1$.  The coupling Hamiltonian takes the form:
\beq
	H_{\rm full} = \frac{i}{\sqrt{\d t}} \sum_m{(M_m b_m^\dagger - M_m^\dagger b_m)} + H \label{eq:04-hfull}
\eeq
The Wigner method for input-output relations is easiest derived in the $S = 0$ case, so we focus on that to begin.  At the start of the interaction $t$, the Wigner function may be written as $W_{\rm sys}(x) W_\beta(\beta)$, where $\beta$ is the external field.  We will propagate the Wigner function forward to time $t+\d t$, obtaining a mixed Wigner function.

Using the formalism from the last chapter, we can derive a Fokker-Planck equation for the joint Wigner function over the interval $[t, t+\d t]$.  The joint system undergoes purely Hamiltonian evolution, governed by Eq.~(\ref{eq:04-hfull}).  Assuming that the Hamiltonian is ``nice'', as discussed in the previous section -- that is, assuming that the following factorization holds,
\beq
	\left([(AB)_{\rm sym}, H_{\rm full}]\right)_p = \left(\left(A\right)_p\left([B, H]\right)_p + \left([A, H]\right)_p\left(B\right)_p\right)_{\rm sym} \label{eq:04-identity-h}
\eeq
where $A, B \in \{X_1, \ldots, X_n, b, b^\dagger\}$, then this Hamiltonian the flow satisfies Liouville's theorem and $C^{(2)} = 0$, as shown in Sec.~\ref{sec:04-ham}.  The Wigner function at time $t+\d t$ can be sampled by solving the following ODE's on the interval $[t, t+\d t]$:
\bea
	\frac{\d  x_k}{\d t} & = & -i[X_k, H]_p + \frac{1}{\sqrt{\d t}}\left([X_k, M_m] b_m^\dagger + [M_m^\dagger, X_k] b_m\right)_p \\
	\frac{\d \beta_m}{\d t} & = & \frac{1}{\sqrt{\d t}}(M_m)_p
\eea
\begin{figure}[tbp]
\begin{center}
\includegraphics[width=0.80\textwidth]{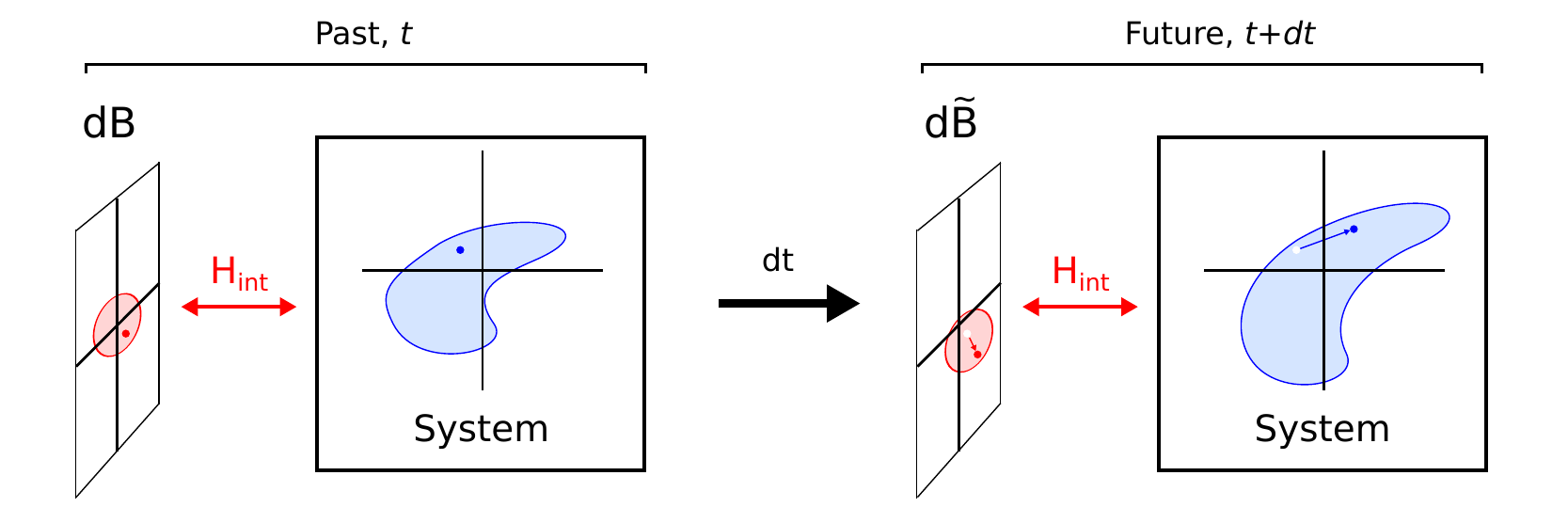}
\caption{2-part Wigner function (left) before interaction, (right) after, for (blue) system and (red) one input-output slice.  The Wigner function is sampled by random points (dot in figure), which move as the system evolves.}
\label{default}
\end{center}
\end{figure}
These can be converted in to It\^{o} SDEs.  The result looks very much like the QSDEs, but all the variables are c-numbers.
\bea
	\d x_k & = & -i[X_k, H]_p \d t + \d\beta_m^* [X_k, M_m]_p + [M_m^\dagger, X_k]_p \d\beta_m \\
	\d\tilde{\beta}_m & = & \d\beta_m + (M_m)_p \d t
\eea
Constant scattering terms are easily included.  Using the Gough-James circuit algebra, we know that if $S$ is a constant, then the SLH model may be written as a series product:
\beq
	(S, L, H) = (S, \_, \_) \triangleleft (1, S^{-1}L, H)
\eeq
This just corresponds to obtaining the SDEs for $(1, S^{-1}L, H)$ (i.e. setting $M = S^{-1}L$) and then scattering the output fields $\d\tilde{\beta} \rightarrow S\,\d\tilde{\beta}$.  The result is:
\begin{empheq}[box=\fbox]{align}
	\d x_k &\ =\ -i[X_k, H]_p \d t + \d\beta_m^* S^\dagger_{mn}[X_k, L_n]_p + [L_n^\dagger, X_k]_p S_{nm} \d\beta_m \label{eq:04-wig-qsde1} \\
	\d\tilde{\beta}_m &\ =\ S_{mn} \d\beta_n + (L_m)_p \d t \label{eq:04-wig-qsde1}
\end{empheq}
These are the Wigner SDEs, with input-output relations, for a Wigner function that encompasses \textit{both} system and bath.  The $\d\beta$'s are the c-number analogue to the quantum processes $\d B$.  They are commuting processes, and obey the same statistics:
\beq
	\d\beta\,\d\beta^* = \frac{1}{2}\avg{\d B\,\d B^\dagger + \d B^\dagger \d B} = \frac{1}{2}\d t
\eeq
All of the noise in the system comes from these inputs.  The noise in each input comes from the fact that each input time-slice $\d B$ is a quantum field, living in the ground state, and sampling from its Wigner function means picking a point $\d\beta$ with statistics $\avg{\d\beta\,\d\beta^*} = \d t/2$.

\subsection{Homodyne Detection}

What happens when we do homodyne detection on the outputs?  Recall that, before the system and bath interact, the joint Wigner function has the form:
\beq
	W[x_1, \ldots, x_n, \d\beta(t_1), \ldots, \d\beta(t_N)] = W_0(x) \prod_i W_{\ket{0}}(\d\beta(t_i)/\d t)
\eeq
and we can sample from this distribution with a state vector
\beq
	\bigl(x_1(0), \ldots, x_n(0), \d\beta(t_1), \ldots, \d\beta(t_N)\bigr)
\eeq
where $(x_1, \ldots, x_n)$ are sampled according to $W_0$, and the $\d\beta_i$ are sampled according to the ground-state Wigner function, i.e.\ Gaussian with $\avg{\d\beta(t_i) \d\beta(t_i)^*} = \d t/2$.

After the interaction, the Wigner distribution is all jumbled up, but we can efficiently sample from it by looking at the final state vector
\beq
	\bigl(x_1(T), \ldots, x_n(T), \d\tilde{\beta}(t_1), \ldots, \d\tilde{\beta}(t_N)\bigr)
\eeq
All of the outputs $\d\tilde{\beta}(t_i)$ are sent into a homodyne detector.  Homodyne detection is just measuring $X = (\d B + \d B^\dagger)/\d t$.  The probability density is the marginalized Wigner function, and the vector $(\d\tilde{\beta}(t_1), \ldots, \d\tilde{\beta}(t_n))$ samples from that distribution.

It follows that the homodyne signal for any trajectory is just the real part of the trajectory output field, multiplied by two:
\beq
	\boxed{X(t) = \frac{1}{\d t} \left[\d\tilde{\beta}(t)+\d\tilde{\beta}^*(t)\right] = L(t)+L^\dagger(t) + (\beta(t)+\beta^*(t))}
\eeq
Likewise for the P quadrature.
\beq
	\boxed{P(t) = -\frac{i}{\d t} \left[\d\tilde{\beta}(t)-\d\tilde{\beta}^*(t)\right] = \frac{L(t)-L^\dagger(t)}{i} + \frac{\beta(t)-\beta^*(t)}{i}}
\eeq

\subsection{Heterodyne Detection}

In heterodyne detection, each output is split into two, and homodyne detection is performed on each component.  The result of the beam-splitting is additional quantum noise:
\beq
	\d\tilde{\beta}_1 = \frac{\d \tilde{\beta} + \d\gamma^*}{\sqrt{2}},\ \ \ 
	\d\tilde{\beta}_2 = \frac{\d \tilde{\beta} - \d\gamma^*}{\sqrt{2}}
\eeq
(We used $\d\gamma^*$ rather than $\d\gamma$ for convenience; they have the same statistics).  An X measurement is made on the first quadrature, and P on the second.  The results of these measurements are:
\beq
	X = \sqrt{2}\d t^{-1} \mbox{Re}[\d\tilde{\beta}+\d\gamma^*],\ \ \ 
	P = \sqrt{2}\d t^{-1} \mbox{Im}[\d\tilde{\beta}-\d\gamma^*]
\eeq
The heterodyne signal is the combination of these two.  It is equal to the trajectory output, plus some noise:
\beq
	\boxed{\beta_{\rm het} = \frac{X+iP}{\sqrt{2}} = \frac{\d\tilde{\beta} + \d\gamma}{\d t} = L + \frac{\d \beta + \d\gamma}{\d t}}
\eeq
The extra noise, of course, comes from the uncertainty principle -- it is possible to precisely measure a single quadrature in a field, but it is not possible to measure both quadratures at once.

Note that there is no way to set, a priori, the detector output, either homodyne or heterodyne -- these are random variables that are sampled from a distribution.  Conditioning that distribution on some particular output ends up being a hard task -- the whole distribution has to be re-weighted and re-normalized.  Since we do not keep track of the whole distribution but sample from trajectories, conditioning is not possible using the Wigner trajectory method.  Thus, while the Wigner method is a good way to generate sample trajectories, it cannot be used as a filter to learn the state of a quantum system.

Nor is it possible to model photon counting.  Photon counting is a highly nonlinear sort of measurement, which can be used to create states with negative Wigner functions.  For example, photon subtraction of a squeezed state creates a ``Schr\"{o}dinger kitten'' state \cite{Ourjoumtsev2006, Neergaard2006, Wakui2007}, which has negative Wigner function near $\alpha = 0$.  Even in the absence of conditioning, photon-counting detectors allow one to perform {\it boson sampling}, and boson sampling of Gaussian states can be used to calculate molecular vibronic spectra \cite{Huh2015}, a problem that has no efficient classical algorithm.  Thus it should not surprise us that photon counting cannot be modeled with the truncated Wigner method.

\section{Example Systems}

\subsection{Linear Cavity}

Consider the optical cavity from Sec.~\ref{sec:02-cavity}, which has the SLH model:
\beq
	\left(1_{N\times N},\ \ \left[\sqrt{\kappa_1}e^{i\psi_1}, \ldots, \sqrt{\kappa_N}e^{i\psi_N}\right],\ \ 
      \Delta a^\dagger a + i(E^*a - E a^\dagger)\right)
\eeq
This has the following Wigner SDEs:
\bea
	\d\alpha & = & \left[(-i\Delta - \kappa/2)\alpha - E\right]\,\d t - \sum_i \sqrt{\kappa_i}e^{-i\psi_i} \d\beta_i \\
	\d\tilde{\beta}_i & = & \d\beta_i + \sqrt{\kappa_i}e^{i\psi_i} a\,\d t
\eea
These SDEs are exact.  It happens that, when the Hamiltonian is quadratic, all third- and higher-order derivatives vanish from the Wigner PDE, and it becomes an exact Fokker-Planck equation.

\subsection{Kerr Cavity}
\label{sec:05-kerr}

Recall from Sec.~\ref{sec:02-multikerr} that the most general SLH model for a Kerr cavity is:
\beq
	G = \left(1,\ \ \ \sqrt{\beta} \sum_{ij} \Lambda_{m,ij} a_i a_j,\ \ \ \frac{1}{2}\chi \sum_{ijkl} \Psi_{ijkl}a_i^\dagger a_j^\dagger a_k a_l\right)
\eeq
This gives the following Wigner SDEs:
\bea
	\d\alpha_i & = & (-i\chi - \beta)\sum_{jkl}\Psi_{ijkl}\left[(\alpha_j^\ast\alpha_k - \delta_{jk})\alpha_l\right]\d t - 2\sqrt{\beta} \sum_m \Lambda_{m,ij}^\ast \alpha_j^\ast \d\beta_m \label{eq:04-daikerr} \\
	\d\tilde{\beta}_m & = & \d\beta_m + \sum_m \Lambda_{m,ij} \alpha_i\alpha_j
\eea
This model is discussed at length in a paper with our HP colleagues \cite{Santori2014, Santori2014-CLEO}.  One can use Kerr cavities to construct an SR latch \cite{Mabuchi2011b}, from which one can build up a whole zoo of digital components \cite{Horowitz1989}.  The stochastic terms give rise to spontaneous switching events in the latch, which propagate errors down digital circuits like optical counters.

One can use truncated Wigner theory for Kerr systems whenever $\chi, \beta \ll 1$ \cite{Santori2014}.  Note that (\ref{eq:04-daikerr}) contains both a nonlinear term and also an additional linear dispersion / absorption (the $\delta_{jk}$ term).  This term is small since $\chi$ and $\beta$ are small, and is usually dropped (though more rigorously it can be absorbed into the cavity detuning $\Delta$ and loss $\kappa$).

\subsection{Nondegenerate OPO}
\label{sec:04-ndopo}

A nondegenerate OPO cavity has three resonant modes -- satisfying $\omega_a + \omega_b = \omega_c$.  All of these modes can interact with input-output fields.  This gives an SLH model of the following form:
\beq
	G = \left(1,\ \ \begin{bmatrix} \sqrt{\kappa_a} a \\ \sqrt{\kappa_b} b \\ \sqrt{\kappa_c} c \end{bmatrix},\ \ 
          \Delta_a a^\dagger a + \Delta_b b^\dagger b + \Delta_c c^\dagger c
          + \frac{\epsilon^* a b c^\dagger - \epsilon a^\dagger b^\dagger c}{2i} \right)
\eeq
The master equation can be converted into a PDE for the Wigner function:
\bea
	\frac{\partial W[\rho]}{\partial t} & = & 
	-\frac{\partial}{\partial\alpha} \biggl((-\kappa_a/2 - i\Delta_a)\alpha + \frac{1}{2}\epsilon\,\beta^\ast \gamma W[\rho]\biggr)
	-\frac{\partial}{\partial\beta} \biggl((-\kappa_b/2 - i\Delta_b)\alpha + \frac{1}{2}\epsilon\,\alpha^\ast \gamma W[\rho]\biggr) \nonumber \\
	& & - \frac{\partial}{\partial\gamma} \left((-\kappa_b/2 - i\Delta_b)\alpha - \frac{1}{2}\epsilon^\ast\,\alpha\beta W[\rho]\right)
-\frac{\partial}{\partial\alpha^\ast} (...)
-\frac{\partial}{\partial\beta^\ast} (...)
-\frac{\partial}{\partial\gamma^\ast} (...) \nonumber \\
& & + \frac{\partial^2}{\partial\alpha\partial\alpha^\ast} \left(\frac{1}{2}\kappa_a W[\rho]\right)
+ \frac{\partial^2}{\partial\beta\partial\beta^\ast} \left(\frac{1}{2}\kappa_b W[\rho]\right)
+ \frac{\partial^2}{\partial\gamma\partial\gamma^\ast} \left(\frac{1}{2}\kappa_c W[\rho]\right)
\nonumber \\
	& & + \frac{1}{8}\left(\epsilon \frac{\partial^3 W[\rho]}{\partial\alpha\partial\beta\partial\gamma^\ast} + \epsilon^\ast \frac{\partial^3 W[\rho]}{\partial\alpha^\ast\partial\beta^\ast\partial\gamma}\right)
\eea
The triple-derivative terms come from the $\chi^{(2)}$ nonlinearity.  They can generally be omitted as long as $\left|\alpha\beta\right|, \left|\alpha\gamma\right|, \left|\beta\gamma\right| \gg 1$.  Eliminating these terms turns this into a Fokker-Planck equation, which gives us the following SDEs:
\bea
	\d\alpha & = & \left[(-\kappa_a/2 - i\Delta_a)\alpha + \frac{1}{2}\epsilon\,\beta^\ast\gamma\right] \d t - \sqrt{\kappa_a} \d\beta_{a} \\
	\d\beta & = & \left[(-\kappa_b/2 - i\Delta_b)\beta + \frac{1}{2}\epsilon\,\alpha^\ast\gamma\right] \d t - \sqrt{\kappa_b} \d\beta_{b} \\
	\d\gamma & = & \left[(-\kappa_c/2 - i\Delta_c)\gamma - \frac{1}{2}\epsilon^\ast\,\alpha\beta\right] \d t - \sqrt{\kappa_c} \d\beta_{c} \\
	\d\tilde{\beta}_{a} & = & \d\beta_{a} + \sqrt{\kappa}\,\alpha\,\d t \\
	\d\tilde{\beta}_{b} & = & \d\beta_{b} + \sqrt{\kappa}\,\beta\,\d t \\
	\d\tilde{\beta}_{c} & = & \d\beta_{c} + \sqrt{\kappa}\,\gamma\,\d t
\eea

\subsection{Degenerate OPO}

A degenerate OPO cavity has two resonant modes -- an $\omega$ mode $a$ and a $2\omega$ mode $b$.  Both of these modes can interact with input-output fields.  This gives an SLH model of the following form:
\beq
	G = \left(1,\ \ \begin{bmatrix} \sqrt{\kappa_a} a \\ \sqrt{\kappa_b} b \end{bmatrix},\ \ 
          \Delta_a a^\dagger a + \Delta_b b^\dagger b 
          + \frac{\epsilon^* a^2 b^\dagger - \epsilon (a^\dagger)^2 b}{2i} \right)
\eeq
Using the Wigner function rules, the master equation
\beq
	\frac{\d \rho}{\d t} = -i[H, \rho] + \left(L \rho L^\dagger - \frac{1}{2}(L^\dagger L \rho + \rho L^\dagger L)\right)
\eeq
can be converted into a PDE for the Wigner function:
\bea
	\!\!\!\!\!\!\!\frac{\partial W[\rho]}{\partial t} & = & 
	-\frac{\partial}{\partial\alpha} \biggl((-\kappa_a/2 - i\Delta_a)\alpha + \epsilon\,\alpha^\ast \beta W[\rho]\biggr)
	-\frac{\partial}{\partial\alpha^\ast} (...) \nonumber \\
	& & -\frac{\partial}{\partial\beta} \left((-\kappa_b/2 - i\Delta_b)\alpha - \frac{1}{2}\epsilon^\ast\,\alpha^2 W[\rho]\right)
	-\frac{\partial}{\partial\beta^\ast} (...) \nonumber \\
	& & + \frac{\partial^2}{\partial\alpha\partial\alpha^\ast} \left(\frac{1}{2}\kappa_a W[\rho]\right)
+ \frac{\partial^2}{\partial\beta\partial\beta^\ast} \left(\frac{1}{2}\kappa_b W[\rho]\right)
+ \frac{1}{8}\left(\epsilon \frac{\partial^3 W[\rho]}{\partial\alpha^2\partial\beta^\ast} + \epsilon^\ast \frac{\partial^3 W[\rho]}{(\partial\alpha^\ast)^2\partial\beta}\right)
\eea
The triple-derivative terms come from the $\chi^{(2)}$ nonlinearity.  They can generally be omitted as long as $\left|\alpha\right|^2 \gg 1$ and $\left|\alpha\beta\right| \gg 1$, since in this case the first- and second-derivative terms are much larger.  Eliminating these terms turns this into a Fokker-Planck equation, which gives us the following SDEs:
\bea
	\d\alpha & = & \left[(-\kappa_a/2 - i\Delta_a)\alpha + \epsilon\,\alpha^\ast\beta\right]\d t - \sqrt{\kappa_a} \d\beta_{a} \\
	\d\beta & = & \left[(-\kappa_b/2 - i\Delta_b)\alpha - \frac{1}{2}\epsilon^\ast\,\alpha^2\right]\d t - \sqrt{\kappa_b} \d\beta_{b} \\
	\d\tilde{\beta}_{a} & = & \d\beta_{a} + \sqrt{\kappa}\,\alpha\,\d t \\
	\d\tilde{\beta}_{b} & = & \d\beta_{b} + \sqrt{\kappa}\,\beta\,\d t
\eea

\section{Atom Cavity}


The atom cavity was discussed in Sec.~\ref{sec:02-atom}.  It consists of $N$ two-level atoms, with Pauli operators $\sigma_{\pm,i}, \sigma_{z,i}$, coupled to a single cavity mode $a$.  There are three environmental couplings -- cavity loss $\kappa$, atomic spontaneous emission $\gamma_{\parallel}$, and non-radiative decay $\gamma_{nr}$.  This has the following SLH model:
\begin{eqnarray}
	S & = & 1 \\
	L & = & \begin{bmatrix} \sqrt{\kappa} a \\ \sqrt{\gamma_{\parallel}}\;\sigma_{-,i} \\ \sqrt{\gamma_{nr}/2}\;\sigma_{z,i} \end{bmatrix} \\
	H & = & \Delta_c a^\dagger a + \frac{1}{2}\Delta_a \sum_k\sigma_{z,k} + i g_0 \sum_k (a^\dagger \sigma_{-,k} - a \sigma_{+,k})
\end{eqnarray}
Following Lugiato \cite{Lugiato1978, Lugiato1982}, define $X = (a, a^\dagger, \sigma_-, \sigma_+, \sigma_z)$ as the operator algebra basis, where $\sigma = \sum_i \sigma_i$ is the total spin.  It can be shown, using the adjoint equations, that this basis is closed under time evolution.  Likewise, define $x = (\alpha, \alpha^*, v, v^*, m)$ as the c-number Wigner basis, and define the polarization decay $\gamma_\perp = \gamma_{nr} + \gamma_{\parallel}/2$.  Applying the method in Sec.~\ref{sec:04-procedure}, we arrive at the following Fokker-Planck coefficients:
\beq
	C^{(1)}(x) = \begin{bmatrix} \left(-i\Delta_c - \kappa/2\right)\alpha + g_0 v \\
	    \left(i\Delta_c - \kappa/2\right)\alpha^* + g_0 v^* \\
	    \left(-i\Delta_a - \gamma_\perp\right)v + g_0 \alpha m \\
	    \left(i\Delta_a - \gamma_\perp\right)v^* + g_0 \alpha^* m \\
	    -\gamma_{\parallel}(m+N) - 2g_0(\alpha v^* + \alpha^* v) \end{bmatrix}
\eeq
Second-order moments:
\beq
	C^{(2)}(x) = \begin{bmatrix}
		0 & \kappa/2 & 0 & 0 & 0 \\
		\kappa/2 & 0 & 0 & 0 & 0 \\
		0 & 0 & 0 & N\gamma_\perp & v \gamma_{\parallel} \\
		0 & 0 & N\gamma_\perp & 0 & v^* \gamma_{\parallel} \\
		0 & 0 & v\gamma_{\parallel} & v^*\gamma_{\parallel} & 2(N+m)\gamma_{\parallel} \end{bmatrix}
\eeq
Generally speaking, the Fokker-Planck approximation is only valid in the limit of many atoms, $N \gg 1$, and probably many photons as well.

The SDEs may be written:
\bea
	\d\alpha & = & \bigl[\left(-i\Delta_c - \kappa/2\right)\alpha + g_0 v\bigr] \d t - \sqrt{\kappa}\,\d\beta \\
	\d v & = & \bigl[\left(-i\Delta_a - \gamma_\perp\right)v + g_0 \alpha m\bigr] \d t + \d\xi_v \\
	\d m & = & \bigl[-\gamma_{||}(m+N) - 2g_0(\alpha v^* + \alpha^* v)\bigr] \d t + \d\xi_m \\
	\d\tilde{\beta} & = & \d\beta + \sqrt{\kappa}\alpha\,\d t
\eea
The optical noise term $\d\beta$ is the standard optical input field.  The other noise terms $\d\xi = (\d\xi_v, \d\xi_v^*, \d\xi_m)$ have the following noise matrix:
\beq
	\d\xi\,\d\xi^T = \begin{bmatrix} 
		0 & N\gamma_\perp & v \gamma_{||} \\
		N\gamma_\perp & 0 & v^* \gamma_{||} \\
		v\gamma_{||} & v^*\gamma_{||} & 2(N+m)\gamma_{||} \end{bmatrix}
\eeq
This is satisfied for:
\bea
	\d\xi_v & = & \sqrt{\frac{N\gamma_\perp}{2}} \left(\d w_1 + i\,\d w_2\right) \\
	\d\xi_m & = & \sqrt{\frac{2\gamma_{||}^2}{N\gamma_{\perp}}} \left(\mbox{Re}[v]\d w_1 + \mbox{Im}[v]\d w_2 + \sqrt{\frac{N\gamma_{\perp}}{\gamma_{||}}(m+N) - v^*v}\right)
\eea
where the $\d w_i$ are Wiener processes.  Note that this only works for $v^*v \leq \frac{\gamma_\perp}{\gamma_\parallel}N(m+N)$.  When this condition is not satisfied, the covariance matrix is not positive definite and the Wigner method cannot be used.

\begin{figure}[tbp]
\begin{center}
\includegraphics[width=0.8\textwidth]{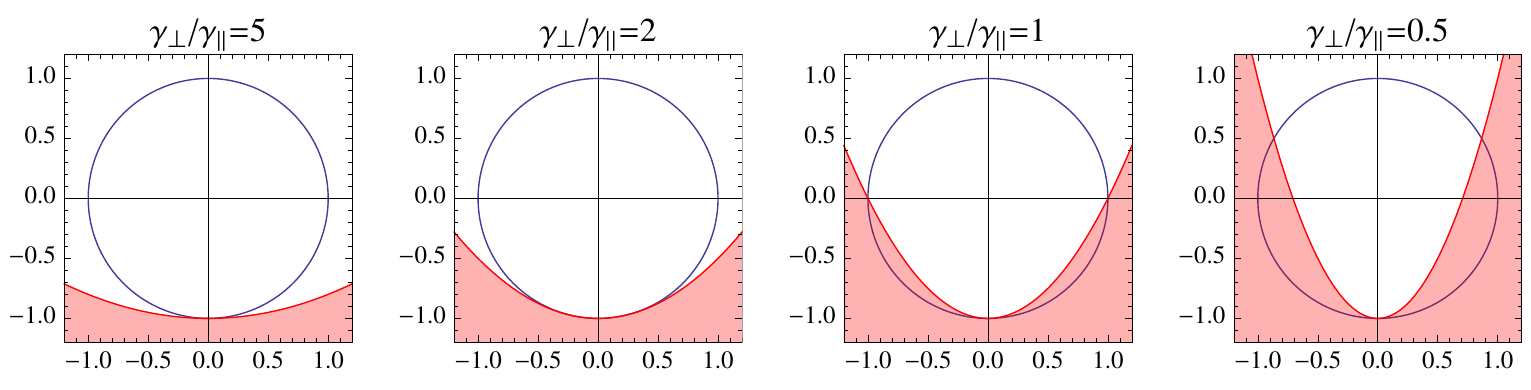}
\caption{Plots of a cross-section of the Bloch sphere (axes are $(v/N, m/N)$) and the ``forbidden region'' $|v/N|^2 > (\gamma_\perp/\gamma_\parallel)(m/N+1)$.}
\label{fig:04-f3}
\end{center}
\end{figure}

The forbidden region is shown in Figure \ref{fig:04-f3}.  Note that it depends strongly on the ratio of polarization decay to excitation decay, $\gamma_\perp/\gamma_\parallel$.  The stronger the non-radiative decay (i.e. the more ``incoherent'' the atoms are), the smaller this region.  Clearly the Wigner method will not work for the highly coherent case where $\gamma_\parallel \gg \gamma_{nr}$ ($\gamma_\perp/\gamma_\parallel = 1/2$), since large, frequently accessed patches of the Bloch sphere are forbidden.  But in the opposite limit, which is often the case for quantum dots, NV-centers or other artificial atoms, the Wigner theory can work quite well.

\subsection{Steady-State Limit}

Coupling to an ensemble of atoms introduces an effective optical nonlinearity, since the atoms are highly nonlinear systems.  The strength of this nonlinearity can be compared to others like the Kerr and free-carrier effects by considering the steady-state behavior.  While this says nothing about the noise or dynamics of the system, it is a good first-order way to compare optical nonlinearities.

To obtain the steady-state limit, assume a constant input $\d\beta = \beta_{\rm in}\d t$, ignore all other noise terms and set the time derivatives to zero:
\bea
	\frac{\d \alpha}{\d t} & = & \left(-i\Delta_c - \kappa/2\right)\alpha + g_0 v - \sqrt{\kappa} \beta_{\rm in} = 0 \\
	\frac{\d  v}{\d t} & = & \left(-i\Delta_a - \gamma_\perp\right)v + g_0 \alpha m = 0 \\
	\frac{\d  m}{\d t} & = & -\gamma_{\parallel}(m+N) - 2g_0(\alpha v^* + \alpha^* v) = 0 \\
	\beta_{\rm out} & = & \sqrt{\kappa} \alpha + \beta_{\rm in}
\eea
The $v$ and $m$ equations are linear in $(v, m)$; this gives rise to a matrix equation for the triplet $(v, v^*, m)$ in terms of $\alpha$.  One can solve this matrix equation, obtaining:
\beq
	v = -\frac{N g_0}{\gamma_\perp + i \Delta_a} \left[1 + \frac{4}{N\gamma_\parallel} \mbox{Re}\left[\frac{N g_0^2}{\gamma_\perp+i\Delta_a}\right]\alpha^*\alpha\right]^{-1}\alpha
\eeq
This gives the following equation for $\alpha$:
\beq
\left(-i\Delta_c-\frac{\kappa}{2}-\frac{N g_0^2}{\gamma_{\perp}+i\Delta_a}\left[1+\frac{4}{N \gamma_{\parallel}}\text{Re}\left[\frac{N g_0^2}{\gamma_{\perp}+i\Delta_a}\right]\alpha^*\alpha \right]^{-1}\right)\alpha - \sqrt{\kappa}\beta_{\text{in}} = 0
\eeq
Taking the absolute value, one obtains a nonlinear relation between the input power and the internal photon number.  As with the Kerr cavity, this can be used to determine when the system exhibits amplification and bistability.

\subsection{Adiabatic Elimination}

In the limit of rapid dephasing, $\gamma_\perp \gg \gamma_\parallel, \kappa$, the atomic polarization $v$ evolves much faster than either the optical field $\alpha$ or the excitation number $m$.  The standard procedure for adiabatic elimination is to replace $v$ with its steady-state value:
\beq
	v \rightarrow \frac{1}{\gamma_\perp + i\Delta_a}(g_0 m \alpha + \xi_v)
\eeq
The equations of motion are best expressed in terms of $\alpha$ and the number of excited atoms, $z \equiv (m+N)/2$.  In addition, it will be useful to define a linear absorption and excitation-dependent dispersion, as follows:
\beq
	\frac{1}{2}\eta + iN \delta \equiv \frac{g_0^2 N}{\gamma_\perp + i\Delta_a}
\eeq
The equations of motion are:
\begin{empheq}[box=\fbox]{align}
	\d\alpha & = \left(-i\Delta_c - \frac{\kappa}{2}\right)\alpha\,\d t + \left(\left[-\frac{\eta(1-2z/N)}{2} - i\delta z\right]\alpha\,\d t + \sqrt{\eta}\,\d\beta_v\right) - \sqrt{\kappa}\,\d\beta_{\rm in} \nonumber \\
	& =  \left[-i(\Delta_c+\delta z) - \frac{\kappa+\eta(1-2z/N)}{2}\right]\alpha\,\d t + \sqrt{\eta}\,\d\beta_v - \sqrt{\kappa}\,\d\beta_{\rm in} \\
	dz & = -\gamma_\parallel z\,\d t - 2\mbox{Re}\left[\alpha^*\left(\left[-\frac{\eta(1-2z/N)}{2} - i\delta z\right]\alpha\,\d t + \sqrt{\eta}\,\d\beta_v\right)\right] + \sqrt{\gamma_\parallel z}\, \d w
\end{empheq}
There are two effects at play here: saturable absorption prevents the excitation number $z$ from exceeding $N/2$, since the absorption saturates at that level.  Dispersion has a Kerr-like nonlinear effect on the field.  When we derive the Wigner equations for the free carrier nonlinearity in the next chapter, we will see that they take a very similar form.


\section{Linearized Systems}
\label{sec:04-linear}

The truncated Wigner method converts the quantum equations of motion into a set of semiclassical SDEs.  One obtains a semiclassical ABCD model, analogous to Eqs.~(\ref{eq:04b-abcd-1}-\ref{eq:04b-abcd-2}), with additional noise terms:
\bea
	\d\dbl{x} & = & \dbl{A}\,\dbl{x}\,dt + \dbl{B}\,\d\dbl{\beta}_{\rm in} + \dbl{F}\,\d w \label{eq:04-linabcd1} \\
	\d\dbl{\beta}_{\rm out} & = & \dbl{C}\,\dbl{x}\,dt + \dbl{D}\,\d\dbl{\beta}_{\rm in} \label{eq:04-linabcd2}
\eea
Here $\dbl{x}$ is the c-number state vector, $\d\dbl{\beta}$ are the input/output fields, and $\d w$ is an additional noise vector with statistics $\d w_i \d w_j = \delta_{ij}\d t$.

\subsection{Basic Theory}

\subsubsection{Moment Equations}

The Gaussian moment equations for $\dbl{\mu} \equiv \langle \dbl{x} \rangle$, $\dbl{\sigma} \equiv \langle \dbl{x}\dbl{x}^* \rangle$ (see Eqs.~(\ref{eq:04b-mom1}-\ref{eq:04b-mom2}, \ref{eq:04b-mom3})) become:
\beq
	\frac{\d\dbl{\mu}}{\d t} = \dbl{A}\dbl{\mu},\ \ \ 
	\frac{\d\dbl{\sigma}}{\d t} = \dbl{A}\dbl{\sigma} + \dbl{\sigma}\dbl{A}^\dagger + \frac{1}{2} \dbl{B} \dbl{B}^\dagger + \dbl{F} \dbl{F}^\dagger \label{eq:04-mom}
\eeq
The internal state at $t \rightarrow \infty$ can be solved by setting $\d\dbl{\sigma}/\d t = 0$, giving a Lyapunov equation for $\dbl{\sigma}$; compare (\ref{eq:04b-lyapunov}).
	
\subsubsection{Circuit Algebra}

The concatenation product $G_1 \boxplus G_2$ is (compare Eq.~(\ref{eq:04b-concat})):
\begin{align}
	\dbl{a} &= P_a\begin{bmatrix} \dbl{a}^{(1)} \\ \dbl{a}^{(2)} \end{bmatrix}, 
	&\dbl{A} &= P_a\begin{bmatrix} \dbl{A}^{(1)} & 0 \\ 0 & \dbl{A}^{(2)} \end{bmatrix}P_a^{-1},
	&\dbl{B} &= P_a\begin{bmatrix} \dbl{B}^{(1)} & 0 \\ 0 & \dbl{B}^{(2)} \end{bmatrix}P_{\d B}^{-1}, \nonumber \\ 
	\d\dbl{B} &= P_{\d B}\begin{bmatrix} \d\dbl{B}^{(1)} \\ \d\dbl{B}^{(2)} \end{bmatrix},
	&\dbl{C} &= P_{\d B}\begin{bmatrix} \dbl{C}^{(1)} & 0 \\ 0 & \dbl{C}^{(2)} \end{bmatrix}P_a^{-1},
	&\dbl{D} &= P_{\d B}\begin{bmatrix} \dbl{D}^{(1)} & 0 \\ 0 & \dbl{D}^{(2)} \end{bmatrix}P_{\d B}^{-1}, \nonumber \\
	\dbl{F} &= P_a\begin{bmatrix} \dbl{F}^{(1)} & 0 \\ 0 & \dbl{F}^{(2)} \end{bmatrix}
\end{align}
The series product $G_2 \triangleleft G_1$ is (compare Eq.~(\ref{eq:04b-series})):
\begin{align}
	\dbl{A} &= P_a\begin{bmatrix} \dbl{A}_1 & 0 \\ \dbl{B}_2 \dbl{C}_1 & \dbl{A}_2 \end{bmatrix}P_a,
	&\dbl{B} &= P_a\begin{bmatrix} \dbl{B}_1 & \dbl{B}_2 \dbl{D}_1 \end{bmatrix},\nonumber \\
	\dbl{C} &= \begin{bmatrix} \dbl{D}_2 \dbl{C}_1 \\ \dbl{C}_2 \end{bmatrix}P_{a},
    &\dbl{D} &= \dbl{D}_2 \dbl{D}_1,
    &\dbl{F} &= P_a\begin{bmatrix} \dbl{F}_1 & 0 \\ 0 & \dbl{F}_2 \end{bmatrix}
\end{align}
The feedback operator $[G]_{k\rightarrow l}$ is realized by (compare Eq.~(\ref{eq:04b-feedback})):
\begin{align} 
	\dbl{A} &= \dbl{A} + \dbl{B}_{:,l}(1 - \dbl{D}_{kl})^{-1}\dbl{C}_{k,:} 
	&\dbl{B} &= \dbl{B}_{:,!l} + \dbl{B}_{:,l}(1 - \dbl{D}_{kl})^{-1}\dbl{D}_{k,!l} \nonumber \\
	\dbl{C} &= \dbl{C}_{!k,:} + \dbl{D}_{!k,l}(1 - \dbl{D}_{kl})^{-1}\dbl{C}_{k,:} 
	&\dbl{D} &= \dbl{D}_{!k,!l} + \dbl{D}_{!k,l}(1 - \dbl{D}_{kl})^{-1}\dbl{D}_{k,!l}
\end{align}
and $F$ remains the same.

\subsubsection{Adiabatic Elimination}

Adiabatic elimination of the linear system is realized by (compare Eq.~(\ref{eq:04b-ad-el})):
\beq
	\d\dbl{\beta}_{\rm out} = \left(\dbl{D} - \dbl{C} \dbl{A}^{-1} \dbl{B}\right) \d\dbl{\beta}_{\rm in} + \left(\dbl{C}\dbl{A}^{-1}\dbl{F}\right)\,dw 
	+ (\dbl{c} - \dbl{C}\dbl{A}^{-1}\dbl{a}) \d t
	\label{eq:04b-ad-el}
\eeq

\subsubsection{Input-Output Relations}

Following Sec.~\ref{sec:04b-inout}, we can model input-output behavior in the Wigner picture in terms of a transfer function $\dbl{T}(\omega)$.  We will also need a noise matrix $\dbl{N}(\omega)$.  To start, we define doubled-up frequency-domain input-output fields $\dbl{\beta}_\omega = (\beta_\omega,\ \beta_{-\omega}^*)$ (see Eq.~(\ref{eq:04b-adbl})).

The ABCD equations, in the frequency domain, become (compare Eqs.~(\ref{eq:04b-abcdf1}-\ref{eq:04b-abcdf2})):
\bea
	-i\omega \dbl{a}_\omega & = & \dbl{A}\dbl{a}_\omega + \dbl{B}\dbl{b}_{{\rm in},\omega} + \dbl{F} w_\omega \\
	\dbl{b}_{{\rm out},\omega} & = & \dbl{C}\dbl{a}_\omega + \dbl{D}\dbl{b}_{{\rm in},\omega}
\eea
The input and output are related by a matrix and some noise (compare Eq.~(\ref{eq:04b-tf})):
\beq
	\dbl{\beta}_{{\rm out},\omega} = \underbrace{\left[D + C \frac{1}{-i\omega - A}B\right]}_{\dbl{T}(\omega)}\dbl{\beta}_{{\rm in},\omega} + \underbrace{C\frac{1}{-i\omega - A} F}_{\dbl{N}(\omega)} w_\omega
\eeq
Note that these matrices have the doubled-up structure (compare Eq.~(\ref{eq:04b-tstruc})):
\beq
    \dbl{T}(\omega) = \begin{bmatrix} T_{-}(\omega) & T_{+}(\omega) \\ 
        T_{+}(-\omega)^* & T_{-}(-\omega)^* \end{bmatrix},\ \ \ 
    N(\omega) = \begin{bmatrix} N_+(\omega) \\ N_-(\omega)\end{bmatrix}
\eeq
The amplitude-gain relations, Sec.~\ref{sec:04b-gain}, carry over unchanged.

The squeezing spectrum is defined in terms of the $\dbl{M}(\omega)$ and $\dbl{N}(\omega)$; see Sec.~\ref{sec:04b-noise}, which are given by:
\begin{eqnarray}
    \mathcal{N} + \frac{1}{2} & \!\!\!=\!\!\! & \frac{1}{2}\left[\frac{|T_{-}(\omega)|^2 + |T_{+}(\omega)^2| + |T_{+}(-\omega)^2| + |T_{-}(-\omega)^2|}{2} + |N_+(\omega)|^2 + |N_-(\omega)|^2
    \right] \\
    \mathcal{M} & \!\!\!=\!\!\! & \left[\frac{T_{-}(\omega)T_{+}(-\omega) + T_{+}(\omega)T_{-}(-\omega)}{2} + N_+(\omega)N_-(\omega)^\dagger
    \right]
\end{eqnarray}

\subsection{Example: Kerr Cavity}
\label{sec:04b-kerr}

The Kerr cavity is the simplest such linearized system.  All of its degrees of freedom are bosonic, so its linearized form corresponds to an actual quantum model.

Following Sec.~\ref{sec:05-kerr}, the Wigner SDEs for the most general Kerr cavity are:
\beq
	\d\alpha_i = (-i\chi-\beta)\sum_{jkl} \bigl[\alpha_j^*\alpha_k\alpha_l\bigr] \d t - 2\sqrt{\beta}\sum_m \Lambda_{m,ij}^*\alpha_j^*\d\beta_{{\rm 2PA},m} + \mbox{(linear terms)}
\eeq
The single-mode case is the most ubiquitous.  Adding linear terms for cavity detuning and mirror losses, it becomes:
\bea
	\d\alpha & = & \left[\left(-i\Delta - \frac{1}{2}\kappa\right) + (-i\chi-\beta)\alpha^*\alpha^2\right]\,\d t - \sum_i\sqrt{\kappa_i}\d\beta_{{\rm in}, i} - 2\sqrt{\beta} \alpha^*\d\beta_{{\rm 2PA},m} \\
	\d\beta_{{\rm out}, i} & = & \d\beta_{{\rm in}, i} + \sqrt{\kappa_i}\alpha
\eea
Linearizing these around a given state, $\alpha \rightarrow \alpha + \delta\alpha$, one finds the following equations of motion:
\bea
	d(\delta\alpha) & = & \left[\left[\left(-i\Delta - \frac{1}{2}\kappa\right) + (-i\chi-\beta)\alpha^*\alpha\right]\delta\alpha + \left[(-i\chi-\beta)\alpha^2\right]\delta\alpha^*\right]\,\d t \nonumber \\
	& & - \sum_i\sqrt{\kappa_i}\d\beta_{{\rm in}, i} - 2\sqrt{\beta} \alpha^*\d\beta_{{\rm 2PA},m} \\
	\d\beta_{{\rm out}, i} & = & \d\beta_{{\rm in}, i} + \sqrt{\kappa_i}\alpha
\eea
In doubled-up notation, with a single input-output port, we get the ABCD equations:
\bea
	d(\delta\dbl{\alpha}) & = & \begin{bmatrix} \begin{array}{c} (-i\chi-\beta)\alpha^*\alpha \\ \ +(-i\Delta - \kappa/2)\! \end{array} & (-i\chi-\beta)\alpha^2 \\ \left[(-i\chi-\beta)\alpha^2\right]^* & \begin{array}{c} \bigl[(-i\chi-\beta)\alpha^*\alpha \\ \ +(-i\Delta - \kappa/2)\bigr]^*\! \end{array} \end{bmatrix}\delta\dbl{\alpha}\,\d t - \sqrt{\kappa} \begin{bmatrix} 1&0\\0&1 \end{bmatrix} \d\dbl{\beta}_{\rm in} + \sqrt{\beta}\begin{bmatrix} \alpha^* & i\alpha^* \\ \alpha & -i\alpha \end{bmatrix}\d w \nonumber \\ \
	\d\dbl{\beta}_{\rm out} & = & \sqrt{\kappa}\begin{bmatrix} 1&0\\0&1 \end{bmatrix}\delta\dbl{\alpha}\,\d t + \begin{bmatrix} 1&0\\0&1 \end{bmatrix}\d\dbl{\beta}_{\rm in} \label{eq:04-kerr-abcd}
\eea
This is equivalent to a single-mode OPO (Sec.~\ref{sec:04b-dopo}) if we make the following substitutions:
\bea
	\Delta + 2\chi\alpha^*\alpha & \leftrightarrow & \Delta_{\rm opo} \nonumber \\
	\kappa & \leftrightarrow & \kappa_{\rm opo}\ \mbox{(coupling in $B$, $C$)} \nonumber \\
	\kappa + 2\beta\alpha^*\alpha & \leftrightarrow & \kappa_{\rm opo}\ \mbox{(loss in $A$)} \nonumber \\
	(-i\chi - \beta)\alpha^2 & \leftrightarrow & \epsilon_{\rm opo} \label{eq:04-kerrcorr}
\eea
There are extra noise terms, and corresponding loss, due to two-photon absorption.  In the $\beta = 0$ limit, this goes away but the nonlinearity, so we get perfect squeezing in a model identical to the OPO.

\subsubsection{Squeezing of Internal State}

For a given Kerr cavity, and variable input, where is squeezing the strongest?  Assuming a small $\chi$, both $\epsilon_{\rm opo}$ and $\kappa_{\rm opo}$ are slowly varying functions of $\alpha$.  But $\Delta_{\rm opo}$ is not, passing through zero at $\alpha^*\alpha = -\Delta/2\chi$.  Maximum squeezing will happen at approximately this point.

Not all OPOs are stable, and nor are all Kerr cavities.  The stability of a Kerr solution depends on whether or not the linearized model, i.e.\ the OPO, is stable.  This gives the criterion:
\beq
	3(\chi^2 + \beta^2)(\alpha^*\alpha)^2 + (4\chi\Delta + \beta\kappa)(\alpha^*\alpha) + (\Delta^2+\kappa^2/4) \geq 0
\eeq
This is a convex quadratic function, of the form $ax^2 + bx + c$ so it is stable for all values of $\alpha$ if $b^2/4a + c \geq 0$.  This gives a global stability criterion for Kerr cavities:
\beq
	\left(\Delta^2 + \frac{\kappa^2}{4}\right) - \frac{4(\chi\Delta + \beta\kappa/4)^2}{3(\chi^2+\beta^2)} \geq 0
\eeq
This ends up being equivalent to $\Delta \geq -\sqrt{3/4}\,\chi$ if there isn't any two-photon absorption.  For parameters outside this range, the Kerr cavity goes bistable.

\begin{figure}[bt]
	\centering
	\includegraphics[width=1.00\textwidth]{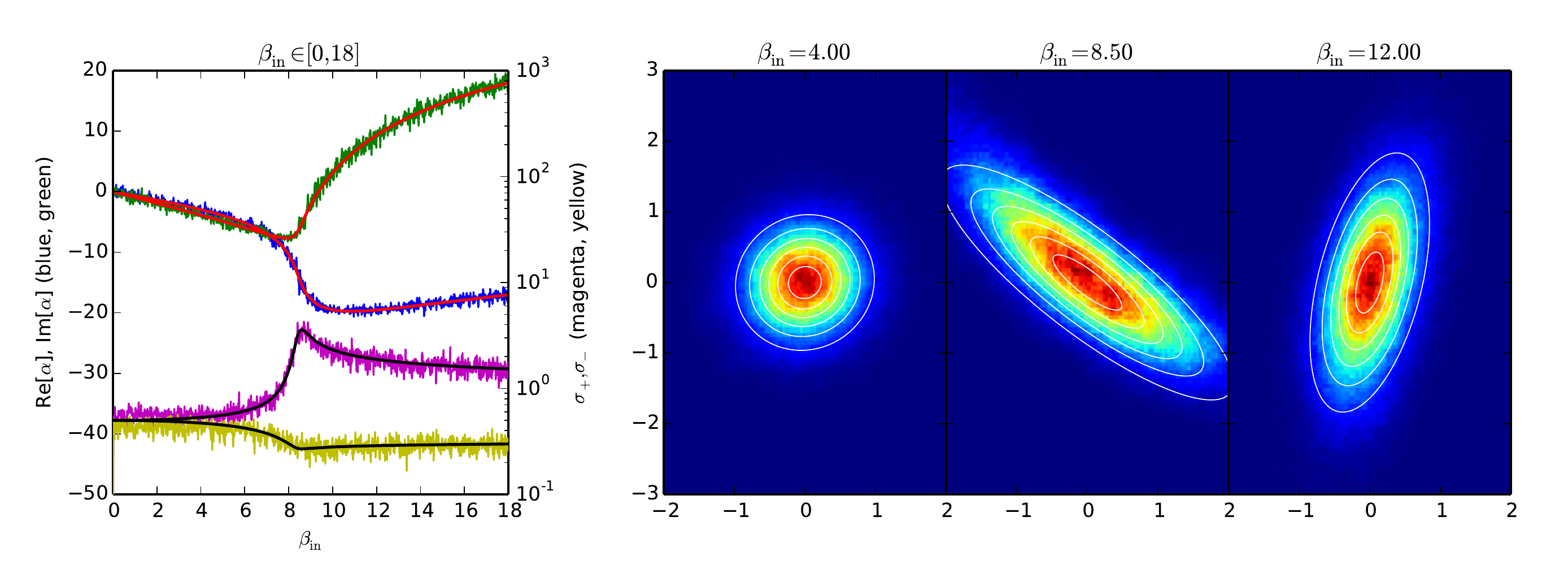}
	\caption{Left: internal state of the Kerr cavity $\alpha$ (blue, green), and eigenvalues $\sigma_+, \sigma_-$ of 	\label{fig:04b-f3}
the covariance matrix (magenta, yellow) as a function of input field $\beta_{\rm in}$.  Right: Simulation of Wigner function for internal state (equilibrium value subtracted) at $\beta_{\rm in} = 4.0,\ 8.5,\ 12.0$ (colored) compared to ABCD model prediction (white contours)}
\end{figure}

Figure \ref{fig:04b-f3} shows the internal state of the Kerr cavity as a function of bias field.  Not surprisingly, one quadrature is highly amplified when the device amplification is greatest (around $\beta_{\rm in} = 8.5$).  But since the Kerr cavity also acts as a squeezer, the noise in the opposite quadrature is suppressed.  In the strong-amplification case, though, the Wigner function deviates slightly from the ideal Gaussian.  This is due to nonlinearities in the Kerr model.  If we were to take the limit $\chi \rightarrow 0$ with $\beta_{\rm in}/\sqrt{\chi}$ fixed, these nonlinearities would go away and the Kerr model would behave exactly like an OPO.

\subsubsection{Output Spectrum}

Since the Kerr cavity resembles an OPO, we expect to see squeezing in the output light.  One output quadrature ends up squeezed, while the other is anti-squeezed.  The bandwidth of the squeezing should be related to the bandwidth of the cavity.

\begin{figure}[tb]
	\centering
	\includegraphics[width=1.00\textwidth]{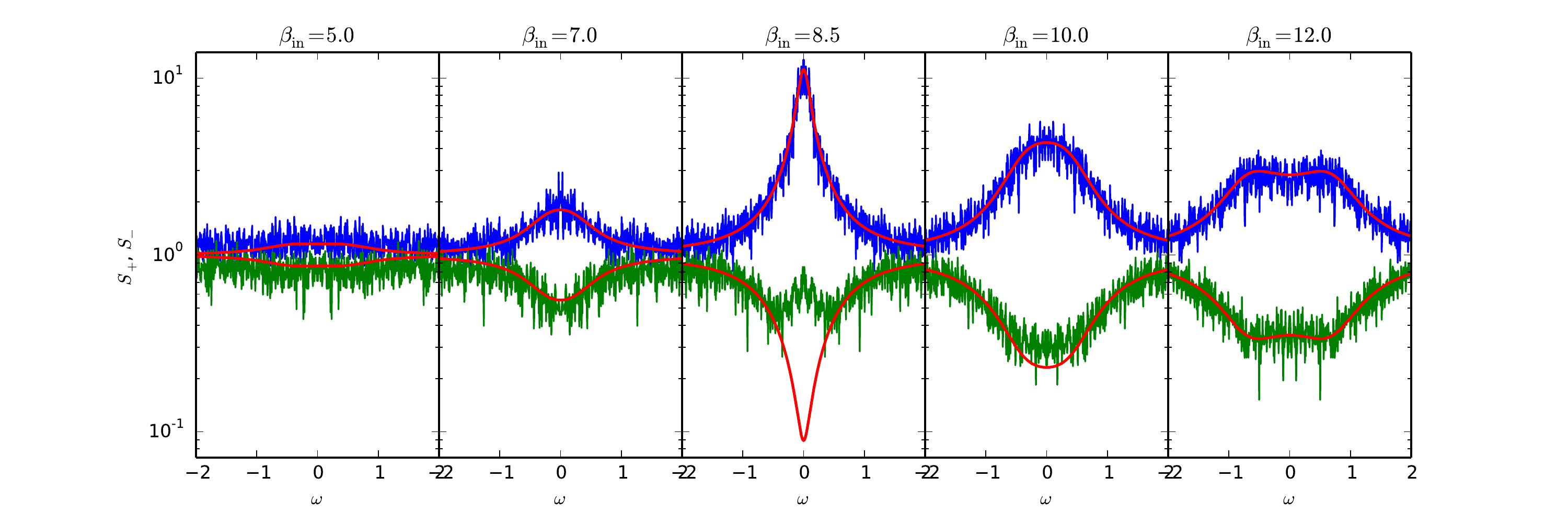}
	\caption{Output noise spectra for biased Kerr cavity.}
	\label{fig:04b-f6}
\end{figure}

However, the Kerr cavity is a nonlinear system.  If that nonlinearity is strong enough, it will cause deviations from the ideal OPO spectrum.  This is especially true in the strong-amplification case.  When the Kerr cavity is set to amplify strongly, the internal field is very spread out along one direction (see center-right plot in Fig.~\ref{fig:04b-f3}).  Such an elongated state is very sensitive to nonlinearities, particularly along the axis being squeezed.  As a result, the actual squeezing will deviate from its predicted value in the strong-squeezing case.  This is seen in the central plot of Fig.~\ref{fig:04b-f6}

\ifstandalone{\bibliography{NoteRefs}{}
\bibliographystyle{alpha}
}
\ifdefined\multidoc\else\input{Header}\fi

%

\ifstandalone{\setcounter{chapter}{5}}
\chapter{Free-Carrier Theory}
\label{ch:05b}

The next two chapters are based on the following paper:

\begin{itemize}
	\item \href{http://dx.doi.org/10.1103/PhysRevA.92.023819}{R.~Hamerly and H.~Mabuchi, ``Quantum Noise of Free-Carrier Dispersion in Semiconductor Optical Cavities'', Physical Review A 92, 023819 (2015)}
\end{itemize}

Optical logic requires a platform that is fast, low-power and scalable to compete with electronics.  In the past decade, nano-photonics has advanced to the point where optical cavities of size $\lesssim (\lambda/n)^3$ and $Q$ factors $\gtrsim 10^4$ can be fabricated with standard techniques \cite{Notomi2010, Notomi2011}.  The hope is that these cavities can be used to amplify the optical nonlinearity of materials or defects and perform all-optical logic for communications and computing at speeds and energy scales comparable to electronics.

Free-carrier dispersion is a promising nonlinearity for low-power optical logic.  The effect arises in all semiconductors.  In a semiconductor, there is a filled valence band and an empty conduction band, and when photons are absorbed, they excite electrons from the valence band to the conduction band.  Each absorption creates two free carriers -- an electron and a hole -- which evolve independently and decay on some timescale set by the material and its geometry.  The carriers provide feedback to the optical field by altering the absorption of the material (free-carrier absorption) or its refractive index (free-carrier dispersion).  On timescales long compared to the free-carrier lifetime, it acts as an effective optical nonlinearity and can be used to construct switches, amplifiers and other logic elements.

Accurate, semiclassical models for free-carrier effects already exist, and these are valid when the carrier and photon number are very large \cite{LundstromBook, Bennett1990}.  However, the real promise of free-carrier effects lies in their application to low-power photonic computing.  In some materials, free-carrier effects are strong enough that switching can be achieved with as few as 100 photons per cavity.  In this regime, quantum effects become important and place fundamental limits on device performance.  For example, quantum fluctuations in the photon number add noise to quantum amplifiers \cite{Caves1982} and lead to spontaneous switching in optical memories \cite{Kerckhoff2011b, Mabuchi2011a}.  This motivates the need to develop a quantum model for the free-carrier nonlinearity that works at low photon numbers, similar to the models that exist for cavity quantum electrodynamics (QED) and $\chi^{(3)}$ (Kerr) systems \cite{Kimble1998, Agrawal1979}.

In this chapter, I derive a quantum-mechanical model for the free-carrier nonlinearity, following the standard open quantum systems formalism (Ch.~\ref{ch:01}) used for cavity QED, optical parametric oscillators (OPOs) and $\chi^{(3)}$ systems \cite{Gardiner1985}.  However, simulating even a single cavity in this model is not practical, since the large number of available carrier modes makes the full Hilbert space exponentially large.  Using a method based on the Wigner function (Ch.~\ref{ch:04}), one can reduce the master equation to a set of c-number Langevin equations that are simple to simulate \cite{Santori2014, Lugiato1978, Gardiner1988}.  These equations bear resemblance to semiconductor laser rate equations and Bloch equations found in the literature \cite{Haug1969, AgrawalDutta, Lindberg1988}.  An adiabatic elimination reduces the model further, giving a set of stochastic differential equations (SDEs) for the field, electron number, and hole number in the cavity.  The deterministic part of these equations matches the classical models found in the previous literature, but the noise terms are new -- and have a quantum origin.

Section~\ref{sec:05b-model} introduces the quantum model for the free-carrier cavity.  In Section~\ref{sec:05b-wigner}, I introduce the Wigner formalism and apply it to this model, deriving a set of stochastic differential equations (SDEs) which can be simplified by invoking a weak-doping, fast-dephasing limit.  The key result of this chapter, summarized in Equations (\ref{eq:05b-csde-1}-\ref{eq:05b-csde-4}) and (\ref{eq:05b-sde-sc1}-\ref{eq:05b-sde-sc2}), resembles the equations of motion for the Kerr cavity derived in \cite{Santori2014}, but there are extra noise terms.

A proof of closedness of the operator algebra is given in Appendix~\ref{sec:05b-fullsdes}.  Extensions to the free-carrier model, incorporating two-photon absorption and free-carrier absorption, are treated in Appendix~\ref{sec:05b-related}.  Appendix~\ref{sec:05b-materials} shows how the key parameters in (\ref{eq:05b-csde-1}-\ref{eq:05b-csde-4}), the coupling constant and carrier-dependent detuning, can be derived from measured material properties.  Plasma dispersion in silicon and band-filling in III-V materials are used as examples.

Future chapters will make use of these results.  In Chapter \ref{ch:06b}, I apply (\ref{eq:05b-csde-1}-\ref{eq:05b-csde-4}) to study the steady-state behavior, and correspondence to the Kerr cavity, of the free-carrier system.  Next, the free-carrier SDEs are applied to simulate two devices: a phase-sensitive amplifier in Section~\ref{sec:06b-amp}, and an all-optical SR-latch in Section~\ref{sec:06b-latch}.  For the amplifier, the free-carrier device does not show squeezing, whereas its Kerr analog does.  For the latch, the spontaneous switching rate is larger for the free-carrier device, and the discrepancy grows as the latch's bistable states become more widely separated.  Chapter \ref{ch:07} studies a Hopf bifurcation in free-carrier cavities, comparing the quantum noise in such devices to a quantum-limited linear amplifier.

\section{Quantum Model}
\label{sec:05b-model}

\begin{figure}[tbp]
\begin{center}
\includegraphics[width=0.50\textwidth]{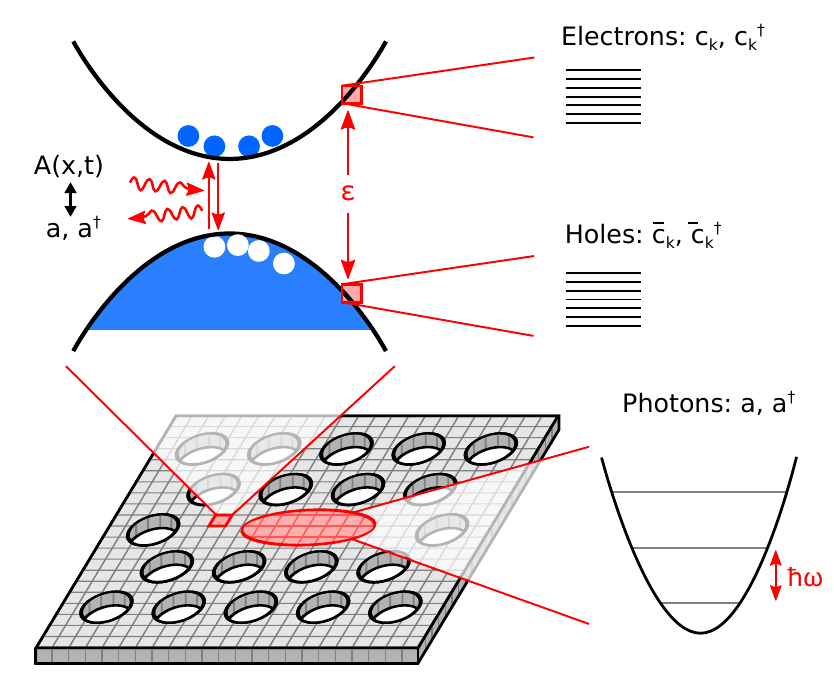}
\caption{Electronic and optical modes of a semiconductor cavity.}
\label{fig:05b-f1}
\end{center}
\end{figure}

Consider a single-mode optical cavity fabricated from an undoped semiconductor.  The optical degree of freedom can be represented as a harmonic oscillator, with the creation / annihilation operators $a, a^\dagger$.  In band theory, the electronic degree of freedom is represented by many uncoupled fermion modes.  For a two-band model, we have a single electron band and hole band.  Each mode has its own (fermionic) creation / annihilation operators -- $c_\k, c_\k^\dagger$ for electrons, $\bar{c}_\k, \bar{c}_\k^\dagger$ for holes, where $\k$ is the mode index.  The operator algebra is:
\begin{eqnarray}
	[a, a^\dagger] = 1,\ \ 
	\{c_\k, c_\l^\dagger\} = \{\bar{c}_\k, \bar{c}_\l^\dagger\} = \delta_{\k\l} \label{eq:05b-opalg}
\end{eqnarray}
The operator algebra is key to the Wigner analysis in Section~\ref{sec:05b-wigner}.  In short, one can define a generalized Wigner function for the quantum system if one can find a ``closed'' set of operators $\{X_i\}$, where the bath-averaged time derivatives $(\d X_i/\d t)_{\rm ad}$, $(\d (X_iX_j)/\d t)_{\rm ad}$, defined in Section \ref{sec:05b-wigner}, are always functions of the $\{X_i\}$.  However, Wigner functions for fermionic operators require the use of Grassmann variables \cite{Cahill1999}, for which the analogy to classical phase space is less intuitive.  Thus, we identify fermion pairs and perform the following bosonization:
\begin{eqnarray}
	\sigma_{-\k} & = & c_\k \bar{c}_\k \label{eq:05b-bos1} \\
	\sigma_{+\k} & = & \bar{c}_\k^\dagger c_\k^\dagger \label{eq:05b-bos2} \\
	n_\k & = & c_\k^\dagger c_\k \label{eq:05b-bos3} \\
	\bar{n}_\k & = & \bar{c}_\k^\dagger \bar{c}_\k \label{eq:05b-bos4} \\
	Q_\k & = & n_\k\bar{n}_\k = \sigma_{+\k}\sigma_{-\k} = c_\k^\dagger\bar{c}_\k^\dagger \bar{c}_\k c_\k \label{eq:05b-bos5}
\end{eqnarray}
This is similar to the operator algebra in an ensemble of two-level atoms \cite{Lugiato1978, Lugiato1982}, but there are some extra terms.  Analogous to the atom ensemble, the electronic polarization is given by $\sigma_{-\k}$.  However, the free-carrier system contains two number operators $n_{\k}$, $\bar{n}_{\k}$ rather than one, as well as a pairing operator $Q_{\k}$.  These arise because the electrons and holes in the free-carrier system have more freedom of movement: in an ensemble of atoms, each electron is confined to its parent atom and $n_\k = \bar{n}_\k = Q_\k$, while in a semiconductor these three quantities are no longer equal, since electrons and holes freely scatter between modes $\k$.  Operators (\ref{eq:05b-bos1}-\ref{eq:05b-bos5}) are bosonic because they are products of an even number of fermionic operators.  Note that several bosonic operators, namely $c_\k \bar{c}_\k^\dagger$ and $\bar{c}_\k c_k^\dagger$, are not included in (\ref{eq:05b-bos1}-\ref{eq:05b-bos2}) -- this is because we are interested in systems that respect charge conservation, while $c_\k \bar{c}_\k^\dagger$ and $\bar{c}_\k c_\k^\dagger$ violate it.

The bosonized operators are closed under commutation, with the following nonzero commutators:
\begin{eqnarray}
	{[}\sigma_{\pm \k}, n_\l] & = & \mp\delta_{\k\l}\sigma_{\pm \k} \\
	{[}\sigma_{\pm \k}, \bar{n}_\l] & = & \mp\delta_{\k\l}\sigma_{\pm \k} \\
	{[}\sigma_{\pm \k}, \sigma_{\mp \l}] & = & \pm\delta_{\k\l} (n_\k+\bar{n}_\k-1) \\
	{[}\sigma_{\pm \k}, Q_\l] & = & \mp \delta_{\k\l} \sigma_{\pm \k}
\end{eqnarray}
Note that these are all commutators, rather than anti-commutators, because the operators have been bosonized.

\subsection{Hamiltonian}

The Hamiltonian consists an an optical part $H_{\rm ph}$ which resembles a harmonic oscillator, an electronic part $H_{\rm el}$ given by the electronic band structure, and an interaction part $H_{\rm int}$ due to the $A\cdot p$ light-matter interaction.  It can be written as:
\beq
	H = \underbrace{\vphantom{\frac{n_k}{2}}\Delta_c a^\dagger a}_{H_{\rm ph}} + \sum_\k \Bigl[\underbrace{\Delta_\k \frac{n_\k + \bar{n}_\k}{2}}_{H_{\rm el}} + \underbrace{\vphantom{\frac{n_k}{2}}i g_\k (a^\dagger \sigma_{-\k} - a \sigma_{+\k})}_{H_{\rm int}}\Bigr]
	\label{eq:05b-ham}
\eeq
where $\Delta_c = \hbar(\omega_c - \omega_{\rm ph})$ is the cavity resonance detuning, $\Delta_\k = E_{\k,c} - E_{\k,v} - \hbar\omega_{ph}$ is the detuning of the transition, and $g_\k$ is the atom-photon coupling.  The coupling can be expressed in terms of material parameters as shown in Appendix \ref{sec:05b-materials}.

Hamiltonian (\ref{eq:05b-ham}) resembles the cavity QED Hamiltonian.  This is because both systems contain an optical term, and electronic term, and a light-matter interaction of the $A\cdot p$ form.  Thus, it should not be surprising if free-carrier cavities exhibit many of the same phenomena observed in cavity QED, e.g.\ bistability, amplification, limit cycles \cite{Kwon2013}.

\subsection{External Interactions}
\label{sec:05b-l}

In the cavity, the optical field is relatively well isolated from its environment.  The two primary interactions are optical absorption, which gives rise to particle-hole pairs and is treated through Eq.~(\ref{eq:05b-ham}), and coupling to the external waveguide.  Because these couplings are usually quite weak, the optical field tends to retain its coherence in spite of them.

The same is not true for the carriers.  Many forces act to dephase, thermalize, and scatter the free carriers on very quick timescales (typically around 10--100 fs) \cite{LundstromBook, Sabbah2002, Lin1987}.  Even for very poor cavities with $Q \lesssim 1000$, this is much faster than the photon lifetime.  The practical upshot of this will be that, on optical timescales, the ``coherent'' part to the carrier fields $\sigma_{\pm\k}$ can be adiabatically eliminated and only the ``slowly-varying'' carrier numbers $n_\k, \bar{n}_\k, Q_\k$ remain relevant to the system.

\begin{figure}[tbp]
\begin{center}
\includegraphics[width=0.60\textwidth]{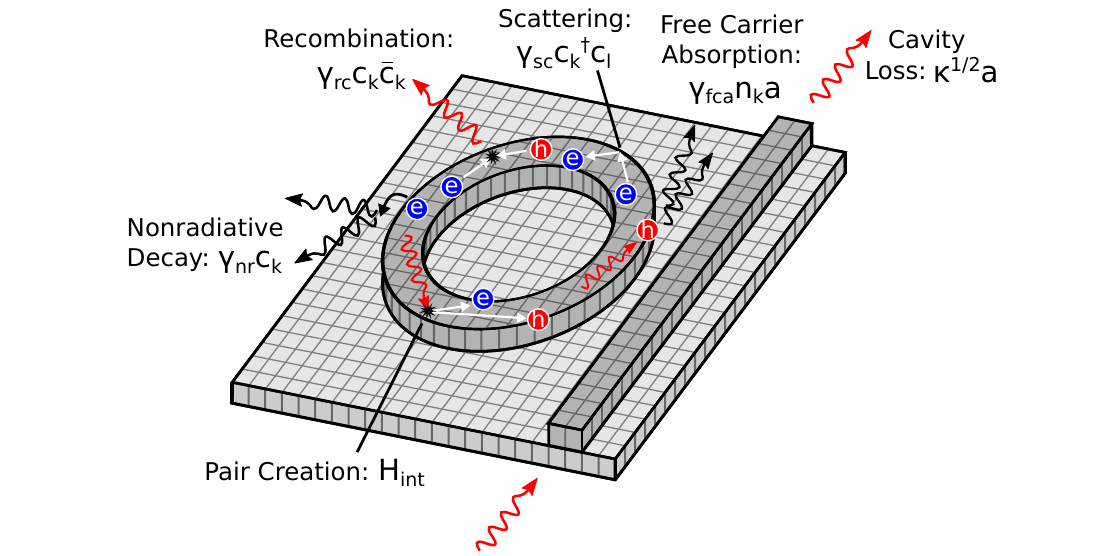}
\caption{Major free-carrier effects in an optical cavity}
\label{fig:05b-f2}
\end{center}
\end{figure}

For a bosonic, Markovian bath, external interactions can be treated by adding extra Lindblad terms to the Master equation \cite{Gardiner1985}.  The main external processes are given in Figure \ref{fig:05b-f2} above.  As Lindblad terms, they are:

\begin{itemize}
	\item Cavity Loss, mediated by
		\beq
			L_{\rm cav} = \sqrt{\kappa}\;a
		\eeq
	\item Recombination, mediated by
		\beq
			L_{\rm rc,\k} =
				\sqrt{\gamma_{rc,\k}}\;\sigma_{-\k}
		\eeq
	\item Nonradiative Decay, mediated by
		\beq
			L_{{\rm nr},{\k}} =
				\sqrt{\gamma_{nr,\k}}\;c_{\k},\ \
				\sqrt{\bar{\gamma}_{nr,\k}}\;\bar{c}_{\k}
		\eeq
	\item Scattering / Dephasing, mediated by
		\beq
			L_{\rm sc,\k\rightarrow \l} =
				\sqrt{\gamma_{\k\rightarrow\l}}\,c_{\l}^\dagger c_{\k},\ \
				\sqrt{\bar{\gamma}_{\k\rightarrow\l}}\,\bar{c}_{\l}^\dagger \bar{c}_{\k}
		\eeq
\end{itemize}

Some materials also have significant free-carrier absorption and two-photon absorption.  For simplicity, these are not treated presently, but are discussed in Appendix~\ref{sec:05b-otherprocess}.

\subsection{Single-Mode Theory}
\label{sec:05b-singlemode}

To make the computation more tractable, assume for now that all carrier modes $\k$ are identical.  This is not always a realistic assumption, and in Appendix \ref{sec:05b-manymodes}, I extend the result to non-identical modes.  But assuming identical modes for now, $\Delta_{\k}, g_{\k},$ and all the $\gamma_{\k}$'s become independent of $\k$.  Suppose that there are $N$ modes.  One can now define mode-summed operators:
\beq
\sigma_{\pm} = \sum_{\k} \sigma_{\pm \k},\ \mbox{etc.} \label{eq:05b-sumop}
\eeq
This reduces the dimensionality of the state space from $5N+2$ to $7$.  In terms of these, the Hamiltonian and interaction terms are:
\bea
	\!\!\!\!H & \!=\! & \Delta_c a^\dagger a + \frac{1}{2}\Delta_e (n + \bar{n}) + i g (a^\dagger \sigma_{-} \!- a \sigma_{+}) \label{eq:05b-slh-1} \\
	\!\!\!\!L_{\rm cav} & \!=\! & \sqrt{\kappa}\;a \\
	\!\!\!\!L_{\rm rc,\k} & \!=\! &
				\sqrt{\gamma_{rc}}\;\sigma_{-\k} \\
	\!\!\!\!L_{{\rm nr},\k} & \!=\! &
				\sqrt{\gamma_{nr}}\;c_{\k},\ \
				\sqrt{\bar{\gamma}_{nr}}\;\bar{c}_{\k} \\
	\!\!\!\!L_{\rm sc,\k\rightarrow \l} & \!=\! &
				\sqrt{\frac{\gamma_{\rm sc}}{2}}\;c_{\l}^\dagger c_{\k},\ \
				\sqrt{\frac{\bar{\gamma}_{\rm sc}}{2}}\;\bar{c}_{\l}^\dagger \bar{c}_{\k} \label{eq:05b-slh-n}
\eea
Note that the sum operators used are bosonic, not fermionic.  The single-mode theory would not work if one had started with the fermionic operators.

\section{Wigner Function and SDEs}
\label{sec:05b-wigner}

\subsection{Derivation from Quantum Model}

Under the quantum model described in Section \ref{sec:05b-model} above, the state of the cavity is given by a density matrix $\rho$ and evolves according to the master equation:
\beq
	\frac{\d \rho}{\d t} = -i[H, \rho] + \frac{1}{2} \sum_k \left(2L_k \rho L_k^\dagger - L_k^\dagger L_k \rho - \rho L_k^\dagger L_k\right) \label{eq:05b-master}
\eeq
Unfortunately, with an exponentially large Hilbert space, it is not practical to compute $\rho$ or its evolution.  To get around this problem, I follow the approach of Chapter~\ref{ch:04}.  Expressing $\rho$ in terms of a generalized Wigner function, one computes the equations for the Wigner function, shows that they can be approximated by a Fokker-Planck equation, and solves the Fokker-Planck equation stochastically using SDEs.

This approach was derived by Carter \cite{Carter1995} for optical fibers with a $\chi^{(3)}$ interaction; the same approach can be applied to optical cavities or cavity-based photonic circuits \cite{Santori2014}.  In both cases, there is an injective linear mapping between optical states $\rho$ and Wigner functions $W(\alpha, \alpha^*)$.  Gronchi and Lugiato~\cite{Lugiato1978} extended this method to weakly-coupled many-atom cavity QED.  In this case, in addition to an optical degree of freedom, one must also keep track of $N$ two-level atoms.  The procedure is to define a closed Lie algebra $\{ X_i \}$ of both optical and atomic operators, and a semiclassical phase-space that with c-number variables $\{ x_i \}$.  The generalized Wigner function is defined analogous to the optical function -- the Fourier transform of a characteristic function:
\beq
	W(x,t) = \int{\d ^ny\,e^{-i\sum_k x_k y_k} {\rm Tr}\left[e^{i\sum_k X_k y_k}\rho(t)\right]}
\eeq
In our case, the operator algebra consists of the optical and single-mode carrier operators (\ref{eq:05b-sumop}) and is given as follows:
\beq
	X = \left[a, a^\dagger, \sigma_{-}, \sigma_{+}, n, \bar{n}, Q\right]
\eeq
This is a $7$-dimensional, operator-valued vector.  The Wigner function thus lives is a $7$-dimensional phase space, defined over the c-number variables:
\beq
	x = \left[\alpha, \alpha^*, v, v^*, m, \bar{m}, q\right]
\eeq
Under certain closedness conditions discussed in Appendix \ref{sec:05b-closedness}, the Wigner function satisfies a generalized Fokker-Planck equation.  Truncating at second order, this reduces to a Fokker-Planck equation.  The validity of this truncation generally requires that nonlinear coupling constants be weak \cite{Santori2014}, and for two-level systems, that the number of atoms (carrier modes in this case) be large \cite{Lugiato1978}.  Both assumptions are true here.  As a solution to a Fokker-Planck equation, the Wigner function can be sampled stochastically by solving the following SDEs:
\beq
	dx_m = \mu_m\,\d t + \sum_n R_{mn}\,\d w_n \label{eq:05b-fpe-sde}
\eeq
with $dw_n$ a Wiener process and
\bea
	\mu_m & = & C_m^{(1)}(x) \equiv \left(\frac{\d  X_m}{\d t}\right)_p \label{eq:05b-cum1} \\
	(RR^T)_{mn} & = & C_{mn}^{(2)}(x) \equiv \frac{1}{2}\left(\frac{\d (X_mX_n-X_nX_m)}{\d t}\right)_p \nonumber \\
	& & - x_m \left(\frac{\d  X_n}{\d t}\right)_p - \left(\frac{\d  X_m}{\d t}\right)_p x_n \label{eq:05b-cum2}
\eea
where the time-derivatives are computed using the adjoint to (\ref{eq:05b-master})
\beq
	\left.\frac{\d A}{\d t}\right|_{\rm ad} \equiv -i[A, H] + \frac{1}{2}\sum_k{2L_k^\dagger A L_k - L_k^\dagger L_k A - A L_k^\dagger L_k} \label{eq:05b-ad}
\eeq
and $(\cdots)_p$ is defined so that normally ordered products return simple c-number polynomials, e.g.\ $(a)_p = \alpha$, $\frac{1}{2}(a^\dagger a + a a^\dagger)_p = \alpha^*\alpha$, $(a^\dagger a)_p = \alpha^*\alpha - \frac{1}{2}$, etc.  Compare Eqs.~(\ref{eq:04-c1-corr}-\ref{eq:04-c2-corr})

Computing the cumulant matrices $C^{(1)}$ and $C^{(2)}$ from the quantum model (\ref{eq:05b-slh-1}-\ref{eq:05b-slh-n}) is straightforward but very tedious, so I used {\it Mathematica} to derive the result.  The details are given in Appendix \ref{sec:05b-fullsdes}; the SDEs are:
\begin{eqnarray}
	\d\alpha & = & \left[\left(-\frac{\kappa}{2} - i\Delta_c \right)\alpha + g v\right]\d t + \d\xi_1 \label{eq:05b-sde1} \\
	\d\alpha^* & = & \left[\left(-\frac{\kappa}{2} + i\Delta_c\right)\alpha^* + g v^*\right]\d t + \d\xi_2 \\
	\d v & = & \Bigl[-\alpha (N-m-\bar{m}) g \nonumber \\
	& & + \left(-\frac{\gamma_{tot}}{2} - i\Delta_e\right) v\Bigr]\d t + \d\xi_3 \\
	\d v^* & = & \Bigl[-\alpha^* (N-m-\bar{m}) g \nonumber \\
	& & + \left(-\frac{\gamma_{tot}}{2} + i\Delta_e\right) v^*\Bigr]\d t + \d\xi_4 \\
	\d m & = & \left[-g(\alpha v^* + v\alpha^*) - \gamma_{nr} m - \gamma_{rc} q\right]\d t + \d\xi_5 \\
	\d\bar{m} & = & \left[-g(\alpha v^* + v\alpha^*) - \bar{\gamma}_{nr} \bar{m} - \gamma_{rc}q\right]\d t + \d\xi_6 \\
	\d q & = & -\gamma_{sc} (q - m\bar{m}/N)\d t + \d\xi_7 \label{eq:05b-sdeN}
\end{eqnarray}
where $\gamma_{tot} = \gamma_{sc} + \gamma_{rc} + \gamma_{nr} + \bar{\gamma}_{nr}$ and the noise processes $\d\xi_i$ have the covariance matrix:
\beq
	\d\xi \d\xi^T = C^{(2)}\d t
\eeq
where $C_{\rm cav}^{(2)} + C_{\rm int}^{(2)} + C_{\rm rc}^{(2)} + C_{\rm nr}^{(2)} + C_{\rm sc}^{(2)}$ is the sum of the terms in Eqs.~(\ref{eq:05b-app1}-\ref{eq:05b-app2}).  Note how Equations (\ref{eq:05b-sde1}-\ref{eq:05b-sdeN}) resemble both the Maxwell-Bloch equations and the Langevin equations for many-atom cavity QED derived by Gronchi and Lugiato~\cite{Lugiato1978}.  However, because of scattering between carrier modes, we need to keep track of $m$, $\bar{m}$ and $q$ separately.

\subsection{Approximations}
\label{sec:05b-approx}

\subsubsection{Fast Dephasing, Nondegenerate Excitation}

Three approximations make these equations more tractable: fast-dephasing, nondegenerate excitation and the single-carrier approximation.  {\it Fast dephasing} assumes that the scattering rate $\gamma_{sc}$ and detuning $\Delta_e$ are faster than any other timescale in the system, thus
\beq
	\gamma_{nr}, \gamma_{rc}, \gamma_{fc} \ll \gamma_{sc}, \Delta_e \label{eq:05b-fastdeph}
\eeq
This is related to the relaxation-time approximation that holds for most semiconductors \cite{LundstromBook}.  In useful, optimized free-carrier devices, all of the carrier timescales -- $\gamma_{nr}, \gamma_{rc}, \gamma_{fc}$ -- are of order the photon lifetime.  To achieve strong carrier effects, we generally have cavities with $Q \gtrsim 1000$, giving a photon lifetime of $\tau_{ph} \gtrsim \mbox{ps}$.  Ultrafast studies show that inter-mode scattering takes place on times of order 10--100 fs \cite{Sabbah2002, Lin1987}, giving scattering rates at least 10--100 times faster than any other timescale in the system.

Fast dephasing leads to an adiabatic elimination of the dipole terms $(v, v^*)$ and the pair density $q$.  These variables will be replaced by their steady-state values, and a new set of SDEs are obtained for the reduced basis $(\alpha, \alpha^*, m, \bar{m})$.

{\it Nondegenerate excitation} assumes that the number of carriers is much less than the number of carrier modes -- in other words, the valence and conduction bands are far from being degenerately filled with electrons or holes.  This approximation is invoked by setting
\beq
	m, \bar{m}, q \ll N
\eeq
This simplifies the equations of motion by discarding effects like absorption saturation that are negligible for low-power, high-$Q$ optical logic devices.  The resulting equations of motion are:
\begin{empheq}[box=\fbox]{align}
	\d\alpha & = \left[-\frac{\kappa+\eta}{2} - i\bigl(\Delta_c + \delta_c m+ \bar{\delta}_c\bar{m}\bigr)\right]\alpha\,\d t + \d\xi_\alpha \label{eq:05b-csde-1} \\
	\d m & = \left[\eta\,\alpha^*\alpha - \gamma_{nr}m - \gamma_{rc}m\bar{m}\right]\d t + \d\xi_m \label{eq:05b-csde-3}  \\
	\d\bar{m} & = \left[\eta\,\alpha^*\alpha - \bar{\gamma}_{nr}\bar{m} - \gamma_{rc}m\bar{m}\right]\d t + \d\xi_{\bar{m}} \label{eq:05b-csde-4}
\end{empheq}
with noise terms
\begin{empheq}[box=\fbox]{align}
	\d\xi_\alpha & = -\sqrt{\kappa}\,\d \beta_{\rm in} - \sqrt{\eta}\,\d \beta_\eta \\
	\d\xi_m & = 2\sqrt{\eta}\,\mbox{Re}[\alpha^* \d\beta_\eta] + \sqrt{\gamma_{nr} m}\,\d w_m + \sqrt{\gamma_{rc} m\bar{m}}\,\d w_{rc} \\
	\d\xi_{\bar{m}} & = 2\sqrt{\eta}\,\mbox{Re}[\alpha^* \d\beta_\eta] + \sqrt{\gamma_{nr} \bar{m}}\,\d w_{\bar{m}} + \sqrt{\gamma_{rc} m\bar{m}}\,\d w_{rc}
\end{empheq}
where the $\d\beta$'s are complex vacuum Wiener processes, e.g.\ $\d\beta_\eta^*\d\beta_\eta = \frac{1}{2}\d t$, and the $\d w$'s are real Wiener processes, e.g. $dw_m^2 = \d t$.

In the equations above, we rescaled $\gamma_{rc}$ and defined a bandfilling carrier-dependent detuning $\delta_c$ and linear absorption $\eta$:
\bea
	\delta_c = \bar{\delta}_c & = & \frac{g^2}{\Delta_e - \frac{1}{2}i \gamma_{sc}} \label{eq:05b-delta-bf} \\
	\eta & = & 2N\mbox{Im}[\delta_c] \label{eq:05b-eta-bf}
\eea
Since $\Delta_e = \omega_e - \omega$, this function has one pole (for $\omega$) in the lower half-plane, $\omega = \omega_e - \frac{1}{2}i\gamma_{sc}$.  As a result, its real and imaginary parts satisfy the Kramers-Kronig relations.  The carrier-dependent dispersion and absorption are given by the real and imaginary parts of $\delta_c$, respectively.  Around $1/\delta_c$ carriers are needed to shift the cavity resonance by one linewidth; since this quantity is much smaller than $N$ under the nondegenerate approximation, it follows that $\mbox{Im}[\delta_c] \ll \mbox{Re}[\delta_c]$.  For a pure bandfilling effect, we can generally neglect the imaginary part.

The $\d\beta_{\rm in}$ in (\ref{eq:05b-csde-1}-\ref{eq:05b-csde-4}) is the vacuum noise of the input field and $\d\beta_\eta$ is the noise due to linear absorption; each behaves as a vacuum Wiener process $\d\beta^* \d\beta = \frac{1}{2}\d t$ \cite{Santori2014}; compare Sec.~\ref{sec:04-inout}.  The $dw_m$, $dw_{\bar{m}}$ and $dw_q$ are real-valued noises due to carrier loss and recombination, and go as $dw^2 = \d t$.

The noise term for $\alpha$ is fairly standard for open quantum systems: a sum of two vacuum noises.  The noise terms for $m$ and $\bar{m}$ have Poisson statistics: for each process with rate $R\,\d t$, there is a corresponding noise term $\sqrt{R}\,\d w$.  Since carrier generation involves photon absorption, one should not be surprised by the Poisson noise on this signal.  Likewise, since the carrier number is quantized and carrier decay is a random process, there should also be Poisson noise on the decay terms.

\subsubsection{Single-Carrier Approximation}
\label{sec:05b-sca}

In many cases, the equations (\ref{eq:05b-csde-1}-\ref{eq:05b-csde-4}) can be reduced further by positing that $m = \bar{m}$ and introducing an {\it effective carrier number} $N_c$ equal to this quantity.  For example, it will hold if only one of the carrier species is relevant (for instance in silicon, where $\delta_c \gg \bar{\delta}_c$ due to the plasma effect \cite{Bennett1990}), if the recombination process $\gamma_{rc}$ is dominant, or if the number of recombination sites is limited (much smaller than the number of carriers) so that electrons and holes tend to decay together.  If any of these cases hold true, equations (\ref{eq:05b-csde-1}-\ref{eq:05b-csde-4}) become:
\begin{empheq}[box=\fbox]{align}
	\d\alpha & = \left[-\frac{\kappa+\eta}{2} - i(\Delta_c + \delta_c N_c)\right]\alpha\,\d t + \d\xi_\alpha \label{eq:05b-sde-sc1} \\
	\d N_c & = \left[\eta\,\alpha^*\alpha - \gamma_{nr}N_c - \gamma_{rc}N_c^2\right]\d t + \d\xi_N \label{eq:05b-sde-sc2}
\end{empheq}
with noise terms
\begin{empheq}[box=\fbox]{align}
	\d\xi_\alpha & = -\sqrt{\kappa}\,\d \beta_{\rm in} - \sqrt{\eta}\,\d \beta_\eta \\
	\d\xi_N & = 2\sqrt{\eta}\,\mbox{Re}[\alpha^* \d\beta_\eta] + \sqrt{\gamma_{nr} N_c}\,\d w_{nr} + \sqrt{\gamma_{rc} N_c^2}\,\d w_{rc}
\end{empheq}

Eqs.~(\ref{eq:05b-csde-1}-\ref{eq:05b-csde-4}) and (\ref{eq:05b-sde-sc1}-\ref{eq:05b-sde-sc2}) are the key results from this chapter.  To recapitulate, I introduced a method to simulate optical cavities where free-carrier dispersion is the dominant nonlinearity.  This method is based on deriving an approximate Fokker-Planck equation for the Wigner function, the approximation being valid in the weak-coupling limit where the detuning per carrier is much smaller than the cavity linewidth and the mean photon number is large.  Importantly, this allows us to keep track of the dominant quantum effects (vacuum noise in the optical field, Poisson noise in the carrier excitation and decay) without running a full quantum simulation.  

The following chapters will study this nonlinearity from a device perspective.  In Chapter \ref{ch:06b}, I apply (\ref{eq:05b-sde-sc1}-\ref{eq:05b-sde-sc2}) to simulate an optical amplifier and an SR-latch.  Because the semiclassical properties of these devices are well known, our interest lies in the quantum noise in the free-carrier amplifier and latch.  Since the free-carrier dispersion creates an effective $\chi^{(3)}$ nonlinearity, one important question is how free-carrier devices line up against analogous Kerr devices, for which the quantum model is well known.

\section*{Appendix}

\renewcommand\thesection{\arabic{chapter}.\Alph{appsection}}
\renewcommand\thesubsection{\arabic{chapter}.\Alph{appsection}.\arabic{subsection}}

\setcounter{appsection}{1}\section{Closedness of Operator Algebra}
\label{sec:05b-closedness}

\subsection{Closedness and the Wigner Function}

The single-mode model of Sec.~\ref{sec:05b-singlemode} reduces the number of phase-space dimensions from $5N+2$ to $7$, but the operator algebra $X = [a, a^\dagger, \sigma_+, \sigma_-, n, \bar{n}, Q]$ is a very restricted basis set.  Many degrees of freedom cannot be expressed in terms of the $X_i$.  However, if certain closedness conditions are satisfied, operators in the algebra stay in the algebra under time evolution.  Since the Wigner function is tied to expectations of operator products, this allows us to set up a PDE for the Wigner function.  In essence, the degrees of freedom contained in the single-mode Wigner function exactly ``decouple'' from the other degrees of freedom in the system, and the single-mode model is valid.

$X$ is closed under commutation and thus forms a valid basis for an algebra $\mathcal{B}$ -- a vector space spanned by the $X_i$ and their products, e.g.\ $X_i X_j, X_i X_j X_k$, etc.  We say that $\mathcal{B}$ is {\it closed under time evolution} if the time derivative of every element of $\mathcal{B}$ is in $\mathcal{B}$:
\beq
	B \in \mathcal{B}\ \ \Rightarrow\ \ \left.\frac{\d B}{\d t}\right|_{\rm ad} \in \mathcal{B} \label{eq:05b-closed}
\eeq
If (\ref{eq:05b-closed}) holds, then for every c-number product $x_m\ldots x_p$, there exists a polynomial $M_{i\ldots p}(x)$ such that
\beq
	\left(\frac{\d (X_m \ldots X_p)_{sym}}{\d t}\right)_{\rm p} = M_{m\ldots p}(x)
\eeq
and from the correspondence between Wigner moments and operator products, we obtain an equation of motion for the Wigner function's moments:
\beq
	\frac{\d }{\d t}\langle x_m \ldots x_p \rangle_{W} = \langle M_{m\ldots p}(x) \rangle_{W} \label{eq:05b-moments}
\eeq
It is a well-known result in stochastic calculus that we can recast (\ref{eq:05b-moments}) as a generalized Fokker-Planck equation for $W(x, t)$, where the moments are replaced by cumulants.  Eqs.~(\ref{eq:05b-fpe-sde}-\ref{eq:05b-cum2}) arise when this equation is truncated to second order.

\subsection{Proof of Closedness}

We will prove closedness for a relatively broad class of Hamiltonians and Lindblad terms.  To start, define a boson space $\mathcal{B}_\k$, a restricted fermion space $\mathcal{F}_\k$, and sum-operator spaces $\mathcal{B}^{(n)}$ (note no index $\k$):
\bea
	\mathcal{B}_\k & \equiv & \mbox{span}(\sigma_{+\k}, \sigma_{-\k}, n_{\k}, \bar{n}_{\k}, Q_{\k}) \\
	\mathcal{F}_\k & \equiv & \{x_1 c_\k + x_2 \bar{c}_\k^\dagger\ |\ x_1,x_2\in \mathcal{B}_\k\} \\
	\mathcal{B}^{(n)} & \equiv & \mbox{span}(X_{i_1}\ldots X_{i_m}, m \leq n),\ \ \mathcal{B} \equiv \mathcal{B}^{\infty}
\eea
For example, $c_\k \in \mathcal{F}_\k$, $Q \in \mathcal{B}^{(1)}$, $\sigma_+^2 \in \mathcal{B}^{(2)}$.  Below, we prove several lemmas about the ordering of bosonic and fermionic operators.

{\it Lemma 1.}  If $f \in \mathcal{F}_\k$ and $b \in \mathcal{B}^{(1)}$, then $f b = b f + f'$ and $f^\dagger b = b f^\dagger + (f'')^\dagger$, where $f', f'' \in \mathcal{F}_\k$.

{\it Proof. } The case for $f$ is proved by a search of all relevant cases.  $\mathcal{B}$ and $\mathcal{F}_\k$ have 7 and 10 basis vectors, respectively, so this is 70 commutators to check (most are zero).  Given this, the $f^\dagger$ case holds because $[f^\dagger, b]^\dagger = -[f, b^\dagger]$.

{\it Lemma 2.}  If $f \in \mathcal{F}_\k$ and $b \in \mathcal{B}^{(n)}$, then $f b = b f + \sum_i b'_i f'_i$ and $f^\dagger b = b f^\dagger + \sum_i b''_i (f''_i)^\dagger$, where $f_i', f_i'' \in \mathcal{F}_\k$ and $b_i', b_i'' \in \mathcal{B}^{(n-1)}$.

{\it Proof. } Induction on $n$.  The $n=1$ case is proved in Lemma 1.  Assuming it holds for $n-1$, write $b = \sum_j b_{1,j} b_{n-1,j}$ with $b_{1,j} \in \mathcal{B}^{(1)}, b_{n-1,j} \in \mathcal{B}^{(n-1)}$.  Using both Lemma 1 and the $n-1$ case, we move the fermionic operator from the left to the right side of the expression (summation signs omitted for claity):
\bea
	f b & = & f b_{1,j} b_{n-1,j} = (b_{1,j} f + f'_j) b_{n-1,j} \nonumber \\
	& = & b_{1,j}(b_{n-1,j} f + b'_{n-2,jk} f''_{jk}) + (b_{n-1,j} f'''_j + b''_{n-2,jk} f''''_{jk}) \nonumber \\
\eea 
With appropriate index renaming, this takes the desired form.  The $f^\dagger$ case is analogous.
	
{\it Lemma 3.}  If $f_\k, f'_\k \in \mathcal{F}_\k$, then $f_\k^\dagger f'_\k \in \mathcal{B}_\k$ and $\sum_\k f_\k^\dagger f'_\k \in \mathcal{B}^{(1)}$.

{\it Proof. } Done by a search of all relevant cases -- 100 in all since $\mathcal{F}_\k$ has 10 basis vectors.  

{\it Theorem 1.}  The operator algebra $\mathcal{B}$ is closed under (\ref{eq:05b-ad}) if the Hamiltonian is in $\mathcal{B}$ and the Lindblad terms take the following form: $L \sim b f_{\k_1} \ldots f_{\k_n}$, with $b \in \mathcal{B}$ and either $f_{\k_i} \in \mathcal{F}_{\k_i}$ or $f_{\k_i}^\dagger \in \mathcal{F}_{\k_i}$, or $L \sim b g_\k$, for $b \in \mathcal{B}, g \in \mathcal{B}_\k$.  There must be one $L$ for each multi-index $\k_i$.

{\it Proof. } To prove closedness, we must show that (\ref{eq:05b-ad}) is in $\mathcal{B}$ for all $A\in\mathcal{B}$.  This is the sum of a Hamiltonian and Lindblad terms.  The Hamiltonian term $-i[A, H]$ is obvious since both $A$ and $H$ are in the algebra of $X$, which is closed under commutation.

The Lindblad term is $\frac{1}{2}(2L^\dagger A L - L^\dagger L A - A L^\dagger L)$.  We first use Lemma 2 to move the indexed parts $f_{\k_1} \ldots f_{\k_n}$ to the same side of the expression; for instance, for $L^\dagger A L$, we find
\begin{align}
	& (f'_{\k_n})^\dagger \ldots (f'_{\k_1})^\dagger b^\dagger A b f_{\k_1} \ldots f_{\k_n} \nonumber \\
	& \rightarrow \sum_{i} b'_{i_1\ldots i_n, j_1\ldots j_n} (f'_{i_n,\k_n})^\dagger \ldots (f'_{i_1,\k_1})^\dagger f_{j_1,\k_1} \ldots f_{j_n,\k_n}
\end{align}
To each term in this sum, we apply Lemma 3 to combine the fermionic operators into bosonic operators.  
\begin{align}
	& \sum_{\k_1} (f'_{i_n,\k_n})^\dagger \ldots (f'_{i_1,\k_1})^\dagger f_{j_1,\k_1} \ldots f_{j_n,\k_n} \nonumber \\
	& \ \ \ \rightarrow (f'_{i_n,\k_n})^\dagger \ldots (f'_{i_2,\k_2})^\dagger b_1 f_{j_2,\k_2} \ldots f_{j_n,\k_n}
\end{align}
Summation over $\k$ is critical here; without it $b_1$ would not be a bosonic sum-operator in $\mathcal{B}^{(1)}$.  Thus, the algebra $\mathcal{B}$ is not closed for a Lindblad term with just a single $\k$ -- we must sum over all the $\k$'s in order to recover closedness.

Now we use Lemma 3 to move the bosonic operator $b$ to the left, recombine the operators with index $\k_2$, and repeat until all fermionic operators have been combined.  This gets rid of all the indices $\k_i$, resulting in an operator that lives in $\mathcal{B}$.  The terms $L^\dagger L A$ and $A L^\dagger L$ are done the same way.  It follows that the Lindblad term in (\ref{eq:05b-ad}) lives in $\mathcal{B}$.  As before, the action of a single Lindblad term breaks closedness, but when we sum over $\k$, it is recovered.

The result for $L = b g_\k$ can be shown without Lemmas 1--3.  We just use the commutation relations of the $X_\k$ to move the all the indexed terms to the same side, where they can be combined and summed into a term in $\mathcal{B}$.

This theorem encompasses all the quantum models studied in this paper.  A few examples of things it does {\it not} apply to would be index-dependent effects, say $H \sim E_\k n_\k$, or certain effects that violate charge conservation, such as $L \sim c_\k \bar{c}_\k^\dagger$.

\newpage

\setcounter{appsection}{2}\section{Full Wigner SDEs}
\label{sec:05b-fullsdes}

The cumulants $C^{(1)}$ and $C^{(2)}$ are computed from Eqs. (\ref{eq:05b-cum1}-\ref{eq:05b-cum2}) using {\it Mathematica}.  The terms are separated by physical origin in the sections below.

\subsection{Uncoupled Cavity, Carrier Terms}

In this case, $H = \Delta_c a^\dagger a + \Delta_e(n + \bar{n})/2$ and $L + \sqrt{\kappa}\,a$.  It is easy to show that:
\beq
	C^{(1)} = \begin{bmatrix} (-i\Delta_c-\frac{\kappa}{2})\alpha \\ (i\Delta_c-\frac{\kappa}{2})\alpha^* \\
	-i\Delta_e v \\ i\Delta_e \bar{v} \\ 0 \\ 0 \\ 0 \end{bmatrix},\ \ \
	C^{(2)} = \begin{bmatrix}
		 0 & \frac{\kappa}{2} & 0 & 0 & 0 & 0 & 0 \\
		\frac{\kappa}{2} & 0 & 0 & 0 & 0 & 0 & 0 \\
		 0 & 0 & 0 & 0 & 0 & 0 & 0 \\
		 0 & 0 & 0 & 0 & 0 & 0 & 0 \\
		 0 & 0 & 0 & 0 & 0 & 0 & 0 \\
		 0 & 0 & 0 & 0 & 0 & 0 & 0 \\
		 0 & 0 & 0 & 0 & 0 & 0 & 0
	\end{bmatrix} \label{eq:05b-app1}
\eeq

\subsection{Photon-Carrier Interaction}

Here, $H = ig (a^\dagger \sigma_{-} - a\sigma_{+})$, and there are no environment couplings.  There is no noise term here.
\bea
	C^{(1)} & \!\!=\!\! & g\begin{bmatrix} v & v^* & \!-\alpha(N\!-\!m\!-\!\bar{m})\! & \!-\alpha^*(N\!-\!m\!-\!\bar{m})\! &
		\!-(\alpha v^* \!+\! v\alpha^*)\! & \!-(\alpha v^* \!+\! v\alpha^*)\! & \!-(\alpha v^* \!+\! v\alpha^*)
		\end{bmatrix}^{\rm T} \nonumber \\
	C^{(2)} & \!\!=\!\! & 0
\eea


\subsection{Free-Carrier Dispersion / Absorption}

Here, $L_\k = \sqrt{\gamma_{fca}}\,n_\k a, \sqrt{\bar{\gamma}_{fca}}\,n_\k a$ and $H = a^\dagger a(\delta_{fcd} n + \bar{\delta}_{fcd} \bar{n})$.  Define $\delta_{fc} = \delta_{fcd} - i\gamma_{fca}/2$.  Considering only electrons ($\delta_{fcd}, \gamma_{fcd}$), the cumulants become:
\beq
	C^{(1)} = \begin{bmatrix}
		-i\delta_{fc}m\alpha \\
		i\delta_{fc}^*m\alpha^* \\
		-i \delta_{fc} (\alpha^*\alpha - \frac{1}{2}) v \\
		i \delta_{fc}^* (\alpha^*\alpha - \frac{1}{2}) v^* \\ 0 \\ 0 \\ 0 \end{bmatrix},\ \ \
	C^{(2)} = \begin{bmatrix}
		0 & \tfrac{1}{2}\gamma_{fca}m\!\!\!\! & 0 & \!\!\!\!\gamma_{fca}v^*\alpha\!\!\!\! & 0 & 0 & 0 \\
		* & 0 & \!\!\gamma_{fca}v\alpha^*\!\!\!\! & 0 & 0 & 0 & 0 \\
		0 & * & 0 & \!\!\!\!\!\!\!\!\!\!\begin{array}{c}\gamma_{fca}(\alpha^*\alpha-\frac{1}{2})\\\times(N\!+\!2q\!-\!m\!-\!\bar{m})\end{array}\!\!\! & 0 & 0 & 0 \\
		* & 0 & * & 0 & 0 & 0 & 0 \\
		0 & 0 & 0 & 0 & 0 & 0 & 0 \\
		0 & 0 & 0 & 0 & 0 & 0 & 0 \\
		0 & 0 & 0 & 0 & 0 & 0 & 0
		\end{bmatrix} \label{eq:05b-fca-c12}
\eeq
Holes are included by replacing $\delta_{fc} \rightarrow \bar{\delta}_{fc}$, $\gamma_{fca} \rightarrow \bar{\gamma}_{fca}$.  The total $C^{(1)}$, $C^{(2)}$ is the sum of the two.

Equation~(\ref{eq:05b-fca-c12}) has a nontrivial noise matrix.  However, this can be greatly simplified in the non-degenerate, fast-dephasing limit usually taken.

\subsection{Recombination}

This is mediated by the term $L = \sqrt{\gamma_{rc}}\,\sigma_-$.  Recombination only takes place when an electron and hole occupy the same state $\k$, so the rate goes as the pair density $q$, not as the carrier density $m+\bar{m}$.
\beq
	C^{(1)} = \gamma_{rc} \begin{bmatrix} 0 \\ 0 \\ -\frac{1}{2}v \\ -\frac{1}{2}v^* \\ -q \\ -q \\ -q
		\end{bmatrix},\ \ \
	C^{(2)} = \begin{bmatrix}
		0 & 0 & 0 & 0 & 0 & 0 & 0 \\
		0 & 0 & 0 & 0 & 0 & 0 & 0 \\
		0 & 0 & 0 & \!\!\!\!\!\!\!\!\!\!\frac{1}{2}(N+2q-m-\bar{m}) & \frac{1}{2} v & \frac{1}{2} v & \frac{1}{2} v \\
		0 & 0 & \frac{1}{2}(N+2q-m-\bar{m})\!\!\!\!\!\!\!\!\!\! & 0 & \frac{1}{2}v^* & \frac{1}{2}v^* & \frac{1}{2}v^* \\
		0 & 0 & \frac{1}{2} v & \frac{1}{2} v^* & q & q & q \\
		0 & 0 & \frac{1}{2} v & \frac{1}{2} v^* & q & q & q \\
		0 & 0 & \frac{1}{2} v & \frac{1}{2} v^* & q & q & q
		\end{bmatrix}
\eeq

\subsection{Nonradiative Decay / Excitation}

Nonradiative decay is mediated through a term of the form $L = \sqrt{\gamma_{nr}}\,c_k,\ \sqrt{\bar{\gamma}_{nr}}\,\bar{c}_k$.  Strictly speaking, one must include write $L = \sqrt{\gamma_{nr}} c_k r_l^\dagger$, etc.\ where $r_l$ is the electronic mode into which the carrier decays, to make the $L$ operator bosonic.  However, if there are many more recombination sites than carriers, this mode's dynamics are not relevant and the fermionic $L$ gives the right result.  Considering only $\gamma_{nr}$ terms, the cumulants are:
\beq
	C^{(1)} = \begin{bmatrix}
		0 \\ 0 \\ \frac{1}{2}\gamma_{nr} v \\
		\frac{1}{2}\gamma_{nr}v \\ -\gamma_{nr}m \\ 0 \\
		-\gamma_{nr}q
		\end{bmatrix},\ \ \
	C^{(2)} = \begin{bmatrix}
		0 & 0 & 0 & 0 & 0 & 0 & 0 \\
		0 & 0 & 0 & 0 & 0 & 0 & 0 \\
		0 & 0 & 0 & \!\!\tfrac{1}{2}\gamma_{nr}(N-\bar{m})\!\! & \frac{1}{2}\gamma_{nr} v & 0 & \frac{1}{2}\gamma_{nr}v \\
		0 & 0 & \!\!\tfrac{1}{2}\gamma_{nr}(N-\bar{m})\!\! & 0 & \frac{1}{2}\gamma_{nr} v^* & 0 & \frac{1}{2}\gamma_{nr}v^* \\
		0 & 0 & \frac{1}{2}\gamma_{nr} v & \frac{1}{2}\gamma_{nr} v^* & \gamma_{nr}m & 0 & \gamma_{nr}q \\
		0 & 0 & 0 & 0 & 0 & 0 & 0 \\
		0 & 0 & \frac{1}{2}\gamma_{nr} v & \frac{1}{2}\gamma_{nr} v^* & \frac{1}{2}\gamma_{nr} q & 0 & \gamma_{nr}q
		\end{bmatrix}
\eeq
The $\bar{\gamma}_{nr}$ is found by replacing $\gamma_{nr}\rightarrow\bar{\gamma}_{nr}$, $m\rightarrow\bar{m}$ and permuting rows and columns 5 and 6.  The total cumulant is the sum of the two.

\subsection{Scattering}

The scattering terms are $L = \sqrt{\gamma_{sc}/2N}\,c_\k^\dagger c_\l,\ \sqrt{\bar{\gamma}_{sc}/2N}\, \bar{c}_\k^\dagger \bar{c}_\l$.  Defining an average scattering rate by $\frac{1}{2}(\gamma_{sc} + \bar{\gamma}_{sc}) \rightarrow \gamma_{sc}$, we have:
\beq
	C^{(1)} = \gamma_{sc}\begin{bmatrix} 0 \\ 0 \\ -\frac{1}{2} v \\ -\frac{1}{2} v^* \\ 0 \\ 0 \\
		\frac{m\bar{m}}{N}-q \end{bmatrix},\ \ \
	C^{(2)} = \gamma_{sc}\begin{bmatrix}
			0 & 0 & 0 & 0 & 0 & 0 & 0 \\
			0 & 0 & 0 & 0 & 0 & 0 & 0 \\
			0 & 0 & \frac{1}{N} v^2 & \frac{1}{2}N\,e_1 & 0 & 0 & \frac{1}{2} v\,e_2 \\
			0 & 0 & \frac{1}{2}N\,e_1 & \frac{1}{N} v^2 & 0 & 0 & \frac{1}{2} v^*e_2 \\
			0 & 0 & 0 & 0 & 0 & 0 & 0 \\
			0 & 0 & 0 & 0 & 0 & 0 & 0 \\
			0 & 0 & \frac{1}{2} v\,e_2 & \frac{1}{2} v^*\,e_2 & 0 & 0 & q + \frac{m\bar{m}}{N} + e_3
		\end{bmatrix} \label{eq:05b-app2}
\eeq
where $e_1 = 1-(m+\bar{m})/N+2m\bar{m}/N^2$, $e_2 = 1 + (2q-m-\bar{m}-2/3)/N$, and $e_3 = q(q-2(m+\bar{m}))/N$.  In Section \ref{sec:05b-approx}, we take the limit $m, \bar{m}, q \ll N$.  In this limit, $e_1 = e_2 = 1$, $e_3 = 0$.

\setcounter{appsection}{3}\section{Related Models}
\label{sec:05b-related}

Equations (\ref{eq:05b-csde-1}-\ref{eq:05b-csde-4}) are the simplest free-carrier model: identical modes, no two-photon absorption, no interaction between carriers, no excitons.  In many ways it is unrealistic.  However, it forms the basis for generalized models that include these effects and better approximate the real system.

\subsection{Non-Identical Modes}
\label{sec:05b-manymodes}

The most obvious generalization is to include many non-identical carrier modes.  This means that, rather than grouping all of the modes together into $(m, \bar{m})$, they are binned into spectrum of modes $(m_\a, \bar{m}_\a)$.  A similar binning technique is used in many-atom cavity QED when the atomic couplings are not equal \cite{Kwon2013}.  The equations are a straightforward generalization of (\ref{eq:05b-csde-1}-\ref{eq:05b-csde-4}):
\bea
	\d\alpha & = & \Bigl[-\frac{\kappa+\eta}{2} - i\Delta_c - i\sum_\a(\delta_\a m_\a+\bar{\delta}_\a \bar{m}_\a)\Bigr]\alpha\,\d t - \d\xi_\alpha \\
	\d m_\a & = & \Bigl[\eta_\a\,\alpha^*\alpha - \gamma_{nr,\a}m_\a - \gamma_{rc,\a}m_\a \bar{m}_\a + \sum_\b{(\gamma_{\b\rightarrow\a} m_\b - \gamma_{\a\rightarrow\b} m_\a)} \Bigr]\d t + \d\xi_{m,\a} \\
	\d\bar{m}_\a & = & \Bigl[\eta_\a\,\alpha^*\alpha - \bar{\gamma}_{nr,\a}\bar{m}_\a - \gamma_{rc,\a}m_\a \bar{m}_\a + \sum_\b{(\bar{\gamma}_{\b\rightarrow\a} \bar{m}_\b - \bar{\gamma}_{\a\rightarrow\b} \bar{m}_\a)}\Bigr]\d t + \d\xi_{\bar{m},\a} \label{eq:05b-db2}
\eea
The only change here is the introduction of indices and the cross-scattering terms $\gamma_{\a\rightarrow\b}$.  These terms, like the other carrier excitation / decay terms, have Poisson statistics.  The Poisson statistics of different modes are, of course, correlated just as the flows are -- this conserves total carrier number in the scattering processes.

If scattering between bins $m_\a$ is fast compared to carrier excitation or decay, we can replace the mode occupations by the thermal average $m_\a = f_{e,\a}(T) m, \bar{m}_\a = f_{h,\a}(T) \bar{m}$, where $f_{e,\a}, f_{h,\a}$ are normalized Boltzmann distributions.  In terms of the total carrier numbers $m = \sum_\a m_\a$ and $\bar{m} = \sum_\a \bar{m}_\a$, we recover Equations (\ref{eq:05b-csde-1}-\ref{eq:05b-csde-4}), with the effective rates:
\begin{align}
	\delta & = \sum_\a \delta_\a f_{\a}(T) & \bar{\delta} & = \sum_\a \delta_\a \bar{f}_{\a}(T) \nonumber \\
	\gamma_{nr} & = \sum_\a \gamma_{nr,\a} f_{\a}(T) & \bar{\gamma}_{nr} & = \sum_\a \bar{\gamma}_{nr,\a} \bar{f}_{\a}(T) \nonumber \\
	\gamma_{rc} & = \sum_\a \gamma_{rc,\a} f_\a(T) \bar{f}_\a(T)
\end{align}

\subsection{Other Processes: Kerr, TPA, FCA}
\label{sec:05b-otherprocess}

A host of additional processes may be relevant in semiconductor cavities: among the most important are the Kerr effect, two-photon absorption (TPA), and free-carrier absorption (FCA).  Thermal effects and excitonic effects, while very important for some systems, are beyond the scope of this paper.

\subsubsection{TPA and Kerr}

In indirect-gap materials, like silicon, the linear absorption is not an effective pathway for carrier generation.  Instead, two photon absorption is the dominant excitation process.  Typically, two-photon absorption also comes with a dispersive (Kerr) effect.  In other cases, the band gap is tuned to be very close to the photon energy, and both processes are important.  Unlike linear absorption, which tends to create carriers very close to the band gap, two-photon absorption tends to create highly excited carriers with excess kinetic energy.  After excitation, these carrier quickly thermalize and subsequently decay.

We can model this with the following Hamiltonian and decay process:
\bea
	H & = & \frac{1}{2}\Delta_\x (n_\x + \bar{n}_\x) + i g \sum_\x \bigl((a^\dagger)^2 \sigma_{-\x} - a^2 \sigma_{+\x}\bigr) \nonumber \\
	\\
	L_{\x\rightarrow\k} & = & \sqrt{\gamma_{th}} c_\x c_\k^\dagger,\ \ \sqrt{\bar{\gamma}_{th}} \bar{c}_\x \bar{c}_\k^\dagger
\eea
where the new modes $c_\x, \bar{c}_\x$ defined for the highly excited carriers.  Note that, as these modes are highly excited, there is no process $L_{\k\rightarrow\x}$.

Since the excited state is so short-lived, it can be adiabatically eliminated.  For on-resonant transitions $\Delta_\x = 0$ this gives a two-photon absorption term $\beta$; in the off-resonant case $\Delta_\x \neq 0$, one finds two photon absorption plus a dispersive $\chi^{(3)}$ (Kerr) term.

These effects add the following terms to the Wigner equations:
\bea
	\!\!\!\Delta(\d \alpha) & = & (-i\chi - \beta) (\alpha^*\alpha)\alpha\,\d t - 2\sqrt{\beta}\alpha^* \d\beta_\beta \label{eq:05b-tpk1} \\
	\!\!\!\Delta(\d m) & = & \beta(\alpha^*\alpha)^2 \d t + \sqrt{\beta}\left((\alpha^*)^2\d\beta_\beta + \alpha^2 \d\beta_\beta^*\right) \label{eq:05b-tpk2} \\
	\!\!\!\Delta(\d \bar{m}) & = & \beta(\alpha^*\alpha)^2 \d t + \sqrt{\beta}\left((\alpha^*)^2\d\beta_\beta + \alpha^2 \d\beta_\beta^*\right) \label{eq:05b-tpk3}
\eea
Note how Equations (\ref{eq:05b-tpk1}-\ref{eq:05b-tpk3}) predict that single electron-hole pair is created for every {\it two} photons absorbed, in contrast to linear absorption (\ref{eq:05b-csde-1}-\ref{eq:05b-csde-4}), where the ratio is one-to-one.

\subsubsection{Free-Carrier Absorption}

In some materials, including silicon, free carriers can increase the absorption of the medium, an effect known as free-carrier absorption.  In addition, for indirect band-gap materials, the free-carrier dispersion is larger than the band-filling result (\ref{eq:05b-delta-bf}) predicts, due to the collective response of the free-carrier plasma \cite{Bennett1990}.  These effects can be accounted for by adding the phenomenological terms:
\bea
	H & = & -i(\delta_{\rm fcd} n + \bar{\delta}_{\rm fcd} \bar{n}) a^\dagger a \\
	L & = & \sqrt{\gamma_{\rm fca} n}\, a,\ \ \sqrt{\bar{\gamma}_{\rm fca} \bar{n}}\, a
\eea
This can be accommodated in the model (\ref{eq:05b-csde-1}-\ref{eq:05b-csde-4}) if the substitution $\delta_c \rightarrow \delta_c + \delta_{\rm fcd} - i\gamma_{\rm fca}/2$ is made and an extra noise is included:
\beq
	\Delta(\d \alpha) = -\sqrt{\gamma_{\rm fca} m + \bar{\gamma}_{\rm fca} \bar{m}}\,\d \beta_{\rm fca}
\eeq
where $\d\beta_{\rm fca}$ is another vacuum Wiener process.

Altogether, the Wigner equations for the free-carrier cavity, including $\chi^{(3)}$, two-photon absorption, FCD and FCA, take the form:
\begin{align}
	\d\alpha & = \left[-\frac{\kappa}{2} - i(\Delta_c + \delta_c\,m+\bar{\delta}_c\,\bar{m})\right]\alpha\,\d t + \underbrace{\left[-\frac{\eta}{2}\alpha\,\d t - \sqrt{\kappa} \d\beta_\eta\right]}_{(\d \alpha)_\eta} \nonumber \\
	& \quad+ \underbrace{\left[(-i\chi-\beta)(\alpha^*\alpha)\alpha\,\d t - 2\sqrt{\beta}\,\alpha^*\d\beta_\beta\right]}_{(\d \alpha)_\beta} - \sqrt{\gamma_{\rm fca} m + \bar{\gamma}_{\rm fca} \bar{m}}\,\d \beta_{\rm fca} \label{eq:05b-sde-full1} \\
	\d m & = \bigl[-\gamma_{nr}m\,\d t - \sqrt{\gamma_{nr} m}\,\d w_{m}\bigr] + \bigl[-\gamma_{rc}m\bar{m}\,\d t + \sqrt{\gamma_{rc}m\bar{m}}\,\d w_{rc}\bigr] \nonumber \\
	& \quad - \left[\alpha^*(\d \alpha)_\eta + \alpha(\d \alpha)_\eta^*\right]
		- \frac{\alpha^*(\d \alpha)_\beta + \alpha(\d \alpha)_\beta^*}{2} \label{eq:05b-sde-full2} \\
	\d\bar{m} & = \bigl[-\bar{\gamma}_{nr}\bar{m}\,\d t - \sqrt{\bar{\gamma}_{nr} \bar{m}}\,\d w_{\bar{m}}\bigr] + \bigl[-\gamma_{rc}m\bar{m}\,\d t + \sqrt{\gamma_{rc}m\bar{m}}\,\d w_{rc}\bigr] \nonumber \\
	& \quad - \left[\alpha^*(\d \alpha)_\eta + \alpha(\d \alpha)_\eta^*\right]
		- \frac{\alpha^*(\d \alpha)_\beta + \alpha(\d \alpha)_\beta^*}{2} \label{eq:05b-sde-full3}
\end{align}

\subsection{Single-Carrier Approximation}

The single-carrier approximation assumes $m = \bar{m} = N$, and replaces $\delta_c + \bar{\delta_c} \rightarrow \delta_c$, $\gamma_{\rm fca} + \bar{\gamma}_{\rm fca} \rightarrow \gamma_{\rm fca}$.  The equations reduce to:
\begin{align}
	\d\alpha & = \left[-\frac{\kappa}{2} - i(\Delta_c + \delta_c N_c) \right]\alpha\,\d t + \underbrace{\left[-\frac{\eta}{2}\alpha\,\d t - \sqrt{\kappa} \d\beta_\eta\right]}_{(\d \alpha)_\eta} \nonumber \\
	& \quad + \underbrace{\left[(-i\chi-\beta)(\alpha^*\alpha)\alpha\,\d t - 2\sqrt{\beta}\,\alpha^*\d\beta_\beta\right]}_{(\d \alpha)_\beta} - \sqrt{\gamma_{\rm fca} N_c}\,\d \beta_{\rm fca} \label{eq:05b-sde-sc1-b} \\
	\d N_c & = \bigl[-\gamma_{nr}N_c\,\d t - \sqrt{\gamma_{nr} N_c}\,\d w_{nr}\bigr] + \bigl[-\gamma_{rc}N_c^2\,\d t + \sqrt{\gamma_{rc}N_c^2}\,\d w_{rc}\bigr] \nonumber \\
	& \quad - \left[\alpha^*(\d \alpha)_\eta + \alpha(\d \alpha)_\eta^*\right]
		- \frac{\alpha^*(\d \alpha)_\beta + \alpha(\d \alpha)_\beta^*}{2} \label{eq:05b-sde-sc2-b}
\end{align}

\setcounter{appsection}{4}\section{Carrier Detuning in terms of Material Properties}
\label{sec:05b-materials}

In this section we derive expressions for the coupling constant $g_\k$ and the carrier-dependent detuning $\delta_\k$ as a function of material properties.  This is important because it allows one to match the results from this work to the semiclassical treatment of FCD found elsewhere in the literature.

In standard single-particle electrodynamics, to first order in the optical field the light-matter coupling goes as:
\beq
	H_{\rm int} = \frac{e \vec{A} \cdot \vec{p}}{m_0} \label{eq:05b-eap}
\eeq
This can be generalized to many-particle systems by ``second-quantizing'' the Hamiltonian in terms of fermionic creation / annihilation operators $f_\k, f_\k^\dagger$ \cite{KiraKoch2012}:
\beq
	H_{\rm int} \rightarrow \frac{e}{m_0} \sum_{\k,\l\in\rm \{states\}}{f_\k^\dagger \bracket{\k}{A \cdot p}{\l} f_\l} \label{eq:05b-hint}
\eeq
Consider a two-band model.  The $f_\k$ here represent both valence-band and conduction-band states.  If the field $A$ is driving at optical frequencies, only transitions between the valence band and conduction band need be considered -- for these, the Hamiltonian becomes:
\bea
	H_{\rm int} & \rightarrow & \frac{e}{m_0} \sum_{\k} \Bigl[c_{\k}^\dagger \bar{c}_{\k}^\dagger \bbracket{\k,c}{\vec{A}(x,t) \cdot \vec{p}}{\k,v} \nonumber \\
	& & \qquad\qquad + \bar{c}_{\k} c_{\k} \bbracket{\k,v}{\vec{A}(x,t) \cdot \vec{p}}{\k,c}\Bigr] \label{eq:05b-hint2}
\eea
For a resonant structure, $E(x, t)$ and $B(x, t)$ depend on the normal-mode fields $E_\omega(x)$ and their time-dependent amplitude $a_\omega(t)$ (which becomes the photon annihilation operator when the system is quantized).  Working in the Coulomb gauge $\phi(x,t) = 0$, $E = -\partial A/\partial t$ and $A(x, t)$ is given by:
\beq
	\vec{A}(x, t) = \mbox{Re}\left[\sum_\omega -i \sqrt{2\hbar/\omega\epsilon_0} a_\omega \vec{E}_\omega(x) e^{-i\omega t}\right]
\eeq
Here $E_\omega$ is normalized so that $\int{n(x)^2 |E_\omega|^2 d^3x} = 1$, and $a_\omega$ is the photon annihilation operator.  For a good resonator, typically only one frequency $\omega$ is relevant (though multiple frequencies is a simple extension of this work), so hereafter we replace $a_\omega \rightarrow a$.  Going into the interaction picture and neglecting rotating-wave terms and adding an arbitrary phase shift to the $c_i$ to fix the sign of $g_\k$, we find:
\beq
	H_{\rm int} = \sum_{\k} ig_{\k} \left(a^\dagger c_{\k}\bar{c}_{\k} - a\,\bar{c}_{\k}^\dagger c_{\k}^\dagger\right)
\eeq
with coupling constant $g_{\k}$ given by:
\beq
	g_{\k} = \frac{e}{m_0} \sqrt{\frac{\hbar}{2\omega\epsilon_0}}\ \Bigl| \vec{E}_\omega(x) \cdot \sbracket{\k,c}{\vec{p}}{\k,v} \Bigr|
\eeq
The electronic and photon parts to the Hamiltonian take their canonical forms.  The end result is (\ref{eq:05b-ham}).

Having derived the coupling $g_\k$, we proceed to express the carrier-dependent detuning in (\ref{eq:05b-delta-bf}) in terms of actual material properties.  The carrier-dependent detuning is what fundamentally limits the performance of a free-carrier device -- it sets the minimum number of carriers needed to switch by one linewidth, the energy figure of merit for a photonic switch.  It is given by:
\beq
	\Delta(m,\bar{m}) = \sum_\k(\delta_\k m_\k + \bar{\delta}_\k \bar{m}_\k)
\eeq
with $\delta_\k = g^2/(\Delta_\k-\frac{1}{2}i\gamma_{sc})$, as in (\ref{eq:05b-delta-bf}).

This section considers two common cases: a III-V semiconductor near the band gap, where band filling is dominant, and silicon far from the band gap, where the plasma effect dominates.  These effects are well studied in bulk materials; the point of this section is to translate them to the optical resonator picture used in this paper.

\subsection{III-V Semiconductor near Band Gap}

Here, the dominant effect comes from band-filling dispersion.  We assume that all modes have roughly the same energy, $E_g$, and that the optical field is at $E = x E_g$, where $x < 1$.  If $x \approx 1$, then one can show that the carrier-dependent detuning takes the form:
\bea
	\Delta(m,\bar{m}) & \approx & \sum_\k{\delta_{\k}(m_\k+\bar{m}_\k)} \nonumber \\
	& \approx & \sum_\k\frac{\hbar^2 e^2 |\vec{E}_\omega(x_\k) \cdot \vec{p}_{cv}|^2}{m_0^2 \epsilon_0 E_g^2} \frac{1}{2x(1-x)}(m_\k + \bar{m}_\k) \nonumber \\
	& = & \frac{\hbar^2 e^2 |E_\omega(x_\k)|^2 |p_{cv}|^2}{m_0^2 \epsilon_0 E_g^2} |\hat{E}_\omega(x_\k)\cdot\hat{p}_{cv}|^2 \frac{1}{2x(1-x)} (m_\k + \bar{m}_\k)
\eea
where $m_0$ is the bare electron mass and $\vec{p}_{cv}$ is the matrix element $\bra{\k,c} \vec{p} \ket{\k, v}$ between conduction- and valence-band states

This is a two-band calculation, which only includes transitions from a single valence band.  Adding a second valence band doubles the effect of the electrons -- since each electron ``blocks'' two transitions, one from each valence band, its bandfilling effect is doubled (Figure \ref{fig:05b-f3}).  This does not happen for holes, since each hole only ``blocks'' the one transition to the conduction band.  Thus the correct carrier-dependent detuning is:
\bea
	\Delta(m,\bar{m}) & \approx & \sum_\k \frac{3\hbar^2 e^2 |E_\omega(x_\k)|^2 |p_{cv}|^2}{m_0^2 \epsilon_0 E_g^2} |\hat{E}_\omega(x_\k)\cdot\hat{p}_{cv}|^2 \frac{1}{2x(1-x)} \frac{2m_\k + \bar{m}_\k}{3}
\eea
To get a sense of scaling, we replace $|E_\omega|^2 \rightarrow |\tilde{E}_\omega|^2/(n_0^2 V)$.  Here, $\tilde{E}_\omega$ is designed to have near-unit amplitude within the cavity, and $V$ is the mode volume.  Unlike $E_\omega$, $\tilde{E}_\omega$ is not normalized (its integral is not one), but having near unit-amplitude is what matters here.  The mode volume is defined in terms of a normalized quantity, $\tilde{V} = V/(\lambda/n)^3$, which is $O(1)$ for photonic crystals and $O(10)$ for rings.  Instead of looking at $\Delta$, we look at $\Delta/\omega$, since this is unitless, and we are well aware that $\Delta(m,\bar{m})/\omega \sim 1/Q$ means that enough carriers have been injected to move the cavity one linewidth.
\beq
	\frac{\Delta(m,\bar{m})}{\omega} = \frac{3e^2 n_0 |p_{cv}|^2}{8\pi^3 m_0^2 \hbar c^3 \epsilon_0 \tilde{V}} \frac{x}{2(1-x)} \sum_\k{|\tilde{E}(x_\k)\cdot\hat{p}_{cv}|^2 \frac{2m_\k + \bar{m}_\k}{3}} \label{eq:05b-del}
\eeq

\begin{figure}[tbp]
\begin{center}
\includegraphics[width=0.45\textwidth]{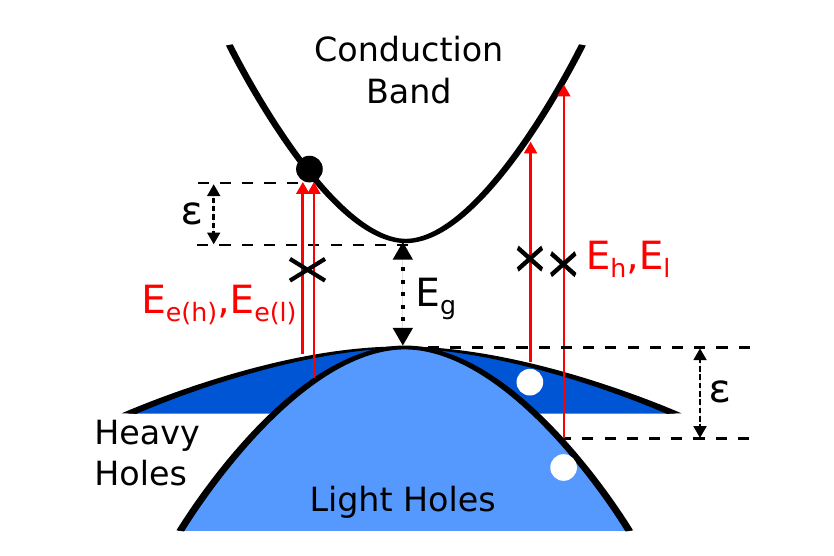}
\caption{Bandfilling in a direct-gap III-V semiconductor.  Carriers block certain optical transitions, changing the absorption spectrum, which in turn alters the index of refraction.}
\label{fig:05b-f3}
\end{center}
\end{figure}

This is a product of four terms.  (1) The first is a bunch of fundamental constants, plus material and cavity parameters like the cavity size and the index of refraction, and the magnitude of the matrix element $|p_{cv}|$.  These constants depend only on the device, not on the wavelength operated at or the particular carrier mode being excited.  (2) Next we have a term that depends on the closeness to the band edge: $x/2(1-x)$.  In practice, we will want $x$ to be as close to one as possible in order to maximize this quantity -- however, as $x \rightarrow 1$ linear absorption limits the cavity $Q$, so there is a tradeoff here.  (3) Next is a position term that depends on the field strength at $x_\k$, the location of the carrier (assuming carriers localized to well below a wavelength here).  (4) Finally, the carrier number.

When carrier thermalization and diffusion is fast compared to the decay processes, one can write this as an effective single-mode model, where the per-carrier detuning is given by the thermal average:
\begin{align}
	\frac{\Delta(m,\bar{m})}{\omega}
	& \ \rightarrow\ \frac{3e^2 n_0 |p_{cv}|^2}{8\pi^3 m_0^2 \hbar c^3 \epsilon_0 \tilde{V}} \frac{x}{2(1-x)} \avg{|\tilde{E}(x_\k)\cdot\hat{p}_{cv}|^2}_\k \frac{2m + \bar{m}}{3} \nonumber \\
	& \ =\ \frac{e^2 n_0 |p_{cv}|^2}{8\pi^3 m_0^2 \hbar c^3 \epsilon_0 \tilde{V}} \frac{x}{2(1-x)} \langle|\tilde{E}(x_\k)|^2\rangle_\k \frac{2m + \bar{m}}{3} \label{eq:05b-bfm}
\end{align}
In the limit $x \approx 1$, this is consistent with previous derivations of the band-filling dispersion \cite{Bennett1990, Said1992}, under the replacements $|p_{cv}|^2 \rightarrow E_g m_0^2/2m_e$ and $\langle|\tilde{E}(x)|^2\rangle_\k \rightarrow 1$ (this is always $O(1)$ and the equality can be imposed by scaling $\tilde{V}$).  Because of the rotating-wave approximation taken in this paper, it will not be valid when $x$ deviates far from 1.  However, optimized devices exploiting band-filling always operate near the band gap.

\subsection{FCD in Silicon}

In silicon, the indirect band gap makes the band-filling effect very weak.  Instead, free-carrier dispersion is dominated by the plasma effect \cite{Bennett1990}.  Consider a simple Drude model with a carrier density $N$.  The index of refraction is modified as follows:
\beq
	n^2 \rightarrow n_0^2\left(1 - \frac{n_c e^2/m_c n_0^2 \epsilon_0}{\omega^2 + i\omega/\tau}\right)
\eeq
where $n_c = n, p$ are the densities and $m_c = m_e, m_h$ are the masses for electrons and holes.  In the high-frequency limit $\omega \gg \omega_p$, the real part dominates and this becomes:
\beq
	\Delta n = -\frac{\hbar^2e^2}{2n_0\epsilon_0 E^2} \left[\frac{n}{m_e} + \frac{p}{m_h}\right]
\eeq
Assuming the carriers are confined to a volume $V$, and defining the dimensionless $\tilde{V} = V/(\lambda/n)^3$ as above, and using standard coupled-mode theory to convert $\Delta n$ to a detuning, we find:
\beq
	\Delta(m,\bar{m}) = \frac{e^2 n_0}{16\pi^3 \hbar c^3 \epsilon_0 \tilde{V}} \left[\frac{m}{m_e} + \frac{\bar{m}}{m_h}\right]
\eeq
A more detailed treatment shows that the dependence is linear for electrons, but nonlinear for holes \cite{Soref1987}.  This nonlinearity can be treated phenomenologically in (\ref{eq:05b-csde-1}-\ref{eq:05b-csde-4}); the quantum noise terms derived in this section do not change.

\renewcommand\thesection{\arabic{chapter}.\arabic{section}}
\renewcommand\thesubsection{\arabic{chapter}.\arabic{section}.\arabic{subsection}}

\ifstandalone{}
\ifdefined\multidoc\else\input{Header}\fi

%

\ifstandalone{\setcounter{chapter}{6}}
\chapter{Free-Carrier Amplifiers and Latches}
\label{ch:06b}

This and the previous chapter are based on the following paper:

\begin{itemize}
	\item \href{http://dx.doi.org/10.1103/PhysRevA.92.023819}{R.~Hamerly and H.~Mabuchi, ``Quantum Noise of Free-Carrier Dispersion in Semiconductor Optical Cavities'', Physical Review A 92, 023819 (2015)}
\end{itemize}

In the previous chapter, I derived a semiclassical model for an optical cavity with a free-carrier nonlinearity.  That derivation was done from first principles, and although the resulting model was semiclassical, it incorporated the main quantum effects -- shot noise in the photon and free-carrier numbers.  This is a compromise between a full quantum treatment, which is needed when the photon or carrier number is very low but is impractical, and the noiseless, classical coupled-mode approach used in the literature.  It should be an accurate description of the system for intermediate photon and carrier numbers $20 \lesssim N_{ph}, N_c \lesssim 1000$, where quantum noise is a relevant, but not dominant, effect.

This chapter takes this model and shows how free-carrier cavities can be designed to implement optical logic.  This includes digital logic as well as nonlinear analog devices such as amplifiers, oscillators and spike generators.  While digital logic is an important goal in photonics, the vast zoo of non-digital free-carrier devices is interesting in its own right, especially since these devices are often more robust to imperfections and can operate at lower energies.  

First, the basic free-carrier model is quoted.  Analytic expressions are derived in limiting cases where the carrier lifetime is much longer (or shorter) than the photon lifetime, and I argue that the most efficient devices will always live between these extremes.  Steady state solutions are studied because these give important insight into the amplification, switching and self-oscillation discussed later.  Next, I look at two devices in detail -- the amplifier and the switch.  The former is key to feedback control; the latter is a building block for digital logic.

In later chapters, more complex devices -- limit-cycle oscillators, relays, and Ising machines -- are studied, but those results will build off of the intuition developed in this chapter.

\section{Steady-State Behavior}
\label{sec:06b-ss}

\begin{figure}[btp]
\begin{center}
\includegraphics[width=0.50\textwidth]{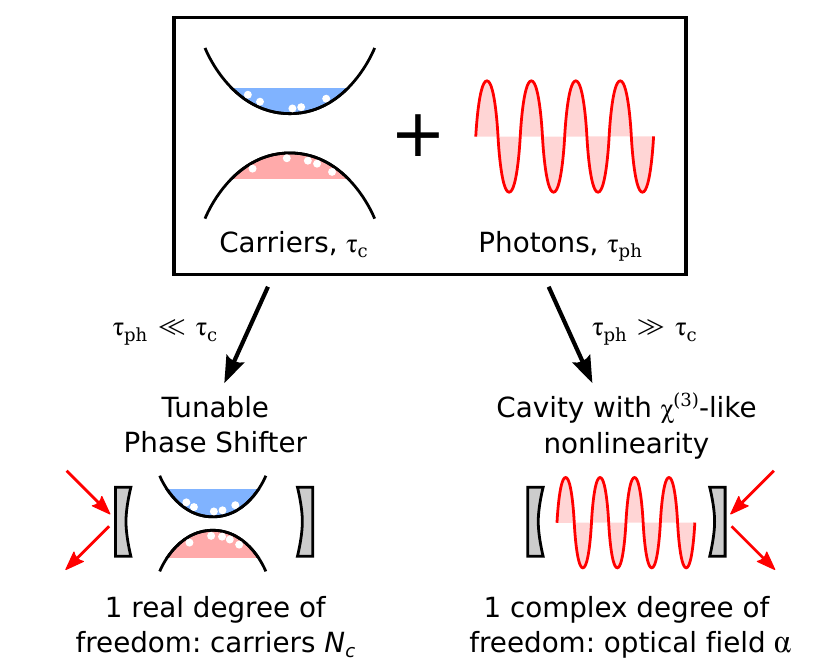}
\caption{Adiabatic elimination of a free-carrier device into a tunable phase shifter (left) and a Kerr-like nonlinear cavity (right).}
\label{fig:06b-f1}
\end{center}
\end{figure}

Consider now the case where (\ref{eq:05b-sde-sc1}-\ref{eq:05b-sde-sc2}) hold.  Suppose also that direct recombination is negligible ($\gamma_{rc} = 0$).  This is the limit to which III-V photonic crystals operated near the band edge, which have the best performance to date \cite{Nozaki2010}, belong (the parameters in Table \ref{tab:06b-t1} are less than a factor of 5 from the state of the art).  In this limit, a free-carrier cavity modeled by equations (\ref{eq:05b-sde-sc1}-\ref{eq:05b-sde-sc2}) has two timescales -- an optical lifetime $\tau_{ph} \equiv 1/(\kappa+\eta)$ and a free-carrier lifetime $\tau_{c} \equiv 1/\gamma_{nr}$.

First, a steady-state limit is discussed.  This is the case when \textit{both} the carrier and photon lifetimes are much shorter than the relevant timescales.  Questions of thermal stability, for instance, can be treated in the steady-state limit.  Next the limiting case of $\tau_{ph} \gg \tau_c$, where the carrier population varies much faster than the photon population, is treated and we show that the free-carrier model reduces to a Kerr model with extra noise terms.

In this section, we work in normalized units by setting $k \equiv \kappa+\eta \rightarrow 1$.  Rates, time constants and powers are scaled by appropriate powers of $k$.  This allows our results to generalize to a wide range of systems spanning orders of magnitude in speed and size.

\subsection{Steady-State Limit}

\begin{table}[tbp]
\begin{center}
\begin{tabular}{c|ll|c}
\hline\hline
Name & Description & Reference & Value (this chapter) \\ \hline
	$k$ & Photon Decay, $k \equiv \kappa + \eta = \omega/Q$ & & 0.42 ps$^{-1}$ \\
	$\kappa$ & Output Coupling & & $0.8k$ \\
	$\eta$ & Linear Absorption, $\eta = \frac{2\omega}{n}\mbox{Im}(n) = \frac{c\alpha}{n}$ & Eqs.~(\ref{eq:01b-lin1}-\ref{eq:01b-lin2}) & $0.2k$ \\
	$\beta$ & 2PA, $\beta \!=\! \bigl(1.54 \!\times\! 10^{-7}\bigr) \frac{\omega x^4 \sqrt{E_g/\text{eV}}}{n_0\tilde{V}\sqrt{m_e/m_0}} \mbox{Im}(f_\chi)$ & Eq.~(\ref{eq:01b-chidim}) & $\approx 0$ \\
	$\chi$ & Kerr, $\chi \!=\! \bigl(1.54 \!\times\! 10^{-7}\bigr) \frac{\omega x^4 \sqrt{E_g/\text{eV}}}{n_0\tilde{V}\sqrt{m_e/m_0}} \mbox{Re}(f_\chi)$ & Eq.~(\ref{eq:01b-chidim}) & $\approx 0$ \\
	$\delta$ & FCD, $\delta \!=\! \bigl(3.62 \!\times\! 10^{-10}\bigr) \frac{n_0(E_g/\text{eV})}{\tilde{V}(m_e/m_0)} \frac{x}{1-x^2}\omega$ & Eq.~(\ref{eq:01b-dk}) & $0.014k$ \\
	$\delta_T$ & Thermal, $\delta_T = -0.411 \frac{(dE_g/dT)/\text{eV}}{(E_g/\text{eV})^{3/2} n_0} f_{bgs}(E/E_g)$ & Eqs.~(\ref{eq:01a-bgs}, \ref{eq:01b-lin1}) & $-1.1k$ \\
	$\gamma_{nr}$ & Carrier Decay, typically $O(D/L^2)$ & & $1/(2\text{ps}) = 1.2k$ \\
	$\gamma_{rc}$ & Recombination & & $\approx 0$ \\
	$\gamma_{th}$ & Thermal Decay, $O(D_{th}/L^2)$, geometric & &  $1/(1.6\text{ns})=0.0014k$ \\
	$r$ & $\Delta T$/Photon, $r = \bigl(8.40 \times 10^{-8}\bigr) \frac{(x E_g/\text{eV})^4 n_0^3}{\tilde{V}_{th} C/(\text{J/cm}^3\text{K})}$ & & $2.7\times 10^{-5}$ \\ \hline
	$\tau_c$ & Carrier Lifetime, $1/\gamma_{nr}$ & & 2 ps \\
	$\tau_{ph}$ & Photon Lifetime, $1/k$ & & 2.4 ps \\
	$\tau_{th}$ & Thermal (phonon) Lifetime, $1/\gamma_{th}$ & & 1.6 ns \\
	$\chi_c$ & FCD ``$\chi^{(3)}$'' coefficient, $\chi_c = \eta\delta/\gamma_{nr}$ & & $0.0022k$ \\
	$\chi_{th}$ & Thermal ``$\chi^{(3)}$'' coefficient, $\chi_{th} = \eta r \delta_T/\gamma_{th}$ & & $-0.0042k$ \\
	$N_{sw,c}$ & Carriers needed to switch cavity, $k/\delta$ & & $70$ \\
	$N_{sw,ph}$ & Photons needed to switch cavity, $k/\chi_c$ & & $450$ \\ 
	$N_{sw,th}$ & Thermal switching energy in units of $\hbar\omega$, $k/r\delta_T$ & & $34000$ \\
	\hline\hline
\end{tabular}
\caption{System parameters used for the simulations in this section.  Based on GaAs photoic-crystal cavity, with $x = 0.98$, $\tilde{V} = 0.25$, $\tilde{V}_{th} = 0.25$, $Q = 5000$ ($Q_{\rm unloaded} = 25000$).  Of the parameters above, $\kappa, \eta, \beta, \delta, \delta_T, \gamma_T,$ and $\gamma_{th}$ have units of s$^{-1}$.  We normalize them by writing them in terms of $k \equiv \omega/Q$.}
\label{tab:06b-t1}
\end{center}
\end{table}

In the steady-state case, we set all noise terms to zero and solve for $\dot{N}_c = \dot{\alpha} = 0$.  Solving for $\dot{\alpha} = 0$, the steady-state internal field $\bar{\alpha}$ can be related to $N_c$ and the input field $\beta_{in}$ as follows:
\beq
	\alpha = \frac{-\sqrt{\kappa} \beta_{in}}{\frac{\kappa+\eta}{2} + i(\Delta_c + \delta_c N_c)} \label{eq:06b-ab}
\eeq
This is the familiar formula for the field in a resonant cavity, where the detuning $\Delta_c + \delta_c N_c$, depends on the free-carrier number.  Solving the $\dot{N}_c = 0$ equation gives $N_c = (\eta/\gamma_{rc})\alpha^*\alpha$.  This can be rearranged into a polynomial equation for $\alpha^*\alpha$:
\beq
	\kappa\,\beta_{in}^* \beta_{in} = (\alpha^*\alpha) \left[(k/2)^2 + \bigl(\Delta_c + (\eta\delta_c/\gamma_{nr}) (\alpha^*\alpha) \bigr)^2\right] \label{eq:06b-cubic}
\eeq

When the external power $P = \beta_{in}^*\beta_{in}$ is set, this is a cubic equation for the internal photon number $N_{ph} = \alpha^*\alpha$.  It is the same optical bistability cubic as the Kerr cavity \cite{Agrawal1979, Yurke2006}, with the effective Kerr nonlinearity:
\beq
	\chi_{\rm eff} = \frac{\eta\delta_c}{\gamma_{nr}}
\eeq
One can solve the cubic (\ref{eq:06b-cubic}) to obtain $\alpha^*\alpha$; it is not always uniquely defined.  Just like Kerr cavities and atom cavities, free-carrier cavities exhibit hysteresis and bistability, with both ``low'' and ``high'' intensity states being allowed for the same input power.  Figure \ref{fig:06b-f2} shows the stable lower- and upper states, and an unstable middle-state, for varying values of $\Delta_c$.

The intuition behind this bistability is that, when the cavity is off resonance and a sufficiently large number of carriers are injected, it will shift back on resonance.  If there is a strong enough input, then a large power builds up inside the cavity and this large carrier population can be maintained through absorption, giving rise to the high state.  On the other hand, if there are no carriers to begin with, the cavity stays off resonance and there is never enough power in the cavity to raise the carrier number -- hence the low state.  Analytically, one can show that the bifurcation sets in when:
\beq
	\Delta_c < -\sqrt{\frac{3}{4}}\,(\kappa+\eta)
\eeq
Much of our intuition behind free-carrier nonlinearities comes from this steady-state picture.  It does not include any quantum effects or even any dynamics, but the shapes of the curves in Figure~\ref{fig:06b-f2} suggest that the device could be used as an amplifier or a switch.  We will show in the next chapter that free-carrier cavities can do much more than this, but that will build on the fundamentals discussed here.

The steady-state picture has been amply discussed in the literature \cite{Agrawal1979, Yurke2006, Kwon2013}, so it is not worth describing in more detail here.  Rather, we now proceed to look at the quantum noise and dynamics of these systems.

\begin{figure}[tbp]
\begin{center}
\includegraphics[width=0.58\textwidth]{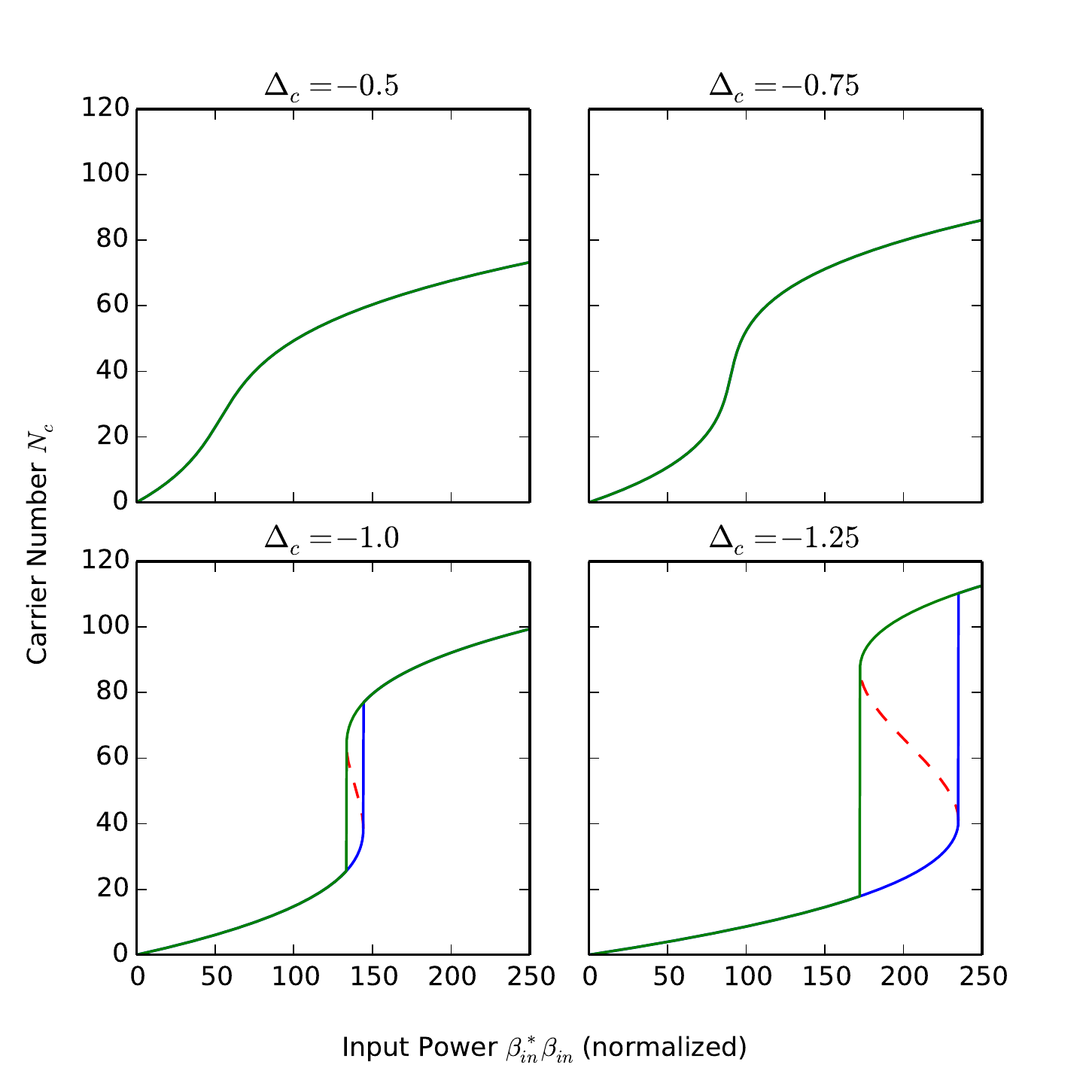}
\caption{Steady-state solutions to $N_c$ for optical free-carrier cavity at different detunings.}
\label{fig:06b-f2}
\end{center}
\end{figure}

\subsection{Effective $\chi^{(3)}$ Model}

Next, we go from the steady-state picture to the limit of short carrier lifetime.  In this opposite limit, $\tau_{ph} \gg \tau_c$.  Typical devices do not realize this limit, but it is useful because it enables an apples-to-apples comparison between the free-carrier and Kerr effects.

To adiabatically eliminate the carrier number, one replaces $N_c$ with its steady-state value:
\beq
	N_c\,dt \rightarrow \frac{\eta(\alpha^*\alpha)dt + d\xi_N}{\gamma_{nr}}
\eeq
This gives the following SDE for the relevant dynamical variable, $\alpha$:
\beq
	d\alpha = \left[-\frac{\kappa + \eta}{2} - i\left(\Delta_c + \frac{\eta\delta_c}{\gamma_{nr}}(\alpha^*\alpha)\right)\right]\alpha\,dt - \sqrt{\kappa} d\beta_{in} + d\xi_\alpha' \label{eq:06b-kerrsde}
\eeq
where the $d\xi_\alpha'$ is a new noise term that depends both on the $d\xi_\alpha$ and $d\xi_N$.  As before, the analogy to the Kerr model is clear: Equation~(\ref{eq:06b-kerrsde}) is very close to the Wigner equations for the Kerr cavity \cite{Santori2014}, but the noise term is different.  The effect of this noise term will be discussed in the following sections, where the performance of Kerr- and free-carrier based amplifiers and switches is analyzed.

\section{Amplifier}
\label{sec:06b-amp}

Figure \ref{fig:06b-f2} shows that, for certain detunings, the state of the cavity changes very rapidly with a change in input power.  One can imagine using such a device to amplify differential signals: if the input signal is perturbed, that perturbation will be multiplied by some gain factor in the output.

\begin{figure}[tb]
\centering
\includegraphics[width=0.45\textwidth]{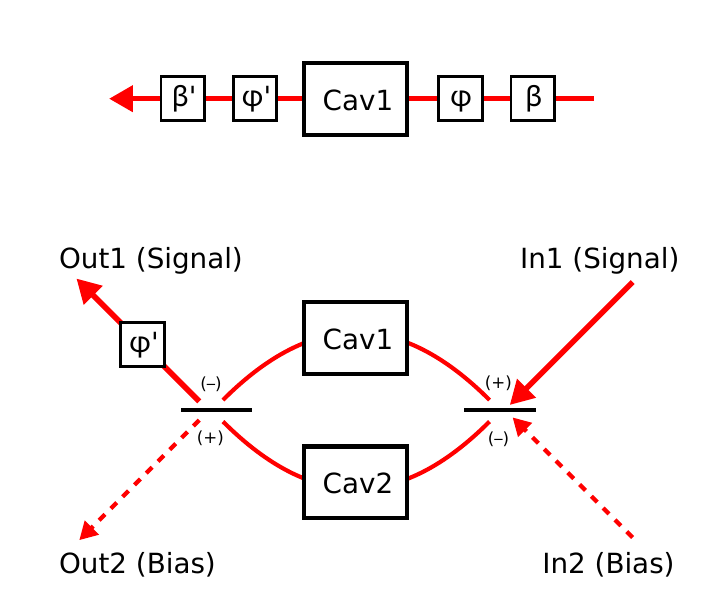}
\label{fig:06b-f3}
\caption{Left: Simple single-cavity amplifier, a cavity sandwiched between phase shifters and displacements $G = L(\beta') \triangleleft e^{i\phi'} \triangleleft \mbox{(Cav)} \triangleleft e^{i\phi} \triangleleft L(\beta)$.  Right: Symmetric two-cavity amplifier.}
\end{figure}

The real picture is actually a bit more complicated, since the input field has two quadratures.  In the Kerr cavity, one of the quadratures is amplified while the other is de-amplified \cite{Yurke2006}.  This gives rise to {\it phase-sensitive amplification} which, since there is no additional noise in the Kerr system, also squeezes the quantum noise of one quadrature below the vacuum level.

Key to an optical amplifier are its gain $G(\omega)$, its noise spectrum $S(\omega)$, and the scale on which nonlinear effects take over.  The gain and noise can be predicted by linearizing equations of motion (\ref{eq:05b-sde-sc1}-\ref{eq:05b-sde-sc2}) around the steady-state value.  This takes the general form (Eqs.~(\ref{eq:04-linabcd1}-\ref{eq:04-linabcd2})):
\begin{eqnarray}
    d\bar{x} & = & \bar{A}\bar{x}\,dt + \bar{B}\,d\bar{\beta}_{\rm in} + \bar{F}\,dw \\
    d\bar{\beta}_{\rm out} & = & \bar{C}\bar{x}\,dt + \bar{D}\,d\bar{\beta}_{\rm in}
\end{eqnarray}
where $\bar{x}$ and $\bar{\beta}$ are {\it doubled-up} state vectors, which include the complex field operators and their conjugates \cite{Gough2010b}, as well as the (real) carrier number: $\bar{x} = (\delta\alpha,\ \delta\alpha^*,\ \delta N_c)$, $d\bar{\beta} = (d\beta, d\beta^*)$ (removing any constant coherent input), and $\alpha$, $N_c$ are the steady-state values.

Linearization is key because many general results of stochastic systems theory only apply to linear or approximately linear systems \cite{AstromBook}.  For example, in a linearized system, the output squeezing spectrum can be computed exactly for Gaussian inputs \cite{WallsMilburn, Crisafulli2013}.  Many results in quantum feedback control theory are also restricted to linear systems \cite{Nurdin2009, Hamerly2012}.

With a linearized model in hand, it is a simple matter to compute the internal state covariance $\bar{\sigma}$, the transfer and noise matrix $\dbl{T}(\omega)$, $\dbl{N}(\omega)$, and the frequency-domain input-output relation \cite{Gough2010b, Hamerly2013}, see Sec.~\ref{sec:04-linear}:
\begin{align}
	& \bar{A}\bar{\sigma} + \bar{\sigma}\bar{A}^\dagger + \frac{1}{2}\bar{B}\bar{B}^\dagger + \bar{F}\bar{F}^\dagger = 0 \label{eq:06b-lyap} \\
	& \bar{\beta}_{{\rm out},\omega} = \underbrace{\left[\bar{D} + \bar{C} \frac{1}{-i\omega - \bar{A}}\bar{B}\right]}_{\dbl{T}(\omega)} \bar{\beta}_{{\rm in},\omega} +
		\underbrace{\bar{C} \frac{1}{-i\omega - \bar{A}} \bar{F}}_{\dbl{N}(\omega)} w_\omega \label{eq:06b-dblt}
\end{align}
Unfortunately, because the doubled-up matrices here are 3-by-3 rather than 2-by-2, the analytic results are rather cumbersome and therefore not reproduced here.  Instead, in this section I compute these quantities numerically and compare the results to the Kerr system.  The results here are compared against a Kerr cavity with the same effective nonlinearity, $\chi = \eta\delta_c/\gamma_{nr}$.

\subsection{Gain}

The gain is computed from the singular values of the doubled-up transfer function $T(\omega)$.  If both singular values are the same, the device is a phase-insensitive amplifier.  Both the Kerr and free-carrier cavities, however, only amplify one quadrature.  As Figure~\ref{fig:06b-f4} shows, they de-amplify the other quadrature as well.

At and below the ideal input $\beta_{\rm in} \approx 9$, the Kerr and free-carrier cavities seem to amplify in the same way.  For over-driven cavities, the behavior is very different.  The free-carrier cavity becomes very efficient at amplifying off-resonance, whereas the Kerr cavity hardly amplifies at all.

Gain is maximized when the system is very close to instability -- that is, when at least one of the eigenvalues of $\dbl{A}$ is very close to the imaginary axis.  From (\ref{eq:06b-dblt}), an eigenvalue decomposition of $\dbl{A}$ gives the transfer function the following form:
\beq
	T = \dbl{D} + \sum_i \frac{v_i u_i^T}{-i\omega - \lambda_i}
\eeq
where $v_i, u_i$ are related to $B$, $C$ and the eigenvectors and $\lambda_i$ are the eigenvalues of $A$.  Since $B, C \sim O(\sqrt{\kappa}$), the numerator term is proportional to $\kappa$.  Near the resonance, the sum is dominated by the eigenvalue closest to zero, $\lambda_{\rm max}$.  The maximum gain should intuitively take the form:
\beq
	G \sim \frac{O(\kappa)}{|-i\omega - \lambda_{\rm max}|}
\eeq
This is a Lorentzian with a peak at $\mbox{Im}(\lambda_{\rm max})$ and bandwidth of $\Delta\omega = -\mbox{Re}(\lambda_{\rm max})$.  The peak gain is thus $G_{\rm max} = -O(\kappa)/(\mbox{Re}(\lambda_{\rm max}))$.  This gives us a gain-bandwidth relation:
\beq
	G_{\rm max} \Delta\omega = O(\kappa)
\eeq
The greater the amplifier gain, the slower it responds and the narrower its bandwidth.

\begin{figure}[tb]
\centering
\includegraphics[width=0.66\textwidth]{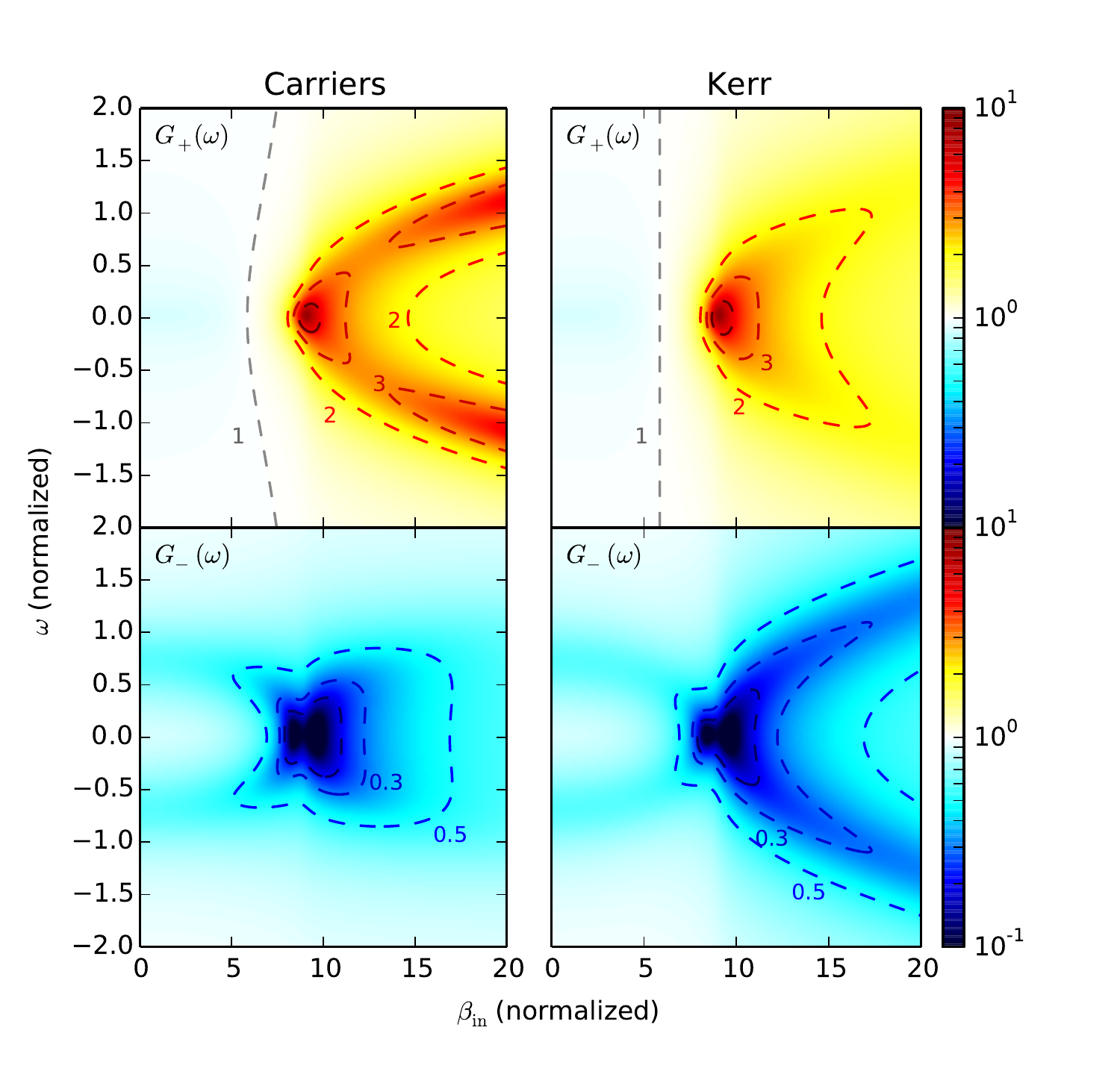}
\label{fig:06b-f4}
\caption{Plot of the maximum and minimum gain $G_+(\omega)$, $G_-(\omega)$ for free-carrier cavity (left) and Kerr cavity (right).  Parameters are from Table \ref{tab:06b-t1}, with $\Delta_c = -0.7$.}
\end{figure}

\subsection{Internal State}

\begin{figure}[p]
\begin{center}
\includegraphics[width=0.66\textwidth]{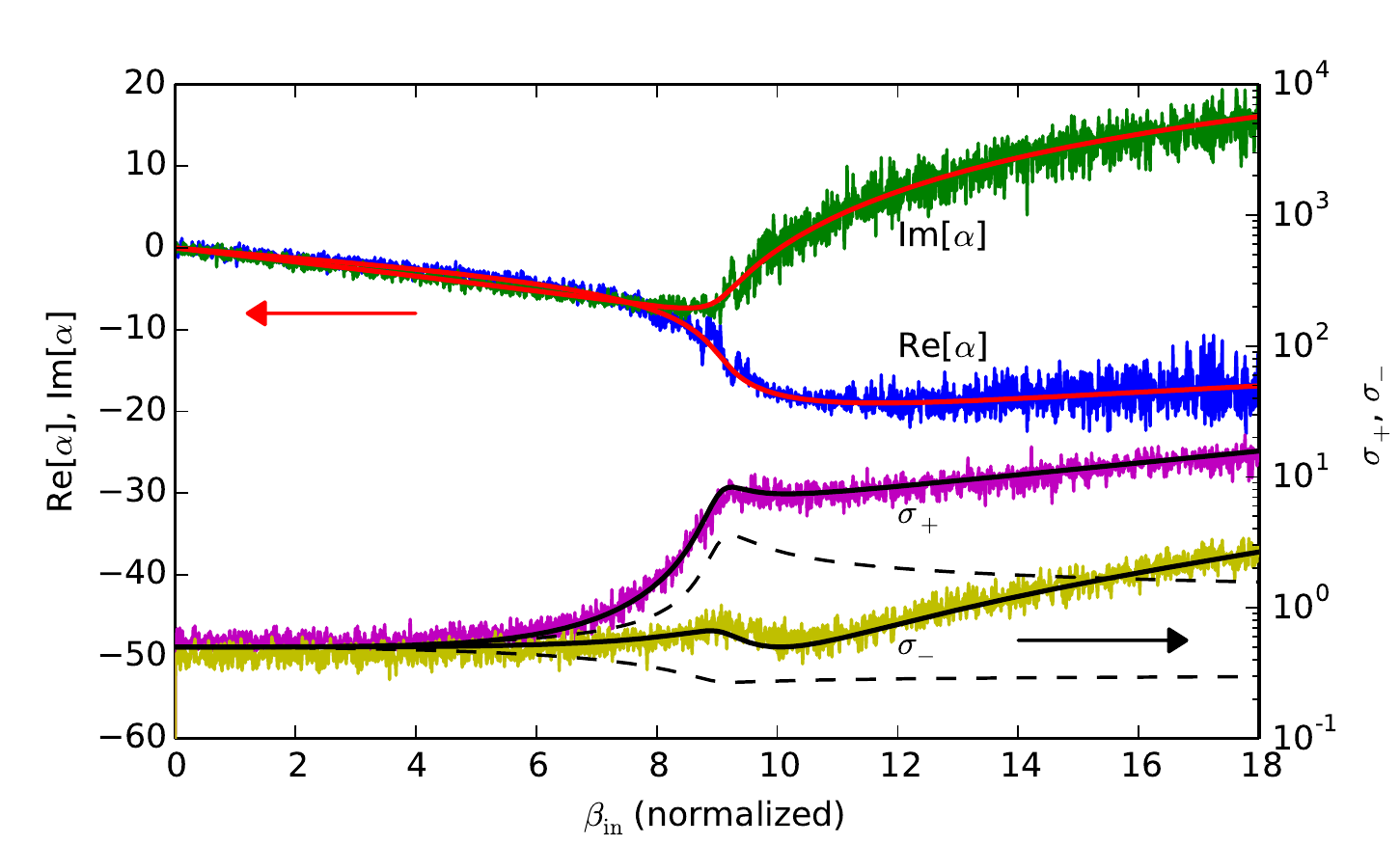}
\caption{Internal state of free-carrier cavity, simulated eigenvalues of $\sigma$ ($\sigma_+$ and $\sigma_-$, the larger and smaller eigenvalue, respectively) compared to analytic result (solid lines).  The dashed line is the analytic result for an equivalent Kerr cavity.  $\Delta_c = -0.7$}
\label{fig:06b-f5b}
\end{center}
\end{figure}

\begin{figure}[p]
\begin{center}
\includegraphics[width=0.66\textwidth]{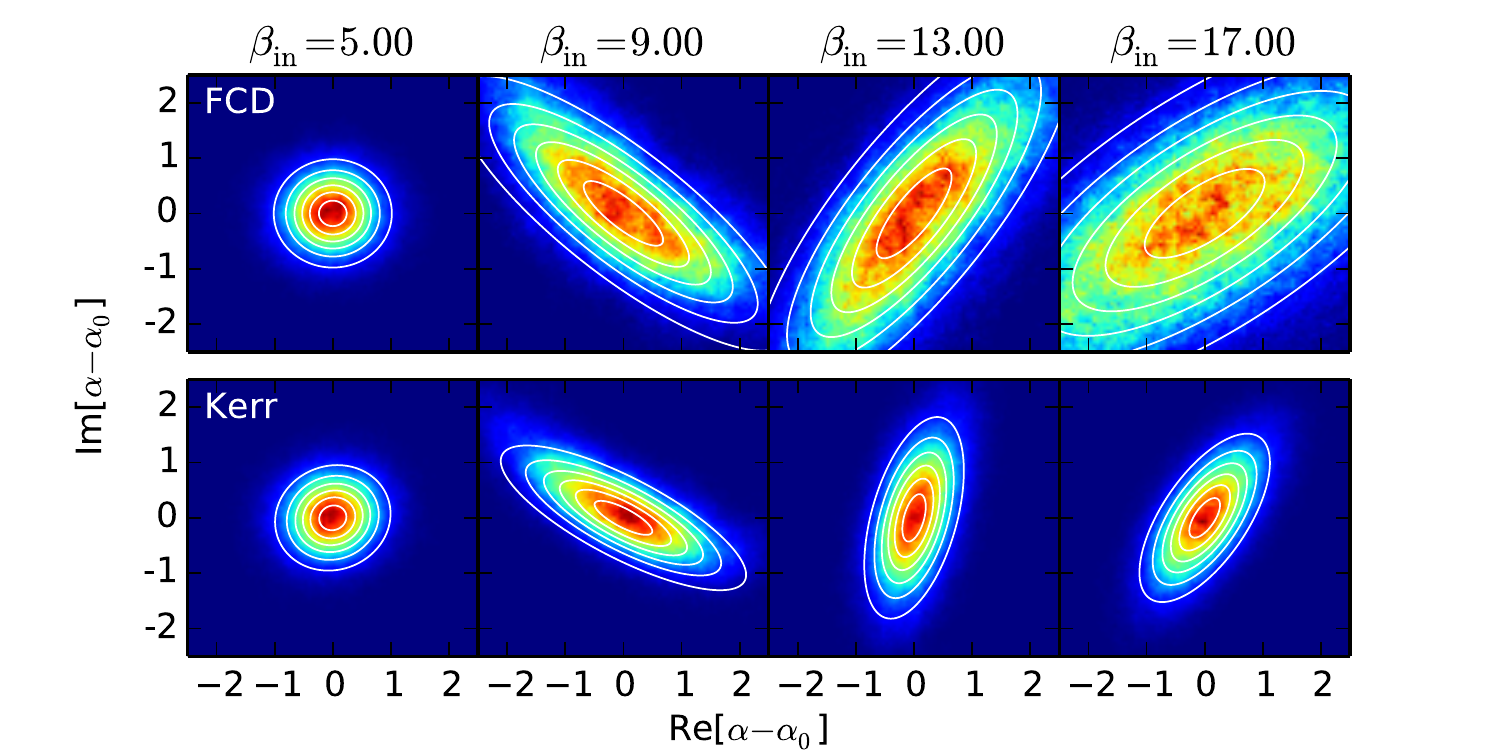}
\caption{Simulated Wigner functions for free-carrier (top) and Kerr (bottom) cavities with the same effective $\chi^{(3)}$.  Analytic approximation for linearized model given in white contours.  $\Delta_c = -0.7$}
\label{fig:06b-f5a}
\end{center}
\end{figure}

The internal state is computed using the Lyapunov equation (\ref{eq:06b-lyap}).  This time, the Wigner equations contain additional noise terms, which make the state noisier than the state of an equivalent Kerr cavity.  This is plotted in the Figures~\ref{fig:06b-f5b}-\ref{fig:06b-f5a}.  The state remains roughly Gaussian, but the size of the Gaussian is larger than in the Kerr case, especially above the inflection point.

Unlike in the Kerr case, the mode in the free-carrier cavity is never squeezed.  As seen in Figure~\ref{fig:06b-f5b}, the eigenvalues $\sigma_+, \sigma_-$ of the covariance matrix $\sigma$ are always $\geq \frac{1}{2}$, ensuring that the state is always ``classical'' in the sense that it has a valid $P$ representation.  Given that the carrier excitation and decay process is highly incoherent, it should not be too surprising that the cavity always remains in a classical state.  But it is a clear departure from the Kerr model, and this classicality could conceivably be used to distinguish between the two in an experiment.

Also note that the noise grows linearly with the input field at high powers.  This happens because the free-carrier number is constantly fluctuating, being driven by excitation and decay events that mimic a Poisson process.  At high carrier numbers, this means that the cavity detuning and consequently the cavity field become very noisy.  This does not happen in the Kerr cavity, where the nonlinearity is mediated by virtual transitions which do not add any noise to the system.  It is a peculiar consequence of the incoherence of the free-carrier mechanism.

\subsection{Output Noise Spectrum}

\begin{figure}[b!]
\begin{center}
\includegraphics[width=0.66\columnwidth]{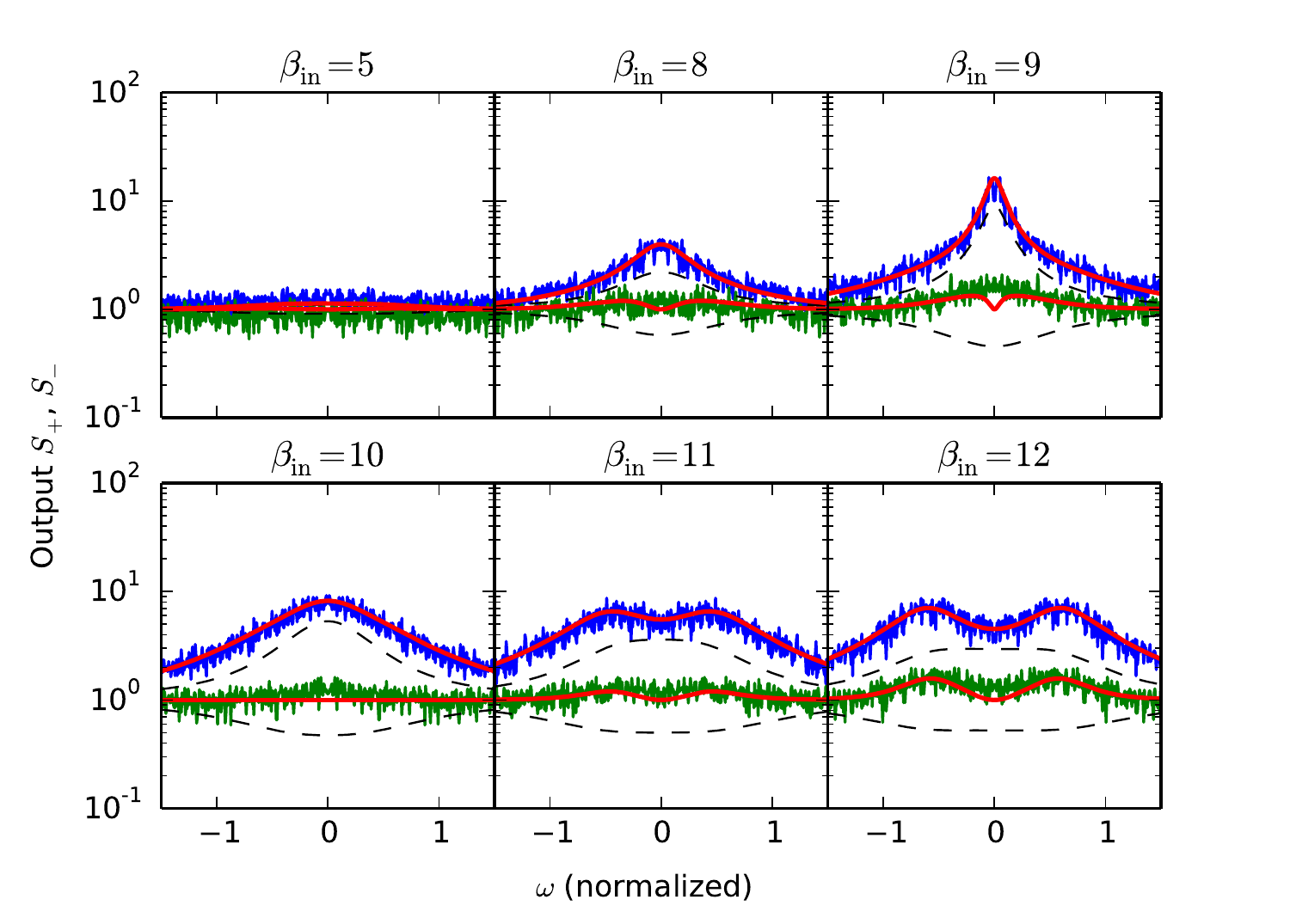}
\caption{Noise spectrum modes $S_+$, $S_-$ for the a free-carrier cavity with $\Delta_c = -0.7$ at various pump powers.  Green and blue lines are numerical simulations; red solid line is the prediction from the linearized ABCD model.  The dashed lines are the prediction from the Kerr model.}
\label{fig:06b-f6}
\end{center}
\end{figure}

\begin{figure}[t!]
\begin{center}
\includegraphics[width=0.66\textwidth]{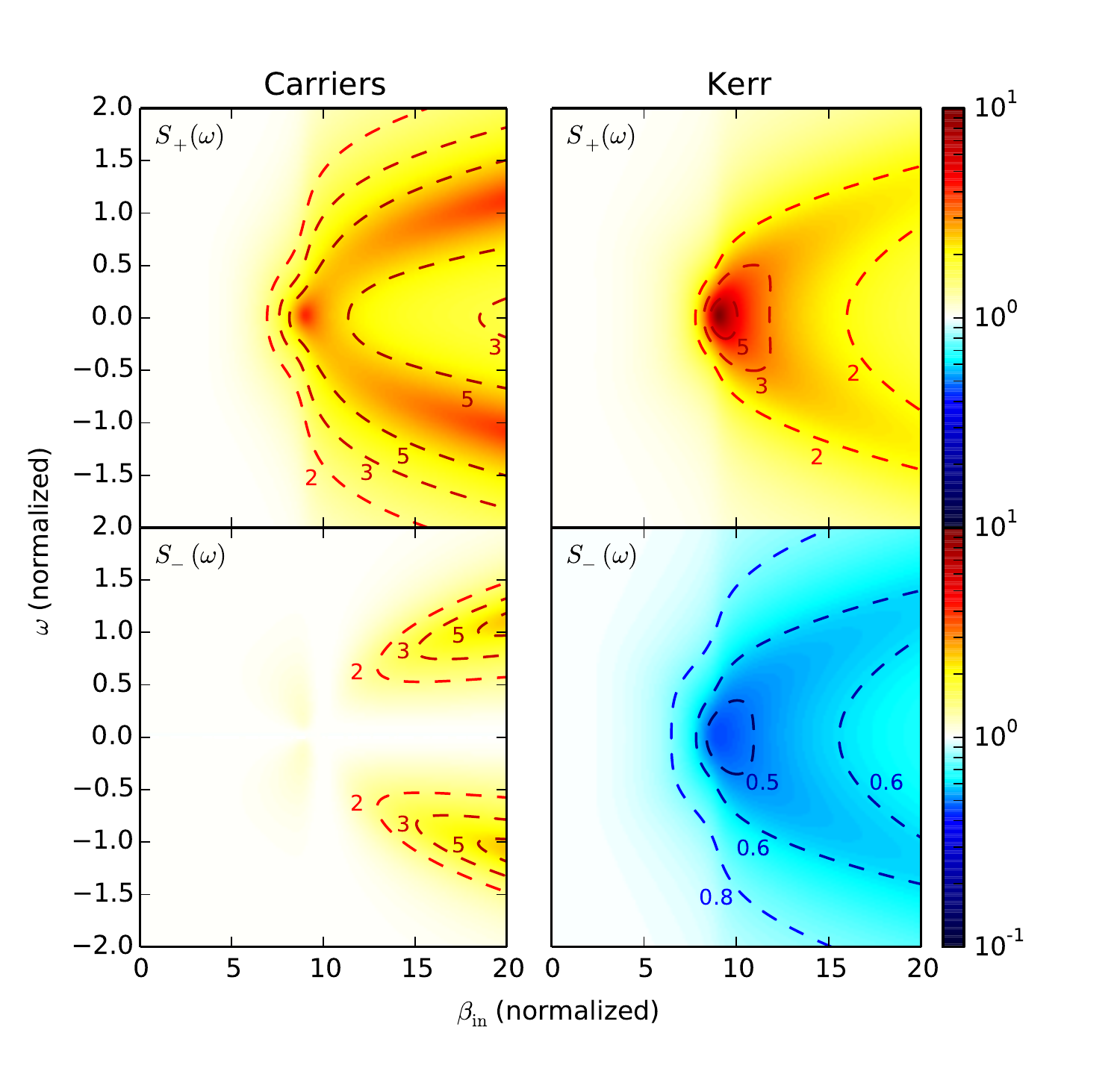}
\caption{Noise spectrum $S_+$, $S_-$ as a function of input $\beta_{\rm in}$ and frequency $\omega$ for Kerr and free-carrier models.  $\Delta_c = -0.7$}
\label{fig:06b-f7}
\end{center}
\end{figure}

Given a linearized input-output model, we can compute the squeezing spectrum (noise spectrum) for the cavity output field \cite{WallsMilburn, Gough2009c}, see Sec.~\ref{sec:04-linear}.  The squeezing spectrum for quadrature $\theta$ is defined as the power spectral density of a homodyne measurement of $\beta_{\rm out}(t)$.  That is, for the following homodyne signal,
\beq
	j_\theta(t) = e^{-i\theta}\beta_{\rm out}(t) + e^{i\theta} \beta_{\rm out}^*(t)
\eeq
the squeezing spectrum is:
\beq
	S_\theta(\omega) = \sqrt{2 P_\theta(\omega)},\ \ P_\theta(\omega) = \frac{\avg{j_\theta(\omega)^*j_\theta(\omega')}}{2\pi\delta(\omega-\omega')}
\eeq
$S_{\theta}(\omega)$ is normalized so that the coherent state has $S_\theta(\omega) = 1$.  For general states, $S_\theta(\omega)$ depends on $\theta$.  The maximum and minimum of $S_\theta(\omega)$, with respect to $\theta$, are denoted $S_+(\omega)$ and $S_-(\omega)$, respectively.

The squeezing spectrum of the Kerr cavity can be computed analytically \cite{Yurke2006}.  By contrast, since the free-carrier squeezing spectrum involves the inverse of a $3\times3$ matrix, it is unlikely that a simple expression can be found.  However, it is not difficult to compute numerically.

In Figure~\ref{fig:06b-f6}, the noise spectrum is obtained in two separate ways: first, simulating the full system in the time domain and taking the Fourier transform of the homodyned output (blue, green curves); and second, from the analytic predictions of the linearized ABCD model.  These agree everywhere except for very large pump powers, where the system approaches a bifurcation.

Figure~\ref{fig:06b-f7} displays the noise spectrum for the whole range $0 \leq \beta_{\rm in} \leq 20$, for both Kerr and free-carrier devices.  Two things are obvious.  First, the noise curve (at least for the $S_+$ component) matches the general form of the gain curve in Figure~\ref{fig:06b-f4}.  This is of course necessary because there must be noise wherever there is gain.  The free-carrier cavity, unlike the Kerr cavity, amplifies not only at $\omega = 0$ near the point of maximum gain, but also for $\omega \neq 0$ for $\beta_{\rm in}$ above that point.

Unlike the Kerr cavity, the free-carrier cavity does not squeeze the output field.  Regardless of the parameters, regardless of the pump power, both $S_+$ and $S_-$ are always above the vacuum level, indicating that this is a classical field with no squeezing.  The Kerr cavity, on the other hand, squeezes light over a broad range of the spectrum.  This is in agreement with the results of the previous section, which showed that the internal field of the free-carrier cavity was classical.  If the input and intracavity field are in a classical state, so is the output.

\section{Spontaneous Switching in SR-Latch}
\label{sec:06b-latch}

\begin{figure}[bp]
\begin{center}
\includegraphics[width=0.52\textwidth]{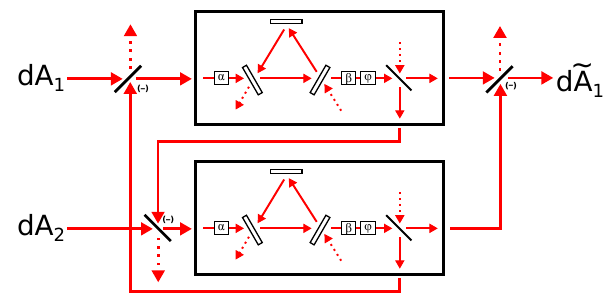}
\caption{Circuit diagram for a photonic SR-Latch}
\label{fig:06b-f8}
\end{center}
\end{figure}

It is also possible to construct a switching device using only amplifiers, provided the amplification is large enough \cite{Mabuchi2011b}.  The circuit in Figure~\ref{fig:06b-f8} uses two identical amplifiers in a feedback loop.  Suppose that each amplifier has a gain $G$.  Consider the fate of a perturbation in the top amplifier.  An input $\delta\beta$ is amplified to $G\,\delta\beta$.  This amplifier has a fan-out of 2, so $(G/\sqrt{2})\delta\beta$ passes to the right and exits the system, while $(G/\sqrt{2})\delta\beta$ passes to the lower amplifier.

In the lower amplifier, it grows to $(G^2/2)\delta\beta$, is fed back into the original amplifier.  After passing through this loop, the signal strength has grown to $(G^2/2)\delta\beta$.  This leads to a latching instability if the gain is sufficiently large:
\beq
	G > \sqrt{2}
\eeq
Symmetry gives the latch some very desirable properties.  Unlike the single-cavity switch, the two states here are symmetric.  Thus, there is less worry about finding the right bias field to ``balance'' the low and high state, and transitions between the states look the same.  But this comes at the cost of the added complexity of two cavities, plus the extra connections.

\begin{figure}[tbp]
\begin{center}
\includegraphics[width=0.66\textwidth]{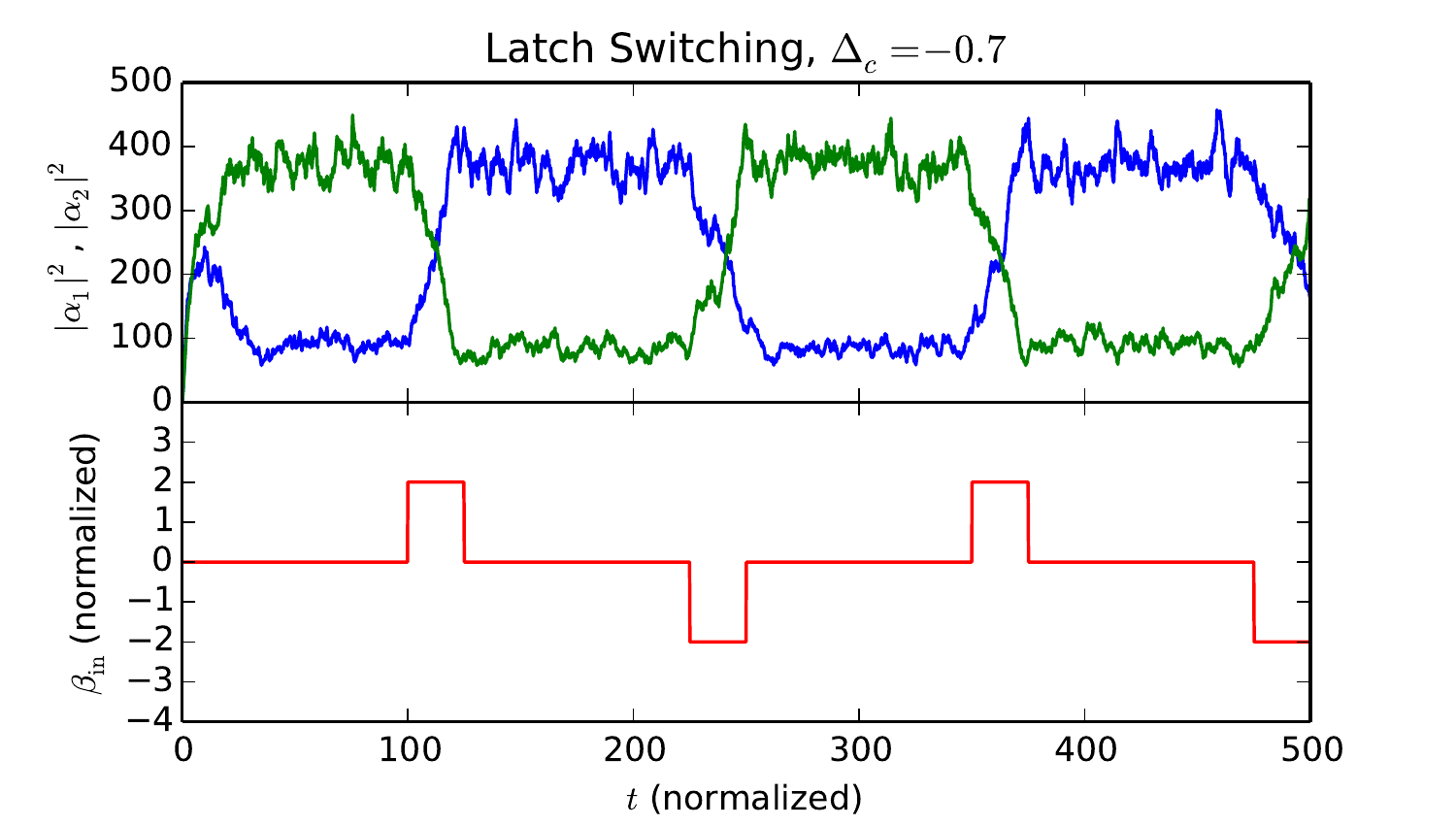}
\caption{Top: time series of the latch internal state.  Bottom: input field.}
\label{fig:06b-f9}
\end{center}
\end{figure}

Figure \ref{fig:06b-f9} shows a latch simulation for the same parameters used in the previous section.  Here, the detuning is set to $\Delta_c = -0.7$, large enough to realize a large gain, but not large enough make an individual cavity bistable.  The symmetry between the two states is very clear.

Externally driven switching in the latch is {\it good}, because it allows the user to set the state of the latch, which becomes a memory element.  But thanks to quantum noise, Kerr and free-carrier devices also undergo {\it spontaneous switching}.  This is generally {\it bad}, because it limits the lifetime of a carrier-based memory.

In the Kerr case, spontaneous switching is driven by vacuum fluctuations \cite{Santori2014}.  In the free-carrier case, vacuum fluctuations combine with stochastic carrier excitation and decay to drive the switching process.  Because there are more fluctuations, we naturally expect the free-carrier cavity to spontaneously switch at a higher rate than the Kerr cavity.

\begin{figure}[tbp]
\begin{center}
\includegraphics[width=0.66\textwidth]{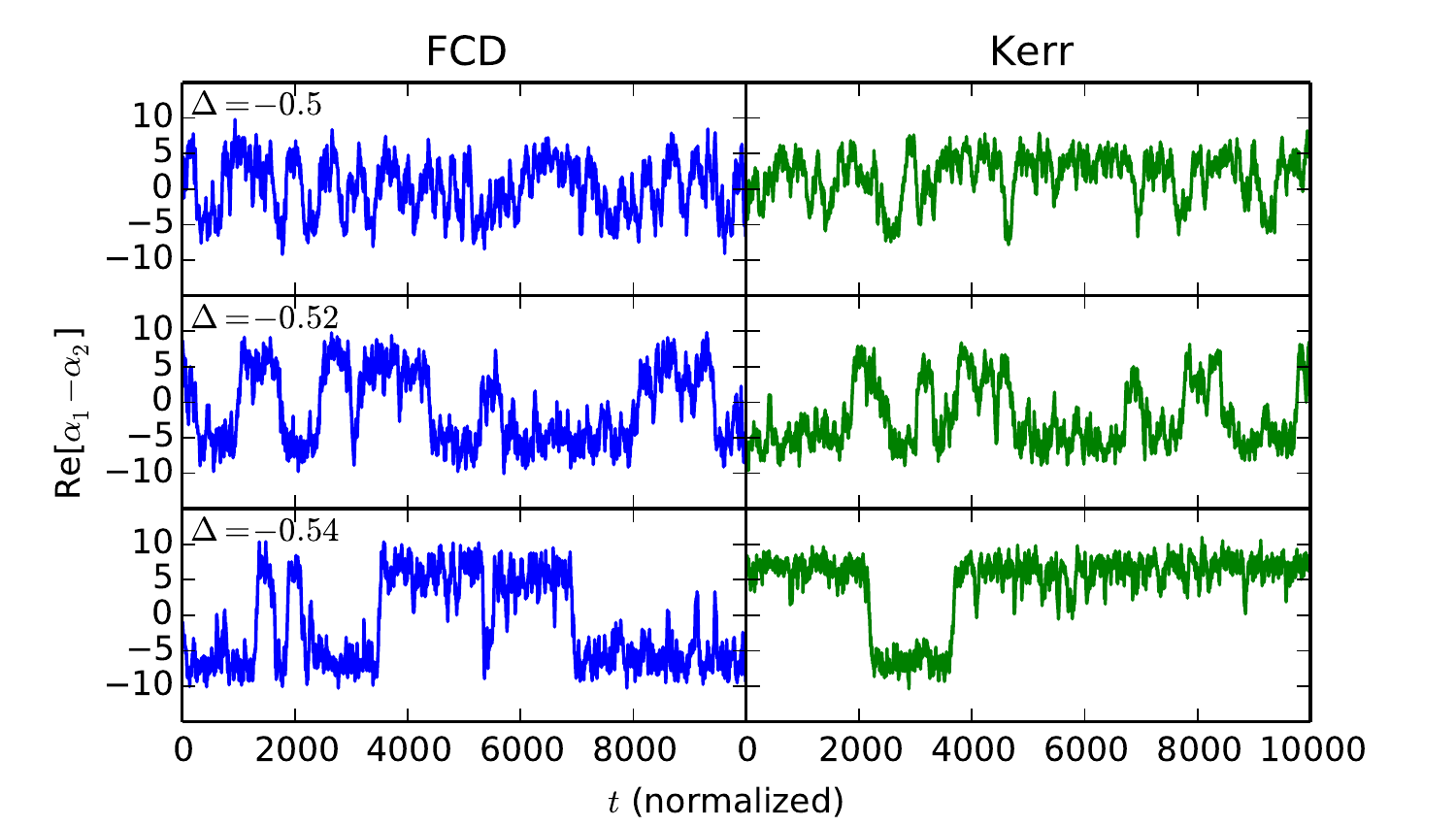}
\caption{Asymmetric part of the latch state Re[$\alpha_1 - \alpha_2$] for free-carrier based latch (left) and Kerr-based latch of the same $\chi^{(3)}$.  Cavity detuning set to $\Delta_c = 0.50, 0.52, 0.54$.}
\label{fig:06b-f10}
\end{center}
\end{figure}

When the switching rate is low, the switching process is well described by a two-state Markov chain.  In a two-state Markov chain, there are two states $a$ and $b$, with jump probabilities
\beq
	P(a \rightarrow b) = \gamma_{a} dt,\ \ \
	P(b \rightarrow a) = \gamma_{b} dt
\eeq
In the latch, the states are symmetric, so $\gamma_a = \gamma_b \equiv \gamma_{sw}$.  The probability of being in a given state evolves as:
\bea
	\frac{dP_a}{dt} & = & -\gamma_{sw} P_a + \gamma_{sw} P_b \nonumber \\
	\frac{dP_b}{dt} & = & \gamma_{sw} P_a - \gamma_{sw} P_b
\eea
Solving this linear system, one finds that the system reverts to its equilibrium distribution with a characteristic time $\tau_{sw} = 1/(2\gamma_{sw})$.  This time can be measured from simulations of the latch by looking at the autocorrelation function $R(\tau)$, which decays exponentially for the Markov process:
\beq
	R(\tau) = \frac{\avg{\alpha(t)\alpha(t-\tau)^*}}{\avg{\alpha(t)\alpha(t)^*}} \rightarrow e^{-\tau/\tau_{sw}}
\eeq
Figure \ref{fig:06b-f10} shows time traces of the asymmetric field $\alpha_1 - \alpha_2$ as the latch detuning is varied from $-0.50$ to $-0.54$, about where the latching transition occurs.  Larger negative detunings correspond to higher gain (see Fig.~\ref{fig:06b-f2}), and likewise stronger latching.  However, for a fixed detuning, the free-carrier cavity has a shorter spontaneous switching lifetime.

\begin{figure}[tbp]
\begin{center}
\includegraphics[width=0.66\textwidth]{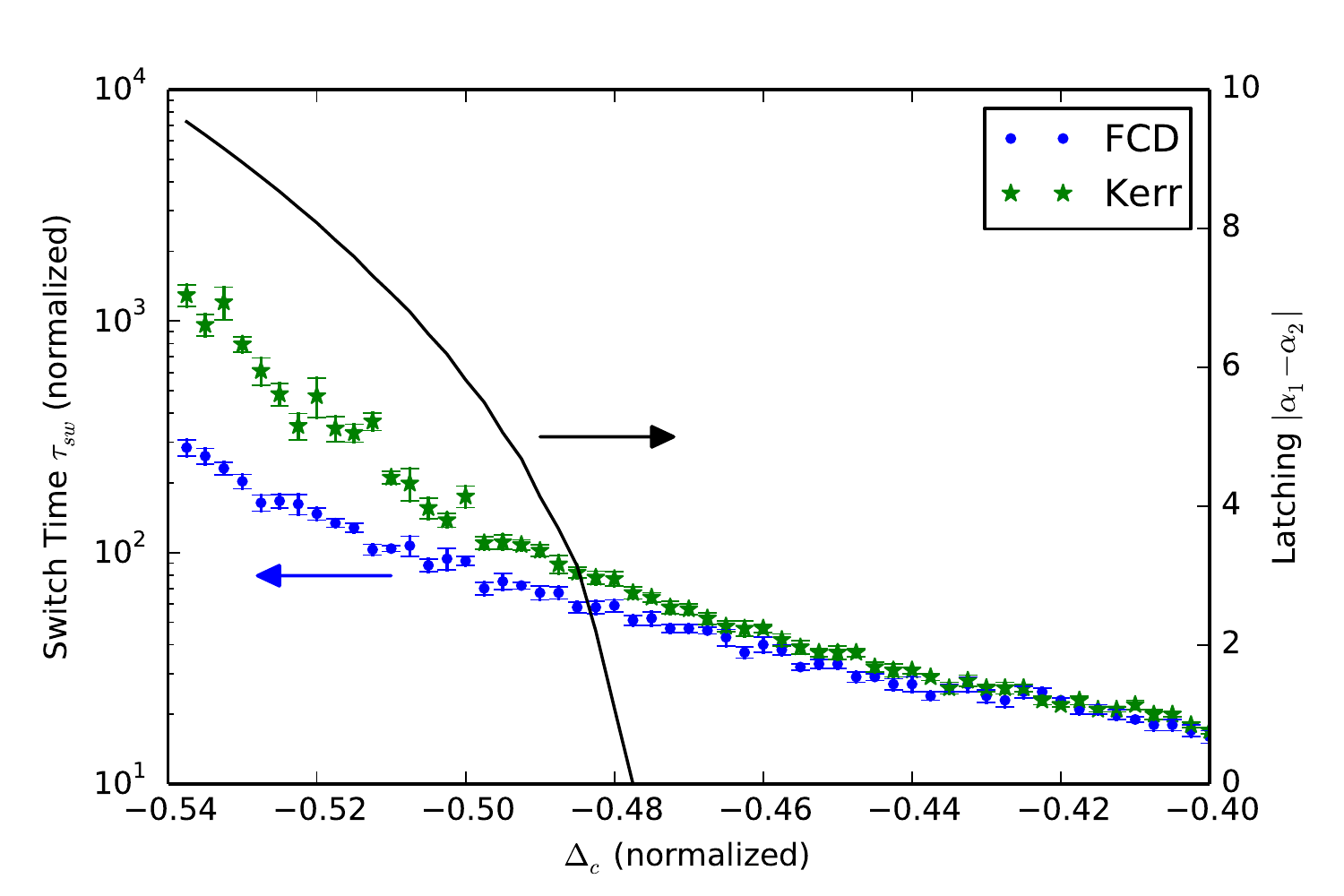}
\caption{Spontaneous switching lifetimes $\tau_{sw}$ for the Kerr and free-carrier latch as a function of detuning.}
\label{fig:06b-f11}
\end{center}
\end{figure}

This is also seen in Figure \ref{fig:06b-f11}, which plots $\tau_{sw}$ for the free-carrier and equivalent Kerr latches.  Because the free-carrier cavity has more quantum noise than the Kerr cavity, its spontaneous switching rate is higher.  The effect becomes noticeable once the latching transition sets in, and grows as the latching grows stronger.

\section{Bifurcation Analysis of SR-Latch}

The SR-latch is a simple circuit that uses feedback to create new dynamics.  Start with two nonlinear amplifiers.  By definition, an amplifier cannot store information -- its state and output are fully determined by the input.  However, if two amplifiers are placed in a feedback loop so that the output of one applies a negative signal to its partner, the latch becomes {\it bistable} and can be used to store a bit of information.  What's more, the configuration is highly symmetric.  This symmetry separates the pump degrees of freedom, which are needed to provide the energy for latching, from the signal degrees of freedom, which set and read out the state.

This chapter studies the phase space and bifurcations of a free-carrier latch.  This is important, because very general arguments can be made about when and where latching happens, as well as the nature of the latching bifurcation.  The latch has two bifurcations -- a low-energy latching instability and a high-energy limit cycle.  I study both of these are studied in turn, paying attention to the differences between this device and both a single-cavity system and a $\chi^{(3)}$ latch.

Understanding the dynamics of the latch from this high-level perspective may open many doors to future work.  For example, if we want to build an integrated Ising machine (Ch.~\ref{ch:09}), a network of latches may be a good alternative to OPOs \cite{Wang2013}.  It may be possible to construct an optical ``relay'' from latches, which could be used for photonic decoding of LDPC codes \cite{Pavlichin2014}.  These are things I pondered in the later years of graduate school, but never had the time to work out.

\subsection{Photonic Design with Ring Cavities}

\begin{figure}[tbp]
\begin{center}
\includegraphics[width=0.8\textwidth]{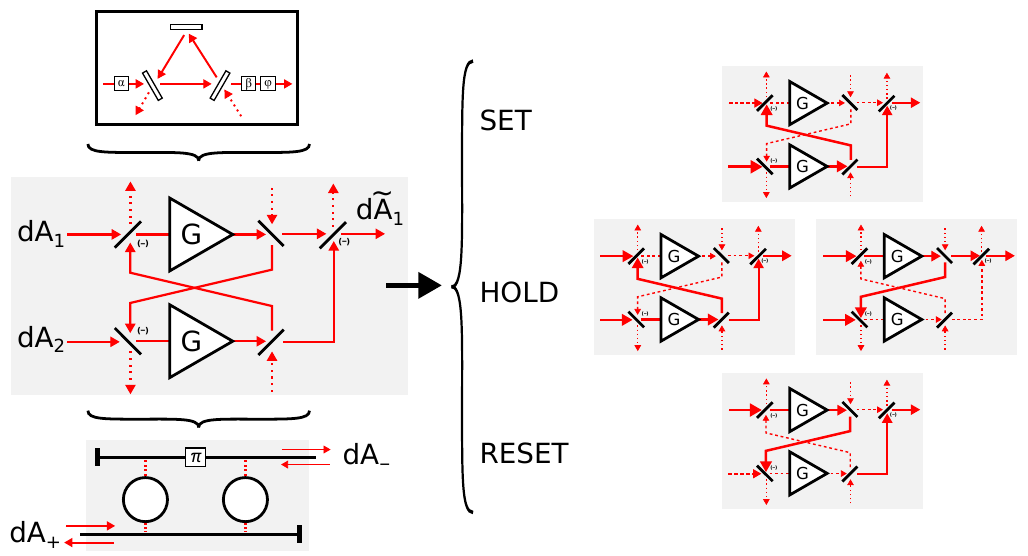}
\caption{Left: Amplifier element (top), placed in a feedback loop to form a latch circuit (middle), along with a possible photonic implementation (bottom).  Right: Depending on the driving, the latch either has a single stable state (top, bottom) or is bistable (middle).}
\label{fig:06b-f12}
\end{center}
\end{figure}

It is straightforward to take an optical nonlinearity like $\chi^{(3)}$ or free-carrier dispersion and build a phase-sensitive amplifier out of it (Sec.~\ref{sec:06b-amp}).  Two such amplifiers can be placed in a negative-feedback loop like Fig.~\ref{fig:06b-f12} to form the latch.

Feedback-control theory provides a very qualitative explanation for the latching behavior.  In the SET and RESET configurations, the input field is stronger than the internal feedback, which has the effect of forcing the system into its desired state regardless of the internal dynamics.  In the HOLD state, if the top amplifier is ``on'', the beam from the top to bottom amplifier interferes with the input, forcing the bottom amplifier into the ``off'' state.  The lack of output from the bottom amplifier likewise ensures the top amplifier stays on.  The same line of argument shows that the opposite state, with the top amplifier off and the bottom one on, is also stable.

Dynamical systems theory allows us to treat this problem more quantitatively.  In the Wigner formalism, a single Kerr cavity has two independent degrees of freedom $x = (\alpha, \alpha^*)$ (three degrees of freedom if one includes free carriers).  The latch consists of two cavities, so its phase space is spanned by the variables $(x_1, x_2)$.  Due to the symmetry between cavities, a better basis to use is $x_\pm = (x_1 \pm x_2)/2$.  If the system is driven symmetrically, there will always be a stable fixed point at $x_1 = x_2$, that is, $x_- = 0$.  Latching happens when this fixed point goes unstable in the $x_-$ variable.

To see this, consider the SLH model for the latch \cite{Tezak2012}.  In addition to single-cavity terms, there is an inter-cavity Hamiltonian, and most important, $L$ terms that go as $k_{1i} a_1 + k_{2i} a_2$ for some $k_{1i}, k_{2i}$.  This can actually be rewritten as:
\beq
	L_1 = \sqrt{\kappa_+/2} (a_1 + a_2),\ \ \ L_2 = \sqrt{\kappa_-/2} (a_1 - a_2)
\eeq
for different $k_1, k_2$.

\subsubsection{SLH Model}

A simple photonic design, shown in the Figure~\ref{fig:06b-f12} (lower left) realizes this same latching behavior.  In this design, two cavities are sandwiched between a pair of waveguides, with the lower-waveguide coupling stronger than the upper one.  Each cavity has a single standing-wave mode, which couples equally to waveguide signals propagating left and right.  Cavity 1 has the SLH model:
\beq
	{\rm Cav}_1 = \left(1_{4\times 4},\ \ \begin{bmatrix} \sqrt{\kappa_+/8}\,a_1 \\ \sqrt{\kappa_+/8}\,a_1 \\ \sqrt{\kappa_-/8}\,a_1 \\ \sqrt{\kappa_-/8}\,a_1 \end{bmatrix},\ \ H_{\rm int}\right) \equiv 1_{\rm int} \boxplus 1_{+,L} \boxplus 1_{+,R} \boxplus 1_{-,L} \boxplus 1_{-,R} 
\eeq
The $\pi$ phase shift on the top waveguide has the effect of shifting the sign of the $L$ terms that couple to the asymmetric input $dA_-$.  Thus, Cavity 2 has the SLH model:
\beq
	{\rm Cav}_2 = \left(1_{4\times 4},\ \ \begin{bmatrix} \sqrt{\kappa_+/8}\,a_2 \\ \sqrt{\kappa_+/8}\,a_2 \\ \sqrt{\kappa_-/8}\,a_2 \\ \sqrt{\kappa_-/8}\,a_2 \end{bmatrix},\ \ H_{\rm int}\right) \equiv 2_{\rm int} \boxplus 2_{+,L} \boxplus 2_{+,R} \boxplus 2_{-,L} \boxplus 2_{-,R} 
\eeq
Tracing the optical paths, one finds that that latch has the following Gough-James expression:
\beq
    {\rm Latch} = 1_{\rm int} \boxplus 2_{\rm int} \boxplus \left(1_{+L} \triangleleft 2_{+L} \triangleleft 2_{+R} \triangleleft 1_{+R}\right) \boxplus \left(1_{-L} \triangleleft 2_{-L} \triangleleft 2_{-R} \triangleleft 1_{-R}\right)
\eeq
The components $1_{\rm int}$, $2_{\rm int}$ give the internal, nonlinear dynamics.  The two series products are for the top and bottom couplings.  The bottom waveguide couples the two cavities in phase, so signals will set or read $a_1 + a_2$.  Due to a phase lag, the top waveguide couples them out of phase, allowing the difference to be read out.  In latching mode, a constant input is applied to the bottom waveguide, while the top waveguide is used for setting or readout.
\begin{eqnarray}
    1_{+L} \triangleleft 2_{+L} \triangleleft 2_{+R} \triangleleft 1_{+R} & = &  
    \left(1, \sqrt{\kappa_+/8}a_1, 0\right) \triangleleft \left(1, \sqrt{\kappa_+/8}a_2, 0\right) \triangleleft \left(1, \sqrt{\kappa_+/8}a_2, 0\right) \triangleleft \left(1, \sqrt{\kappa_+/8}a_1, 0\right) \nonumber \\
    & = & \left(1, \sqrt{\kappa_+/2}(a_1 + a_2), 0\right) \\
    1_{-L} \triangleleft 2_{-L} \triangleleft 2_{-R} \triangleleft 1_{-R} & = &  
    \left(1, \sqrt{\kappa_-/8}a_1, 0\right) \triangleleft \left(1, -\sqrt{\kappa_-/8}a_2, 0\right) \triangleleft \left(1, -\sqrt{\kappa_-/8}a_2, 0\right) \triangleleft \left(1, \sqrt{\kappa_-/8}a_1, 0\right) \nonumber \\
    & = & \left(1, \sqrt{\kappa_-/2}(a_1 - a_2), 0\right)
\end{eqnarray}
This yields the SLH model:
\beq
	{\rm Latch} = \left(1_{2\times 2},\ \ \begin{bmatrix} \sqrt{\kappa_+/2}\,(a_1 + a_2) \\ \sqrt{\kappa_-/2}\,(a_1 - a_2) \end{bmatrix},\ \ H_{{\rm int},1} + H_{{\rm int},2}\right)
\eeq

\subsubsection{Equations of Motion}

For a latch based on free-carrier cavities, this results is the following Wigner equations of motion:
\begin{eqnarray}
    d\alpha_1 & = & (d\alpha_1)_{\rm int} - \frac{\kappa_+ (\alpha_1 + \alpha_2) + \kappa_-(\alpha_1 - \alpha_2)}{4} dt - \sqrt{\kappa_+/2}\,d\beta_{+,\rm in} - \sqrt{\kappa_-/2}\,d\beta_{-,\rm in} \\
    dN_1 & = & (dN_1)_{\rm int} \\
    d\alpha_2 & = & (d\alpha_2)_{\rm int} - \frac{\kappa_+ (\alpha_1 + \alpha_2) - \kappa_-(\alpha_1 - \alpha_2)}{4} dt - \sqrt{\kappa_+/2}\,d\beta_{+,\rm in} + \sqrt{\kappa_-/2}\,d\beta_{-,\rm in} \\
    dN_2 & = & (dN_2)_{\rm int} \\
    d\beta_{+,\rm out} & = & d\beta_{+,\rm in} + \sqrt{\kappa_+/2}(\alpha_1 + \alpha_2)dt \\
    d\beta_{-,\rm out} & = & d\beta_{-,\rm in} + \sqrt{\kappa_-/2}(\alpha_1 - \alpha_2)dt    
\end{eqnarray}
The $(d\alpha)_{\rm int}$, $(dN)_{\rm int}$ depend on the cavity parameters; the rest of the dynamics is determined solely by the circuit layout.  If there are additional degrees of freedom (temperature, excitons), these can be accounted for as well.  However, for steady-state determination, the additional degrees of freedom do not matter -- one only needs the effective optical nonlinearity, adiabatically eliminating the non-optical modes.

Define the following symmetric and antisymmetric modes:
\beq
    \alpha_\pm = \frac{\alpha_1 \pm \alpha_2}{\sqrt{2}},\ \ \ N_{\pm} = \frac{N_1 \pm N_2}{\sqrt{2}}
\eeq
I put a $\sqrt{2}$ in the denominator to make this an orthogonal transformation: that way, the $a_\pm$ are properly normalized fields that satisfy the canonical commutation relations: $[a_+, a_+^\dagger] = 1$, $[a_-, a_-^\dagger] = 1$.  This isn't necessary for the $N_\pm$, which are classical variables, but it helps the notation to be consistent.  The Wigner SDEs become:
\begin{eqnarray}
    d\alpha_+ & = & \frac{(d\alpha_1)_{\rm int} + (d\alpha_2)_{\rm int}}{\sqrt{2}} - \frac{1}{2}\kappa_+ \alpha_+ dt - \sqrt{\kappa_+}\,d\beta_{+,\rm in} \\
    dN_+ & = & \frac{(dN_1)_{\rm int} + (dN_2)_{\rm int}}{\sqrt{2}} \\
    d\alpha_- & = & \frac{(d\alpha_1)_{\rm int} - (d\alpha_2)_{\rm int}}{\sqrt{2}} - \frac{1}{2}\kappa_- \alpha_- dt - \sqrt{\kappa_-}\,d\beta_{-,\rm in} \\
    dN_- & = & \frac{(dN_1)_{\rm int} - (dN_2)_{\rm int}}{\sqrt{2}} \\
    d\beta_{+,\rm out} & = & \sqrt{\kappa_+}\alpha_+ dt + d\beta_{+,\rm in} \\
    d\beta_{-,\rm out} & = & \sqrt{\kappa_-}\alpha_- dt + d\beta_{-,\rm in}
\end{eqnarray}
where the internal dynamics are given by:
\begin{eqnarray}
    (d\alpha_i)_{\rm int} & = & \left[-\frac{\eta}{2} - (\beta+i\chi)\alpha_i^*\alpha_i - i(\Delta + \delta N_i)\right]\alpha_i\,dt + d\xi_{\alpha,i} \\
    (dN_i)_{\rm int} & = & \left[\eta\alpha_i^*\alpha_i + \beta(\alpha_i^*\alpha_i)^2 - \gamma N_i\right]\,dt + d\xi_{N,i}
\end{eqnarray}

\subsubsection{Normalized Coordinates}

To make the results as general as possible, choose to work in ``normalized'' coordinates.  These reduce the number of free parameters in the problem from 11 to 8, of which 6 are constants set by the material or cavity geometry.  Defining $k = \kappa_+ + \eta$, we set:
\beq
	t \rightarrow \frac{\bar{t}}{k},\ \ \ 
	N \rightarrow \frac{\bar{N}}{\delta/k},\ \ \ 
	\alpha \rightarrow \frac{\bar{\alpha}}{\sqrt{\beta/k}},\ \ \ 
	\beta_{\rm in} \rightarrow \frac{\bar{\beta}_{\rm in}}{\sqrt{\beta/k^2}} \label{eq:06b-normcoords}
\eeq
Intuitively, time $\bar{t}$ is scaled so that the cavity photon lifetime is one (for the $\alpha_+$ mode, which decays fastest).  The carrier number is scaled so that $\bar{N} = 1$ shifts the cavity by one linewidth.  Both $\alpha$ and $\beta_{\rm in}$ are scaled by the two-photon absorption: $|\bar{\alpha}| = 1$ means that single- and two-photon loss processes are equally likely.

The reduced equations take the form:
\begin{eqnarray}
    d\bar{\alpha}_+ & = & \frac{(d\bar{\alpha}_1)_{\rm int} + (d\bar{\alpha}_2)_{\rm int}}{\sqrt{2}} - \frac{1}{2}\bar{\kappa}_+ \bar{\alpha}_+ d\bar{t} - \sqrt{\bar{\kappa}_+}\,d\bar{\beta}_{+,\rm in} \label{eq:06b-rm1} \\
    d\bar{N}_+ & = & \frac{(d\bar{N}_1)_{\rm int} + (d\bar{N}_2)_{\rm int}}{\sqrt{2}} \\
    d\bar{\alpha}_- & = & \frac{(d\bar{\alpha}_1)_{\rm int} - (d\bar{\alpha}_2)_{\rm int}}{\sqrt{2}} - \frac{1}{2}\bar{\kappa}_- \bar{\alpha}_- d\bar{t} - \sqrt{\bar{\kappa}_-}\,d\bar{\beta}_{-,\rm in} \\
    d\bar{N}_- & = & \frac{(d\bar{N}_1)_{\rm int} - (d\bar{N}_2)_{\rm int}}{\sqrt{2}} \\
    d\bar{\beta}_{+,\rm out} & = & \sqrt{\bar{\kappa}_+}\,\bar{\alpha}_+ d\bar{t} + d\bar{\beta}_{+,\rm in} \\
    d\bar{\beta}_{-,\rm out} & = & \sqrt{\bar{\kappa}_-}\,\bar{\alpha}_- d\bar{t} + d\bar{\beta}_{-,\rm in}
\end{eqnarray}
with
\begin{eqnarray}
    (d\bar{\alpha}_i)_{\rm int} & = & \left[-\left(\bar{\eta}/2+\bar{\delta}_{\rm fca}\bar{N}_i\right) - (1 + i\bar{\chi})\bar{\alpha}_i^*\bar{\alpha}_i - i\left(\bar{\Delta}+\bar{N}_i\right)\right]\bar{\alpha}_i\,d\bar{t} + d\bar{\xi}_{\alpha,i} \\
    (d\bar{N}_i)_{\rm int} & = & \left[\bar{\mu} (\bar{\alpha}^*\bar{\alpha}) + \bar{\zeta} (\bar{\alpha}^*\bar{\alpha})^2 - \bar{\gamma}\bar{N}\right]d\bar{t} + d\bar{\xi}_{N,i} \label{eq:06b-rmF}
\end{eqnarray}
\begin{table}[bp]
\begin{centering}
\begin{tabular}{cc|cc}
	\hline\hline
	Reduced Parameter           & Formula                   & Si $\mu$-ring (TPA) & GaAs PhC (TPA+LA) \\ \hline
	$\bar{\delta}_{\rm fca}$    & $\delta_{\rm fca}/\delta$ & $-0.071$            & $0$               \\
	$\bar{\mu}$                 & $\delta\eta/k\beta$       & $0$                 & $17.1$            \\
	$\bar{\zeta}$               & $\delta/\beta$            & $151.3$             & $34.1$            \\
	$\bar{\chi}$                & $\chi/\beta$              & $0$                 & $0$               \\
	$\bar{\gamma}$              & $\gamma/k$                & $1.0$               & $1.2$             \\
	$\bar{\eta}$                & $\eta/k$                  & $0.5$               & $0$               \\
	$\bar{\kappa}_+$            & $\kappa_+/k$              & $1.0$               & $0.5$             \\
	$\bar{\kappa}_-$            & $\kappa_-/k$              & variable            & variable          \\
	$\bar{\Delta}$              & $\Delta/k$                & variable            & variable          \\ \hline\hline
\end{tabular}
\caption{Reduced parameters for simulations in this section.  Si $\mu$-ring: Q = 2.5--$5\times 10^5$, $\tilde{V}$ = 5--20, $\beta$ = $3.9 \times 10^{-6}$, $\delta$ = $(5.6 - 0.4i)\times 10^{-4}$, $\gamma = 1$ \cite{JohnsonThesis, Johnson2006}.  GaAs PhC: similar to Table~\ref{tab:06b-t1}}
\label{tab:06b-t2}
\end{centering}
\end{table}

There are 8 parameters in Eqs.~(\ref{eq:06b-rm1}-\ref{eq:06b-rmF}), listed below.
\beq
	\underbrace{\bar{\Delta} = \frac{\Delta}{k},}_{\rm Tunable}\ \ \ 
	\underbrace{\bar{\kappa}_+ = \frac{\kappa_+}{k},\ \ \ 
	\bar{\kappa}_- = \frac{\kappa_+}{k},\ \ \ 
	\bar{\eta} = \frac{\eta}{k},\ \ \ 
	\bar{\gamma} = \frac{\gamma}{k},}_{\rm Cavity\;Design}\ \ \ 
	\underbrace{\bar{\delta}_{\rm fca} = \frac{\delta_{\rm fca}}{\delta},\ \ \ 
	\bar{\mu} = \frac{\delta\eta}{k\beta},\ \ \ 
	\bar{\zeta} = \frac{\delta}{\beta},\ \ \ 
	\bar{\chi} = \frac{\chi}{\beta}}_{\rm Material\;Properties}
\eeq
Of these, four are material constants.  Of the three that depend on cavity design, three add up to one ($\bar{\eta} + \bar{\kappa}_+ + \bar{\kappa}_- = 1$) and we require $\bar{\kappa}, \bar{\gamma} \sim O(1)$ to have efficient coupling between the pump, cavity modes and carriers.  We will see below that the latch functions best when $\bar{\kappa}_- < \bar{\kappa}_+$, so that parameter can be assumed to vary from 0 to $\bar{\kappa}_+$.  Only the detuning $\bar{\Delta}$ and the inputs $\bar{\beta}_{\pm,\rm in}$, can be varied dynamically.  

Thus, rescaling and material constraints significantly constrains the parameter space, making it easier to make universal statements that apply to all optical latches.

\begin{figure}[b!]
\begin{center}
\includegraphics[width=1.00\textwidth]{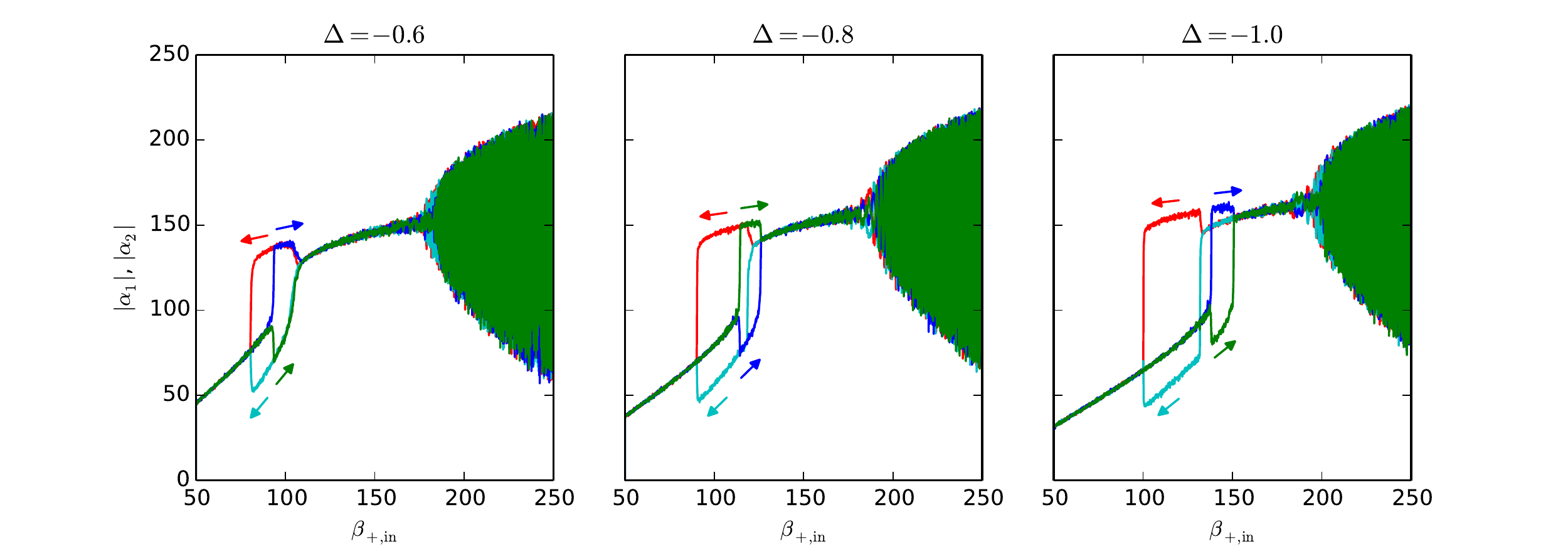}
\caption{Cavity fields $\alpha_1, \alpha_2$ for the latch with the Si $\mu$-ring (TPA) parameters.  The input $\beta_{\rm in}$ is swept from $50$ to $250$ (blue, green curves) and back (red, cyan curves)}
\label{fig:06b-f13}
\end{center}
\end{figure}

\subsection{Phase Space and Bifurcations}

Compared to a single cavity, the latch has a more complex phase space.  Recall that both the Kerr cavity and free-carrier cavity had a pitchfork bifurcation resulting in optical bistability (Sec.~\ref{sec:06b-ss}).  The free-carrier cavity also has a Hopf bifurcation at high powers, giving rise to a limit cycle (Ch.~\ref{ch:07}).  Since the latch is formed from two coupled cavities, we expect to see all these effects and perhaps some more.

Figure \ref{fig:06b-f13} shows three simulations of a free-carrier cavity based on the silicon parameters in the table above.  In this plot, the pump field $\beta_{+\rm in}$ is swept up and down, and the internal cavity modes $a_1$, $a_2$ are shown.  Like a single cavity, one sees a latching bifurcation at low powers, and a limit cycle at high powers.  However, there is an added hysteresis in the latching, so that for some configurations, three or more states are stable.

The goal of this section is to construct a {\it phase diagram} for the latch, in terms of the parameters $\beta_{\rm in}$, $\Delta$ and $\kappa_-$, that explains all the behavior in Figure \ref{fig:06b-f13}.

\subsubsection{Stability of Symmetric Fixed Points}

Suppose that $(\bar{x}_1, \bar{x_2})$ is a fixed point of the latch, where $\bar{x} = (\bar{\alpha}, \bar{\alpha}^*, \bar{N})$ is the state of one cavity.  Linearizing (\ref{eq:06b-rm1}-\ref{eq:06b-rmF}) about this point, and throwing away the noise terms:
\bea
	\frac{d}{dt} \begin{bmatrix} \delta\bar{x}_+ \\ \delta\bar{x}_- \end{bmatrix} & = & 
		\begin{bmatrix} \frac{J_1 + J_2}{2} - \frac{1}{2}\kappa_+ K & \frac{J_1 - J_2}{2} \\
			\frac{J_1 - J_2}{2} & \frac{J_1 + J_2}{2} - \frac{1}{2} \kappa_- K \end{bmatrix}
		\begin{bmatrix} \delta\bar{x}_+ \\ \delta\bar{x}_- \end{bmatrix} 
		- \begin{bmatrix} \sqrt{\kappa_+} K & 0 \\ 0 & \sqrt{\kappa_-} K \end{bmatrix}
		\begin{bmatrix} \bar{\beta}_{+, \rm in} \\ \bar{\beta}_{-, \rm in} \end{bmatrix} \label{eq:06b-lin} \\
	\begin{bmatrix} \bar{\beta}_{+, \rm out} \\ \bar{\beta}_{-, \rm out} \end{bmatrix} & = &
		\begin{bmatrix} \sqrt{\kappa_+} K & 0 \\ 0 & \sqrt{\kappa_-} K \end{bmatrix}
		\begin{bmatrix} \bar{x}_+ \\ \bar{x}_- \end{bmatrix} + 
		\begin{bmatrix} \bar{\beta}_{+, \rm in} \\ \bar{\beta}_{-, \rm in} \end{bmatrix}
\eea
where $J_i$ is the Jacobian at $\bar{x}_i$, and $K$ is a projector matrix, as follows (compare Eq.~(\ref{eq:07b-linabcd})):
\bea
	J_i & = & \begin{bmatrix}
		(J_i)_{11} & -(1 + i\bar{\chi}) \bar{\alpha}_i^2 & -i\bigl(1 - i\bar{\delta}_{\rm fca}\bigr)\bar{\alpha}_i \\ 
		\left[-(1 + i\bar{\chi}) \bar{\alpha}_i^2\right]^* & (J_i)_{11}^* & \left[-i\bigl(1 - i\bar{\delta}_{\rm fca}\bigr)\bar{\alpha}_i\right]^* \\
		\bigl(\bar{\mu} + 2\bar{\zeta}|\bar{\alpha}_i|^2\bigr)\bar{\alpha}_i^* & 
		\bigl(\bar{\mu} + 2\bar{\zeta}|\bar{\alpha}_i|^2\bigr)\bar{\alpha}_i & -\bar{\gamma}
		\end{bmatrix} \nonumber \\
		& & {\rm with}\ \ (J_i)_{11} = \left(-\frac{\bar{\eta}}{2} - i\bar{\Delta}\right) - i\bigl(1 - i\bar{\delta}_{\rm fca}\bigr) \bar{N}_i - 2(1 + i\bar{\chi})|\bar{\alpha}_i|^2 \\
	K & = & \begin{bmatrix} 1 & 0 & 0 \\ 0 & 1 & 0 \\ 0 & 0 & 0 \end{bmatrix} 
\eea
The symmetric state is a special case.  Here, $\bar{x}_1 = \bar{x}_2$ and $J_1 = J_2 \equiv J$, so the modes decouple and the equations of motion become:
\beq
	\frac{d\bar{x}_\pm}{dt} = J \pm \frac{1}{2}\kappa_\pm K \label{eq:06b-lin-red}
\eeq

\begin{figure}[tbp]
\begin{center}
\includegraphics[width=0.8\textwidth]{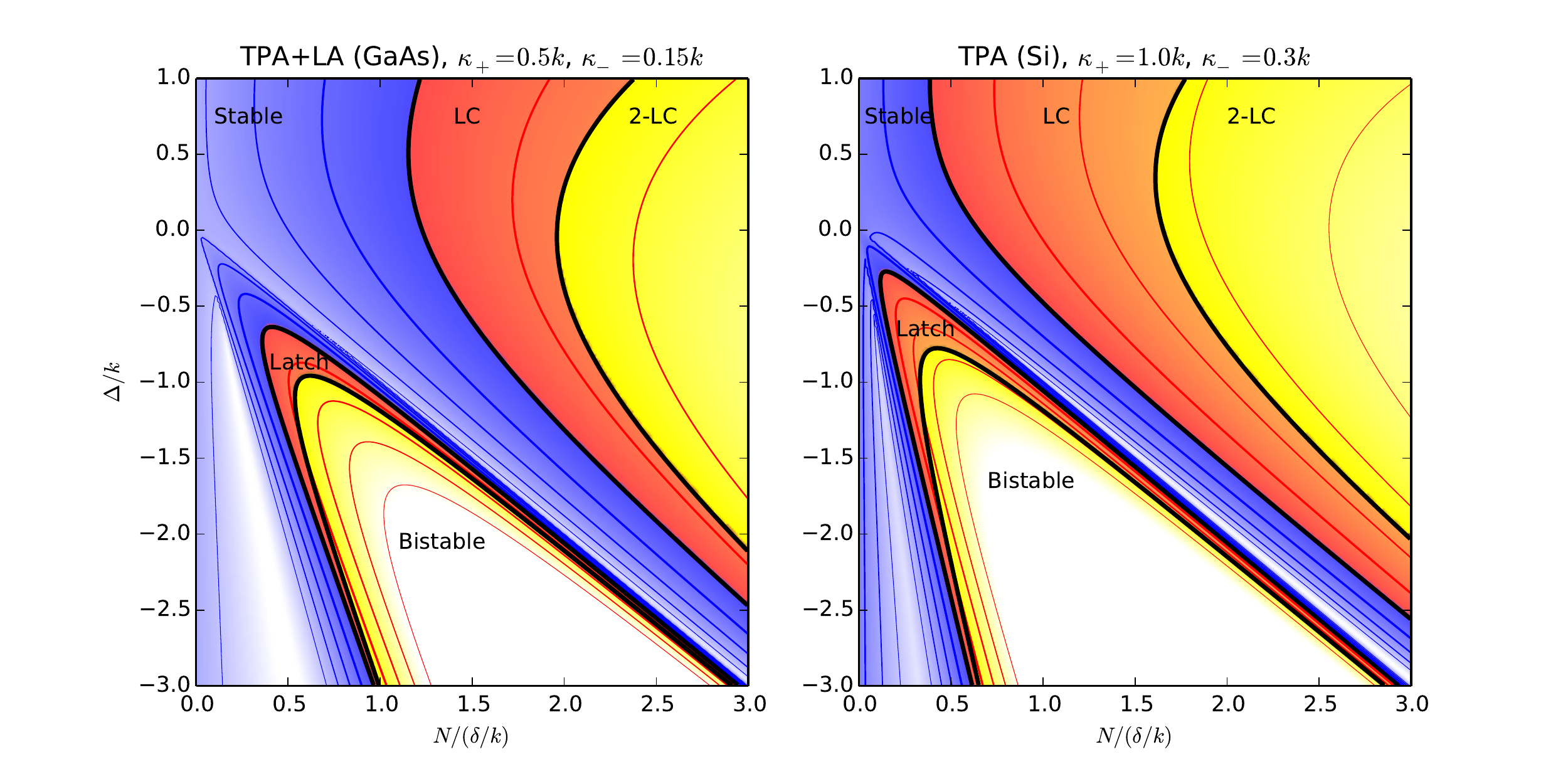}
\caption{Stability of the symmetric state $\alpha_1 = \alpha_2$, for GaAs cavity parameters (left) and Si cavity parameters (right).  Plotted in terms of normalized coordinates (\ref{eq:06b-normcoords}).  Red region corresponds to instability in the $\alpha_-$ mode (standard latching or limit cycle behavior); yellow region corresponds to instability in both $\alpha_+$ and $\alpha_-$ modes.  Lines are contours of the $\alpha_-$ eigenvalue.}
\label{fig:06b-f14}
\end{center}
\end{figure}

Since $\kappa_- < \kappa_+$ by design, Eq.~(\ref{eq:06b-lin-red}) says that the asymmetric mode $\alpha_-$ always goes unstable {\it before} the symmetric mode.  This is the mode that gives rise to the ``latch'' states $(\bar{x}_H,\bar{x}_L)$, $(\bar{x}_L,\bar{x}_H)$.  An unstable $\alpha_+$ mode, by contrast, would give rise to optical ``bistability'' states $(\bar{x}_H,\bar{x}_H)$, $(\bar{x}_L,\bar{x}_L)$.  The equation above shows that the latching region should be larger than the bistability region, and in principle both can coexist.  This is confirmed in the traces in Figure~\ref{fig:06b-f13}.

This is also seen in Figure \ref{fig:06b-f14}.  In this figure, the stability of both modes $\delta\alpha_+, \delta\alpha_-$ is plotted in terms of the normalized carrier number $\bar{N}$ and detuning $\bar{\Delta}$.  There are four distinct regions here: stable in both $\delta\alpha_+, \delta\alpha_-$ (blue), unstable in $\alpha_-$ only (red), unstable in both (yellow).

\subsubsection{Phase Diagram}

\begin{figure}[tbp]
\begin{center}
\includegraphics[width=1.00\textwidth]{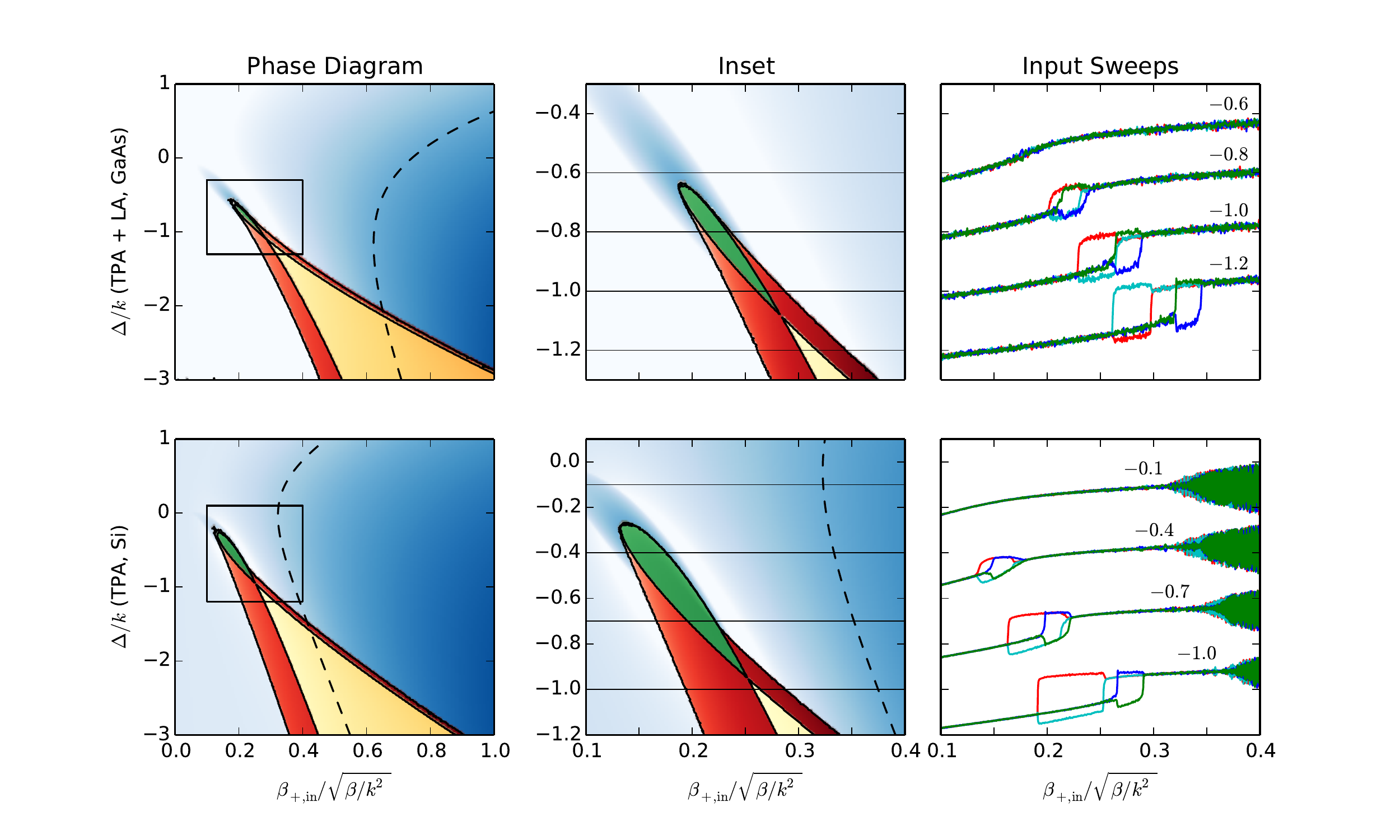}
\caption{Phase diagram of the latch.  Left: Phase diagram for the interval $\bar{\beta}_{+,\rm in} \in [0, 1]$, $\bar{\Delta} \in [-3, 1]$.  Blue is non-latching, green is bistable to latching, red is twistable (two latching states plus a symmetric state), yellow is tetrastable (two latching states plus two symmetric states).  Center and right: Inset and simulations sweeping the input power in this region.}
\label{fig:06b-f15}
\end{center}
\end{figure}

One can numerically solve for the fixed points of the latch using standard gradient-descent methods.  In this way, I construct the phase diagram in Figure~\ref{fig:06b-f15}.  Note that this roughly follows the pattern of Figure~\ref{fig:06b-f10}, in that both a pitchfork (solid lines) and Hopf (dashed line) bifurcation are present.  However, because of the latching mechanism, this diagram is more complex than that of the single cavity.

The latch supports up to four stable states.  In the blue region in Figure~\ref{fig:06b-f15}, there is only one such state.  This is the symmetric state, which becomes a limit cycle above the Hopf bifurcation (dashed line in the figure).  The ``latching'' instability, where the symmetric state $(\alpha_s, \alpha_s)$ goes unstable and two stable fixed points $(\alpha_L, \alpha_H)$, $(\alpha_H, \alpha_L)$ are formed, may be either subcritical or supercritical, depending on the parameters.  In the supercritical case, the system passes directly from the blue region to the green region, where only the latching states are stable.  In the subcritical case, there is a hysteresis region (red in figure) where both the symmetric and latching states are stable.  There is also a region where there are two stable latching states {\it and} two stable symmetric states, yellow in the figure.

These bifurcations can also be seen from simulations of the internal state as the input $\beta_{+,\rm in}$ is scanned up and down (right pane of the figure).  Hysteresis is present when the cavity has a subcritical bifurcation.

\ifstandalone{\bibliography{NoteRefs}{}
\bibliographystyle{alpha}
}
\ifdefined\multidoc\else\input{Header}\fi

\newcommand{\binA}{\beta_{{\rm in}}}
\newcommand{\boutA}{\beta_{{\rm out}}}
\newcommand{\bin}[1]{\beta_{{\rm in},{#1}}}
\newcommand{\bout}[1]{\beta_{{\rm out},{#1}}}
\newcommand{\bbinA}{\bar{\beta}_{{\rm in}}}
\newcommand{\bboutA}{\bar{\beta}_{{\rm out}}}
\newcommand{\bbin}[1]{\bar{\beta}_{{\rm in},{#1}}}
\newcommand{\bbout}[1]{\bar{\beta}_{{\rm out},{#1}}}
\newcommand{\Amat}{\bar{A}}

\ifstandalone{\setcounter{chapter}{6}}

\chapter{Free-Carrier Limit-Cycle Oscillators}
\label{ch:07}

This chapter is based on the following paper:

\begin{itemize}
	\item \href{http://dx.doi.org/10.1103/PhysRevApplied.4.024016}{R.~Hamerly and H.~Mabuchi, ``Optical Devices Based on Limit Cycles and Amplification in Semiconductor Optical Cavities'', Physical Review Applied 4, 024016 (2015)}
\end{itemize}

Many problems in simulation, optimization and machine learning are analog in nature and mapping them onto a digital processor incurs significant overhead.  As a result, there has been a recent revival of interest in analog or ``neuromorphic'' computing systems \cite{Utsunomiya2011, Tezak2015}.  Devices that can spontaneously oscillate are a key component in this neuromorphic architecture.  Such devices can function as an analog memory \cite{Tezak2015}, a phase-insensitive amplifier \cite{Kwon2013, KwonThesis}, or a complex-valued neuron \cite{HiroseBook}, among other things.  In addition, large networks of such oscillators can be applied to complex optimization and machine learning tasks, such as Ising problems \cite{Utsunomiya2011}.

In most dynamical systems, spontaneous oscillations arise from a Hopf bifurcation \cite{StrogatzBook}.  In optics, the simplest such system is the non-degenerate $\chi^{(2)}$ optical parametric oscillator (OPO), which behaves as a quantum-limited amplifier below threshold \cite{Yamamoto1990} and has a symmetric limit cycle above \cite{Reid1988}.  In addition, cavity quantum electrodynamics (QED) systems can self-oscillate in the right conditions \cite{Armen2006, Kwon2013}.  However, nanofabrication with $\chi^{(2)}$ materials such as KTP and LiNbO$_3$ is still in its infancy \cite{Poberaj2012}, and most implementations of cavity QED -- trapped atoms, quantum dots, NV centers -- are not scalable with current technology.  To realize neuromorphic computing with photonics, there is an unfulfilled need for self-oscillating photonic devices based on a scalable technology.

Free-carrier dispersion can fulfill this unmet need.  This effect is present in silicon and all III-V semiconductors, and is scalable and low-power \cite{Notomi2010}.  Previous work by Malaguti et al.~\cite{Malaguti2011, Malaguti2013} and Chen et al.~\cite{Chen2012} showed that when the photon and carrier lifetime are comparable, an optical cavity can pass through a Hopf bifurcation and undergo self-oscillation.  However, these studies focused on the many-photon classical limit, where quantum fluctuations can be ignored.  If such a device is optimized for low power, quantum fluctuations in the photon and carrier number may substantially alter the dynamics and limit the performance of real devices.

In Chapters \ref{ch:05b}-\ref{ch:06b}, I derived a set of stochastic equations for free-carrier optical cavities that  model these quantum fluctuations, and applied them to study phase-sensitive amplifiers and latches \cite{Hamerly2015-1}.  Here, I apply those equations to study the effects of quantum noise on the free-carrier Hopf bifurcation.

Sections \ref{sec:07b-conditions} and \ref{sec:07b-sims} discuss the general theory of the oscillations, which arise from an instability in the linearized model around the system's fixed point.  Because this is done in a general, scale-invariant way, it should be possible to observe these oscillations in a wide range of systems spanning orders of magnitude in speed, size and energy.  Next, we consider the equations of motion close to the bifurcation point and show that the bifurcation resembles the non-degenerate OPO at threshold with some extra noise.  Section \ref{sec:07b-below} models the device below threshold: it functions as a phase-insensitive linear amplifier with noise $\sim 5\times$ above the Caves bound \cite{Caves1982}.  The near-threshold behavior, which follows the critical exponents of the Hopf bifurcation, is discussed in Section \ref{sec:07b-near}.

The above-threshold case is covered in Section \ref{sec:07b-above}.  Like the non-degenerate OPO, the free-carrier cavity has a limit cycle in this regime.  The above-threshold OPO can be considered a ``quantum-optimal'' limit cycle in the sense that it can function as an optimal homodyne detector.  By comparison, the free-carrier limit cycle is $\sim 10\times$ noisier than the OPO.  This difference is due to the incoherent nature of carrier excitation and decay.

Limit-cycle devices can be very useful in optimization and machine learning.  In Section \ref{sec:07b-ising}, I propose and simulate an Ising machine based on the free-carrier limit cycle, which should be several orders of magnitude faster and less power-consuming than a supercomputer.  In addition, Section \ref{sec:07b-relay} discusses an all-optical XOR gate based on the limit-cycle effect.

\section{Conditions for Self-Oscillation}
\label{sec:07b-conditions}

\subsection{Equations of Motion}

A single-mode free-carrier optical cavity has three degrees of freedom: two field quadratures $(\alpha, \alpha^*)$ and the free carrier number $N$.  Typically, the following effects are relevant:

\begin{enumerate}
	\item Cavity-waveguide coupling.  This gives rise to a linear loss $\kappa$ in the cavity field.
	\item Linear and two-photon absorption.  The former dominates for near-bandgap operation of direct-gap semiconductors; the latter for indirect-gap systems.  Gives rise to a linear loss term $\eta$ and a quadratic loss term $\beta$.  Both act as source terms for the carrier number.
	\item Free-carrier dispersion / absorption.  The cavity detuning shifts as a function of the carrier number: $\Delta \rightarrow \Delta + \delta_c N$.  If $\delta_c = \delta_1 - i\,\delta_2$ is complex, this accounts for free-carrier absorption as well.
	\item Carrier decay.  Typically due to recombination at surface sites or diffusion out of the cavity.  This gives rise to a linear loss term $\gamma$ for $N$.
\end{enumerate}

In this text, I ignore the following effects:

\begin{enumerate}
	\item Excitons, which tend to be the dominant effect only at low temperatures or in exotic materials.
	\item Thermo-optic effect.  Temperature changes much more slowly than the photon or carrier number, so does not typically play a role in the fast dynamics of the device.  It may, however, lead to stability issues, which are not the focus of this paper \cite{VanVaerenbergh2012, JohnsonThesis, Johnson2006}.
	\item Optomechanical effects, which are negligible unless a cavity has been specifically engineered to probe them.
\end{enumerate}

Under these assumptions, the device can be modeled as an open quantum system that couples to a Markovian bath; see generally \cite{Gardiner1985, WallsMilburn, GardinerBook}.  The full quantum theory is quite involved and is discussed earlier in the thesis.  In short, starting from a quantum model with a bosonic photon mode and many fermionic carrier modes, one can construct a generalized Wigner function in terms of a set of bosonized operators and derive a Fokker-Planck equation for this function using the truncated Wigner method \cite{Santori2014, Lugiato1978}.  This can be recast as a set of stochastic differential equations (SDEs) which sample from the Wigner function as a probability distribution.  Assuming that dephasing and thermalization are much faster than the photon or carrier lifetimes, one obtains the following stochastic equations of motion (Eqs.~(\ref{eq:05b-sde-sc1-b}-\ref{eq:05b-sde-sc2-b})):
\begin{eqnarray}
    \d\alpha & = & \left[-\frac{\kappa+\eta}{2} - (\beta+i\chi)\alpha^*\alpha - i(\Delta + N \delta_c)\right]\alpha\,\d t - \sqrt{\kappa}\, \d\binA \nonumber \\
    & & + \underbrace{\left[-\sqrt{\eta}\, \d\beta_\eta - 2\sqrt{\beta}\alpha^*\d\beta_\beta - \sqrt{2N \delta_2}\, \d\beta_{fca}\right]}_{\d\xi_\alpha} \label{eq:07b-eom1} \\
    \d N & = & \left[\eta\,\alpha^*\alpha + \beta(\alpha^*\alpha)^2 - \gamma N\right]\d t \nonumber \\
    & & + \underbrace{\left[\sqrt{\eta}(\alpha^*\d\beta_\eta + \alpha\,\d\beta_\eta^*) + \sqrt{\beta}\bigl((\alpha^*)^2 \d\beta_\beta + \alpha^2 (\d\beta_\beta)^*\bigr) + \sqrt{\gamma N} \d w_\gamma\right]}_{\d\xi_N} \label{eq:07b-eom2}
\end{eqnarray}
and the output optical field is:
\beq
	\d\boutA = -\sqrt{\kappa}\,\alpha\,\d t + \d\binA
\eeq
In these equations, $\d\binA$ is a complex Wiener process representing the input field, which for vacuum input has the It\^{o} rule $\d\binA \d\binA^* = \d t/2$.  The processes $\d\beta_\eta$, $\d\beta_\beta$ and $\d\beta_{fca}$ correspond to linear, two-photon and free-carrier absorption respectively, and also have vacuum statistics.  The $\d w_\gamma$ is a real Wiener process satisfying $\d w_\gamma^2 = \d t$, giving the Poisson statistics of carrier decay.  The real and imaginary parts of $\delta_c$ are $\delta_c = \delta_1 - i\delta_2$.  Typical values for the parameters in (\ref{eq:07b-eom1}-\ref{eq:07b-eom2}) are given in Table \ref{tab:07b-t1}.

These equations resemble the coupled-mode equations used to analyze semiconductor microcavities elsewhere in the literature \cite{Malaguti2011, Malaguti2013, Chen2012}.  Unlike the equations used elsewhere, (\ref{eq:07b-eom1}-\ref{eq:07b-eom2}) include quantum-noise terms.  As a result, these equations allow us to model the quantum behavior of devices previously only discussed classically, and study the fundamental quantum limits to device performance.

We can analyze optical bistability and self-oscillation by linearizing these equations of motion about their equilibrium point.  Defining the doubled-up vector $\bar{x} = (\delta\alpha,\delta\alpha^*,\delta N)$, the equations of motion take the following form:
\begin{align}
    & \underbrace{\d\!\begin{bmatrix} \delta\alpha \\ \delta\alpha^* \\ \delta N \end{bmatrix}}_{\d\bar{x}} \!=\!
    \underbrace{\begin{bmatrix} -\frac{\eta+\kappa}{2} - i(\Delta + N\delta_c) - 2(\beta+i\chi)\alpha^*\alpha\!\!\!\!\!\!\!\!\!\!\!\!\!\! & -(\beta+i\chi)\alpha^2 & -i\delta_c\alpha \\ -\bigl((\beta+i\chi)\alpha^2\bigr)^* & \!\!\!\!\!\!\!\!\!\!\!\!\!\!\bigl(-\frac{\eta+\kappa}{2} - i(\Delta + N\delta_c) - 2(\beta+i\chi)\alpha^*\alpha\bigr)^*\!\!\!\!\! & (-i\delta_c\alpha)^* \\ (\eta + 2\beta\alpha^*\alpha)\alpha^* & (\eta + 2\beta\alpha^*\alpha)\alpha & -\gamma \end{bmatrix}\ \!\!\!\!\begin{bmatrix} \delta\alpha \\ \delta\alpha^* \\ \delta N \end{bmatrix} \d t}_{\bar{A} \bar{x}\,\d t} \nonumber \\
    & + \underbrace{\begin{bmatrix} -\sqrt{\kappa} & 0 \\ 0 & -\sqrt{\kappa} \\ 0 & 0 \end{bmatrix} \!\!\begin{bmatrix} \d\binA \\ \d\binA^* \end{bmatrix}}_{\bar{B}\,\d\bbinA}
    + \underbrace{\begin{bmatrix}
    	-\sqrt{\eta}\,\d\beta_\eta - 2\sqrt{\beta}\alpha^*\d\beta_\beta - \sqrt{2N \delta_2}\, \d\beta_{fca} \\
    	-\sqrt{\eta}\,\d\beta_\eta^* - 2\sqrt{\beta}\alpha\, \d\beta_\beta^* - \sqrt{2N \delta_2}\, \d\beta_{fca}^* \\
		\sqrt{\eta}(\alpha^*\d\beta_\eta + \alpha\,\d\beta_\eta^*) + \sqrt{\beta}((\alpha^*)^2 \d\beta_\beta + \alpha^2 (\d\beta_\beta)^*) + \sqrt{\gamma N} \d w_N \end{bmatrix}}_{\bar{F} \d w} \label{eq:07b-linabcd}
\end{align}
Likewise, the output can be related to the input and internal state by:
\beq
	\underbrace{\begin{bmatrix} \d\boutA \\ \d\boutA^* \end{bmatrix}}_{\d\bboutA} =
	\underbrace{\begin{bmatrix} \sqrt{\kappa} & 0 & 0 \\ 0 & \sqrt{\kappa} & 0 \end{bmatrix}
		\begin{bmatrix} \delta\alpha \\ \delta\alpha^* \\ \delta N \end{bmatrix} \d t}_{\bar{C}\bar{x}\d t} +
	\underbrace{\begin{bmatrix} 1 & 0 \\ 0 & 1 \end{bmatrix}
    	\begin{bmatrix} \d\binA \\ \d\binA^* \end{bmatrix}}_{\bar{D}\d\bbinA} \label{eq:07b-linabcd2}
\eeq
Together, Eqs.~(\ref{eq:07b-linabcd}-\ref{eq:07b-linabcd2}) may be written formally as:
\bea
	\d\bar{x} & = & \bar{A}\bar{x}\,\d t + \bar{B}\,\d\bbinA + \bar{F}\,\d w \label{eq:abcd-1} \\
	\d\bboutA & = & \bar{C}\bar{x}\,\d t + \bar{D}\,\d\bbinA \label{eq:abcd-2}
\eea
which is the standard form for a linear stochastic input-output system (Sec.~\ref{sec:04-linear}).

Equation (\ref{eq:07b-linabcd}) separates the dynamics into three parts: a deterministic term $\bar{A}\bar{x}\d t$, noise due to quantum fluctuations of the input $\bar{B}\,\d\bbinA$, and additional free-carrier noise $\bar{F}\d w$.  (Here, $\d w$ is a vector Wiener process constructed from the real and imaginary parts of the noise terms $\d\beta_\eta, \d\beta_\beta, \d\beta_{fca}, \d w_\gamma$, and normalized to satisfy the It\^{o} table $\d w_i \d w_j = \delta_{ij}\d t$; the matrix $\bar{F}$ is constructed so that (\ref{eq:07b-linabcd}) is satisfied).

The matrix $\Amat$ has three eigenvalues.  Due to its doubled-up structure, complex eigenvalues must come in conjugate pairs.  Thus, $\Amat$ can either have three real eigenvalues or one real eigenvalue and one complex conjugate pair.  If the equilibrium is stable, all three eigenvalues must have a negative real part.

There are two ways for an equilibrium to go unstable.  First, a negative real eigenvalue can cross zero and turn positive.  Since only a single direction goes unstable, the equilibrium point bifurcates into two stable equilibria.  This is the standard cusp catastrophe of optical bistability in Kerr and cavity QED systems \cite{Agrawal1979}.  The previous chapter discussed it in the context of carrier-based switches and amplifiers.  By calculating the determinant of $\Amat$, we can catch this instability -- for stable equilibrium, $\det \Amat < 0$, but if the equilibrium transitions to unstable, $\det \Amat$ will become positive.

\textit{Self-oscillation} takes place when a conjugate pair of eigenvalues cross the imaginary axis.  In this case, two directions go unstable, so the equilibrium point bifurcates into a ring of steady states, or more often, a limit cycle.  The determinant will remain negative, but the product
\beq
L(\Amat) \equiv \left(\text{tr}(\Amat)^2 - \text{tr}(\Amat^2)\right)\text{tr}(\Amat) - 2\det(\Amat)
\eeq
changes sign at this bifurcation.  To see why, suppose that the matrix $\Amat$ has eigenvalues $\lambda, \mu, \mu^*$.  Then for some transformation $P$,
\beq
P^{-1} \Amat P  = \begin{bmatrix} \lambda & & \\ & \mu & \\ & & \mu^* \end{bmatrix}
\eeq
By the cyclic property of traces and determinants, $L(\Amat) = L(P^{-1} \Amat P)$, and the latter evaluates to:
\beq
L(\Amat) = L(P^{-1} \Amat P) = 4|\lambda + \mu|^2 \text{Re}(\mu)
\eeq
This will change sign from negative to positive when passing through a Hopf bifurcation.

\begin{table}[b!]
\centering
\begin{tabular}{c|l|cc}
Name & Description & GaAs PhC & Si $\mu$-ring \\ \hline
$k$            & $\kappa+\eta$          & $0.42$ ps$^{-1}$      & $0.31$ ns$^{-1}$ \\
$\kappa$       & I/O Coupling           & $k/2$\footnote{All dimensional quantities in this table are scaled to the linear loss $k$.}                 & $k$ \\
$\eta$         & LA                     & $k/2$                 & $0$ \\
$\beta$        & TPA                    & $7.9 \times 10^{-5}k$ & $3.7 \times 10^{-6}k$ \\
$\chi$         & Kerr                   & $0$\footnote{Negligible, as dispersive effect is dominated by free carriers.}                   & $0$ \\
$\delta$       & FCD                    & $2.7 \times 10^{-3}k$ & $(5.6-0.4i)\times 10^{-4}k$ \\
$\gamma$       & Carrier Decay          & $1.2k$                & $1.0k$ \\ \hline
$\bar{\delta}$ & $\delta_2/\delta_1$    & $0$                   & $0.07$ \\
$\bar{\zeta}$  & $\delta_1/\beta$       & $34$                  & $150$ \\
$\bar{\chi}$   & $\chi/k$               & $0$                   & $0$ \\
$\bar{\gamma}$ & $\gamma/k$             & $1.2$                 & $1.2$ \\
$\bar{\kappa}$ & $\kappa/k$             & $0.5$                 & $1.0$ \\
$\bar{\eta}$   & $\eta/k$               & $0.5$                 & $0$ \\
$\bar{\Delta}$ & $\Delta/k$             & varies                & varies \\ \hline\hline
\end{tabular}
\caption{Cavity parameters.  GaAs PhC: $\hbar\omega = 0.9 E_g$, $\tilde{V} = 0.25$, $Q = 5000$, $\tau_{fc} = 2$ ps; compare \cite{Nozaki2010}.  Si $\mu$-ring: $\lambda = 1.5\mu$ m, $\tilde{V} = 40$, $Q = 4 \times 10^5$, $\tau_{fc} = 3$ ns; see \cite{JohnsonThesis}.  Compare Table~\ref{tab:06b-t1}}
\label{tab:07b-t1}
\end{table}

\subsection{Scaling Laws}
\label{sec:07b-scaling}

\begin{figure}[tbp]
\begin{center}
\includegraphics[width=0.66\textwidth]{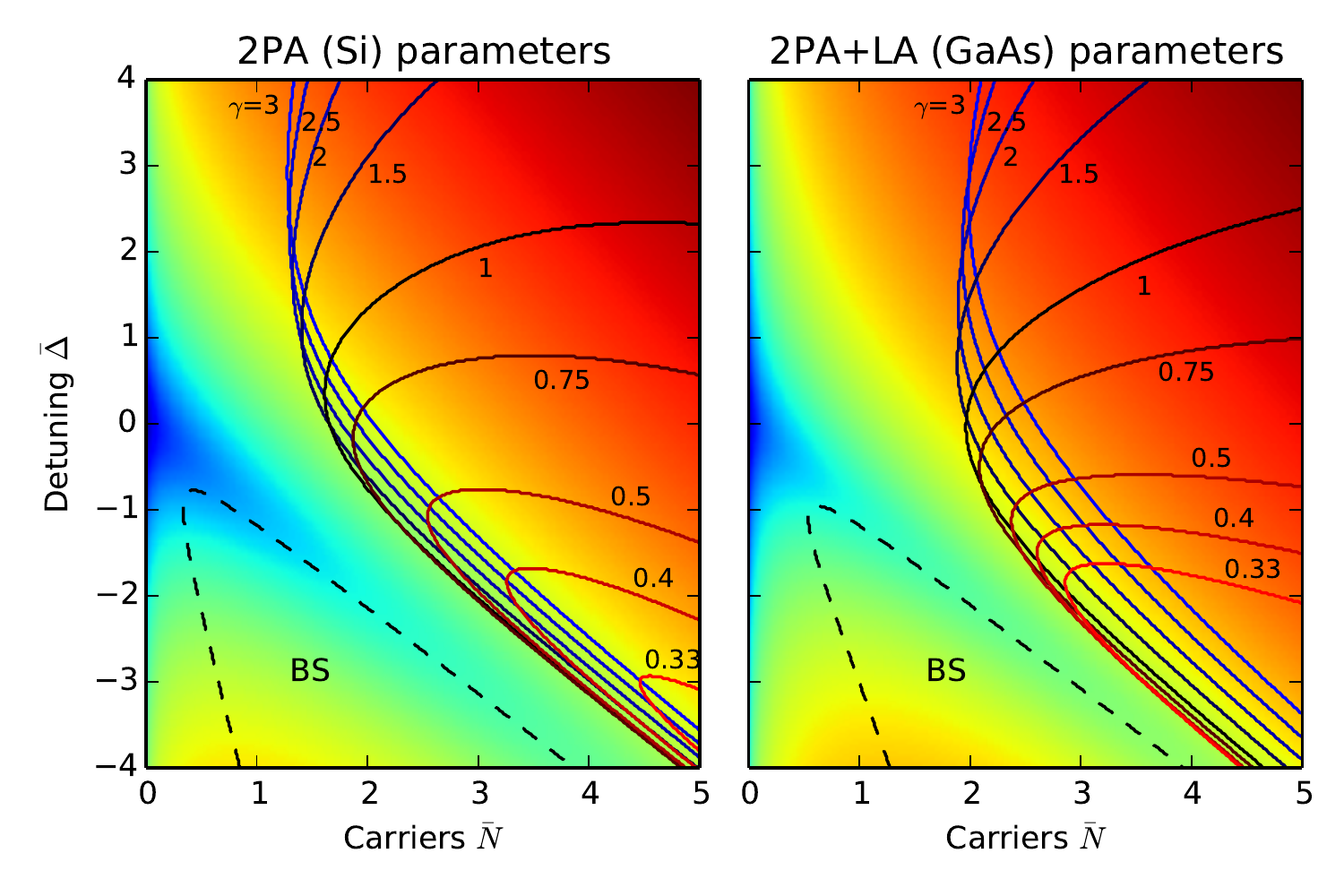}
\caption{Oscillation region as a function of cavity parameters.  Two materials are shown: Si at 1.5 $\mu$m (left) and GaAs near the band edge (right).  Oscillations occur to the right of the solid curves.  Curves represent different values of $\bar{\gamma}$, from 0.33 to 3.0.  Optical bistability occurs in the dashed region.  Color represents the steady-state input power.}
\label{fig:07b-f1}
\end{center}
\end{figure}

Equations (\ref{eq:07b-eom1}-\ref{eq:07b-eom2}), and the resulting matrix $\Amat$, have 8 free parameters.  That's a lot.  Naively, searching for oscillating conditions would appear difficult because of all the parameters one must consider.  However, several scaling laws let us reduce this to 6 ``normalized'' parameters, of which 3 are material constants.

Start with equations of motion (\ref{eq:07b-eom1}-\ref{eq:07b-eom2}).  Let $k = \kappa + \eta$ be the total cavity linear loss.  Scale time, the electric field, the input field, and the carrier number as follows:
\beq
	t \rightarrow \frac{\bar{t}}{k},\ \ \
	\alpha \rightarrow \frac{\bar{\alpha}}{\sqrt{\beta/k}},\ \ \
	\binA \rightarrow \frac{\bar{\beta}_{\rm in}}{\sqrt{\beta/k^2}},\ \ \
	N \rightarrow \frac{\bar{N}}{\delta_1/k} \nonumber
\eeq
Intuitively, time $\bar{t}$ is scaled so that the cavity photon lifetime is one.  The carrier number is scaled so that $\bar{N} = 1$ shifts the cavity by one linewidth.  The intracavity field $\bar{\alpha}$ and input field $\bar{\beta}_{\rm in}$ are scaled to the two-photon absorption: $|\bar{\alpha}| = 1$ means that the single- and two-photon loss processes are equally strong.

The reduced equations take the following form:
\begin{eqnarray}
\d\bar{\alpha} & = & \left[-\left(1/2+\bar{\delta}\bar{N}\right) - \bar{\alpha}^*\bar{\alpha} - i\left(\bar{\Delta}+\bar{N}\right)\right]\bar{\alpha}\,\d\bar{t} \nonumber\\
& &- \sqrt{\bar{\kappa}}\bar{\beta}_{\rm in} \d\bar{t} + \frac{\sqrt{\beta} F_\alpha}{k} \d\bar{w} \\
\d\bar{N} & = & \left[\bar{\eta}\bar{\zeta} (\bar{\alpha}^*\bar{\alpha}) + \bar{\zeta} (\bar{\alpha}^*\bar{\alpha})^2 - \bar{\gamma} \bar{N}\right]\d\bar{t} + \frac{\delta_1 F_N}{k^{3/2}} \d\bar{w}
\end{eqnarray}
In the absence of noise, these equations have 6 independent parameters:

\begin{align}
	\bar{\delta} = \frac{\delta_{2}}{\delta_1},\ \
	\bar{\zeta} = \frac{\delta_1}{\beta},\ \
	\bar{\chi} = \frac{\chi}{\beta} \quad
	& \biggr\} \quad \begin{array}{c} {\rm Material} \\ {\rm Properties} \end{array} \nonumber \\
	\bar{\gamma} = \frac{\gamma}{k},\ \
	\bar{\kappa} = 1-\bar{\eta} = \frac{\kappa}{k} \quad
	& \biggr\} \quad \begin{array}{c} {\rm Cavity} \\ {\rm Design} \end{array} \nonumber \\
	\bar{\Delta} = \frac{\Delta}{k} \quad
	& \biggr\} \quad {\rm Tunable}
\end{align}
where $k = \kappa + \eta$ and $\delta_c = \delta_1 - i\delta_2$.

\begin{figure}[tbp]
\begin{center}
\includegraphics[width=0.66\textwidth]{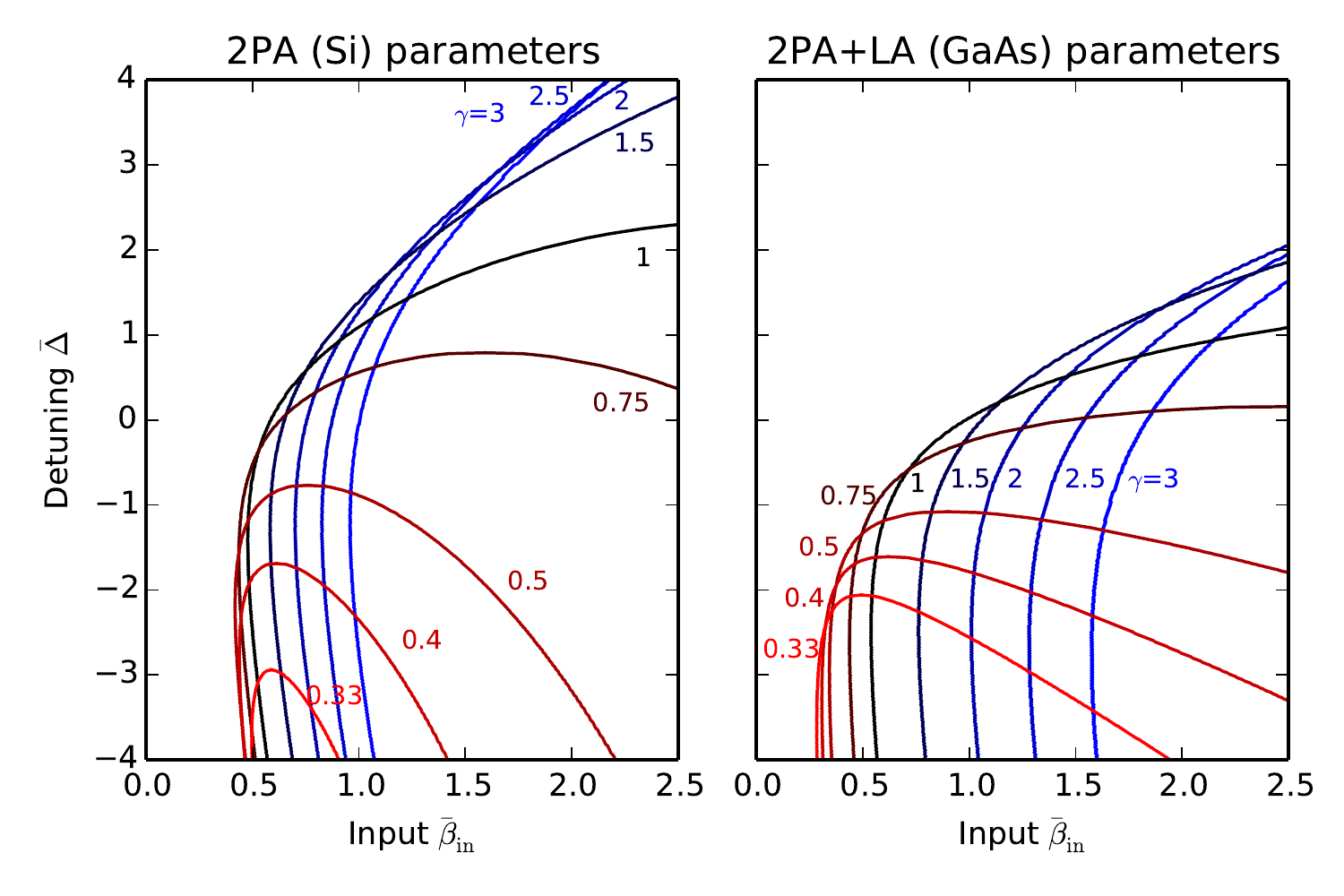}
\caption{Oscillation region as a function of cavity parameters.  Here, the x-axis is normalized input field rather than normalized $N$.}
\label{fig:07b-f2}
\end{center}
\end{figure}

Once a material and laser wavelength are picked, only three parameters can be varied.  The relative linear absorption $\bar{\eta} = 1 - \bar{\kappa}$ typically cannot vary much -- in a linear-absorption cavity it should be $O(1)$ to maximize the nonlinearity, and in TPA materials like silicon it is zero.  The ratio of optical to free-carrier lifetimes, $\bar{\gamma}$, can vary by several orders of magnitude, depending on the cavity geometry and $Q$.  For instance, it is easy to make low-$Q$ cavities with a very small $\bar{\gamma}$.  State-of-the-art micro-rings have $Q \sim 10^6$ and $\tau_c \sim $ns and consequently $\gamma/k \sim 1$.  Coincidentally, photonic crystals tend to have a similar ratio, though the carrier decay mechanism (\d iffusion) is different.  It is also possible to make large cavities with very high $Q$ and large $\bar{\gamma}$.

Obviously, both the input power and detuning can also be varied.  For a given material, these quantities exhaust the parameter space.  By plotting the self-oscillating regions as a function of $\bar{\Delta}$ and $\bar{N}$ (a function of the input), for reasonable values of $\bar{\gamma}$, we are essentially plotting the entire parameter space.  As shown in Figure \ref{fig:07b-f1}, in a large fraction of the parameter space, the cavity should self-oscillate.

Figure \ref{fig:07b-f2} shows the self-pulsing region as a function of input field and detuning.  This is generally similar to Figure \ref{fig:07b-f1}, although the low-$\gamma$ regions appear more accessible because, although the internal carrier number is high, the carriers are long-lived and the cavity requires less optical power.  However, these cavities are complicated by optical bistability (which occurs in the same region), and the slow response time is generally not desirable.  The most desirable conditions seem to occur when the photon and carrier lifetimes are comparable, and the cavity is driven with a slightly detuned pump.

\begin{figure}[tbp]
\begin{center}
\includegraphics[width=0.66\textwidth]{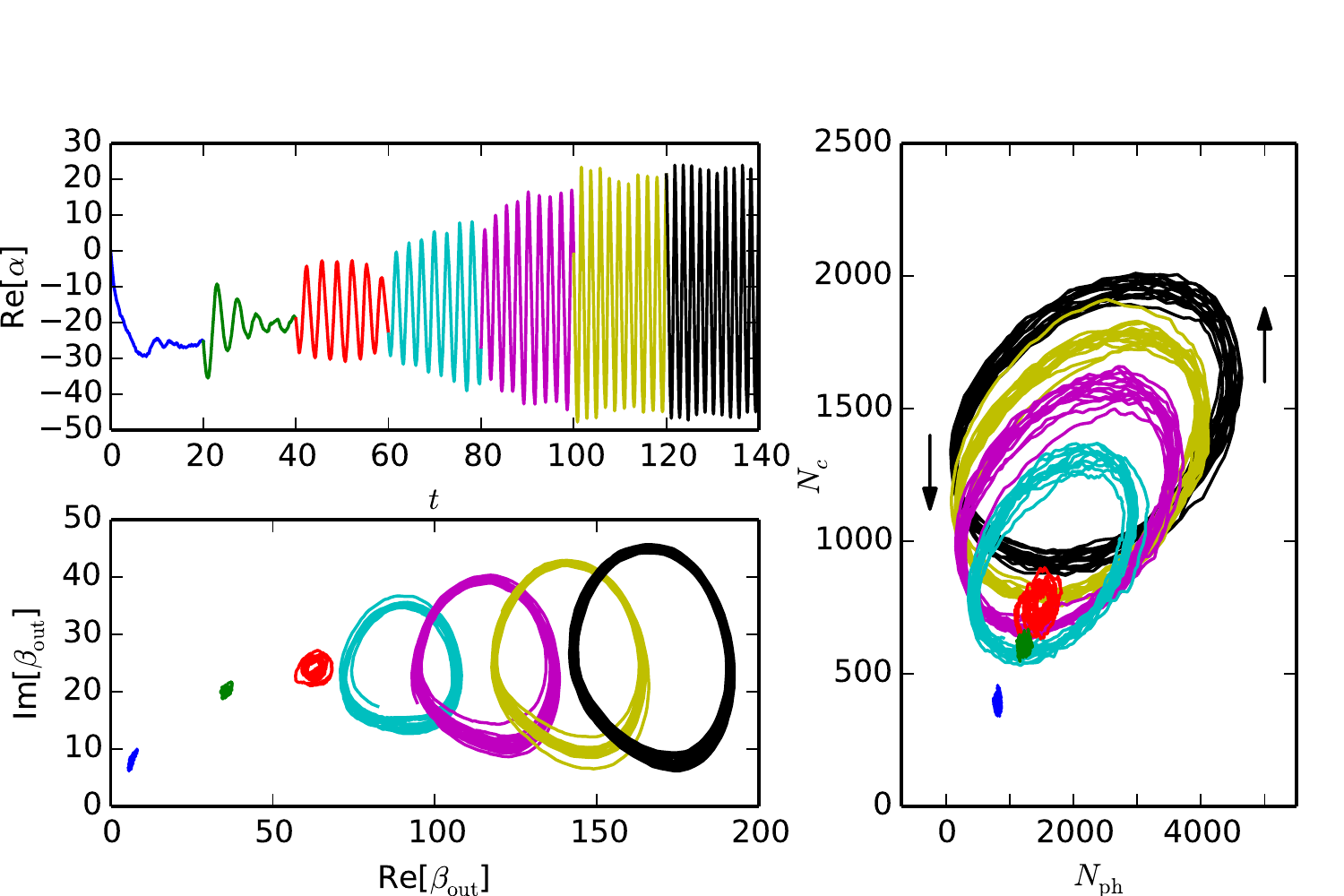}
\caption{Top: time trace of Re[$\alpha(t)$] as the input field is stepped from $\binA = 25$ through $175$.  Bottom: Output field quadratures at these input powers.  Right: Oscillation between photons and carriers.}
\label{fig:07b-f3}
\end{center}
\end{figure}

\section{Semiclassical Simulations}
\label{sec:07b-sims}

Quantum simulations (in the semiclassical Wigner picture) add noise to this model.  For concreteness, in this section and the sections that follow, we consider a GaAs photonic-crystal cavity with parameters given in Table \ref{tab:07b-t1}; however, our results are applicable to a range of devices.  Quantities with units of time or inverse time ($t$, $\Delta$, etc.) will be normalized to the cavity lifetime $1/k$.

Figure \ref{fig:07b-f3} shows simulations for a detuning $\Delta = -0.8$.  The input field is stepped from $\binA = 25$ (blue) to $175$ (black) in increments of $25$.  The top plot shows a typical time trace.  Oscillations clearly set in at around $\binA = 75$.  In addition to the amplitude, the oscillation frequency also increases with pump power.

The right panel of Figure \ref{fig:07b-f3} plots internal photon number (horizontal) against carrier number (vertical).  This provides a qualitative picture of the oscillations: when the photon number is high, more photons are absorbed and the free carrier number increases.  Eventually the carrier number becomes so high that the cavity shifts off-resonance, reducing the cavity's effective driving strength and consequently the photon number.  Once the photon number falls, the carrier number falls because fewer photons are being absorbed, but eventually this brings the cavity back on resonance, increasing the photon number and repeating the cycle.

\begin{figure}[tbp]
\begin{center}
\includegraphics[width=0.66\textwidth]{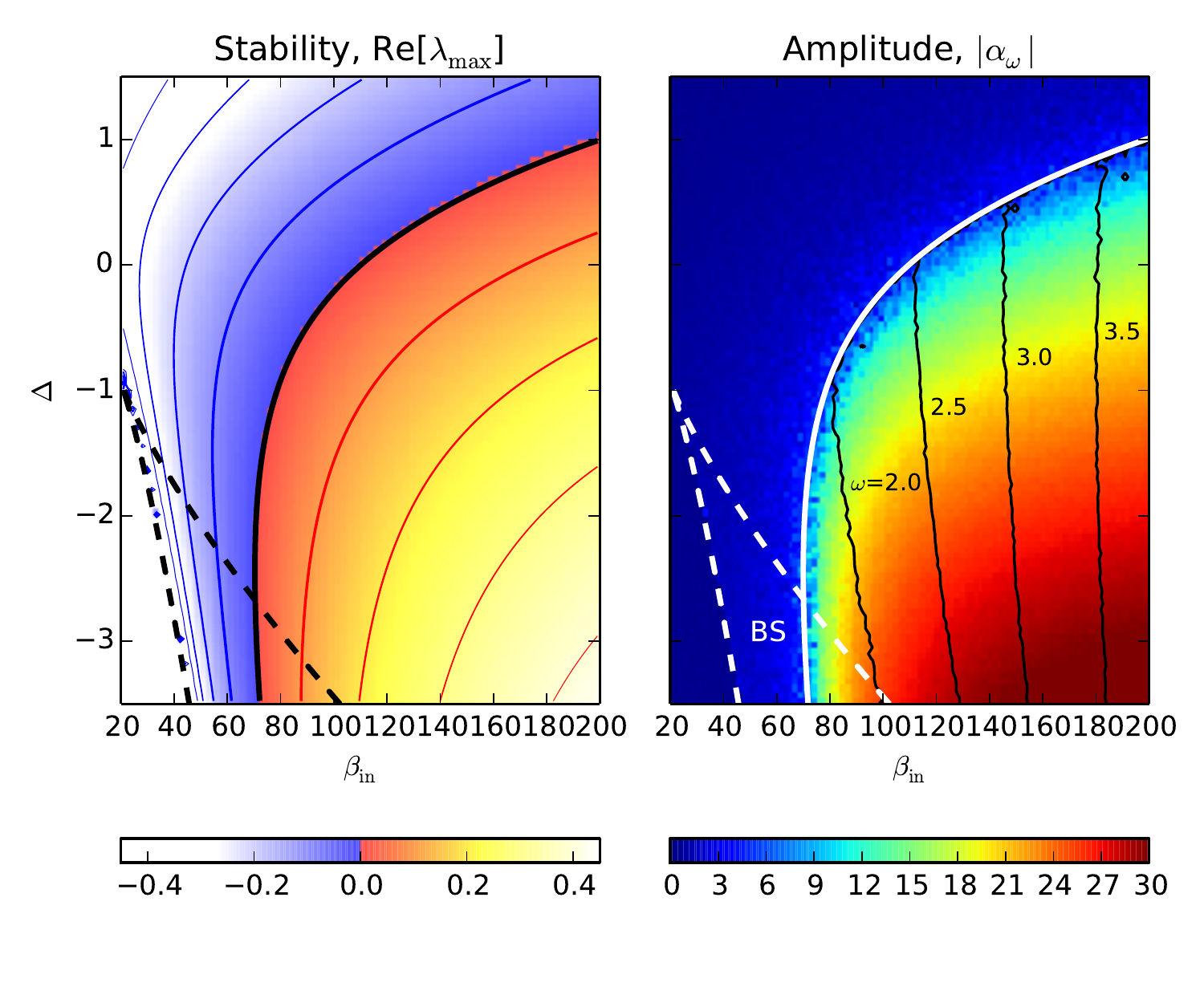}
\caption{Left: Stability of equilibrium point, measured by the real part of the largest eigenvalue of $A$.  Right: Amplitude of limit cycle, with contours designating the limit cycle frequency.}
\label{fig:07b-f4}
\end{center}
\end{figure}

To get a more general picture, consider all possible pump powers and detunings for this system.  If a limit cycle forms, we are interested in its amplitude and frequency.  The amplitude should be large, so that a significant fraction of the pump is converted to photons at the limit-cycle frequency.  The frequency should be large enough that the pump and limit cycle fields can be easily demultiplexed with a cavity.  Figure \ref{fig:07b-f4} plots both of these figures of merit.  As expected, the amplitude at $\omega$ only becomes nonzero in the unstable region where $\text{Re}[\lambda_{\rm max}] > 0$.  The frequency also grows with pump power, starting at $\omega \approx 1.7$ and growing to $\omega \approx 4$; this is probably a nonlinear effect of the strong pumping.

Two other figures of merit are the limit cycle ``efficiency'' and the gain.  Efficiency is defined in terms of the output and absorbed power:
\beq
    \eta \equiv \frac{P_{\omega,\rm out}}{P_{\omega,\rm out} + P_{\rm abs}}
\eeq
Efficiency is defined this way rather than output over input because much of the input power is not consumed by the device; it is just a constant bias that can be recycled.  If there finite conversion to $\omega$ and no absorption, we say the efficiency is 1; if no conversion, it is obviously zero.  The left panel of Figure~\ref{fig:07b-f5} plots efficiency as a function of detuning and input field.  While not close to 100\%, the efficiency is not too small, either -- peaking at around 20\%.

If we drive the device with a sinusoidal field whose frequency is close to the limit-cycle frequency, that field should be amplified.  In this way, the free-carrier cavity acts as a phase-insensitive amplifier.  The amplitude gain $G(\omega) = \beta_{\omega,\rm out}/\beta_{\omega,\rm in}$ is plotted at $\omega = 1.7$ in the right panel of Figure~\ref{fig:07b-f5}.

\begin{figure}[tbp]
\begin{center}
\includegraphics[width=0.66\textwidth]{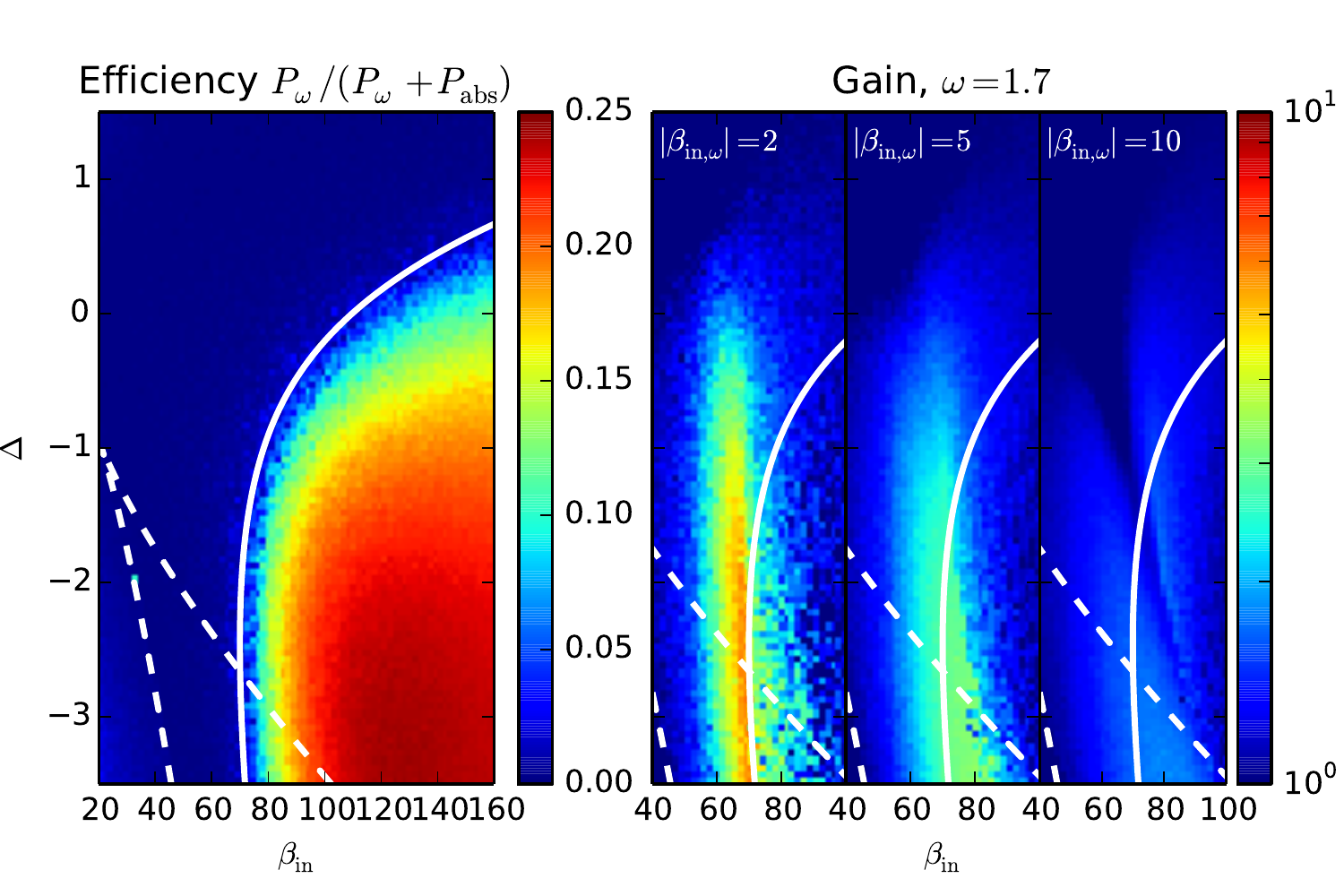}
\caption{Left: Photon conversion efficiency, the ratio of limit cycle photons emitted to photons absorbed.  Right: Amplitude gain $\bout{\omega}/\bin{\omega}$ at $\omega = 1.7$ for different values of seed amplitude $\bin{\omega} = 2, 5, 10$.}
\label{fig:07b-f5}
\end{center}
\end{figure}

\section{Below Threshold: Linear Amplification}
\label{sec:07b-below}

Below the Hopf bifurcation, a complex pair of eigenvalues approach the imaginary axis.  The corresponding eigenvectors span a plane in phase space; since motion tangent to this plane is only marginally stable, perturbations will be strongly amplified.  Since this plane is two-dimensional, we expect linear, phase-insensitive amplification of both quadratures of the input field \cite{KwonThesis, Wiesenfeld1986}.

For any mesoscopic linear amplifier, an important question to ask is: how much noise does the amplifier have?  Quantum mechanics sets a strict bound on the noise of a quantum linear amplifier \cite{Caves1982}, and this bound is realized with the non-degenerate OPO \cite{Yamamoto1990}.  Since free carriers are excited incoherently, one expects an amplifier driven by carriers to be noisier than a quantum-limited amplifier; however, if the difference is not too large, the free-carrier amplifier may still be preferred because of material, power, or footprint considerations.

\subsection{Nondegenerate OPO}
\label{sec:07b-ndopo}

Although this chapter is about free-carrier effects, it is helpful to introduce the non-degenerate OPO here as a ``benchmark'' system because it is a well-studied system that saturates the Caves bound.  It can be modeled as a quantum input-output system \cite{Gough2009b, Gardiner1985} with three fields: signal $a$, idler $b$ and pump $c$ (Sec.~\ref{sec:02-opo}).  The internal Hamiltonian is
\beq
	H = \Delta_a a^\dagger a + \Delta_b b^\dagger b + \Delta_c c^\dagger c + \frac{\epsilon^*a b c^\dagger - \epsilon a^\dagger b^\dagger c}{2i}
\eeq
and input-output couplings
\bea
	L_1 & = & \sqrt{\kappa_a} a \nonumber \\
	L_2 & = & \sqrt{\kappa_b} b \nonumber \\
	L_3 & = & \sqrt{\kappa_c} c
\eea
Following the Wigner method of \cite{Santori2014}, one can convert the master equation into a PDE for the Wigner function, and truncating higher-order terms, this PDE becomes a Fokker-Planck equation.  This can then be converted into an SDE, and solving the SDE produces trajectories that sample from the Wigner function \cite{WallsMilburn}.  Adiabatically eliminating the pump field and setting $\Delta_a = -\Delta_b \equiv \Delta$, $\kappa_a = \kappa_b \equiv \kappa$ (symmetric doubly-resonant cavity), one obtains the following equations of motion:
\begin{eqnarray}
	\d\alpha_1 & = & \left[\left(-i\Delta - \frac{\kappa + \beta\,\alpha_2^*\alpha_2}{2}\right) \alpha_1 + \epsilon\, \alpha_2^*\right]\,\d t - \sqrt{\kappa}\, \d\bin{1} - \sqrt{\beta}\,\alpha_2^* \d\bin{3} \\
	\d\alpha_2 & = & \left[\left(i\Delta - \frac{\kappa + \beta\,\alpha_1^* \alpha_1}{2}\right) \alpha_2 + \epsilon\, \alpha_1^*\right]\,\d t - \sqrt{\kappa}\, \d\bin{2} - \sqrt{\beta}\,\alpha_1^* \d\bin{3} \\
	\d\bout{1} & = & \sqrt{\kappa}\,\alpha_1 \d t + \d\bin{1} \\
	\d\bout{2} & = & \sqrt{\kappa}\,\alpha_2 \d t + \d\bin{2}
\end{eqnarray}
where $\beta = \epsilon^*\epsilon/\kappa$ is the intrinsic coupling strength of the OPO.

Here, $\alpha_1$ and $\alpha_2$ are the signal and idler, which have the same lifetime but opposite detunings.  The pump does not resonate.  These equations are symmetric with respect to $\alpha_1 \leftrightarrow \alpha_2^*$.  Because of the symmetry, the dynamics can be decomposed into a ``symmetric'' mode $\alpha_+ = (\alpha_1 + \alpha_2^*)/2$ and an ``antisymmetric'' mode $\alpha_- = (\alpha_1 - \alpha_2^*)/2$ (and likewise for the $\d\beta_\pm$).  In addition, define $\d w_1, \d w_2$ as quadratures of the pump noise, $\d\bin{3} = (\d w_1 + i\,\d w_2)/2$.  The equations of motion become:
\bea
    \d\alpha_\pm & = & \left[(-i\Delta -\kappa/2 \pm \epsilon) \alpha_\pm - \frac{\beta}{2} \bigl(\alpha_\pm^2 - \alpha_\mp^2)\alpha_\pm^*\right]\d t \nonumber \\
    & & - \sqrt{\kappa}\,\d\bin{\pm} \mp \frac{1}{2}\sqrt{\beta}\left(\alpha_\pm \d w_1 - i \alpha_\mp \d w_2\right)
\eea
The symmetric mode $\alpha_+$ has gain (a $+\epsilon$ term) while the antisymmetric mode has additional loss.  As a result, at near- or above-threshold pumping, $\alpha_+$ can become very large, but $\alpha_-$ always stays near zero.  In the weakly coupled case ($\beta \ll 1$), we can throw away the terms that couple $\alpha_+$ and $\alpha_-$ in the equation above, and combine the noise terms, giving:
\beq
    \d\alpha_+ =  \left[(-i\Delta -\kappa/2 + \epsilon) \alpha_+ - \frac{\beta}{2} \left|\alpha_+\right|^2 \alpha_+\right]\d t - \sqrt{\kappa/2} \d\beta_+ - \frac{1}{2}\sqrt{\beta}\alpha_+ \d w_1 \label{eq:daplus}
\eeq
Linearizing about the fixed point $\alpha_1 = \alpha_2 = 0$, and transforming into the frequency domain, we arrive at the input-output relation:
\beq
	\bout{1}(\omega) = \underbrace{\frac{\left|(\omega - \Delta) + i\kappa/2\right|^2 + (\epsilon/2)^2}{\left(-(\omega - \Delta) + i\kappa/2\right)^2 + (\epsilon/2)^2}}_{e^{i\phi} \cosh\eta} \bin{1}(\omega) + \underbrace{\frac{2(\kappa/2)(\epsilon/2)}{\left(-(\omega - \Delta) + i\kappa/2\right)^2 + (\epsilon/2)^2}}_{e^{i\psi} \sinh\eta} \bin{2}^*(-\omega)
\eeq
For phase-insensitive amplification, the gain $G$ an noise $S$ at frequency $\omega$ may be defined as:
\bea
	G(\omega) & \equiv & \left|\frac{\bout{1}(\omega)}{\bin{1}(\omega)}\right| \label{eq:g} \\
	S(\omega) & \equiv & \sqrt{2P(\omega)},\ \ P(\omega) = \frac{\langle \bout{1}(\omega)^*\bout{1}(\omega')\rangle}{\delta(\omega-\omega')} \label{eq:s}
\eea
In terms of $\eta$, they are:
\beq
	G(\omega) = \cosh\eta,\ \ S(\omega) = \sqrt{2\cosh^2\eta - 1} \label{eq:gw}
\eeq
Note that this $S(\omega)$ is different from the squeezing spectrum of \cite{WallsMilburn, Gough2009c}; rather, it is a measure of the electromagnetic energy at frequency $\omega$.  The squeezing spectrum, by contrast, is a power spectrum of a homodyne measurement.

From (\ref{eq:gw}) one sees that the non-degenerate OPO saturates the Caves bound for phase-insensitive amplifiers \cite{Caves1982}:
\beq
S(\omega) \geq \sqrt{2G(\omega)^2 - 1}
\eeq

\begin{figure}[tbp]
\begin{center}
\includegraphics[width=0.66\textwidth]{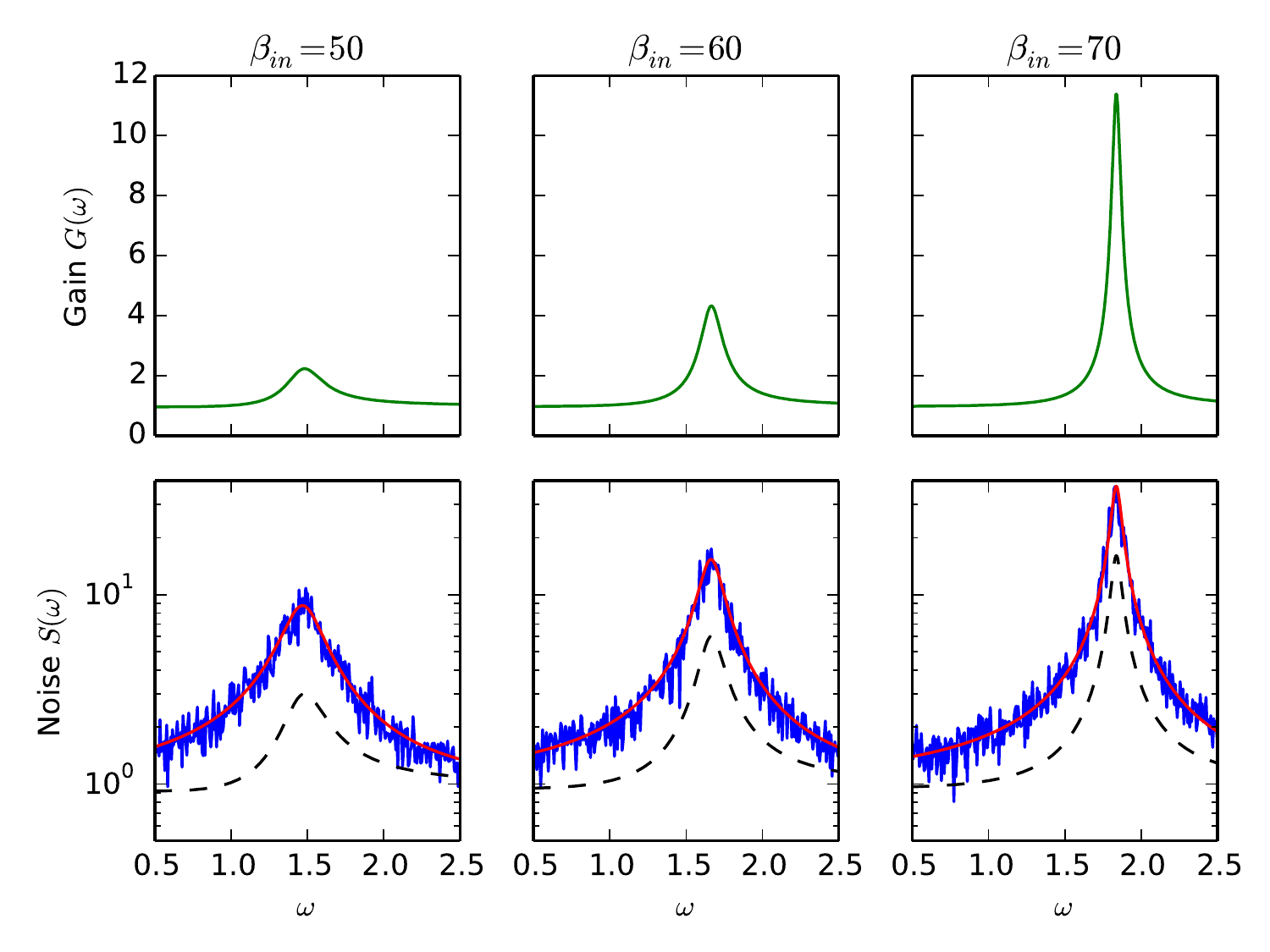}
\caption{Plots of the amplitude gain (top) and noise (bottom) for free-carrier cavity with $\Delta = -1.0$ approaching the Hopf bifurcation.  In the lower graph, the blue line comes from numerical simulation, the red curve is the analytic linearized model, and the black dashed curve is the Caves bound.}
\label{fig:07b-f6}
\end{center}
\end{figure}

\subsection{Free-Carrier Amplifier}

Turning to the free-carrier amplifier, first transform Equations~(\ref{eq:abcd-1}-\ref{eq:abcd-2}) to the frequency domain:
\bea
	-i\omega \bar{x}(\omega) & = & \bar{A}\bar{x}(\omega) + \bar{B}\bbinA(\omega) + \bar{F}w(\omega) \\
	\bboutA(\omega) & = & \bar{C}\bar{x}(\omega) + \bar{D}\bbinA(\omega)
\eea
with state $\bar{x}(\omega) = \bigl(\alpha(\omega), \alpha^*(-\omega), \dbl{N}(\omega)\bigr)$ and input-output field $\bar{\beta}(\omega) = \bigl(\beta(\omega), \beta^*(-\omega)\bigr)$.  This is the standard frequency-domain form for doubled-up variables \cite{Gough2010b}.

Solving for $\bar{x}$, this becomes a linear input-output relation with a transfer function and a noise matrix:
\beq
	\bboutA(\omega) = \underbrace{\left[\bar{D} + \bar{C} \frac{1}{-i\omega - \bar{A}} \bar{B}\right] \bbinA(\omega)}_{\dbl{T}(\omega) \bbinA(\omega)} + \underbrace{\left[\bar{C} \frac{1}{-i\omega - \bar{A}} \bar{F}\right] w(\omega)}_{\dbl{N}(\omega) w(\omega)}
\eeq
Applying the definitions of $G$ and $S$ in Eqs.~(\ref{eq:g}-\ref{eq:s}), we find:
\beq
	G(\omega) = |\dbl{T}(\omega)_{11}|,\ \
	\frac{S(\omega)^2}{2} = \left[\frac{\dbl{T}(\omega)\dbl{T}(\omega)^\dagger}{2} + \dbl{N}(\omega)\dbl{N}(\omega)^\dagger\right]_{11}
\eeq
Unlike the OPO, the free-carrier amplifier does not have a simple expression for $G(\omega)$ or $S(\omega)$.  However, they are straightforward to evaluate numerically, and can be compared to a full nonlinear simulation.

Figure \ref{fig:07b-f6} shows the gain and noise for the cavity studied in Section \ref{sec:07b-sims}, with $\Delta = -1.0$.  Far from the limit-cycle frequency, there is no gain and the output noise matches that of the vacuum.  As the power is increased and the system approaches the Hopf bifurcation, the gain and noise at the resonance obviously diverge.  But the noise always remains a factor of $\sim$2--3 above the Caves bound (in terms of noise power, a factor of $\sim$5 above the bound).  This is due to the incoherent nature of the free-carrier nonlinearity.

\section{Near Threshold: Critical Exponents}
\label{sec:07b-near}

Near the bifurcation point, the system transitions from a stable fixed point to a stable limit cycle.  Dynamical systems exhibit universal behavior near this bifurcation, in the sense that every system with a Hopf bifurcation can be transformed into the same normal form \cite{StrogatzBook, WigginsBook}.  The same is not true when one adds noise and quantum effects.  Two systems with the same semiclassical equations of motion can behave very differently once quantum noise is added.  Nevertheless, all systems will show the same {\it qualitative} behavior near a bifurcation point.

Before discussing the free-carrier oscillations, consider the non-degenerate OPO near threshold.  Below threshold, there is a stable fixed point at $\alpha_+ = \alpha_- = 0$.  Above threshold, there is a limit cycle at:
\beq
	\left|\alpha_+\right| = \sqrt{\frac{2\epsilon-\kappa}{\beta}}
\eeq
Thus, if we smoothly vary the parameter $\epsilon$ near the bifurcation point, $\epsilon = \kappa/2 + \delta\epsilon$, the limit cycle amplitude goes as $\sqrt{\epsilon}$.  This is a universal feature.  However, not all OPOs are equal up to a transformation -- the behavior of the quantum states depends strongly on the value of $\beta$.  For $\beta \ll 1$, dissipation is dominant and the system stays in a classical state with a positive Wigner function.  For $\beta \gg 1$, the Wigner formalism breaks down.  (This is true for OPOs in general.  It is known that in this regime the {\it degenerate} OPO can access ``highly quantum'' states with non-positive Wigner function such as number states and cat states \cite{Wolinsky1988, Mirrahimi2014, Mabuchi2012}.)

\begin{figure}[tbp]
\begin{center}
\includegraphics[width=0.66\textwidth]{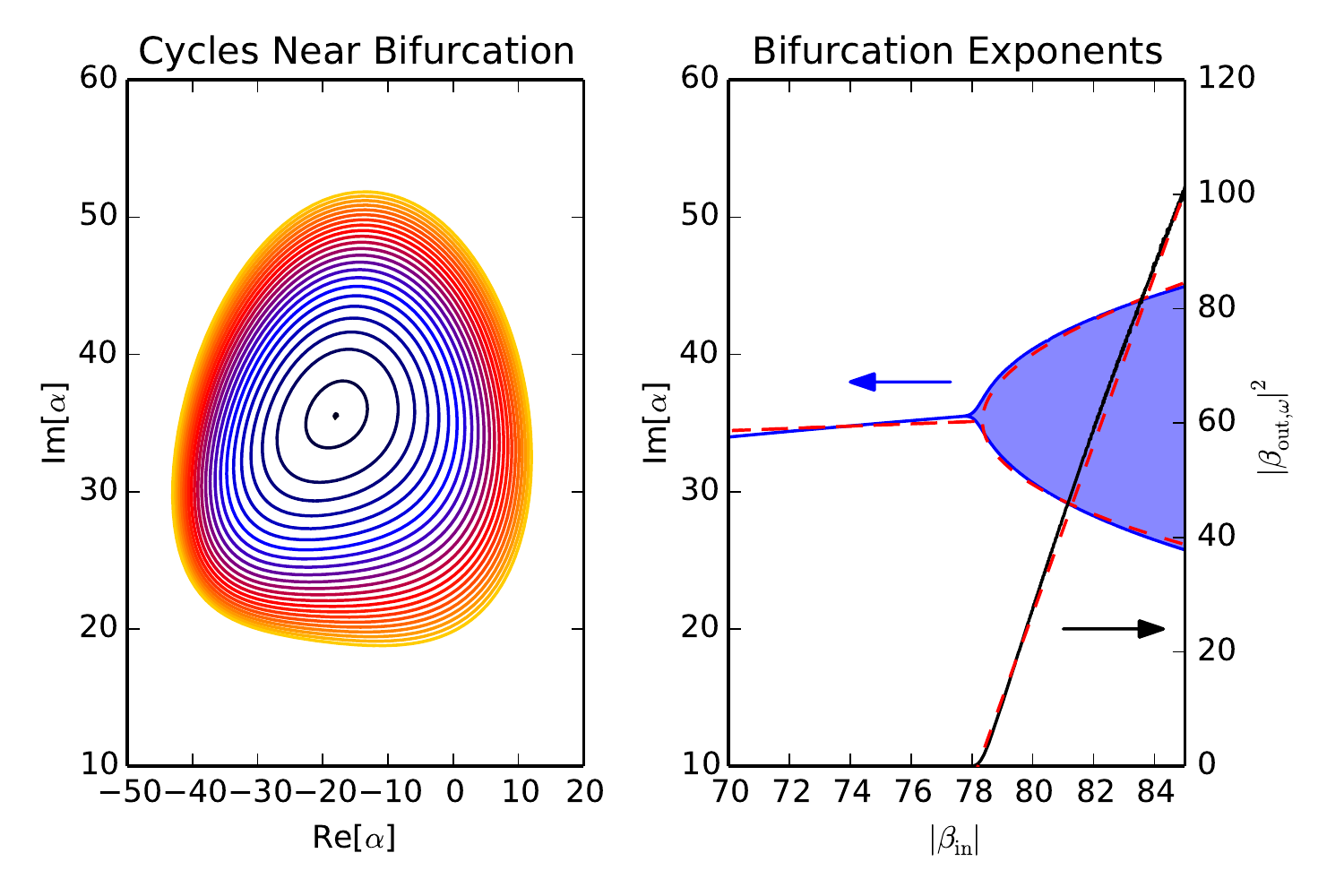}
\caption{Left: Free-carrier limit cycles just above the bifurcation point (noiseless simulation), for evenly spaced $\binA = 78, 79, 80, \ldots$  Right: Size of the limit cycle in terms of $\alpha$ (blue) and $|\boutA(\omega)|^2$ (black), and the critical exponents $\alpha \sim \boutA(\omega) \sim \sqrt{\delta \binA}$}
\label{fig:07b-f7}
\end{center}
\end{figure}

The fixed-point eigenvalues near the bifurcation are: $\lambda = (\epsilon - \kappa/2) \pm i\Delta$, and therefore:
\beq
    \left|\alpha_+\right|	\sim \sqrt{\frac{{\rm Re}[\lambda]}{\beta/2}}\ \ \Leftrightarrow \ \
    	\beta \sim \frac{{\rm Re}[\lambda]}{|\alpha_+|^2} \label{eq:07b-beta}
\eeq
In classical dynamical systems theory, we can freely transform the system variable $\alpha$, so the parameter $\beta$ can be rescaled to 1.  This is part of the process of transforming to the normal coordinate frame.  Classically, $\alpha$ is dimensional and therefore $\beta$ is not universal in any way.  But in quantum mechanics, there {\it is} a universal scale for $\alpha$: the single-photon scale.  Because of this, $\beta$ becomes a universal parameter, and is related to the ``quantumness'' of the bifurcation.

Figure \ref{fig:07b-f7} shows that the free-carrier Hopf bifurcation satisfies the same critical exponent as the non-degenerate OPO: in terms of the input power $\binA$, the average oscillating field goes as $|\alpha| \sim \delta \binA^{1/2}$.  One can calculate the effective $\beta$ for this bifurcation using Eq.~(\ref{eq:07b-beta}): fitting to the figures, it works out to $\beta \sim 0.0002$, well in the semiclassical regime.

\begin{figure}[tbp]
\begin{center}
\includegraphics[width=0.66\textwidth]{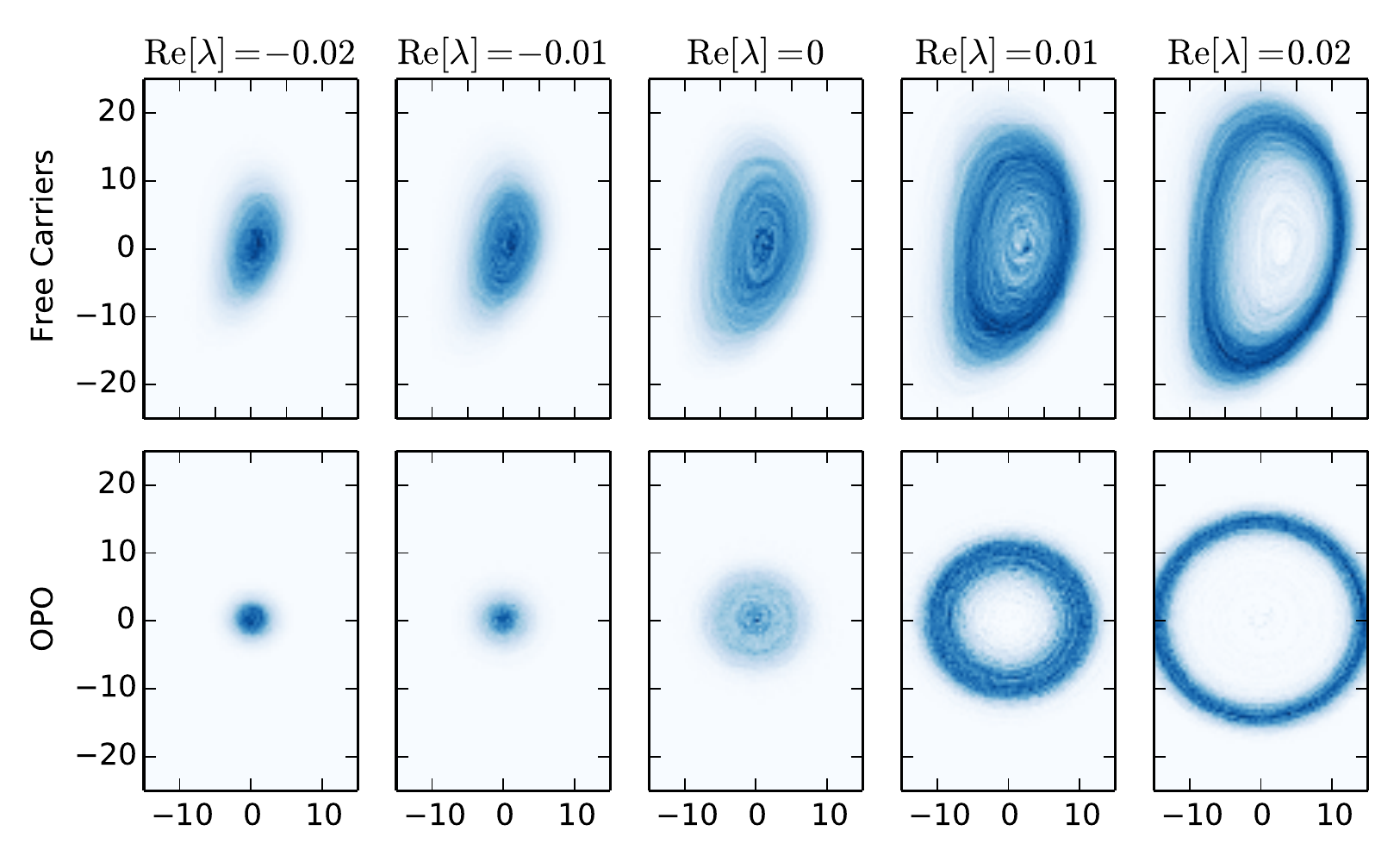}
\caption{Phase plots (axes are Re[$\alpha$], Im[$\alpha$]) of the limit cycles for free carriers ($\Delta = -1.0, a_{\rm in} = 72.5$ through $84.5$) and the non-degenerate OPO ($\beta = 0.0002, \epsilon = 0.48$ through $0.52$).}
\label{fig:07b-f8}
\end{center}
\end{figure}

Even after accounting for $\beta$, the free-carrier and OPO Hopf bifurcations are not equivalent up to a transformation, as they would be in classical bifurcation theory.  Again, the culprit is quantum mechanics: the incoherent process of carrier excitation and decay adds extra quantum noise, making the free-carrier limit cycle ``fuzzier'' than its OPO counterpart.  This is shown in Figure~\ref{fig:07b-f8}.

\section{Above Threshold: Limit Cycle}
\label{sec:07b-above}

Above threshold, we classically expect a limit cycle.  Quantum noise will blur this out to some degree, but sufficiently far above threshold, the cycle should be clear.

Limit cycles are a classic topic in dynamical systems; some key results are reviewed in Appendix~\ref{sec:07b-04b-lc}.  To summarize the important points:  For an $n$-dimensional phase space, there is a function $(\xi, \vec{u}) \rightarrow \mathbb{R}^n$, that maps the limit cycle phase $\xi$ and local perturbations $\vec{u}$ onto a portion of the phase space.  When the perturbations are small compared to the limit cycle, they can be ignored entirely, reducing the dimensionality of the system from $n$ to 1.  This reduced system has the following equation of motion:
\beq
	\d\xi = \omega\,\d t + \sum_i{\rm Re}[B_i(\xi)^* \d\bin{i}]\d t + F(\xi) \d w \label{eq:07b-lcxi}
\eeq
Here, $B_i(\xi)$ is the response to an external perturbation $\d\bin{i}$ and $F(\xi) \d w$ is the intrinsic limit cycle noise.

Any limit-cycle system can be used as a homodyne detector.  To see why, consider a coherent input $\bin{i} = \avg{\beta_i}e^{-i\omega_c t} + \bin{i}^{\rm (vac)}$, where $\omega_c$ is the limit cycle frequency.  Averaging over many cycles, this input changes the limit-cycle phase as follows:
\bea
	\Delta\xi - \omega t & = & \int_0^T{\sum_i{\rm Re}[B_i(\xi)^* \d\bin{i}]} + \int_0^T{F(\xi) \d w} \nonumber \\
	& \sim & N\left(T\sum_i {\rm Re}\left[\mu_{\xi,i} \avg{\beta_i}\right],\ \
		D_\xi T\right) \label{eq:homodyne}
\eea
That is, the phase change has a normal distribution, with mean and variance given by the drift and diffusion constants:
\bea
	\mu_{\xi,i} & = & \avg{B_i(\xi)^* e^{-i\xi}}_\xi \label{eq:mu} \\
	D_\xi & = & \frac{1}{2}\avg{|B_i(\xi)|^2}_\xi + \avg{|F(\xi)|^2}_\xi \label{eq:dxi} \\
	& & \left(\mbox{where}\ \avg{\ldots}_\xi \equiv \frac{1}{2\pi} \int_0^{2\pi} {(\ldots)\d\xi}\right) \nonumber
\eea
The drift term $\mu_{\xi,i}$ governs the {\it response rate} of the limit cycle to an external stimulus (in this case, the field).  The diffusion term $D_\xi$ tells us how quickly the limit-cycle phase diffuses in the absence of a stimulus (assuming coherent inputs).  Both terms show up in the homodyne measurement (\ref{eq:homodyne}).  The standard quantum limit \cite{Yamamoto1990} bounds the accuracy of this measurement: in terms of the $\mu_{\xi,i}$ and $D_\xi$, this gives rise to a {\it drift-diffusion inequality}:
\beq
	D_\xi \geq \frac{1}{4} \sum_i |\mu_{\xi,i}|^2 \label{eq:diffeq}
\eeq
This relation holds for all limit cycles.  One can also derive it from Eqs.~(\ref{eq:mu}-\ref{eq:dxi}) by applying the Schwarz inequality.  Equality holds only for special, ``quantum-limited'' limit cycles where $F(\xi) = 0$ and $B_i(\xi) \sim e^{-i\xi}$.  In the sections below, we compare the performance of the non-degenerate OPO and the free-carrier limit cycle using this metric, and show that the OPO saturates the drift-diffusion inequality, while the free-carrier device does not.

\subsection{Non-degenerate OPO}

Again, it will be important to contrast the results obtained here with the non-degenerate OPO; as we will show, this device can function as a quantum-limited homodyne detector for signal and idler fields.  Because it is quantum-limited, no other limit-cycle device will beat the OPO at this task, just like no other linear amplifier can beat the non-degenerate OPO below threshold.

As we show in Appendix~\ref{sec:07b-04b-lc}, the non-degenerate OPO has a limit cycle with $|\alpha_+| = \sqrt{(2\epsilon-\kappa)/\beta}$ and a phase that evolves as:
\bea
	\d\xi & = & \Delta\,\d t + {\rm Re}\,\left[\frac{-i\sqrt{\kappa}}{\alpha_+}\d\bin{+}\right] \nonumber \\
	& = & \Delta\,\d t + {\rm Re}\,\left[\frac{-i\sqrt{\kappa}}{2\alpha_+}\d\bin{1} + \frac{i\sqrt{\kappa}}{2\alpha_+^*}\d\bin{2}\right]
\eea
so that for signal and idler fields varying as $\beta_1 e^{-i\Delta t}$, $\beta_2 e^{i\Delta t}$, the drift-diffusion terms are:
\bea
	\mu_{\xi,1} & = & -i\frac{\sqrt{\kappa}}{2|\alpha_+|} \label{eq:muopo1} \\
	\mu_{\xi,2} & = & -i\frac{\sqrt{\kappa}}{2|\alpha_+|} \label{eq:muopo2} \\
	D_\xi & = & \frac{\kappa}{8|\alpha_+|^2} \label{eq:dopo}
\eea
It is not difficult to see from (\ref{eq:muopo1}-\ref{eq:dopo}) that the drift-diffusion inequality (\ref{eq:diffeq}) is saturated.  In this limit, the non-degenerate OPO functions as an optimal, quantum-limited homodyne detector.
\begin{figure}[tbp]
\begin{center}
\includegraphics[width=0.6\textwidth]{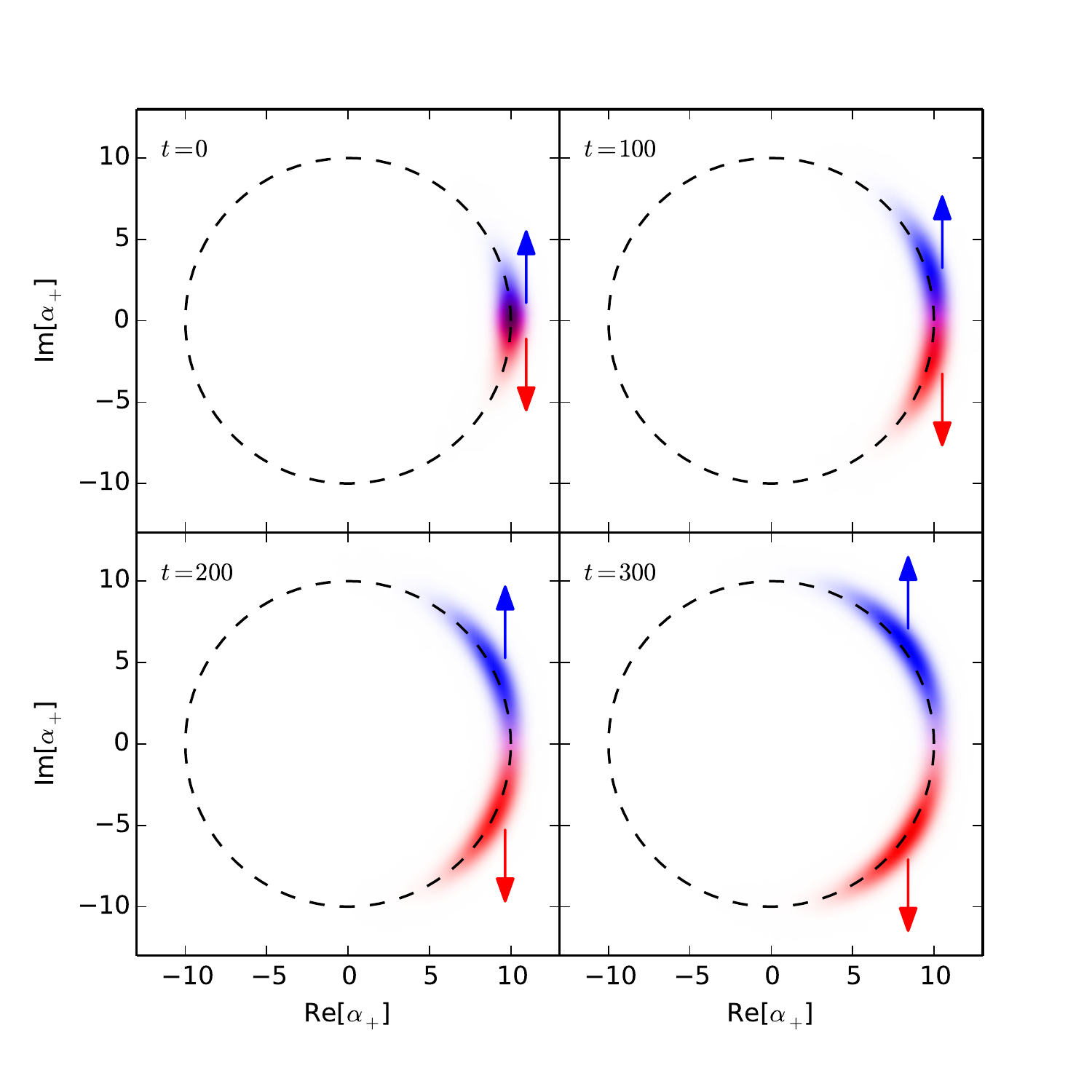}
\caption{Wigner function of the nondegenerate OPO ($\eta = 1.0, \beta = 0.01$) subject to a bias $\beta_1 = 0.15i$ (red) and $-0.15i$ (blue).  The state $\xi(t)$ for $t > 0$, which can be accurately read out with either homodyne or heterodyne detection, effectively encodes a measurement of the $p$-quadrature of the input, ${\rm Im}[\bar{\beta}_1]$.}
\label{fig:07b-f9a}
\end{center}
\end{figure}
This is sketched in Figure \ref{fig:07b-f9a}.  Here, a non-degenerate OPO with $\Delta = 0$ is used to measure the $p$ quadrature of a signal field.  Depending on the sign of the field, the state either drifts to the top or the bottom, and the diffusion incurred is due to the quantum uncertainty of the homodyne measurement.

\subsection{Free-Carrier Cavity}

Since the equations of motion for the free-carrier cavity are more complicated, a simple analytic expression for $\mu_\xi$ and $D_\xi$ does not exist.  However, these can be computed numerically.  Following the results of Section~\ref{sec:07b-below}, it is reasonable to expect diffusion rates 5--10 times faster than for the non-degenerate OPO, the extra diffusion due to incoherent processes involving free carriers.

\begin{figure}[tbp]
\begin{center}
\includegraphics[width=0.66\textwidth]{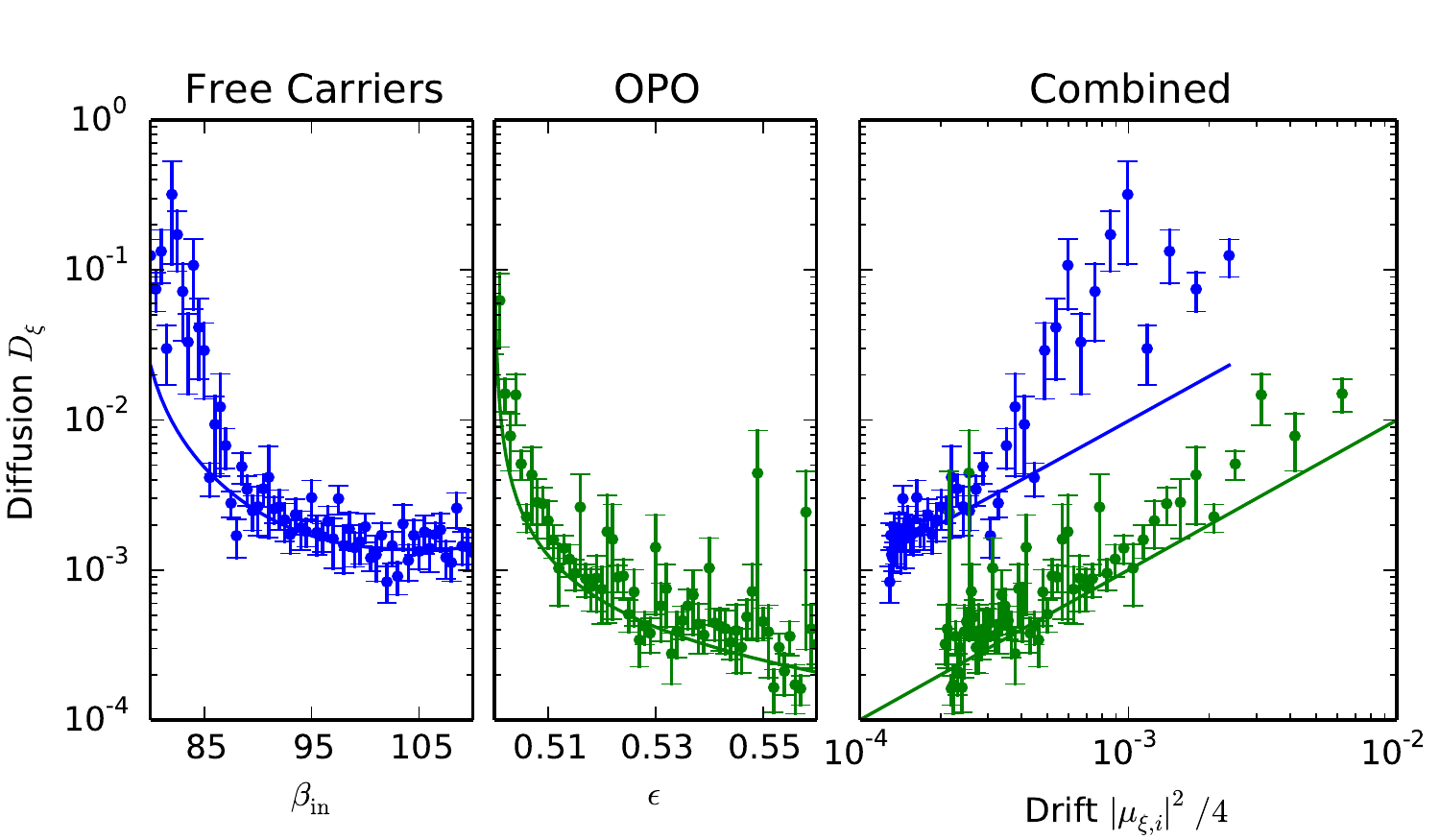}
\caption{Left: Limit-cycle phase diffusion for free-carrier cavity, $\Delta = -1.0$, as a function of input field.  Center: Phase diffusion for non-degenerate OPO, $\beta = 0.0002$, as a function of pump.  Right: Combined, where the drift term $\frac{1}{4}\sum_i |\mu_{\xi,i}|^2$is the common $x$ axis.}
\label{fig:07b-f9b}
\end{center}
\end{figure}

Figure \ref{fig:07b-f9b} plots the simulated phase diffusion constant $D_\xi$ for both the OPO and the free-carrier limit cycle.  As one approaches the bifurcation, the diffusion rate increases and diverges from the linearized result (\ref{eq:dxi}), solid curves in the figure.  However, far from the bifurcation, the linearized model agrees with the full simulation for both the OPO and free carriers.

To compare the OPO and free-carrier cavity on equal footing, the right panel of Figure \ref{fig:07b-f9b} plots the diffusion $D_\xi$ against the right-hand side of (\ref{eq:diffeq}): $\frac{1}{4}\sum_i|\mu_{\xi,i}|^2$.  The OPO simulations, at least for large $|\alpha_+|$, lie on the line $D_\xi = \frac{1}{4}\sum_i|\mu_{\xi,i}|^2$ (green line), while the free-carrier simulations lie a factor of $\sim 10$ above.

\subsection{Entrainment}
\label{sec:07b-driving}

\newcommand{\omegaC}{\omega_c}
\newcommand{\omegaIn}{\omega_{\rm in}}

If the system is driven with a periodic seed field whose frequency $\omegaIn$ does not exactly match the limit-cycle frequency $\omegaC$, the limit cycle may or may not lock to the seed ({\it entrainment}), depending on its amplitude.  To study this effect conceptually, assume a symmetric, noiseless limit-cycle model with a periodic drive $\binA + \bin{\omega} e^{-i\omega t}$, and transform to comoving coordinates $\zeta = \xi - \omegaIn t$.  Equation (\ref{eq:07b-lcxi}) takes the form \cite{StrogatzBook}:
\beq
	\frac{\d\zeta}{\d t} = (\omegaC-\omegaIn) - \left|\bin{\omega} B\right| sin(\zeta) \label{eq:07b-lczeta}
\eeq

For frequencies $|\omegaC - \omegaIn| < |B \bin{\omega}|$, there is a fixed point at $\zeta = \sin^{-1}((\omegaC-\omegaIn)/|\bin{\omega}B|)$, so the oscillator will lock to the seed.  If we plot $\omegaIn$ on the $x$-axis and $\bin{\omega}$ on the $y$ axis, this phase locking will happen in a vertical cone centered at $(\omegaC, 0)$.  Full free-carrier cavity simulations also show this effect.  Figure~\ref{fig:07b-f10} shows results for a $\Delta = -1.0$ cavity with pump $\binA = 100$, which naturally oscillates at $\omegaC = 2.27$.  On top of this, an oscillating field $\bin{\omega} e^{-i\omegaIn t}$ drives the cavity.

\begin{figure}[tbp]
\begin{center}
\includegraphics[width=0.7\textwidth]{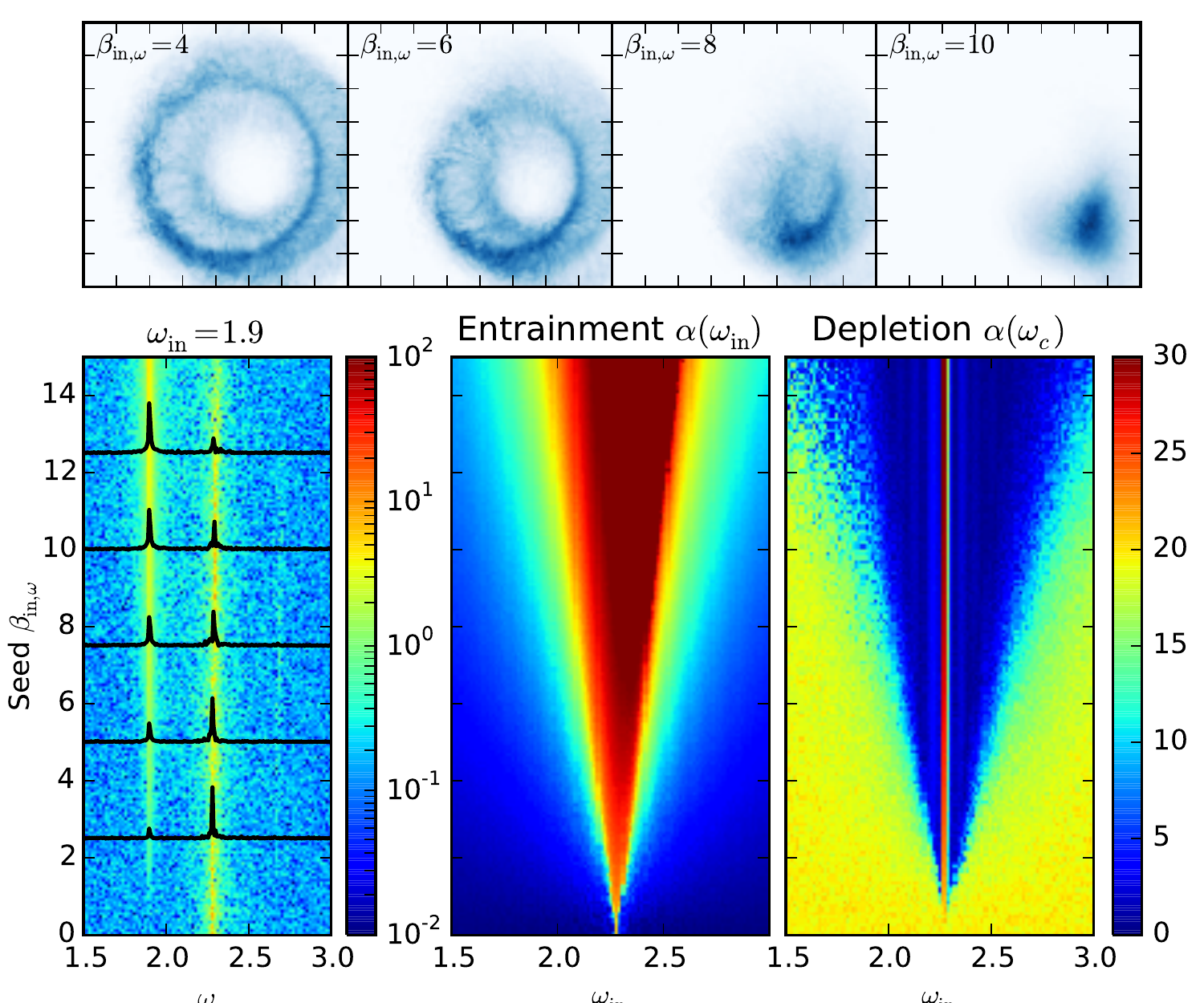}
\caption{Entrainment of free-carrier limit cycle, $\Delta = -1.0$, $\binA = 100$.  Top: Phase plots of the output field in a rotating wave frame, $e^{-i\omegaIn} \boutA$ (mean subtracted).  For large seed inputs, the device clusters to one side of the diagram, indicating phase locking.  Bottom left: output spectrum as a function of seed power, at $\omegaIn = 1.9$.  Bottom right: Entrainment cone.  Plots of $\alpha(\omegaIn)$ and $\alpha(\omegaC)$ (intracavity amplitude at seed and natural frequency, respectively) as a function of seed frequency and amplitude.}
\label{fig:07b-f10}
\end{center}
\end{figure}

The top pane in Figure~\ref{fig:07b-f10} shows the real and imaginary quadratures of the output field in a rotating-wave frame: $\tilde{\beta} e^{i\omegaIn t}$.  This is for seed frequency $\omegaIn = 1.9$ and cavity frequency $\omegaC = 2.3$, so $|\omegaIn - \omegaC| \approx 0.4$, or about 16\%.  For weak seed fields, the rotated output makes loops about the origin -- the phase is not locked.  However, around $\bin{\omega} = 10$, it clusters in a given direction -- indicating locking.

The bottom-left plot shows the output spectrum $\boutA(\omega)$ as a function of $\omega$ and the seed amplitude.  One sees two peaks, one at the limit-cycle frequency $\omegaC$ and one at the seed frequency $\omegaIn$.  The peak at the natural frequency $\omegaC$ is strongest when the pump is weak, and eventually goes away for strong pumping.  Conversely, the peak at the drive frequency $\omegaIn$ is absent for weak pumping, and grows with the pump strength.

This is seen more clearly in the bottom-right plots.  Instead of confining ourselves to $\omegaIn = 1.9$, in these plots we vary both the amplitude $\bin{\omega}$ and frequency $\omegaIn$ of the pump.  The left plot shows the power at the input frequency, while the right plot shows the power at the original frequency.  Inside the {\it entrainment cone}, the oscillator locks and the former dominates; outside the cone, the oscillator is unable to lock and the natural frequency is dominant.

From the shape of the entrainment cone, we estimate $B \approx 0.04$ for this set of parameters.

\subsection{Impulse Response}

\begin{figure}[b!]
\begin{center}
\includegraphics[width=0.66\textwidth]{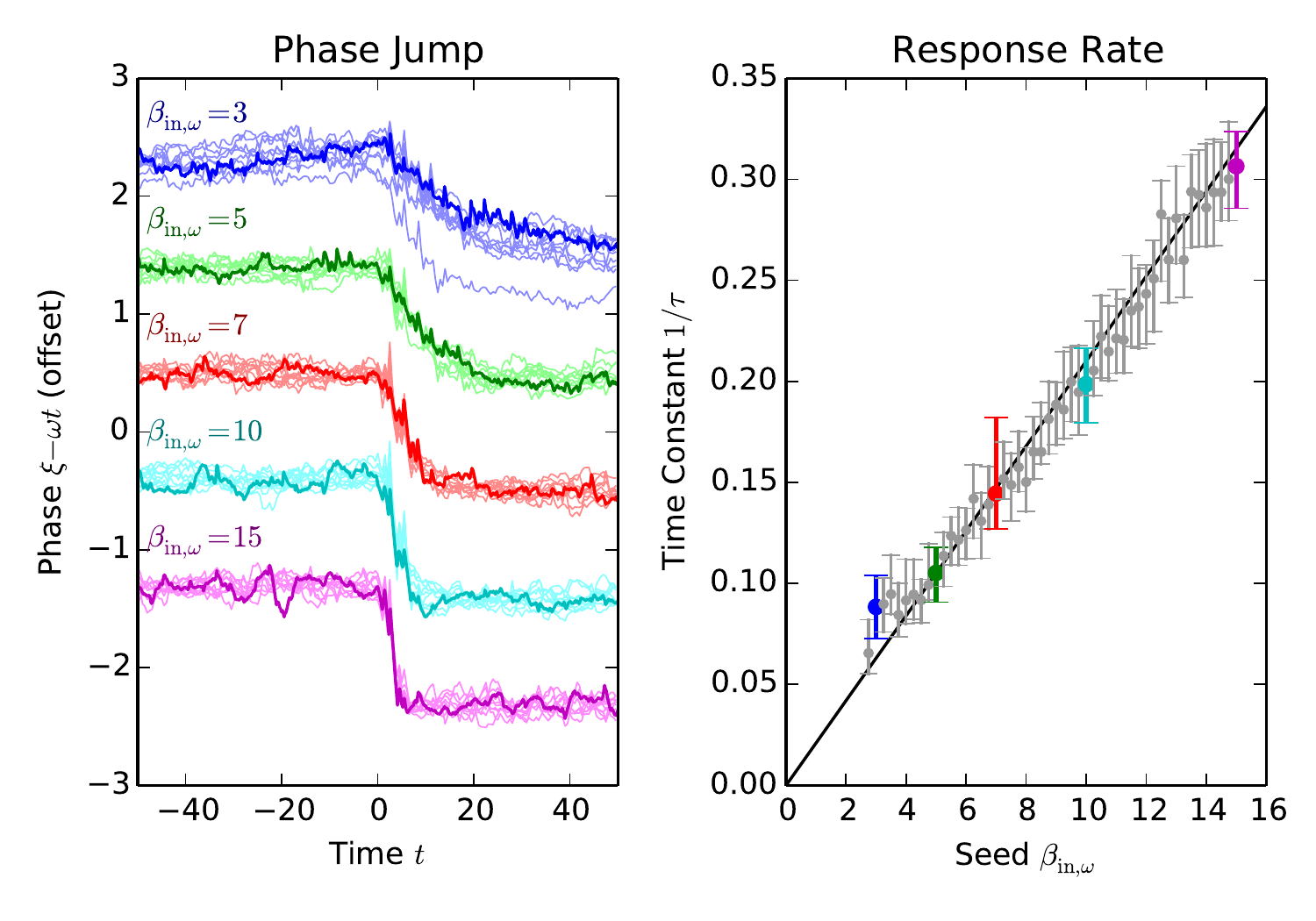}
\caption{Left: Time traces of the limit-cycle phase $\xi$ for a driven system where the seed phase jumps by one radian at $t = 0$.  Right: Response rate $1/\tau$, obtained by exponential fitting, as a function of seed amplitude $a_\omega$.  Parameters: $\Delta = -1.0, a_{\rm in} = 100$.}
\label{fig:07b-f11}
\end{center}
\end{figure}

Suppose that the oscillator has been locked to an external field and now the phase of that field is changed.  The oscillator should follow that phase, but there will be a time lag.  From Eq.~(\ref{eq:07b-lczeta}) we can estimate this time lag to be of order:
\beq
	\tau \sim \frac{1}{|\bin{\omega} B|} \label{eq:07b-tau}
\eeq
In Figure \ref{fig:07b-f11}, the same free-carrier system is simulated with a seed field $\omegaIn = \omegaC = 2.27$.  However, at time $t = 0$, the phase of the input shifts by 1 radian.  For seed amplitudes $\bin{\omega} \gtrsim 3$, the system quickly realigns to the new phase, with a time-constant given by (\ref{eq:07b-tau}).  From this, we can estimate $B \approx 0.02$.  This agrees with the entrainment-cone estimate to within a factor of 2; the lack of exact agreement is due to the circular cycle assumption that underlies (\ref{eq:07b-lczeta}, \ref{eq:07b-tau}).

\section{Applications}

\subsection{Ising Machine}
\label{sec:07b-ising}

Many optimization problems can be recast as Ising problems, which involve finding the minimum of the Ising Hamiltonian: $H = \sum_{ij} J_{ij} \vec{\sigma}_i \cdot \vec{\sigma}_j$.  If $\sigma$ is constrained to lie on the $xy$-axis the problem is called an XY model, the each spin maps onto an angle $\sigma_i = (\cos\zeta_i, \sin\zeta_i)$ and the Hamiltonian becomes:

\beq
	U[\zeta] = \sum_{ij} J_{ij} \cos(\zeta_i - \zeta_j) \label{eq:ising-h}
\eeq

The general Ising problem for arbitrary $J_{ij}$ is NP-hard \cite{Barahona1982}.

Ising problems map naturally onto oscillator networks.  Let each Ising spin be mapped onto an oscillating free-carrier cavity.  Let each oscillator have multiple independent input and output ports.  This can be accomplished using the ``railroad topology'' of Figure \ref{fig:07b-f14c}.  Suppose that an output of cavity $j$ is fed into an input of cavity $i$.  Assuming all cavities have the same limit-cycle frequency, under the assumptions of Section \ref{sec:07b-driving}, the phase of cavity $i$ evolves as:

\begin{figure}[tbp]
\begin{center}
\includegraphics[width=0.66\textwidth]{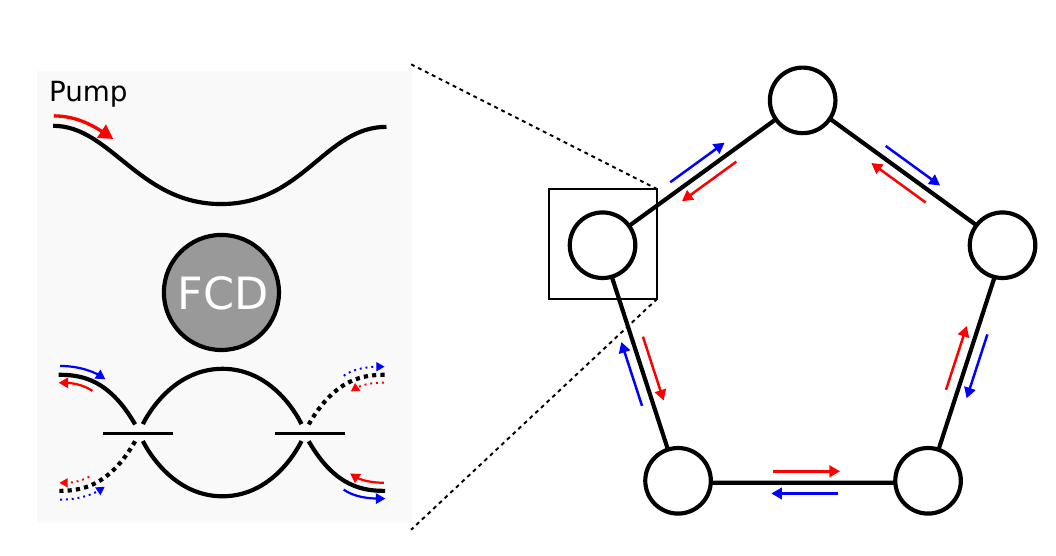}
\caption{Optical free-carrier cavity used as a node in an Ising machine.}
\label{fig:07b-f14c}
\end{center}
\end{figure}

\beq
	\d\zeta_i = -J_{ij} \sin(\zeta_i - \zeta_j) \label{eq:dzeta}
\eeq
where $J_{ij}$ depends on the waveguide coupling, the phase of the connection, and the limit-cycle amplitude.  It is not difficult to see that, with the appropriate connections, one can realize a cavity network that minimizes (\ref{eq:ising-h}) by the steepest-descent method.

A full discussion of optical Ising machines is beyond the scope of this chapter.  The concept was proposed by Utsunomiya et al. \cite{Utsunomiya2011}, who suggested implementing it using injection-locked lasers.  Recent theoretical work \cite{Wang2013} and experiments with 4-bit \cite{Marandi2014} and 16-bit \cite{KentaThesis} Ising machines using a time-multiplexed pulsed OPO show that the device matches or surpasses classical algorithms in accuracy.  However, free-carrier oscillations may be a preferable platform for Ising machines because of their low power requirements and compatibility with existing fabrication processes.

Figure \ref{fig:07b-f14b} shows the simulated Ising-machine performance for antiferromagnetic couplings on five graphs: pair, triangle, square, pentagon and tetrahedron.  Of these, the pair and square have zero-energy configurations, while the rest are frustrated systems.  The square and tetrahedron were studied with an OPO Ising machine in \cite{Marandi2014}.

Larger networks also show convergence in reasonable time.  In Figure \ref{fig:07b-f14b}, we plot the performance of a 16-spin network, both with a nearest-neighbor interaction and with a cross-interaction (which shows frustration).  These are the graphs studied in the OPO network of \cite{KentaThesis}.  As long as it does not get trapped in local minima, the device converges to the minimum of $U[\zeta]$ in $50-100$ cavity lifetimes.

\begin{figure}[tbp]
\begin{center}
\includegraphics[width=0.66\textwidth]{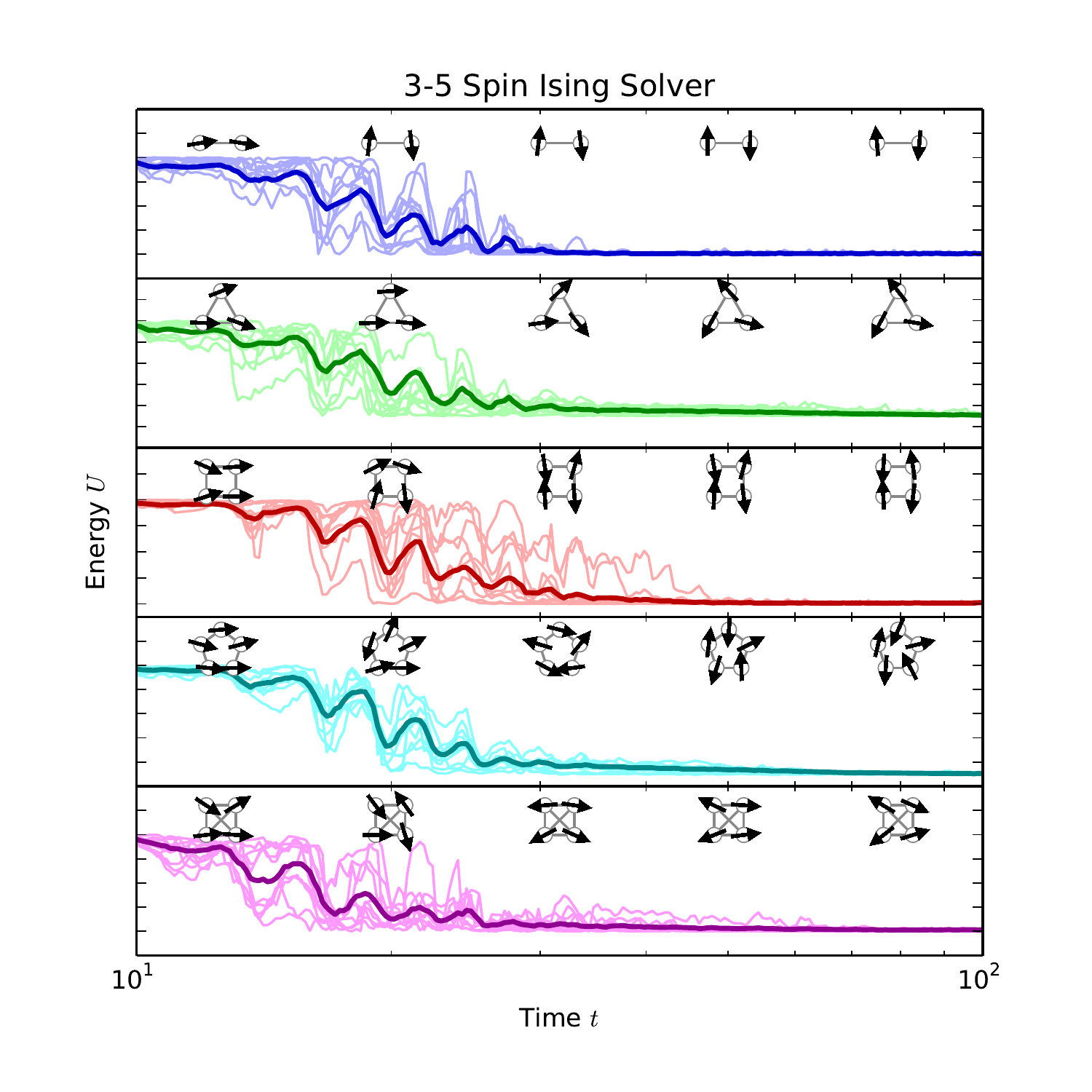}
\caption{Ising machine performance for small graphs.  Top to bottom: pair, triangle, square, pentagon, and tetrahedron.}
\label{fig:07b-f14a}
\end{center}
\end{figure}

\begin{figure}[tbp]
\begin{center}
\includegraphics[width=0.66\textwidth]{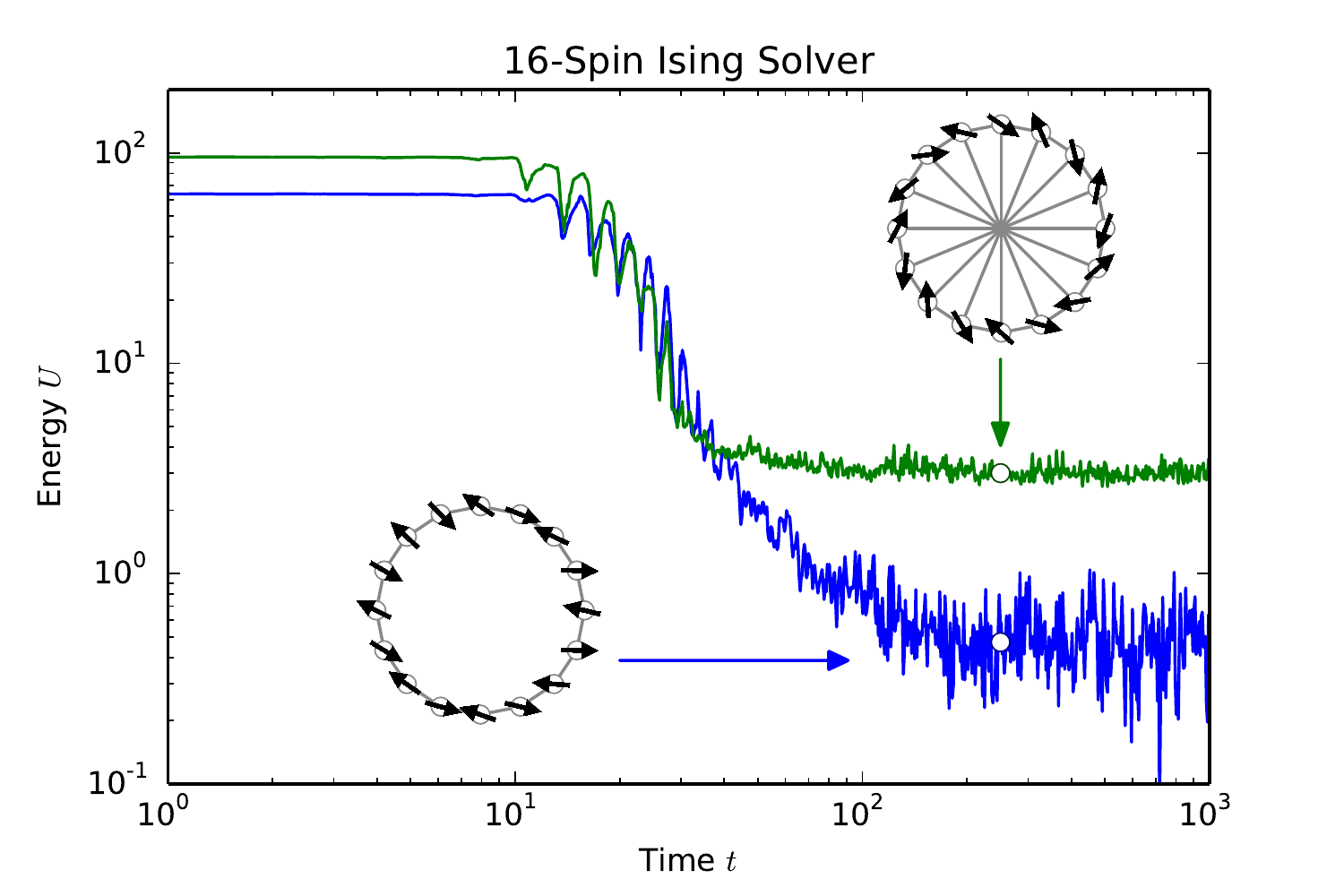}
\caption{Ising machine performance for 16-gon and frustrated 16-gon with cross-couplings.}
\label{fig:07b-f14b}
\end{center}
\end{figure}

Because the free-carrier Ising machine maps the optimization directly onto the hardware dynamics, it can achieve a per-watt performance orders or magnitude greater than a microprocessor solving the same problem.  For the network used in Figure \ref{fig:07b-f14b} (see Sec.~\ref{sec:07b-sims} for cavity parameters), during oscillation each cavity consumes $\sim 2000$ photons, or about 0.5 fJ, per cavity lifetime and takes $\sim100$ lifetimes to converge, an energy cost of $\sim 50$ fJ per spin and a computation time of $\sim 300$ ps.  A microprocessor using steepest-descent or stimulated annealing will also take $\sim 100$ steps to converge, but be required to compute (\ref{eq:dzeta}) at each step.  Since (\ref{eq:dzeta}) involves computing a trigonometric function, it will take $\sim 50$ flops and $\sim 100$ clock cycles per step \cite{AgnerFogNote}, or $\sim 5000$ flops per spin overall.  As of 2015, the most energy-efficient supercomputer was the L-CSC at GSI, Darmstadt, which runs at 3 GHz and requires $0.2$ nJ per flop \cite{Green500}, giving a simulation time of $\sim$3 $\mu$s and energy cost of $\sim$1 $\mu$J per spin.  On the basis of this rough calculation, the free-carrier Ising machine should perform $\sim$$10^4\times$ faster and consume $\sim$$10^7\times$ less energy.

\subsection{Free-Carrier Relay}
\label{sec:07b-relay}

\begin{figure}[b!]
\begin{center}
\includegraphics[width=0.66\textwidth]{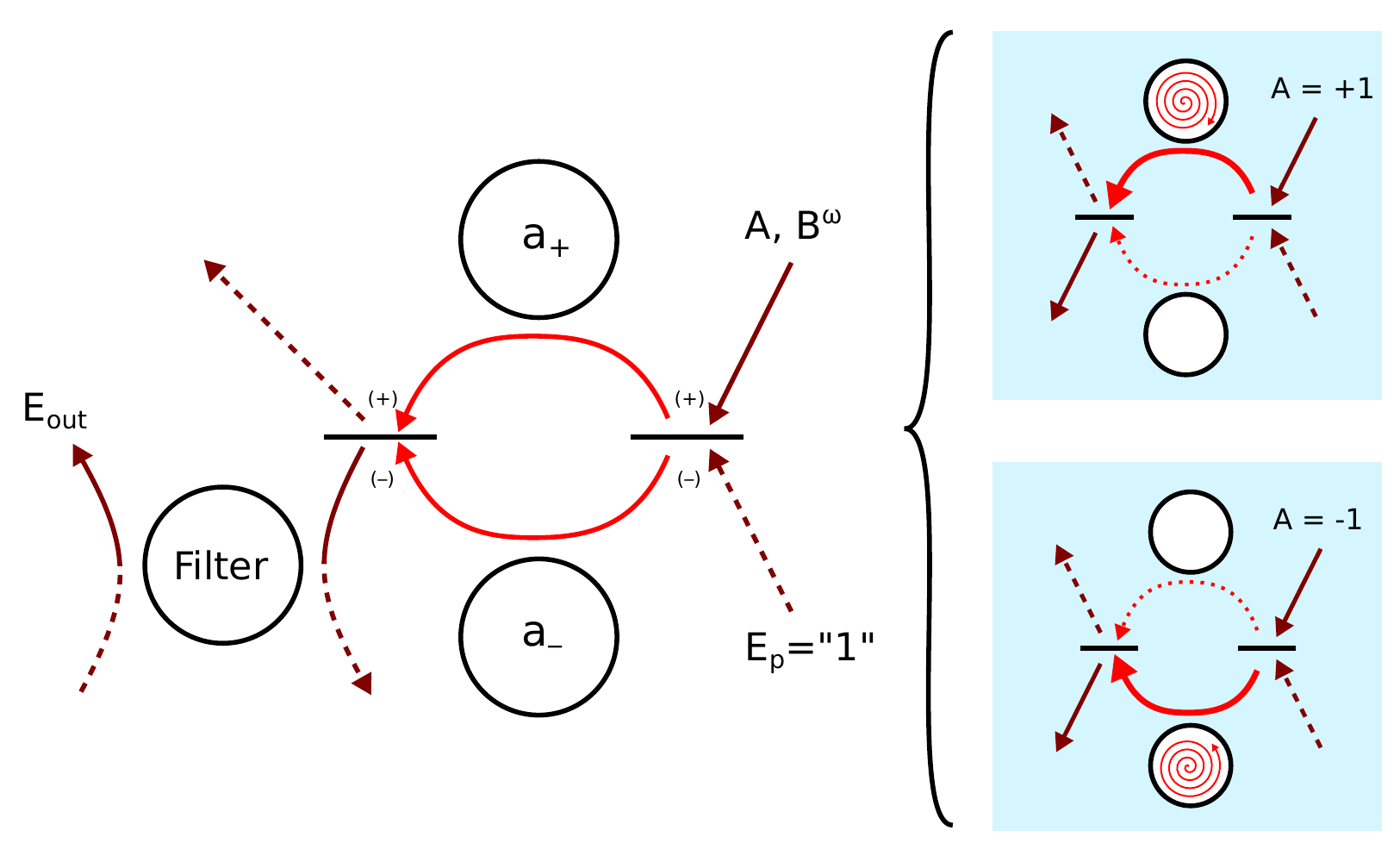}
\caption{Left: Layout of the free-carrier relay.  Right: Relay behavior when control bit $A$ is set to $+1$ (left) or $-1$ (right).}
\label{fig:07b-f12}
\end{center}
\end{figure}

In a previous sections, we showed that free-carrier cavities can undergo spontaneous self-oscillation if driven hard enough.  Here we show that this can be used to construct a free-carrier ``relay''.  Such a device has many logic applications, including message passing algorithms for error correction \cite{Pavlichin2014}.  A relay acts like a classical CNOT gate: if the digital inputs $A, B \in \{-1, 1\}$, then the relay maps these to:
\beq
    (A,\ B) \stackrel{\rm Relay}{\longrightarrow} (A,\ A B)
\eeq
That is, output $B$ is flipped if $A = -1$.

The relay is a circuit with two free-carrier cavities, arranged as in Figure~\ref{fig:07b-f12}.  The inputs $A$ and $B$ arrive on the same channel, but are offset in frequency.  Data is encoded on the {\it phase} of the inputs ($0$ or $\pi$), not the amplitude; thus, for a fixed field amplitude $|A|$, a $1$ corresponds to $+|A|$, while $-1$ corresponds to $-|A|$.

The input is mixed with a pump field on a beamsplitter, so that the field entering cavity $a_\pm$ is:
\beq
    \bin{\pm} = \frac{A \pm E_p}{\sqrt{2}} + \frac{B e^{-i\omega t}}{\sqrt{2}}
\eeq
A free-carrier cavity will self-oscillate if the input field is stronger than some threshold: $|\binA| > \beta_{\rm th}$.  Let:
\beq
    |A| - |E_p| < \beta_{\rm th} < |A| + |E_p|.
\eeq

If $A = +1$, then the top resonator is above threshold and self-oscillates at $\omega$, while the bottom resonator does not self-oscillate.  For the $B$ field at this frequency, this means that the top channel has more gain than the bottom channel.  When these are interfered on a beamsplitter, the output at this frequency is $\frac{1}{2}(G_{\rm high} - G_{\rm low}) B e^{-i\omega t}$.  Since $G_{\rm high} > G_{\rm low}$, the phase of $B$ does not change.

On the other hand, if $A = -1$, the lower channel has higher gain.  When recombined on the beamsplitter, the output is $-\frac{1}{2}(G_{\rm high} - G_{\rm low}) B e^{-i\omega t}$ -- the phase of $B$ does flip.  This is shown in Figure \ref{fig:07b-f12}.  Thus, the relay realizes the CNOT map $(A, B) \rightarrow (A, AB)$.

\begin{figure}[tbp]
\begin{center}
\includegraphics[width=0.66\textwidth]{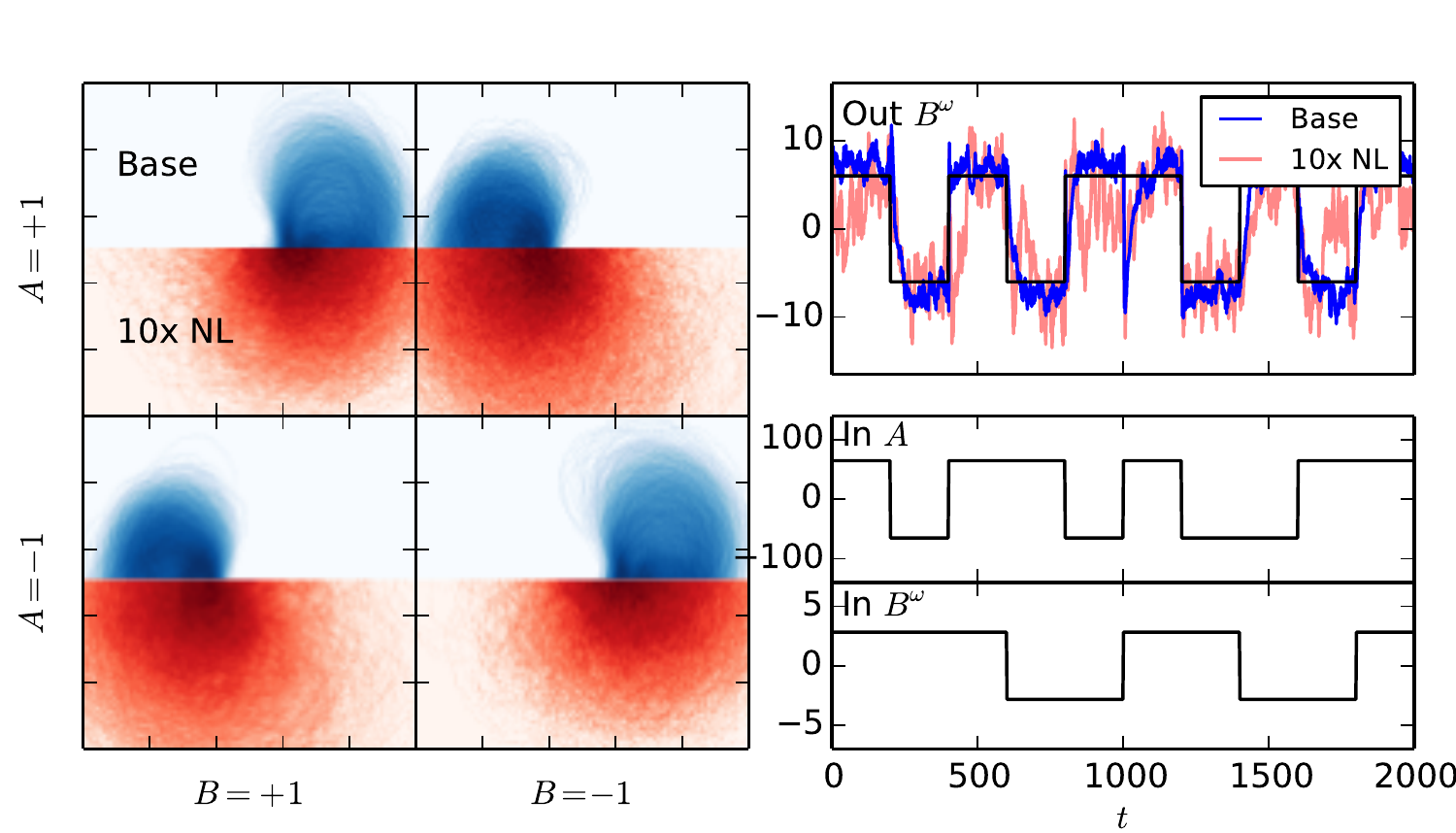}
\caption{Left: Plots of the real and imaginary parts of the rotating-frame output $B^\omega$, as a function of the input $A$ and $B^\omega$.  Right: Time trace of the relay output (top), where the inputs $A$ and $B^\omega$ are switched regularly (bottom).  Both the base (blue) and 10x NL (red) scenarios are shown.  Outputs are scaled by $\sqrt{10}$ for the 10x NL case.}
\label{fig:07b-f13}
\end{center}
\end{figure}

Figure \ref{fig:07b-f13} demonstrates the relay operation.  Two results are plotted: a ``base'' case with the same cavity parameters used elsewhere in the paper (blue in figure) and a hypothetical ``10x NL'' case where the nonlinearity (parameters $\delta, \beta$) has been increased by a factor of ten.  Both cavities have a detuning $\Delta = -2.0$.  In order to control the phase of the beam at $\omega$, the input $A$ must be fairly large ($A = \pm 65$ was used here, scaled by $\sqrt{10}$ for the 10x NL case).  However, the input $B$ at $\omega$ can be quite small; in the simulation taking a value of about 3.  Since the output amplitude is around 7, this provides an XOR with enough gain for a fanout of 4-5.

Both relays display the same overall behavior, but because the cavity in the 10x NL relay has a stronger nonlinearity, it operates at a lower photon number and thus the photon shot noise is more significant.  This degrades the performance of the XOR gate.  Ultimately, there is tradeoff between gate fidelity and energy consumption for free-carrier based systems.  Since this tradeoff arises from quantum mechanics, it cannot be avoided by choosing different materials or cavity designs.  The benefit of our SDE approach (\ref{eq:07b-eom1}-\ref{eq:07b-eom2}) is that it reveals not only the classical behavior of the relay, but also this basic quantum limit to its performance.

\section{Conclusion}

Systems with a Hopf bifurcation can perform a wide range of useful tasks with applications in sensing and photonic logic.  In this paper, we have studied the supercritical Hopf bifurcation in a semiconductor optical cavity where the dominant optical nonlinearity is due to free carrier dispersion.  Following the previous chapter, I simulated the dynamics of a the free-carrier cavity using Wigner SDEs that capture both the semiclassical motion and the quantum fluctuations in photon and carrier number.

Below the bifurcation, the free-carrier optical cavity acts as a phase-insensitive amplifier.  This device is the basis for heterodyne detection, where both quadratures of the field are simultaneously measured with an added noise penalty.  The Caves bound places a lower limit on the noise, and this limit is satisfied in the non-degenerate OPO.  By contrast, the free-carrier cavity has $\sim 5\times$ more noise in the output, an effect we attribute to the incoherent nature of carrier excitation and decay.

Above the bifurcation, the device has a limit cycle.  Quantum fluctuations cause the phase of this cycle to diffuse, and the diffusion rate can be computed by linearizing the SDEs in a normal coordinate frame centered on the limit cycle.  In this limit, one can use the device to store a continuous number in the range $[0, 2\pi)$, or alternately, to perform a homodyne measurement on signals at the limit-cycle frequency.  Limits on the efficiency of homodyne measurement lead to a quantum lower bound on the limit-cycle diffusion rate.  This bound is saturated by the non-degenerate OPO, while the diffusion rate of the free-carrier cavity is $\sim 10\times$ larger.  Again, this is due to the incoherent carrier excitation and decay processes.

Limit-cycle systems are useful in logic and computing because they can be locked to external signals, and their outputs can in turn be used to lock other limit cycles.  While an analysis such large-scale networks is beyond the scope of this paper, we have explored the basic phenomenon that underlies this: entrainment in an external field.  Utilizing entrainment, I showed that the free-carrier cavity can be used to construct a coherent Ising machine that finds the minimum of a preprogrammed cost function.  With reasonable cavity parameters, such a coherent Ising machine could run $\sim 10^4\times$ faster with $\sim 10^7\times$ less energy than a comparable algorithm on a supercomputer.  In addition, we showed that entrainment can be used to construct a limit-cycle ``relay'' -- an all-optical classical CNOT gate, which has applications in message-passing schemes.

Although the free-carrier cavity is noisier and performs more poorly than quantum-limited systems like the non-degenerate OPO, it is much more convenient to build.  Free-carrier optical cavities can be built from silicon or III-V materials, which have mature and scalable fabrication processes.  In addition, the per-photon effect is much stronger, enabling operation at lower powers.  When it comes to building an actual device, these practical concerns may prevail over the theoretical elegance of quantum-limited systems.

\section*{Appendix}

\renewcommand\thesection{\arabic{chapter}.\Alph{appsection}}
\renewcommand\thesubsection{\arabic{chapter}.\Alph{appsection}.\arabic{subsection}}

\setcounter{appsection}{1}\section{Limit Cycles and $k$-dimensional Attractors}
\label{sec:07b-04b-lc}

\begin{figure}[b!]
\begin{center}
\includegraphics[width=0.60\textwidth]{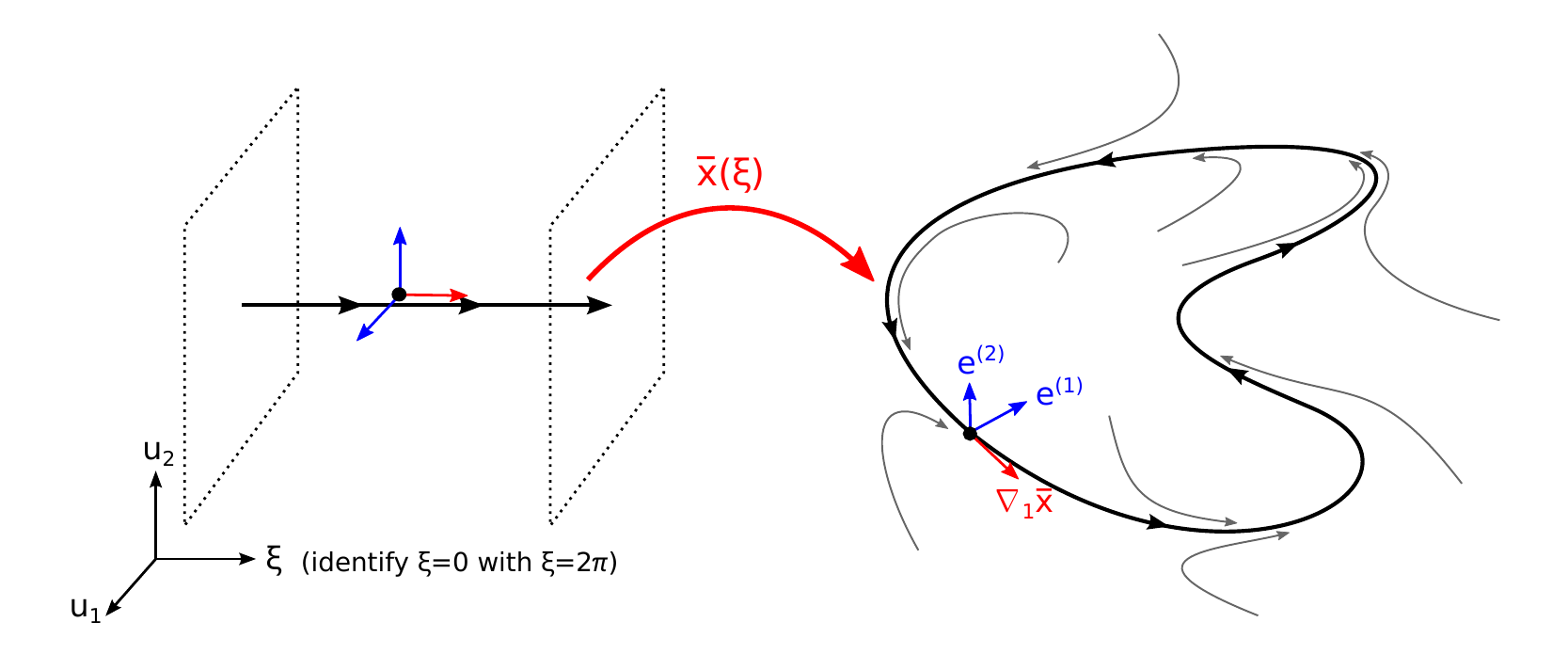}
\caption{Diagram of a limit cycle in a normal coordinate frame (left) and in the actual phase space (right), along with the transverse $e^{(i)}$ (blue) and longitudinal $\nabla_i\bar{x}$ (red) vectors.}
\label{fig:04b-f7}
\end{center}
\end{figure}

Many dynamical systems do not have a fixed point.  Instead, they have a stable limit cycle, or more generally, a stable $k$-dimensional attractor.  The $k = 1$ case corresponds to a limit cycle.  The cycle may be parameterized as follows:
\beq
	x(t) = \bar{x}(\omega t)
\eeq
where $\omega$ is the oscillation frequency.  The map $\bar{x}: \mathbb{R} \rightarrow \mathbb{R}^n$ defines the attractor's manifold, and is sufficient if we are only interested in how the system behaves without forcing.  However, the map tells us nothing about forcing or deviations from the attractor.  When noise and forcing are present, these perturbations become relevant, and we need more information about the system to handle them.

\subsection{Linearization About Attractor}

Consider a nonlinear system of differential equations of the most general form:
\beq
    dx_i = \left[f_i(x) + F_i(x, t)\right]\d t + g_{ij}(x)\d w_j(t) \label{eq:07b-lc}
\eeq
Here, $x$ is the state of the system, $f_i(x)$ is its natural (unforced) derivative, $g_{ij}$ is the noise coupling (to Wiener process $\d w_j$) and $F_i(x,t)$ is the external forcing.  In the absence of forcing, let's suppose that Equation~(\ref{eq:07b-lc}) gives rise to a stable attractor $\bar{x}(\omega t)$.  This has natural period $T = 2\pi/\omega$, so $\bar{x}(\xi + 2n\pi) = \bar{x}(\xi)$ for integers $n$.  Deviations from this cycle are given by: $x(t) = \bar{x}(\omega t) + \delta x(t)$.  In the absence of noise or external forcing, the perturbations evolve as follows:
\beq
    d(\delta x_i) = \frac{\partial f_i}{\partial x_j} \delta x_i \equiv A_{ij}(\bar{x}_\xi) \delta x_j
\eeq
where $A_{ij}(x) = \partial f_i/\partial x_j$ is the Jacobian of the dynamical system; see (\ref{eq:07b-linabcd}), and $\xi$ is the attractor phase, with $\bar{x}_\xi \equiv \bar{x}(\xi)$.

The key trick is to perform a coordinate transformation that separates the $dx$, and $n$-dimensional vector, into $1$ longitudinal perturbation and $n-1$ transverse perturbations.  The longitudinal perturbation keeps the system on the limit cycle, and therefore does not decay.  The transverse perturbations deviate from the limit cycle, and decay to zero as $t \rightarrow \infty$.  We denote these by $v_\xi$ and $e^{(i)}_\xi$, as follows:
\beq
	\delta x(t) = \delta\xi(t) v_\xi + \sum_{i=0}^{n-1} u_i(t) e^{(i)}_\xi \label{eq:07b-lcparam}
\eeq
Here we have traded an $n$-dimensional state vector $x(t)$ for $n-1$ transverse variables $u_i(t)$ and one longitudinal variable $\delta\xi_i$.

Applying (\ref{eq:07b-lcparam}) to the equations of motion with noise and forcing, we obtain:
\beq
	d(\delta\xi) v_\xi + du_i e^{(i)}_\xi = \Bigl(\underbrace{\left[A(\bar{x}_\xi) v_\xi - \omega\frac{\d v_\xi}{\d\xi}\right]}_{D_t v_\xi = 0} \delta\xi + \underbrace{\left[A(\bar{x}_\xi) e^{(i)}_\xi - \omega\frac{\d e^{(i)}_\xi}{\d\xi}\right]}_{D_t e^{(i)}_\xi} u_i\Bigr)\d t + \left(F(\bar{x}_\xi, t)\d t + g(\bar{x}_\xi) \d w\right) \label{eq:07b-lcpt}
\eeq
(implicit summation over $i$)

The covariant derivative $D_t$ of a $\xi$-dependent vector is defined as
\beq
	D_t q_\xi \equiv A(\bar{x}_\xi)q_\xi - \omega \frac{\d  q_\xi}{\d\xi}
\eeq
This derivative accounts for both the equations of motion and our parameterization near the limit cycle.  It is similar to the covariant derivative in Riemannian geometry \cite{WaldBook}.  Because the tangent vector $v_\xi$ always transforms into itself when propagated around the manifold, its covariant derivative is zero.  Likewise, because the transverse vectors always decay to zero, they cannot evolve into $v_\xi$; thus $D_t e_\xi^{(i)}$ has no $v_\xi$ component.

In matrix form, Equation (\ref{eq:07b-lcpt}) is:
\beq
	\begin{bmatrix} v_\xi & e_\xi \end{bmatrix} \begin{bmatrix} d(\delta\xi(t)) \\ du(t) \end{bmatrix} =
	\begin{bmatrix} 0 & D_t e_\xi \end{bmatrix} \begin{bmatrix} \delta\xi(t) \\ u(t) \end{bmatrix}\d t + F(\bar{x}_\xi,t)\d t + g(\bar{x}_\xi)\d w
\eeq
This becomes a matrix ODE:
\bea
	\d\begin{bmatrix} \delta\xi(t) \\ u(t) \end{bmatrix} & = & \left(\begin{bmatrix} v_\xi & e_\xi \end{bmatrix}^{-1} \begin{bmatrix} 0 & D_t e_\xi \end{bmatrix}\right)\begin{bmatrix} \delta\xi(t) \\ u(t) \end{bmatrix} \d t + \begin{bmatrix} v_\xi & e_\xi \end{bmatrix}^{-1} \left(F(\bar{x}_\xi,t)\d t + g(\bar{x}_\xi)\d w\right) \label{eq:07b-lcmat} \nonumber \\
	& \equiv & \begin{bmatrix} 0 & 0 \\ 0 & A_{T} \end{bmatrix} \begin{bmatrix} \delta\xi(t) \\ u(t) \end{bmatrix}
	+ \begin{bmatrix} B_L \\ B_T \end{bmatrix} \left(F(\bar{x}_\xi,t)\d t + g(\bar{x}_\xi)\d w\right) \label{eq:07b-lcmat2}
\eea
In the equations above, $\xi(t) = \omega t$ has a fixed time-dependence.  The dynamical variable $\delta\xi(t)$ adds a perturbation to this $\xi$.  We can roll $\delta\xi$ into $\xi$, {\it turning $\xi$ into a dynamical variable}, so the state vector becomes:
\beq
    x(t) = \bar{x}(\xi(t)) + \sum_{j=0}^{n-1}{u_j(t) e^{(j)}_{\xi(t)}}
\eeq
The matrix ODE becomes:
\begin{align}
	\d\xi(t) & = \omega + B_L(\xi) \left(F(\bar{x}_\xi,t)\d t + g(\bar{x}_\xi)\d w\right) \label{eq:07b-lc-ode1} \\
	du(t) & = A_T(\xi)u(t)\d t + B_T(\xi) \left(F(\bar{x}_\xi,t)\d t + g(\bar{x}_\xi)\d w\right) \label{eq:07b-lc-ode2}
\end{align}
This equation captures our intuition regarding limit cycles and attractors.  External forces ($F$, $g$) can give rise to two kinds of perturbations: longitudinal (encoded in changes to $\xi$) and transverse ($u$).  Because of our choice of coordinates, {\it the perturbations evolve independently}.  The $A_T$ matrix causes transverse perturbations to decay as $t \rightarrow \infty$, while longitudinal perturbations do not.  Often, we are only interested in the longitudinal perturbations; in this case we can ignore the $u(t)$ altogether.

Altogether, we can arrive at (\ref{eq:07b-lc-ode1}-\ref{eq:07b-lc-ode2}) for an arbitrary limit cycle by following these four steps:

\begin{enumerate}
	\item Get equations of motion $dx = \left[f(x) + F(x, t)\right]\d t + g(x)\d w$
	\item Get limit cycle $\bar{x}(\xi)$ and the tangent vector $v_\xi$
	\item Find a set of vectors $e^{(i)}_\xi$ at each point $\xi$ that satisfy the following:
    	\begin{enumerate}
    		\item $\{e^{(i)}_\xi, v_\xi\}$ spans the whole vector space $\mathbb{R}^n$
    		\item Perturbations along the $\delta x \sim e^{(i)}$ eventually go to zero as $t \rightarrow \infty$
		\end{enumerate}
	\item Compute $A_T, B_L, B_T$ in Eqs.~(\ref{eq:07b-lcmat}-\ref{eq:07b-lcmat2})
\end{enumerate}

\subsection{Non-degenerate OPO}

Now we apply this to the non-degenerate OPO introduced in Section \ref{sec:07b-ndopo}.  The equations of motion are reproduced below:
\bea
    \d\alpha_\pm & = & \left[(-i\Delta -\kappa/2 \pm \epsilon) \alpha_\pm - \frac{\beta}{2} \left(\alpha_\pm^*\alpha_\pm \alpha_\pm - \alpha_\pm^* \alpha_\mp \alpha_\mp\right)\right]\d t \nonumber \\
    & & - \sqrt{\kappa} \d\bin{\pm} \mp \frac{1}{2}\sqrt{\beta}\left(\alpha_\pm \d w_1 - i \alpha_\mp \d w_2\right)
\eea
The limit cycle occurs at:
\beq
	|\alpha_+| = \sqrt{\frac{\epsilon - \kappa/2}{\beta/2}}
\eeq
Following the procedure above, we first find a mapping from $[0, 2\pi]$ to the limit cycle.  This is easy: $\alpha_+(\xi) = |\alpha_+| e^{-i\xi}, \alpha_-(\xi) = 0$.  Next, one needs the $v_\xi$ and $e^{(i)}$.  In terms of the basis $(\alpha_+, \alpha_-)$, a good choice is:
\beq
	\nabla_1 \bar{x}_\xi = \begin{bmatrix} -i\alpha_+ \\ 0 \end{bmatrix},\ \ \
	e^{(1)} = \begin{bmatrix} \alpha_+ \\ 0 \end{bmatrix},\ \ \
	e^{(2)} = \begin{bmatrix} 0 \\ 1 \end{bmatrix},\ \ \
	e^{(3)} = \begin{bmatrix} 0 \\ i \end{bmatrix}
\eeq
One can check that these are linearly independent (in doubled-up space) and span the whole space.  Plus, due to the symmetry of the problem, it should be pretty clear that perturbations orthogonal to the limit cycle ($e^{(1)}$) or perturbations to the $\alpha_-$ mode ($e^{(2)}, e^{(3)}$) always decay to zero.

In this case we are not concerned about deviations from the limit cycle, so there is no need to calculate the $A_T$ (which depends on covariant derivatives $D_e e_\xi$).  All we need to find is $B_L$.  At the end of the day we get the following equation of motion:
\beq
	\dot{\xi} = \Delta + {\rm Re}\,\left[\frac{-i\sqrt{\kappa}}{\alpha_+}\bin{+}\right]
\eeq
If the inputs $\bin{1}, \bin{2}$ are vacuum noise, the noise term on the right becomes
\beq
	\d\xi = \Delta\,\d t + \sqrt{\frac{\beta}{8} \frac{\kappa/2}{\epsilon - \kappa/2}}\,\d w \label{eq:04b-ndopo-diff}
\eeq

\renewcommand\thesection{\arabic{chapter}.\arabic{section}}
\renewcommand\thesubsection{\arabic{chapter}.\arabic{section}.\arabic{subsection}}

%

\ifstandalone{\bibliography{NoteRefs}{}
\bibliographystyle{alpha}
}
%

\ifdefined\multidoc\else\input{Header}\fi

\ifstandalone{\setcounter{chapter}{8}}
\chapter{1D and 2D Pulsed Ising Machines}
\label{ch:09}

This chapter is based on the following paper:

\begin{enumerate}
	\item R.~Hamerly, K.~Inaba, T.~Inagaki, H.~Takesue, Y.~Yamamoto and H.~Mabuchi, ``Topological defect formation in 1D and 2D spin chains realized by network of optical parametric oscillators.''  International Journal of Modern Physics B (submitted), arXiv:\href{http://arxiv.org/abs/1605.08121}{1605.08121}
\end{enumerate}

Many important problems in computer science can be solved by message-passing algorithms.  In such algorithms, information lives on the nodes of a graph, while computation consists of updating the values of the nodes by passing ``messages'' along the graph's edges.  Examples of such algorithms include neural networks\cite{Izhikevich2007}, probabilistic graphical models\cite{Koller2009}, low-density parity check codes\cite{Pavlichin2014} and topological surface codes\cite{Fujii2014}.  Message-passing algorithms are advantageous because they are intrinsically parallel, making them straightforward to implement on multi-core architectures.

As digital microprocessors reach their physical limits, there has been a surge of research into special-purpose hardware for various message-passing algorithms.  In electronics, examples include CMOS artificial neural networks\cite{Benjamin2014,Merolla2014,Misra2010,Schemmel2010} and CMOS chips for simulated-annealing\cite{Yoshimura2015}.  Quantum annealers have a similar graphical architecture, with data stored at the vertices (qubits), while pairwise couplings along the edges transmit information along the graph.

This chapter focuses on a coherent {\it optical} network, which functions as a message-passing algorithm to solve the {\it Ising problem} and the related {\it XY problem}.  These problems consist of finding the global minimum of the Ising potential $\min_\sigma[U(\sigma)]$, where
\beq
	U(\sigma) = -\frac{1}{2}\sum_{ij} {J_{ij} \vec{\sigma}_i \cdot \vec{\sigma}_j} \label{eq:09-u}
\eeq
In (\ref{eq:09-u}), $J_{ij}$ is the coupling between spins $\vec{\sigma}_i$, $\vec{\sigma}_j$.  The spins $\vec{\sigma} \in \mathbb{R}^d$ have unit norm $|\vec{\sigma}| = 1$.  For the Ising problem $d = 1$ and $\vec{\sigma} \in \mathbb{Z}_2 = \{-1, 1\}$; for the XY problem $d = 2$ and $\vec{\sigma} \in S_1 = U(1)$.  Higher-dimensional problems ($d = 3,4,5\ldots$) can also be defined, but will not be considered here.

The general Ising problem is NP-hard\cite{Barahona1982}, but algorithms based on convex relaxation or heuristics can give approximate solutions in polynomial time.  A number of schemes have been studied to map such algorithms directly onto electronic\cite{Yoshimura2015} or photonic circuits\cite{Hamerly2015-2}.

\begin{figure}[tbp]
\begin{center}
\includegraphics[width=1.00\textwidth]{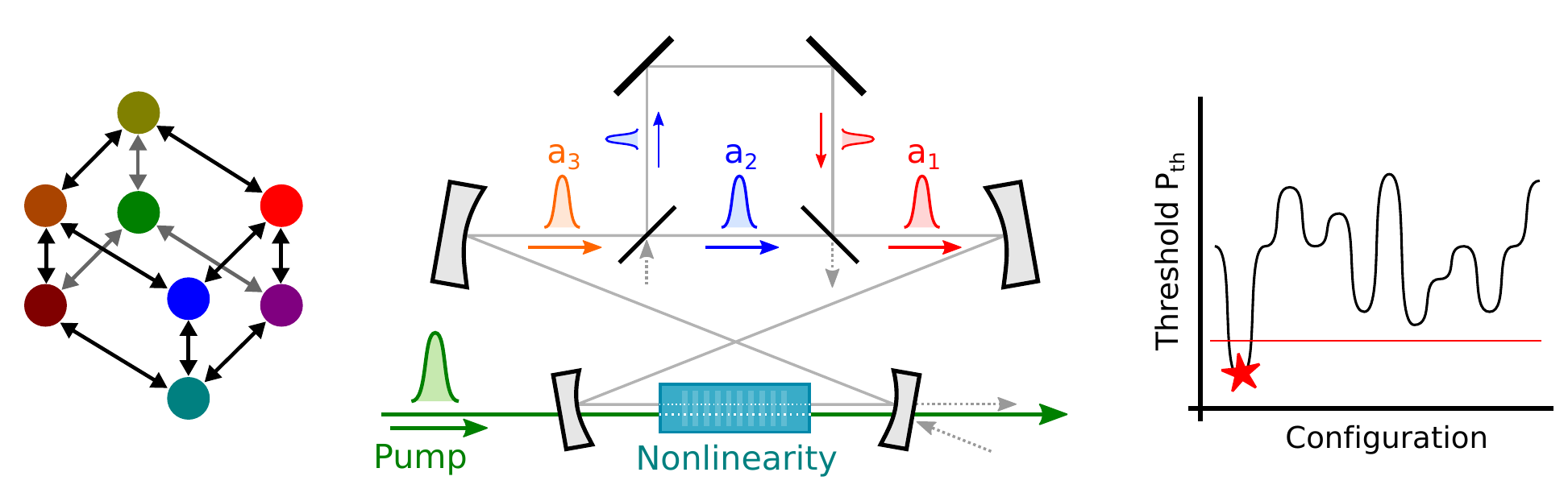}
\caption{Left: Ising machine consisting of optical gain elements (circles) with injection coupling (arrows), cubic graph.  Center: time-multiplexed implementation.  Right: illustration of the minimum-gain principle.}
\label{fig:09-f1}
\end{center}
\end{figure}

The coherent Ising machine is a network of identical nonlinear gain elements symmetrically coupled by optical injection that solves the Ising problem by a minimum-gain principle\cite{Haribara2016,Wang2013,Marandi2014}.  According to this principle, if the couplings are chosen to implement the potential $J_{ij}$, the configuration that oscillates should minimize the potential (\ref{eq:09-u}).  For the nonlinear gain, an injection-locked laser or an optical parametric oscillator (OPO) can be used.  In practice, the spins in the machine are time-multiplexed as pulses in a synchronously-pumped laser or OPO and couplings are realized by delay lines that couple pulses at different locations in the cavity (Fig.~\ref{fig:09-f1}).

The Ising machine was proposed as an injection-locked laser network\cite{Takata2012,Utsunomiya2011}.  Later, the theory was extended to OPOs\cite{Wang2013} and simulations showed promising performance on MAX-CUT Ising problems of size $N \leq 20$.  Experimental results followed for an $N=4$ OPO network\cite{Marandi2014} and an $N = 16$ network\cite{KentaThesis,Takata2016}, as well as simulations for G-set graphs\cite{Haribara2015} up to $N = 20000$.

In this chapter, I analyze the OPO Ising machine for solving the simplest class of Ising problems: 1D and 2D ferromagnetic chains.  Although these problems are trivial in the sense that the solutions are well-known, the analytic theory one can derive gives the reader a more lucid understanding of how the Ising machine actually works.  Because of their simplicity, 1D and 2D models may serve as a good way to ``benchmark'' the performance of different Ising machines.  Moreover, they are one of the simplest systems to realize in the laboratory, requiring only one delay line, allowing for direct comparison between the theory and currently realizable experiments.  As a model experimental system, I use the four-wave mixing fiber OPO implemented in our previous paper\cite{Inagaki2016}.

Section \ref{sec:theory} covers the theory of the time-multiplexed OPO Ising machine.  Based on this theory, I derive semiclassical equations of motion for the OPO pulse amplitudes.  In the original formulation of the Ising machine as a network of continuous-wave OPOs, these are stochastic differential equations\cite{Wang2013}, but for the pulsed case we show that they become {\it difference equations}, relating the pulse amplitudes between successive round trips.

These equations are solved in Sec.~\ref{sec:09-coll}, where I show that the dynamics breaks down into two stages: a {\it growth stage} where the field amplitudes are well below threshold and growth is linear, and a {\it saturation stage} where the OPO amplitudes saturate, giving rise to nonlinear dynamics defined by domains and domain walls.  Using this picture, Sec.~\ref{sec:statistics} derives expressions for the correlation length, domain-wall density and domain-length histogram for Ising machine solution.  This is compared to experimental data from the fiber OPO of Inagaki et al.\cite{Inagaki2016}; we show that our theory matches the experimental results, while a simple thermal Ising model does not.

Sections \ref{sec:09-2d}-\ref{sec:xy1d2d} explore more complex systems that have not yet been realized in OPO experiments.  In Sec.~\ref{sec:09-2d}, the two-dimensional lattice is treated.  The same growth / saturation stage picture applies, but during the latter we find 2D domains separated by 1D domain walls which move towards their center of curvature and collapse in a time quadratic in the domain size.  XY models are treated in Sec.~\ref{sec:xy}-\ref{sec:xy1d2d}, where the basic equations are introduced and applied to 1D and 2D systems.  Instead of domains, the XY model gives winding-number states for the 1D chain and vortices for 2D.  These vortices resemble those from Berezinskii-Kosterlitz-Thouless theory\cite{Berezinskii1971,Kosterlitz1974,Kosterlitz1973}, but they are generated by a non-thermal mechanism, and so their distribution is also athermal.

\section{Fiber OPO Theory}
\label{sec:theory}

First, we derive equations of motion for the pulse amplitudes in the cavity.  For concreteness, consider the case of a singly resonant $\chi^{(3)}$ fiber OPO (typical parameters, following Inagaki et al.\cite{Inagaki2016} are given in Table \ref{tab:09-t1}).  In this system, a narrowband filter ensures that the signal $a_i(t)$ is resonant, while the pump fields $b_i(t)$, $c_i(t)$ are not.  The nonlinearity is provided by the degenerate four-wave mixing process $2\omega_a \leftrightarrow \omega_b + \omega_c$ in the nonlinear fiber.

\begin{table}[tb]
\begin{center}
\begin{tabular}{c|cc}
\hline\hline
Term & Value & Description \\
\hline 
$\lambda_a$ & 1541 nm & Signal wavelength \\
$\lambda_b, \lambda_c$ & 1552 nm, 1531 nm & Pump wavelengths \\
$\gamma$ & $21\;\mbox{W}^{-1}\mbox{km}^{-1}$ & Fiber nonlinearity \\
$G_0$ & 7 dB & Fiber gain at threshold \\
$r, t$ & $1/\sqrt{2}$ & Delay mirror coefficients, $r^2 + t^2 = 1$ \\
$f$ & 2 GHz & Pulse frequency (time between pulses is $1/f$) \\
$\tau$ & 60 ps & Pulse width \\
$N$ & 10000 & Number of pulses \\ \hline\hline
\end{tabular}
\caption{Typical parameters for a pulsed four-wave mixing fiber OPO Ising machine.}
\end{center}
\label{tab:09-t1}
\end{table}

If the OPO network is viewed as a computer, the ``memory'' is stored in the signal pulse amplitudes $a_i(t)$, $i \in \{0, 1, 2, \ldots, N-1\}$ is the pulse index and $t \in \{0, 1, 2, \ldots\}$ is the round-trip number, which serves as a discretized time.  The ``processor'' consists of the $\chi^{(3)}$ fiber, a nonlinear map which acts on each pulse independently; and the delay line(s), which create a linear coupling between the pulses.  The ``inputs'' are the amplitudes of the pump pulses $b_i(t)$, $c_i(t)$, which can be programmed with an amplitude modulator placed in front of the pump laser.

Each round trip can be modeled as a cascade of three operations: nonlinear gain, coupling, and linear loss.  Ignoring vacuum noise, this gives the following map:
\beq
	a_i(t) \stackrel{\rm Fiber}{\longrightarrow} \frac{F[a_i(t)]}{e^{\alpha L_{\rm eff}/2}}
	\stackrel{\rm Loss}{\longrightarrow} \frac{F[a_i(t)]}{\sqrt{G_0}} 
	\stackrel{\rm Coupling}{\longrightarrow} \underbrace{\sum_j C_{ij}\frac{F[a_j(t)]}{\sqrt{G_0}}}_{a_i(t+1)} \label{eq:09-diffeq}
\eeq
Equation (\ref{eq:09-diffeq}) relates $a_i(t+1)$ to $a_i(t)$, giving us an equation of motion for the OPO network.  In the sections below, we obtain the nonlinear gain function $F[a_i(t)]$ and the coupling matrix $C_{ij}$, that form the core of (\ref{eq:09-diffeq}).  Once these are known, Ising machines of arbitrary complexity can be simulated.

\subsection{Nonlinear Fiber}

In the highly nonlinear fiber, the $\chi^{(3)}$ term gives rise to self-phase modulation (SPM) cross-phase modulation (XPM), and degenerate four-wave mixing (DFWM).  In the limit $|b|^2 \ll |c|^2$ with $c$ a flat-top pulse, SPM and XPM give constant phase shifts and can be cancelled by the appropriate phase matching\cite{AgrawalBook}, leaving only the DFWM term.  In this chapter I assume that the pulses are sufficiently long that the pulse amplitude is a constant (in time) and dispersion can be neglected; in this case the fields $a, b, c$ depend only on the distance $z$ the fiber, and the fiber field equations are\cite{AgrawalBook,BoydBook}:

\bea
	\frac{\d a}{\d z} & = & \gamma a^* b c - \frac{1}{2}\alpha a \\
	\frac{\d b}{\d z} & = & -\frac{1}{2}\gamma a^2 c^* - \frac{1}{2}\alpha b \\
	\frac{\d c}{\d z} & = & -\frac{1}{2}\gamma a^2 b^* - \frac{1}{2}\alpha c		
\eea

One can rescale the dependent variables $(a, b, c)$ to eliminate the constant $\gamma$; likewise, one can transform the independent variable $t$ to get rid of the linear absorption term.  With the field rescaling $x = (\gamma L_{\rm eff})^{-1/2}e^{-\alpha z/2} \bar{x}$ ($x = a, b, c$, $L_{\rm eff} = (1-e^{-\alpha L})/\alpha$) and length scaling $s = (1 - e^{-\alpha z})/(1 - e^{-\alpha L})$, the equations simplify to
\beq
	\frac{\d \bar{a}}{\d s} = \bar{a}^*\bar{b}\bar{c},\ \ \ 
	\frac{\d \bar{b}}{\d s} = -\frac{1}{2} \bar{a}^2 \bar{c}^*,\ \ \ 
	\frac{\d \bar{c}}{\d s} = -\frac{1}{2} \bar{a}^2 \bar{b}^* \label{eq:feom}
\eeq
and are solved on the interval $s \in [0, 1]$.  Gain occurs when $\bar{a}, \bar{b}, \bar{c}$ satisfy the correct phase relation.  Up to a global phase shift, this requires that $\bar{a}, \bar{b}, \bar{c}$ all be real and positive.  Taking $c(s) = c_{\rm in}$ constant since $c \gg a, b$, and using the constant of motion $B = \bar{b}^2 + \bar{a}^2/2$ from detailed balance, one derives:
\beq
	\frac{\d \bar{a}}{\d s} = c_{\rm in} \bar{a} \sqrt{B^2 - \bar{a}^2/2} \label{eq:09-dads}
\eeq

The fiber output is $a(z=L) = e^{-\alpha L_{\rm eff}/2} \bar{a}(s=1)$.  If the total cavity loss is $G_0$, then the field passes through an additional loss term $\sqrt{G_0 e^{\alpha L_{\rm eff}}}$, giving $a_{\rm out} = G_0^{-1/2} \bar{a}(1)$.  We find
\beq
	\bar{a}(1) = \bar{a}(0) e^{B \bar{c}_{\rm in}} \left[1 + (e^{2B\bar{c}_{\rm in}} - 1)\frac{1 - \sqrt{1 - \bar{a}(0)^2/2B^2}}{2}\right]^{-1} \label{eq:09-a1}
\eeq

The strong pump $c$ has a fixed amplitude, while the weak pump $b$ can be varied.  Define $b_0$ as the cavity threshold in the absence of coupling.  Linearizing (\ref{eq:09-a1}) in the limit $a \ll b$, we find that threshold is achieved when $e^{\bar{b}_0 \bar{c}_{\rm in}} = G_0^{1/2}$.  In terms of the $G_0$, the fiber input-output relation including both gain and loss is:
\beq
	a_{\rm out} = a_{\rm in} G_0^{\frac{1}{2}\bigl(\sqrt{(b_{\rm in}^2 + a_{\rm in}^2/2)}/b_0 - 1\bigr)}
	\left[1 + \bigl(G_0^{\sqrt{(b_{\rm in}^2 + a_{\rm in}^2/2)}/b_0} - 1\bigr) \frac{1 - \sqrt{2b_{\rm in}^2/(a_{\rm in}^2 + 2b_{\rm in}^2)}}{2}\right]^{-1} \label{eq:09-aout}
\eeq
Equation (\ref{eq:09-aout}) has two limits.  When $a_{\rm in} \ll b_{\rm in}$, the terms in the square brackets can be ignored and the field experiences linear gain: $a_{\rm out} = G_0^{\frac{1}{2}(b/b_0 - 1)} a_{\rm in}$.  Thus, the (power) gain for the fiber above threshold is $G_0^{b/b_0}$, and when $b > b_0$ this exceeds the cavity loss.  On the other hand, when $a_{\rm in} \gg b_{\rm in}$, the exponential inside the square brackets dominates and the field is substantially reduced.  This is the DFWM process working in reverse.

From quantum mechanics we know that the field is not defined by a scalar variable $a_i(t)$ but by a state in a harmonic potential.  In the truncated Wigner picture this gives rise to vacuum noise in the signal and pump fields\cite{Carter1995,Kinsler1991,Santori2014}.  To treat this, we need to add fluctuations to the fields $b, c$ before they are inserted: $b_{i,\rm in} \rightarrow b_{i,\rm in} + w^{(b)}_i$, $c_{i,\rm in} \rightarrow c_{i,\rm in} + w^{(c)}_i$, where $w^{(b,c)}_i$ are complex Gaussians that satisfy $\langle w^* w\rangle = \tfrac{1}{2}$, $\langle w \rangle = \langle w^2 \rangle = 0$.  This is the discrete-time analogue of vacuum noise.

To account for the quantum noise in $a_i$, it is easiest to assume that the loss happens in a lumped element after the fiber, rather than concurrently with the gain.  Near threshold, this is a reasonable approximation; elsewhere the noise is larger by a constant $O(1)$ factor.  Making use of this assumption, one must add vacuum fluctuations $a_{i,\rm out} \rightarrow a_{i,\rm out} + \sqrt{1 - 1/G_0} w^{(a)}_i$ to the signal.

All of these results can be applied to $\chi^{(2)}$ OPOs because the strong pump $c$ was presumed constant.  Removing it from Eqs.~(\ref{eq:feom}), one recovers the standard SHG equations, with $\epsilon = \gamma c$ as the $\chi^{(2)}$ parameter.

For a more realistic treatment of the pulsed OPO, one must abandon the continuous-wave picture in Eqs.~(\ref{eq:feom}) and treat the pulse shape itself as a dynamical variable.  The result is a ``multimode'' theory of the OPO, where the actual pulse is a weighted sum of normal modes.  This is a topic unto itself, which we have treated at length in the following chapter; a key finding is that if the cavity dispersion is large enough, or a sufficiently narrowband filter is inserted in the cavity, only a single normal mode resonates, and multimode effects can be ignored.  Although the multimode theory changes the exact expression $F(a_{\rm in})$, this ultimately does not matter.  We show in subsequent sections that the performance of the Ising machine depends only on the general form of $F(a_{\rm in})$: the gain at threshold and the near-threshold saturation (which goes as $O(a_{\rm in}^3)$).

\subsection{Coupling}

This section considers inter-pulse couplings mediated by delay lines and beamsplitters (Fig.~\ref{fig:09-f2}).  Recent experiments all use delay-line couplings\cite{Inagaki2016,KentaThesis,Marandi2014}, although it poses difficulties when many delay lines are involved.  A $d$-bit delay has five parameters: $r, t, r', t', \phi$, where $r^2 + t^2 = 1$, $(r')^2 + (t')^2 = 1$.  With fast modulators, in principle one can make all of these parameters (except $d$) pulse-dependent, giving them an index $i$.  Tracing the paths in Figure \ref{fig:09-f2}, and including the vacuum that enters through the lower-left beamsplitter, the input-output relation for a single delay is:
\beq
	a_i \rightarrow t'_i t_i a_i + r'_i r_{i-d} e^{i\phi_i} a_{i-d} + \left(t'_i r_i w^{(J)}_i + r'_i t_{i-d} e^{i\phi}w^{(J)}_{i-d}\right) \label{eq:09-ai-delay}
\eeq
where the $w_i^{(J)}$ are vacuum processes with $\langle w^* w\rangle = \tfrac{1}{2}$.  One must be careful to avoid negative indices: for instance $a_{-1}(t)$ maps to $a_{N-1}(t-1)$.

\begin{figure}[tb]
\begin{center}
\includegraphics[width=0.9\textwidth]{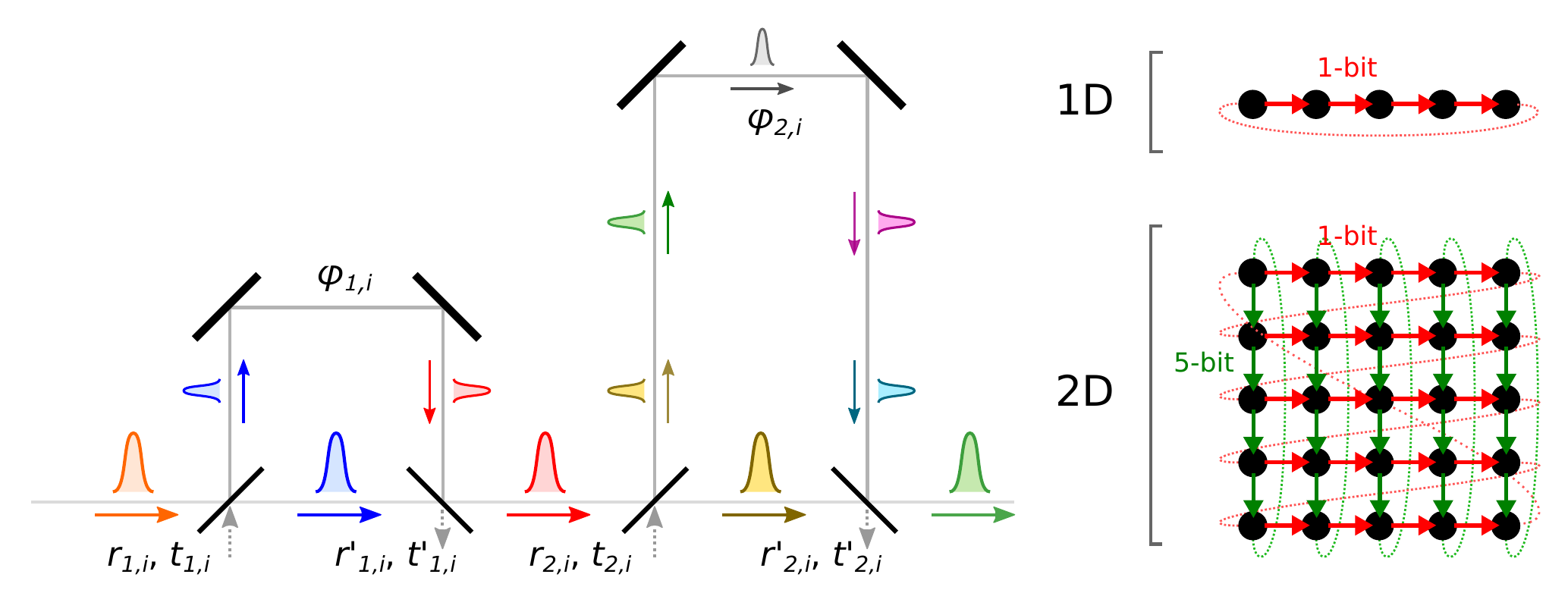}
\caption{Left: schematic a 1-bit and 5-bit delay line.  Right: Ising graphs implemented $N = 5$ with 1-bit delay (top), $N=25$ with 1-bit and 5-bit delay (bottom).}
\label{fig:09-f2}
\end{center}
\end{figure}

If the delays are static and $t = t'$, then (\ref{eq:09-ai-delay}) takes the simplified form:
\beq
	a_i \rightarrow t^2 a_i + r^2 e^{i\phi} a_{i-d} + tr \left(w^{(J)}_i + e^{i\phi}w^{(J)}_{i-d}\right) \label{eq:09-ai-delay2}
\eeq
This section will focus on the static-delay limit, since the experiments to date use static delays.  But the theory and code can accommodate the arbitrary case.

By relabeling the paths so that the long path is the ``cavity'' path and the short path is the ``delay'', and swapping $r \leftrightarrow t$, a delay can be converted into an ``advance'', which mixes $a_i$ with $a_{i+d}$ (again one must be careful with labeling; $a_N(t)$ corresponds to $a_0(t+1)$).  The cavity is enlarged by $d$, so $N \rightarrow N + d$.  Thus it is possible to engineer symmetric length-$d$ couplings using two identical $d$-bit delays.

A 1-bit delay implements the nearest-neighbor coupling of a 1D Ising chain.  To implement a 2D $m \times n$ lattice, one needs a 1-bit delay for the horizontal coupling and an $m$-bit delay for the vertical.  This gives a lattice with periodic but ``offset'' boundary conditions, as shown in Figure \ref{fig:09-f2}.  To implement the lattice without the offsets requires three delays, with time dependence; that case is not treated here.

\subsection{Linear and Near-Threshold Limits}
\label{sec:09-limits}

The fiber OPO has two analytically tractable limits: the linear case $a \ll b$ and the near-threshold case $b \approx b_0$.  These limits arise when we expand the fiber input-output relation (\ref{eq:09-aout}) to third order in $a_{\rm in}$:
\beq
	a_{\rm out} = a_{\rm in} \sqrt{G/G_0} \left[1 - \frac{G - (1 + \log G)}{8} (a_{\rm in}/b)^2 + O\left((a_{\rm in}/b)^4\right)\right] \label{eq:09-inout3}
\eeq
where
\beq
	G = G_0^{b/b_0} \label{eq:09-gain}
\eeq
The {\it linear limit} applies when $a \ll b$.  Taking only the linear term in (\ref{eq:09-inout3}) and combining it with (\ref{eq:09-ai-delay2}), one finds (for a single $d$-bit delay):
\beq
	a_i(t+1) = G_0^{\tfrac{1}{2}(b/b_0 - 1)} \left[t^2 a_i + r^2 a_{i-d}\right] + \mbox{(noise terms)} \label{eq:09-linear}
\eeq
The {\it near-threshold limit} applies when $b \approx b_0$.  In this case, (\ref{eq:09-inout3}) is expanded in powers of $(b - b_0)$:
\beq
	a_{\rm out} = a_{\rm in} + \left[\frac{\log G_0}{2}(b/b_0 - 1) - \frac{G_0 - (1 + \log G_0)}{8} (a_{\rm in}/b_0)^2\right] a_{\rm in} \label{eq:09-nt0}
\eeq
Combining this with (\ref{eq:09-ai-delay2}) and noting that $a_{\rm out} \approx a_{\rm in}$, we get a difference equation for $a_i(t)$.  Below it is written for a single $d$-bit delay:
\bea
	a_i(t+1) - a_i(t) & = & \left[\frac{\log G_0}{2}(b/b_0 - 1) - \frac{G_0 - (1 + \log G_0)}{8} (a_i(t)/b_0)^2\right] a_i(t) \nonumber \\
	& & \qquad +\ r^2 (a_{i-d}(t) - a_i(t)) + \mbox{(noise terms)} \label{eq:09-nt1}
\eea
Near threshold, the field $a_i(t)$ tends to vary slowly in both position and time.  This justifies replacing $a_i(t)$ with a smoothly-varying function $a(x, t)$ and swapping (\ref{eq:09-nt1}) with a PDE.  Ignoring the noise terms, it is:
\bea
	\frac{\partial a}{\partial t} + \frac{1}{2}\frac{\partial a^2}{\partial t^2} & = & \left[\frac{\log G_0}{2}(b/b_0 - 1) - \frac{G_0 - (1 + \log G_0)}{8} \frac{a^2}{b_0^2}\right] a \nonumber \\
	& & \qquad -\ r^2 d \frac{\partial a}{\partial x} + \frac{r^2 d^2}{2} \frac{\partial^2 a}{\partial x^2} \label{eq:09-nt2}
\eea

Steady-state solutions will drift with a speed $v_d = r^2 d$.  Substituting $x = \xi + v_d t$, one obtains a driftless equation of motion which, upon neglecting higher-order time-derivative terms ($\partial_t^2 a, \partial_t\partial_\xi a \ll \partial_t a$), yields:
\beq
	\frac{\partial a}{\partial t} = \left[\frac{\log G_0}{2}(b/b_0 - 1) - \frac{G_0 - (1 + \log G_0)}{8} \frac{a^2}{b_0^2}\right] a + \frac{r^2(1-r^2) d^2}{2} \frac{\partial^2 a}{\partial \xi^2} \label{eq:09-nt3}
\eeq
Although less tractable numerically, the steady state of (\ref{eq:09-nt3}) can be found analytically, yielding helpful insights about domain walls as discussed in the next section.

\section{Collective Dynamics of 1D Chain}
\label{sec:09-coll}

For the fiber OPO Ising machine, the evolution of the 1D chain is a two-stage process: in the {\it growth stage}, the field is weak compared to the saturation value, pump depletion can be ignored and the signal grows exponentially from the vacuum.  Because of inter-pulse coupling, different (Fourier) modes will grow at different rates, the ferromagnetic mode growing fastest.  This lasts for a time $T$, which is logarithmic in the saturation power and inversely proportional to the normalized pump amplitude.

In the {\it saturation stage}, the field saturates to one of two values: $a \rightarrow \pm a_0$.  The sign depends on the sign of the field after the growth stage.  Different regions will have different signs, called {\it domains} in analogy to the classical ferromagnet, and these domains will be separated by topological defects (domain walls).  The domain walls are not fixed, and their mutual attraction can cause some of the smaller domains to annihilate.

\subsection{Growth Stage}
\label{sec:09-growth}

In the growth stage, the field $a_i(t)$ follows Eq.~(\ref{eq:09-linear}).  Restricting attention to the 1D chain using a single delay line, this becomes:
\beq
	a_i(t+1) = G_0^{\frac{1}{2}(b/b_0 - 1)} \left[t^2 a_i + r^2 a_{i-1}\right] + \mbox{(noise terms)} \label{eq:09-linear2}
\eeq
The linear map (\ref{eq:09-linear2}) is diagonalized by going to the Fourier domain $a_i \rightarrow \tilde{a}_k$.  For small $k$, the result is:
\bea
	\tilde{a}_k(t+1) & = & G_0^{\frac{1}{2}(b/b_0-1)}\left(t^2 + r^2 e^{2\pi i k/N}\right) \tilde{a}_k(t) \nonumber \\
	& \approx & \underbrace{G_0^{\frac{1}{2}(b/b_0-1)} e^{-2t^2r^2(\pi k/N)^2}}_{\rm gain}\; \underbrace{\vphantom{G_0^{\frac{1}{2}(b/b_0-1)}} e^{ik(2\pi r^2/N)}}_{\rm drift} \; \tilde{a}_k(t) \label{eq:09-grev}
\eea

\begin{figure}[tbp]
\begin{center}
\includegraphics[width=1.00\textwidth]{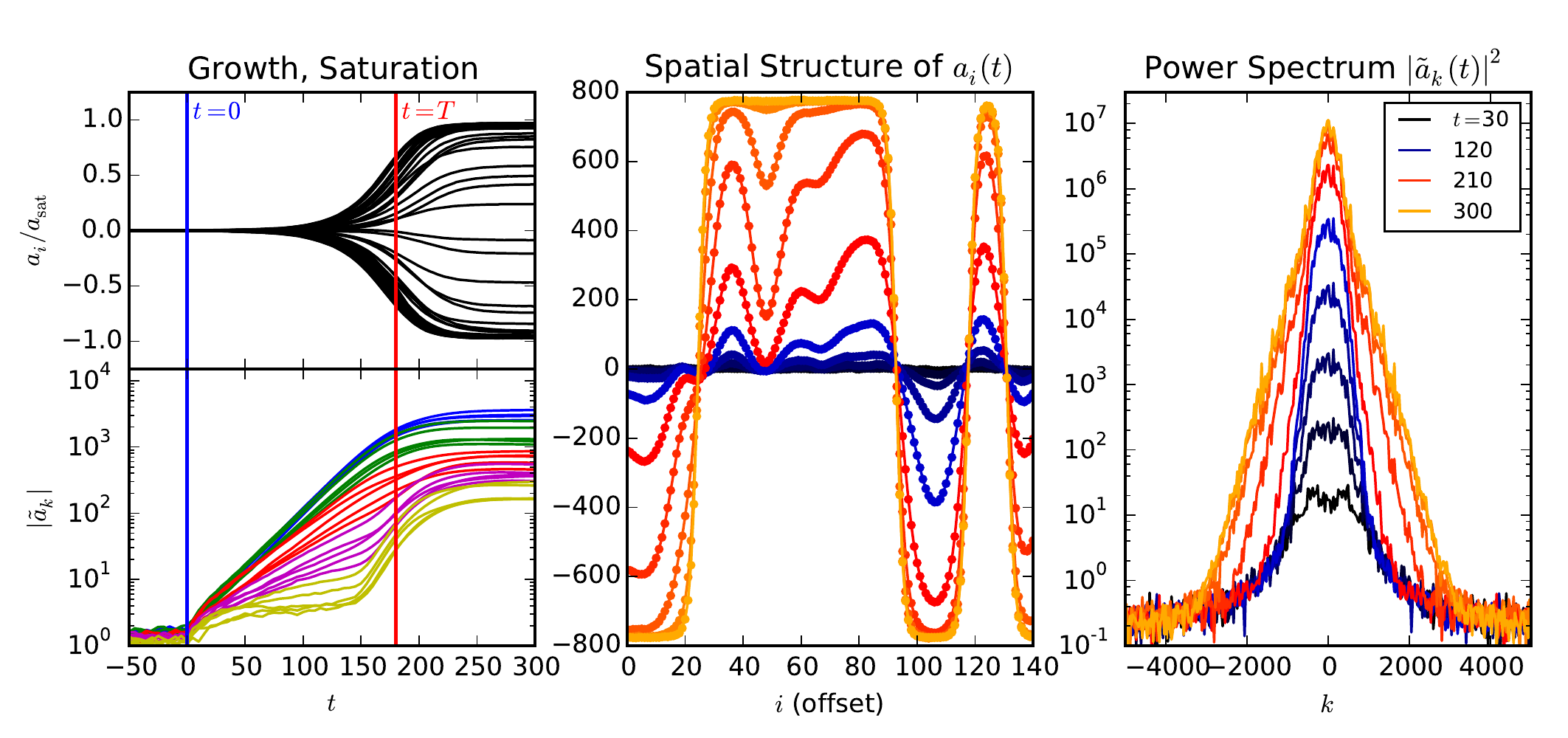}
\caption{Left: growth of OPO pulse amplitudes  $a_i$ (top) and Fourier modes $\tilde{a}_k$ (bottom).  Center: fields $a_i(t)$ for 1D chain at times $t = 30, 60, \ldots 300$ ($x$-axis shifted to cancel drift term).  Right: power spectrum $|\tilde{a}_k|^2$ at times $t = 30, 60, \ldots 300$.  Parameters: $G_0 = 7\;\mbox{dB}$, $b/b_0 = 1.05$}
\label{fig:09-f3}
\end{center}
\end{figure}

The two effects: gain and drift, are separated in Eq.~(\ref{eq:09-grev}).  Drift is a result of the unidirectional coupling.  For a single delay line, the drift speed is $v_d = r^2$.  The gain term depends on $k$, so different modes are amplified at different rates.  This amplification stops when the fields reach their saturation value.  If $N_{\rm sat}$ is the photon number at saturation and we start from vacuum noise, it takes approximately $\log(N_{\rm sat})/\log(G/G_0)$ round trips to reach saturation, that is:
\beq
	T = \frac{1}{b/b_0 - 1} \frac{\log (N_{\rm sat})}{\log(G_0)}
\eeq
Since $T$ depends only logarithmically on $N_{\rm sat}$, which is $O(10^5-10^7)$ in fiber OPOs, factors of two or three are not significant, so we can estimate $N_{sat} \rightarrow b_0^2$, the pump energy at threshold.

Starting with vacuum and propagating the growth equation (\ref{eq:09-grev}) $T$ time steps, we find that at the end of the growth stage the Fourier modes will be distributed as follows:
\beq
	\tilde{a}_k(T) \sim \sqrt{N_{sat}} e^{-2r^2t^2 T(\pi k/N)^2} \label{eq:09-akdist}
\eeq

The modes with smaller $k$ have larger amplitudes, suggesting that the nearest-neighbor interaction forms some kind of short-range order.  A good measure of this is the autocorrelation function $R(x)$.  Before saturation, $R(x)$ is also a Gaussian:
\beq
	R(x) \sim \langle a_i a_{i+x}\rangle = \sum_k e^{2\pi ikx/N} \langle \tilde{a}_k^*\tilde{a}_k \rangle
		\sim e^{-x^2/2x_0^2},\ \ \ x_0 \equiv \sqrt{2T}\,rt \label{eq:09-autocorr}
\eeq

\subsection{Saturation Stage}
\label{sec:09-crst}

In the next stage, pump depletion sets in and the fields inside the OPOs saturate.  The simplest way to model this is to assume that the interaction term $J$ is negligible at this stage.  Under this {\it simple saturation assumption} (SSA), the field in each OPO grows independently until it reaches one of two saturation values: $\pm \sqrt{G/\beta}$.  The {\it sign} of the initial field $c_i(T)$ is preserved, and all its amplitude information is lost.  This can be achieved with a sign function:

\beq
	a_i(\infty) = a_{\rm sat}\,\mbox{sign}[a_i(T)] \label{eq:09-sign}
\eeq

Rather than collapsing into a single ferromagnetic state, the system forms domains of fixed spin, separated by fixed domain walls.  This can be seen in the center plot of Figure \ref{fig:09-f3}.

However, Figure \ref{fig:09-f3} also reveals that the domain walls are not necessarily abrupt phase jumps as (\ref{eq:09-sign}) would have.  Depending on the coupling and pump strength, domain walls can be quite wide.  Near threshold, the shape admits an analytic solution via (\ref{eq:09-nt2}).  Replacing $a_i(t) \rightarrow a(x, t)$ as in Sec.~\ref{sec:09-limits}, a change of variables reduces (\ref{eq:09-nt2}) to the canonical form
\bea
	\frac{\partial\bar{a}}{\partial\bar{t}} & = & (1 - \bar{a}^2)\bar{a} + \frac{1}{2} \frac{\partial^2 \bar{a}}{\partial\bar{x}^2} \label{eq:09-static-norm} \\
	& & \left(\bar{a} = a_0^{-1} a,\ \ \ 
	\bar{x} = (x - vt)/\ell,\ \ \ 
	\bar{t} = t/\tau\right) \label{eq:09-static-norm-units}
\eea
where
\beq
	a_0 = 2b_0 \sqrt{\frac{(b/b_0-1)\log G_0}{G_0 - (1 + \log G_0)}},\ \ \ 
	\ell = \sqrt{\frac{2 r^2(1-r^2)}{(b/b_0-1)\log G_0}},\ \ \ 
	v = r^2,\ \ \ 
	\tau = \frac{2}{(b/b_0-1)\log G_0} \label{eq:09-wallparams}
\eeq
are the saturation field, domain wall length, drift speed, and relaxation time, respectively.  Equation (\ref{eq:09-static-norm}) has an analytic solution: $\bar{c} = \pm \tanh(\bar{x} - \bar{x}_{w})$.  This is the domain wall.  

The left plot of Figure \ref{fig:09-f5b} zooms in on a domain wall.  As the pump grows, the wall gets sharper, its width decreasing as $(b/b_0-1)^{-1/2}$ given in (\ref{eq:09-wallparams}).  If the pump is very strong or the coupling is weak, $\ell \lesssim 1$ and the smoothly-varying field assumption behind (\ref{eq:09-nt2}) breaks down.  However, it seems to hold quite well for the values chosen here (the solid lines in the figure are the $\tanh$ solution).

Domain walls are dynamic objects.  In the presence of a perturbation, they move.  Performing perturbation theory about the $\tanh$ solution, one finds that the Hessian is singular: most of its eigenvalues are $O(1)$ or larger, but for the vector $\partial c/\partial x$, it is zero.  While other perturbations are strongly confined, perturbations along the $\partial c/\partial x$ direction are unimpeded.  These correspond to moving the domain wall left or right.  We can deduce the {\it domain-wall velocity} by taking the inner product (the eigenvalues are orthogonal):
\beq
	\bar{v}_w = -\left[\int{\frac{\partial\bar{a}}{\partial\bar{x}} \frac{\partial\bar{a}}{\partial\bar{x}}d\bar{x}}\right]^{-1} \int{\frac{\partial\bar{a}}{\partial\bar{t}} \frac{\partial\bar{a}}{\partial\bar{x}}d\bar{x}} = -\frac{3}{4} \int{\mbox{sech}^2(\bar{x} - \bar{x}_w) \frac{\partial\bar{a}}{\partial\bar{t}}\,d\bar{x}} \label{eq:09-vw}
\eeq

Consider a function $\bar{a}(x, t)$ with two domain walls at $\pm\bar{L}/2$.  The precise way they are ``glued together'' at $\bar{x} \approx 0$ only matters to second order in the perturbation theory; $a = \tanh(\bar{L}/2 - |\bar{x}|)$ is a valid solution.  Applying (\ref{eq:09-vw}), one finds the following domain-wall speed and collision time:
\beq
	\bar{v}_w = \frac{3}{2}\mbox{sech}^4(\bar{L}/2),\ \ \ 
	\bar{T}_{\bar{L}} = \frac{1}{48}e^{2\bar{L}} \label{eq:09-tcoll}
\eeq

As the domain walls move, smaller domains will evaporate while large domains remain unaffected.  All the domains that survive after a time $t$ have a size $\bar{L} \geq (1/2)\log(48\bar{t})$.

\begin{figure}[tb]
\begin{center}
\includegraphics[width=1.00\textwidth]{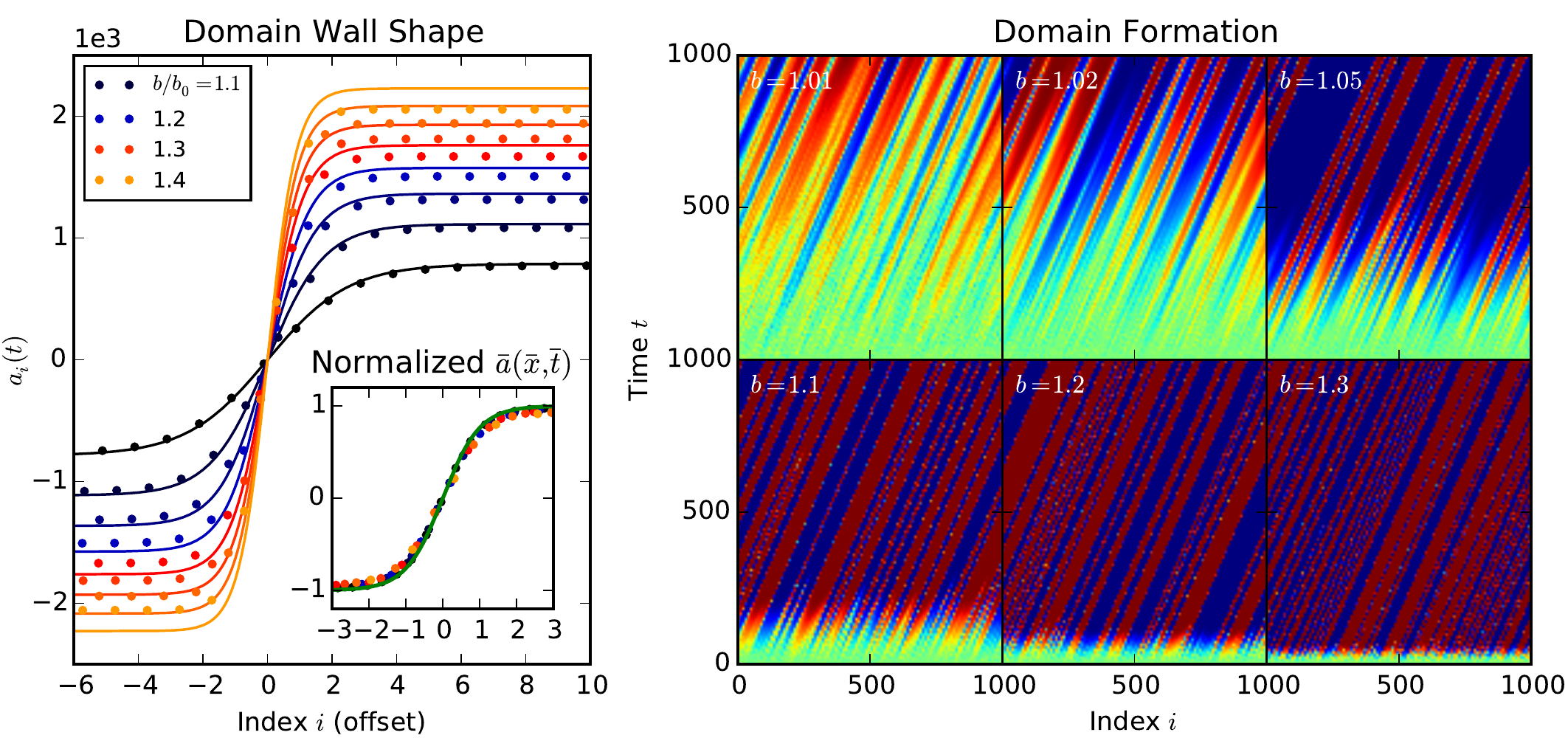}
\caption{Pulse amplitudes $a_i(t)$ near a domain wall as pump is swept slowly from $b/b_0 = 1.0$ to $1.4$ (normalized units in inset).  Right: color plot of pulse amplitudes $a_i(t)$ as function of index $i$ (horizontal) and time $t$ (vertical).  Pump values $b/b_0$ range from $1.01$ to $1.30$.}
\label{fig:09-f5b}
\end{center}
\end{figure}

The right plot in Fig.~\ref{fig:09-f5b} shows the formation of domain walls as a color plot in both the pulse index $i$ and time $t$.  The domain drift is obvious here.  In addition, the average domain size clearly shrinks the further the system is from threshold.  Looking closely, one also sees events where domain walls collide and annihilate some of the smaller domains -- but in general this is rare, because the domains that form by time $T$ tend to be moderate in size, and the lifetime (\ref{eq:09-tcoll}) can be quite long.

\section{Final-State Statistics}
\label{sec:statistics}

From the linear- and saturation-stage theory from Section \ref{sec:09-coll}, we can calculate statistical properties of the final-state ($t \rightarrow \infty$) system.  These properties are of interest because can be used to benchmark the performance of different Ising machines, or to compare the Ising machine against other optimizers.  In this section, we compute the autocorrelation function, defect density, success probability and domain-length histogram for the 1D Ising machine.  These are measurable quantities, allowing for a direct comparison between theory and experiment.

\subsection{Autocorrelation Function}

The autocorrelation function, given by $R(x) = \langle a_i a_{i+x} \rangle / \langle a_i^2 \rangle$, is a key quantity in statistical mechanics.  For the thermal Ising model with $H = -\tfrac{1}{2} J \sum_i \sigma_i \sigma_{i+1}$, it falls off exponentially with distance in one dimension, $R(x) = \tanh(\beta J)^x$.

Since the Ising machine is not in thermal equilibrium, we do not expect {\it a priori} that $R(x)$ will be exponential.  Indeed, at the end of the growth stage, Eq.~(\ref{eq:09-autocorr}) shows that $R(x)$ is a Gaussian.  The easiest way to compute $R(x)$ as $t \rightarrow \infty$ is to assume the simple saturation approximation (\ref{eq:09-sign}).  Replacing $a_i \rightarrow \mbox{sign}(a_i)$, the autocorrelation at $t \rightarrow \infty$ is found to be:
\beq
	R(x;\infty) = 1 - 2P(a_{i}(T)a_{i+x}(T) < 0) \label{eq:09-autocorr2}
\eeq
where $T$ is the saturation time.  Since the evolution in $t < T$ is approximately linear, the probability distribution of $a(T)$ is a two-dimensional Gaussian.  Its covariance is related to the autocorrelation at time $T$, $R(x;T) = e^{-x^2/2x_0^2}$:
\beq
	\sigma_{i,i+x} = \begin{bmatrix} 1 & e^{-x^2/2x_0^2} \\ e^{-x^2/2x_0^2} & 1 \end{bmatrix}
\eeq
Following (\ref{eq:09-autocorr2}), the autocorrelation may be expressed as an integral over a Gaussian with linear constraints:
\beq
	R(x) = 1 - 4 \int_{\mathcal{Q}} \frac{1}{2\pi\sqrt{\det\sigma}} e^{-\frac{1}{2}a^T\sigma^{-1}a} da_i da_{i+x} \label{eq:09-autocorr3}
\eeq
where $\mathcal{Q} = \{(a_i, a_{i+x}): a_i < 0, a_{i+x} > 0\}$ is the upper-left quadrant in $(a_i, a_{i+x})$.  To solve this, perform a linear transformation that diagonalizes the quadratic form in the exponent; $\mathcal{Q}$ is deformed to a pie slice, and the resulting integral is proportional to its angle.  The autocorrelation becomes:
\beq
	R(x) = 1 - \frac{4}{\pi} \tan^{-1}\sqrt{\tanh(x^2/4x_0^2)} \label{eq:09-rx-final}
\eeq

To compute $R(x)$ from the experimental data, one must first reconstruct the pulse amplitudes $a_i(t)$ from the measurement record.  In Inagaki et al.\cite{Inagaki2016}, no local oscillator is present, so the signal is passed through a Mach-Zehnder with a delay line, measuring the quantities $I_{1,i} = |a_i + a_{i+1}|^2$, $I_{2,i} = |a_i + a_{i+1}|^2$.  If the pulse energy $|a_i|^2$ is the same for each pulse, the angle between neighboring pulses is given by $\cos(\Delta\theta_i) = (I_{1,i}-I_{2,i})/(I_{1,i}+I_{2,i})$.  A negative value of $\cos(\Delta\theta_i)$ indicates a phase flip.  This is plotted in the upper-left panel of Fig.~\ref{fig:09-f6}.  Taking $a_i$ to be real for the degenerate OPO, we can invert the relation between the $a_i$ and the $I_{1,i}, I_{2,i}$ to reconstruct the original amplitude sequence $a_i(t)$.  It is then straightforward to compute the autocorrelation function and the correlation length.

The right plot of Fig.~\ref{fig:09-f6} shows the autocorrelation length as a function of pump amplitude, obtained by fitting experimental data to (\ref{eq:09-rx-final}).  The experimental $x_0$ agree with Eq.~(\ref{eq:09-autocorr}), with a particular fit for $b/b_0 \approx 1.4$ shown in the inset.

Although, Eq.~(\ref{eq:09-rx-final}) looks like an exponential to the unaided eye, plotting them on top of each other, the former is a much better fit to the experimental data, as shown in the inset plot.  However, it turns out that the best exponential fit to (\ref{eq:09-rx-final}) is $R(x) = e^{-x/x'_0}$, with $x'_0 = 1.00463x_0$.  Thus, we can obtain $x_0$ from experimental data by fitting the autocorrelation to an exponential.  The right plot in \ref{fig:09-f6} shows this for a variety of pump powers.  The agreement with experimental data is reasonably good.

\begin{figure}[tbp]
\begin{center}
\includegraphics[width=1.0\textwidth]{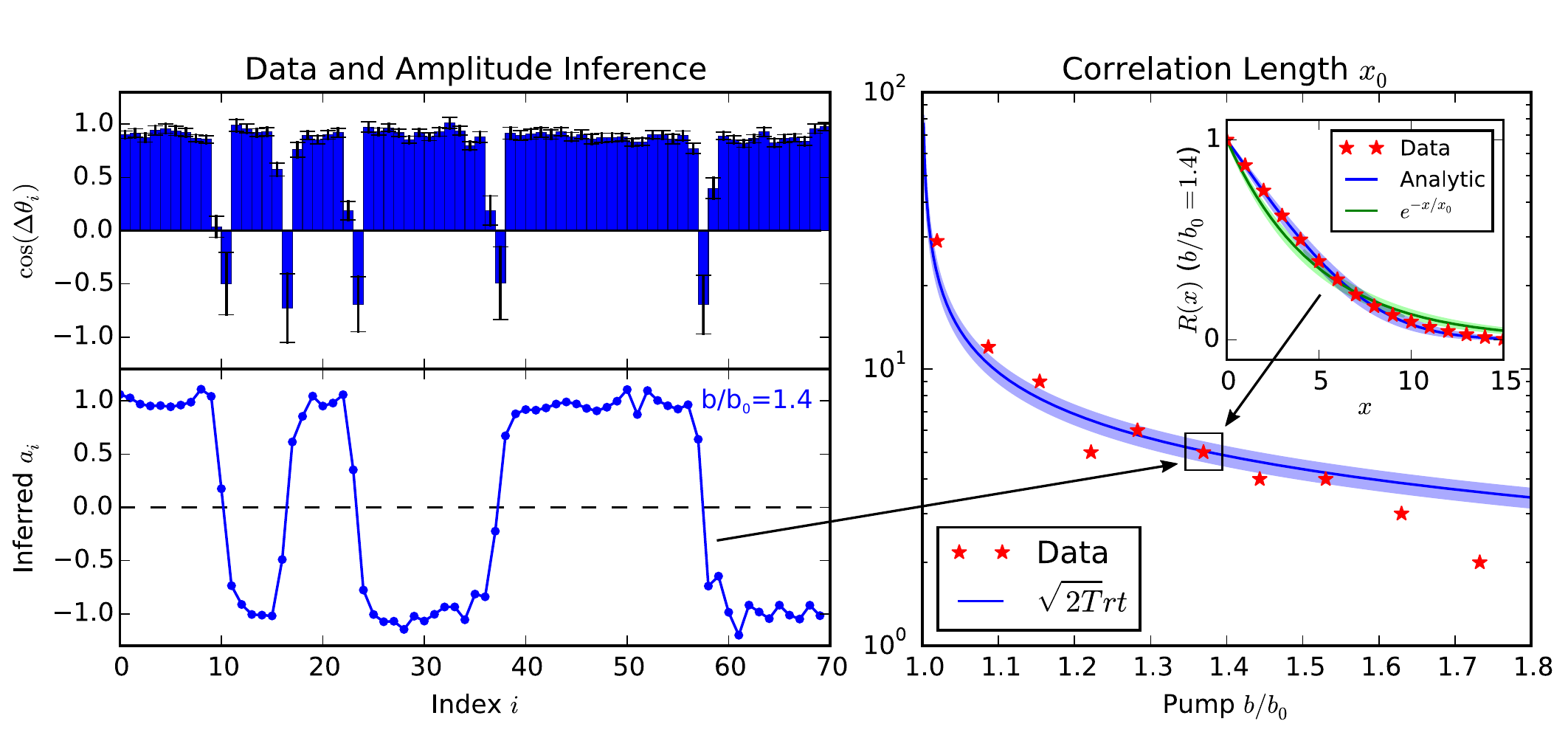}
\caption{Left: fiber OPO data for $\cos(\Delta\theta_i)$ (top) and reconstructed amplitude $a_i$ (bottom).  Right: autocorrelation length as a function of pump $b/b_0$, compared to Eq.~(\ref{eq:09-autocorr}).  Inset: autocorrelation $R(x)$ and analytic fits: form (\ref{eq:09-rx-final}) in blue, exponential in green.  Red stars are experimental data.  Shaded regions show sensitivity of the analytic curves to $N_{\rm sat}$ when varied from $4 \times 10^5$ to $4 \times 10^7$.}
\label{fig:09-f6}
\end{center}
\end{figure}

\subsection{Defect Density}

Another key statistic is the {\it defect (domain wall) density}.  This is the average number of domain walls divided by the size of the chain $n_d = N_d/N$.  The average domain length is then $L_d = 1/n_d$.  For a thermal Ising model with $H = -\tfrac{1}{2} J \sum_i \sigma_i\sigma_{i+1}$, one has $n_d = (1+e^{\beta J})^{-1}$.

Since $a_i(\infty)$ has fixed amplitude, one can compute $n_d$ from the autocorrelation function: $n_d = (1 - R(1))/2$.  For $x_0 \gtrsim 10$, $R(x)$ may be linearized about $x = 0$, giving the result:

\beq
	n_d = \frac{1}{\pi x_0},\ \ \ L_d = \pi x_0,\ \ \ x_0 = \sqrt{2T}\,rt \label{eq:09-nda}
\eeq

Figure \ref{fig:09-f7} (left) compares experimental data from Inagaki et al.\cite{Inagaki2016} (Fig. 3) to both Eq.~(\ref{eq:09-nda}) and numerical simulations.  The data match the simulations when $t \rightarrow \infty$, but deviate from Eq.~(\ref{eq:09-nda}).  This suggests that the full numerical model works well, but Eq.~(\ref{eq:09-nda}), which relies on the simple saturation assumption (\ref{eq:09-sign}), is inaccurate.  This is the result of domain-wall motion and collision in the saturation stage, which reduces the number of defects as $t \rightarrow \infty$.

\begin{figure}[tbp]
\begin{center}
\includegraphics[width=1.00\textwidth]{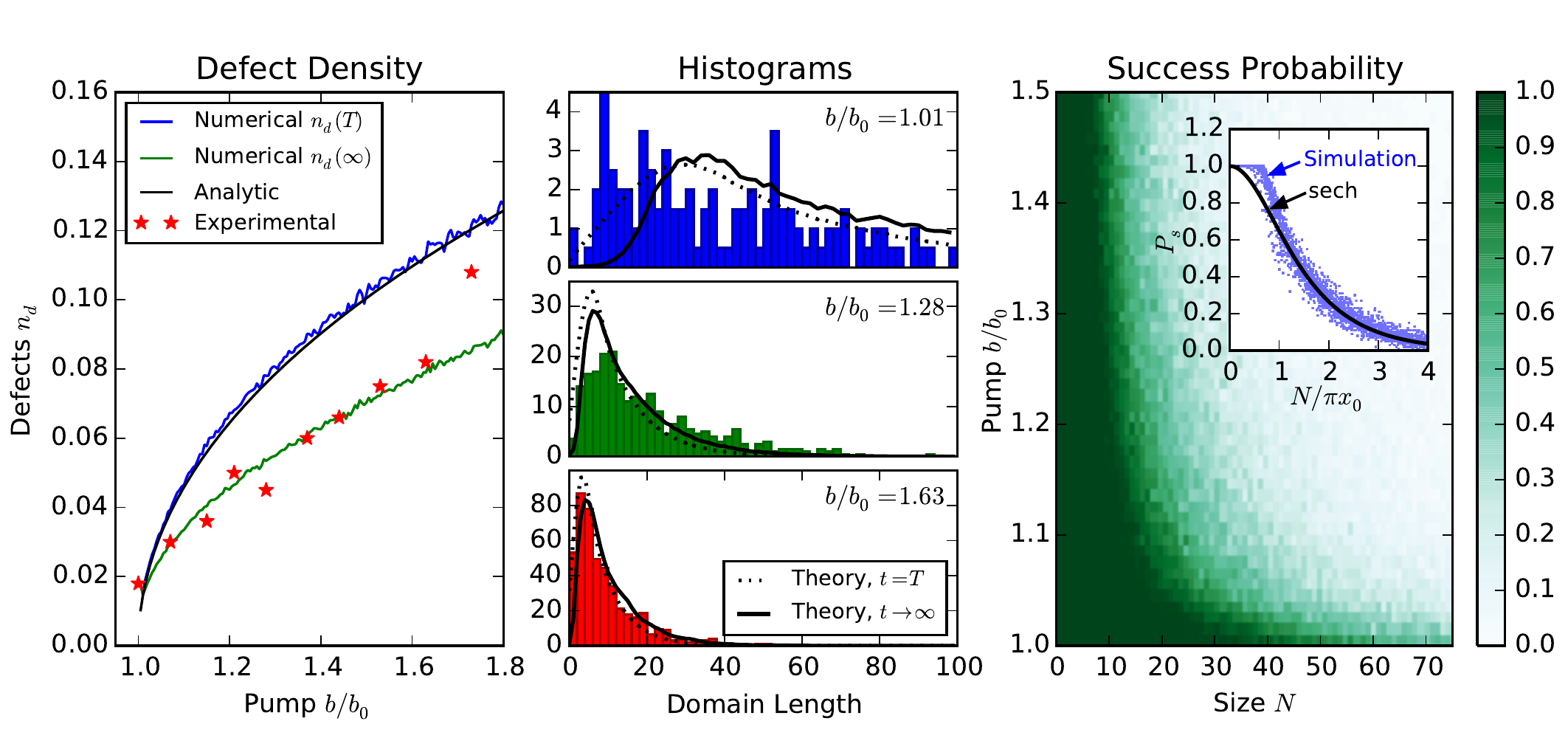}
\caption{Left: plot of defect density as a function of pump $b/b_0$, numerical and analytic models (Eq.~\ref{eq:09-nda}) compared to experimental data.  Center: domain length histograms for $b/b_0 = 1.01$, $1.28$ and $1.63$.  Bars denote experimental data.  Right: success probability $P_s$ as a function of system size $N$ and pump $b/b_0$.}
\label{fig:09-f7}
\end{center}
\end{figure}

\subsection{Domain Length Histograms}

Experimental data for the domain-length distribution $P(\ell)$ is plotted in Fig.~\ref{fig:09-f7} (center).  There is a reasonable fit between the data and numerical simulations as $t \rightarrow \infty$.  Note, however, that the calculated histogram at $t \rightarrow \infty$ differs from that at $t \rightarrow T$.  This difference reflects the domain-wall dynamics in the saturation phase.  In particular, since small domains evaporate faster than large domains, the population of small domains is depleted, and the average domain length grows.  Since $N_d L_d = N$, an increase in domain length results in a decrease in defect density, giving rise to the difference between the $t = T$ and $t = \infty$ lines in the left plot.

In a thermal Ising model, the probability distribution of spin $\sigma_{i+1}$ depends only on its nearest neighbors; mathematically this makes it a Markov chain in $i$.  Thus, the distribution $P(\ell)$ should be exponential in $L$: $P(\ell) \sim e^{-n_d \ell}$.  The histograms in Fig.~\ref{fig:09-f7} have exponential tails, but are clearly not exponential for $\ell$ near zero.  This means that the Ising machine never reaches thermal equilibrium, even when $t \rightarrow \infty$.  Rather, it ``freezes out'' fluctuations accumulated during the linear growth stage, through a highly nonlinear process involving domain wall motion and collisions.  Only if one waits an exponentially long time will the larger domains evaporate, bringing the machine to the ground state.

\subsection{Success Probability}

\begin{table}[bt]
\begin{center}
\begin{tabular}{p{0.2\textwidth}|p{0.25\textwidth}|p{0.25\textwidth}|p{0.2\textwidth}}
\hline\hline
  & Thermal 
  & CIM Theory 
  & Experiment \\
\hline
Mechanism 
  & Thermally-activated flips create a Boltzmann distribution.
  & Linear growth of OPO amplitudes, followed by saturation.
  & \\ \hline
Correlation $R(x)$
  & $e^{-x/x_0}$
  & See Eq.~(\ref{eq:09-rx-final})
  & Matches CIM \\
Corr.\ length $x_0$
  & $-1/\log(\tanh(\beta J/2))$
  & $\sqrt{2T} rt$
  & Matches CIM \\
Defect density $n_d$
  & $1/(1+e^{\beta J})$ 
  & $1/\pi x_0$
  & Matches CIM \\
Length dist.\ $P(\ell)$
  & $(1+e^{-\beta J})^{-\ell}$
  & Non-exponential, Fig.~\ref{fig:09-f7}
  & Matches CIM \\
Success probability
  & $\mbox{sech}(N e^{-\beta J})$
  & See Fig.~\ref{fig:09-f7}
  & $\approx 0$ for $N = 10000$ \\
\hline\hline
\end{tabular}
\end{center}
\caption{Comparison of thermal Ising model and the final state in the coherent Ising machine.}
\label{tab:09-t2}
\end{table}

The success probability $P_s$ of the Ising machine is defined as the probability that it reaches the ground state at some time $T_{\rm final}$.  The chosen $T_{\rm final}$ depends on experimental parameters, should be large compared to the saturation time $T$, but not exponentially large (since this would always give the ground state).  Figure \ref{fig:09-f7} (right) plots the success probability (numerically computed for $t \rightarrow \infty$) as a function of system size and pump power.  As expected, the probability is greatest near threshold for small systems, where the average defect number $N/\pi x_0$ is small.

If we assume the final state is thermal, the success probability can be calculated analytically.  For an $N$-spin ring with $H = -\tfrac{1}{2} J \sum_i \sigma_i\sigma_{i+1} + N/2$ and periodic boundary conditions, the partition function is:
\beq
	Z = \sum_n (1 + (-1)^n) e^{-n \beta J} = (1 + e^{-\beta J})^N + (1 - e^{-\beta J})^N
\eeq
The success probability is the ground-state probability for the system.  The ground state has energy zero and degeneracy 2, so $P_s = 2/Z$.  For low defect densities, $e^{-\beta J} \ll 1$, and the success probability becomes:
\beq
	P_s = \mbox{sech}\left(N e^{-\beta J}\right) \label{eq:ps-sech}
\eeq
Note that $e^{-\beta J}$ is the approximate defect density (for $e^{-\beta J} \ll 1$).  Thus, in analogy to the thermal model, we suspect that the success probability of the 1D Ising machine should depend on the defect density as well.  Using the relation $n_d \approx 1/\pi x_0$, in the inset figure, $P_s$ is plotted against $N/\pi x_0$ for all $(N, b/b_0)$ values shown in the larger plot.  Like the thermal model, the full Ising machine success probability falls off exponentially for high $N$, fitting reasonably well to the form $P_s = \mbox{sech}(N/\pi x_0)$.

\section{2D and Frustrated Systems}
\label{sec:09-2d}

\subsection{2D Square Lattice}
\label{sec:09-2dlat}

The two-dimensional Ising lattice exhibits richer physics than its 1D counterpart.  In particular, the thermal 2D system has a phase transition at finite temperature with long-range order below the transition temperature\cite{Onsager1944}.  Likewise, we suspect that a mechanism must exist to ensure long-range order in the 2D Ising machine.

Referring back to Figure \ref{fig:09-f2}, an $m \times n$ Ising lattice can be realized in an OPO network using a 1-bit and $m$-bit delay.  This implements the couplings $a_{i,i} \rightarrow a_{i,i+1}$, $a_{i,i}\rightarrow a_{i+1,i}$.  If the spins are serialized in C order $a_{i,j} \leftrightarrow a_{mi+j}$, then periodic boundary conditions $a_{i,n+1} = a_{i,1}$; $a_{m+1,j} = a_{a,j+1}$ are enforced.  There is a slight vertical offset compared to standard periodic boundary conditions (see Fig.~\ref{fig:09-f2}) but in the limit $m, n \gg 1$ with ferromagnetic couplings, this offset is negligible.

As before, the dynamics are described by a {\it growth stage} and a {\it saturation stage}.  In the growth stage, the Fourier modes are amplified independently, in analogy to Eq.~(\ref{eq:09-akdist}) we have:
\beq
	\tilde{a}_k(T) \sim \sqrt{N_{sat}} e^{-2r^2t^2 T \pi^2 (k_x^2+k_y^2)/N^2} \label{eq:09-akdist2}
\eeq

This gives the same autocorrelation function, generalized to two dimensions: $R(x) = e^{-(x^2+y^2)/2x_0^2}$, with $x_0 = \sqrt{2T}\,rt$.  Here $T = (b/b_0-1)^{-1} \log(N_{\rm sat})/\log(G_0)$ is the saturation time; see Sec.~\ref{sec:09-growth}.

\begin{figure}[tbp]
\begin{center}
\includegraphics[width=1.0\textwidth]{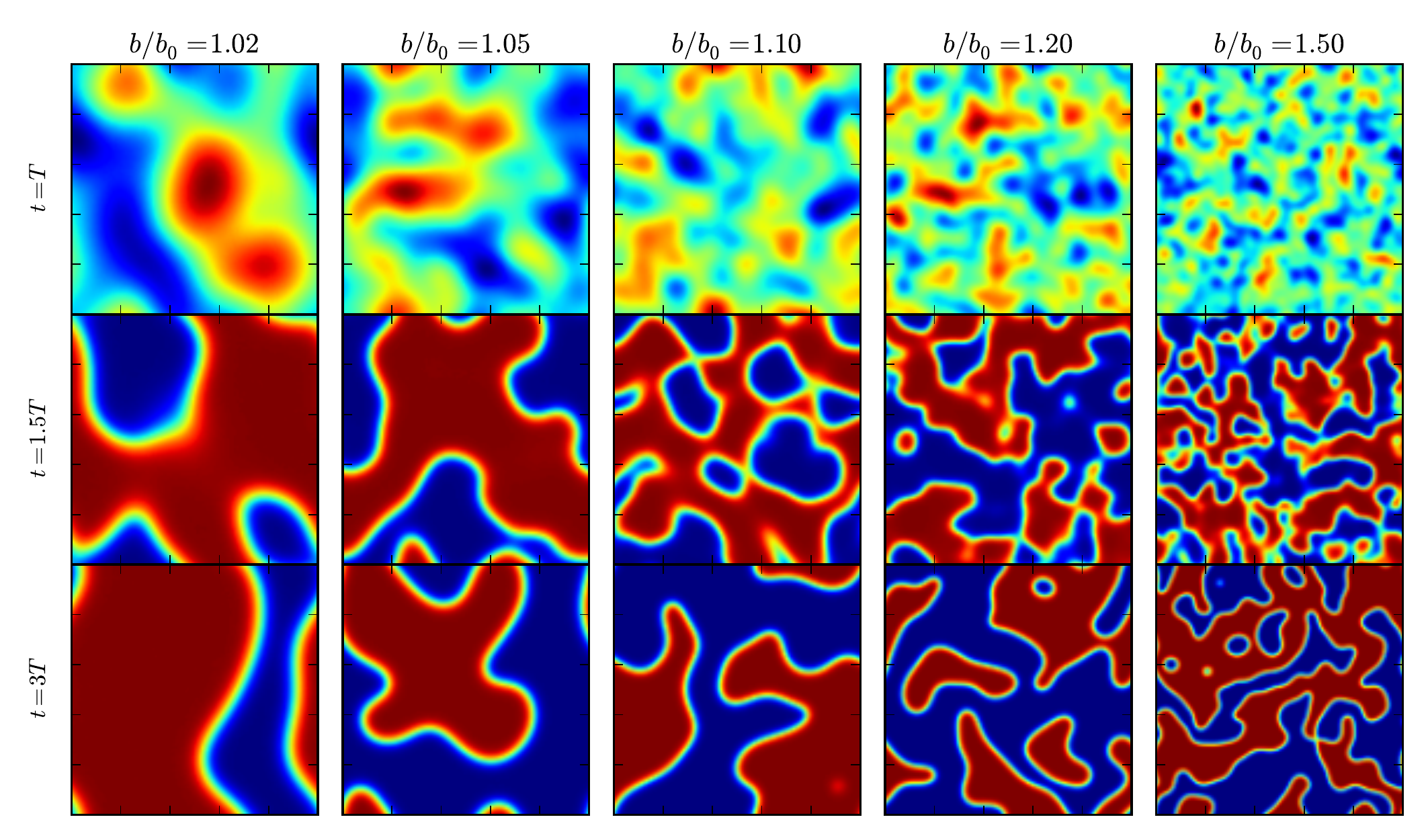}
\caption{Simulation of 2D OPO Ising machine, $100\times 100$ grid.  Pump ranges from $b/b_0 = 1.02$ to $1.50$.}
\label{fig:09-f9}
\end{center}
\end{figure}

Growth-stage fluctuations are imprinted on the domain structure of the OPO, and persist for some time.  Since these fluctuations are longer-range the larger the saturation time $T$, the Ising machine displays longer-range order when the pump is closer to threshold, just like the 1D case.  Figure \ref{fig:09-f9} shows the state of the machine for five different pump powers $b/b_0$.  The larger $b/b_0$, the smaller the domains that form.

After saturation, we can proceed analytically as long as the pump is near threshold.  Invoking the limit (\ref{eq:09-nt0}) and inserting both horizontal and vertical delays to obtain the two-dimensional analog of (\ref{eq:09-nt2}):
\bea
	a_{i,j}(t+1) - a_{i,j}(t) & = & \left[\frac{\log G_0}{2}(b/b_0 - 1) - \frac{G_0 - (1 + \log G_0)}{8} (a_{i,j}(t)/b_0)^2\right] a_{i,j}(t) \nonumber \\
	& & + \left[r^4 a_{i-1,j-1}(t) + r^2t^2 (a_{i-1,j}(t)+a_{i,j-1}(t)) - t^4 a_{i,j}(t)\right] \label{eq:09-nt4}
\eea

Near threshold, the field $a_{ij}(t)$ tends to vary slowly in both position and time.  Following the same procedures used to obtain (\ref{eq:09-nt3}), replaces the discrete increments with derivatives and drops higher-order $\partial^2a/\partial t^2, \partial^2a/\partial x\partial t, \partial^2a/\partial y\partial t$ terms, obtaining:
\beq
	\frac{\partial a}{\partial t} = \left[\frac{\log G_0}{2}(b/b_0 - 1) - \frac{G_0 - (1 + \log G_0)}{8} \frac{a^2}{b_0^2}\right] a + \frac{r^2(1-r^2)}{2} \left[\frac{\partial^2 a}{\partial \xi_x^2} + \frac{\partial^2 a}{\partial \xi_y^2}\right] \label{eq:09-nt5}
\eeq
where $\xi_x = x - r^2 t$, $\xi_y = y - r^2 t$ are the comoving coordinates.  Setting $\bar{a} = a_0^{-1} a$, $\bar{x} = (x-vt)/\ell$, $\bar{y} = (y-vt)/\ell$, $\bar{t} = t/\tau$, Eq.~(\ref{eq:09-nt5}) is converted to its canonical form (the 2D version of (\ref{eq:09-static-norm}))
\beq
	\frac{\partial\bar{a}}{\partial\bar{t}} = (1 - \bar{a}^2)\bar{a} + \frac{1}{2} \left(\frac{\partial^2 \bar{a}}{\partial\bar{x}^2} + \frac{\partial^2 \bar{a}}{\partial\bar{y}^2}\right) \label{eq:09-static-norm2}
\eeq
with $a_0$, $\ell$, $v$ and $\tau$ given in (\ref{eq:09-wallparams}).  The steady-state solutions to (\ref{eq:09-static-norm2}), $\bar{a} = \tanh(\bar{x}\cos\theta + \bar{y}\sin\theta)$, are linear domain walls.

\begin{figure}[btp]
\begin{center}
\includegraphics[width=1.00\textwidth]{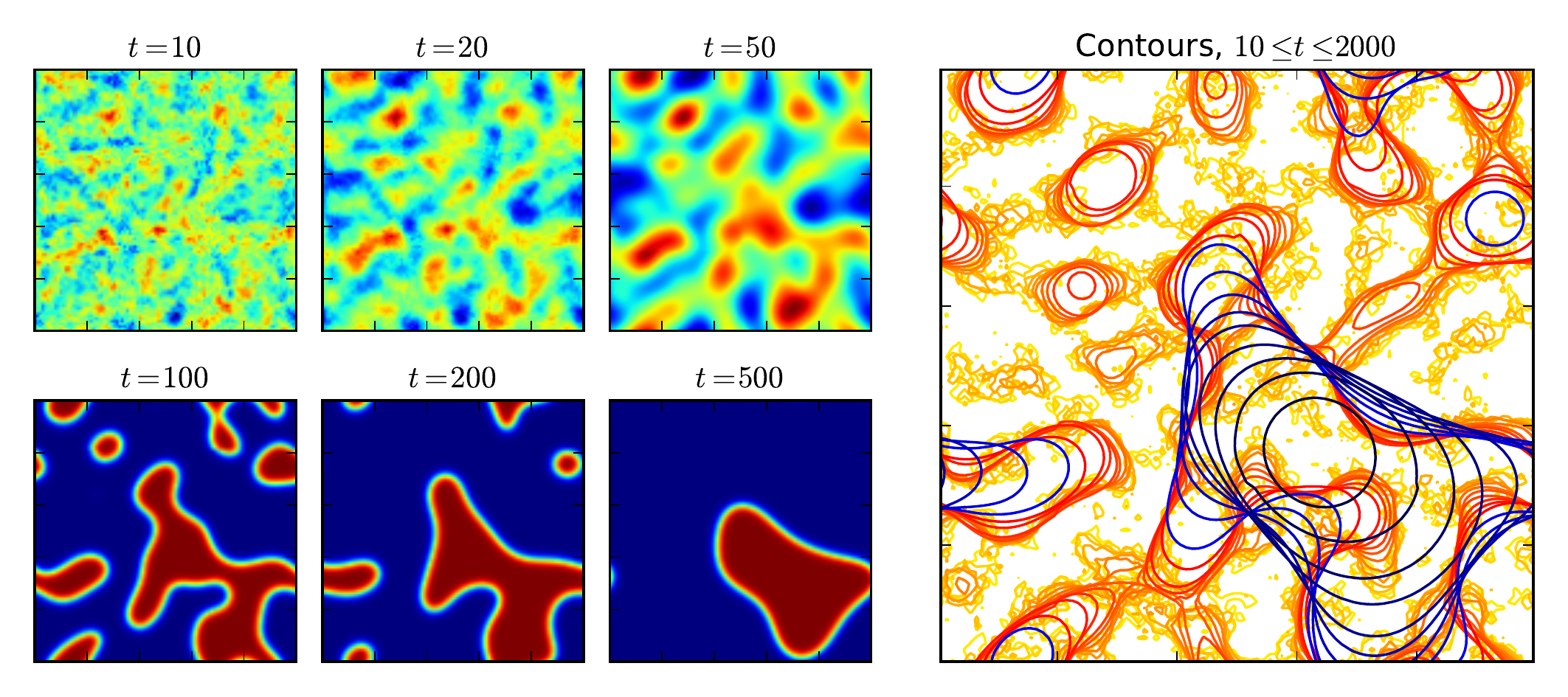}
\caption{Left: Simulation of 2D OPO Ising machine, $100\times 100$ grid, $b/b_0 = 1.1$.  Right: Location of domain walls for $10 \leq t \leq 2000$ (larger $t$ values are darker).}
\label{fig:09-f8}
\end{center}
\end{figure}

Curved domain walls will move towards the center of curvature at a rate proportional to $1/R$.  This can be seen intuitively if we imagine each spin on the wall picking a sign based on a majority vote of its neighbors.  The rate can be computed by considering the special case of a circular domain.  Working in cylindrical coordinates, (\ref{eq:09-static-norm2}) becomes:
\beq
	\frac{\partial\bar{a}}{\partial\bar{t}} = (1 - \bar{a}^2)\bar{a} + \frac{1}{2} \frac{\partial^2 \bar{a}}{\partial\bar{r}^2} + \frac{1}{2\bar{r}} \frac{\partial \bar{a}}{\partial\bar{r}} \label{eq:09-static-norm3}
\eeq

Here, $(2r)^{-1} \partial\bar{a}/\partial\bar{r}$ is a perturbation to the 1D equation (\ref{eq:09-static-norm}).  Applying the same results used to compute the attraction of neighboring walls (Eq.~(\ref{eq:09-vw})), the drift velocity is
\beq
	\bar{v}_w = -\frac{1}{2\bar{r}} \label{eq:09-vw-2d}
\eeq

For a circular domain of size $\bar{r}$, this gives $d\bar{r}/dt = -1/2\bar{r}$, which implies a collapse time of $\bar{T}_{\bar{r}} = \bar{r}^2$.  This time only scales quadratically with the domain size -- unlike the 1D case, where domains of size $L$ live for $(1/48)e^{2\bar{L}}$ time.  As a result, the 2D Ising machine on an $m\times n$ lattice should reach the ground state with high probability if allowed to run for $O(m^2, n^2)$ time.

Figure \ref{fig:09-f8} shows a simulation for $b/b_0 = 1.1$ ($T \approx 100$).  The top-left plots correspond to linear growth, which by $t = 100$ has saturated into domains.  Locally, the domain walls migrate towards their center of curvature, which the more tightly curved parts moving faster, following (\ref{eq:09-vw-2d}).  This can also be seen in the right plot, which superimposes the domain boundaries 23 time slices in $10 \leq t \leq 2000$.  Just after $t = 2000$, the system collapses into the ferromagnetic state.

\subsection{Frustrated Chains and Lattices}

In frustrated Ising models, different couplings compete and the resulting spin structure can be much richer than simple (anti-)ferromagnetism.  Most systems in classical and quantum physics involve frustration to some degree.  Moreover, frustration is a intimately connected to computational complexity; while non-frustrated Ising problems are trivial to solve, frustration makes the problem NP-hard in general\cite{Barahona1982}.

\begin{figure}[bp]
\begin{center}
\includegraphics[width=1.0\textwidth]{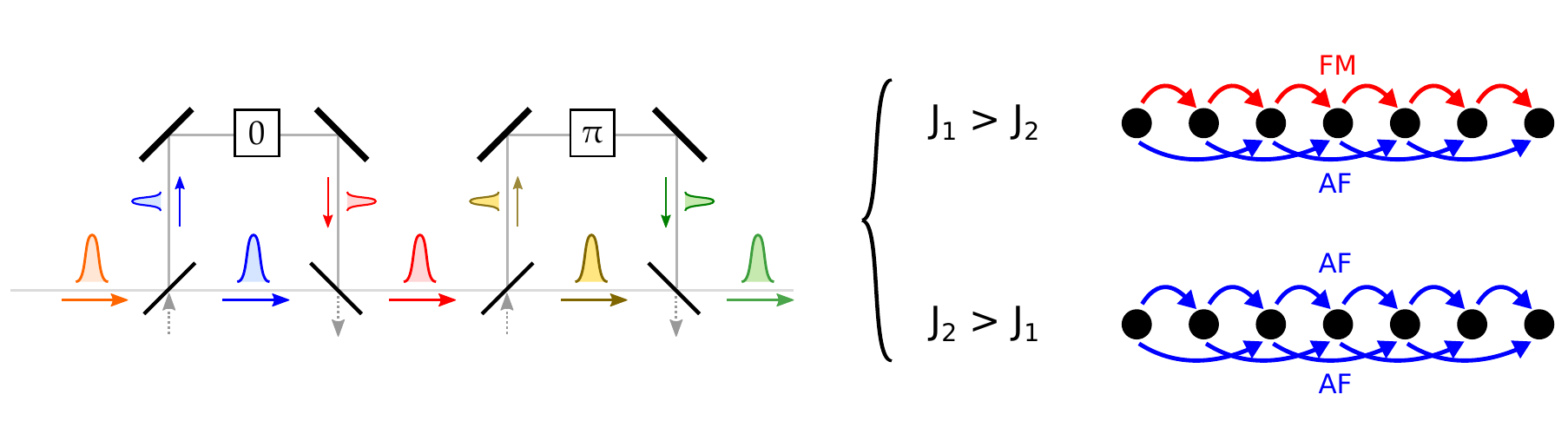}
\caption{Cascade of ferro- and antiferromagnetic couplings creates a frustrated spin chain.}
\label{fig:09-f11}
\end{center}
\end{figure}

The simplest way to introduce frustration to the 1D Ising chain is to cascade two 1-bit delays, one with phase 0 (beamsplitter $r = \sqrt{J_1}$) and one with phase $\pi$ ($r = \sqrt{J_2}$).  (This requires tunable beamsplitters, but the tuning only needs to happen on slow timescales.)  During the linear growth stage, the round-trip gain is:
\beq
    a(t+1) = G_0^{\frac{1}{2}(b/b_0-1)} \left[a_i(t) + (J_1 - J_2) a_{i-1}(t) - J_1 J_2 a_{i-2}(t)\right] \label{eq:09-fr-dadt}
\eeq

The nearest-neighbor coupling is ferromagnetic if $J_1 > J_2$, antiferromagnetic if $J_2 > J_1$, but in either case it wants to align next-nearest neighbors.  This conflicts with the next-nearest term in (\ref{eq:09-fr-dadt}), causing frustration.  (The case $J_1 = J_2$ is special because the nearest-neighbor term cancels out.  In this case, the even and odd spins decouple, so the chain can be ``unwrapped'' into two independent (antiferromagnetic) chains of size $N/2$).

Because of the time-invariant couplings, the eigenvectors will be Fourier modes.  For Fourier mode $k$, we have:
\beq
	G_k \equiv \left| \frac{\tilde{a}_k(t+1)}{\tilde{a}_k(t)} \right|^2 = G_0^{b/b_0-1} 
    \left[1 - 4J_1(1-J_1)\sin^2(k/2N)\right]\left[1 - 4J_2(1-J_2)\cos^2(k/2N)\right]
\eeq
There are three distinct possibilities:

\begin{enumerate}
    \item $G_k$ decreasing for all $k \in [0, \pi]$.  Maximum at $k = 0$.  Ferromagnetic order at growth stage.
    \item $G_k$ increasing on $k \in [0, \pi]$.  Maximum at $k = N\pi$.  Antiferromagnetic order at growth stage.
    \item $G_k$ non-monotonic.  Maximum for some $k \in (0, N\pi)$.  Frustrated system at growth stage.
\end{enumerate}
Examining the first derivatives of $G_k$ at $x = \{0, 1\}$, we deduce that:
\begin{eqnarray}
    J_2(1-J_2) < \frac{J_1(1-J_1)}{1 + 4J_1(1-J_1)} & \Leftrightarrow & \mbox{Ferromagnetic} \nonumber \\
    J_1(1-J_1) < \frac{J_2(1-J_2)}{1 + 4J_2(1-J_2)} & \Leftrightarrow & \mbox{Antiferromagnetic} \label{eq:09-phasediag}
\end{eqnarray}
Figure \ref{fig:09-f11} (left plot) illustrates the phase diagram defined by (\ref{eq:09-phasediag}).  

For weak couplings, there is a clear boundary between ferro- and antiferromagnetic behavior.  But for strong couplings, we get this interesting ``frustrated'' regime.  It's not hard to show that, in the frustrated regime, the $k$ with maximum gain is:
\beq
    k_{\rm max} = \cos^{-1}\left(\frac{J_1(1-J_1) - J_2(1-J_2)}{4J_1(1-J_1)J_2(1-J_2)}\right)
\eeq
Note that this is only defined in the frustrated region; elsewhere $k_{\rm max}$ is $0$ or $\pi$ depending on whether the dominant coupling is ferro- or antiferromagnetic.

Frustration increases the threshold beyond $b_0$, the uncoupled OPO threshold.  The gain at pump $b$ for the dominant mode is calculated to be:

\beq
	G_{\rm max} = \left\{\begin{array}{ll}
		G_0^{b/b_0-1} \left[1 - 4J_2(1 - J_2)\right] & \mbox{FM} \\
		G_0^{b/b_0-1} \left[1 - 4J_1(1 - J_1)\right] & \mbox{AF} \\
		G_0^{b/b_0-1} \frac{\left(J_1(1-J_1) + J_2(1-J_2) - 4J_1(1-J_1)J_2(1-J_2)\right)^2}{4J_1(1-J_1)J_2(1-J_2)} & \mbox{Frustrated} \end{array}\right.
\eeq

The gain still varies exponentially with $b$.  One can define the frustrated threshold $b'_0$ such that $G_{\rm max}(b'_0) = 1$, and threshold gain $G'_0 = G_{\rm max}(0)^{-1}$.  In terms of these quantities, the gain at $k_{\rm max}$ varies as $G_{\rm max} = (G'_0)^{b/b'_0-1}$, in analogy to (\ref{eq:09-gain})
		
\begin{figure}[tbp]
\begin{center}
\includegraphics[width=1.00\textwidth]{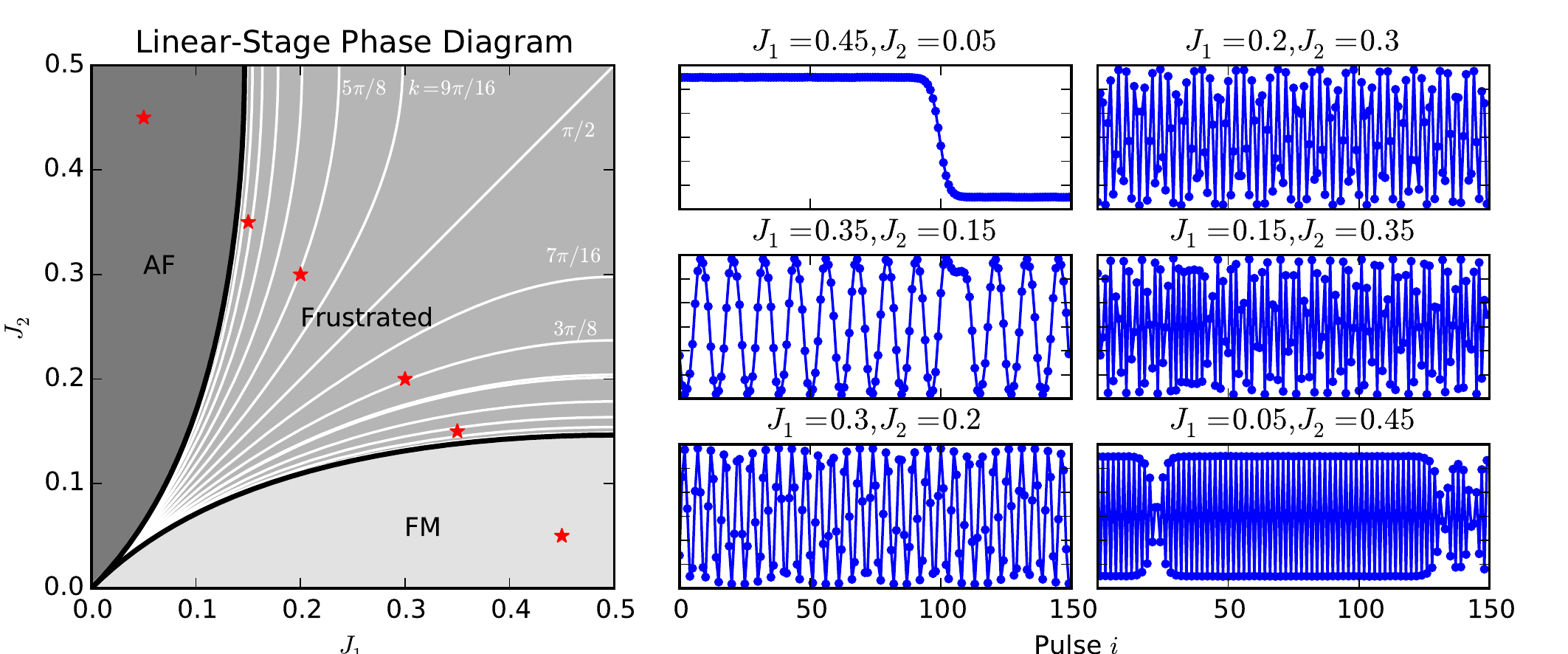}
\caption{Left: phase diagram for the frustrated chain.  Contours of $k_{\rm max}$ shown in white.  Red stars correspond to plots on the right.  Right: Ising machine output as the couplings $J_1$ and $J_2$ are varied.}
\label{fig:09-f11}
\end{center}
\end{figure}

The right plots in Figure \ref{fig:09-f11} show the transition from ferromagnetic to antiferromagnetic order as one passes through the frustration region in parameter space.  First, the ferromagnetic domains give way to an oscillatory order parameter, whose wavelength decreases until it starts to approximate antiferromagnetic order.  Eventually this leads to the antiferromagnetic domains in the lower-right plot.  The $J_1 > J_2$ and $J_2 > J_1$ regimes are related by a symmetry: replacing $J_1 \leftrightarrow J_2$ and $a_i \rightarrow (-1)^i a_i$, the equations of motion are unchanged.

\begin{figure}[p]
\begin{center}
\includegraphics[width=1.00\textwidth]{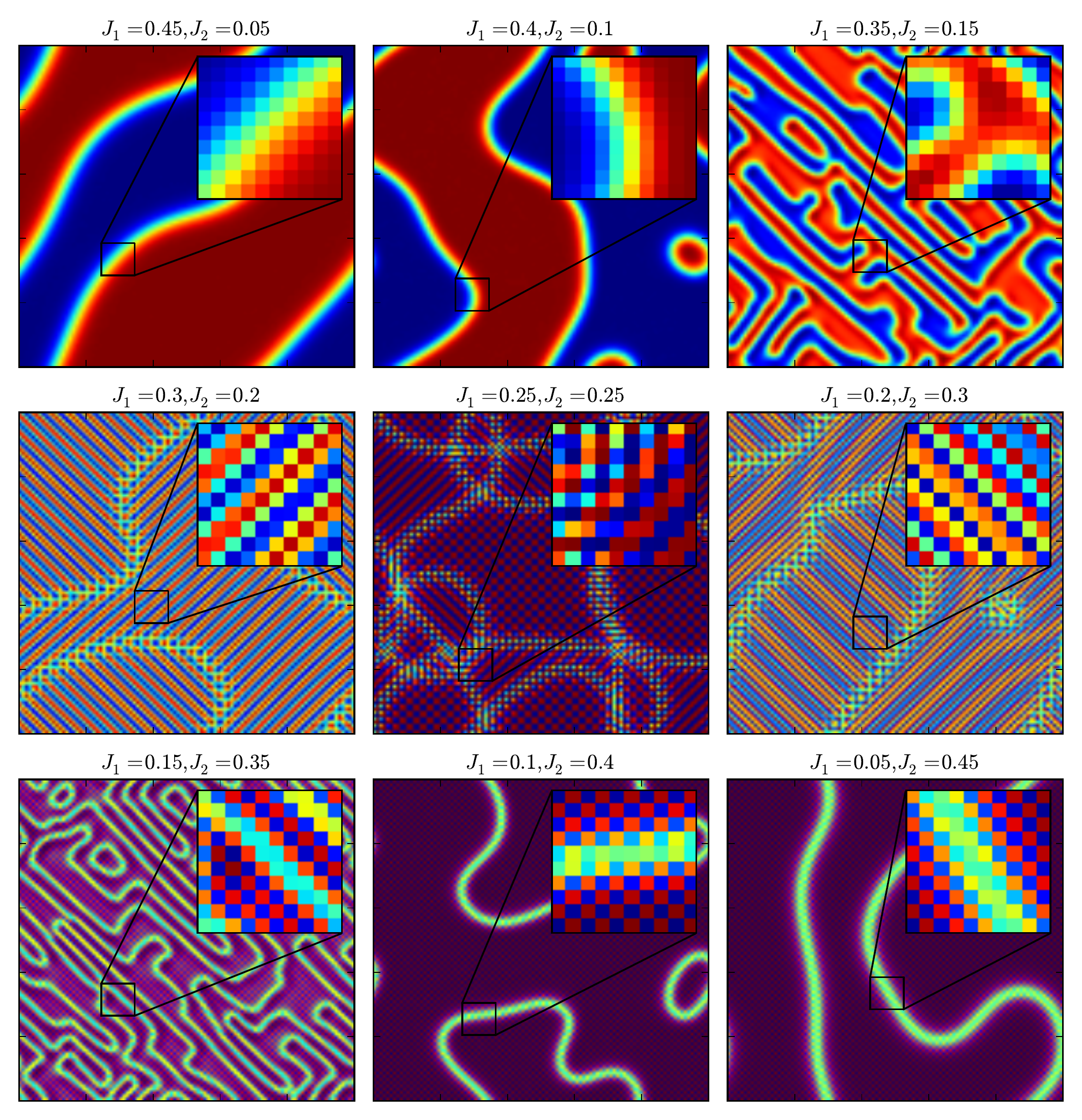}
\caption{Ising machine simulations ($b/b'_0 = 1.02$, $t=1500$) for frustrated system with $J_1, J_2$ ranging from mostly ferromagnetic (upper left) to mostly antiferromagnetic (bottom right)}
\label{fig:09-f12}
\end{center}
\end{figure}

Most of the theory developed above carries over to frustrated 2D lattices.  The phase diagram in Fig.~\ref{fig:09-f11} is unchanged, but now the Fourier modes in the growth stage have two wavenumbers $k_x, k_y$.  It is not hard to show that the mode gain is:
\begin{eqnarray}
G_k & = & G_0^{b/b_0-1} \left[1 - 4J_1(1-J_1)\sin^2(k_x/2N)\right]
    \left[1 - 4J_2(1-J_2)\cos^2(k_x/2N)\right] \nonumber \\
    & & \qquad \times\left[1 - 4J_1(1-J_1)\sin^2(k_y/2N)\right]
    \left[1 - 4J_2(1-J_2)\cos^2(k_y/2N)\right]
\end{eqnarray}

There are four frequencies that maximize $G_k$ are $(k_{\rm max}, k_{\rm max})$, $(-k_{\rm max}, -k_{\rm max})$, $(k_{\rm max}, -k_{\rm max})$, $(-k_{\rm max}, k_{\rm max})$.  These create upper and lower diagonal ``stripes'' (see Fig.~\ref{fig:09-f12}).  These striped domains compete with each other and form domains with domain walls in the frustrated region.

In addition to the stripes, Figure \ref{fig:09-f12} shows some interesting behavior in and near the frustrated zone.  At $J_1 = J_2$, one finds a doubling of the unit cell and three domain types appear to exist: stripes with $k=(\pm\pi/2, \pm\pi/2)$ and checkerboards. One should not read too much into this, because there is no nearest-neighbor coupling in this system, meaning it is equivalent to four $50\times 50$ lattices with antiferromagnetic coupling, interleaved in both $x$ and $y$ directions.

Near the frustration transition, the lattice forms filaments of opposite phase.  The ferromagnetic domain walls are subject to two nearly-equal opposing forces: nearest-neighbor interactions want to shrink and circularize the walls, as per Fig.~\ref{fig:09-f8}; on the contrary, the next-nearest neighbor effect wants to create striped order in the system.  The resulting structure is not unlike that of the manganites, where opposite phases coexist and percolate into each other\cite{Dagotto2013,Nagaev2002}.

\section{XY Machine Based on OPO}
\label{sec:xy}

\begin{figure}[b]
\begin{center}
\includegraphics[width=0.90\textwidth]{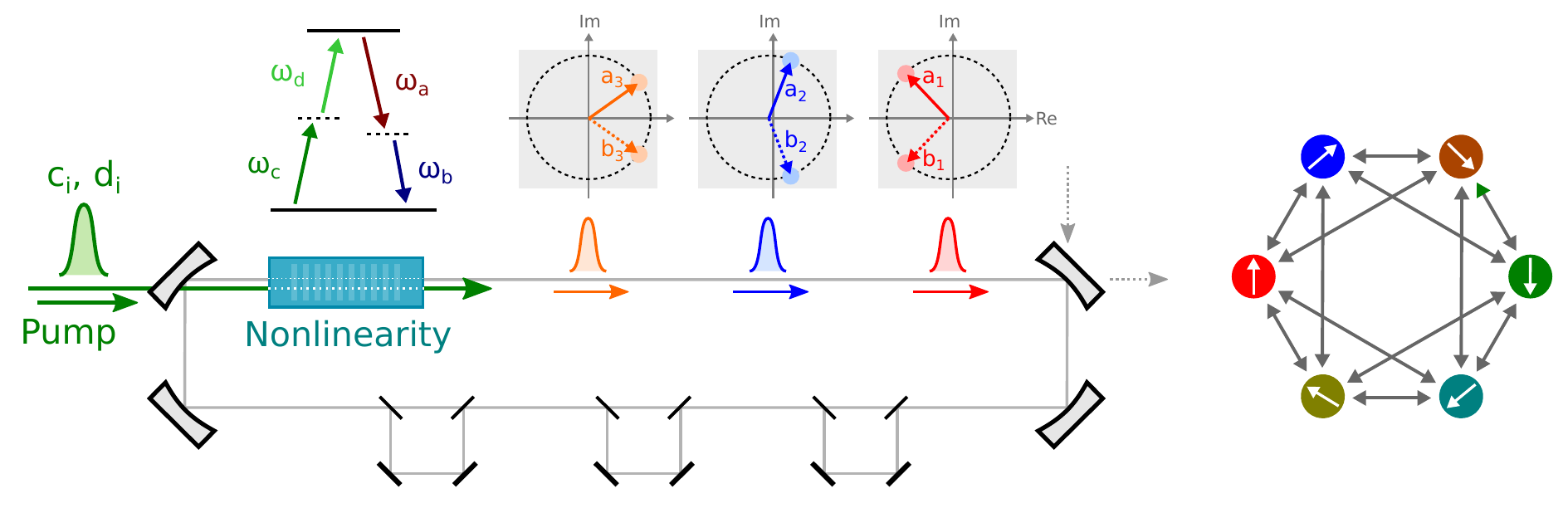}
\caption{Diagram of a time-multiplexed XY machine based on nondegenerate four-wave mixing.}
\label{fig:09-f13}
\end{center}
\end{figure}

In statistical physics, the XY model refers to a network of spins $\vec{\sigma}_i \in \mathcal{R}^2$, $|\vec{\sigma}_i^2| = 1$, with the Hamiltonian $U(\sigma) = -\tfrac{1}{2} \sum_{ij} J_{ij}\vec{\sigma}_i \cdot \vec{\sigma}_j$.  Each spin has a continuous $U(1)$ degree of freedom, rather than being discrete-valued.  It could equivalently be formulated in terms of angles, with $\vec{\sigma}_i = (\cos\phi_i, \sin\phi_i)$ living in a potential
\beq
	U(\phi) = -\frac{1}{2} \sum_{ij}{J_{ij} \cos(\phi_i - \phi_j)} \label{eq:09-uxy}
\eeq
Nondegenerate OPOs also have a $U(1)$ degree of freedom above threshold.  As a result, it is natural to map XY dynamics onto a nondegenerate OPO.  

\subsection{Gain Equations}

Consider a non-degenerate fiber OPO based on the four-wave mixing process $\omega_a + \omega_b \leftrightarrow \omega_c + \omega_d$ (Fig.~\ref{fig:09-f13}).

\begin{align}
	\frac{\d a}{\d z} = \frac{1}{2}\gamma b^* c d - \frac{1}{2}\alpha a & & 
		\frac{\d c}{\d z} = -\frac{1}{2}\gamma ab d^* - \frac{1}{2}\alpha c \\		
	\frac{\d b}{\d z} = \frac{1}{2}\gamma a^* cd - \frac{1}{2}\alpha b & & 
		\frac{\d d}{\d z} = -\frac{1}{2}\gamma ab c^* - \frac{1}{2}\alpha d
\end{align}

The pump fields $c$ and $d$ do not resonate.  We can assume without loss of generality that they are real.  The signal experiences gain when the phases of $a$ and $b$ are equal and opposite, that is $ab \in \mathbb{R}$.  If $(a, b)$ is a steady state in the OPO, so is $(a\,e^{i\phi}, b\,e^{-i\phi})$.  Thus the nondegenerate OPO has a ring of steady states, each with its own phase.  The spin $\vec{\sigma}_i$ is represented with this phase.

Rescaling $a, b, c, d, z$ to eliminate $\gamma$ and $\alpha$, the field equations are reduced to their canonical form.  Assuming $ab \in \mathbb{R}$ and real $c, d$:

\beq
	\frac{\d |\bar{a}|}{\d s} = |\bar{b}| \bar{c}\bar{d},\ \ \ 
	\frac{\d |\bar{b}|}{\d s} = |\bar{a}| \bar{c}\bar{d},\ \ \ 
	\frac{\d \bar{c}}{\d s} = -|\bar{a}||\bar{b}| \bar{d},\ \ \ 
	\frac{\d \bar{d}}{\d s} = -|\bar{a}||\bar{b}| \bar{c} \label{eq:09-xy-eom1}
\eeq

To proceed further, one assumes that one of the pump fields is much stronger than the other: $|d| \gg |c|$.  This allows us to ignore depletion in $d$ and treat $\bar{d}$ as a constant.  It is worth noting that the resulting system (in the $\alpha = 0$ limit) becomes equivalent to a $\chi^{(2)}$ OPO, with $\tfrac{1}{2}\gamma d \rightarrow \epsilon$ the $\chi^{(2)}$ coupling parameter.

The general three- and four-wave mixing problems can be solved analytically in terms of Jacobi elliptic functions\cite{Armstrong1962,Chen1989-2,Chen1989}.  Generally, two limits are of interest for OPOs: singly- and doubly-resonant.

\subsubsection{Singly Resonant Case}

For the singly resonant OPO, the initial idler amplitude is zero.  Following Armstrong et al.\cite{Armstrong1962} (Eq.~6.13) and including fiber and additional linear losses, the output signal is:
\beq
	a_{\rm out} = a_{\rm in} G_0^{-1/2} \sqrt{1 + \frac{c_{\rm in}^2}{a_{\rm in}^2}\left[1 - \mbox{cd}^2\left(\frac{1}{2}\gamma d_{\rm in} L_{\rm eff} \sqrt{a_{\rm in}^2 + c_{\rm in}^2}; \frac{c_{\rm in}^2}{a_{\rm in}^2 + c_{\rm in}^2}\right)\right]}
\eeq
where $\mbox{cd}(x;v)$ is a Jacobi elliptic function.  This linearizes for $a \ll c, d$ to $a_{\rm out} = G_0^{-1/2} \cosh(\tfrac{1}{2}\gamma c_{\rm in} d_{\rm in} L_{\rm eff}) a_{\rm in}$.  Since cavity losses are included here, the threshold $c = c_0$ is defined so that $a_{\rm out} = a_{\rm in}$.  The gain above threshold is:
\beq
	G = \cosh\left(\frac{c}{c_0} \cosh^{-1}(G_0^{1/2})\right)^2
\eeq

\subsubsection{Doubly Resonant Case}

If the signal and idler frequencies are similar enough and we don't filter one of them out, they will propagate through the cavity with the same $Q$ factor.  As a result, $a$ and $b$ will have the same magnitude.  If, furthermore, the overall phase is stabilized, then we have $a = b^*$.  All modes orthogonal to the $a = b^*$ subspace experience loss in the gain medium, and can be ignored.

Setting $b_{\rm in} = a_{\rm in}^*$ amounts to equating the constants of motion $A, B$.  Equations (\ref{eq:09-xy-eom1}) reduce to $\d \bar{a}/\d s = \bar{d}_{\rm in}\,\bar{a}\sqrt{\bar{c}_{\rm in}^2 + \bar{a}_{\rm in}^2 - \bar{a}^2}$.  This matches Eq.~(\ref{eq:09-dads}) up to scaling factors, so the input-output relation is analogous.  Converting to the form (\ref{eq:09-aout}), we have:
\beq
	a_{\rm out} = a_{\rm in} G_0^{\frac{1}{2}\sqrt{a_{\rm in}^2+c_{\rm in}^2}/c_0} \left[1 + (G_0^{\sqrt{a_{\rm in}^2+c_{\rm in}^2}/c_0} - 1) \frac{1 - \sqrt{1 - (a_{\rm in}/\sqrt{a_{\rm in}^2+c_{\rm in}^2})^2}}{2}\right]^{-1} \label{eq:09-inout-xy}
\eeq

Linearizing (\ref{eq:09-inout-xy}) for small input fields, we find $a_{\rm out} = G_0^{c/2c_0} a_{\rm in}$.  Thus the (power) gain for the waveguide pumped above threshold is the same as in the degenerate case:
\beq
	G = G_0^{c/c_0}
\eeq
Going to third order in $a_{\rm in}$, it is not hard to derive the XY version of Eq.~(\ref{eq:09-inout3}), valid when the pump is near threshold:
\beq
	a_{\rm out} = a_{\rm in} \sqrt{G/G_0} \left[1 - \frac{G - (1 + \log G)}{4} \frac{|a_{\rm in}|^2}{c^2} + O\left((a_{\rm in}/b)^4\right)\right] \label{eq:09-inout3-xy}
\eeq

As in the degenerate case, quantum noise can be modeled by adding vacuum fluctuations to the input pump fields $c_{i,\rm in} \rightarrow c_{i,\rm in} + w_i^{(c)}$, $d_{\rm in} \rightarrow d_{i,\rm in} + w_i^{(d)}$ (and idler $b_{\rm in} \rightarrow b_{i,\rm in} + w_i^{(b)}$, if the system is singly resonant) and treating the loss in signal as a lumped element after the gain medium: $a_{i,\rm out} \rightarrow a_{i,\rm out} + \sqrt{1 - 1/G_0}w_i^{(a)}$ (plus $b_{i,\rm out} \rightarrow b_{i,\rm out} + \sqrt{1 - 1/G_0}w_i^{(b)}$ if doubly resonant).  As before, $w$ is a discrete-time noise process with vacuum statistics: $\langle w^* w \rangle = \tfrac{1}{2}$.

In the rest of this chapter, I assume a doubly-resonant OPO for concreteness.  Because the signal and idler amplitudes are equal, the results are analytically more tractable.  But it is worth noting that the same calculations could be done using the singly-resonant results above.

\subsection{Couplings}

Tracing the paths in the canonical delay-line diagram (Fig.~\ref{fig:09-f2}), vacuum enters the cavity through the first beamsplitter.  The transmitted beam passes along the cavity without delay, while the reflected beam is delayed by one pulse spacing, contributing to $a_{i+d}$ instead.  There are five parameters: $r, t, r', t', \phi$, which can in principle vary in time.

\beq
	a_i \rightarrow t_i t'_i a_i + r_i r'_{i-d} e^{i\phi_i} a_{i-d} + \left(t_i r_i w^{(J)}_i + r_i t_{i-d} e^{i\phi}w^{(J)}_{i-d}\right) \label{eq:09-ai-delay-xy}
\eeq

Couplings will be more difficult to implement in the doubly-resonant regime because both signal and idler fields propagate with separate parameters $r, t, r', t', \phi$.  To maintain the condition $a = b^*$ the beamsplitter coefficients must be the same and the phases must be opposite:

\bea
	a_i & \rightarrow & t_i t'_i a_i + r_i r'_{i-d} e^{i\phi_i} a_{i-d} + \left(t_i r_i w^{(J,a)}_i + r_i t_{i-d} e^{i\phi}w^{(J,a)}_{i-d}\right) \label{eq:09-ai-delay-xy1} \nonumber \\
	b_i & \rightarrow & t_i t'_i b_i + r_i r'_{i-d} e^{-i\phi_i} b_{i-d} + \left(t_i r_i w^{(J,b)}_i + r_i t_{i-d} e^{-i\phi}w^{(J,b)}_{i-d}\right) \label{eq:09-ai-delay-xy2}
\eea

\section{1D and 2D XY Models}
\label{sec:xy1d2d}

Like the Ising machine, the XY machine is a dynamical system whose motion can be divided into two stages.  In the {\it growth stage}, quantum fluctuations are amplified from the vacuum.  The modes that are amplified the most are the largest eigenvalues of the coupling matrix $J_{ij}$.  The growth stage runs for a time $T = O((c-c_0)^{-1})$; the longer $T$, the more the state is resolved to the largest eigenvectors.  In the {\it Kuramoto stage}, the spin amplitude saturates and the system follows Kuramoto-model dynamics, which may be highly nonlinear.  After a while, it relaxes to a local minimum of the potential (\ref{eq:09-uxy}).

Both 1D and 2D XY models are intimately connected to the topology of $U(1)$.  Since the first homotopy group of the $d$-dimensional torus is $\pi_1(T_d) = \mathbb{Z}^d$, local minima are given by {\it winding states} which can be characterized by $d$ winding numbers $w_1, w_2, \ldots w_d$ for the $d$ dimensions\cite{NakaharaBook}.  In addition, in $d \geq 2$ dimensions, topologically protected vortices can form, which in thermal systems give rise to the Berezinskii-Kosterlitz-Thouless (BKT) vortex-pair transition\cite{Kosterlitz1973}.

\subsection{1D Chain}
\label{sec:xy1d}

A ferromagnetic 1D chain is realized with a single delay line of phase 0; see Sec.~\ref{sec:09-coll}.  The linear dynamics of $a_i(t)$ are the same as for the Ising model: working in the Fourier basis $\tilde{a}_k(t)$, the system of difference equations diagonalizes.  The initial quantum noise is amplified to macroscopic values.  At the saturation time $T = (c/c_0-1)^{-1} \log(N_{\rm sat})/\log(G_0)$, these Fourier modes have mean amplitude:
\beq
	\sqrt{\langle |\tilde{a}_k(T)|^2 \rangle} = \sqrt{N_{\rm sat}} e^{-2r^2 t^2 T(\pi k/N)^2} \label{eq:09-ak-xy}
\eeq

This has a correlation length $x_0 = \sqrt{2T}\,rt$.  The only difference here is both quadratures of $a$ experience gain in the XY model.  After the growth stage, the amplitude $a_i$ quickly saturates, but the phase is still free to move.  Assuming $a_i(t) = a_{\rm sat} e^{i\phi_i(t)}$, the phase is found to follow the difference equation:
\beq
	\phi_i(t+1) = \left[t^2 \phi_i + r^2 \phi_{i-1}\right] \label{eq:09-km-xy}
\eeq
Equation (\ref{eq:09-km-xy}) is a linear equation with the boundary condition $\phi_{N} = \phi_0 + 2m\pi$.  As in the growth stage, the best way to solve it is to use a Fourier series:
\beq
	\phi_x = \frac{m x}{N} + \sum_k \phi_k e^{2\pi ikx/N}
\eeq
Note that (\ref{eq:09-km-xy}) and (\ref{eq:09-linear2}) are the same up to the constant gain term.  Thus the eigenvalues for the $\phi_k$ will be:
\beq
	\lambda_k = \frac{\phi_i(t+1)}{\phi_i(t)} = t^2 + r^2 e^{-2\pi ik/N} = \underbrace{e^{-\frac{1}{2}(r t)^2(2\pi k/N)^2}}_{\rm diffusion}\, \underbrace{e^{ir^2(2\pi k/N)}}_{\rm drift} 
\eeq

\begin{figure}[tbp]
\begin{center}
\includegraphics[width=1.00\textwidth]{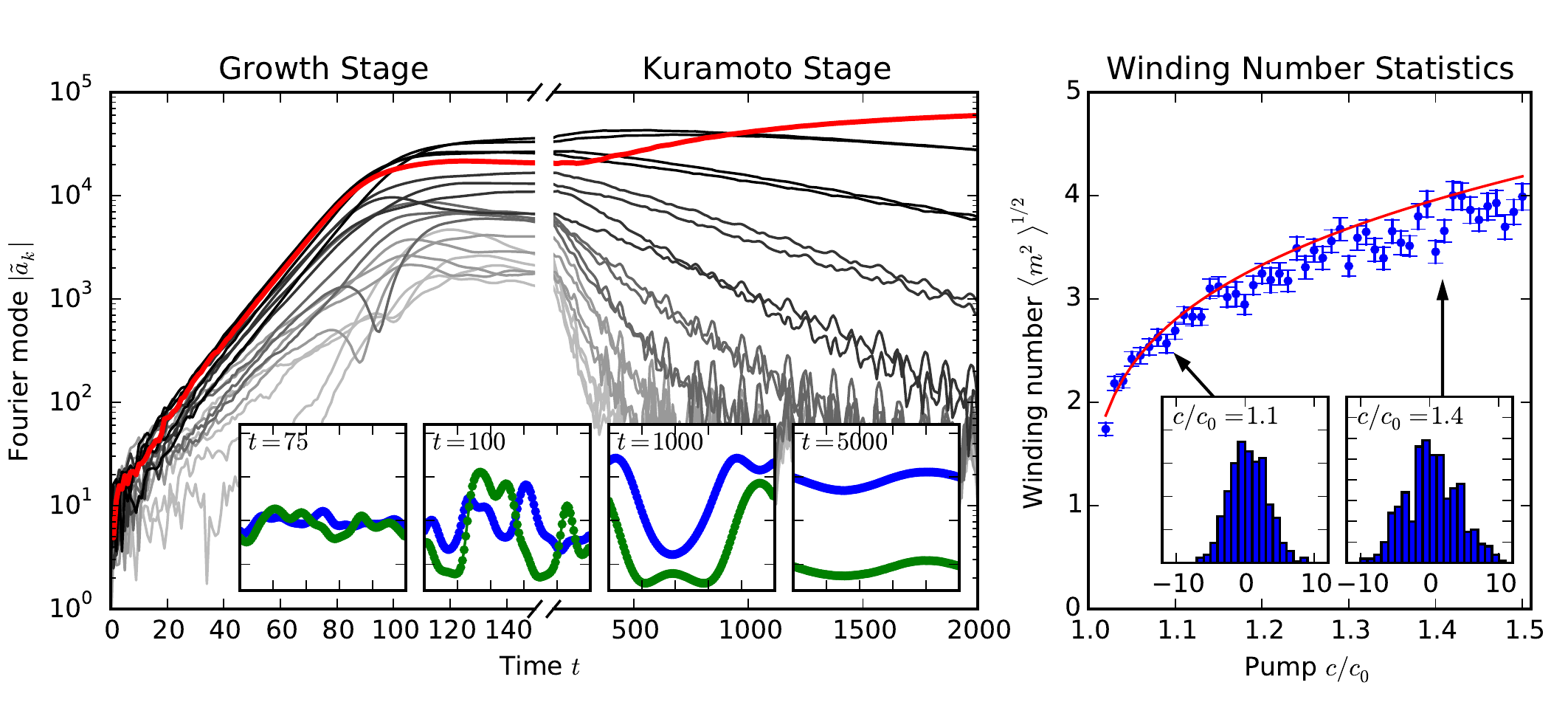}
\caption{Left: Fourier modes $|\tilde{a}_k(t)|$ for a 100-spin XY machine simulation.  Right: average winding number $\langle m^2\rangle^{1/2}$, with the fit $0.232\sqrt{N/x_0}$.}
\label{default}
\end{center}
\end{figure}

The steady state will be a state of constant winding $\phi_x = mx/N$.  For $m = 0$ this is the global minimum for the XY potential, for $m \neq 0$ an excited local minimum.  For sufficiently long chains, where $N \gg x_0$, the parts of the chain separated by $\gtrsim x_0$ are uncorrelated, so on these length-scales, the phase executes a random walk.  Thus the average number of windings is normally distributed about zero, with a standard deviation that goes as $\langle m^2\rangle^{1/2} \approx 0.232 \sqrt{N/x_0}$ (the constant must be determined numerically).

Note the two timescales in this problem.  The first is the growth-stage time.  If we want to reach the global minimum, the growth stage must be long enough for $x_0 \approx N$.  Since $x_0 = \sqrt{2T} rt$, this constrains the growth-stage time to be $T \gtrsim (N/rt)^2$. On the other hand, just to reach a local minimum, we must wait long enough in the Kuramoto stage for the phase excitations $\phi_k$ to decay to zero -- this takes $O(N/rt)^2$ time as well.  So no matter what kind of minimum we want, global or local, we must wait $O(N/rt)$ time, but to get the global minimum, this must happen in the growth stage, when the field is weak compared to saturation.

Another important thing to note is that the precise form of the nonlinear input-output map $a_{\rm in} \rightarrow a_{\rm out}$ does not matter.  In the growth stage, this map is linearized so all that matters is the gain, which determines the saturation time.  In the Kuramoto stage, since the amplitude saturates much more rapidly than the phase dynamics, the equation for $\phi_i$ does not even depend on the gain element.  This seems to suggest that all XY machines are equivalent when it comes to solving the 1D Ising problem.

\subsection{2D Lattice}

As far as local minima are concerned, the 2D lattice is just like a 1D chain in two directions.  The spins are indexed by two coordinates $a_{i,j}$ with the connections $a_{i,i}\rightarrow a_{i,i+1}$, $a_{i,i}\rightarrow a_{i+1,i}$ (Sec.~\ref{sec:09-2dlat}) and the equilibrium solutions are states of constant winding number: $a_{x,y} = e^{i(m_x x + m_y y)}$.  The growth stage is also analogous:  the Fourier amplitudes grow according to (\ref{eq:09-ak-xy}) so that the autocorrelation is $e^{-(x^2+y^2)/2x_0^2}$.

Having saturated the amplitude and thus reached the Kuramoto stage, the 2D model becomes quite different.  Topological {\it vortex defects} form and the dynamics are dominated by inter-vortex interactions.

\subsubsection{Vortex Shape, Frequency}

For an infinite lattice, an isolated vortex is a stable solution to the round-trip equations of the OPO.  Following the analysis leading to (\ref{eq:09-nt3}) and (\ref{eq:09-nt5}), which is applicable in the near-threshold limit, the round-trip equations can be rewritten as a nonlinear PDE with gain and diffusion:
\beq
	\frac{\partial a}{\partial t} = \left[\frac{\log G_0}{2}(c/c_0 - 1) - \frac{G_0 - (1 + \log G_0)}{4} \frac{|a|^2}{c_0^2}\right] a + \frac{r^2(1-r^2)}{2} \left[\frac{\partial^2 a}{\partial \xi_x^2} + \frac{\partial^2 a}{\partial \xi_y^2}\right] \label{eq:09-nt6}
\eeq
where $\xi_x = x-r^2 t$, $\xi_y = y-r^2t$.  This differs from (\ref{eq:09-nt5}) only in that $a$ is complex-valued here.  Setting $\bar{a} = a_0^{-1} a$, $\bar{x} = (x-vt)/\ell$, $\bar{y} = (y-vt)/\ell$, $\bar{t} = t/\tau$, this equation is converted to its canonical form:
\beq
	\frac{\partial\bar{a}}{\partial\bar{t}} = (1 - |\bar{a}|^2)\bar{a} + \frac{1}{2} \left(\frac{\partial^2 \bar{a}}{\partial\bar{x}^2} + \frac{\partial^2 \bar{a}}{\partial\bar{y}^2}\right) \label{eq:09-static-norm2}
\eeq
with the constants
\begin{align}
	a_0 &= c_0 \sqrt{2\frac{(c/c_0-1)\log G_0}{G_0 - (1 + \log G_0)}},
	& \ell &= \sqrt{\frac{2 r^2(1-r^2)}{(c/c_0-1)\log G_0}}, \nonumber \\
	v &= r^2,
	& \tau &= \frac{2}{(c/c_0-1)\log G_0} \label{eq:09-wallparams2}
\end{align}
Going to polar coordinates $(r, \phi)$, the vortex is the solution $A(r) e^{\pm i\phi}$ with $A(r)$ satisfying the differential equation:
\beq
	\frac{1}{2} \left(A'' + \frac{1}{r} A'\right) + \left(1 - A^2 - \frac{1}{2r^2}\right)A = 0 \label{eq:09-vorta}
\eeq

It turns out that $A(r) \approx \tanh (r)$ is a good approximation for the amplitude.  This comes from the fact that $\tanh (r)$ is a solution to (\ref{eq:09-vorta}) if the $A'/2r$ and $A/2r^2$ terms are ignored, and these terms nearly cancel out for the solution $\tanh(r)$.

To calculate the number of vortices at time $T$, one finds the probability that the phase winds $2\pi$ around one unit cell of the lattice.  Defining $a_{\rm sq} = [a_P, a_Q, a_R, a_S]$ as the pulse amplitudes for four corners of any lattice cell, clockwise ordered, the joint probability is a Gaussian with the covariance matrix:
\beq
	\avg{a_{\rm sq}a_{\rm sq}^T} = \begin{bmatrix} 1 & e^{-1/2x_0^2} & e^{-1/x_0^2} & e^{-1/2x_0^2} \\
		e^{-1/2x_0^2} & 1 & e^{-1/2x_0^2} & e^{-1/x_0^2} \\
		e^{-1/x_0^2} & e^{-1/2x_0^2} & 1 & e^{-1/2x_0^2} \\
		e^{-1/2x_0^2} & e^{-1/x_0^2} & e^{-1/2x_0^2} & 1 \end{bmatrix}
\eeq
In the near-threshold limit, $x_0 \gg 1$.  Conditioned on the mean value $\mu_{\rm sq} = \tfrac{1}{4}(a_P+a_Q+a_R+a_S)$, $\hat{a}_{\rm sq} \equiv \sqrt{2}\,x_0 a$ is distributed as:
\beq
	P(\hat{a}_{\rm sq} | \mu) = N\left(\sqrt{2}\,x_0\mu \begin{bmatrix} 1 \\ 1 \\ 1 \\ 1 \end{bmatrix},\ 
		\begin{bmatrix} 1 & 0 & -1 & 0 \\ 0 & 1 & 0 & -1 \\ -1 & 0 & 1 & 0 \\ 0 & -1 & 0 & 1 \end{bmatrix}\right)
\eeq
up to terms small in the expansion in $1/x_0$.  The probability $P({\rm vortex})$ depends only on $\xi \equiv |\sqrt{2}\,x_0\mu|$ and is maximal for $\xi = 0$, decaying to zero as $\xi \rightarrow \pm\infty$.  The vortex density is thus 
\beq
	n_v \equiv P({\rm vortex}) = \int{P({\rm vortex}|\sqrt{2}\,x_0\mu) P(\mu) \d^2\mu} \approx \frac{1}{2x_0^2} \int{P({\rm vortex}|\xi) \xi\,\d\xi} \approx \frac{0.159}{x_0^2} \label{eq:09-nvort}
\eeq
The constant $0.159$ in (\ref{eq:09-nvort}) must be determined numerically.  The total number of vortices will be $N n_v$, where $N$ is the size of the lattice.

\begin{figure}[tbp]
\begin{center}
\includegraphics[width=1.0\textwidth]{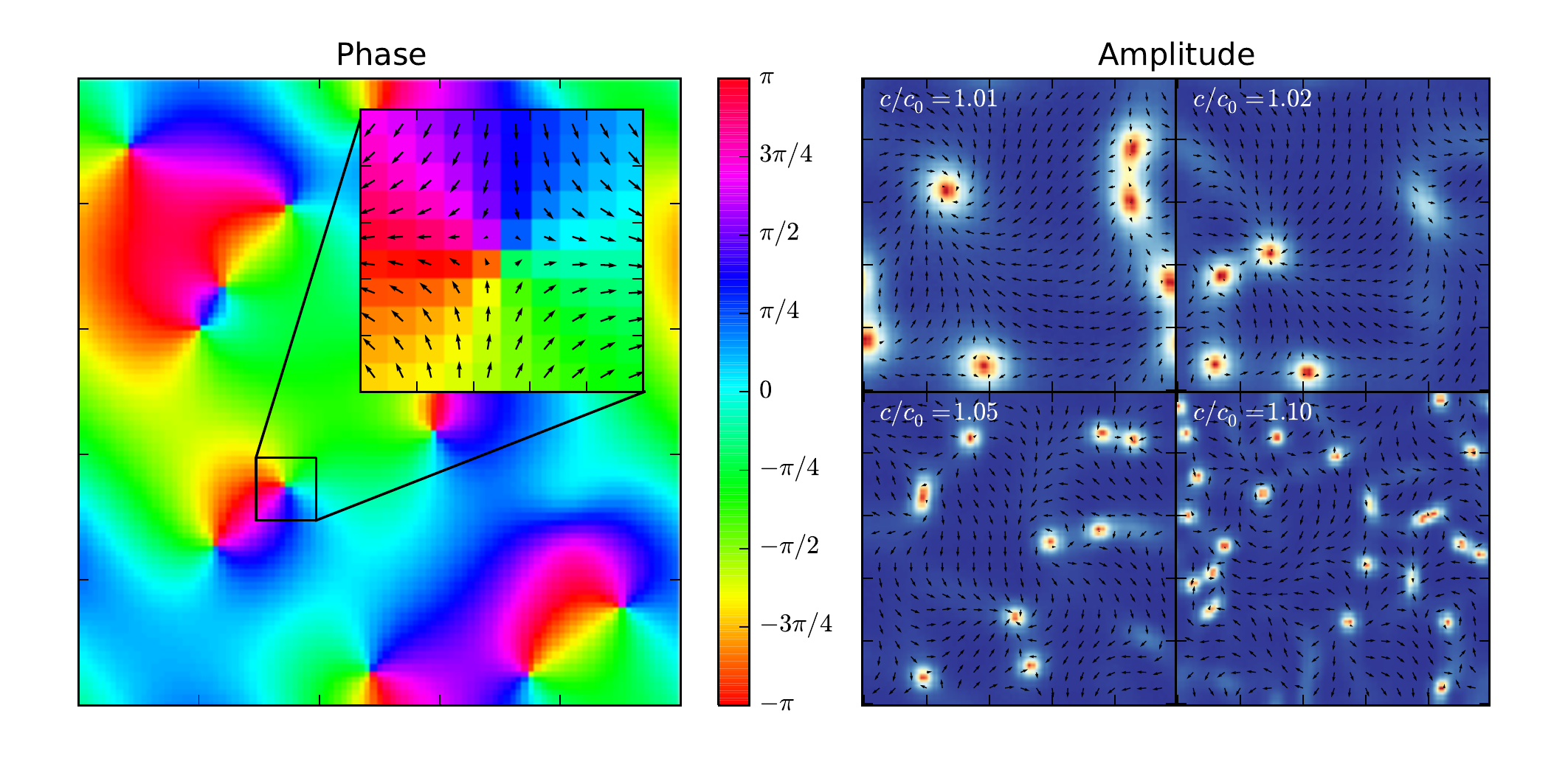}
\caption{Phase (left) and amplitude (right) for 2D XY model with vortices.}
\label{fig:09-f15}
\end{center}
\end{figure}

\subsubsection{Vortex Interactions}

A single vortex is a stable fixed point of the near-threshold equations of motion (\ref{eq:09-static-norm2}).  But if two vortices are placed together, the solution is no longer stable.  Far from the vortex cores $|r| \gtrsim \ell$, the field amplitude is constant and only its phase varies: $A(x, y) = a_{\rm sat} e^{i\phi(x, y)}$.  The equation of motion for $\phi(\bar{x}, \bar{y})$ is:
\beq
	\frac{\partial\phi}{\partial\bar{t}} = \frac{1}{2}\nabla^2 \phi
\eeq

One finds the vortex attraction in a manner analogous to Eq.~(\ref{eq:09-vw}) (for 1D domain-wall attraction).  For the solution $\phi_z(\bar{x}, \bar{y}) = \mbox{Im}[\log((\bar{x}-\bar{z}) + i\bar{y}) - \log((\bar{x}+\bar{z}) + i\bar{y})]$, which is parameterized by the separation $2z$, all perturbations decay rapidly except the translation modes $\partial\phi_z/\partial\bar{x}$, $\partial\phi_z/\partial\bar{y}$, and the attraction mode $\partial\phi_z/\partial \bar{z} = \bar{x}\,\partial\phi_z/\partial \bar{x}$.  The vortex attraction can be computed:
\beq
	\bar{v}_{\rm vort} = -\left[\int\frac{\partial\phi}{\partial\bar{z}}\frac{\partial\phi}{\partial\bar{z}}d\bar{x}d\bar{y}\right]^{-1}
		\int\frac{\partial\phi}{\partial\bar{t}}\frac{\partial\phi}{\partial\bar{z}}d\bar{x}d\bar{y} \label{eq:09-vvort}
\eeq
The first term is an inertial term.  One can compute it by noting that the ``inertia'' of two vortices is roughly twice that of a single vortex, and for a single vortex,
\beq
	\int{\frac{\partial\phi}{\partial\bar{z}}\frac{\partial\phi}{\partial\bar{z}}d\bar{x}d\bar{y}} = \frac{1}{2}\int{(\nabla\phi)^2 d\bar{x}d\bar{y}} = \pi \ln(R/r_0)
\eeq

This is infinite for a single vortex when $R \rightarrow \infty$, consistent with the well-known fact that individual vortices have infinite energy in the XY model.  For a vortex pair at $(z, -z)$ one can set $R \approx z$.  The denominator $r_0$ is set by the lattice size in the classical XY model, of the vortex size $\ell$ here.  The full inertial term will match up to a numerical factor: $A\pi\ln(z/r_0)$.

\begin{figure}[tbp]
\begin{center}
\includegraphics[width=1.00\textwidth]{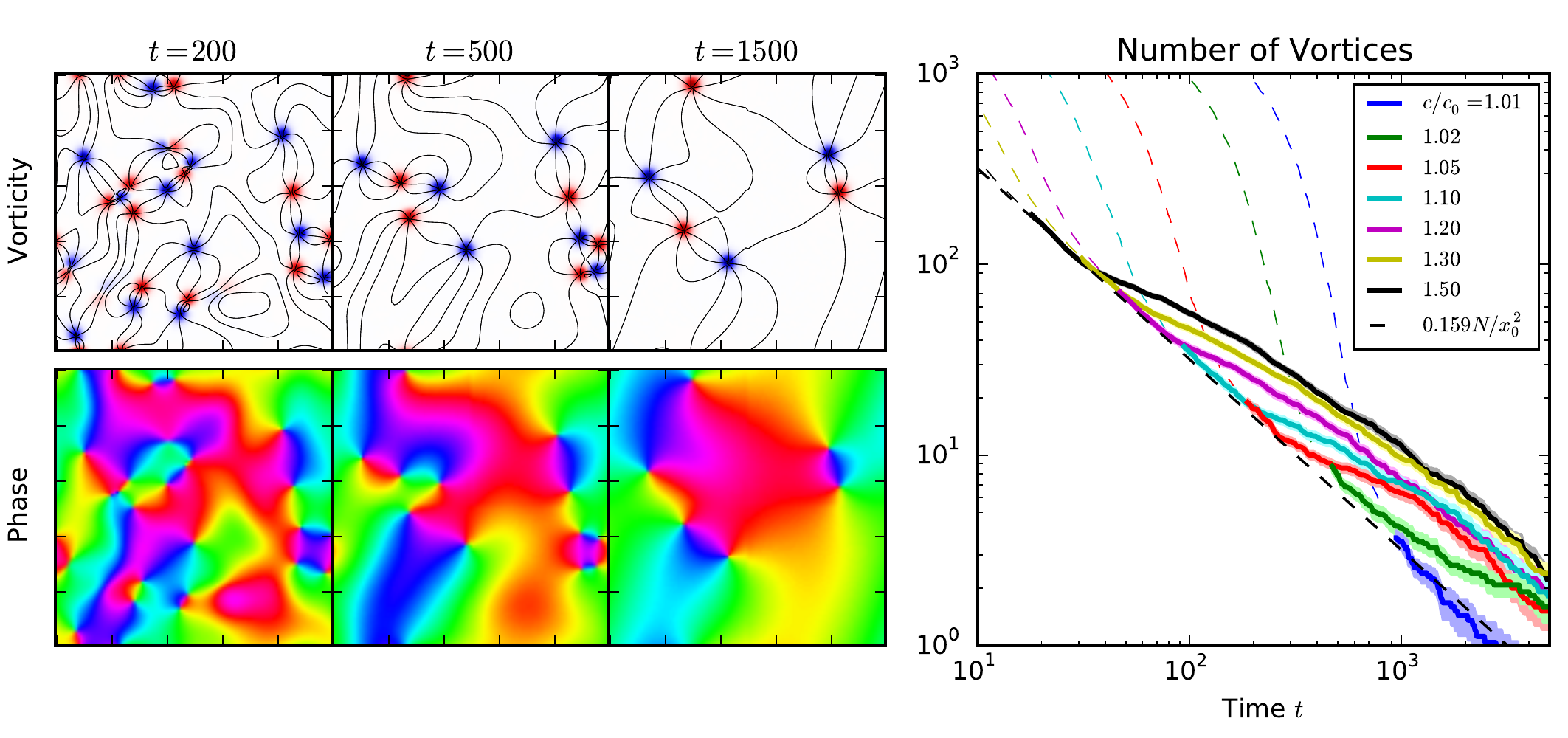}
\caption{Left: plots of vorticity and phase as a function of time (simulation used $c/c_0 = 1.1$).  Right: vortex number as a function of time and pump power ($t > T_{\rm sat}$ in bold).}
\label{fig:09-f16}
\end{center}
\end{figure}

The second term in (\ref{eq:09-vvort}) is a forcing term.  Since $\partial\phi/\partial\bar{t} = \tfrac{1}{2}\nabla^2\phi$, one can rewrite it in terms of a potential, which may be calculated by methods of complex analysis; see Kosterlitz \& Thouless\cite{Kosterlitz1973}:
\beq
	\int\frac{\partial\phi}{\partial\bar{t}}\frac{\partial\phi}{\partial\bar{z}}d\bar{x}d\bar{y} = 
		-\frac{1}{4}\frac{\d}{\d z}\int{(\nabla\phi)^2 d\bar{x}d\bar{y}} = -\frac{1}{4}\frac{\d}{\d\bar{z}}\left[4\pi \ln\frac{2\bar{z}}{r_0}\right] = -\frac{\pi}{\bar{z}}
\eeq
Thus, the attraction speed for a vortex pair at $(z, -z)$, and the lifetime for such a pair, is:
\beq
	\bar{v}_{\rm vort} \approx \frac{1}{A\bar{z}\ln (\bar{z}/r_0)},\ \ \ 
	\bar{T} = \frac{A}{2}\bar{z}^2\left(\log(\bar{z}/r_0) - \tfrac{1}{2}\right) \label{eq:vwxy}
\eeq
In real units, $T = \tau\bar{T}, z = \ell\bar{z}$, so a coefficient $\tau/\ell^2$ appears on the right-hand side.  This evaluates to:
\beq
	v_{\rm vort} = \frac{r^2(1-r^2)}{A} \frac{1}{z \ln(z/r_0\ell)},\ \ \ 
	T = \frac{A}{r^2(1-r^2)} z^2\left(\log(z/r_0\ell) - \tfrac{1}{2}\right) \label{eq:09-tvar}
\eeq
Unlike for domain walls, the vortex collision time scales only logarithmically with the pump, in that $\ell$ renormalizes the cutoff length $r_0$.  It also scales polynomially with $z$, suggesting that for an $L\times L$ lattice the system should reach the ground state (up to a winding number) in $O(L^2 \log L)$ time, similar to the $O(L^2)$ scaling found for 1D chains in Sec.~\ref{sec:xy1d}.  One can run 2D XY-machine simulations with two-vortex initial conditions; the vortex attraction roughly follows (\ref{eq:09-tvar}) with the parameters $A \approx 1.80$, $r_0 \approx 0.25$.

Figure \ref{fig:09-f16} illustrates the vortex interactions, albeit qualitatively.  For a complex field of arbitrary amplitude, one can define the vorticity as $\nabla a^* \times \nabla a$.  The winding number around a loop (for which the field has constant amplitude) equals the integral of the vorticity inside the loop.  This vorticity is plotted as a function of position and time, and the regions of nonzero vorticity correspond to regions where the phase wraps by $2\pi$.  Following the plots from left to right, one sees that vortices of opposite vorticity are attracted to each other and eventually annihilate, consistent with the vortex interaction picture sketched above.

The right plot shows the average number of vortices on a $100\times 100$ XY model for pump amplitudes ranging from $c/c_0 = 1.01$ to $1.50$.  In this plot, the ``number of vortices'' was defined as the number of unit cells in which the phase winds by $\pm 2\pi$.  Well below saturation, when the field amplitudes are random, the number of such ``vortices'' is very high.  By the end of the growth stage, the vortex count stabilizes at $0.159N/x_0^2$, consistent with Eq.~(\ref{eq:09-nvort}).  Thereafter the system enters the Kuramoto stage and its dynamics are driven by vortex-vortex interactions.  From (\ref{eq:09-tvar}), we expect that the number of vortices should scale as $N_v \sim T^{-1}$ up to a logarithmic term; this explains the near-linear falloff of all of the curves on the log-log scale.

\section{Conclusion}

Although the Ising problem is quite old, the OPO-based Ising solver is a new idea.  This chapter presents the first comprehensive treatment of 1D and 2D ferromagnetic Ising and XY machines based on this mechanism.  The Ising machine differs from simulated and quantum annealing in that the ``spins'' are not bits or qubits, but rather optical states in an OPO.  The dynamics of this OPO network can be simulated using semiclassical equations derived from the truncated Wigner method.

Previous papers modeled the Ising machine as a network of coupled cavities and derived continuous-time equations of motion for the state\cite{Haribara2015,Utsunomiya2011,Wang2013}.  In the time-multiplexed picture (Fig.~\ref{fig:09-f1}), that approach is only valid when the cavity has high finesse ($G_0 \approx 1$) and the round-trip coupling between pulses is weak ($r \ll 1$).  Thus, the coupled-cavity model is not accurate for high-gain systems like those at RIKEN\cite{KentaThesis,Takata2016}, NTT\cite{Inagaki2016} and Stanford\cite{McMahon2016}.  On the other hand, high-gain systems are advantageous because they are faster and more resilient to experimental noise and loss.

In this chapter, we derive a more general approach which holds for cavities of arbitrary finesse and coupling.  The truncated Wigner picture is used and the state is described by semiclassical pulse amplitudes $a_i(t)$, where $i$ is the pulse index and $t$ is the {\it discrete} time (round-trip number).  This state satisfies a set of {\it difference equations} (Eqs.~(\ref{eq:feom}, \ref{eq:09-aout}, \ref{eq:09-ai-delay2})) that relate $a_i(t+1)$ to $a_i(t)$.  These equations reduce to the continuous-time equations in the high-finesse limit $G_0 \rightarrow 1$, $r \rightarrow 0$.

Both 1D and 2D Ising chains were simulated using this model.  The dynamics of the Ising machine can be broken into two stages: a {\it growth stage} (Sec.~\ref{sec:09-growth}) in which the field amplitudes are far below saturation, and a {\it saturation stage} (Sec.~\ref{sec:09-crst}), by which most of the OPO amplitudes have saturated.  During the growth stage, the OPO amplitudes start from random values and grow linearly, with longer-wavelength Fourier modes growing the fastest.  This induces correlations between nearby OPOs, forming ferromagnetic domains after saturation.  During the saturation stage, these domains evolve, and the attraction of nearby domain walls causes smaller domains to annihilate.

We used this model to compute basic statistical quantities in 1D: the correlation function $R(x)$, correlation length $x_0$, defect density $n_d$, domain length distribution $P(\ell)$, and success probability $P_s$.  In the Ising machine, these all depend on the time to saturation $T$ (which is a function of pump rate) and the coupling mirror parameters $r, t$; for the thermal model they are functions of the coupling $J$ and effective temperature $1/\beta$.  Experimental data from Inagaki et al.\cite{Inagaki2016} match closely with our numerical predictions.

The dynamics depend strongly on dimension.  For the 1D chain, the domain lifetime scales exponentially with domain size, so one can say that after the growth stage, the domain structure ``freezes out'', and will not relax to the ferromagnetic ground state unless one waits an exponentially long time.  Conversely, in the 2D case this lifetime scales as the size squared, since domain walls are curved and always move towards their center of curvature (Sec.~\ref{sec:09-2dlat}).  Thus, long-range order is established in $O(L^2)$ time for an $L \times L$ lattice, and all domain walls are eventually destroyed.

Ising simulations for frustrated 1D and 2D systems were also studied.  In this case, the Fourier modes with maximum gain have nonzero $k$, giving rise to periodic order in the final state (Fig.~\ref{fig:09-f11}).  In 2D, one finds two competing phases of periodic order: up- and down-diagonal stripes, which compete with each other, analogous to the competition between up- and down-states in the Ising model (Fig.~\ref{fig:09-f12}).

We also studied a related OPO network, the coherent XY machine.  This device uses a network of {\it non}-degenerate OPOs to find the ground state of the XY potential.  If the XY machine is based on pulses in a high-gain cavity with strong couplings (Fig.~\ref{fig:09-f13}), one obtains a similar set of difference equations, this time for both signal and idler fields.  As before, if we take the limit $G_0 \rightarrow 1$, $r \rightarrow 0$, this reduces to the continuous-time coupled-cavity model studied elsewhere.

For 1D XY systems, the only possible topological defect is the winding number.  In $O(N^2)$ time, the system always relaxes to a state with constant winding.  Before this ``smoothing out'', winding can be treated as a random walk per unit length, and the winding number has a Gaussian with standard deviation that goes as $N^{1/2}$.  For 2D systems, vortex defects form, analogous to the BKT transition\cite{Kosterlitz1973}.  In contrast to BKT, vortices in the XY machine are formed through the OPO growth / saturation process, not thermally; thus their distribution is very different.

It is hoped that our results for these simple models will shed insight into Ising machines more generally.  From the results above, a few things stand out:

\begin{enumerate}
\item As an ``algorithm'', the Ising machine is behaving like a convex relaxation technique.  Dividing the dynamics into growth and saturation stages makes this more obvious.  During the growth stage, the eigenmodes of the coupling matrix $C_{ij}$ grow at different rates and the machine tends toward the dominant eigenmode.  This is solving the maximum eigenvalue problem, which has a single local minimum and is solvable in polynomial time (although it is not technically convex).  However, this eigenmode may not be a valid Ising state, so during the saturation stage, the system relaxes into a valid state as the pulse amplitudes saturate.
\item Unlike simulated annealing, randomness does not appear to play a major role in this algorithm.  While random noise seeds the initial state, most of the subsequent dynamics is deterministic because the field amplitudes are far above the quantum level.  When the system reaches a local minimum in the saturation stage, it is unable to ``tunnel'' out (in either a classical or quantum sense) because the photon number is so high.
\item Even ``trivial'' problems can have long-lived metastable states (e.g.\ domain walls) or local minima (winding numbers).  The current machine does not have a way to escape these minima, since the noise is so small compared to the coherent amplitude at saturation.  However, it is equally worth mentioning that simulated annealing is not very efficient on the 1D chain, requiring at least $O(N^2)$ time to converge.  For the Ising machine, the convergence time is also $O(N^2)$, if this time is spent during the growth stage.
\end{enumerate}

While so far only the 1D chain has studied experimentally, by adding extra delay lines, it is straightforward to extend current work to the 2D and frustrated cases.  Moreover, the groups at Stanford\cite{McMahon2016} and NTT\cite{Inagaki2016b} are working towards machines with ``all-to-all'' connectivity via injection and measurement feedback\cite{Haribara2016}.  The measurement-feedback theory is probably a straightforward extension of this work, with additional stochastic terms for detector, ADC/DAC and injection noise.  Beyond the scope of this work, the measurement-feedback approach is promising because it can handle arbitrary spin networks, not just the 1D and 2D lattices of this chapter.

\ifstandalone{\bibliography{NoteRefs}{}
\bibliographystyle{alpha}
}
\ifdefined\multidoc\else\input{Header}\fi

\ifstandalone{\setcounter{chapter}{9}}
\chapter{Reduced Models for Pulsed OPOs}
\label{ch:10}

This chapter is based on the following paper:

\begin{itemize}
\item R.~Hamerly. A.~Marandi, M.~Jankowski, M.~M.~Fejer, Y.~Yamamoto and H.~Mabuchi, ``Reduced models and design principles for half-harmonic generation in synchronously-pumped optical parametric oscillators'' (in preparation)
\end{itemize}

The optical parametric oscillator (OPO) is an indispensable tool in nonlinear optics.  As a lightsource, it benefits from the broadband $\chi^{(2)}$ nonlinearity, allowing it to produce light at near- and mid-IR frequencies \cite{Marandi2016}, an essential resource for molecular spectroscopy \cite{Parker2012}, high-harmonic generation \cite{Popmintchev2012} and dielectric laser accelerators \cite{Peralta2013}.  From an optical logic standpoint, since the $\chi^{(2)}$ effect is much stronger than the $\chi^{(3)}$ effect, nonlinearity (and thus computation) can be achieved with much lower powers.  Recently, networks of OPOs have been proposed as tools for combinatorial optimization \cite{Wang2013, Marandi2014} and machine learning \cite{Tezak2015}.  Integrated $\chi^{(2)}$ photonics is rapidly maturing and recent success with LiNbO$_3$ waveguides \cite{Jackel1991, Korkishko1998, Iwai2003, Roussev2004, Chang2016} and microstructures \cite{Poberaj2012, Rabiei2014, Guarino2007, Lin2015} in particular suggest that large-scale, integrated OPO systems are feasible in the near future.

Since optical nonlinearities are most pronounced at strong field intensities, and field intensity is enhanced in pulsed mode, there has been a growing interest in the synchronously-pumped OPO (SPOPO), in which the pump is a train of ultrashort pulses synchronized to the round-trip time of the cavity \cite{VanDriel1995}.  Highly nonlinear effects can take place at modest average powers.  SPOPOs are used for numerous applications including pulse compression \cite{Khaydarov1994, Marandi2015}, frequency-domain entanglement generation \cite{Roslund2014}, cluster-state preparation \cite{Yokoyama2013} and coherent computing \cite{Marandi2014, KentaThesis}.  On the other hand, SPOPOs have far more degrees of freedom than their continuous-wave counterparts, so modeling them and predicting their behavior is a challenge.

This chapter discusses computationally efficient schemes for modeling degenerate SPOPOs.  Pulse dynamics in a SPOPO is a competition between three effects: $\chi^{(2)}$ nonlinearity, dispersion, and group-velocity mismatch (temporal walkoff).  Section \ref{sec:10-intro} introduces the physical system and its equations of motion.  These equations can be solved numerically using a split-step Fourier method (which can easily be scaled to multicore / GPU architectures for performance), giving rise to a discrete round-trip Ikeda-like map for the pulse amplitude \cite{Ikeda1979}.  While this numerical model is accurate and agrees with experiments, it is computationally costly to run, particularly for guided-wave systems with large temporal walkoff.

Sections \ref{sec:10-linear}-\ref{sec:10-boxpulse} derive approximate, physically-motivated {\it reduced models} for the SPOPO system.  These models reduce the OPO simulation time by several orders of magnitude, but within their respective regimes of operation, give steady-state pulse shapes and dynamical behavior that match the full numerical model.  The resulting computational speedup is particularly useful for large simulations of many OPOs in parallel -- for example, large-scale Ising or XY machines based on time-multiplexed OPO networks \cite{Takata2016, Inagaki2016, Hamerly2016-2}.  Moreover, these models facilitate device optimization and robustness studies, by allowing the designer to simulate a SPOPO with a wide range of test parameters.  Finally, these models shed analytic and physical insight into the dynamics of SPOPOs.

In Section \ref{sec:10-linear}, I derive a linearized model based on an eigenmode expansion.  The eigenmodes and their eigenvalues are computed, and related to analytic formulae that reveal a power-law scaling in the steady-state signal pulse width as a function of pump pulse width, dispersion and single-pass gain.  Section \ref{sec:10-nonlinear} extends this model by treating pump depletion to first order in perturbation theory, leading to equations with cubic terms that resemble the Langevin equations for continuous-wave OPOs \cite{Kinsler1991}.  This model accurately predicts the oscillation threshold, power efficiency, signal pulse shape, and stability for the SPOPO near threshold.

An ansatz based on the simulton solution in a $\chi^{(2)}$ waveguide \cite{Akhmanov1968, Trillo1996} is presented in Section \ref{sec:10-simulton}.  By postulating a sech-shaped signal pulse, effects of the pump shape, dispersion, and nonlinearity all map onto a set of ODE's for the amplitude, centroid and width of the sech pulse.  This ansatz restricts the range of validity compared to Sec.~\ref{sec:10-nonlinear} (although it can also be valid well above threshold, where the eigenmode treatment fails \cite{Jankowski2016}), but it is physically more intuitive and sheds more light into the pulse dynamics.

In the opposite regime well above threshold, Section \ref{sec:10-boxpulse} obtains an analytic form by ignoring dispersion.  The result is a box-shaped pulse whose width is a function of the pump amplitude and whose spectrum approximates a sinc-function.  We note that this section is a generalization of \cite{Becker1974} to the case of nonzero walkoff.

While the results of this chapter are general and apply to any degenerate SPOPO with dispersion and temporal walkoff, for concreteness we consider a guided-wave PPLN OPO with a fiber cavity, implemented in \cite{Marandi2015, McMahon2016}, as an example system.

\section{The Synchronously Pumped OPO}
\label{sec:10-intro}

\begin{figure}[tbp]
\begin{center}
\includegraphics[width=0.50\columnwidth]{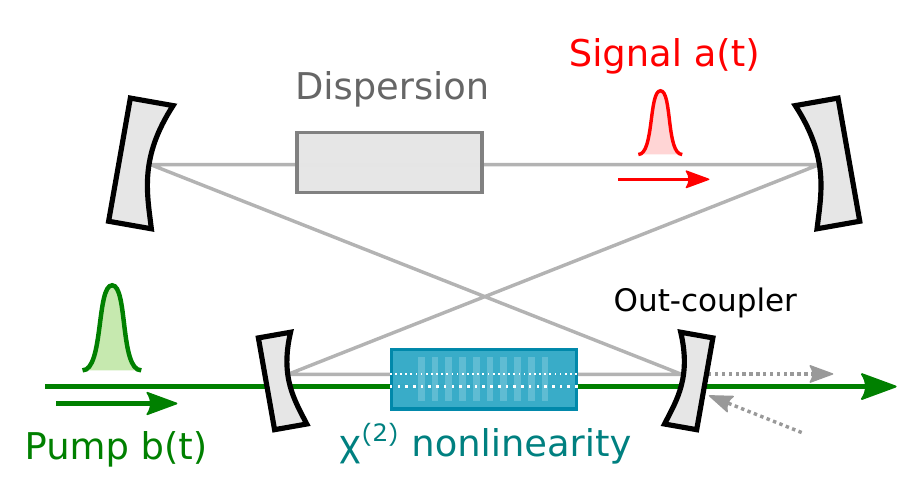}
\caption{Typical synchronously pumped OPO design.  }
\label{fig:10-f1}
\end{center}
\end{figure}

Figure \ref{fig:10-f1} sketches the design.  The degenerate, synchronously pumped OPO consists of a cavity with a nonlinear $\chi^{(2)}$ medium, an output coupler, and a lumped dispersion element (for all dispersion excluding the $\chi^{(2)}$ medium).  In isolation, the $\chi^{(2)}$ medium is an amplifier, and the feedback loop created by the cavity turns it into an oscillator.  As a concrete example, in the fiber-coupled OPO in \cite{Marandi2015}, the $\chi^{(2)}$ medium is a PPLN waveguide and the dispersive element is the optical fiber.

\subsection{Equations of Motion}

Propagation through the OPO is a two-step process: (1) nonlinear $\chi^{(2)}$ medium and (2) linear dispersion element.  The waveguide dynamics are governed by a pair of PDE's.  To derive these equations, first write the electric field in terms of slowly-varying amplitudes \cite{BoydBook, AgrawalBook}
\bea
    \vec{E}(z, t) & = & \mathcal{E}_a \vec{E}_{T,a}(x, y) e^{i(\bar{\beta}_a z - \bar{\omega} t)} a(z, t) \nonumber \\
    & & - i\mathcal{E}_b\vec{E}_{T,b}(x, y) e^{i(\bar{\beta}_b z - 2\bar{\omega} t)} b(z, t) + \mbox{c.c.} \label{eq:10-svea}
\eea
where $a(z,t)$ and $b(z,t)$ are the envelope functions for the pump and signal.  Here $z$ is the propagation direction and $\vec{E}_{T,a}$, $\vec{E}_{T,b}$ are normalized transverse mode profiles.  The constants $\mathcal{E}_{a,b} = \sqrt{\hbar\omega_{a,b}/2n(\omega_{a,b}) \epsilon_0 c}$ are chosen so that $\int{|a|^2 \d t}$, $\int{|b|^2 \d t}$ correspond to the pump and signal photon number.  Applying Maxwell's equations to (\ref{eq:10-svea}) and adding dispersion and a $\chi^{(2)}$ nonlinearity, the envelope functions evolve as follows:\bea
    \!\!\frac{\partial a}{\partial z} & \!=\! & \left[-\frac{\alpha_a}{2} - \frac{i\beta_2^{(a)}}{2!} \frac{\partial^2}{\partial t^2} + \frac{\beta_3^{(a)}}{3!} \frac{\partial^3}{\partial t^3} + \ldots\right] a + \epsilon\,a^* b \label{eq:10-at} \\
    \!\!\frac{\partial b}{\partial z} & \!=\! & \left[-\frac{\alpha_b}{2} - u \frac{\partial}{\partial t} - \frac{i\beta_2^{(b)}}{2!} \frac{\partial^2}{\partial t^2} + \frac{\beta_3^{(b)}}{3!} \frac{\partial^3}{\partial t^3} + \ldots\right] b - \frac{1}{2}\epsilon\,a^2 \nonumber \\ \label{eq:10-bt}
\eea
where $\alpha_{a,b}$ are the waveguide losses, $u = (\beta_1^{(b)} - \beta_1^{(a)}) = (v_a - v_b)/v_a v_b$ is the walkoff (group-velocity mismatch), and $\epsilon = \bigl(2\omega \mathcal{E}_b d_{\rm eff}/n(\omega) c\bigr) \int {E_{T,a}^2 E_{T,b} \d x\,\d y}$ is the nonlinear coefficient.  Equations (\ref{eq:10-at}-\ref{eq:10-bt}) reveal that the dynamics is a competition between three effects:

\begin{enumerate}
	\item Nonlinearity: second-harmonic generation and parametric gain when pulses overlap in time
	\item Dispersion: short pulses are spread out and chirped
	\item Walkoff (group velocity mismatch): pump and signal move with respect to each other, limiting the duration of their overlap
\end{enumerate}

Previous studies of this problem have either ignored the walkoff or treated it as a perturbation \cite{Becker1974, Cheung1990}, or have focused on the high-finesse limit when the single-pass PPLN gain is small \cite{Patera2010, DeValcarcel2006, Roslund2014}.  Equations (\ref{eq:10-at}-\ref{eq:10-bt}) generalize these results to the high-gain, large-walkoff case that is more commonplace when long $\chi^{(2)}$ crystals and/or ultrashort pulses are used \cite{Marandi2014, Marandi2016}.

Similar equations can be derived from a quantum model for the $\chi^{(2)}$ system \cite{Raymer1991, Werner1997}.  The procedure is similar to that used for optical fibers \cite{Drummond2001}, but in the resulting equations, the roles of $z$ and $t$ are swapped.  These quantum equations are equivalent to (\ref{eq:10-at}-\ref{eq:10-bt}) under reasonable assumptions.

For very short or high-power pulses, (\ref{eq:10-at}-\ref{eq:10-bt}) become inaccurate and higher-order effects such as $\chi^{(3)}$ and Raman scattering must be included.  Moreover, pulses spanning more than one octave merit special treatment as the slowly-varying envelope approximation breaks down \cite{Phillips2011, PhillipsThesis}; these are beyond the scope of this work.

To solve Eqs.~(\ref{eq:10-at}-\ref{eq:10-bt}), I employ the split-step Fourier method \cite{AgrawalBook}.  First, a sampling window $[0, T]$ is defined, with $T$ is large enough that all of the dynamics happens inside the window.  One can express the field in terms of a Fourier series $a(z, t) = T^{-1/2} \sum_m a_m(z) e^{-im\Omega t}$ (and likewise for $b$), where $\Omega = 2\pi/T$ and $m$ is the Fourier index.  The dispersive terms in (\ref{eq:10-at}-\ref{eq:10-bt}) are propagated in the frequency domain, while the nonlinear terms are propagated in the time domain.  Since most of the computation time is spent performing FFT's to go between time and frequency domains, I implemented the solver in CUDA \cite{CUDAGuide} because of the substantial FFT speedup afforded by modern GPUs \cite{Moreland2003, Sreehari2012}.

The second step, propagation through the dispersive element, is trivial because it is linear.  Since only the signal resonates in the setup (Fig.~\ref{fig:10-f1}), each Fourier component acquires a constant loss and phase shift $a_m \rightarrow G_0^{-1/2} e^{i\phi_m} a_m$, with $\phi_m = \phi_0 + \tfrac{\ell\lambda}{2c} \Omega m + \tfrac{\phi_2}{2!} (\Omega m)^2 + \ldots$

\ctable[caption=Parameters for PPLN waveguide OPO \cite{Marandi2015, McMahon2016} used as example in this chapter,
        label=tab:10-t1,
        pos=t]{ccc}{
\tnote[a]{0.3 dB/cm}
\tnote[b]{LiNbO$_3$, extraordinary polarization}
\tnote[c]{$T_p = L u$, matched to crystal walkoff length}
\tnote[d]{$\epsilon = \sqrt{2\hbar\omega\,\eta}$, where $\eta = 1.0$ W$^{-1}$cm$^{-2}$ is the normalized conversion efficiency \cite{Parameswaran2002, Langrock2007}}
\tnote[e]{5-dB out-coupling loss.}
\tnote[f]{$N_{b,0} = \left[(\alpha_b/4\epsilon)(e^{\alpha_b L/2}-1)^{-1} \log(G_0 e^{\alpha_a L})\right]^2$}
\tnote[g]{$N_{b,0} = T_p b_0^2$}}{
\hline\hline
Term & Meaning & Value \\ \hline
$\lambda_a$ & Pump, Signal $\lambda$ & 1.5 $\mu$m, 0.75 $\mu$m \\
$L$ & Waveguide Length & 40 mm \\
$\alpha_a$, $\alpha_b$ & Waveguide Loss & 0.00691 mm$^{-1}$ \tmark[a] \\
$u$ & Walkoff & $0.329$ ps/mm\tmark[b] \\
$T_p$ & Pump Length & 13.2 ps\tmark[c] \\
$\beta_2^{(a)}$ & Signal GVD & $1.12 \times 10^{-4}$ ps$^2$/mm \\
$\beta_3^{(a)}$ & Signal TOD & $3.09 \times 10^{-5}$ ps$^3$/mm \\
$\beta_2^{(b)}$ & Pump GVD & $4.06 \times 10^{-4}$ ps$^2$/mm \\
$\beta_3^{(b)}$ & Pump TOD & $2.51 \times 10^{-5}$ ps$^3$/mm \\
$\epsilon$ & Nonlinearity & $5.16 \times 10^{-5}$ ps$^{1/2}$/mm\tmark[d] \\
$G_0$ & Cavity loss & 3.33\tmark[e] \\
$N_{b,0}$ & Threshold Photons & $1.94 \times 10^6$ \tmark[f] \\
$b_0$ & Threshold Amplitude & $3.84 \times 10^2$ ps$^{-1/2}$ \tmark[g] \\
\hline\hline
}

The out-coupling loss $G_0$ is the same for all modes, while the dispersion and walkoff terms give different modes different phases.  Here $\ell$ is the cavity length detuning (in units of half-wavelengths); $\phi_0$ and $\ell$ are not independent: $\phi_0 = \pi\ell + \mbox{const}$.  The constant reflects the fact that zero detuning may not correspond to a resonance peak.  For signal pulses much longer than an optical cycle, this constant can be neglected because it corresponds to a small, subwavelength repetition-rate mismatch.

\subsection{Numerical Results}

Figure \ref{fig:10-f2} shows some typical results for the simulations.  The left plot gives the steady-state OPO output power of the $P_{a,\rm out}$, in units of photons per round-trip.  This is proportional to the photon number $N_{a}$.  If the cavity round-trip loss is $O(1)$, the photon number will be different at the beginning and end of the crystal: $N_a\bigr|_{z=L} = G_0 N_a\bigr|_{z=0}$.  The output power, neglecting cavity losses other than the out-coupler and $\chi^{(2)}$ gain medium, is given by $P_{a,\rm out} = (G_0 - 1)N_a\bigr|_{z=0}$.

The figure shows a clear set of resonances called {\it detuning peaks}.  At each detuning peak, the round-trip phase $\phi_0$ is either 0 or $\pi$, since both phases can be amplified by the crystal.  There is an optimal length detuning denoted $\ell = 0$ for which the threshold is the lowest, which is understandable because a nonzero $\ell$ creates a repetition-rate mismatch between the pump and signal, increasing the required pump power.  Adding a nonzero offset to the relation $\phi_0 = \pi\ell + \mbox{const}$ shifts the detuning peaks, but not the envelope; since the envelope is much larger than any peak, this does not have a significant effect on Fig.~\ref{fig:10-f2}.  There is an asymmetry in the plot, where $\ell > 0$ peaks have higher power if the pump is strong enough; this is a result of walkoff and pump depletion that will be explained using sech-pulse theory in Section \ref{sec:10-simulton}.

\begin{figure}[t!]
\begin{center}
\includegraphics[width=1.00\columnwidth]{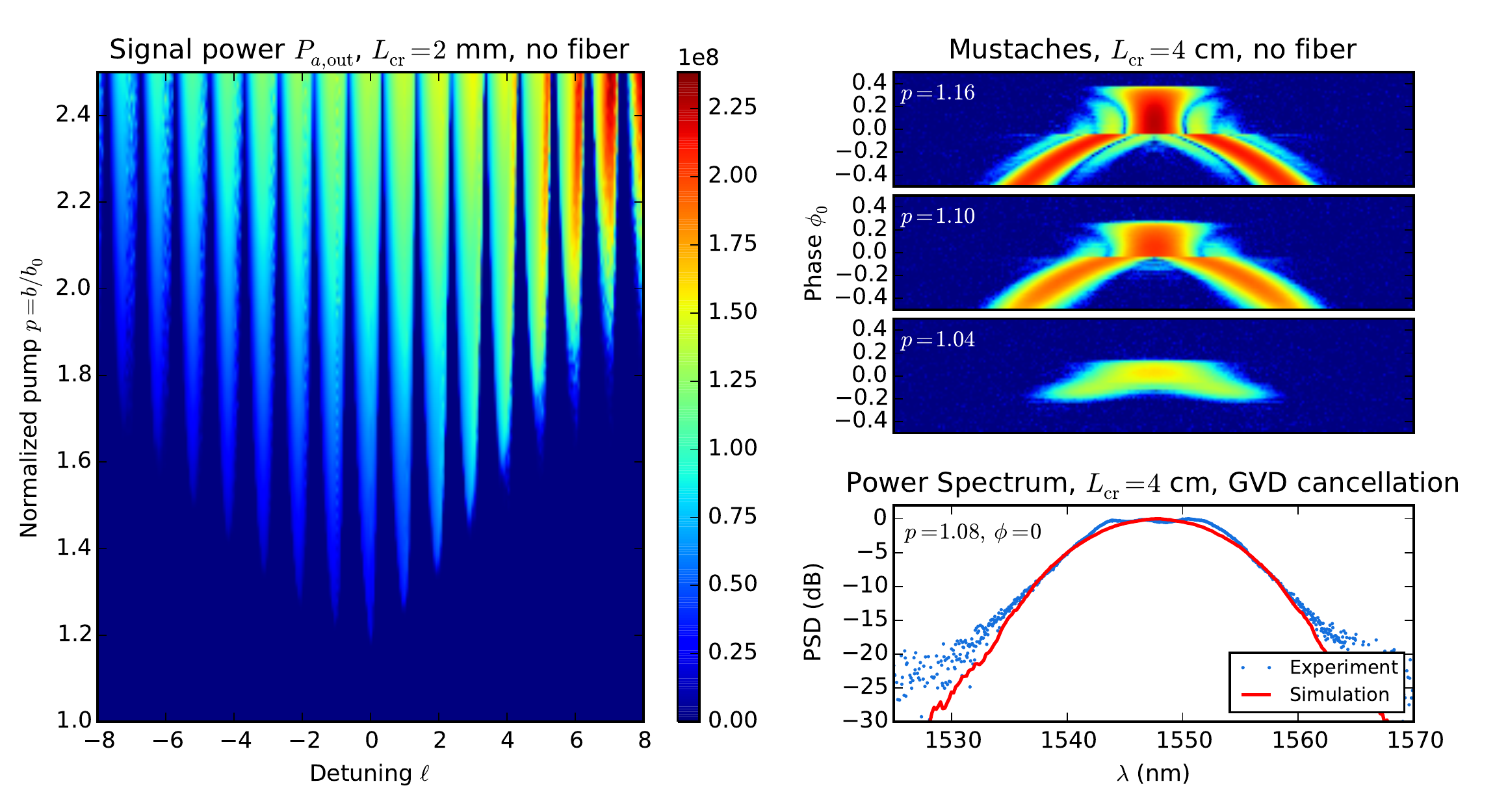}
\caption{Left: Plot of output signal power $P_{b,\rm out}$ (in photons per round-trip) for 2-mm crystal, no dispersion compensation (``free-space'').  Right: Resonance plots of the power spectrum $P(\lambda)$ for 4-cm crystal, no dispersion compensation, with normalized pump values $p = 1.16, 1.10, 1.04$ (top), and spectrum for GVD-compensated cavity at $p = 1.08$ (bottom).}
\label{fig:10-f2}
\end{center}
\end{figure}

Another common figure will be the ``resonance diagrams'' on the top-right plot.  These are plots of the power spectrum $P(\lambda) = |a(\lambda)|^2$ as a function of $\lambda$ and cavity round-trip phase $\phi_0$.  They show how the steady-state spectrum varies within a detuning peak.  As the pump power increases above threshold, the resonance diagrams become increasingly structured.  This structure will be explained later in Sec.~\ref{sec:10-boxpulse} in terms of box-shaped pulses that tend to form well above threshold.

Simulations are performed for many values of $\ell$ in parallel and sweeping $p = b/b_0$, the pump normalized to threshold; the stored output is a 3-dimensional array $a(k,p,\ell)$.  A typical run with 256 parallel simulations of 20000 round-trips each takes 15 hours with an Nvidia Tesla M2070 GPU.  Integrating $|a|^2$ over $k$ gives the power plot in Fig.~\ref{fig:10-f2}.  The resonance diagrams are $p$-slices of $|a|^2$.  Each $\phi$-slice of a resonance diagram is a spectrum.  The lower-right figure shows the simulated power spectrum for a 4-cm PPLN OPO with a GVD-compensated fiber.  Experimental data are in agreement with this result \cite{Marandi2015}.

\section{Linear Eigenmode Theory}
\label{sec:10-linear}

In actively mode-locked lasers, the pulse shape is set by a competition between two forces: a resonant cavity modulation confines the pulse in time, while the finite bandwidth of the gain medium confines it in frequency \cite{Kuizenga1970, Siegman1970, Haus2000}.  These effects give rise to a linear master equation for pulse evolution, which can be solved as an eigenvalue problem, the dominant eigenmode (typically a Gaussian) becoming the lasing mode.

The same story holds for SPOPOs.  In this case, the finite pump length confines the signal in time, while dispersion in the cavity and gain medium confines it in frequency \cite{Becker1974, Khaydarov1994, Khaydarov1995}.  Patera {\it et al.}\ followed a similar procedure for the SPOPO below threshold, linearizing the equations of motion and diagonalizing them to obtain squeezing ``supermodes'' \cite{DeValcarcel2006, Patera2010}.  However, their analysis was restricted to the low-gain, high-finesse case, which is not applicable here.

This section derives an eigenmode expansion that extends the work of Patera et al.\ to the high-gain regime with walkoff, where waveguide-based SPOPOs typically operate.  We do so using a split-step procedure -- a single round trip $a(t;n)\rightarrow a(t;n+1)$ is divided up as follows:

\begin{enumerate}
    \item Continuous-wave step: Solve equations with dispersion terms, but constant pump $b(t) = b_{\rm max}$.  Result: $\tilde{a}(\delta\omega) \rightarrow \Delta(\delta\omega) \tilde{a}(\delta\omega)$ (Sec.~\ref{sec:10-cw})
    \item Dispersionless step: Solve with pulsed pump $b(t) - b_{\rm max}$ (peak value subtracted), and no dispersion terms.  Result: $a(t) \rightarrow \Gamma(t)a(t)$ (Sec.~\ref{sec:10-dispersionless})
\end{enumerate}

This is analogous to the split-step Fourier method used for the nonlinear Schr{\"o}dinger equation \cite{AgrawalBook}.  The key assumption that the pulse shape does not change much during a single step (``gain without distortion ansatz'') is equally necessary here.  This assumption tends to be true unless the pump is far above threshold.

Combining the two steps, the pulse satisfies the following round-trip equation:
\beq
	a(t;n+1) = \Gamma(t)\Delta(i\tfrac{\d}{\d t}) a(t;n)
\eeq
$\Gamma\Delta$ is related to a Hermitian matrix by transformation, so this is diagonalizable and the eigenmodes are found by solving the corresponding eigenvalue equation:
\beq
	\Gamma(t)\Delta(i\tfrac{\d}{\d t}) a_k(t) = \lambda_k a_k(t) \label{eq:10-eig}
\eeq

We can define a {\it gain-clipping function} $G(t) \equiv \log\Gamma(t)$ and a {\it dispersion loss function} $D(\delta\omega) \equiv \log(\Delta(\delta\omega)/\Delta_{\rm max})$, where $\Delta_{\rm max} = \mbox{max}_{\delta\omega}\Delta(\delta\omega)$.  Both of these functions are negative.  Near threshold, $G(t), D(\delta\omega) \ll 1$ and we can obtain a master equation analogous to \cite{Haus2000}:
\beq
	a(t;n+1) = \Delta_{\rm max}\left[1 + G(t) + D(i\tfrac{\d}{\d t})\right] a(t;n) \label{eq:10-nt-rt}
\eeq
Again, one can convert (\ref{eq:10-nt-rt}) into an eigenvalue equation to extract the eigenmodes:
\beq
	\left[g_{\rm cw} + G(t) + D(i\tfrac{\d}{\d t})\right] a_k(t) = g_k a_k(t) \label{eq:10-nt-eig}
\eeq
Here $g_{\rm cw} = \log \Delta_{\rm max}$ is the CW gain and $g_k = \log \lambda_k$ is the eigenmode gain.  Because of the negativity of $G$ and $D$, $g_k \leq g_{\rm cw}$ for all eigenmodes.

\subsection{Continuous Wave Step}
\label{sec:10-cw}

To obtain the CW round-trip gain $\Delta(\delta\omega)$, consider the case of a signal $a_\sig$ at frequency $\omega + \delta\omega$ and idler $a_\idl$ at $\omega - \delta\omega$.  From these we define $a_+ = (a_\sig + a_\idl^*)/2$, $a_- = (a_\sig - a_\idl^*)/2$ (``real'' and ``imaginary'' parts of the field) and use (\ref{eq:10-at}-\ref{eq:10-bt}), excluding pump depletion, to get:
\begin{eqnarray}
    \frac{\d a_\pm}{\d z} & = & \left(-\tfrac{1}{2}\alpha_a \pm \epsilon\,b\right) a_\pm \mp \left(\tfrac{1}{2}\beta_2\delta\omega^2\right) a_\mp \label{eq:10-cw}
\end{eqnarray}

Unless the pump loss $\alpha_b L$ is large, the pump remains relatively constant during the propagation; we can replace it by its average value $b \rightarrow \bar{b} \approx b_{\rm in} e^{-\alpha_b L/4}$.  Equation (\ref{eq:10-cw}) can then be solved by matrix exponentiation.  After exiting the gain medium, the field passes through the dispersion element and is then re-inserted.  There will be additional loss $G_0^{-1/2}$ due to out-coupling, and possibly additional delay and phase due to the cavity detuning.  Thus, the reinserted field is related to the exiting field by: $a_\sig \rightarrow G_0^{-1/2} e^{i(\phi + \psi)}a_\sig$, $a_\idl \rightarrow G_0^{-1/2} e^{i(\phi - \psi)}a_\idl$, where $\phi \equiv \phi_0 + \tfrac{1}{2}\phi_2\delta\omega^2$ is the symmetric phase shift, and $\psi \equiv \pi\ell$ as the asymmetric phase.  The overall round-trip propagation of $a_\pm$ is:
\begin{align}
	&\begin{bmatrix} a_+ \\ a_- \end{bmatrix} \rightarrow 
		G_0^{-1/2} e^{-\alpha_a L/2} e^{i\psi} \nonumber \\
		& \quad \times
		\underbrace{\begin{bmatrix} \cos\phi & -\sin\phi \\ \sin\phi & \cos\phi \end{bmatrix}}_{R(\phi)}
		\underbrace{\exp\left(\begin{bmatrix} \epsilon\,\bar{b} & -\tfrac{1}{2}\beta_2\delta\omega^2 \\ \tfrac{1}{2}\beta_2\delta\omega^2 & -\epsilon\,\bar{b} \end{bmatrix} L\right)}_{M}
		\begin{bmatrix} a_+ \\ a_- \end{bmatrix}
\end{align}

This equation has two eigenvalues: $\lambda_\pm$.  The round-trip gain is the larger of the two.  Note that $\det R(\phi) = \det M = 1$, so the product of the eigenvalues must equal $G_0^{-1} e^{-\alpha_a L}$, which is less than one.  Thus, at most one of the modes experiences gain.  We now assume that the frequency components of the pulse $a(t)$ live primarily in the growing eigenmode, so that we can substitute $\Delta(\delta\omega) \approx \lambda_+(\delta\omega)$.  This eigenvalue is:
\beq
	\Delta(\delta\omega) \approx \lambda_+ = \mbox{sign}(T) G_0^{-1/2} e^{-\alpha_a L/2} e^{i\psi} \left[|T| + \sqrt{T^2 - 1}\right] \label{eq:10-Delta}
\eeq
where $T = \tfrac{1}{2}\mbox{Tr}[R(\phi)M]$.

\begin{figure}[p]
\begin{center}
\includegraphics[width=1.00\textwidth]{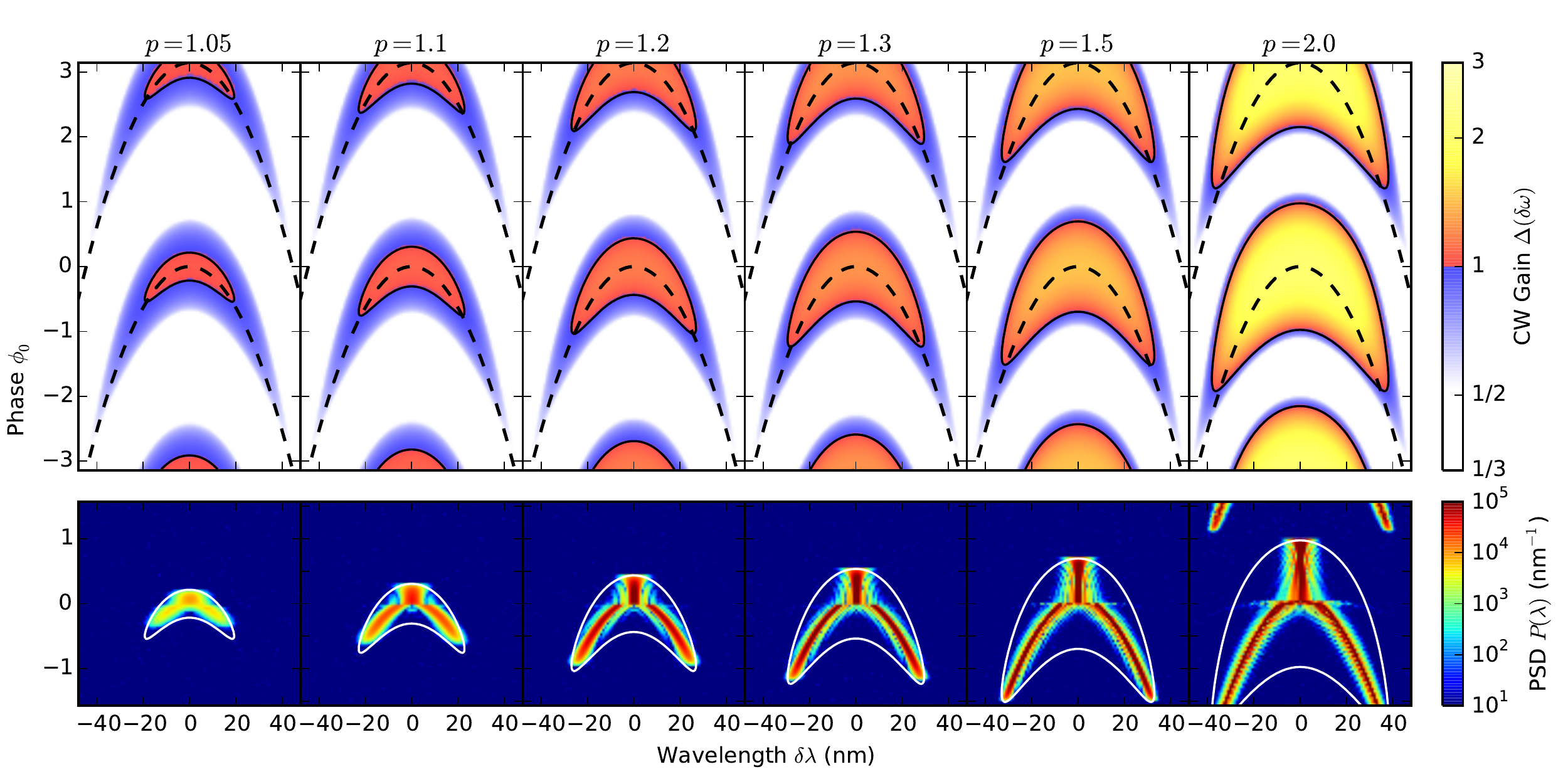}
\caption{Top: CW gain $|\Delta(\delta\omega)|$ as a function of $\delta\lambda = (-\lambda^2/2\pi c) \delta\omega$.  Bottom: plot of power spectral density $P(\lambda)$ (in photons/nm$^2$).  White contour gives the threshold condition $|\Delta| = 1$.  PPLN OPO with $L = 4$ cm, no fiber.}
\label{fig:10-f3a}
\end{center}
\end{figure}

\begin{figure}[p]
\begin{center}
\includegraphics[width=1.00\textwidth]{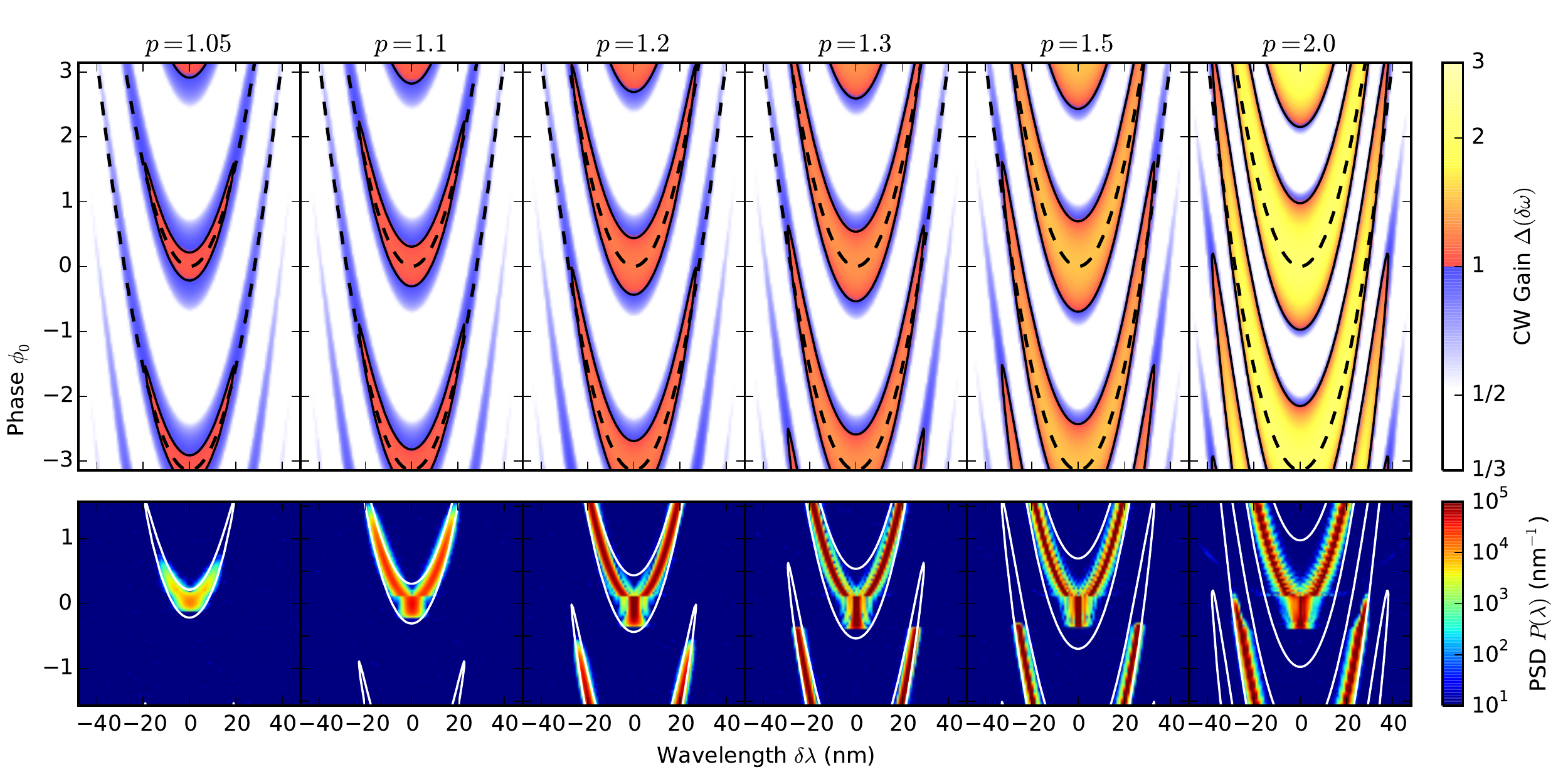}
\caption{PPLN OPO, 1-m SMF-28e fiber ($\beta_2 = -1.58 \times 10^{-26}$ s$^2$/m, $\beta_3 = 1.10 \times 10^{-40}$ s$^3$/m).}
\label{fig:10-f3b}
\end{center}
\end{figure}

\begin{figure}[t]
\begin{center}
\includegraphics[width=1.00\textwidth]{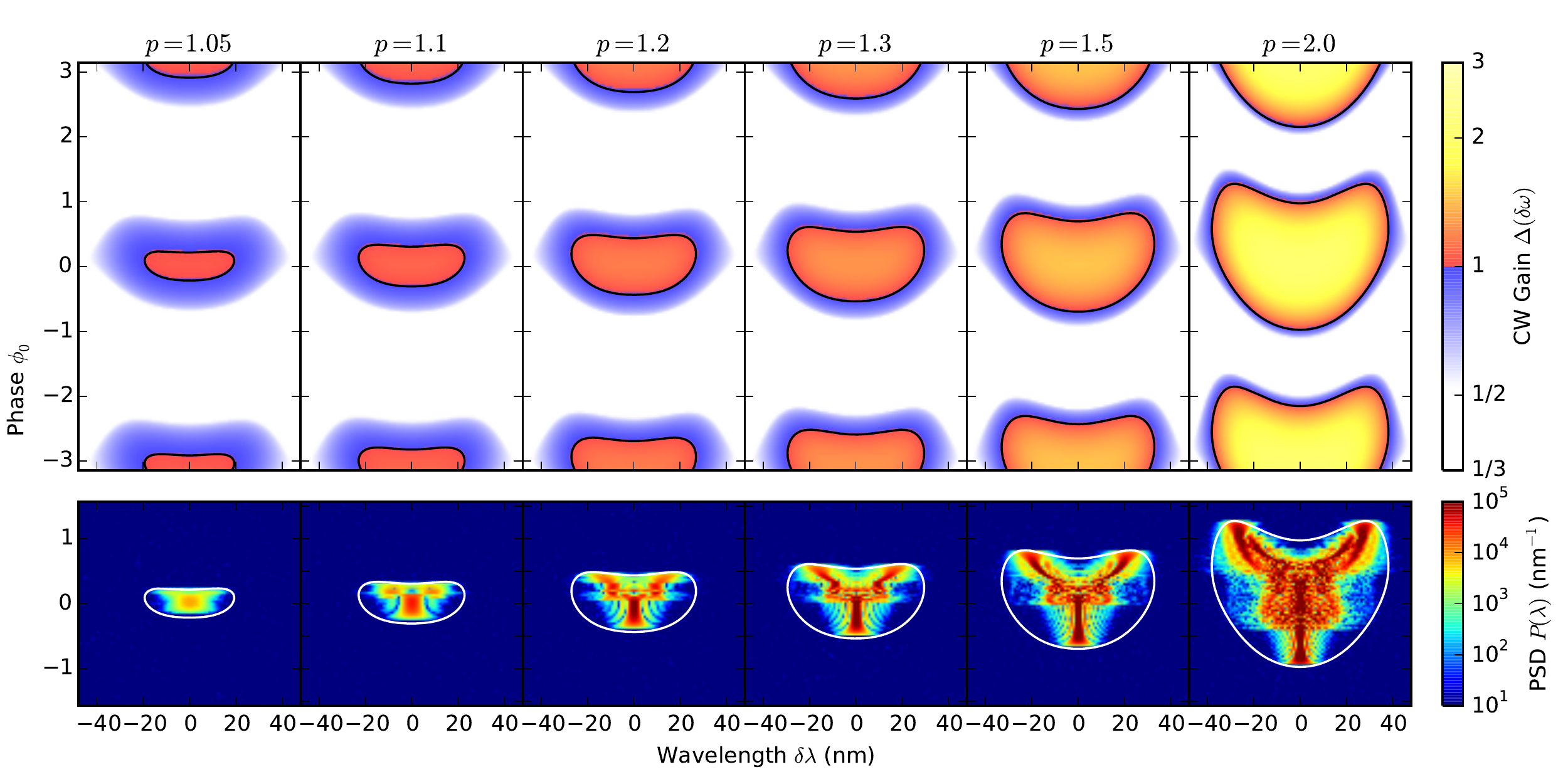}
\caption{PPLN OPO, GVD-compensating fiber ($\phi_2 = -4.49 \times 10^{-27}$ s$^2$, $\phi_3 = 5.14 \times 10^{-41}$ s$^3$).}
\label{fig:10-f3c}
\end{center}
\end{figure}

The pump can be written in terms of its normalized amplitude $p$, where $p = 1$ is the OPO threshold for a CW source with the same peak intensity as $b(t)$.  Since the threshold depends on $\phi_0$, for specificity we take the lowest threshold, when $\phi_0 = 0$, $\delta\omega = 0$:
\beq
	\bar{b} = p\,\bar{b}_0,\ \ \ \bar{b}_{0} = \frac{1}{2L\epsilon} \log(G_0 e^{\alpha_a L})
\eeq
At $p$ times above threshold, the maximum gain is at $\phi_0 = 0$, $\delta\omega = 0$, where dispersion effects disappear:
\beq
	\mbox{max}_{\phi_0, \delta\omega} {\Delta(\delta\omega, \phi_0)} = (G_0 e^{\alpha_a L})^{p-1} \label{eq:10-maxgain}
\eeq

Figures \ref{fig:10-f3a}-\ref{fig:10-f3c} compare the CW gain from Eq.~(\ref{eq:10-Delta}) to numerical spectra.  The frequency content of the OPO signal lives within the frequency-gain window $|\Delta(\delta\omega)| > 1$, as expected, centered on the resonance condition
\beq
	\phi_0 + \tfrac{1}{2}\underbrace{(\phi_2 + \beta_2 L)}_{\phi'_2} \delta\omega^2 = n\pi \label{eq:10-phimatch}
\eeq
which essentially says that the line-of-center phase shift $\phi_0$ must be compensated by the total (waveguide plus fiber) dispersion.  The shape of the spectrum depends on independent factors, which I will revisit in Sec.~\ref{sec:10-boxpulse}.

\subsubsection{Approximate Forms}

Equation (\ref{eq:10-Delta}) gives an accurate model of the CW round-trip gain, but it is cumbersome and it would be helpful to have an approximate form that is easier to work with analytically.  

Naturally, one expects the gain to be maximized when the fiber dispersion compensates the waveguide dispersion, that is: $\phi_0 + \tfrac{1}{2}\phi'_2\delta\omega^2 = n\pi$ (with $\phi'_2 = \phi_2 + \beta_2 L$).  There are two possible limits:

\begin{enumerate}
	\item $\phi_0\phi'_2 \geq 0$.  This is the degenerate limit, because no value of $\delta\omega$ can satisfy the phase relation.  We assume that $a(t)$ is real when it exits the crystal.  This is not exact (Eq.~(\ref{eq:10-cw}) assumes $a(t)$ can have arbitrary phase), but is approximately true because the amplification is phase-sensitive.
	
	Next, we treat the dispersion as a lumped element.  Thus, $a(t)$ entering the cavity has a phase $\phi = \phi_0 + \tfrac{1}{2}\phi'_2\delta\omega^2$.  Since we are only keeping track of the real part of the field as per the first assumption, this amounts to a round-trip gain of:
	\beq
		\Delta(\delta\omega) \approx \Delta_{\rm max} \cos\left(\phi_0 + \tfrac{1}{2}\phi'_2\delta\omega^2\right) \label{eq:10-Delta-app}
	\eeq
	The cosine term can be expanded, giving an approximation for $D(\delta\omega) = \log(\Delta(\delta\omega)/\Delta_{\rm max})$
	\beq
		\quad\quad\quad\ \  D(\delta\omega) \approx -\frac{\phi'_2\tan\phi_0}{2} \delta\omega^2 - \frac{(\phi'_2\sec\phi_0)^2}{8} \delta\omega^4 \label{eq:10-dw-deg}
	\eeq
	
	\item $\phi_0\phi'_2 < 0$.  This is the nondegenerate limit.  We make the same assumptions as before, but this time there exists a $\delta\omega_0 \equiv \sqrt{-2\phi_0/\phi'_2}$ that satisfies the phase relation.  At this frequency, $\Delta(\delta\omega)$ is (approximately) maximized.  Expanding the formula (\ref{eq:10-Delta-app}) about that point, we obtain:
	\beq
		D(\delta\omega) \approx -|\phi_0\phi'_2| (\delta\omega - \delta\omega_0)^2 \label{eq:10-dw-nd}
	\eeq
	Section~\ref{sec:10-emshapes} makes use of Eqs.~(\ref{eq:10-Delta-app}-\ref{eq:10-dw-nd}) to obtain an analytic form for the pulse shape.	
\end{enumerate}

\subsection{Dispersionless Step}
\label{sec:10-dispersionless}

\begin{figure}[b!]
\begin{center}
\includegraphics[width=0.9\columnwidth]{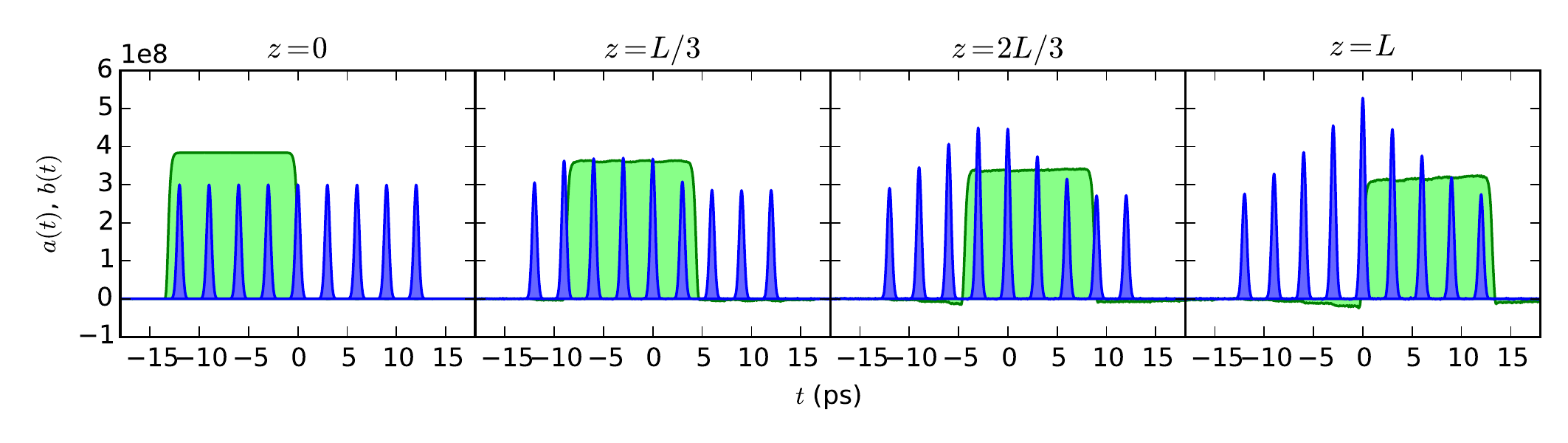}
\caption{Illustration of gain clipping.  A train of femtosecond pulses (blue) are amplified by a picosecond pump pulse (green).}
\label{fig:10-f4}
\end{center}
\end{figure}

The dispersionless step treats Eqs.~(\ref{eq:10-at}-\ref{eq:10-bt}) without the dispersion terms with the residual pump $b_{\rm in}(t) - b_{\rm max}$ (since $b(t) = b_{\rm max}$ was used in the continuous-wave pump, and we need to avoid double-counting the gain).  Since this section is about linear effects, we ignore pump depletion (but see Sec.~\ref{sec:10-nonlinear}), so the pump integrates to $(b_{\rm in}(t - uz) - b_{\rm max})e^{-\alpha_b z/2}$ ($u = v_b^{-1}-v_a^{-1}$ is the temporal walkoff) and Eq.~(\ref{eq:10-at}) becomes:
\beq
    \frac{\partial a(z,t)}{\partial z} =  - \frac{1}{2}\alpha_a a(z,t) + \epsilon\,a(z,t)^* (b_{\rm in}(t - uz) - b_{\rm max}) e^{-\alpha_b z/2} \label{eq:10-da-gc}
\eeq

We assume that $a(z,t)$ is close to real, because the imaginary component experiences loss when propagating through the waveguide.  This is only approximate when there is dispersion ($\beta_2 \neq 0, \phi_2 \neq 0$) or detuning ($\phi_0 \neq 0$).  Integrating (\ref{eq:10-da-gc}) we obtain the input-output map:
\beq
	a(t) \rightarrow \underbrace{\exp\Bigl(\int_0^L{\!\epsilon(b_{\rm in}(t - uz) \!-\! b_{\rm max}) e^{-\alpha_b z/2} \d z}\Bigr)}_{\Gamma(t)} a(t) \label{eq:10-gammat}
\eeq
The {\it gain-clipping function}, defined after Eq.~(\ref{eq:10-eig}) as $G(t) = \log \Gamma(t)$, is:
\beq
	G(t) = \int_0^L{\epsilon(b_{\rm in}(t - uz) - b_{\rm max}) e^{-\alpha_b z/2} \d z} \label{eq:10-gcf}
\eeq
This function is always negative, so the dispersionless step always gives rise to loss. We call this effect ``gain-clipping'' because it results in a temporal localization of gain, and confines the pulse in time.

The concept is illustrated in Figure \ref{fig:10-f4}.  As a signal pulse propagates through the waveguide, it walks through the pump.  The pulse gain depends on the amount of pump that it passes through, which in turn depends on the pulse's position.  Thus $G(t)$ takes the form of an integral.  For box pulses whose duration matches the walkoff time in the crystal ($T_p = L u$), it is given by:
\beq
	G(t) \approx -\frac{\epsilon\,b_{\rm max}}{u}|t| = -p \frac{\log(G_0 e^{\alpha_a L})}{2T_p} |t| \label{eq:10-gain-gc0}
\eeq
The total gain in the split-step approximation is $\Gamma(t)\Delta(k)$.  Assuming a box pump and negligible dispersion, we can replace $\Delta(k) = \Delta_{\rm max}$ with (\ref{eq:10-maxgain}) and thus the gain is
\beq
	\Delta_{\rm max}\Gamma(t) = \exp\left[\frac{\log(G_0 e^{\alpha_a L})}{2} \left((p-1) - p\frac{|t|}{T_p}\right)\right] \label{eq:10-gain-gc}
\eeq

\begin{figure}[tbp]
\begin{center}
\includegraphics[width=1.00\columnwidth]{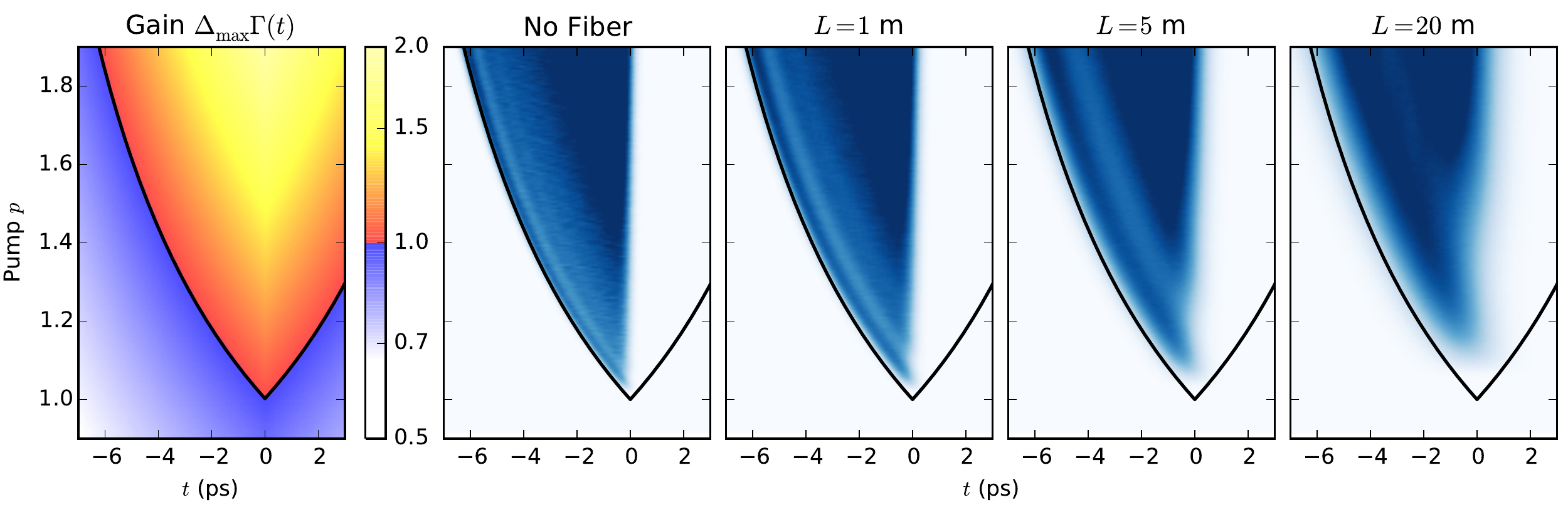}
\caption{Left: Dispersionless round-trip gain $\Delta_{\rm max} \Gamma(t)$ as a function of pump power and time, given by Eq.~(\ref{eq:10-gain-gc}).  Right: Pulse power $|a(t)|^2$ for PPLN waveguide, fiber lengths $L_f = 0$ m and 5 m (which overcompensates the GVD by a factor of 17.6).}
\label{fig:10-f5}
\end{center}
\end{figure}

As Figure \ref{fig:10-f5} shows, the pulse is confined to the positive gain ($\Delta_{\rm max} \Gamma(t) > 1$).  The signal pulses become longer as the pump power is increased, since the gain window becomes larger the larger $p$.  But only the left half of the gain window is filled.  This behavior will be explored in more detail in Sec.~\ref{sec:10-boxpulse}, but in short is a result of walkoff and pump depletion: the right-side region only reaches the pump once it has been depleted by the left side, and is no longer sufficient for amplification.  By this reasoning, the pulse width is derived from (\ref{eq:10-gain-gc}) to be half the gain-window width:
\beq
	T_s = \frac{p-1}{p} T_p
\eeq

This result is consistent with the simulations.  The agreement is strongest when the cavity dispersion is weakest.  As we add dispersion to the cavity, we filter out the high-frequency modes and force $a(t)$ to take a smoother waveform (Fig.~\ref{fig:10-f5}, right plot).  To model the case with dispersion we need both $\Gamma(t)$ and $\Delta(\delta\omega)$ -- this will be done in the following section.

\subsection{Shapes of Eigenmodes}
\label{sec:10-emshapes}

Now that we have the gain-clipping and dispersion terms, Eqs.~(\ref{eq:10-Delta}, \ref{eq:10-gammat}), we are ready to find the eigenmodes.  There are two ways to do this: using Eq.~(\ref{eq:10-eig}) gives $a_k(t)$ exactly, along with the round-trip gain $g_k \equiv \log\lambda_k$; however, this approach must be done numerically.  Alternatively, one can take the near-threshold approximation Eq.~(\ref{eq:10-nt-eig}), and using analytic approximations for $G(t)$, $D(i\tfrac{\d}{\d t})$, obtain analytic expressions for the eigenmodes.  The analytic method is presented first, and compared to Eq.~(\ref{eq:10-eig}) and simulations in the following subsection.

\subsubsection{Analytic Form, Degenerate Case ($\phi_0 \phi'_2 \geq 0$)}

As the resonance diagrams in Figs.~\ref{fig:10-f3a}-\ref{fig:10-f3c} make clear, there are two regimes of interest: degenerate and non-degenerate.  The OPO is degenerate when $\phi'_2\phi_0 > 0$, where $\phi'_2 = \phi_2 + \beta_2 L$ (Eq.~(\ref{eq:10-phimatch})).  In this case, using Eq.~(\ref{eq:10-nt-eig}) and substituting (\ref{eq:10-dw-deg}) and (\ref{eq:10-gain-gc0}) for the $G(t)$ and $D(i\tfrac{\d}{\d t})$ respectively, we find near threshold that $a(t)$ gets mapped after one round trip to
\beq
	\Bigl[g_{\rm cw} \underbrace{-\ \tfrac{\log(G_0 e^{\alpha_a L})}{2T_p} |t|}_{G(t)} + \underbrace{\vphantom{\tfrac{\log(G_0 e^{\alpha_a L})}{2T_p} |t|}\tfrac{\phi'_2\tan\phi_0}{2}\tfrac{\d^2}{\d t^2} - \tfrac{(\phi'_2\sec\phi_0)^2}{8}\tfrac{\d^4}{\d t^4}}_{D(i\tfrac{\d}{\d t})}\Bigr] a(t) \label{eq:10-deg-eig}
\eeq
and thus $[g_{\rm cw} + G(t) + D(i\tfrac{\d}{\d t})] a_k = g_k a_k$ is the eigenvalue equation.

The general case is not solvable analytically, but usually one of the time-derivative terms is much larger than the other, leading to one of two limits:

\begin{enumerate}

\item $\phi_0 \sim O(1)$.  Since $g_k$ is small near threshold, both $G(t)$ and $D(i\tfrac{\d}{\d t})$ must be small, and are typically of the same order.  But if $(\phi'_2\tfrac{\d^2}{\d t^2})a(t) \sim O(g_k) \ll 1$, then $(\phi'_2\tfrac{\d^2}{\d t^2})^2 a(t) \sim O(g_k^2) \ll (\phi'_2\tfrac{\d^2}{\d t^2})a(t)$ and so the fourth-derivative term can be neglected.  In this case (\ref{eq:10-deg-eig}) gives Airy's equation, with the solutions: 
\bea
	a_k(t) & \!\!=\!\! & \mbox{sign}(t)^k \mbox{Ai}\left[\left(\frac{T_p \phi'_2\tan\phi_0}{\log(G_0 e^{\alpha_a L})}\right)^{-1/3} \!|t| - \xi_k \right] \label{eq:10-eig-airy} \\
	g_k & \!\!=\!\! & g_{\rm cw} - \frac{1}{2}\left(\frac{\phi'_2\tan\phi_0}{T_p^2}\log(G_0 e^{\alpha_a L})^2\right)^{1/3} \xi_k \label{eq:10-eig-airy2}
\eea
where $-\xi_k$ are the roots and extrema of the Airy function $\mbox{Ai}(\tau)$ (Table \ref{tab:10-t2}).

\item $\phi_0 \approx 0$.  In this case the second-derivative term is discarded because it goes as $\tan\phi_0$.  The result is a fourth-order analog of Airy's equation: $\d^4y/\d x^4 + xy = 0$, which has two linearly independent solutions that satisfy the boundary conditions at $|t| \rightarrow \infty$: $R_1(\zeta), R_2(\zeta)$ (see Eq.~(\ref{eq:10-rf1}-\ref{eq:10-rf2})). The solution is given by the linear combination
\bea
	a_k(t) & \!\!=\!\! & \mbox{sign}(\zeta)^k \bigl[c_{1,k} R_1(|\zeta| - \zeta_k) + c_{2,k} R_2(|\zeta| - \zeta_k)\bigr]
	 \nonumber \\
	& & \zeta \equiv \left(\tfrac{T_p (\phi'_2)^2}{4\log(G_0 e^{\alpha_a L})}\right)^{-1/5}t \label{eq:10-phizero-ak}  \\
	g_k & \!\!=\!\! & g_{\rm cw} - \frac{1}{2}\left(\frac{\log(G_0 e^{\alpha_a L})^4 (\phi'_2)^2}{4T_p^4}\right)^{1/5}\zeta_k \label{eq:10-eig-hg2}
\eea
that satisfies the differentiability conditions at $t = 0$.  This condition constrains $\zeta_k$ (and thus $g_k$), since these conditions can be reduced to finding a matrix null-space:

\begin{table}[tbp]
\begin{center}
\begin{tabular}{c|cccccccc}
\hline\hline
$k$       & 0    & 1    & 2    & 3    & 4    & 5    & 6    & 7    \\ \hline
$\xi_k$   & 1.02 & 2.34 & 3.25 & 4.09 & 4.82 & 5.52 & 6.16 & 6.79 \\ 
$\zeta_k$ & 0.97 & 2.36 & 3.56 & 4.66 & 5.71 & 6.70 & 7.66 & 8.59 \\ \hline \hline
\end{tabular}
\caption{$\xi_k$ and $\zeta_k$ used in Eqs.~(\ref{eq:10-eig-airy}-\ref{eq:10-eig-hg2})}
\label{tab:10-t2}
\end{center}
\end{table}

\end{enumerate}

\beq
	\underbrace{\begin{bmatrix} R_1'(-\zeta_k) & R_2'(-\zeta_k) \\ R_1'''(-\zeta_k) & R_2'''(-\zeta_k) \end{bmatrix}
	\!\!\begin{bmatrix} c_{1,k} \\ c_{2,k} \end{bmatrix} \!=\! 0}_{k\,=\,0,2,\ldots\ \text{(even solutions)}},\ \
	\underbrace{\begin{bmatrix} R_1(-\zeta_k) & R_2(-\zeta_k) \\ R_1''(-\zeta_k) & R_2''(-\zeta_k) \end{bmatrix}
	\!\!\begin{bmatrix} c_{1,k} \\ c_{2,k} \end{bmatrix} \!=\! 0}_{k\,=\,1,3,\ldots\ \text{(odd solutions)}} 
\eeq
The roots $\zeta_k$ are listed in Table \ref{tab:10-t2}.  For reference, $R_1(\zeta)$ and $R_2(\zeta)$ can be expressed in terms of hypergeometric functions:
\begin{align}
	R_1(\zeta) & = 
			{}_0F_3\bigl(;\tfrac{2}{5}, \tfrac{3}{5},\tfrac{4}{5};\tfrac{-\zeta^5}{625}\bigr) 
		- \tfrac{2\pi}{5^{1/20}\phi^{3/2}\Gamma(\tfrac{1}{5})\Gamma(\tfrac{3}{5})}\; 
			{}_0F_3\bigl(;\tfrac{3}{5},\tfrac{4}{5},\tfrac{6}{5}; \tfrac{-\zeta^5}{625}\bigr) \zeta \nonumber \\
		& \qquad - \tfrac{5^{3/20}\pi}{\phi^{3/2}\Gamma(\tfrac{1}{5})\Gamma(\tfrac{2}{5})}\;
			{}_0F_3\bigl(;\tfrac{4}{5},\tfrac{6}{5},\tfrac{7}{5}; \tfrac{-\zeta^5}{625}\bigr) \zeta^2
		+ \tfrac{5^{3/5}\Gamma(\tfrac{4}{5})}{6\Gamma(\tfrac{1}{5})}\;
			{}_0F_3\bigl(;\tfrac{6}{5},\tfrac{7}{5},\tfrac{8}{5}; \tfrac{-\zeta^5}{625}\bigr) \zeta^3 \label{eq:10-rf1} \\
	R_2(\zeta) & =
			-{}_0F_3\bigl(;\tfrac{3}{5},\tfrac{4}{5},\tfrac{6}{5}; \tfrac{-\zeta^5}{625}\bigr) \zeta
		+ \tfrac{5^{1/5} \phi\,\Gamma(\tfrac{3}{5})}{2\Gamma(\tfrac{2}{5})}\;
			{}_0F_3\bigl(;\tfrac{4}{5},\tfrac{6}{5},\tfrac{7}{5}; \tfrac{-\zeta^5}{625}\bigr) \zeta^2 \nonumber \\
		& \qquad - \tfrac{5^{13/20}\phi^{1/2} \Gamma(\tfrac{3}{5})\Gamma(\tfrac{4}{5})}{12\pi}\;
			{}_0F_3\bigl(;\tfrac{6}{5},\tfrac{7}{5},\tfrac{8}{5}; \tfrac{-\zeta^5}{625}\bigr) \zeta^3 \label{eq:10-rf2}
\end{align}

\comment{
\bea
	\!\!R_1(\zeta) & \!\!=\!\! & 
			{}_0F_3\left(;\tfrac{2}{5}, \tfrac{3}{5},\tfrac{4}{5};\tfrac{-\zeta^5}{625}\right) 
		- \tfrac{2\pi}{5^{1/20}\phi^{3/2}\Gamma(\tfrac{1}{5})\Gamma(\tfrac{3}{5})}\; 
			{}_0F_3\left(;\tfrac{3}{5},\tfrac{4}{5},\tfrac{6}{5}; \tfrac{-\zeta^5}{625}\right) \zeta \nonumber \\
		& & - \tfrac{5^{3/20}\pi}{\phi^{3/2}\Gamma(\tfrac{1}{5})\Gamma(\tfrac{2}{5})}\;
			{}_0F_3\left(;\tfrac{4}{5},\tfrac{6}{5},\tfrac{7}{5}; \tfrac{-\zeta^5}{625}\right) \zeta^2
		+ \tfrac{5^{3/5}\Gamma(\tfrac{4}{5})}{6\Gamma(\tfrac{1}{5})}\;
			{}_0F_3\left(;\tfrac{6}{5},\tfrac{7}{5},\tfrac{8}{5}; \tfrac{-\zeta^5}{625}\right) \zeta^3 \label{eq:10-rf1} \\
	\!\!R_2(\zeta) & \!\!=\!\! & 
			-{}_0F_3\left(;\tfrac{3}{5},\tfrac{4}{5},\tfrac{6}{5}; \tfrac{-\zeta^5}{625}\right) \zeta
		+ \tfrac{5^{1/5} \phi\,\Gamma(\tfrac{3}{5})}{2\Gamma(\tfrac{2}{5})}\;
			{}_0F_3\left(;\tfrac{4}{5},\tfrac{6}{5},\tfrac{7}{5}; \tfrac{-\zeta^5}{625}\right) \zeta^2 \nonumber \\
		& & - \tfrac{5^{13/20}\phi^{1/2} \Gamma(\tfrac{3}{5})\Gamma(\tfrac{4}{5})}{12\pi}\;
			{}_0F_3\left(;\tfrac{6}{5},\tfrac{7}{5},\tfrac{8}{5}; \tfrac{-\zeta^5}{625}\right) \zeta^3 \label{eq:10-rf2}
\eea
}

\subsubsection{Analytic Form, Non-degenerate Case ($\phi_0 \phi'_2 < 0$)}

In the nondegenerate case, most of the frequency content is contained around $\delta\omega_0 = \sqrt{-2\phi_0/\phi'_2}$, which satisfies the phase condition $\phi_0 + \tfrac{1}{2}\phi'_2 \delta\omega_0^2 = 0$.  We thus make the substitution:
\beq
	a(t) = \mbox{Re}\left[\bar{a}(t) e^{-i\,\delta\omega_0 t}\right] \label{eq:10-abar}
\eeq
The eigenvalue equation (\ref{eq:10-nt-eig}) can be solved with the help of (\ref{eq:10-dw-nd}) and (\ref{eq:10-gain-gc0}); neglecting higher-order derivative terms we obtain:
\beq
	\Bigl[g_{\rm cw} \underbrace{-\ \frac{\log(G_0 e^{\alpha_a L})}{2T_p} |t|}_{G(t)} + \underbrace{|\phi'_2\phi_0|\frac{\partial^2}{\partial t^2}}_{D(i\tfrac{\d}{\d t})}\Bigr] \bar{a}_k(t) = g_k \bar{a}_k(t) \label{eq:10-nd-eig}
\eeq
Note that Eq.~(\ref{eq:10-nd-eig}) is the same as (\ref{eq:10-deg-eig}) if we remove the fourth-order derivative and replace $\tfrac{1}{2}\phi'_2\tan\phi_0 \rightarrow |\phi'_2\phi_0|$.  Thus, the solutions are Airy functions:
\bea
	\!\bar{a}_k(t) & \!=\! & \mbox{sign}(t)^k \mbox{Ai}\left[\left(\frac{2T_p |\phi'_2\phi_0|}{\log(G_0 e^{\alpha_a L})}\right)^{-1/3} \!|t| - \xi_k \right] \label{eq:10-nd-airy} \\
	\!g_k & \!=\! & g_{\rm cw} - \frac{1}{2}\left(\frac{2|\phi'_2\phi_0|}{T_p^2}\log(G_0 e^{\alpha_a L})^2\right)^{1/3} \xi_k \label{eq:10-nd-airy2}
\eea

\subsubsection{Full Form}

\begin{figure}[t]
\begin{center}
\includegraphics[width=1.00\columnwidth]{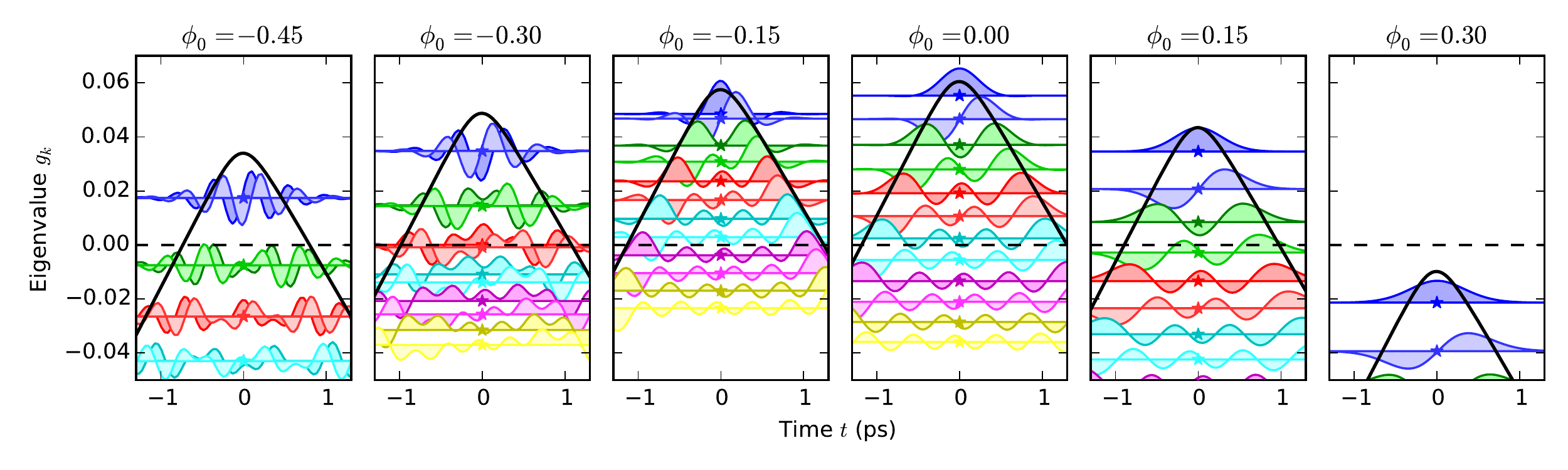}
\caption{Shapes of eigenmodes $a_k(t)$ as a function of $\phi_0$, PPLN OPO with $p = 1.1$ and no fiber.  Dark line is the dispersionless gain $\log(\Delta_{\rm max}) + G(t)$.}
\label{fig:10-f7a}
\end{center}
\end{figure}

\begin{figure}[t]
\begin{center}
\includegraphics[width=1.00\columnwidth]{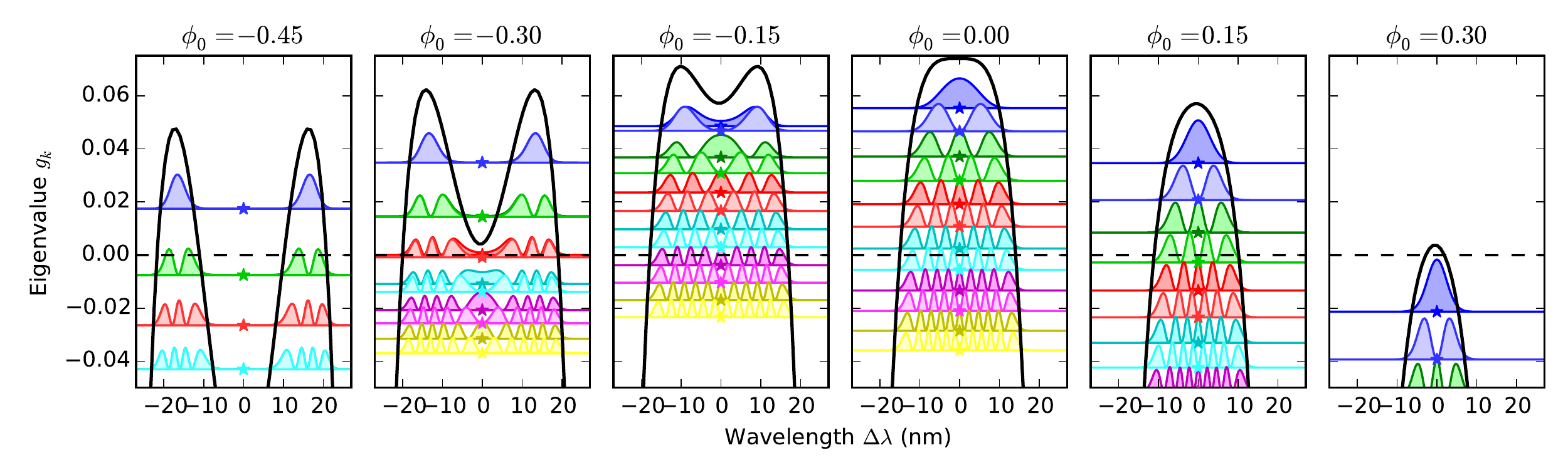}
\caption{Power spectra of eigenmodes $|a_k(\omega)|^2$, dark line is the CW gain $\log(\Delta_{\rm max}) + D(\delta\omega)$.}
\label{fig:10-f7b}
\end{center}
\end{figure}

\begin{figure}[p]
\begin{center}
\includegraphics[width=1.00\columnwidth]{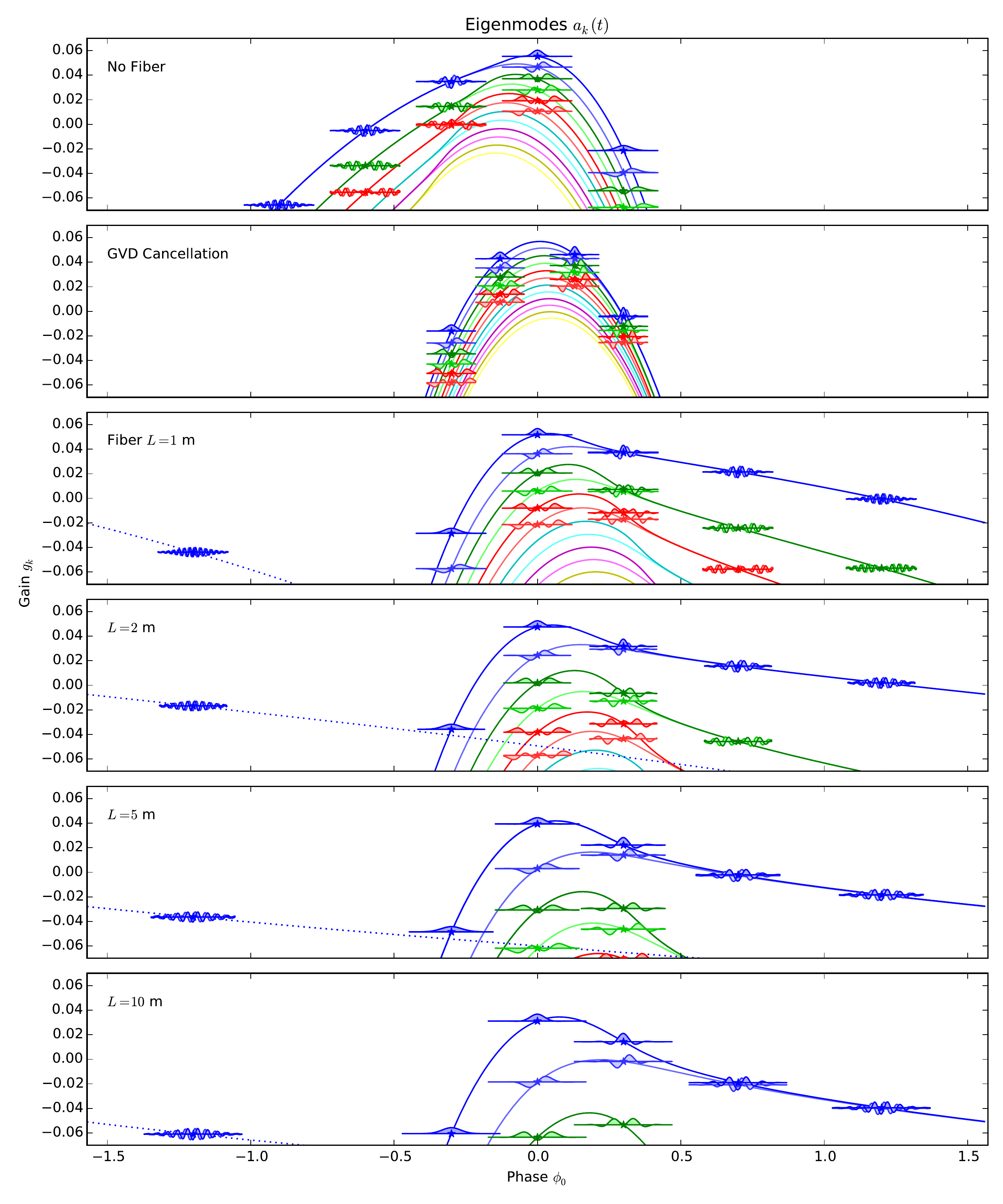}
\caption{Plots of eigenmodes $a_k(t)$ and eigenvalues $g_k$ at pump $p = 1.1$ as a function of cavity phase $\phi_0$ and fiber length $L$.  Pulse widths not to scale between graphs.}
\label{fig:10-f6}
\end{center}
\end{figure}

One can solve the eigenmode equation exactly without resorting to approximations, diagonalizing (\ref{eq:10-eig}) numerically using (\ref{eq:10-Delta}) and (\ref{eq:10-gammat}) for $\Delta(i\tfrac{\d}{\d t})$ and $\Gamma(t)$, respectively.  This approach is necessary in the GVD-compensated case, where the lumped-element approximations (\ref{eq:10-dw-deg}-\ref{eq:10-dw-nd}) break down.  Numerically, it is much easier to diagonalize $\Gamma(t)^{1/2} \Delta(i\tfrac{\d}{\d t}) \Gamma(t)^{1/2}$, which is Hermitian and whose eigenvectors are related to those of $\Gamma(t)\Delta(t)$ by a (nearly constant) function of $t$.

Figures \ref{fig:10-f7a}-\ref{fig:10-f7b} show the temporal and frequency structure of the eigenmodes $a_k(t)$.  The system studied here is the PPLN-waveguide OPO without any fiber.  Like particles in a potential well, each eigenmode wavefunction $a_k(t)$ is largely confined to the region $\log(\Delta_{\rm max} \Gamma(t)) > g_k$, since $\Gamma(t) = e^{G(t)}$ plays the role of the potential here.

The power spectra in Fig.~\ref{fig:10-f7b} show that the OPO smoothly transitions from degenerate to nondegenerate operation as the phase is scanned from positive to negative, consistent with the analysis in the previous sections.  This transition happens because the CW gain function $\Delta(\delta\omega)$ plays the role of a potential here.  This function is quadratic for $\phi_0 > 0$ but transitions to a double-well structure for $\phi_0 < 0$, leading to nondegenerate operation in that regime.

Fiber dispersion is accounted for in Figure \ref{fig:10-f6}.  Here the eigenvalues $g_k$ are plotted against $\phi_0$ for a range of fiber lengths.  As the fiber becomes longer, the spacing between eigenvalues increases, largely consistent with the scaling laws in Eqs.~(\ref{eq:10-eig-airy2}, \ref{eq:10-phizero-ak}, \ref{eq:10-nd-airy}).  

\begin{figure}[t]
\begin{center}
\includegraphics[width=1.00\columnwidth]{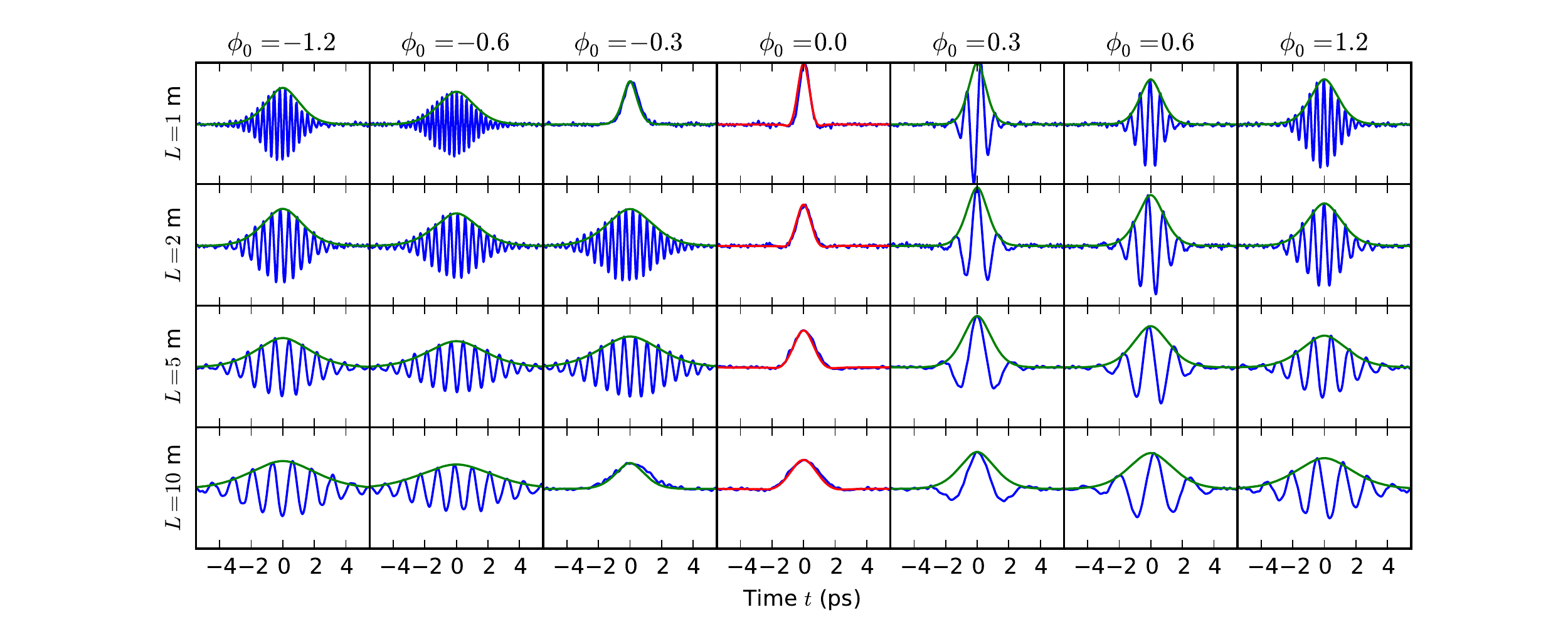}
\caption{OPO steady-state pulse shape just above threshold.  Blue: numerical result.  Green: Airy-function solution, (\ref{eq:10-eig-airy}) for degenerate case and (\ref{eq:10-nd-airy}) for nondegenerate case.  Envelope $\bar{a}_k(t)$ is plotted for nondegenerate case.  Red: hypergeometric result (\ref{eq:10-phizero-ak}).}
\label{fig:10-f8}
\end{center}
\end{figure}

Figure \ref{fig:10-f8} compares the pulse shapes from Eqs.~(\ref{eq:10-eig-airy2}, \ref{eq:10-phizero-ak}, \ref{eq:10-nd-airy}) against simulation data.  The simulation data are taken very close to threshold, so that nonlinear effects do not distort the pulse shape.

In addition to the obvious agreement between theory and simulation, Fig.~\ref{fig:10-f8} shows several important trends in the behavior of pulsed OPOs.  First, the pulses become longer the more fiber is inserted into the OPO ($L = 1$ m already over-compensates the PPLN dispersion).  In addition, the larger one makes $\phi_0$ in the nondegenerate region, the larger the signal-idler splitting, consistent with the signal-idler splitting $\delta\omega = \sqrt{-2\phi_0/\phi'_2}$ (Eq.~(\ref{eq:10-phimatch})).

\subsection{Threshold}

Threshold is both straightforward to measure and easy to derive from the linearized model.  It is the pump power needed to make the principal eigenmode have the highest gain: $g_0 = 0$.  Since the eigenmode gain depends on $\phi_0$, threshold depends on $\phi_0$ as well, giving rise to the detuning peaks in Fig.~(\ref{fig:10-f2}).  For a CW pump at $\phi_0 = 0$, the threshold is clearly $p = 1$.

We can compute thresholds near the center of a detuning peak by inverting the eigenmode gain expression.  Recall from (\ref{eq:10-eig-airy2}, \ref{eq:10-eig-hg2}, \ref{eq:10-nd-airy2}) that the eigenmode gain takes the form:
\beq
	g_k = g_{\rm cw} + g'_k \label{eq:10-gk-zero}
\eeq
where $g'_k$ depends on the differential equation being solved.  Near the center of the detuning peak, the CW gain goes as $\Delta \approx (G_0 e^{\alpha_a L})^{p-1}$ (Eq.~(\ref{eq:10-maxgain})), so we can write $g_{\rm cw}(p) \approx g_{\rm cw}(p=1) + \tfrac{p-1}{2}\log(G_0 e^{\alpha_a L})$.  Setting the gain (\ref{eq:10-gk-zero}) to zero, we obtain an approximate formula for the threshold:
\beq
	p_{\rm th} = 1+\frac{-g_0(p=1)}{\tfrac{1}{2}\log(G_0 e^{\alpha_a L})}
\eeq
This relation is be valid for $|g_0| \ll 1$.  In the same way, we can compute the thresholds for the higher eigenmodes.

\begin{figure}[tbp]
\begin{center}
\includegraphics[width=1.00\columnwidth]{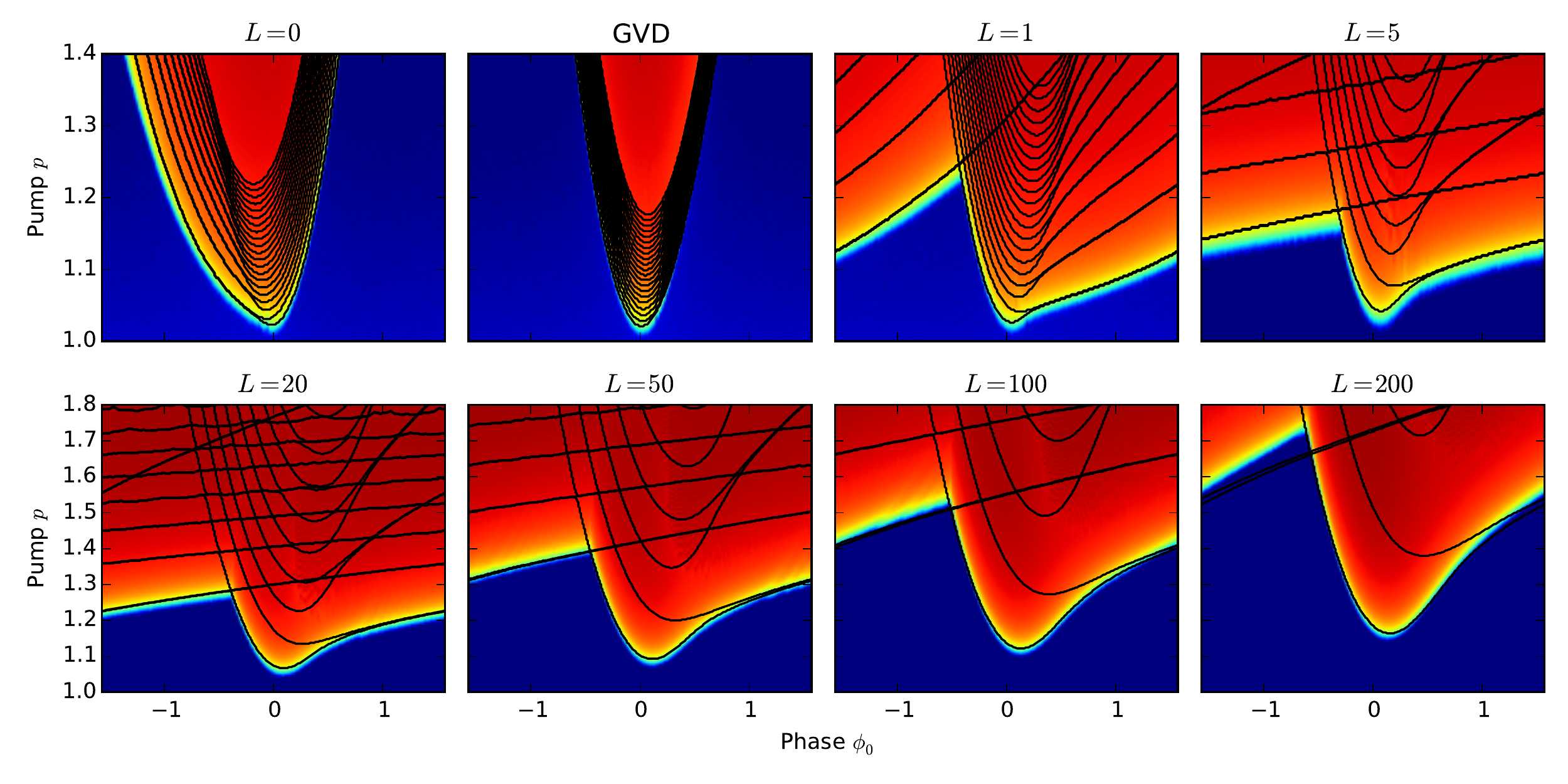}
\caption{Plot of OPO efficiency $\eta = P_{a,\rm out}/P_{b,\rm in}$ in terms of $p$ and $\phi_0$, with cavity dispersion provided by various lengths of fiber $L_f$; GVD refers to a fiber that compensates the dispersion of the $\chi^{(2)}$ medium.  Contours are thresholds for the first 20 eigenmodes $a_k(t)$.}
\label{fig:10-f9}
\end{center}
\end{figure}

By definition, the OPO turns on when the pump power exceeds threshold.  In the simulation results of Fig.~\ref{fig:10-f9}, the OPO efficiency $\eta = P_{a,\rm out}/P_{b,\rm in}$ is plotted against cavity phase and pump power.  In simulations, the OPO turns on right at the point where the highest eigenmode goes above threshold ($g_0 = 0$).  Thus, the eigenmode model should give accurate predictions of pulsed OPO thresholds.

Note that the structure of these thresholds matches that of the eigenmodes.  Consistent with Fig.~\ref{fig:10-f6}, the eigenmodes ``pair up'' in the nondegenerate regime $\phi_0\phi'_2 < 0$.  Also, as the fiber length is increased, the spacing between thresholds increases.

Figure \ref{fig:10-f9} is useful because it tells us when a pulsed OPO is in single-mode operation.  If the pump is below the threshold for the first excited mode $a_1(t)$, then the device behaves like a single-mode OPO.  But once it passes that threshold, multiple modes can oscillate in principle, and the dynamics may become more complex.  Multi-mode effects, coupled with nonlinearity, can give rise to oscillation (Sec.~\ref{sec:10-twomode}), instabilities (Sec.~\ref{sec:10-simstab}), centroid drift (Sec.~\ref{sec:10-ansatz}), and the formation of flat-top pulses (Sec.~\ref{sec:10-boxpulse}).  More complex behavior is possible with multimode OPO networks; recent experiments have hinted towards a multimode description \cite{Takata2016}, and the topic is being actively investigated.

\section{Nonlinear Corrections to Eigenmode Theory}
\label{sec:10-nonlinear}

Now we add nonlinearity to the model.  Nonlinearity is essential to anything above threshold, since it prevents signals from diverging to infinity.  It also makes the otherwise-independent eigenmodes interact.  The resulting pulse shape will depend on OPO parameters like $p$, $\phi_0$.

This section treats nonlinearity as a perturbation to the eigenmode dynamics.  This will only be valid reasonably close to threshold.  Moreover, it is necessary to truncate the nonlinear model by keeping only a finite number of eigenmodes in the basis.  The required number of eigenmodes grows as the pump power increases and more modes go above threshold (Fig.~\ref{fig:10-f9}).  The method described here has $O(N^4)$ complexity, where $N$ is the number of modes, so if too many modes are included it becomes impractical.  However, we will show in this section that a reasonable number ($N \lesssim 20$) gives good agreement with numerical data.  Thus, the nonlinear eigenmode theory is a good alternative ``reduced model'' that captures the dynamics of the full simulations, but takes $10^2$--$10^3$ times less computation time.

In addition to nonlinearity, cavity detuning will be treated in this section.  To treat these two effects, first we introduce the equations of motion and project them onto the eigenmode basis (Sec.~\ref{sec:10-nl-eom}).  Next we discuss the results of an analytic ``two-mode'' model (Sec.~\ref{sec:10-twomode}) which provides insight into pulse stability and dynamics, and finally compare the nonlinear eigenmode model with full simulations (Sec.~\ref{sec:10-nl-num}).

\subsection{Equations of Motion}
\label{sec:10-nl-eom}

The normal modes derived in Section \ref{sec:10-linear} allow us to describe the field of the OPO pulse in terms of a few mode amplitudes rather than hundreds of Fourier components.  This greatly reduces the complexity of the problem, at the cost of having to compute the modes in the first place and being restricted to a subspace spanned by the dominant modes.  Supposing that $a(t;n)$ is the pulse at the $n^{\rm th}$ round trip.  This can be written in terms of the normal modes $a_k(t)$ and their amplitudes $c_k(n)$:
\beq
	a(t;n) = \sum_k a_k(t) c_k(n)
\eeq
In the absence of pump depletion or any other effects, the equation of motion is:
\beq
	c_k(n+1) = e^{g_k} c_k(n) \label{eq:10-dc-disc}
\eeq
In the near-threshold case where $g_k \ll 1$, this can be converted to a differential equation:
\beq
	\frac{\d c_k}{\d n} = g_k c_k \label{eq:10-dc-cont}
\eeq
Pump depletion and cavity length detuning (repetition-rate mismatch) give corrections to the linear model, as described in the sections below.

\subsubsection{Detuning}

When the cavity is detuned by a length $\ell$, the signal picks up a round-trip phase $\pi\ell$ and its envelope shifts by $(\lambda/2c)\ell$:
\beq
	a(t) \rightarrow a(t - \tfrac{\lambda}{2c}\ell) e^{i \pi \ell}
\eeq
The phase shift was accounted for when the normal modes were chosen.  It is easy to account for the envelope shift in the normal-mode picture, using the map
\beq
	c_k \rightarrow S_{kl}(\ell) c_l,\ \ \ S_{kl}(\tau) = \int{a_k(t) a_l(t-\tfrac{\lambda}{2c}\ell)\d t} \label{eq:10-rrrm0}
\eeq
Combining both (\ref{eq:10-dc-disc}) and (\ref{eq:10-rrrm0}), one arrives at the relation $c_k(n+1) =  \sum_l S_{kl} e^{g_l} c_l(n)$.  If the field changes slowly between round trips, e.g. $g_k, S_{k\neq l} \ll 1$, then one has:
\beq
	\underbrace{\frac{\d}{\d n}\begin{bmatrix} c_0 \\ c_1 \\ \vdots \\ c_m \end{bmatrix}}_{\d c/\d n} = 
	\underbrace{\begin{bmatrix} g_0 & \ell J_{01} & \cdots & \ell J_{0m} \\
		-\ell J_{01} & g_1 & \cdots & \ell J_{1m} \\
		\vdots & \vdots & \ddots & \vdots \\
		-\ell J_{0m} & -\ell J_{1m} & \cdots & g_m \end{bmatrix}}_{\ell J + G} 
	\underbrace{\begin{bmatrix} c_0 \\ c_1 \\ \vdots \\ c_m \end{bmatrix}}_{c} \label{eq:10-det-dc}
\eeq
where the coupling matrix $J$ is:
\beq
	J_{kl} = \left.\frac{\d S_{kl}}{\d\ell}\right|_{\ell=0} = -\frac{\lambda}{2c}\int{a_k(t) \frac{\d a_l(t)}{\d t} \d t}
\eeq

Integration by parts shows that $J_{kl}$ is antisymmetric.  Also, it only mixes modes of opposite parity.  The linear dynamics are set by the matrix $G + J$.  This mixes modes of positive and negative eigenvalue.  If the mixing is strong enough, all of the eigenvalues will be negative and the oscillation is suppressed.  Thus the oscillation threshold will increase with increasing $|\ell|$.

\subsubsection{Pump Depletion}

To calculate the effect of pump depletion, go back to Eqs.~(\ref{eq:10-at}-\ref{eq:10-bt}).  During the dispersionless step in Sec.~\ref{sec:10-dispersionless}, we solved these equations in the absence of GVD.  The pump equation can be integrated using the method of characteristics to give:
\beq
    b(z, t)\!=\!b_{\rm in}(t - u z)e^{-\alpha_b z/2} - \frac{\epsilon}{2}\!\int_0^z{\!\!e^{\alpha_b (z'-z)/2} a\!\left(z', t\!+\!u(z'\!-\!z)\right)^2 \!\d z'}
\eeq
We now invoke the ``gain-without-distortion ansatz'' used to derive the linear eigenmode theory.  In this case it takes the form: $a(z', t) \approx G_{z\rightarrow z'} a(z, t)$.  For small $|z'-z|$, say of order one walkoff length, we can expand $G_{z\rightarrow z'}$ in terms of the $z$ coordinate $a(z', t) \approx e^{g(z)(z'-z)/2} a(z, t)$.  When this is so, we can account for the $z'$ dependence in the integral on the right with a factor of $e^{g(z)(z'-z)/2}$,  change the integration variable to $t' = t+u(z'-z)$ and (in the limit that the walkoff length $L u$ is much longer than the signal) set the left bound to $-\infty$, and obtain:
\beq
b(z, t) = b_{\rm in}(t - u z)e^{-\alpha_b z/2} - \frac{\epsilon}{2u} \int_{-\infty}^{t}{e^{(g(z)+\alpha_b/2)(t'-t)/u} a(z, t')^2 \d t'}
\eeq
Substituting this into the differential equation for $a$, we can eliminate the pump and obtain an equation of motion that depends only on the signal:
\begin{align}
    & \!\!\!\!\frac{\partial a(z,t)}{\partial z} = -\frac{1}{2} \alpha_a a(z, t) + \epsilon\,a^*(z,t) b_{\rm in}(t - u z)e^{-\alpha_b z/2} \nonumber \\
    & \qquad \underbrace{-\frac{\epsilon^2}{2u} a^*(z,t)\!\int_{-\infty}^{t}{\!\!e^{(g(z)+\alpha_b/2)(t'-t)/u} a(z, t')^2 \d t'}}_{\partial a/\partial z\bigr|_{\rm NL}} \label{eq:10-intdiff}
\end{align}
Now one can apply the gain without distortion approximation so that $a(z,t)$ can be related to its initial condition, expressing the right-hand side of (\ref{eq:10-intdiff}) in terms of $a_{in}(t)$.  

For a constant pump, $g$ is constant in $z$, but in general it will go as $g = 2G_{0\rightarrow z}^{-1} \d G_{0\rightarrow z}/\d z$.  Although $G_{0\rightarrow z}$ depends on $t$, the dependence is weak in the region where the pulse forms (at least for the waveguide OPOs), so it can be taken to be constant in $t$.  Taking $a_{\rm in}$ to be real, we can integrate through (\ref{eq:10-intdiff}) to obtain the perturbation on $a_{\rm out}$:
\begin{align}
	& a_{\rm out}(t)\bigr|_{\rm NL} = -\frac{\epsilon^2}{2u} G_{0\rightarrow L} \int_0^L {\Bigl[G_{0\rightarrow z}^2 a_{\rm in}(t)} \int_{-\infty}^t{e^{(g(z)+\alpha_b/2)(t'-t)/u} a_{\rm in}(t')^2 \d t'}\Bigr]\d t \label{eq:10-depl-3d}
\end{align}

At threshold, PPLN gain matches cavity loss, so the loss near threshold is approximately $1/G_{0\rightarrow L}$.  This fact combined with (\ref{eq:10-depl-3d}) gives a round-trip equation for $a(t)$.  In terms of the coefficients $c_k$, this may be written as:
\beq
	\Delta c_k\bigr|_{\rm NL} = -2\beta\sum_{lmn} {\Psi_{klmn} c_l c_m c_n} \label{eq:10-3oc}
\eeq
where the $\beta$ (pump back-conversion term) and $\Psi_{klmn}$ are:
\begin{align}
	\beta & = \frac{\epsilon^2}{4u} \int_0^L {G_{0\rightarrow z}^2 \d z} \nonumber \\
	\Psi_{klmn} & = \frac{1}{\int_0^L {G_{0\rightarrow z}^2 \d z}} \int_0^L{\Bigl[G_{0\rightarrow z}^2 \int_{-\infty}^\infty{a_k(t)a_l(t)}} \int_{-\infty}^t{e^{(g(z)+\alpha_b/2)(t'-t)/u} a_m(t') a_n(t')\d t'}\,\d t\Bigr]\d z \label{eq:10-psi0000}
\end{align}
If the gain is constant ($G_{0\rightarrow z} = e^{gz/2}$, $g(z) = g$ constant) and small per walkoff length ($gt/u \ll 1$) then one can simplify this further.  These assumptions generally hold for waveguide OPOs pumped with flat-top pulses.  Using $G_{0\rightarrow L} \approx G_0^{1/2}$ near threshold, one can substitute $g \rightarrow \tfrac{1}{2L}\log(G_0)$; one can then evaluate the integrals in (\ref{eq:10-psi0000}), and applying the formulas in Table \ref{tab:10-t1}, express the remaining constants in terms of the threshold gain $G_0$ and photon number $N_{b,0}$:
\bea
	\beta & \!=\! & \frac{e^{\alpha_b L/2}(G_0-1) \log(G_0 e^{\alpha_a L})^2}{16N_{b,0} \log G_0} \nonumber \\
	\Psi_{klmn} & \!=\! & \int_{-\infty}^\infty{\!a_k(t)a_l(t) \int_{-\infty}^t{\!a_m(t') a_n(t')\d t'}\,\d t} \label{eq:10-betapsi}
\eea
Equation (\ref{eq:10-betapsi}) divides the physics into two terms: $\beta$ is a property of the pump and the waveguide, while $\Psi_{klmn}$ is a geometric factor that depends only on the shape of the normal modes $a_k(t)$.  $\Psi_{klmn}$ also satisfies a few important identities.  Integration by parts gives:
\beq
	\Psi_{klmn} = \delta_{kl}\delta_{mn} - \Psi_{mnkl} \label{eq:10-psi-id1}
\eeq
Typically, the fields $a_k$ have inversion symmetry.  Let's suppose that the $a_k$ are numbered so that the odd-indexed ones are odd and the even-indexed ones are even: $a_k(-t) = (-1)^k a_k(t)$.  Then one finds that exactly half of the $\Psi_{klmn}$ are either zero or a half:
\beq
	\Psi_{klmn} = \frac{1}{2}\delta_{kl}\delta_{mn}\ \ \ (\mbox{if}\ k+l+m+n\ \mbox{even}) \label{eq:10-psi-id2}
\eeq
Combining Equations (\ref{eq:10-det-dc}, \ref{eq:10-3oc}), one has all the physics needed to simulate the OPO near threshold.  Writing these for convenience in continuous-time, the equations of motion are:
\beq
	\frac{\d c_k}{\d n} = g_k c_k + \sum_l J_{kl} c_l - 2\beta \sum_{lmn} \Psi_{jklm} c_l c_m c_n \label{eq:10-dck-qd}
\eeq
In the single-mode limit, this resembles the classic result for a single-mode singly-resonant OPO, with $\Psi$ playing the role of a pump depletion term \cite{Kinsler1991}.  The single-mode theory was extended for high-finesse resonators \cite{DeValcarcel2006, Patera2010}, and the form resembles (\ref{eq:10-dck-qd}).  Note, however, that $c_k$ is constrained to be a real number here, so (\ref{eq:10-dck-qd}) will not capture the squeezing dynamics of the OPO.  A more careful treatment of the eigenmodes, which accounts for both the real and imaginary parts of the field, will be needed to model squeezing.

\subsection{Two-Mode Model}
\label{sec:10-twomode}

Consider a two-mode model.  This model is simple enough that it can be solved analytically, shedding important insight into the bifurcations and stability of the pulsed OPO.  

The time-delay matrix $J_{kl}$ only has two nonzero elements: $J_{10} = -J_{01} \equiv J$.  Most of the values of $\Psi_{klmn}$ are set by identities (\ref{eq:10-psi-id1}-\ref{eq:10-psi-id2}), giving:
\begin{align}
	& \Psi_{0000} = \Psi_{0011} = \Psi_{1100} = \Psi_{1111} = \tfrac{1}{2} \nonumber \\
	& \Psi_{0001} = \Psi_{0010} = -\Psi_{0100} = -\Psi_{1000} \nonumber \\
	& \Psi_{0111} = \Psi_{1011} = -\Psi_{1101} = -\Psi_{1110} \nonumber \\
	& \Psi_{0101} = \Psi_{0110} = \Psi_{1001} = \Psi_{1010} = 0
\end{align}
Putting this all together, we have an equation that depends on 6 parameters $(g_0, g_1, J, \beta, \Psi_{0001}, \Psi_{0111})$:
\bea
	\!\!\dot{c}_0 & \!\!=\!\! & g_0 c_0 \!-\! J\, c_1 \!+\! \beta\left[-(c_0^2 + c_1^2) c_0 \!-\! 2(\Psi_{0001} c_0^2 \!+\! \Psi_{0111} c_1^2) c_1\right] \label{eq:10-dc0}\ \ \ \ \ \ \\
	\!\!\dot{c}_1 & \!\!=\!\! & g_1 c_1 \!+\! J\, c_0 \!+\! \beta\left[-(c_0^2 + c_1^2) c_1 \!+\! 2(\Psi_{0001} c_0^2 \!+\! \Psi_{0111} c_1^2) c_0\right] \label{eq:10-dc1}\ \ \ \ \ \ 
\eea
Since $g_0 > g_1$ are the largest eigenvalues $g_0 \leq 0$ means no signal.  Assuming $g_0$ positive, one can reduce (\ref{eq:10-dc0}-\ref{eq:10-dc1}) by scaling time by $g_0^{-1}$ and the fields by $\sqrt{\beta/g_0}$:
\bea
	\!\!\frac{\d\bar{c}_0}{\d\bar{n}} & \!\!=\!\! & \bar{c}_0 - \bar{J}\, \bar{c}_1 - \left[(\bar{c}_0^2 + \bar{c}_1^2) \bar{c}_0 + 2(\Psi_{0001} \bar{c}_0^2 + \Psi_{0111} \bar{c}_1^2) \bar{c}_1\right]\ \ \ \ \ \  \label{eq:10-dc0-2} \\
	\!\!\frac{\d\bar{c}_1}{\d\bar{n}} & \!\!=\!\! & \bar{g} \bar{c}_1 + \bar{J}\, \bar{c}_0 - \left[(\bar{c}_0^2 + \bar{c}_1^2) \bar{c}_1 - 2(\Psi_{0001} \bar{c}_0^2 + \Psi_{0111} \bar{c}_1^2) \bar{c}_0\right]\ \ \ \ \ \  \label{eq:10-dc1-2}
\eea
Now we only have four parameters ($\bar{J}=J/g_0, \bar{g}=g_1/g_0, \Psi_{0001}, \Psi_{0111}$).  Since the model is two-dimensional, textbook dynamical-systems theory is very useful here \cite{StrogatzBook}.  In particular, we can draw a phase-space diagram and plot the critical points, limit cycles and separatrices.  This can be done by brute force using numerical solvers, but system (\ref{eq:10-dc0-2}-\ref{eq:10-dc1-2}) is simple enough that it has an analytic solution.  Making the substitution $c_1 = c_0\xi$, one can combine the two equations to remove $c_0$, leaving a fourth-order polynomial in $\xi$
\beq
	(1 + \xi^2) \left[\bar{J}\xi^2 + (\bar{g}-1)\xi + \bar{J}\right] + 2(1+\bar{g}\xi^2) \left[\Psi_{0001} + \Psi_{0111}\xi^2 \right] = 0
\eeq
Once this is found, one can plug the result into (\ref{eq:10-dc0-2}) to get $c_0$:
\beq
	c_0^2 = \frac{1 - \bar{J}\xi}{(1 + \xi^2) + 2(\Psi_{0001} + \Psi_{0111}\xi^2)\xi}
\eeq
For given eigenmodes, $\Psi_{0001}$ and $\Psi_{0111}$ are fixed.  As long as the general shape of the eigenmodes remains the same, they will not vary by much.  Thus, the reduced system (\ref{eq:10-dc0-2}-\ref{eq:10-dc1-2}) only has two parameters.  For typical Hermite-Gauss or sech-like eigenmodes, one has $\Psi_{0001} \approx -0.26,\ \Psi_{0111} \approx 0.11$.  

\begin{figure}[tbp]
\begin{center}
\includegraphics[width=1.0\columnwidth]{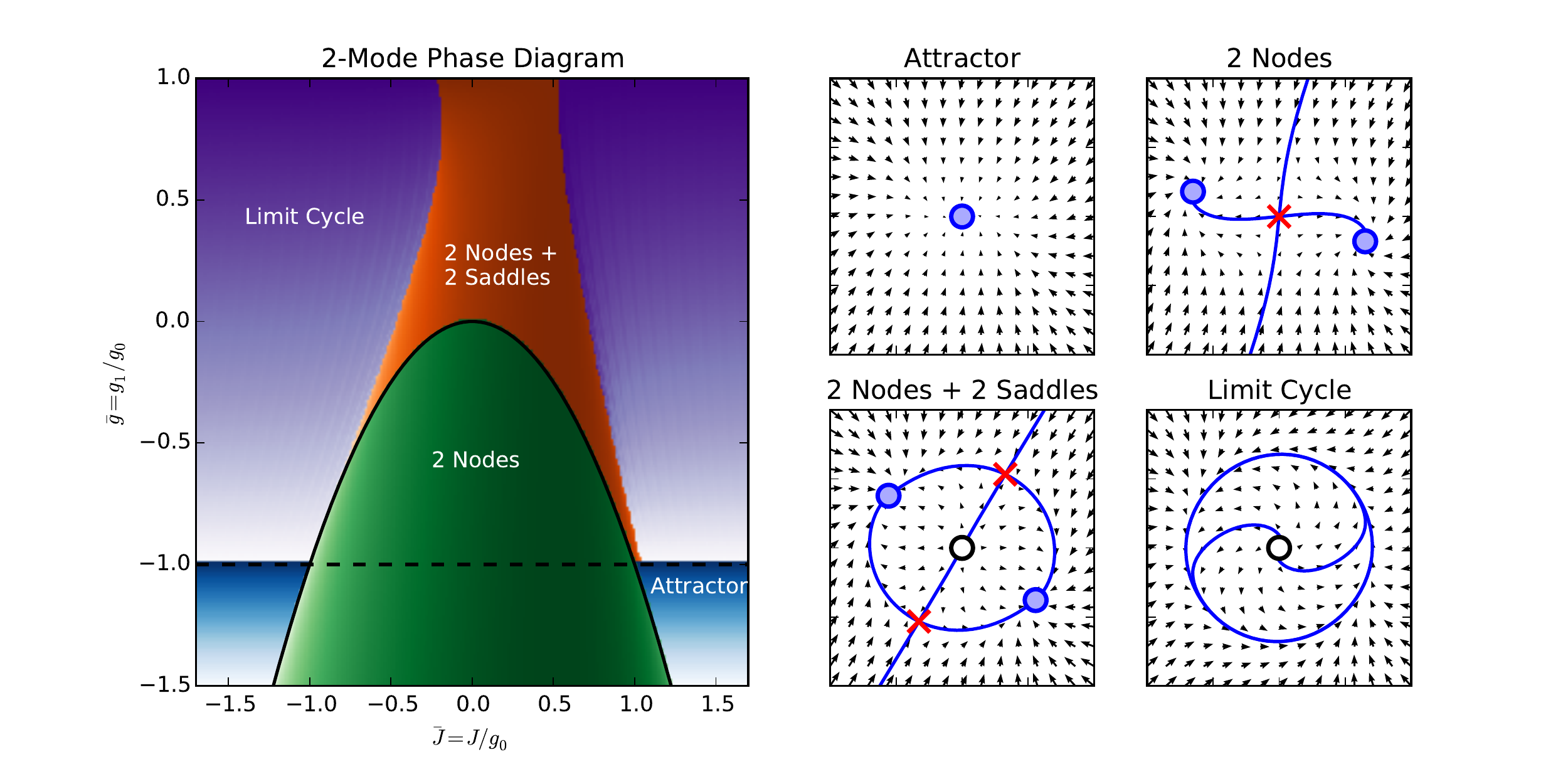}
\caption{Left: phase diagram of the two-mode model in terms of normalized parameters $\bar{J} = J/g_0$, and $\bar{g} = g_1/g_0$.  Right: typical phase-space plots corresponding to the four regions in the diagram.}
\label{fig:10-f10}
\end{center}
\end{figure}

Four types of behavior are possible, as illustrated in Figure \ref{fig:10-f10}.  They are:

\begin{enumerate}
	\item Single attractor.  This occurs if $g_0, g_1 < 0$ or if $g_0 + g_1 < 0$ and $J^2 > -g_0g_1$.  It corresponds to the OPO below threshold.
	\item 2 nodes.  As the pump power is increased, the attractor undergoes a pitchfork bifurcation, creating a saddle point at the origin and two neighboring attractors.  In the limit $g_1/g_0 \rightarrow -\infty$, this reduces to the case of a single-mode OPO above threshold, since the second mode decays too quickly to participate in the dynamics.  In this regime, the OPO behaves qualitatively like the single-mode model.
	\item 2 nodes + 2 saddles.  If the pump increases further, $g_1$ becomes positive and the saddle point at zero splits into two saddles and an unstable node.
	\item Limit cycle.  In the previous picture, nonzero delay causes the attractors and saddle points to move towards each other.  If $|J|$ is large enough, these fixed points annihilate in a saddle-node bifurcation, giving rise to a limit cycle.  Alternatively, one could start in the single-attractor region with sufficiently large $T$, and increasing $g_1$ will lead to the limit-cycle region by way of a Hopf bifurcation.
\end{enumerate}

\begin{figure}[tbp]
\begin{center}
\includegraphics[width=1.0\columnwidth]{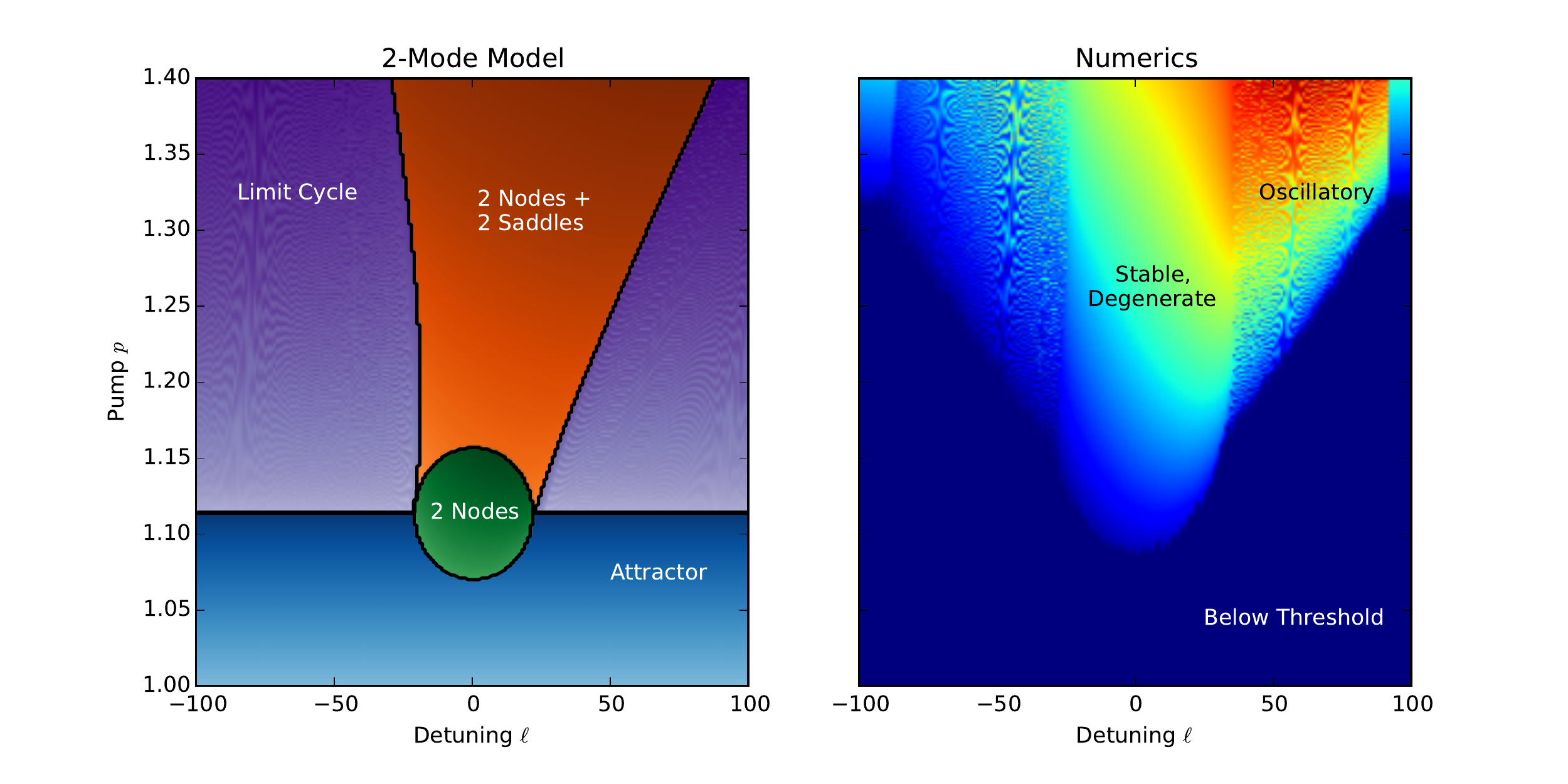}
\caption{Left: 2-mode model phase plot for PPLN OPO with 20-m fiber, $\phi_0 = 0$.  Right: photon number plot for numerical simulation.}
\label{fig:10-f22}
\end{center}
\end{figure}

The OPO pump and detuning are related to the two-mode parameters $\bar{J}, \bar{g}$, so the phase diagram in Fig.~\ref{fig:10-f10} can be mapped onto $(\ell, p)$.  Figure \ref{fig:10-f22} shows the phase diagram as a function of $(\ell, p)$ for an OPO with 20-m of fiber (at the centers of the detuning peaks, $\phi_0 = 0$).  The right plot gives the photon number from a simulation where the pump is swept from $p = 10$ to $p = 1.4$.

Qualitatively, many of the features from the numerical plot agree with the two-mode model.  Near $\ell = 0$, the threshold is lowest, increasing quadratically with $\ell$.  The two-mode model does not predict the threshold correctly for larger $\ell$, since higher-order modes start mixing with $a_0(t), a_1(t)$, raising the threshold still further.

The two-mode model gives a region of stability at low $\ell$, surrounded by a limit-cycle region with no stable fixed points.  The width of this region roughly matches the simulations, although it deviates for large $p$ where higher-order modes become important.  The only way to make the model more accurate is to add more modes; this will be discussed in the next section.

\subsection{Comparison to Numerics}
\label{sec:10-nl-num}

\begin{figure}[b!]
\begin{center}
\includegraphics[width=0.70\columnwidth]{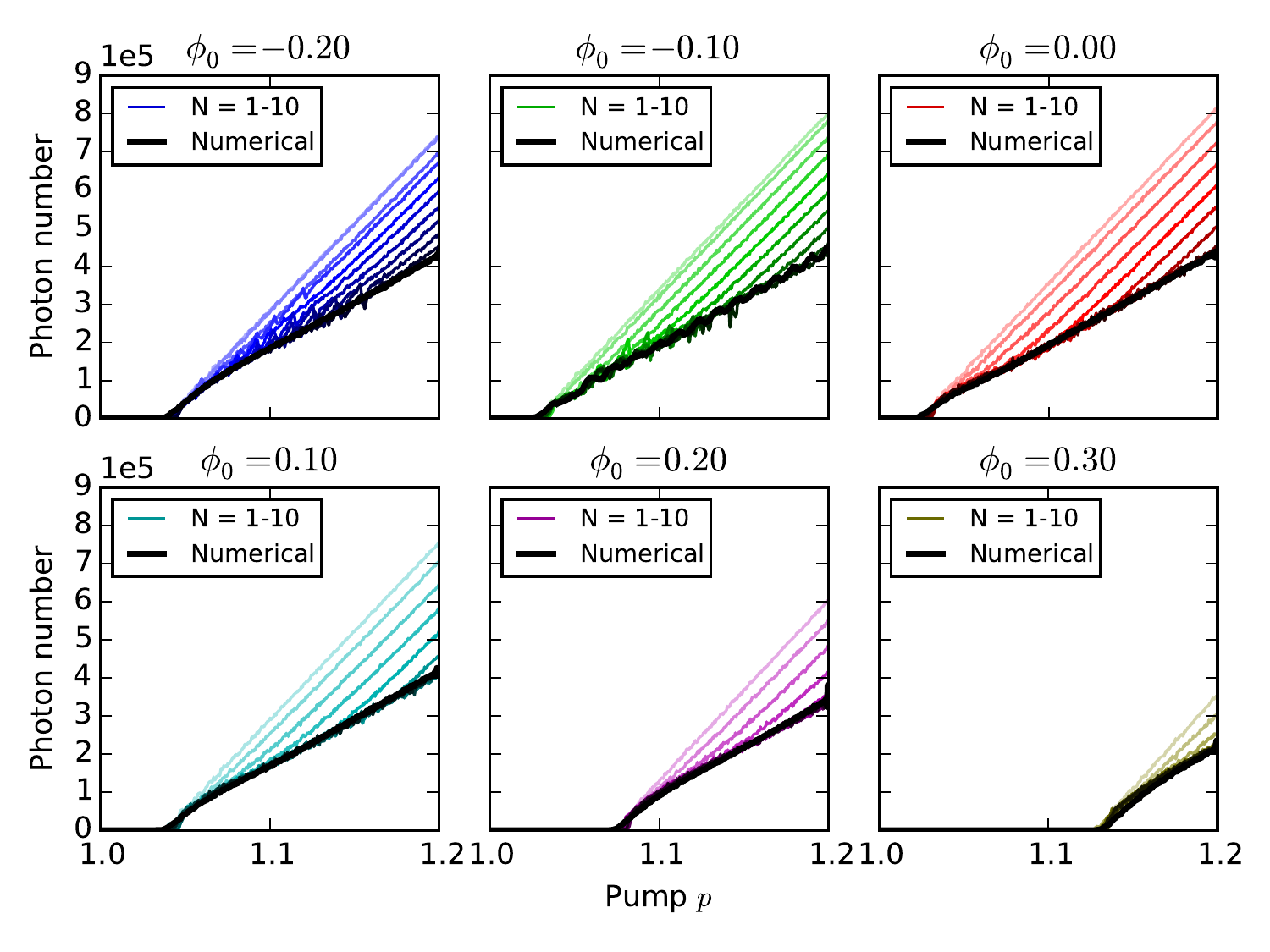}
\caption{Photon number as a function of pump amplitude.  Darker colored lines are eigenmode models with increasing $N$.  Black line is the numerical result.}
\label{fig:10-f11}
\end{center}
\end{figure}

As the number of modes $N$ is increased, the eigenmode model becomes more accurate.  However, the accuracy depends on how far one is from threshold.  The further above threshold, the more modes get excited and the larger $N$ must be to accurately model the OPO.

Figure \ref{fig:10-f11} gives the signal photon number (upon entering the crystal) as a function of pump amplitude.  The colored lines denote results from the eigenmode models, with darker lines for larger values of $N$.  For $N \gtrsim 10$, these lines match the numerical result.

Likewise, the eigenmode model does a good job predicting the steady-state signal pulse shape, provided that enough modes are used.  Figure \ref{fig:10-f12} compares the actual pulse shapes with the eigenmode model.  A linearized treatment would predict a signal centered at the maximum of the gain-clipping function (black curve, left column), but a combination of pump depletion and walkoff push it to the left.  This ``simulton acceleration'' term (see Sec.~\ref{sec:10-simulton}) can be treated to first order in an $N=2$ model, which predicts the centroid drift up to about $p = 1.06$.  Beyond that point, the pulse becomes increasingly elongated and more and more modes must be included to describe it.

\begin{figure}[t!]
\begin{center}
\includegraphics[width=0.70\columnwidth]{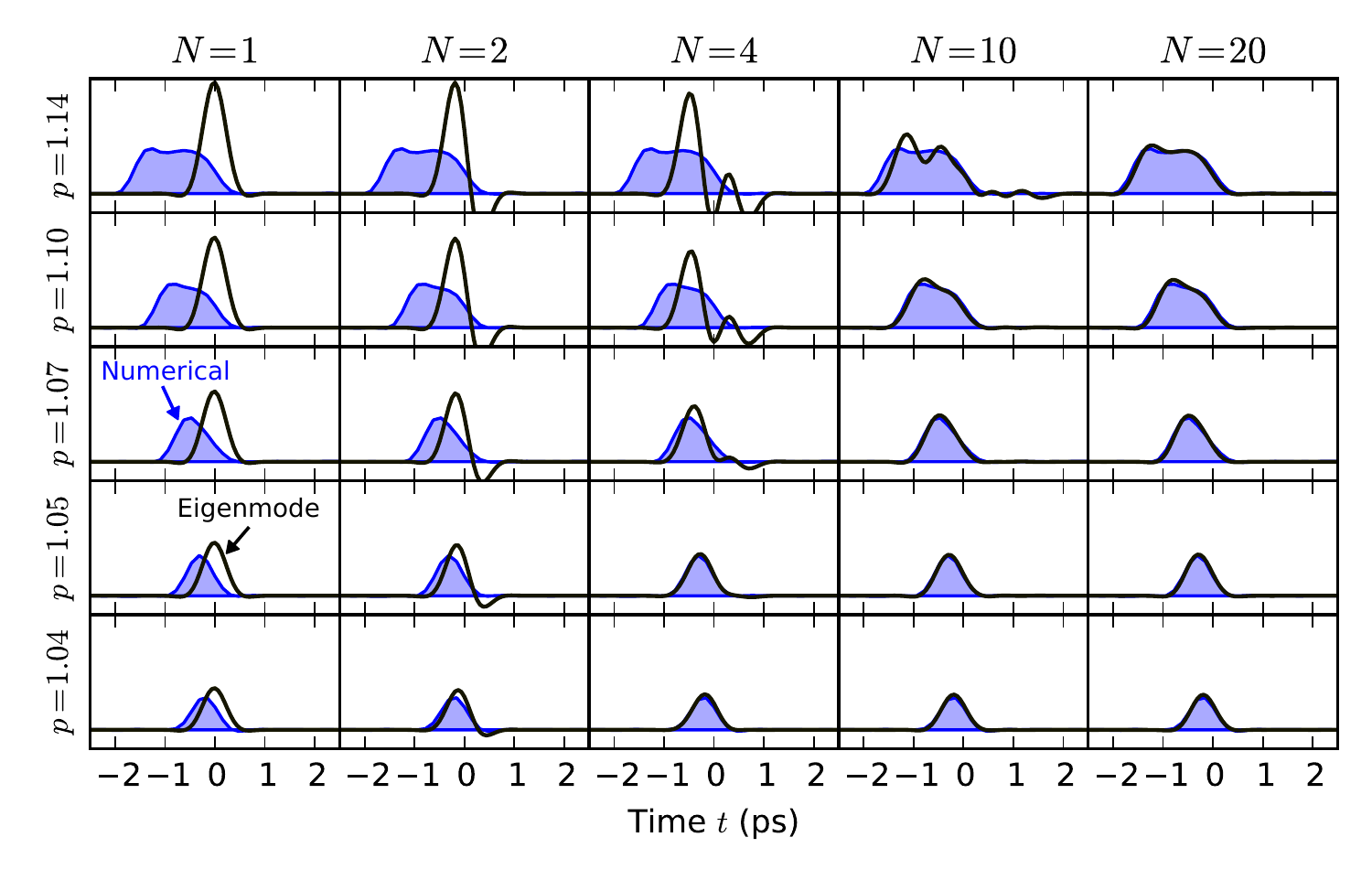}
\caption{Signal pulse shape, computed numerically (filled blue) and with the eigenmode theory (black line).}
\label{fig:10-f12}
\end{center}
\end{figure}

\begin{figure}[t!]
\begin{center}
\includegraphics[width=0.70\columnwidth]{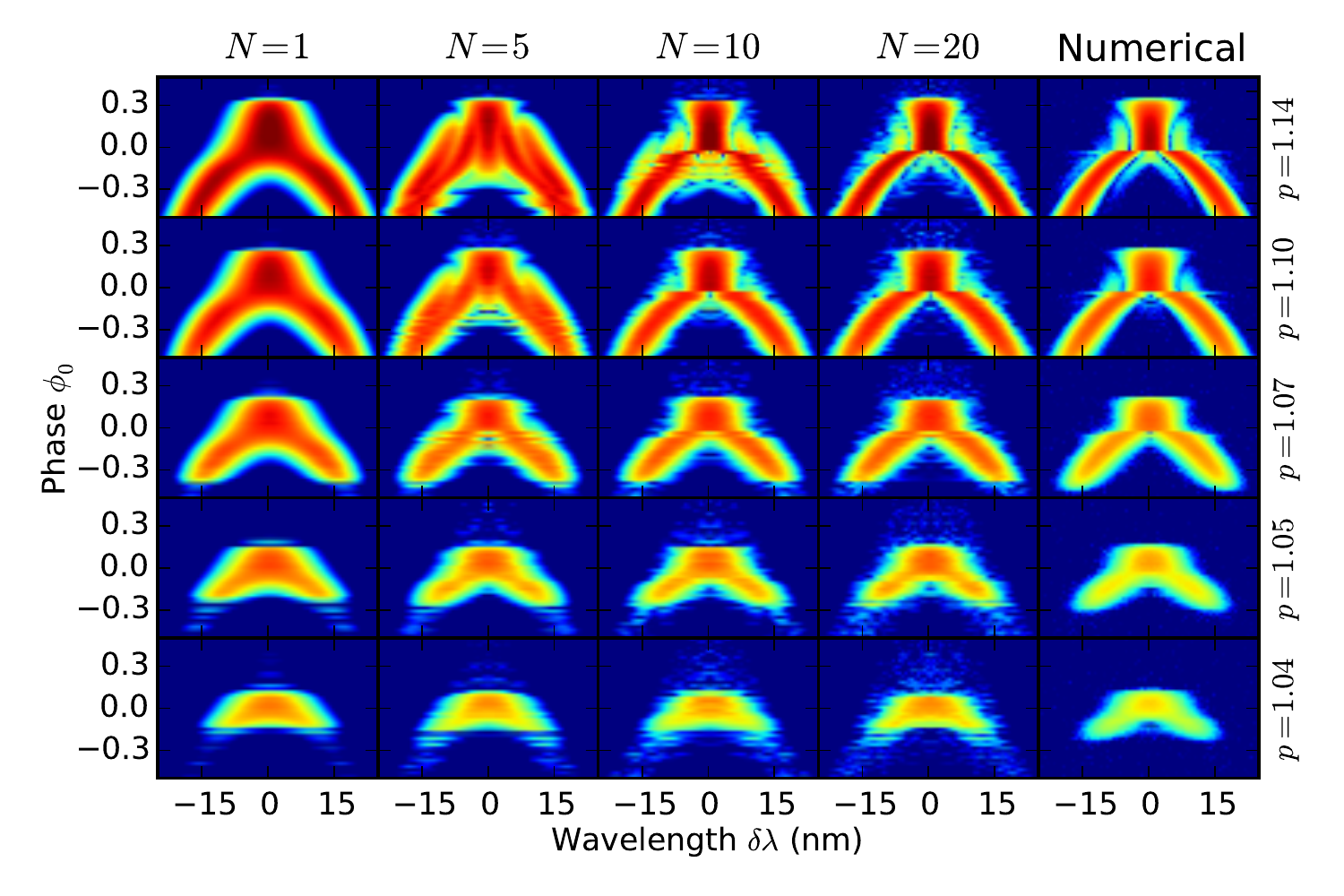}
\caption{Resonance diagrams, computed numerically (right column) and with eigenmode theories of increasing $N$ (left columns).}
\label{fig:10-f13}
\end{center}
\end{figure}

This can also be seen in the resonance diagrams in Fig.~\ref{fig:10-f13}.  As in Fig.~\ref{fig:10-f2}, these give the power spectrum as a function of cavity phase.  All such diagrams show the same general shape, but as the power is increased, the numerical plot acquires finer structure.  This structure is only reproduced if enough modes are kept in the eigenmode expansion, and with insufficient modes, agreement is quite poor.

\section{Sech-pulse Ansatz}
\label{sec:10-simulton}

A common way to model pulse propagation is to assume that the pulse maintains a given shape, and obtain equations of motion for its parameters using manifold projection or Lagrangian techniques \cite{AgrawalBook}.  The eigenmode model of Sec.~\ref{sec:10-nonlinear} is an example of {\it linear} projection, where $a(t)$ is projected onto a linear subspace spanned by the $a_k(t)$.  Unfortunately, this model required many modes in order to reproduce the full OPO dynamics.

This section studies the pulsed OPO using {\it nonlinear} manifold projection onto the space of sech-like pulses

\beq
	a(z,t) = \frac{A(z)}{\sqrt{2\tau}} \mbox{sech}\bigl((t - T(z))/\tau(z)\bigr) \label{eq:10-simulton}
\eeq

The sech pulse is a natural choice because of its relation to the $\chi^{(2)}$ simulton, a bright signal soliton which co-propagates with a dark pump soliton \cite{Akhmanov1968, Trillo1996}.  In fact, if we take, Eq.~(\ref{eq:10-intdiff}) and assume in the weak gain per walkoff length
\bea
    \frac{\partial a(z,t)}{\partial z} & = & -\frac{1}{2} \alpha_a a(z, t) + \epsilon\,a^*(z,t) b_{\rm in}(t - u z)e^{-\alpha_b z/2} \nonumber \\
    & & - \frac{\epsilon^2}{2u} a^*(z,t)\int_{-\infty}^{t}{a(z, t')^2 \d t'} \label{eq:10-intdiff-2}
\eea
then for a flat-top pump, the sech pulse maintains its shape as it propagates through the waveguide.  This observation suggests that, absent other effects, sech pulses should naturally form in PPLN-waveguide SPOPOs, particularly when a flat-top pump is used.  This view is corroborated by the eigenmode model, which gives a nearly sech-shaped pulse in the degenerate regime (Fig.~\ref{fig:10-f14} below) as well as the sech-shaped spectra in experimental data (Fig.~\ref{fig:10-f2}, see also Refs.~\cite{Marandi2016, Jankowski2016})

In this section, we begin with the sech-pulse ansatz (\ref{eq:10-simulton}) and obtain equations of motion for the parameters $A, T, \tau$ (Sec.~\ref{sec:10-ansatz}) and perturbation terms due to gain clipping and dispersion (Sec.~\ref{sec:10-simpert}).  The near-threshold limit is discussed (\ref{sec:10-simgc}) and the sech waveform is compared to first-order eigenmode.  Finally, we compare predictions of the sech-pulse theory to numerical simulations (Sec.~\ref{sec:10-simnum}).

\subsection{Ansatz and Equations of Motion}
\label{sec:10-ansatz}

Assume the simulton-like sech solution (\ref{eq:10-simulton}).  This confines the field $a(t)$ to a 3-dimensional manifold in the state space.  This solution has three free parameters: amplitude $A$ (normalized so that $|A|^2$ is the photon number), centroid $T$ and width $\tau$.  We obtain the reduced model by projecting equation of motion (\ref{eq:10-intdiff-2}) onto the manifold (\ref{eq:10-simulton}).  Projection requires an inner-product, so we use $\ip{f}{g} = \int{f(t)g(t)\d t}$.  Each of the three variables $\xi \in \{A, T, \tau\}$ evolves according to the projection rules:
\beq
	\frac{\d\xi}{\d z} = \frac{\int{\nabla_\xi a\,\partial_z a\,\d t}}{\int{\nabla_\xi a\,\nabla_\xi a\,\d t}} \label{eq:10-manproj}
\eeq
where $\nabla_\xi a$, computed from the ansatz (\ref{eq:10-simulton}), is the tangent vector along $\xi$, and $\partial_z a$ is computed from (\ref{eq:10-intdiff-2}) \cite{Mabuchi2008b, VanHandel2005}.  The equations for $A, T, \tau$ are:

\begin{eqnarray}
    \!\!\frac{\d A}{\d z} & \!\!=\!\! & \left[\int{\gamma(t, z) \frac{\mbox{sech}^2(\tfrac{t-T}{\tau})}{2\tau}\d t} - \frac{\epsilon^2}{4u} A^2\right] A \label{eq:10-sim-da} \\
    \!\!\frac{\d T}{\d z} & \!\!=\!\! & - \tau \frac{\epsilon^2}{4u} A^2 + \frac{3}{2} \int{\gamma(t, z) \mbox{sech}^2(\tfrac{t-T}{\tau})\mbox{tanh}(\tfrac{t-T}{\tau}) \d t} \label{eq:10-sim-dt} \\
    \!\!\frac{\d\tau}{\d z} & \!\!=\!\! & \frac{18}{3+\pi^2} \int{\gamma(t, z) \left[\tfrac{t-T}{\tau} \tanh(\tfrac{t-T}{\tau}) - \tfrac{1}{2}\right] \mbox{sech}^2(\tfrac{t-T}{\tau}) \d t} \nonumber \\ \label{eq:10-sim-dtau}
\end{eqnarray}
where $\gamma(t, z) = \epsilon\,b_{\rm in}(t-uz)e^{-\alpha_b z/2} - \alpha_a/2$.

Three effects come into play here: gain, gain-clipping, and pump depletion.  As in Sec.~\ref{sec:10-linear}, we separate the continuous-wave dynamics from gain-clipping: first we solve the equations of motion assuming a constant-pump gain $\gamma(t, z) \rightarrow \epsilon \bar{b} - \alpha_a/2$, then treat deviations using perturbation theory.  We also add dispersion terms as a perturbations.  The solution will take the form:
\beq
	A = A_0 + \delta A,\ \ \ 
	T = T_0 + \delta T,\ \ \ 
	\tau = \tau_0 + \delta\tau
\eeq
where $A_0, T_0, \tau_0$ satisfy the continuous-wave, lossless equations and $\delta A, \delta T, \delta\tau$ are the gain-clipping and dispersion perturbation terms.

Taking Eqs.~(\ref{eq:10-sim-da}-\ref{eq:10-sim-dtau}) and assuming a constant pump $b(t, z) \rightarrow \bar{b}$, one obtains $\d\tau_0/\d z = 0$ and the equations for $A_0, T_0$:
\beq
    \!\frac{\d A_0}{\d z} = \Bigl[\epsilon\,(\bar{b}_{\rm in} - \tfrac{1}{2}\alpha_a) - \frac{\epsilon^2}{4u} A_0^2\Bigr] A_0,\ \ 
    \frac{\d T_0}{\d z} = -\tau \frac{\epsilon^2}{4u} A_0^2 \label{eq:10-ideal}
\eeq

If the pump field is nearly constant (as is the case with flat pulses or sufficiently long Gaussian pulses) and the waveguide is nearly lossless, $A_0, T_0, \tau_0$ will be a good approximation to the pulse parameters.  The constant pump $\bar{b}$ is chosen to be close to the average value for a CW field of the same peak intensity as $b_{\rm in}(t)$:
\beq
	\bar{b} = \frac{1}{L} \int{b_{\rm max} e^{-\alpha_b z/2} dz} \approx b_{\rm max} e^{-\alpha_b L/4}
\eeq
Solving Eq.~(\ref{eq:10-ideal}) one finds:
\beq
	A_0(z) = \sqrt{\frac{2u(2\epsilon\,\bar{b} - \alpha_a) e^{(2\epsilon\,\bar{b} - \alpha_a)z}}{2u(2\epsilon\,\bar{b} - \alpha_a) + (e^{(2\epsilon\,\bar{b} - \alpha_a)z} - 1) \epsilon^2 A_0(0)^2}}\,A_0(0) \label{eq:10-inout1}
\eeq
At threshold $p = 1$, the constants $\bar{b}$, $u$, $\epsilon$ can be expressed in terms of two experimental parameters: pump intensity $N_{b,0} \approx e^{\alpha_b L/2} \bar{b}_0^2$ and waveguide gain $G_0 = e^{(2\epsilon\bar{b}_0-\alpha_a)L}$ at threshold (Table~\ref{tab:10-t1}).  Above threshold, the pump amplitude scales with $p$, so $\bar{b} = p \bar{b}_0$.  Making the substitutions $N_b = p^2\,N_{b,0}$, $G = G_0^{p} e^{(p-1)\alpha_a L}$, we rewrite Eq.~(\ref{eq:10-inout1}) as:
\beq
	A_0(z) = \left[\frac{G^{z/L}}{1 + \left(G^{z/L} - 1\right) \frac{\log(G e^{\alpha_a L})^2}{\log(G)} \frac{A_0(0)^2}{8 N_{b} e^{-\alpha_b L/2}}}\right]^{1/2} A_0(0) \label{eq:10-aout-1}
\eeq
Combining the first two equations in (\ref{eq:10-ideal}), we can obtain the centroid shift $T$ in terms of the amplitude:
\beq
	T_0(z) = T_0(0) - \tau\,\log\left(\frac{G^{z/2L}}{A(z)/A(0)}\right)
	\label{eq:10-tout-1}
\eeq

Eqs.~(\ref{eq:10-aout-1}-\ref{eq:10-tout-1}) govern the pulse evolution in the presence of a CW pump.  The width $\tau$ does not change.  Note that the pump depletion shifts the centroid of the pulse in addition to reducing its gain.  This {\it simulton acceleration} is caused by pump-signal walkoff: as the pulse walks through the pump, the leading side experiences gain from the undepleted pump while the gain on the trailing side is depleted, shifting the centroid forward.  

\subsection{Perturbations}
\label{sec:10-simpert}

\subsubsection{Gain-Clipping Terms}

Gain clipping gives rise to perturbations in $A$, $T$ and $\tau$.  To find these, we first rewrite (\ref{eq:10-sim-da}-\ref{eq:10-sim-dtau}) as:
\begin{eqnarray}
    \frac{\d(\delta A/A_0)}{\d z} & = & -\frac{\epsilon^2}{2u} A_0^2 (\delta A/A_0) + g(T,\tau,z) \label{eq:10-pert1} \\
    \frac{\d (\delta T)}{\d z} & = & \frac{3\tau_0^2}{2} \frac{\partial g(T_0,\tau_0,z)}{\partial T_0} \label{eq:10-pert2} \\
    \frac{\d (\delta\tau)}{\d z} & = & \frac{18\tau_0^2}{3+\pi^2} \frac{\partial g(T_0,\tau_0,z)}{\partial\tau_0} \label{eq:10-pert3}
\end{eqnarray}
where $g(T,\tau,z)$ is the differential gain-clipping function of the sech-pulse, defined by:
\beq
	g(T, \tau, z) = \int{\epsilon(b_{\rm in}(t-uz)e^{-\alpha_b z/2} - \bar{b}) \frac{\mbox{sech}^2((t-T)/\tau)}{2\tau}\d t}
\eeq
Up to a constant, this is the convolution of the pump $b_{\rm in}(t-uz)$ and sech intensity $(2\tau)^{-1}\mbox{sech}^2((t-T)/\tau)$.

Equations (\ref{eq:10-pert1}-\ref{eq:10-pert3}) can be integrated to give:
\begin{eqnarray}
	\!\!\!\!\!\!\delta A(z) & = & A_0(z) \int_0^z{g(T_0,\tau_0,z') \frac{(A_0(z)/A_0(0))^2}{G^{z/L}}\d z} 
	\label{eq:10-p-da} \\
	\!\!\!\!\!\!\delta T(z) & = & \frac{3\tau_0^2}{2} \int_0^z{\frac{\partial g(T_0,\tau_0,z')}{\partial T_0}\d z'} \label{eq:10-p-dt} \\
	\!\!\!\!\!\!\delta \tau(z) & = & \frac{18\tau_0^2}{3+\pi^2} \int_0^z{\frac{\partial g(T_0,\tau_0,z')}{\partial\tau_0}\d z'} \label{eq:10-p-dtau}
\end{eqnarray}
	
Equation (\ref{eq:10-p-da}) gives the gain-clipping correction to the linear gain.  Although the full form is complicated, it simplifies in the near-threshold regime, where the fraction on the right side of the integral can be ignored.  Equations (\ref{eq:10-p-dt}-\ref{eq:10-p-dtau}) can be simplified if we assume that $T$ and $\tau$ change slowly enough in a single round-trip that we can replace them inside the integral by their initial values.  The input-output relations become:
\begin{eqnarray}
	\delta A(z) & = & A_0(z) G(T_0, \tau_0) \label{eq:10-p2-da} \\
	\delta T(z) & = & \frac{3\tau_0^2}{2} \frac{\partial G(T_0, \tau_0)}{\partial T_0} \label{eq:10-p2-dt} \\
	\delta \tau(z) & = & \frac{18\tau_0^2}{3+\pi^2} \frac{\partial G(T_0, \tau_0)}{\partial\tau_0} \label{eq:10-p2-dtau}
\end{eqnarray}
where
\beq
	G(T, \tau) = \int_0^L {g(T-uz,\tau,z)\d z} \label{eq:10-gttau}
\eeq
is the integrated sech-pulse gain-clipping function.  Up to a constant factor and offset, it is equal to the convolution of the the gain-clipping function $G(t)$ from (\ref{eq:10-gcf}) and the sech waveform.

Combining Eqs.~(\ref{eq:10-aout-1}-\ref{eq:10-tout-1}, \ref{eq:10-p2-da}-\ref{eq:10-p2-dtau}), one obtains the full PPLN input-output relations accounting for both gain-clipping and pump depletion:
\begin{eqnarray}
	\!\!A_{\rm out} & \!=\! & \left[\frac{G}{1 + \left(G - 1\right) \frac{\log(G e^{\alpha_a L})^2}{\log(G)} \frac{A_{\rm in}^2}{8 N_{b} e^{-\alpha_b L/2}}}\right]^{1/2} \nonumber \\
	& & \times \bigl(1 + G(T_{\rm in}, \tau_{\rm in})\bigr)A_{\rm in} \label{eq:10-inout-a} \\
	\!\!T_{\rm out} & \!=\! & T_{\rm in} \!-\! \tau \log\left(\frac{G^{1/2}}{A_{\rm out}/A_{\rm in}}\right) + \frac{3\tau_{\rm in}^2}{2} \frac{\partial G(T_{\rm in},\tau_{\rm in})}{\partial T_{\rm in}}\ \  \label{eq:10-inout-t} \\
	\!\!\tau_{\rm out} & \!=\! & \tau_{\rm in} + \frac{18\tau_{\rm in}^2}{3+\pi^2} \frac{\partial G(T_{\rm in},\tau_{\rm in})}{\partial\tau_{\rm in}} \label{eq:10-inout-tau}
\end{eqnarray}

\subsubsection{Dispersion and Detuning}

Now we add in dispersion.  Following Sec.~\ref{sec:10-linear}, we employ the lumped-element model, since the pulse shape changes only slightly between round trips and dispersion is a linear effect that does not depend on the pulse amplitude.  Restricting ourselves to the degenerate regime $\phi_0\phi'_2 > 0$ where we expect to see simulton-like solutions and following (\ref{eq:10-deg-eig}), we have:
\beq
	\Delta a(t)\bigr|_{\rm dispersion} = \frac{\phi'_2\tan\phi_0}{2} \frac{\d^2 a(t)}{\d t^2} - \frac{(\phi'_2\sec\theta)^2}{8} \frac{\d^4 a(t)}{\d t^4} \label{eq:10-disp}
\eeq
where $\phi_0$ is the round-trip phase and $\phi'_2$ is the total (PPLN plus fiber) dispersion.  We enforce the simulton-like form (\ref{eq:10-simulton}) by projecting (\ref{eq:10-disp}) onto the 3-dimensional sech-pulse manifold.  As before, each of the three variables $A, T, \tau$ changes by Eq.~(\ref{eq:10-manproj}).  Performing the necessary integrals, one finds:
\begin{eqnarray}
	\Delta A & = & \left[-\frac{1}{3} \frac{\phi'_2\tan\phi_0}{2} \tau^{-2} - \frac{7}{15} \frac{(\phi'_2\sec\phi_0)^2}{8} \tau^{-4} \right] A \\
	\Delta \tau & = & \frac{12}{3+\pi^2} \frac{\phi'_2\tan\phi_0}{2} \tau^{-1} + \frac{168}{5(3+\pi^2)}\frac{(\phi'_2\sec\phi_0)^2}{8} \tau^{-3} \nonumber \\
\end{eqnarray}
Higher-order effects such as third-order dispersion and $\chi^{(3)}$ are not included here, but could also be treated with this perturbation theory.  GVD gives no centroid shift.  However, there is a nonzero $\Delta T$ due to cavity detuning: $\Delta T = (\lambda/2c)\ell$.  Combining these with Eqs.~(\ref{eq:10-inout-a}-\ref{eq:10-inout-tau}) and adding a loss $G_0^{-1}$, one obtains round-trip propagation equations for $A, T, \tau$ in the OPO:


\begin{eqnarray}
	A & \rightarrow & \left[1 + G(T, \tau) - \frac{1}{3} \frac{\phi'_2\tan\phi_0}{2} \tau^{-2} - \frac{7}{15} \frac{(\phi'_2\sec\phi_0)^2}{8}\tau^{-4} \right] \left[\frac{G/G_0}{1 + \left(G - 1\right) \frac{\log(G e^{\alpha_a L})^2}{\log(G)} \frac{A^2}{8 N_{b} e^{-\alpha_b L/2}}}\right]^{1/2} A \label{eq:10-prop-a} \\
	T & \rightarrow & T + \frac{\lambda}{2c}\ell - \frac{\tau}{2} \log\left[1 + (G - 1) \frac{\log(G e^{\alpha_a L})^2}{\log(G)} \frac{A^2}{8 N_{b} e^{-\alpha_b L/2}}\right] + \frac{3\tau^2}{2} \frac{\partial G(T,\tau)}{\partial T} \label{eq:10-prop-t} \\
	\tau & \rightarrow & \tau + \frac{18\tau^2}{3+\pi^2} \frac{\partial G(T,\tau)}{\partial\tau_{\rm in}} + \frac{12}{3+\pi^2} \frac{\phi'_2\tan\phi_0}{2} \tau^{-1} + \frac{168}{5(3+\pi^2)} \frac{(\phi'_2\sec\phi_0)^2}{8} \tau^{-3} \label{eq:10-prop-tau}
\end{eqnarray}

\subsection{Near-Threshold Limit}
\label{sec:10-simgc}

Near threshold, the sech-pulse model should match the eigenmode model derived in Sec.~\ref{sec:10-linear}.  In that limit, we can truncate all of the nonlinear gain terms in (\ref{eq:10-prop-a}-\ref{eq:10-prop-tau}) at third order and replace $G \rightarrow G_0$, the at-threshold gain.  In addition, supposing a flat-top pump pulse, the gain-clipping function becomes $G(t) = -\tfrac{1}{2}|t/T_p| \log(G_0 e^{\alpha_a L})$.  Using Eq.~\ref{eq:10-gttau}, $G(T, \tau)$ is:
\beq
	G(T, \tau) = -\frac{\tau}{2T_p} \log(G_0 e^{\alpha_a L})\log\bigl[2\cosh(T/\tau)\bigr]
\eeq
This is maximized for $T = 0$, the trailing edge of the pump (Fig.~\ref{fig:10-f4}).  Since $A, T, \tau$ change slowly on each round trip, we can convert (\ref{eq:10-prop-a}-\ref{eq:10-prop-tau}) to a differential equation analogous to (\ref{eq:10-dc-cont}); performing the near-threshold substitutions, we obtain:

\begin{eqnarray}
	\!\!\!\!\!\frac{\d A}{\d n} & \!\!=\!\! & \left[\frac{p-1}{2}\log(G_0 e^{\alpha_a L}) \!-\! \frac{\log(G_0 e^{\alpha_a L})}{2T_p} \log\bigl[2\cosh(T/\tau)\bigr]\tau \right. \nonumber \\
	& & \ \left.- \frac{1}{3} \frac{\phi'_2\tan\phi_0}{2\tau^{2}} - \frac{7}{15} \frac{(\phi'_2\sec\phi_0)^2}{8\tau^{4}}\right] A - \beta A^3 \label{eq:10-prop-gc-a}\\
	\!\!\!\!\!\frac{\d T}{\d n} & \!\!=\!\! & \ell/c - \tau\beta A^2 - \frac{3\tau^2}{4 T_p} \log(G_0 e^{\alpha_a L}) \tanh(T/\tau) \label{eq:10-prop-gc-t} \\
	\!\!\!\!\!\frac{\d\tau}{\d n} & \!\!=\!\! & \frac{18}{3+\pi^2} \frac{\log(G_0 e^{\alpha_a L})}{2T_p} \left[\log\bigl[2\cosh(\tfrac{T}{\tau})\bigr] \!-\! \tfrac{T}{\tau} \tanh(\tfrac{T}{\tau})\right] \tau^2 \nonumber \\
	& & \ + \frac{12}{3+\pi^2} \frac{\phi'_2\tan\phi_0}{2\tau} + \frac{168}{5(3+\pi^2)} \frac{(\phi'_2\sec\phi_0)^2}{8\tau^3} \label{eq:10-prop-gc-tau}
\end{eqnarray}

Most of these terms make intuitive sense.  For the $A$ equation, the $p-1$ term is the CW gain and the $O(\tau)$, $O(\tau^{-1})$ and $O^(\tau^{-3})$ terms account for gain clipping and dispersion, which reduce the overall gain of the signal.  An $O(A^3)$ term accounts for pump depletion in the near-threshold limit; $\beta$ is given by
\beq
    \beta = \frac{e^{\alpha_b L/2} (G_0 - 1) \log(G_0 e^{\alpha_a L})^2}{16 N_{b,0} \log G_0}
\eeq
which matches Eq.~(\ref{eq:10-betapsi}) from the eigenmode theory.

\comment{

As equation (\ref{eq:10-prop-gc-t}) shows, gain clipping and simulton acceleration both shift the center of the pulse.  In steady state, these effects cancel out, giving a steady-state centroid shift:

\beq
	T = -\tau \tanh^{-1}\left[\frac{4 T_p \beta A^2}{3\tau \log(G_0 e^{\alpha_a L})}\right]
\eeq

The sech pulse is only stable if the argument in the $\tanh^{-1}$ lies in $(-1, 1)$.  Otherwise it will run away and decay to zero.  Not included here is the effect of cavity detuning.  Detuning can either enhance or counteract the simulton acceleration.

}

Equation (\ref{eq:10-prop-gc-tau}) lets us compute the pulse width.  The $O(\tau^2)$ gain-clipping term is compensated by the $O(\tau^{-1})$, $O(\tau^{-3})$ dispersion terms.  Working at $\phi_0 = 0$ and close enough to threshold that the simulton acceleration can be neglected ($T = 0$), one finds the steady-state pulse width:
\beq
	\tau_{\rm sech} = \left(\frac{7}{15} \frac{(\phi'_2)^2 T_p}{\log(G_0 e^{\alpha_a L})\log 2}\right)^{1/5} \label{eq:10-tau-zero}
\eeq
Gain-clipping theory says that signal pulses at $\phi_0 = 0$ are given by combinations of hypergeometric functions (Sec.~\ref{sec:10-emshapes}): $a(t) \sim f(t/\tau_L)$, where $\tau_L = \left((\phi'_2)^2 T_p/4\log(G_0 e^{\alpha_a L})\right)^{1/5}$.  Comparing to (\ref{eq:10-tau-zero}), we find $\tau_{\rm sech} = 1.21\tau_L$.

In the degenerate $\phi_0 \neq 0$ limit, the $\tau^{-1}$ term in (\ref{eq:10-prop-gc-tau}) dominates and the steady-state pulse width is:
\beq
	\tau_{\rm sech} = \left(\frac{2 T_p \phi'_2\tan\phi_0}{3\log(G_0 e^{\alpha_a L})\log 2}\right)^{1/3} \label{eq:10-tau-nz}
\eeq
This should be compared to the eigenmode model, in which the pulse shape is given by an Airy function $\mbox{Ai}(t/\tau_{\rm Ai} + \mbox{const})$, with $\tau_{\rm Ai} = \bigl(T_p \phi'_2\tan\phi_0/\log(G_0 e^{\alpha_a L})\bigr)^{1/3}$.  We find that $\tau_{\rm sech} = 0.987 \tau_{\rm Ai}$.

\begin{figure}[b!]
\begin{center}
\includegraphics[width=0.60\textwidth]{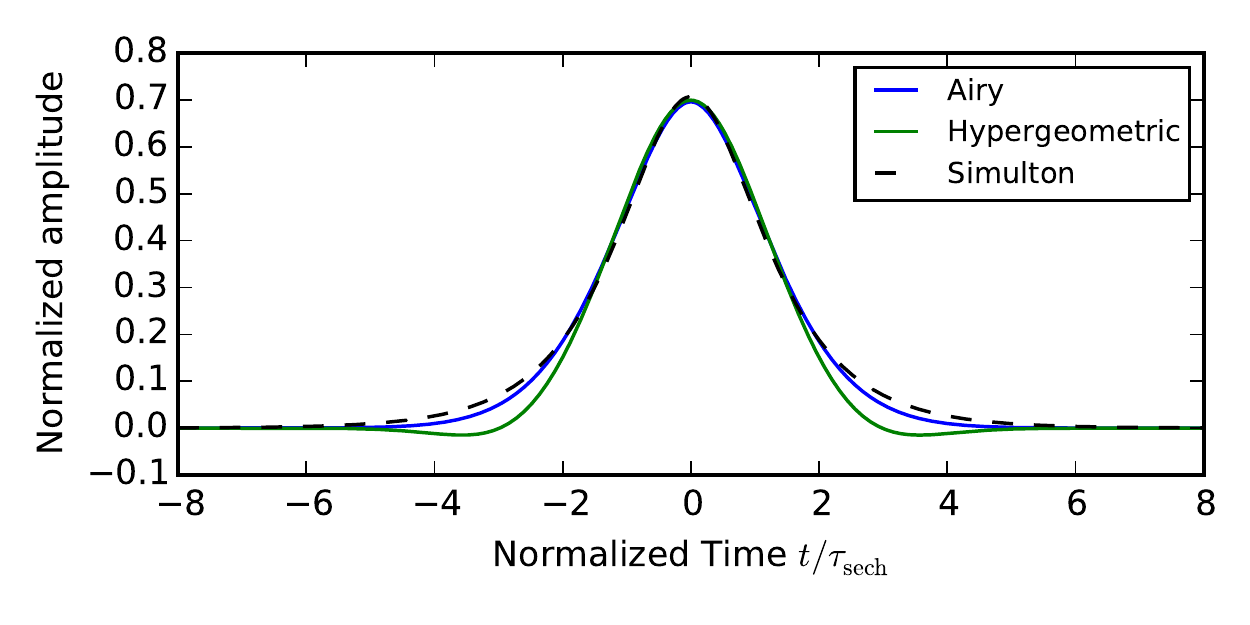}
\caption{Plot of the simulton solution $(2\tau)^{-1/2}\mbox{sech}(t/\tau_{\rm sech})$ against the Airy (Eq.~\ref{eq:10-eig-airy}) and hypergeometric (Eq.~\ref{eq:10-phizero-ak}) eigenfunctions.}
\label{fig:10-f14}
\end{center}
\end{figure}

Although the pulse widths $\tau_{\rm Ai}$, $\tau_{L}$ and $\tau_{\rm sech}$ differ, the respective functions have different shapes, so that the pulse waveforms predicted by eigenmode and simulton theory happen to lie right on top of each other, and their full-width half-maxima agree to a few percent (Fig.~\ref{fig:10-f14}).

\subsection{Comparison to Numerics}
\label{sec:10-simnum}

Numerical simulations for the waveguide OPO show that the simulton model is accurate when the OPO exhibits degenerate, singly-peaked behavior.  This happens in a limited range of circumstances:

\begin{enumerate}
	\item Power: The pulse is sech-shaped near threshold.  Far above threshold, pulses become box-shaped and are better described by the theory in Sec.~\ref{sec:10-boxpulse}.
	\item Phase: One must be near the center of a detuning peak ($\phi_0 \approx 0$) to use the simulton description.  Far from the center for $\phi_0\phi'_2 < 0$, the pulse that resonates starts to resemble a nondegenerate pulse, which is not described by a sech-pulse.
	\item Detuning: The cavity detuning $\ell$ cannot be too large; otherwise the sech-pulse goes unstable and the field amplitude starts to oscillate.
\end{enumerate}

\subsubsection{Steady-State Behavior}

The sech-pulse model does a good job predicting the pulse shape near threshold, provided that the oscillating mode is degenerate.  For $\phi_0 = 0$ or $\phi_0$ sufficiently large, Eqs.~(\ref{eq:10-tau-zero}) and (\ref{eq:10-tau-nz}) can be used to get the pulse width, respectively.  For general $\phi_0$, one must solve for the steady-state of (\ref{eq:10-prop-gc-tau}).  (Near threshold one can take $T \rightarrow 0$ in that equation, resulting in a $5^{\rm th}$-order polynomial in $\tau$.)

However, as Figure \ref{fig:10-f15} shows, one cannot use the sech-pulse model when the OPO oscillates nondegenerately.  Also, it cannot be used when dispersion compensation is used to set $\phi'_2 \rightarrow 0$.  Since dispersion is treated as a lumped element here, this causes the pulse width to shrink to zero (as in Sec~\ref{sec:10-emshapes}).  An OPO with dispersion compensation must be studied numerically or with the eigenmode model, or a more careful approach must be taken, avoiding lumping the dispersion into one element.  In the dispersion-engineered limit where both $\beta_2$ and $\phi_2$ are zero, one must go further and include higher-order dispersion terms.

\begin{figure}[tbp]
\begin{center}
\includegraphics[width=1.0\columnwidth]{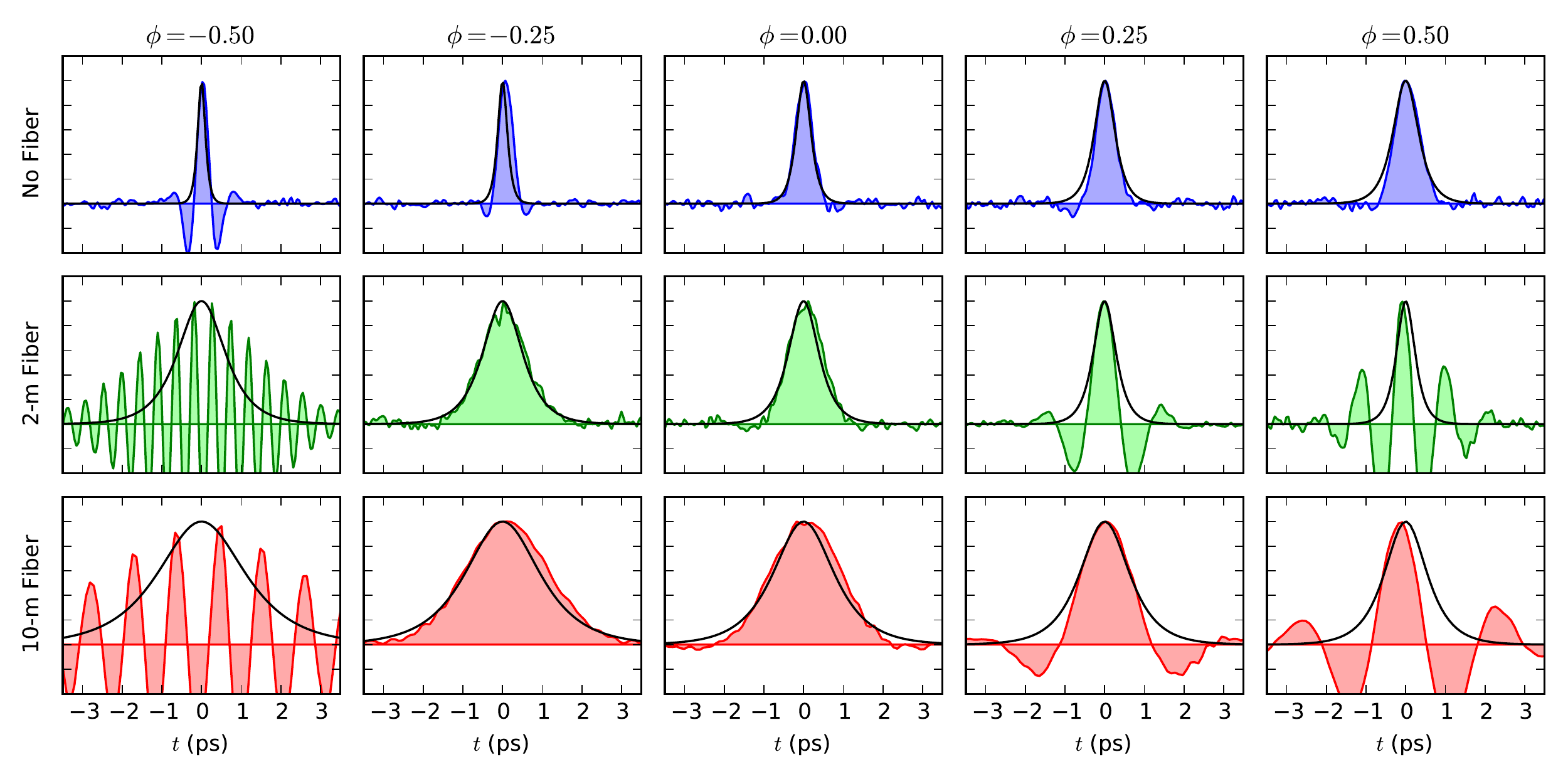}
\caption{Near-threshold pulse shape computed numerically (filled curve), compared to the steady-state sech solution (black line).}
\label{fig:10-f15}
\end{center}
\end{figure}

At threshold, the pulse is centered at the point of maximum gain.  As the pump increases and the amplitude grows, the simulton acceleration causes its centroid to drift towards negative $T$.  In the absence of detuning, a steady state is reached in (\ref{eq:10-prop-gc-t}) when $\beta A^2 = (3\tau^2/4T_p) \log(G_0 e^{\alpha_a L}) \tanh(T/\tau)$.  One can replace $\beta A^2 \rightarrow \tfrac{1}{2}(p-1)\log(G_0 e^{\alpha_a L})$ by making the assumption that those two terms are dominant in the amplitude equation (\ref{eq:10-prop-gc-a}).  Assuming a small $T$ and expanding the hyperbolic tangent, we get:
\beq
	T = \frac{2(p-1) T_p}{3} \label{eq:10-delay}
\eeq
To go beyond this approximation, one must simulate Eqs.~(\ref{eq:10-prop-a}-\ref{eq:10-prop-tau}) or (\ref{eq:10-prop-gc-a}-\ref{eq:10-prop-gc-tau}) numerically.  Figure \ref{fig:10-f16} compares numerical data against the simulton model for two cases: a PPLN OPO without a fiber segment and one with 10 meters of fiber.  The pulse shape matches the sech form well in the linear regime, and continues to match reasonably well as the pulse is displaced from the maximum-gain point.  However, at high pump powers its shape becomes deformed and it begins to resemble a flat-top pulse.

In Sec.~\ref{sec:10-nl-num}, we made a similar comparison with the eigenmode theory.  Figs.~\ref{fig:10-f16} and \ref{fig:10-f12} are computed for the same OPO system, allowing a direct comparison.  We see that for these OPO parameters, the simulton model is accurate up to about $p = 1.10$, does better than the $N=4$ eigenmode model, but not as good as $N=10$.

\begin{figure}[tbp]
\begin{center}
\includegraphics[width=1.0\columnwidth]{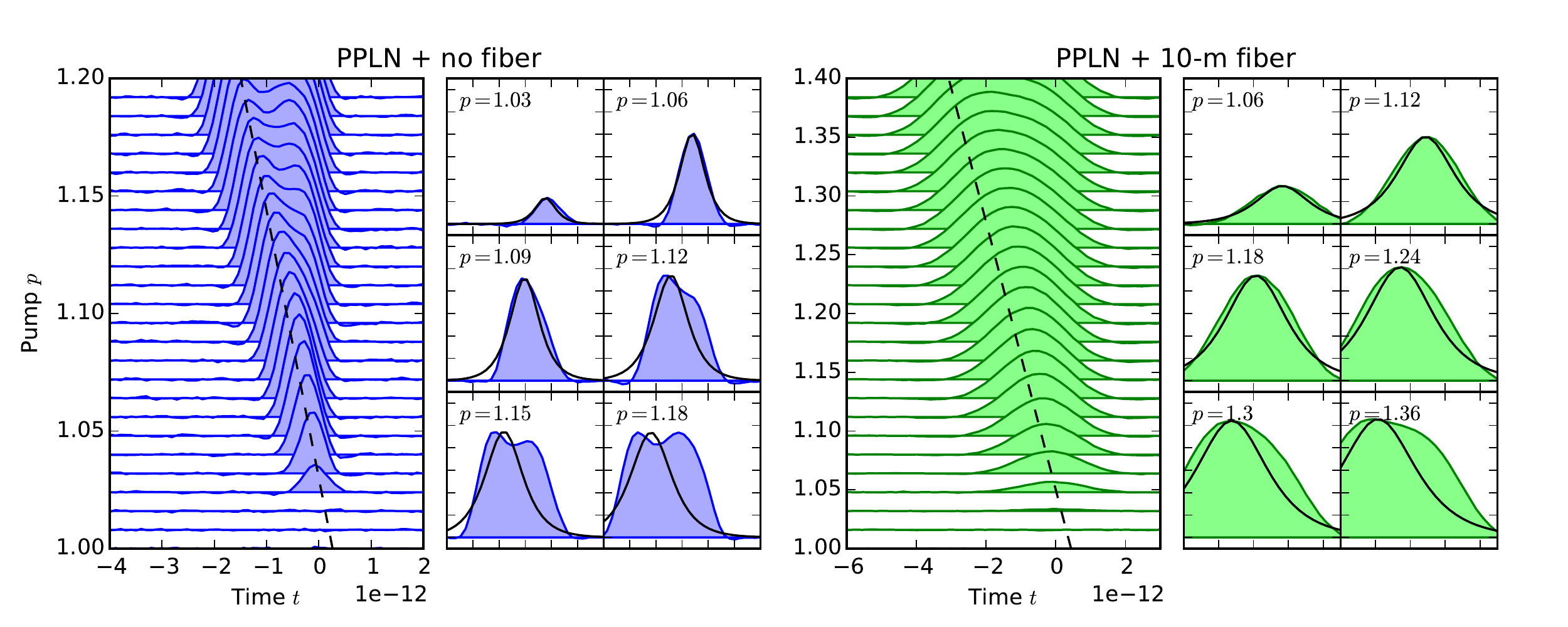}
\caption{Plot of the simulated pulse shape (filled), compared to the simulton steady-state of Eqs.~(\ref{eq:10-prop-a}-\ref{eq:10-prop-tau}) (black line).  Dashed line is the relation (\ref{eq:10-delay}).}
\label{fig:10-f16}
\end{center}
\end{figure}

\subsubsection{Transient Behavior}

Now assume that the pump is turned on abruptly.  In the absence of detuning, the pulse first grows at the maximum-gain point, as per the linear model.  Once pump depletion becomes significant, the pulse shifts forward, reaching an equilibrium when its amplitude saturates.  Both the simulation and simulton model agree here (Fig.~\ref{fig:10-f17-phase}, center-left plot).  This figure visualizes the dynamics with a phase space plot.  The full system is three-dimensional, but the pulse width can be assumed constant, giving a dynamical system with two variables.  This has one attractor, which is a spiral, explaining the initial overshoot in photon number.

Now let's detune the cavity and see what happens.  For negative detuning ($\ell = -7$, left plot), the pulse first grows at $T < 0$ and is shifted further by the simulton acceleration.  In this case, both detuning and simulton acceleration move the pulse in the same direction, away from the maximum-gain point, so its amplitude is reduced.

In contrast, for positive detuning (center-right plot), simulton acceleration opposes the detuning shift.  When the pulse is weak, the latter is dominant, so it grows at $T > 0$, but once pump depletion kicks in, it eventually drifts back to the maximum-gain point, where simulton acceleration and detuning cancel out.  Not surprisingly, photon number is larger than without detuning.

\begin{figure}[tbp]
\begin{center}
\includegraphics[width=0.70\columnwidth]{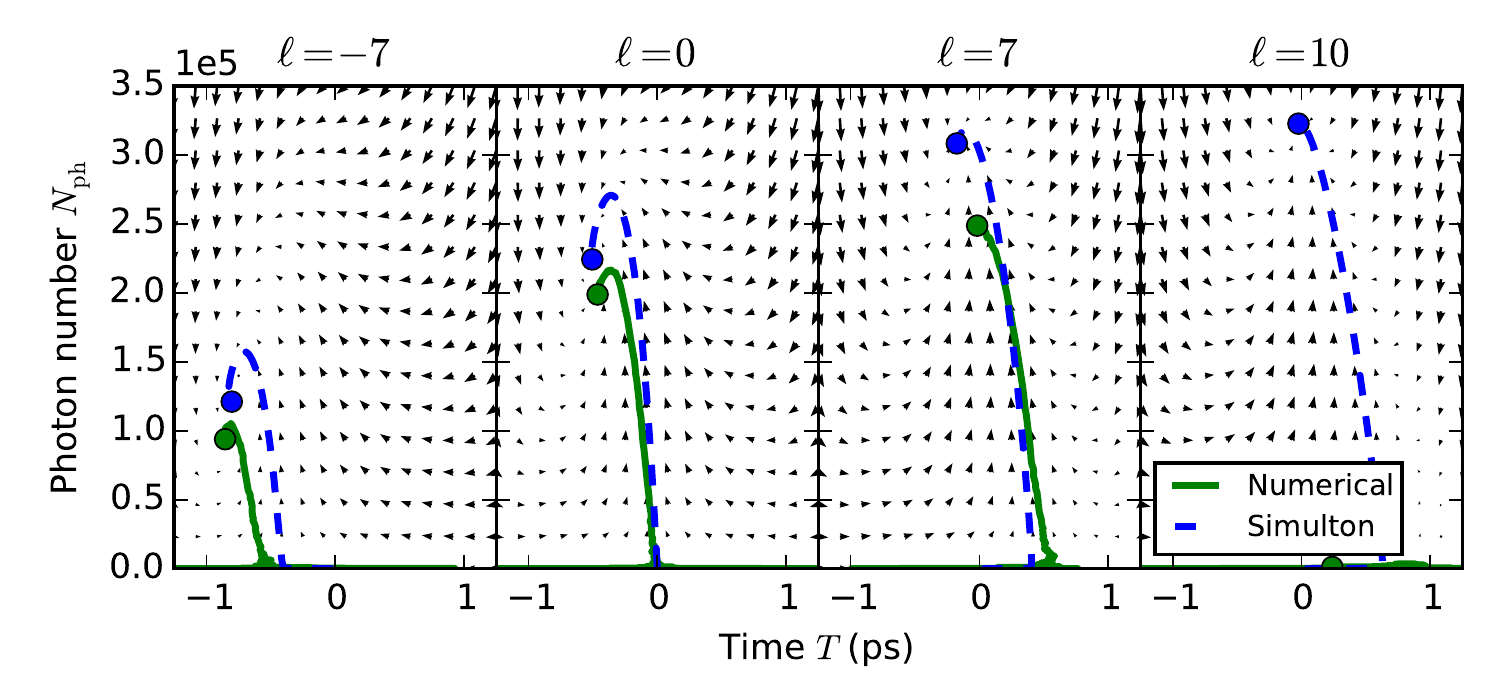}
\caption{Evolution of pulse photon number $N_{\rm ph}(t)$ and centroid $T(t)$ for sech-pulse model (dashed lines) and full numerics (solid).  Four different detuning values shown.}
\label{fig:10-f17-phase}
\end{center}
\end{figure}

For a given pump power, the optimal detuning is the one that cancels the simulton acceleration, so that the pulse can be amplified at the maximum-gain point.  This happens when $T = 0$ is a steady state to (\ref{eq:10-prop-gc-t}).  Applying the same substitution to $\beta A^2$, we find:
\beq
	\frac{\ell_{\rm max}\lambda}{2c} = \frac{p-1}{2}\log(G_0 e^{\alpha_a L})\tau \label{eq:10-ellmax}
\eeq
where $\tau$ is computed from (\ref{eq:10-prop-gc-tau}), which becomes independent of the other variables when $T = 0$.  This depends on the pump power; the larger $p-1$, the larger $\ell$ should be to form the optimal signal pulse.  Overshooting gives rise to weaker signal pulses, and can also cause instabilities that suppress the amplitude and are not captured by the simulton model (Fig.~\ref{fig:10-f17-phase}, right plot).

\subsubsection{Detuning and Stability}
\label{sec:10-simstab}

We can see from Figure \ref{fig:10-f17-phase} that the detuning has a substantial effect on the energy of the pulse that forms.  If $\ell$ is not too large, the numerical result matches the simulton description.

A more complete way to capture this behavior is to look at the pulse properties as a function of both pump $p$ and detuning $\ell$, as shown in Figs.~\ref{fig:10-f18-0}-\ref{fig:10-f18-4}.

\begin{figure}[p]
\begin{center}
\includegraphics[width=0.98\textwidth]{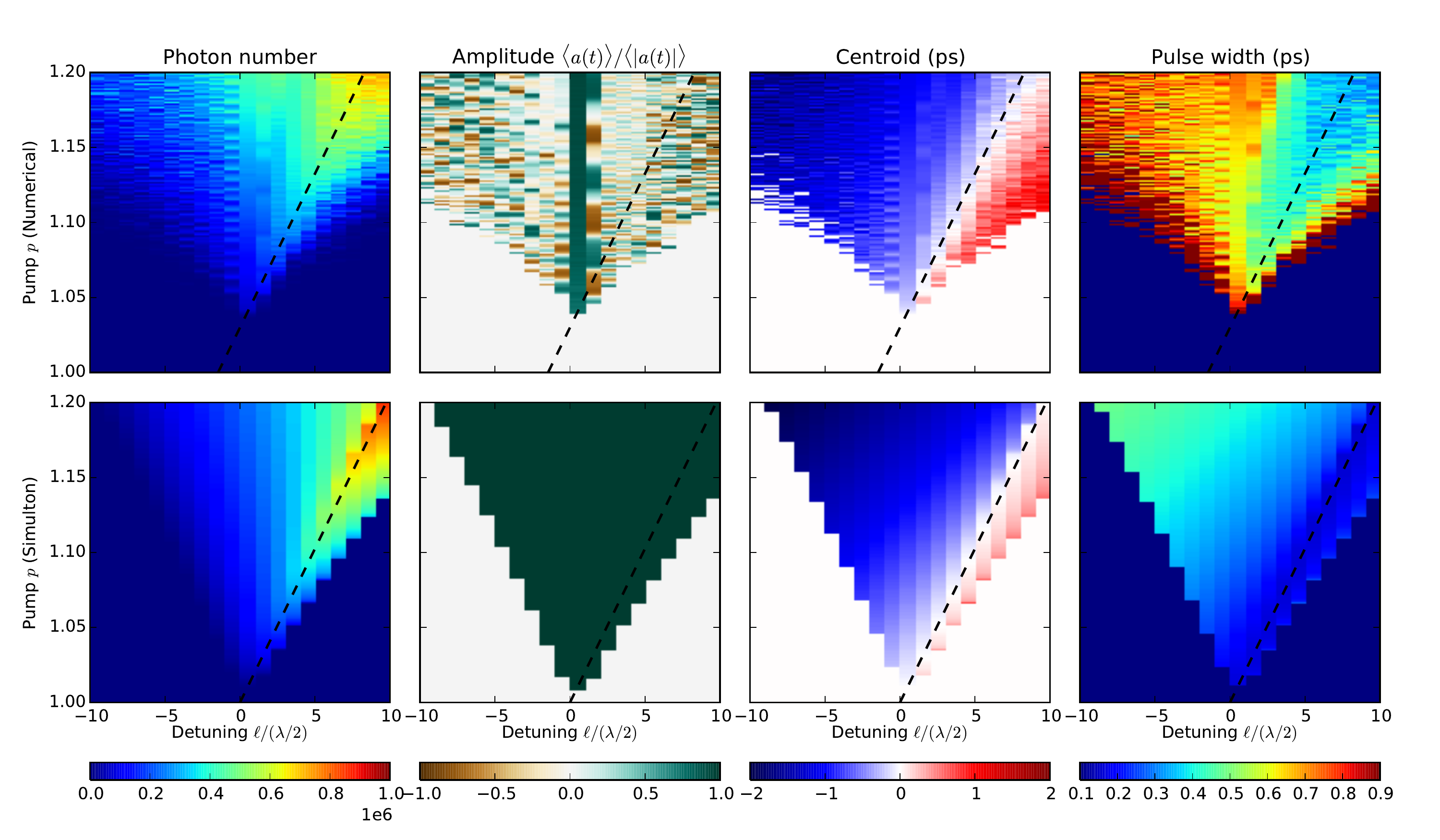}
\caption{Top: numerical simulation, plot of the photon number, normalized amplitude $\langle a(t)\rangle / \langle |a(t)| \rangle$, centroid and pulse width as a function of pump $p$ and detuning $\ell$.  Bottom: predictions from the simulton theory.  Dashed line is Eq.~(\ref{eq:10-ellmax})  PPLN OPO, no fiber.}
\label{fig:10-f18-0}
\end{center}
\end{figure}

\begin{figure}[p]
\begin{center}
\includegraphics[width=0.98\textwidth]{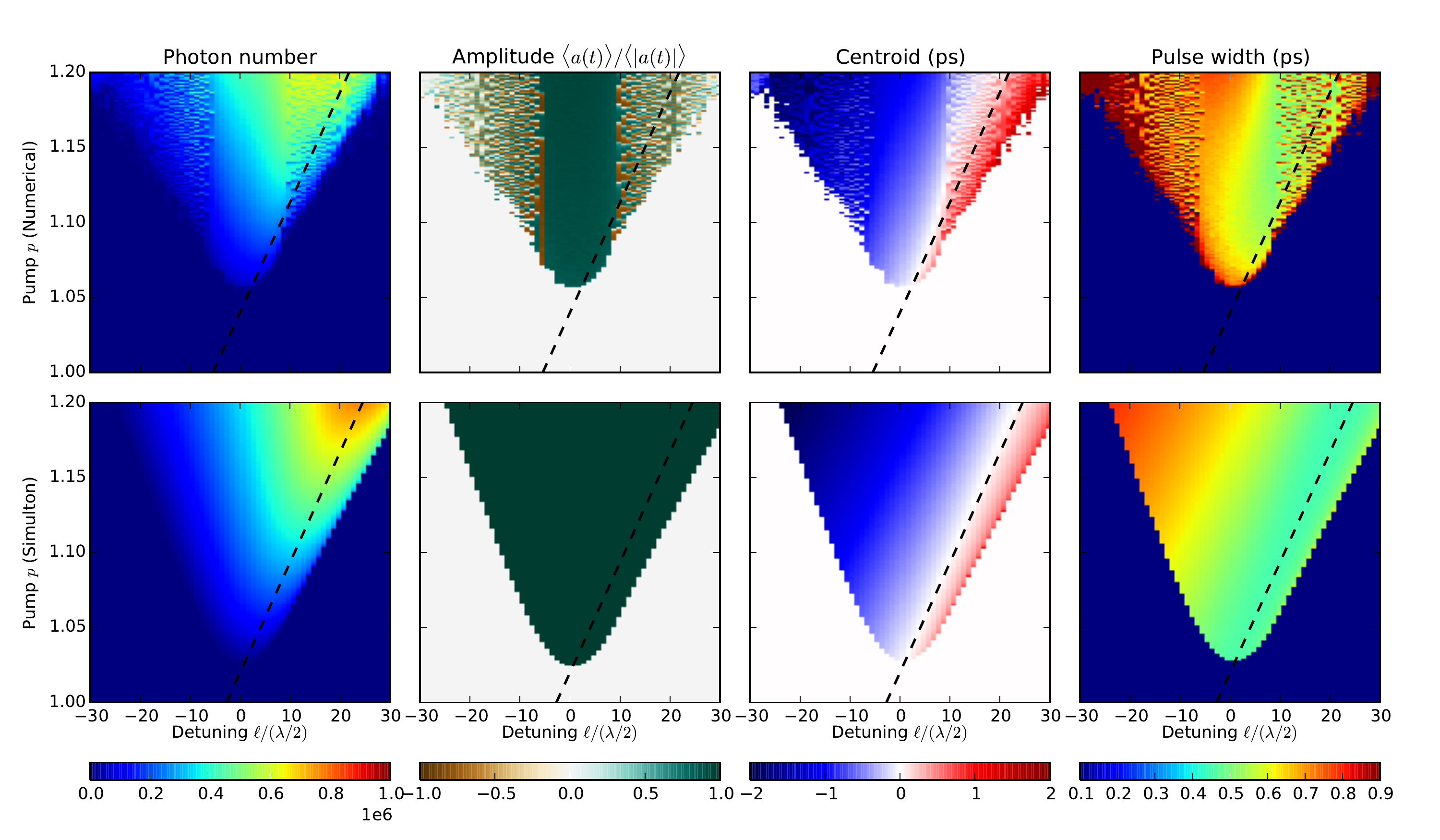}
\caption{Same as Fig.~\ref{fig:10-f18-0}, for PPLN OPO with 4-m fiber.}
\label{fig:10-f18-4}
\end{center}
\end{figure}

In the upper plots in Figs.~\ref{fig:10-f18-0}-\ref{fig:10-f18-4}, several features stand out.  The threshold varies close to linearly with detuning, consistent with the simulton theory (lower plots).  The simulton theory also predicts that when Eq.~(\ref{eq:10-ellmax}) is satisfied, the pulse amplitude is maximized and the pulse width is shortest and the centroid lies at $T = 0$, the trailing edge of the pump.  This is roughly consistent with the data, although there is an overall offset in the thresholds.  The pulse width and photon number also roughly match.  

However, these plots show that the simulton picture is only valid for a limited range of $\ell$.  If $\ell$ is too large, additional effects destabilize the sech-pulse.  Thus, the pulse amplitude $\langle a(t) \rangle / \langle |a(t)| \rangle$, which is constant in the simulton picture, oscillates.  This causes smaller oscillations in the photon number, centroid and pump width.

\section{Box Pulse Theory}
\label{sec:10-boxpulse}

Well above threshold, both the eigenmode and simulton theories fail.  An eigenmode expansion becomes impractical because too many modes need to be used and the computation time scales as $O(N^4)$.  Simulton theory fails because in this regime the pulses are no longer sech-shaped.  We need a new theory that predicts the pulse shapes in this regime.

Simulations show that pulses get longer the further one goes above threshold (Figs.~\ref{fig:10-f5}, \ref{fig:10-f12}, \ref{fig:10-f13}, \ref{fig:10-f16}).  This is a result of the pulse filling the leading side of the positive-gain region $\Delta_{\rm max}\Gamma(t) > 0$ (Sec.~\ref{sec:10-dispersionless}).  Long pulses mean narrow spectra and weak dispersion effects.  The result is a competition between gain and pump depletion, with dispersion playing only a secondary role.

In this section, we ignore dispersion and derive an analytic formula for the pulse shape that is reasonably accurate in this regime.  Dispersion will be treated later, but its main effect will be to add a modulation on the pulse shape when $\phi_0\phi'_2 < 0$, giving rise to a nondegenerate box-like pulse.

\subsection{Degenerate Case $\phi_0 = 0$}
\label{sec:clipping}

First, let's treat the center of the detuning peak $\phi_0 = 0$.  Later on we will treat the general case, but the results are simplest for $\phi_0 = 0$.  Recalling (\ref{eq:10-intdiff}), we drop dispersion terms and invoke the gain-without-distortion ansatz to obtain:
\begin{align}
    & \frac{\partial a(z,t)}{\partial z} = -\frac{1}{2} \alpha_a a(z, t) + \epsilon\,a^*(z,t) b_{\rm in}(t - u z)e^{-\alpha_b z/2} \nonumber \\
    & \quad - \frac{\epsilon^2}{2u} a^*(z,t)\int_{-\infty}^{t}{e^{(g+\alpha_b/2)(t'-t)/u} a(z, t')^2 \d t'} \label{eq:10-intdiff3}
\end{align}
Here $g = \tfrac{1}{L}\log(G_0)$ is the gain per unit length at steady state.  Now make the substitution
\beq
	a(z, t) = e^{gz/2} \bar{a}(z, t)
\eeq
where $\bar{a}(z, t)$ is real and slowly-varying in $z$.  This is valid for flat-top pump pulses, where the gain is roughly constant because the pulse amplitude is constant.  We choose $g$ so that $e^{gL/2}$ is the cavity loss, since in steady state, gain equals loss and thus the single-pass gain should be $e^{gL/2}$.  Deviations will be handled by perturbation theory on $\bar{a}$.  Equation (\ref{eq:10-intdiff3}) becomes:
\begin{align}
    &\frac{\partial\bar{a}(z,t)}{\partial z} = \left[\epsilon\,(b_{\rm in}(t-uz)e^{-\alpha_b z/2} - \bar{b}_0)\right]\bar{a}(z,t) \nonumber \\
    &\qquad - \frac{\epsilon^2 e^{gz}}{2u}\bar{a}(z,t)\int_{-\infty}^t{e^{(g+\alpha_b/2)(t'-t)/u} \bar{a}(z,t')^2\d t} \label{eq:10-nodist}
\end{align}
To obtain the output field, one must integrate (\ref{eq:10-nodist}) from $z=0$ to $L$.  Gain without distortion means that the integrand is close to constant over that interval, so we can approximate the integral with one Picard step, setting $z=0$ everywhere in the integrand.  The evolution over one round-trip is:
\begin{align}
	& \Delta a(t) = a(t)\biggl[\overbrace{\int_0^L{\epsilon(b_{\rm in}(t-uz) e^{-\alpha_b z/2} - \bar{b}_0)\d z}}^{F(t)} \nonumber \\
	& \qquad - \frac{\epsilon^2(e^{gL}-1)}{2gu} \int_{-\infty}^t {e^{(g+\alpha_b/2)(t'-t)/u} a(t')^2 \d t'}\biggr] \label{eq:10-gclip-inout}
\end{align}

\begin{figure}[tbp]
\begin{center}
\includegraphics[width=1.0\columnwidth]{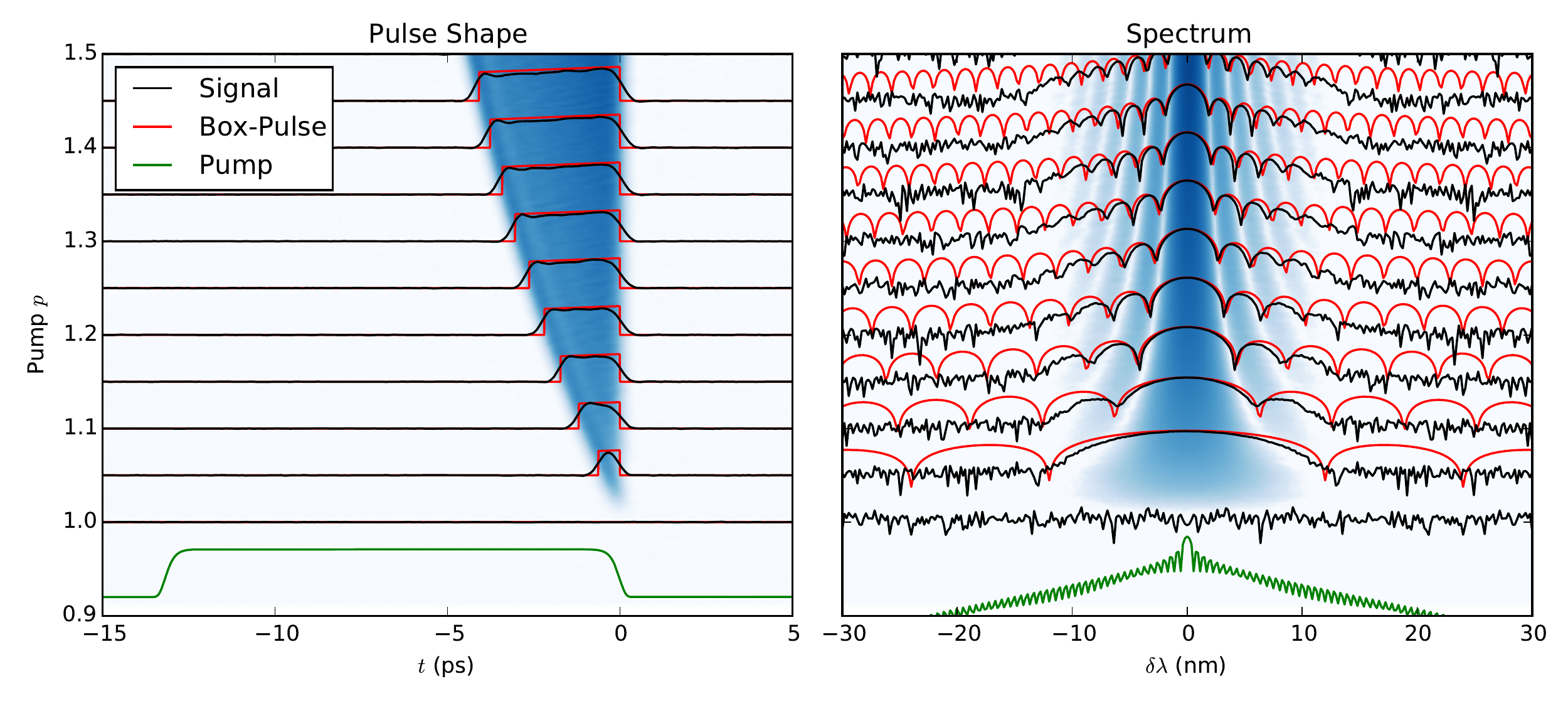}
\caption{Left: plot of the pulse shape for the signal and its box-pulse approximation via Eq.~(\ref{eq:10-box}), for $\phi_0 = 0$, $p \in [1.0, 1.5]$.  Pump is given at the bottom.  Right: power spectrum for the same data, on a log scale.}
\label{fig:10-f19}
\end{center}
\end{figure}

There are two linear terms in (\ref{eq:10-gclip-inout}).  The first is the gain-clipping term, where $F(t)$ is related to $G(t)$ by:

\bea
	F(t) & = & G(t) + \int_0^L {(b_{\rm max} e^{-\alpha_b z/2} - \bar{b}_0)\d z} \nonumber \\
	& = & G(t) + \frac{p-1}{2} \log(G_0 e^{\alpha_a L}) \nonumber \\
	& = & G(t) + \log \bigl[\Delta_{\rm max}(\phi_0 = 0)\bigr] \label{eq:10-boxft}
\eea

In steady state, $a(t)$ stays constant between round trips, so the right-hand side of (\ref{eq:10-gclip-inout}) must equal zero.  There are two ways this can happen:

\begin{enumerate}
\item $F(t) < 0$ or $F(t)$ decreasing.  Since the second integral is always positive and increasing, it is impossible to set the term in square brackets in (\ref{eq:10-gclip-inout}) to zero.  The only way to satisfy the steady-state condition is to set $a(t) = 0$.
\item $F(t) > 0$ and increasing.  In this case, $a(t) \neq 0$ and the terms in the square brackets must cancel out.  Combining (\ref{eq:10-gclip-inout}) with its time derivative (both which must equal zero), we find:
\beq
	a(t)^2 = \frac{2gu}{\epsilon^2(e^{gL}-1)} \left[F'(t) - \frac{g+\alpha_b/2}{u} F(t)\right] \label{eq:10-asq-box}
\eeq
\end{enumerate}

For a flat-top pump pulse, the analytic formula for $G(t)$ (Eq.~\ref{eq:10-gain-gc0}) will suffice; from this we can calculate $F(t) = \tfrac{1}{2}\log(G_0 e^{\alpha_a L}) \bigl[(p-1) - p|t|/T_p\bigr]$.  Using (\ref{eq:10-asq-box}) and substituting $g, \epsilon, u, b_0$ for $G_0, T_p, N_{b,0}$ (Table~\ref{tab:10-t1}) we find the solution
\begin{align}
	& a(t)^2 = \frac{4N_{b,0} e^{-\alpha_b L/2}\log(G_0)}{T_p (G_0 - 1) \log(G_0 e^{\alpha_a L})} \nonumber \\
	& \qquad \times \left[p + \left(\log G_0 + \tfrac{1}{2}\alpha_b L\right) \left((p-1) - p\frac{|t|}{T_p}\right)\right] \label{eq:10-box}
\end{align}
for $-T_p(1-p^{-1}) < t < 0$ (and $a(t) = 0$ otherwise).  This can be integrated to give the total photon number:
\begin{align}
	& N_a = \frac{4N_{b,0} e^{-\alpha_b L/2}\log(G_0)}{T_p (G_0 - 1) \log(G_0 e^{\alpha_a L})} \nonumber \\
	& \qquad \times \left[(p-1) + \frac{(p-1)^2}{2p} \left(\log G_0 + \tfrac{1}{2}\alpha_b L\right)\right]
\end{align}
Figure \ref{fig:10-f19} compares the waveform (\ref{eq:10-box}) and its Fourier transform to full simulations.  The amplitude and the general shape are modeled well by the theory, although it says nothing about the shape of the edges.  As the pulse gets longer with increasing pump power, the spectrum narrows, a fact confirmed in experiments and consistent with previous work \cite{Becker1974}.

\begin{figure}[t]
\begin{center}
\includegraphics[width=1.0\columnwidth]{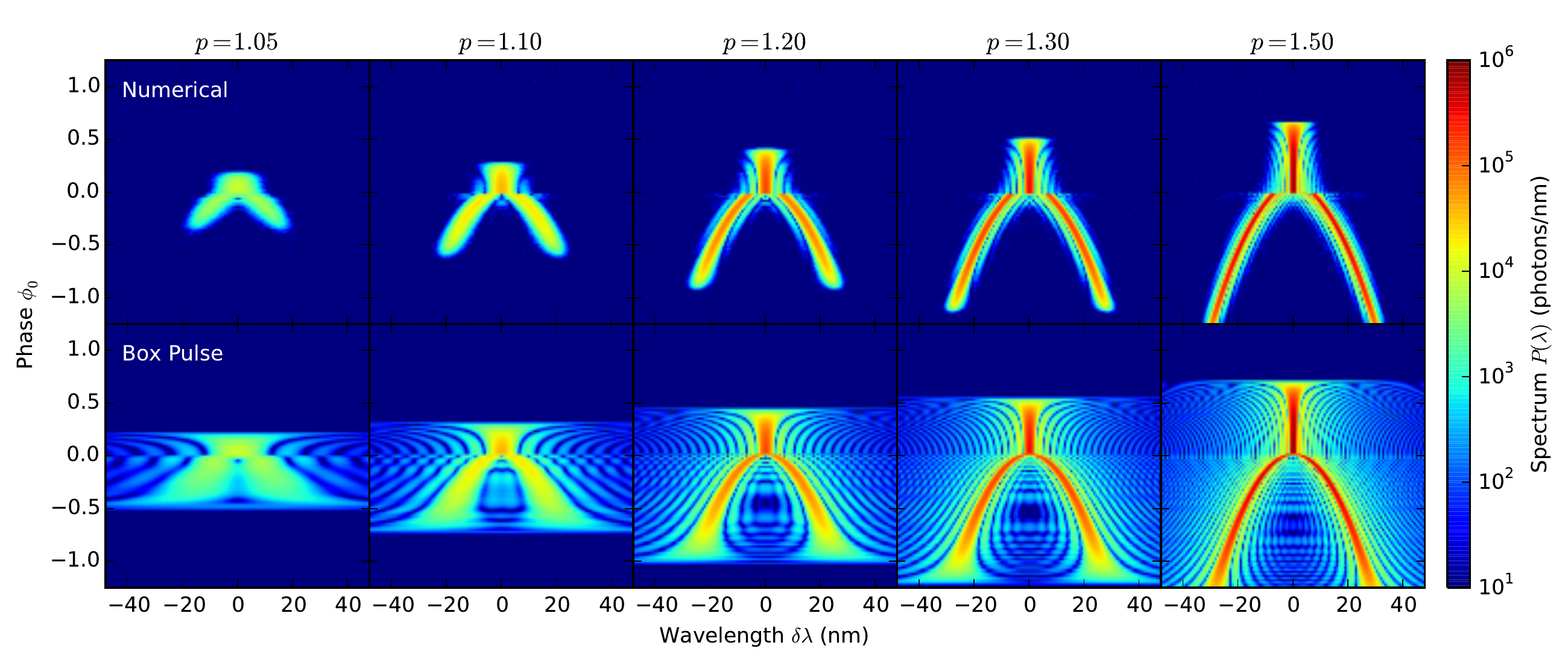}
\caption{Resonance diagrams for the box pulse model (Eq.~\ref{eq:10-box2}) compared to numerical result.}
\label{fig:10-f23}
\end{center}
\end{figure}

\begin{figure}[t]
\begin{center}
\includegraphics[width=1.0\columnwidth]{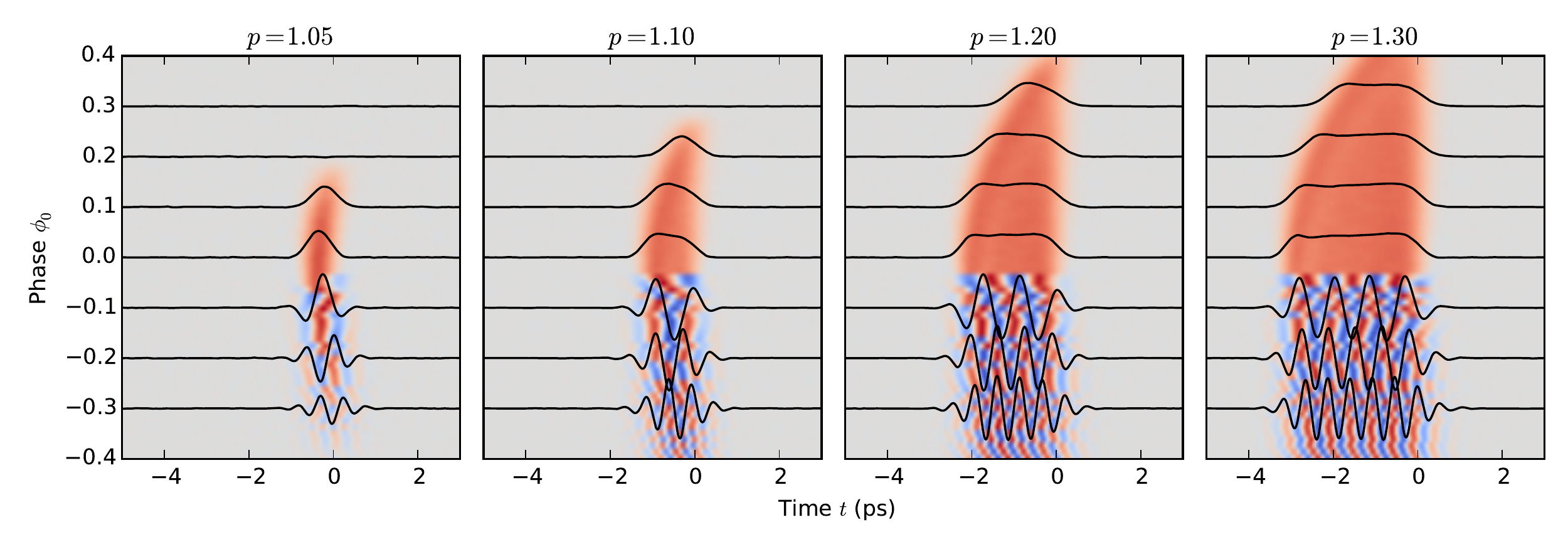}
\caption{Time-domain pulse shapes as function of phase and power, numerical.}
\label{fig:10-f21}
\end{center}
\end{figure}

\subsection{Dispersion}

Gain clipping sets the overall pulse shape, while dispersion evens out the edges and sets the signal-idler splitting.  If $\phi_0 \phi'_2 > 0$, the OPO is degenerate so there is no signal-idler splitting; however, nonzero $\phi_0$ reduces the overall gain, which reduces the signal power.  The most straightforward way to do this is to say that Eq.~(\ref{eq:10-boxft}) should be modified to read
\beq
	F(t) = G(t) + \log\bigl[\Delta_{\rm max}(\phi_0)\bigr]
\eeq
and the rest of the results carry over unchanged.  Eq.~(\ref{eq:10-box}) becomes:
\beq
	a(t)^2 = \frac{4N_{b,0} e^{-\alpha_b L/2}\log(G_0)}{T_p (G_0 - 1) \log(G_0 e^{\alpha_a L})} \biggl[p + \left(\log G_0 + \tfrac{1}{2}\alpha_b L\right) \left(\frac{2\log(\Delta_{\rm max}(p,\phi_0))}{\log(G_0 e^{\alpha_a L})} - p\frac{|t|}{T_p}\right)\biggr] \label{eq:10-box2}
\eeq
For $\phi_0 \phi'_2 < 0$, the pulse is box-shaped but nondegenerate: $a(t) = \mbox{Re}\bigl[\bar{a}(t) e^{-i\,\delta\omega_0 t}\bigr]$, (see Eq.~\ref{eq:10-abar}), and $\bar{a}(t)$ takes the same form as (\ref{eq:10-box2}) but with a $\sqrt{2}$ factor to preserve the overall energy.  

A good way to visualize (\ref{eq:10-box2}) is to plot resonance diagrams for the box-pulse model and compare them to the numerics, as in Fig.~\ref{fig:10-f23}.  The general structure of the resonance plots are the same, but the features on the tails differ, consistent with the smoothing in Fig.~\ref{fig:10-f19}.  However, these tails are suppressed by several orders of magnitude and only show up on the plot because of the log scale.

The Fourier transform of this is given in Fig.~\ref{fig:10-f21}.  The OPO is nondegenerate for $\phi_0 < 0$, but this does not affect the overall shape of the pulse.  Aside from a sinusoidal modulation, it remains box-shaped.

\section{Conclusion}

This chapter has introduced three reduced models that aid the understanding, simulation, and design of synchronously pumped OPOs.  These models are based on mathematical approximations and physical intuition, and show good agreement with numerical simulations for predicting steady-state pulse shapes, transient behavior and stability.  Because the models run several orders of magnitude faster than numerical simulations, they will be a useful tool for simulating large OPO networks, and a guide for device design and optimization.

Near threshold, I derived an eigenmode expansion that predicts the OPO threshold as a function of cavity dispersion and round-trip phase, and gives the correct steady-state pulse shape.  The pulse dynamics is a competition between gain clipping, which shortens the pulse to maximize its overlap with the pump; and dispersion, limits its bandwidth.  We noticed a smooth transition between degenerate and nondegenerate oscillation when the cavity dispersion is not compensated, which could be explained by a simple phase-matching argument.  In both the degenerate and nondegenerate regimes, I obtained analytic formulae for the pulse shape in terms of Airy and hypergeometric functions, which gave analytic expressions for the pulse shape and its threshold.  Moreover, pulse stability could be explained using bifurcation theory with a simple two-mode model.

Far from threshold, the steady-state pulse was found to have a narrow spectrum, and I obtained a box-like pulse shape by solving the equations without dispersion.  In the frequency domain, this appears as a sinc-shaped spectrum which grows narrower the higher the pump relative to threshold.  An analytic expression for the pulse width and amplitude was derived, which agrees with the numerics.

Working between these regimes, I obtained a reduced model based on projection onto a sech-shaped pulse.  This was physically motivated by the ``simulton'' solution in a $\chi^{(2)}$ waveguide, and I accounted for the effects of gain-clipping and dispersion as perturbations to this solution.  While only valid in the degenerate regime close to threshold, this model is helpful because it is fully analytic, and within its regime of validity, agrees with the both the eigenmode model and the numerics.

\ifstandalone{\bibliography{NoteRefs}{}
\bibliographystyle{alpha}
}
\ifdefined\multidoc\else\input{Header}\fi

\ifstandalone{\setcounter{chapter}{11}}
\chapter{Silicon Optical Waveguides}
\label{ch:12}

Shortly after the discovery optical bistability in the 1970's \cite{GibbsBook, Gibbs1976} and demonstrations of ``optical transistors'' in various materials \cite{Miller1979, Miller1981, Jain1976}, optics was touted as a compelling alternative to electronic computing \cite{Smith1986, Smith1984}.  However, the size, quality, and other requirements for an optical transistor \cite{Smith1984, Miller2010} were not realizable with the large, lossy resonators that could be fabricated at the time.  In subsequent decades, due to the exponential ``Moore's Law'' growth in CMOS performance \cite{Moore1965}, optical computing was totally eclipsed by electronics.

Fortuitously, CMOS has turned from competitor to a major driver of photonics research, spawning the field of ``Silicon Photonics.''  High-quality photonic components can be created using CMOS-compatible processes optimized for large-scale integration and high yield, and the fabrication work can be outsourced to commercial foundries \cite{Soref2006}.  Silicon is only ``average'' in terms of its optical properties and state-of-the-art {\it devices} are rarely made from silicon, but due to integration challenges, it is currently the most promising platform for optical {\it circuits}.

Devices and effects that have been realized with silicon include:

\begin{itemize}
	\item Passive components, e.g. add/drop ports, beamsplitters \cite{Shen2016}, polarization rotators \cite{Gao2013}, grating couplers
	\item Electro-optic components, e.g. modulators, photodetectors \cite{Ding2012}
	\item $\chi^{(3)}$ devices, e.g. four-wave mixing \cite{Foster2006, Lin2008, Liu2010}, supercontinuum generation \cite{Leo2015}
	\item Raman scattering, Raman lasers \cite{Boyraz2004, Rong2005}
	\item Induced $\chi^{(2)}$ nonlinearity, e.g. strained silicon \cite{Jacobsen2006}, EFISH \cite{Timurdogan2016}
	\item Free-carrier nonlinearity \cite{Tanabe2005, Xu2006}
	\item Heterogeneous integration \cite{Rabiei2013}
\end{itemize}

The simplest optical component is a waveguide.  Silicon optical waveguides benefit from a high index-contrast ($n = 3.5$ vs.\ $1.5$ or $1.0$), allowing low-loss guided-waves with $\sim$ 0.1$\mu$m cross section \cite{Yamada2005}.  This confinement dramatically enhances the optical nonlinearity in silicon compared to other platforms.  This chapter studies the effects present in silicon optical waveguides, focusing particularly on structures that can be built with established foundries such as IMEC \cite{Selvaraja2014, Lim2014}.

The present chapter covers work done with Dodd Gray (Stanford) and Kambiz Jamshidi (TU-Dresden).  Our lab does not do silicon photonics, but we have recently begun a collaboration with Prof.\ Jamshidi to build and study nonlinear- and potentially quantum-optical circuits.  As such, this is the most open-ended chapter in my thesis, focusing mainly on possibilities rather than results.  It is fitting to end my thesis this way, for it signifies that scientific progress is never ``complete.''  It is always changing, always evolving, a perpetual cycle of striving and discovery.

Works related to this chapter include:

\begin{itemize}
	\item Meysam Namdari, Mahmoud Jazayerifar, Ryan Hamerly, and Kambiz Jamshidi, ``CMOS Compatible Ring Resonators for Phase-Sensitive Optical Parametric Amplification.'' [submitted]
	\item Ryan Hamerly, Levon Mirzoyan, Meysam Namdari, and Kambiz Jamshidi, ``Optical bistability, self-pulsing and soliton formation in silicon micro-rings with active carrier removal.'' [submitted]
\end{itemize}

\section{Waveguide Modes and Dispersion Relation}

To find the propagating modes and their dispersion relation, we need to solve Maxwell's equations in the waveguide.  In a non-magnetic medium, Maxwell's Equations are:
\begin{align}
	\nabla\cdot (n^2 E) & = 0
	& \nabla\times E & = -\mu_0 \frac{\partial H}{\partial t} \nonumber \\
	\nabla\cdot H & = 0
	& \nabla\times H & = n^2 \epsilon_0 \frac{\partial E}{\partial t} \label{eq:12-maxwell}
\end{align}
For a waveguide oriented along the $z$ direction, the index of refraction is a function of $x$ and $y$.  From translation symmetry, we can show that waves propagate along the $z$ direction and take the form:
\beq
	E = E(x, y) e^{i(\beta z - \omega t)},\ \ \ 
	B = B(x, y) e^{i(\beta z - \omega t)} \label{eq:12-eb-wave}
\eeq
Applying (\ref{eq:12-eb-wave}), we get the following equations for the fields $E(x, y)$, $B(x, y)$ \cite{OkamotoBook}:
\begin{align}
	\frac{\partial E_z}{\partial y} - i\beta E_y & = i\omega\mu_0 H_x
	& \frac{\partial H_z}{\partial y} - i\beta H_y & = -i\omega\epsilon_0 n(x,y)^2 E_x \nonumber\\
	i\beta E_x - \frac{\partial E_z}{\partial x} & = i\omega\mu_0 H_y
	& i\beta H_x - \frac{\partial H_z}{\partial x} & = -i\omega\epsilon_0 n(x,y)^2 E_y \nonumber \\
	\frac{\partial E_y}{\partial x} - \frac{\partial E_x}{\partial y} & = i\omega\mu_0 H_z
	& \frac{\partial H_y}{\partial x} - \frac{\partial H_x}{\partial y} & = -i\omega\epsilon_0 n(x,y)^2 E_z 
	\label{eq:12-mode}
\end{align}
Equations (\ref{eq:12-mode}) are an eigenvalue equation: given a frequency $\omega$, we solve for the eigenmode $[E_x,E_y,E_z,H_x,H_y,H_z]$ and its eigenvalue $\beta$.  The modes will be orthonormal in both $E$ and $H$ integrals.  For two modes $(E_i, H_i)$ and $(E_j, H_j)$:
\beq
	\int \epsilon_0 n^2 \bigl(E_i^* \cdot E_j\bigr)\d A = \int \mu_0 \bigl(H_i^* \cdot H_j\bigr)\d A = \delta_{ij}
\eeq
Moreover, the cross-product integral, which is related to the Poynting vector, is orthogonal but with a different normalization constant:
\beq
	\int \bigl(E_i^* \times H_j \bigr)\cdot \hat{z}\,\d A = C_i \delta_{ij}
\eeq
for a given constant $C_i$.

The group velocity can be computed using perturbation theory.  For a general perturbation (to both $\omega$ and $n(x, y)$), the wavenumber changes by \cite{SnyderLove}:
\beq
	\Delta\beta_i = \frac{\int{\left[\Delta(\omega\epsilon_0 n^2) |E_i|^2 + \Delta(\omega \mu_0) |H_i|^2\right]\d A}}{2\int{\mbox{Re}(E_i^* \times H_i)\cdot \hat{z}\,\d A}} \label{eq:12-pert}
\eeq
Note that, when calculating the group velocity, both $\omega$ and $n$ vary in the differentials in Eq.~(\ref{eq:12-pert}).  The {\it effective group index} $n_{g,i} = c/v_{g,i}$ is: :
\beq
	n_{g,i} = \frac{c}{v_{g,i}} = c\frac{\d\beta}{\d\omega} = \frac{\int{c\epsilon_0 n n_g |E|^2 \d A}}{\int{\mbox{Re}(E^*\times H)\cdot \hat{z}\,\d A}} \label{eq:12-ng}
\eeq
Note that the integrals in (\ref{eq:12-ng}) above depend both on the material's phase index $n(x, y)$ and its group index $n_g(x, y)$.  Also note that the numerator and denominator in (\ref{eq:12-ng}) are the energy density (per unit length) and transmitted power:
\beq
	U = \frac{1}{2} \int{\epsilon_0 n n_g |E|^2\d A},\ \ \ 
	P = \frac{1}{2} \int{\mbox{Re}(E^*\times H)\cdot \hat{z}\,\d A},\ \ \ 
	n_{g,\rm eff} = \frac{c U}{P}
\eeq
If the refractive index changes $n \rightarrow n + \Delta n$ (due to heating, etc.), Eq.~(\ref{eq:12-pert}) can be solved and, upon substituting the denominator with (\ref{eq:12-ng}), we obtain:
\beq
	\Delta n_i = n_{g,i} \frac{\int{n\,\Delta n|E|^2 \d A}}{n n_g |E|^2 \d A} \label{eq:12-dbeta}
\eeq
If the waveguide can be split into regions $R_{\textbf{r}}$, each with its own material and its own index $n_{\textbf{r}}$, then (\ref{eq:12-dbeta}) takes a more intuitive form a weighted sum:
\beq
	\Delta n_i = \sum_{\textbf{r}} \frac{n_{g,i}}{n_{g,\textbf{r}}} \Gamma_{\textbf{r}} \Delta n_{\textbf{r}},\ \ \ 
	\Gamma_{\textbf{r}} \equiv \frac{\int_{R_{\textbf{r}}}{n n_g |E|^2\d A}}{\int{n n_g |E|^2\d A}} \label{eq:12-dni}
\eeq
As Eq.~(\ref{eq:12-dni}) shows, the total $\Delta n$ a weighted sum of the $\Delta n_{\textbf{r}}$ for the regions $R_{\textbf{r}}$.  There are two weighting factors: a ratio of group velocities $n_{g,i}/n_{g,\textbf{r}}$, and a filling factor $\Gamma_{\textbf{r}}$.  Since the regions span the whole cross section, the filling factors must sum to unity: $\sum_{\textbf{r}} \Gamma_{\textbf{r}} = 1$

Loss can be modeled with a complex refractive index: $n \rightarrow n + ik$.  In silicon photonics, the loss is typically weak ($k \ll n$) and can be treated using perturbation theory.  The absorption coefficient $\alpha$ is related to $k$ by $\alpha = (2\omega/c)k = (4\pi/c\lambda)k$.  Applying (\ref{eq:12-dni}), we find a similar expression for the waveguide loss $\alpha_i$:
\beq
	\alpha_i = \sum_{\textbf{r}} \frac{n_{g,i}}{n_{g,\textbf{r}}} \Gamma_{\textbf{r}} \alpha_{\textbf{r}} \label{eq:12-wg-abs}
\eeq

\subsection{Slab Waveguide}
\label{sec:12-slab}

\begin{figure}[b!]
\begin{center}
\includegraphics[width=0.5\textwidth]{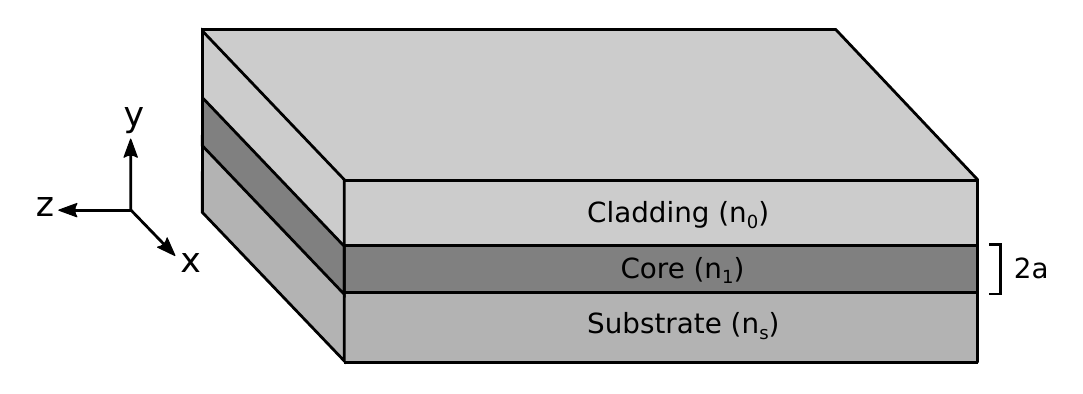}
\caption{Diagram of slab optical waveguide.}
\label{fig:12-f1}
\end{center}
\end{figure}

To find the modes, generally, Eqs.~(\ref{eq:12-mode}) must be solved numerically, but for simple structures (slab waveguide, cylindrical waveguide), one can find analytic solutions.  Reasonable approximate solutions exist for rectangular and rib / ridge waveguides, but these are based on the slab waveguide, so it will be introduced first.

A {\it slab waveguide} is a layer of high-index material, the {\it core}, surrounded by two low-index materials, the {\it cladding} and {\it substrate} (Fig.~\ref{fig:12-f1}).  Technically, this is not a waveguide because it only confines the field in the $y$ direction, so waves can propagate any direction in the $xz$-plane.  If the field propagates along the $z$-axis, $E$ and $H$ have no $y$-dependence.  The field equations become:
\begin{align}
	\frac{\d E_z}{\d y} - i\beta E_y & = i\omega\mu_0 H_x
	& \frac{\d H_z}{\d y} - i\beta H_y & = -i\omega\epsilon_0 n^2 E_x \nonumber\\
	i\beta E_x & = i\omega\mu_0 H_y
	& i\beta H_x & = -i\omega\epsilon_0 n^2 E_y \nonumber \\
    -\frac{\d E_x}{\d y} & = i\omega\mu_0 H_z
	& -\frac{\d H_x}{\d y} & = -i\omega\epsilon_0 n^2 E_z 
	\label{eq:12-mode2}
\end{align}
Inspecting (\ref{eq:12-mode2}), we see that there are two independent modes in the slab waveguide: a {\it transverse-electric (TE)} mode with $E_z = E_y = H_x = 0$, and a {\it transverse-magnetic (TM)} mode with $H_z = H_y = E_x = 0$.  The TE and TM modes satisfy the following equations:

\begin{center}
\begin{tabular}{c|c}
\hline\hline
TE mode & TM mode \\
\hline
$E \sim \hat{x},\ \ H \sim \hat{y},\hat{z}$ &
$E \sim \hat{y},\hat{z},\ \ H \sim \hat{x}$ \\
\hline
$\bigl[\frac{\d^2}{\d y^2} + ((\omega/c)^2 n^2 - \beta^2)\bigr] E_x = 0$
& 
$\bigl[n^2 \frac{\d}{\d y}\bigl(\frac{1}{n^2} \frac{\d}{\d y}\bigr) + \bigl((\omega/c)^2 n^2 - \beta^2\bigr)\bigr] H_x = 0$ \\
$H_y = \frac{\beta}{\omega\mu_0}E_x,\ H_z = \frac{i}{\omega\mu_0} \frac{\d E_x}{\d y}$ & 
$E_y = -\frac{\beta}{\omega\epsilon_0 n^2}H_x,\ E_z = -\frac{i}{\omega\epsilon_0 n^2} \frac{\d H_x}{\d y}$
\\ \hline\hline
\end{tabular}
\end{center}

\begin{figure}[tbp]
\begin{center}
\includegraphics[width=0.8\textwidth]{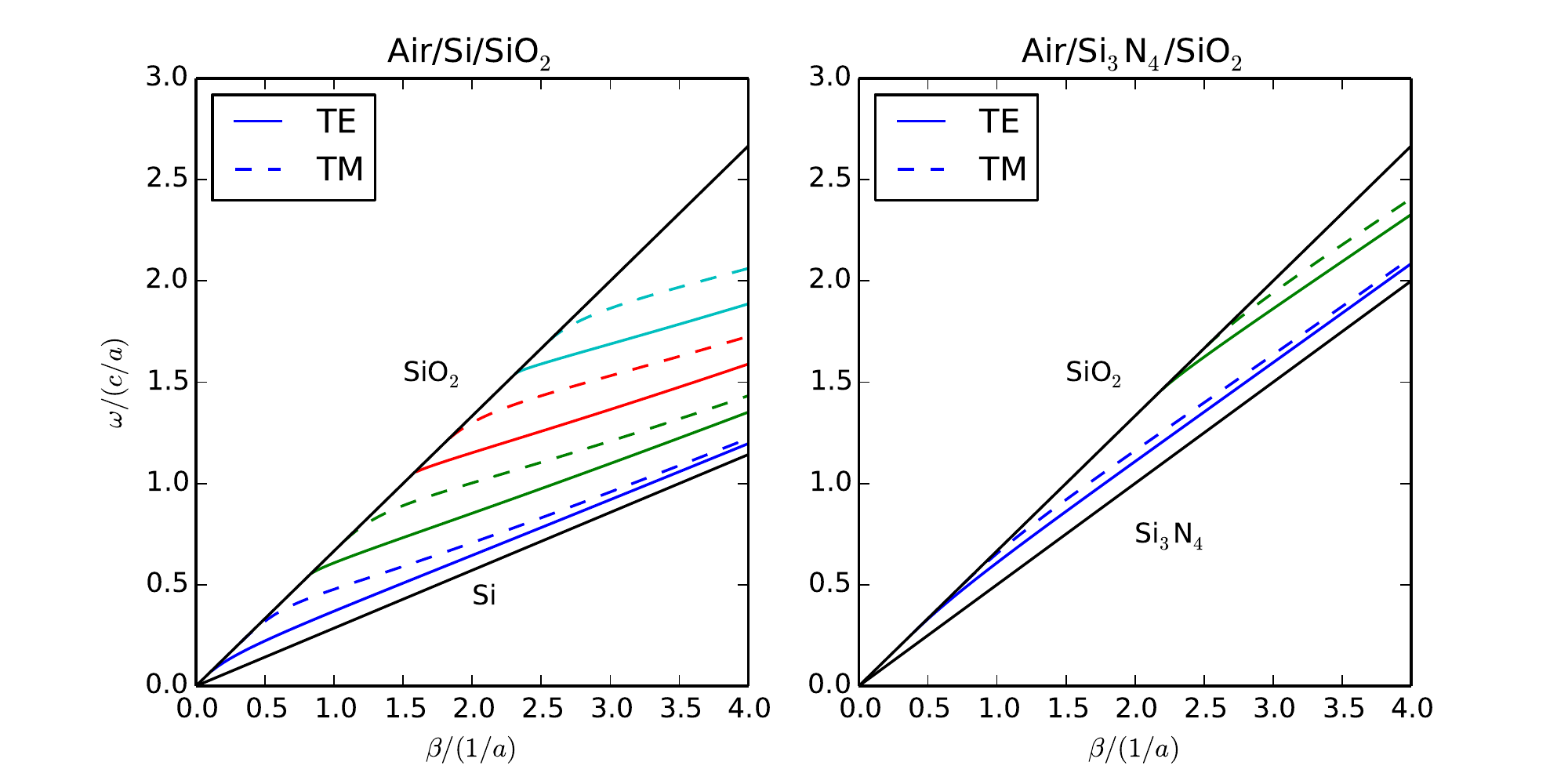}
\caption{Dispersion relation for Air/Si/SiO$_2$ slab waveguide and Air/Si$_3$N$_4$/SiO$_2$ waveguide (right), with $n_{\rm air} = 1, n_{\rm Si} = 3.5, n_{\rm Si_3N_4} = 2.0, n_{\rm SiO_2} = 1.5$.  Black lines give bulk dispersion relations for substrate and cladding.}
\label{fig:12-f2}
\end{center}
\end{figure}

The TE and TM modes have the same form, since the equation for $E_x$ ($H_x$ for TM) is analogous to a Schr\"{o}dinger equation:
\beq
	\left.\begin{array}{cc} E_x & \mbox{(TE)} \\ H_x & \mbox{(TM)} \end{array}\right\} =
	\left\{\begin{array}{ll} \cos(\kappa a - \phi)e^{-\sigma(y-a)} & (y>a) \\
		\cos(\kappa y - \phi) & (-a\leq y\leq a) \\
		\cos(-\kappa a - \phi)e^{\xi(y+a)} & (y<-a) \end{array}\right.
\eeq
For TE, both the field $E_y$ and its first derivative are continuous at the boundaries $y = \pm a$; for TM, the field $H_x$ and the quantity $n^{-2} \d H_x/\d y$ are continuous.  One finds:
\beq
	\kappa = \sqrt{(\omega n_1/c)^2 - \beta^2},\ \ \ 
	\sigma = \sqrt{\beta^2 - (\omega n_0/c)^2},\ \ \ 
	\xi = \sqrt{\beta^2 - (\omega n_s/c)^2} \label{eq:12-ksx}
\eeq
The boundary conditions give the constraint:
\beq
	2\kappa a = m\pi + \tanh^{-1}(\xi/\kappa) + \tan^{-1}(\sigma/\kappa),\ \ \ (m\in\mathbb{Z})
\eeq
Solving for this, after substituting (\ref{eq:12-ksx}), gives the dispersion relation $\beta(\omega)$.  The phase is given by:
\beq
	2\phi = m\pi + \tanh^{-1}(\xi/\kappa) - \tan^{-1}(\sigma/\kappa)
\eeq
Typical dispersion relations are shown in Fig.~\ref{fig:12-f2}.  The guided waves always lie between the bulk dispersion curve for the core ($\omega = (c/n_1)\beta$) and the substrate ($\omega = (c/n_s)\beta$).

\subsection{Rectangular Waveguide (Marcatili Method)}

Consider a waveguide with a rectangular cross section (Fig.~\ref{fig:12-f3}).  The index profile is given by:
\beq
	n(x, y) = \left\{\begin{array}{ll} n_1 & (-d < y < d,\ -a < x < a) \\ n_0 & \mbox{(elsewhere)} \end{array}\right.
\eeq
There is no analytic solution to Eqs.~(\ref{eq:12-mode}) for this waveguide.  In addition, the $(E_x, H_y, H_z)$ and $(H_x, E_y, E_z)$ modes do not decouple, so we do not have strictly TE and TM modes.

An approximate solution was obtained by Marcatili, in which we assume that the field in the corner regions $|x| > a$, $|y| > d$ is small enough to be neglected, since the field decays rapidly outside the waveguide \cite{Marcatili1969, Kumar1983}. 

\begin{figure}[tbp]
\begin{center}
\includegraphics[width=0.7\textwidth]{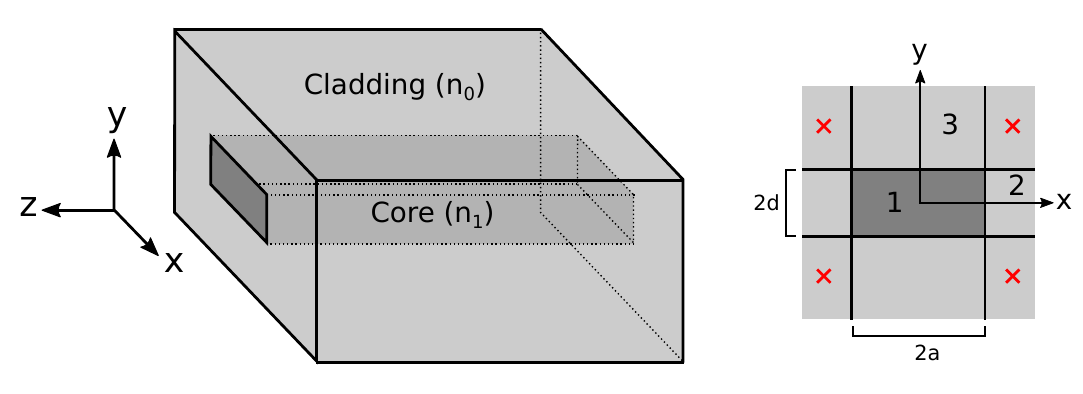}
\caption{Left: rectangular waveguide.  Right: cross-section, with excluded corner-regions in Marcatili's method.}
\label{fig:12-f3}
\end{center}
\end{figure}

Inspired by plane-wave optics, we suppose that in one of the electromagnetic modes, the $E_x$ and $H_y$ terms will be dominant.  Analogous to an $\hat{x}$-polarized plane wave, we will call it the $E_{pq}^x$ mode.  We assume $H_x = 0$ and solve the field equations (\ref{eq:12-mode}) to get:
\beq
	\frac{\partial^2 H_y}{\partial x^2} + \frac{\partial^2 H_y}{\partial y^2} + \bigl((\omega n/c)^2 - \beta^2\bigr) H_y = 0
\eeq
and
\begin{align}
	E_x & = \frac{\omega\mu_0}{\beta} H_y + \frac{1}{\omega\epsilon_0 n^2\beta} \frac{\partial^2 H_y}{\partial x^2}
	& E_y & = \frac{1}{\omega\epsilon_0 n^2\beta} \frac{\partial^2 H_y}{\partial x\partial y} \nonumber \\
	E_z & = \frac{i}{\omega\epsilon_0 n^2}\frac{\partial H_y}{\partial x} 
	& H_z & = \frac{i}{\beta} \frac{\partial H_y}{\partial y} \label{eq:12-expq}
\end{align}
Likewise, the $E_{pq}^y$ mode corresponds to the $\hat{y}$-polarized plane wave.  Here, we asusume that $E_y$ and $H_x$ are dominant and $H_y = 0$, giving the equation
\beq
	\frac{\partial^2 H_x}{\partial x^2} + \frac{\partial^2 H_x}{\partial y^2} + \bigl((\omega n/c)^2 - \beta^2\bigr) H_x = 0
\eeq
and
\begin{align}
	E_x & = -\frac{1}{\omega\epsilon_0 n^2\beta} \frac{\partial^2 H_x}{\partial x\partial y}
	& E_y & = -\frac{\omega\mu_0}{\beta} H_x - \frac{1}{\omega\epsilon_0 n^2\beta} \frac{\partial^2 H_x}{\partial y^2} \nonumber \\
	E_z & = -\frac{i}{\omega\epsilon_0 n^2} \frac{\partial H_x}{\partial y}
	& H_z & = \frac{i}{\beta} \frac{\partial H_x}{\partial x} \label{eq:12-eypq}
\end{align}
In either case, the field ($H_x, H_y$) takes the form:
\beq
	\left.\begin{array}{cc} H_y & (E^x_{pq}) \\ H_x & (E^y_{pq}) \end{array}\right\} =
	\left\{\begin{array}{ll} \cos(k_x x-\phi)\cos(k_y y-\psi) & \mbox{(Region 1)} \\
		\cos(k_x a-\phi)\cos(k_y y-\psi)e^{-\gamma_x(x-a)} & \mbox{(Region 2)} \\
		\cos(k_x x-\phi)\cos(k_y d-\psi)e^{-\gamma_y(y-d)} & \mbox{(Region 3)} \end{array}\right.
\eeq
and the mirror-image regions are filled by symmetry.  The current configuration is symmetric, so the phases are given by: $\phi = (p-1)\pi/2$, $\psi = (q-1)\pi/2$, with $p, q$ positive integers.  The dispersion relation is given by:
\beq
	\beta^2 = (\omega n_1/c)^2 - k_x^2 - k_y^2
\eeq
and $k_x, k_y, \gamma_x, \gamma_y$ are obtained by solving the boundary-value conditions.  For the $E^x_{pq}$ modes:
\begin{align}
	k_x a & = (p-1)\frac{\pi}{2} + \tan^{-1}\left(\frac{n_1^2 \gamma_x}{n_0^2 k_x}\right)
	& k_x^2 + \gamma_x^2 = (\omega/c)^2(n_1^2 - n_0^2) \nonumber \\
	k_y d & = (q-1)\frac{\pi}{2} + \tan^{-1}\left(\frac{\gamma_y}{k_y}\right)
	& k_y^2 + \gamma_y^2 = (\omega/c)^2(n_1^2 - n_0^2)
\end{align}
and the fields are given by:
\begin{align}
	E_x & = \frac{\omega\mu_0}{\beta} H_y + \frac{1}{\omega\epsilon_0 n^2\beta} \frac{\partial^2 H_y}{\partial x^2}
	& E_y & = \frac{1}{\omega\epsilon_0 n^2\beta} \frac{\partial^2 H_y}{\partial x \partial y} \nonumber \\
	E_x & = \frac{i}{\omega\epsilon_0 n^2} \frac{\partial H_y}{\partial x}
	& H_z & = \frac{i}{\beta} \frac{\partial H_y}{\partial y}
\end{align}
Likewise for the $E_y^{pq}$ mode, we solve:
\begin{align}
	k_x a & = (p-1)\frac{\pi}{2} + \tan^{-1}\left(\frac{\gamma_x}{k_x}\right)
	& k_x^2 + \gamma_x^2 = (\omega/c)^2(n_1^2 - n_0^2) \nonumber \\
	k_y d & = (q-1)\frac{\pi}{2} + \tan^{-1}\left(\frac{n_1^2 \gamma_y}{n_0^2 k_y}\right)
	& k_y^2 + \gamma_y^2 = (\omega/c)^2(n_1^2 - n_0^2)
\end{align}

\begin{figure}[tbp]
\begin{center}
\includegraphics[width=0.8\textwidth]{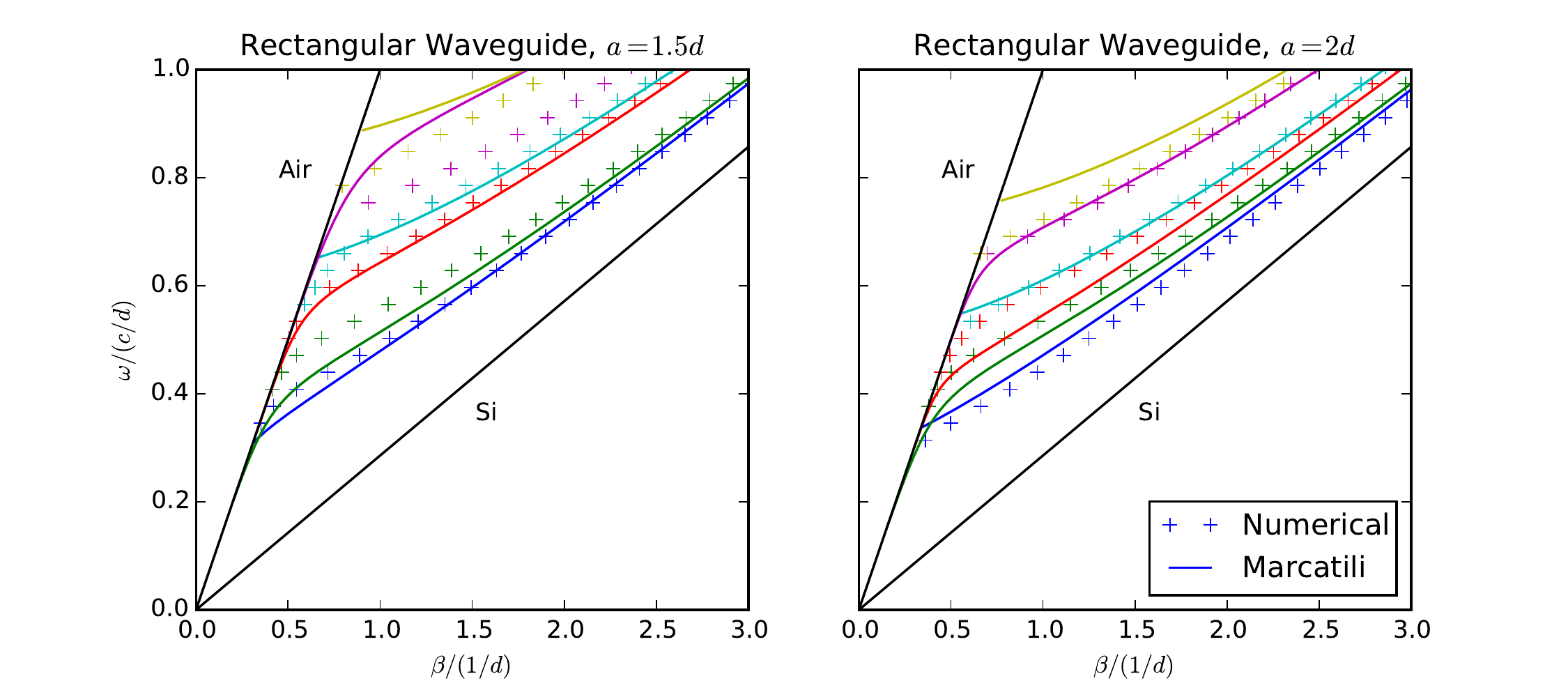}
\caption{Dispersion relation for modes in the rectangular waveguide, $n_1 = 3.5$, $n_0 = 1.0$.  Two aspect ratios are shown: $w/h = 1.5$ and $2.0$.}
\label{fig:12-f4}
\end{center}
\end{figure}

Figure \ref{fig:12-f4} plots the dispersion relation for rectangular waveguides of two aspect ratios: $w/h = 1.5$ and $2.0$.  As with the slab waveguide, all of the dimensions scale, so we plot the curves in dimensionless units, normalized to $d = h/2$.  The modes converge to the light line $\omega = \beta c$ for long wavelengths, suggesting that they are very weakly confined; in this limit the Marcatili method is not accurate.  It is most accurate when the wavelength is short and the modes are strongly confined to the waveguide.

\begin{figure}[tbp]
\begin{center}
\includegraphics[width=1.0\textwidth]{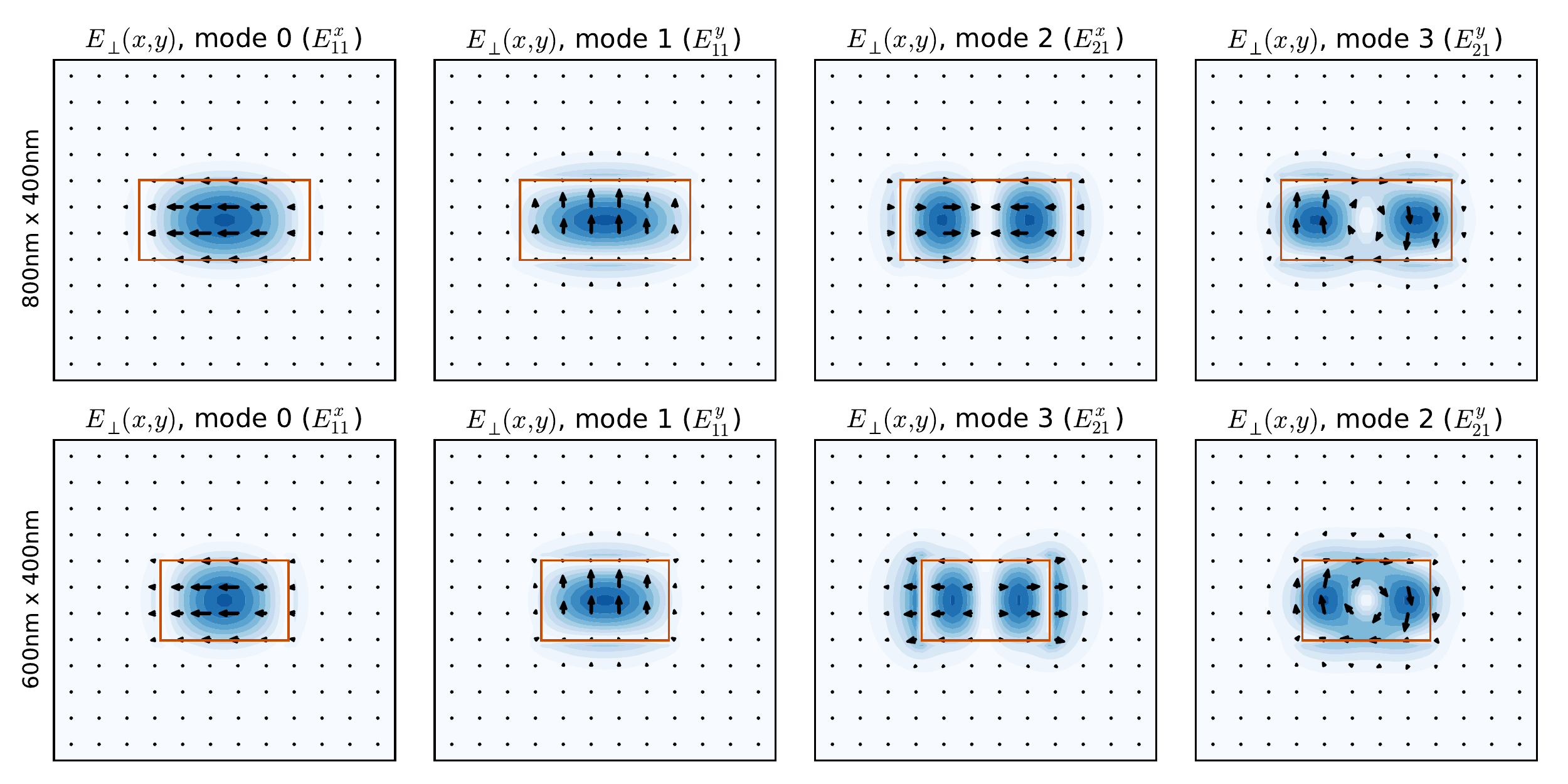}
\caption{Electric field profiles of the lowest 4 modes plotted in Fig.~\ref{fig:12-f4}, at $\omega = 0.837c/d$ (corresponds to $\lambda = 1.5\mu$m for an (800nm $\times$ 400nm) or (600nm $\times$ 400nm) waveguide).}
\label{fig:12-f5}
\end{center}
\end{figure}

The numerically-computed modes are shown in Figure \ref{fig:12-f5}.  We see that the general form -- sinusoidal in the center and exponential outside -- agrees with Marcatili's formulas.  In practice, however, Marcatili's formulas are a very poor approximation for the mode profiles, predicting very high field concentrations outside the waveguide even in the tightly-confined case, and not satisfying the boundary conditions at the surface.  The approximation becomes poorer as the wavelength is increased, a trend evident from Fig.~\ref{fig:12-f4}.

\subsection{Effective Index Method}
\label{sec:12-effind}

Often a waveguide is strongly confining in one direction but weakly confining in the other.  A good example would be a rib waveguide (Fig.~\ref{fig:12-f6}) where the etch depth $H - h$ is much smaller than the rib height $H$.  In this limit, the shape of the fields $E(x, y)$, $H(x, y)$ is only weakly dependent on $x$, suggesting a separation-of-variables solution.

To start, we consider the $E_{pq}^x$ mode, where the $H$-field points primarily along the $\hat{y}$-direction and satisfies the equation:
\beq
	\frac{\partial^2 H_y}{\partial x^2} + \frac{\partial^2 H_y}{\partial y^2} + \left[\bigl(\omega n(x,y)/c\bigr)^2 - \beta^2\right] H_y = 0 \label{eq:12-hy-efi1}
\eeq
Making the separation
\beq
	H_y(x, y) + X(x) Y(x, y)
\eeq
where $Y(x, y)$ satisfies the following eigenvalue equation
\beq
	\frac{\partial^2 Y}{\partial y^2} + (\omega/c)^2 \left[n(x, y)^2 - n_{\rm eff}(x)^2\right] Y \label{eq:12-neff-efi}
\eeq
which is the same ODE used to compute the TE mode in a slab waveguide, but with $n_{\rm eff}$ rather than $\omega$ as an eigenvalue.  The equation depends on $x$ as a parameter; solving for it allows one to compute the {\it effective index} $n_{\rm eff}(x)$.

Assuming weak $x$-confinement, the field $Y(x, y)$ should only depend weakly on $x$.  Mathematically, this means
\beq
	\left|\frac{1}{Y} \frac{\partial Y}{\partial x}\right| \ll \left|\frac{1}{X} \frac{\d X}{\d x}\right|,\ \ \ 
	\left|\frac{1}{Y} \frac{\partial^2 Y}{\partial x^2}\right| \ll \left|\frac{1}{X} \frac{\d^2 X}{\d x^2}\right| \label{eq:12-sepapprox}
\eeq

\begin{figure}[tbp]
\begin{center}
\includegraphics[width=0.7\textwidth]{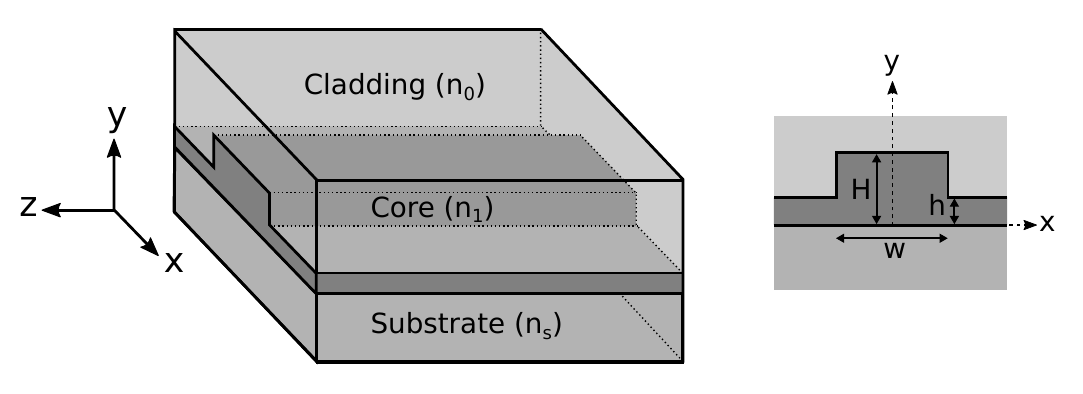}
\caption{Rib waveguide geometry.}
\label{fig:12-f6}
\end{center}
\end{figure}

\begin{figure}[tbp]
\begin{center}
\includegraphics[width=0.9\textwidth]{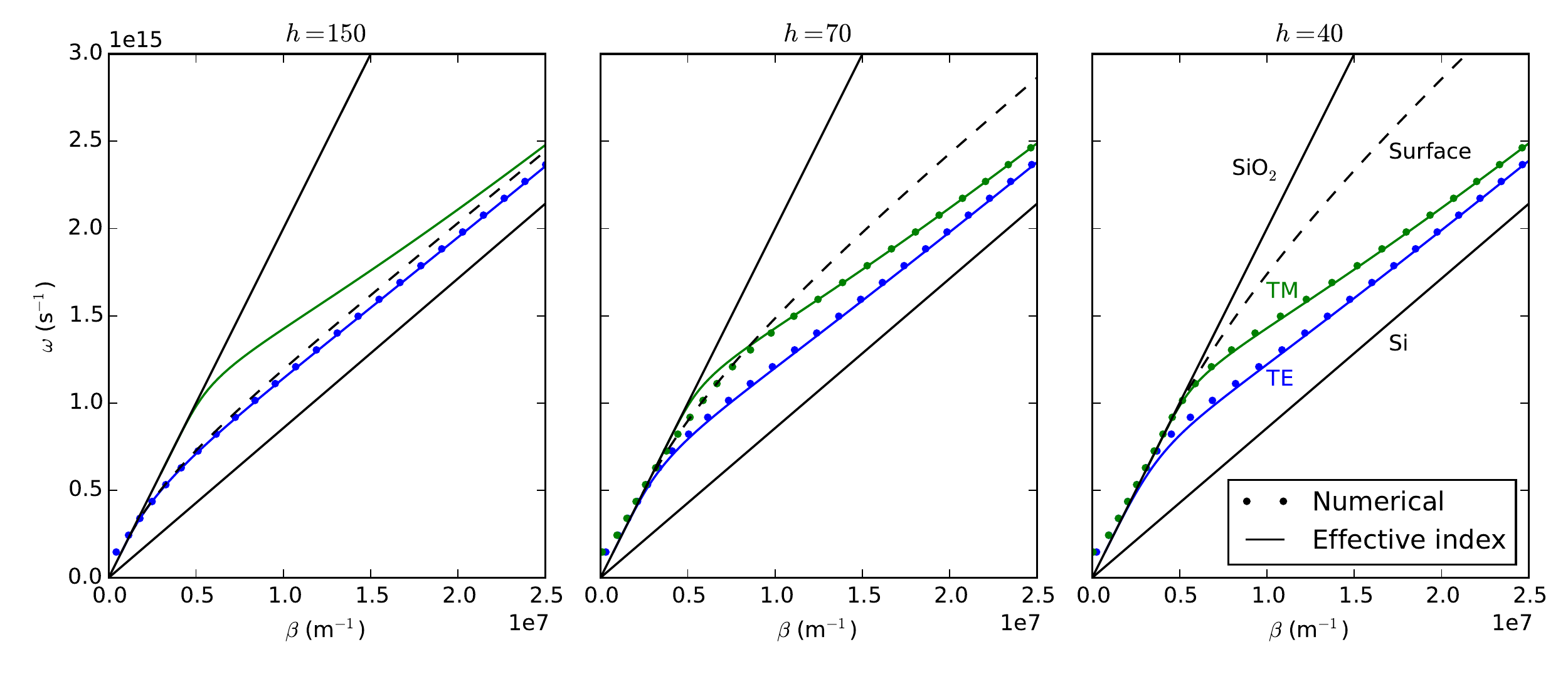}
\caption{Dispersion relation for slab waveguides with varying $h$ ($H = 220$, $w = 450$ nm).  Numerical and effective-index results plotted.}
\label{fig:12-f8}
\end{center}
\end{figure}

In this limit, by substituting Eq.~(\ref{eq:12-neff-efi}) for $Y(x, y)$ into Eq.~(\ref{eq:12-hy-efi1}), we obtain an equation for $X(x)$:
\beq
	\frac{\d^2 X}{\d x^2} + \left[(\omega n_{\rm eff}(x)/c)^2 - \beta^2\right] X = 0
\eeq
The procedure for $E_{pq}^x$ (TE-like) and $E_{pq}^y$ (TM-like) modes works as follows:

\begin{centering}
\begin{tabular}{c|p{0.35\textwidth}|p{0.35\textwidth}}
\hline\hline
Step & $E_{pq}^x$ (TE-like) & $E_{pq}^y$ (TM-like) \\ \hline
1: $Y(x, y)$, $n_{\rm eff}(x)$ 
& Solve slab waveguide equations, TE polarization.  Boundary condition: $\partial Y/\partial y$ is continuous.
& Solve slab waveguide equations, TM polarization.  Boundary condition: $n^{-2} \partial Y/\partial y$ is continuous. \\
2: $X(x)$ 
& Solve slab waveguide equations, TM polarization.  Boundary condition: $n_{\rm eff}^{-2} \d X/\d x$ is continuous.
& Solve slab waveguide equations, TE polarization.  Boundary condition: $\d X/\d x$ is continuous. \\
3: $\vec{E}$, $\vec{H}$
& Get $H_y = X(x)Y(x, y)$.  Compute $\vec{E}$, $\vec{H}$ using Eqs.~(\ref{eq:12-expq}).
& Get $H_x = X(x)Y(x, y)$.  Compute $\vec{E}$, $\vec{H}$ using Eqs.~(\ref{eq:12-eypq}). \\ \hline\hline
\end{tabular}
\end{centering}

While the effective index method is designed for systems with weak horizontal confinement, where $h \approx H$, it gives reasonable answers when $h$ is much smaller.  However, in the limit $h \rightarrow 0$, the $n_{\rm eff}(x)$ may be undefined outside the waveguide because $\omega$ is below the cutoff frequency for a slab of width $h$.  In this case, the best approximation is to assume the wave is confined to the substrate and set $n_{\rm eff} = n_s$.

\begin{figure}[tbp]
\begin{center}
\includegraphics[width=0.8\textwidth]{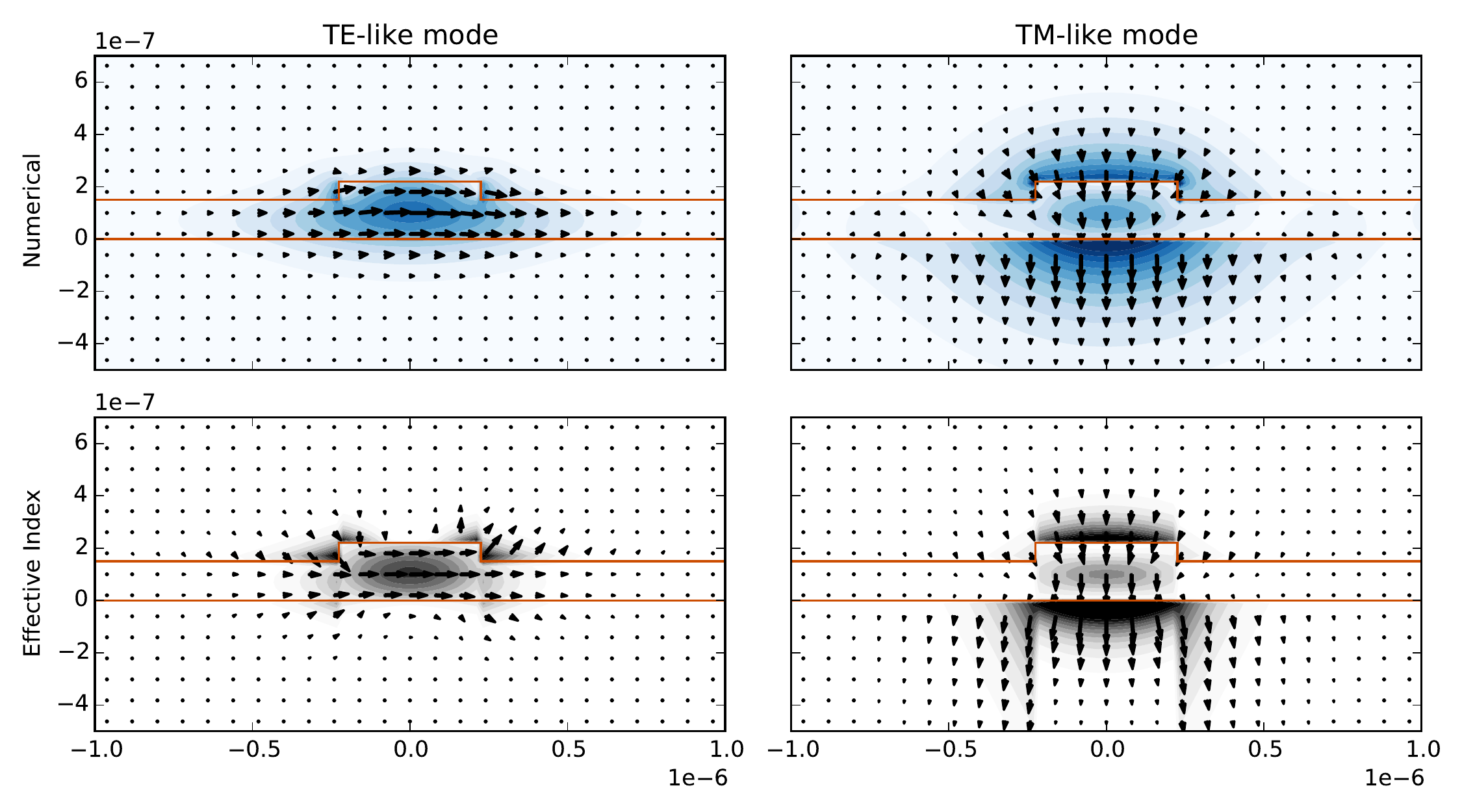}
\caption{Top: TE-like and TM-like modes for rib waveguide with $H = 220$ nm, $h = 150$ nm, $w = 450$ nm.}
\label{fig:12-f7}
\end{center}
\end{figure}

Figure \ref{fig:12-f8} gives the dispersion relation for rib waveguides of varying side heights.  One finds good agreement with the effective index method for the TE and TM modes, even when the ratio $H/h$ is large and the approximation (\ref{eq:12-sepapprox}) breaks down.

In Fig.~\ref{fig:12-f8}, the guided modes lie between the silicon line $\beta = n_1 \omega/c$ and the glass line $\beta = n_s \omega/c$ because the index is always $\leq n_{\rm Si}$, and modes with $\beta < n_s \omega/c$ can leak out through the substrate.  But in the rib waveguide, light can also leak out along the surface, provided that the slab of width $h$ supports a guided mode.  This gives the dashed line in Fig.~\ref{fig:12-f8}.  For large $h$, this line blocks off all modes except the first-order TE-like; only the TE mode is bound.  But for smaller $h$, this line is less restrictive, and both TE and TM modes are bound.

While the dispersion relation is reasonably accurate, the mode profiles one computes are not particularly accurate.  Figure~\ref{fig:12-f7} plots the TE-like and TM-like modes for a $450\times 220$ nm waveguide, with $h = 150$.  While the field inside the waveguide is reasonably accurate, the external field is rather poorly shaped.  However, as the field profile is usually used to compute nonlinear coefficients, which depend only on the interior field, this should not be an issue.

\section{Group Velocity and Dispersion}

Solving for the waveguide modes allows us to reduce the 3+1-dimensional Maxwell equations (\ref{eq:12-maxwell}) to a set of 1+1-dimensional equations for the field $a(z, t)$.  The geometry of the waveguide is distilled into four relevant parameters: the phase-velocity $v_p$, group velocity $v_g$, dispersion $\beta_2$, and nonlinear index $\gamma$ \cite{Lin2007}.

These constants depend on the waveguide's material and, more importantly, its geometry.  Most notably, by choosing the correct waveguide dimensions, one can tailor the dispersion relation of the waveguide.  Such ``dispersion engineering'' has been applied to silicon ridge waveguides \cite{Turner2006}, Si$_3$N$_4$ ridge waveguides \cite{Tan2010} and microresonators \cite{Okawachi2011, Riemensberger2012}, photonic-crystal waveguides \cite{Saynatjoki2007}, and microstructured optical fibers \cite{Ouzounov2001, Knight2000}.  One can use waveguide dispersion to enhance, cancel or invert the material dispersion.

It is useful to express the phase and group velocity in terms of their phase and group indices: $n_{\rm eff} = (\omega/c)/\beta$, $n_g = \beta_1 c$.  The group velocity dispersion is conveniently stated in units of ps$^2$/m; for reference a standard SMF28e fiber has $\beta_2 = -0.016\mbox{ps}^2/\mbox{m}$.

\begin{figure}[p]
\begin{center}
\includegraphics[width=0.90\textwidth]{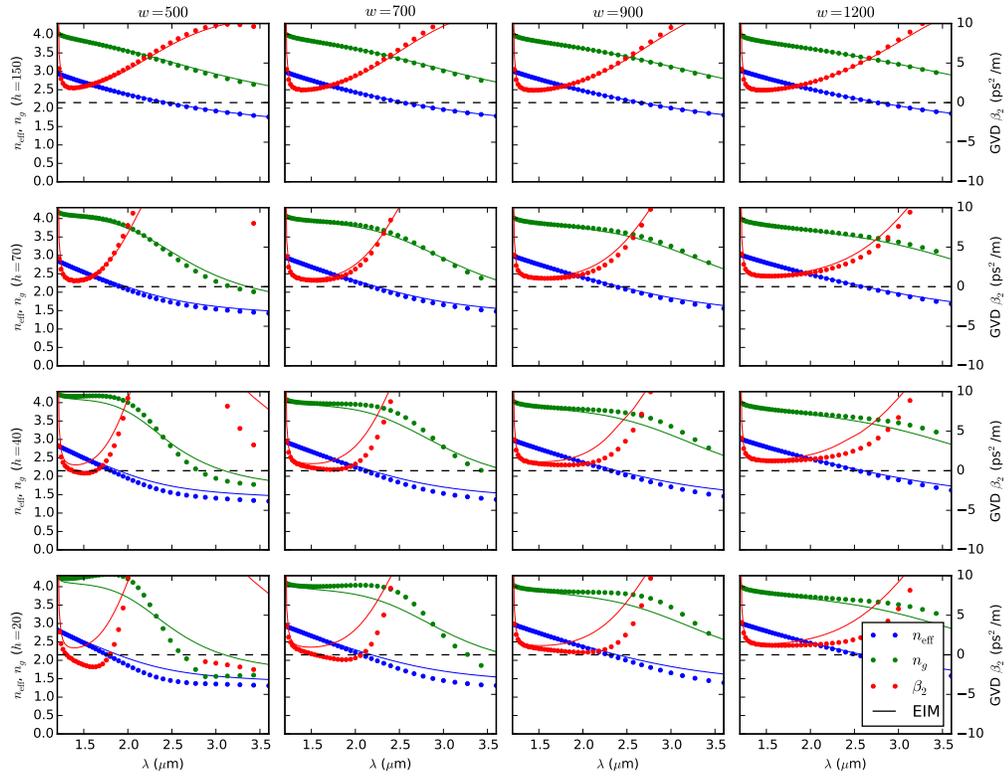}
\caption{Mode index $n_{\rm eff}$, group index $n_g$ and GVD $\beta_2$ for SOI waveguide with $H = 220$.}
\label{fig:12-f16}
\end{center}
\end{figure}

\begin{figure}[p]
\begin{center}
\includegraphics[width=0.90\textwidth]{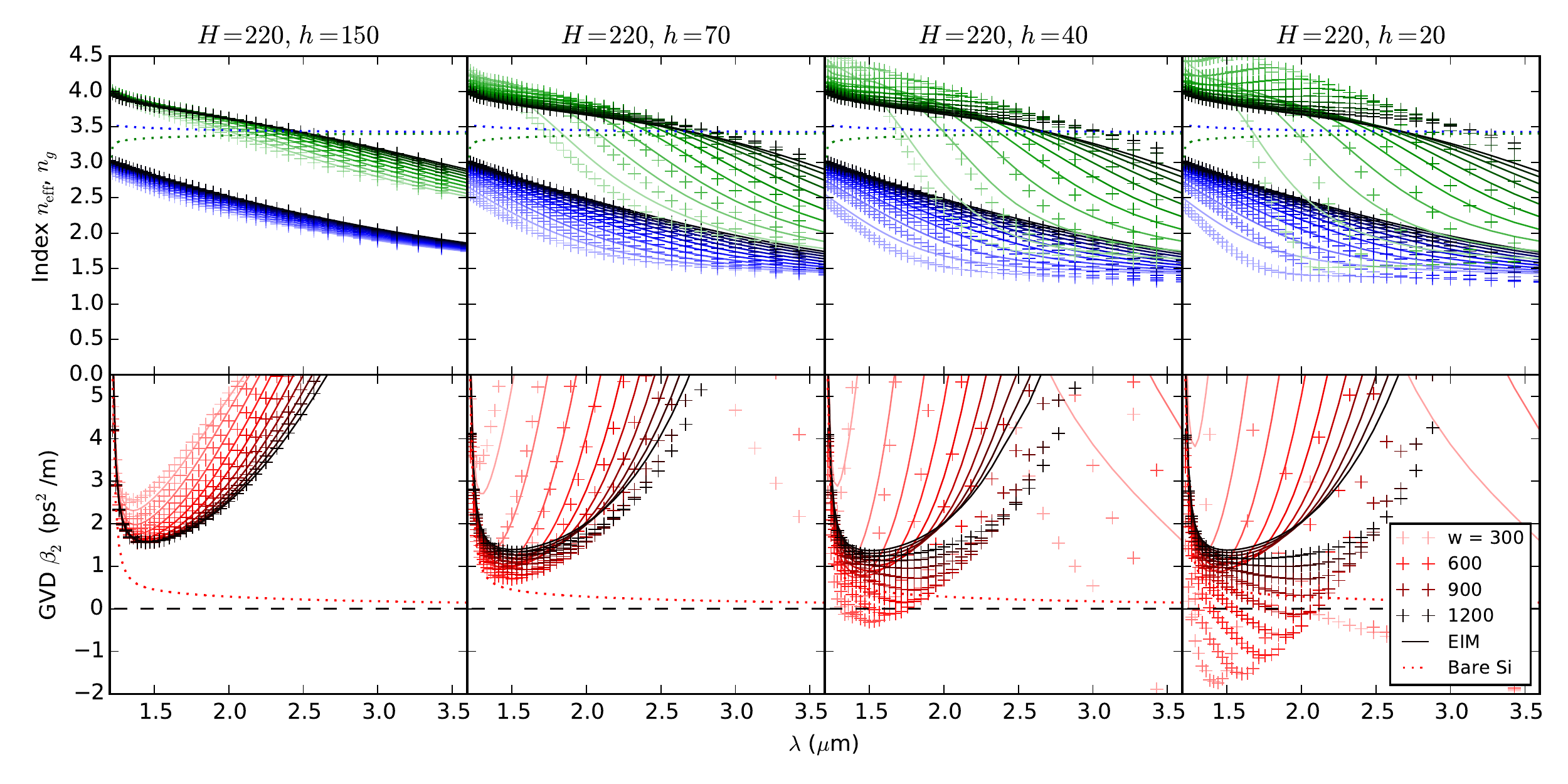}
\caption{Mode index, group index and GVD for SOI rib waveguides.  Bulk silicon values also given.}
\label{fig:12-f17}
\end{center}
\end{figure}

Figs.~\ref{fig:12-f16}-\ref{fig:12-f17} plot the waveguide phase index, group index, and GVD as a function of wavelength for a number of geometries, in the fundamental TE mode.  The top two rows of Fig.~\ref{fig:12-f16} correspond to rib waveguides that can be fabricated at IMEC \cite{Lim2014}: a rib height $H = 220$ with sides $h = 150$ or $70$.  As noted in Sec.~\ref{sec:12-effind}, the effective index method works best when the field is only weakly $x$-dependent -- equivalently, if $H \approx h$.  Thus, it is not surprising to see that effective index theory agrees well with numerical results for the $h = 150$ waveguides.  For $h = 70$ the agreement is still good, although there is some deviation.

For $h = 40$ and $h = 20$ (which cannot presently be fabricated at IMEC), the disagreement becomes more pronounced.  In particular, numerical simulations predict a small range of wavelengths, around 1.8$\mu$m, with anomalous dispersion, while effective index theory predicts normal dispersion for all wavelengths.  It is notable that one can achieve anomalous dispersion in such a thin waveguide, as waveguides used for anomalous dispersion are typically much thicker \cite{Turner2006, Tan2010}.  One should keep in mind, however, that the $h = 20$ or $40$ sidewalls are not currently available at IMEC or many other foundries.

\begin{figure}[tbp]
\begin{center}
\includegraphics[width=1.00\textwidth]{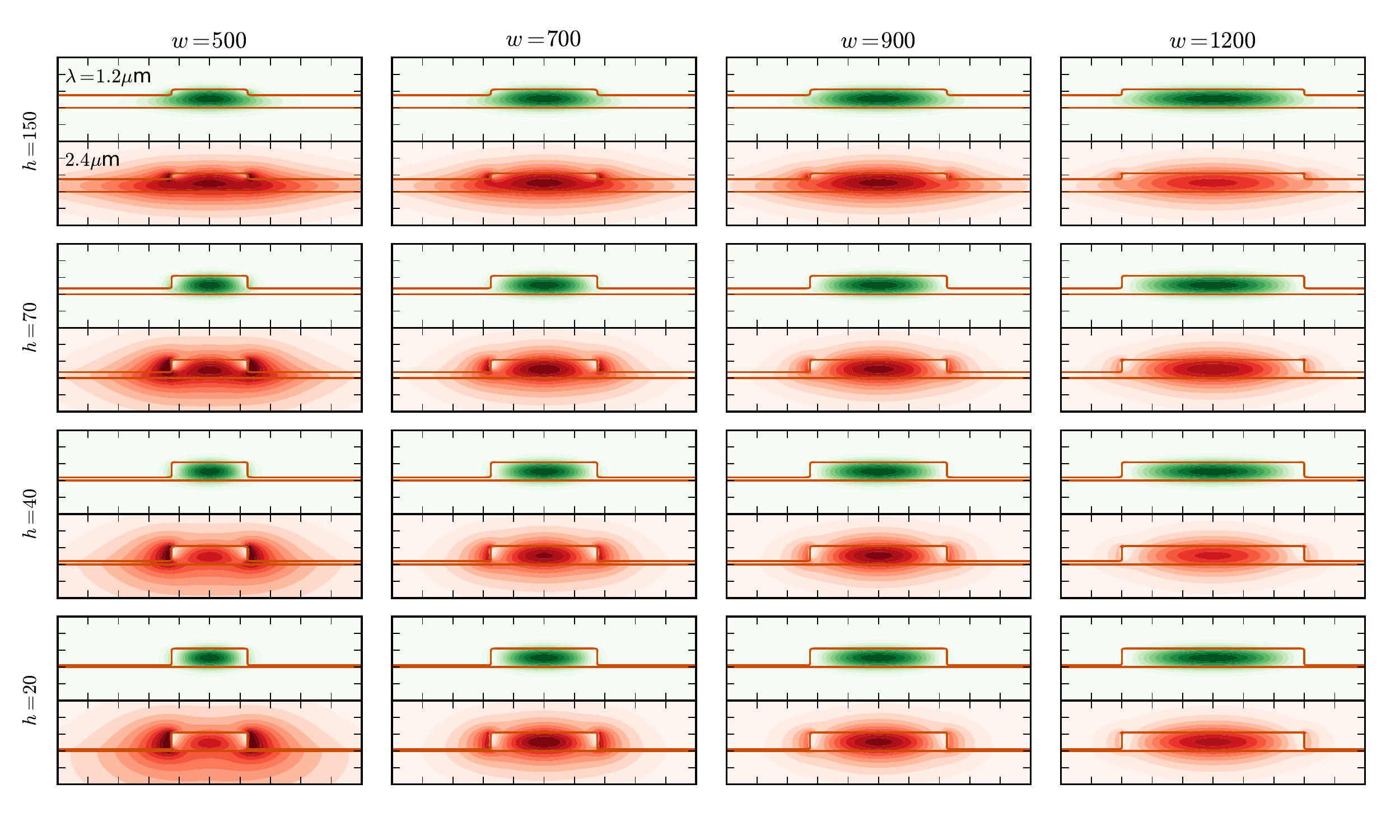}
\caption{Electric field profiles for waveguides shown in Fig.~\ref{fig:12-f16}, sampled at $\lambda = 1.2\mu$m and $2.4\mu$m.}
\label{fig:12-f18}
\end{center}
\end{figure}

Several qualitative features from Figs.~\ref{fig:12-f16}-\ref{fig:12-f17} catch the eye.  First, unlike most materials or weakly-guided optical fibers, the phase and group index differ by $O(1)$ (typically the difference is very small).  This fact indicates that the waveguide dispersion is much more significant than material and fiber dispersion.  

Both the phase and group index tend to increase with decreasing wavelength.  This is due to the increased confinement of light to the high-index silicon core (Fig.~\ref{fig:12-f18}).  The effect is much more pronounced for the smaller waveguides.  However, for the waveguides with thin sidewalls, $n_g$ reaches a maximum and starts decreasing.  From the relation
\beq
	\beta_2 = \frac{1}{c} \frac{dn_g}{d\omega}
\eeq
we can infer that this is the region with anomalous dispersion.

\section{Waveguide Loss}

There are three main loss mechanisms in silicon waveguides, discussed in the sections below:

\begin{enumerate}
	\item Optical absorption -- Si absorbs in the visible and near-IR ($\lambda < 1.1\mu$m), and SiO$_2$ absorbs at longer wavelengths ($\lambda \gtrsim 5\mu$m), giving a device transmission window of 1.1--5$\mu$m.  Free carriers add absorption for all wavelengths, roughly going as $\alpha \sim C_{e,h} n_{e,h} \lambda^2$ in the near IR.
	\item Substrate Loss -- SOI wafers consist of silicon structures on top of a thin (1--3$\mu$m) SiO$_2$ substrate.  Light can leak through the substrate into the bulk silicon beneath.  The rate is exponential in substrate length, approximately $\alpha/\mbox{cm}^{-1} = \exp\bigl[10.6 - 37.9(S/\mu\mbox{m})(\lambda/\mu\mbox{m})^{-1.5}\bigr]$ for a 220-nm slab waveguide.
	\item Scattering Loss -- due to surface roughness, can be calculated using the {\it Payne-Lacey model}, which treats surface defects as antennas and calculates the radiated power \cite{Lacey1990, Payne1994}.
\end{enumerate}

\subsection{Optical Absorption}

\begin{figure}[tbp]
\begin{center}
\includegraphics[width=0.85\textwidth]{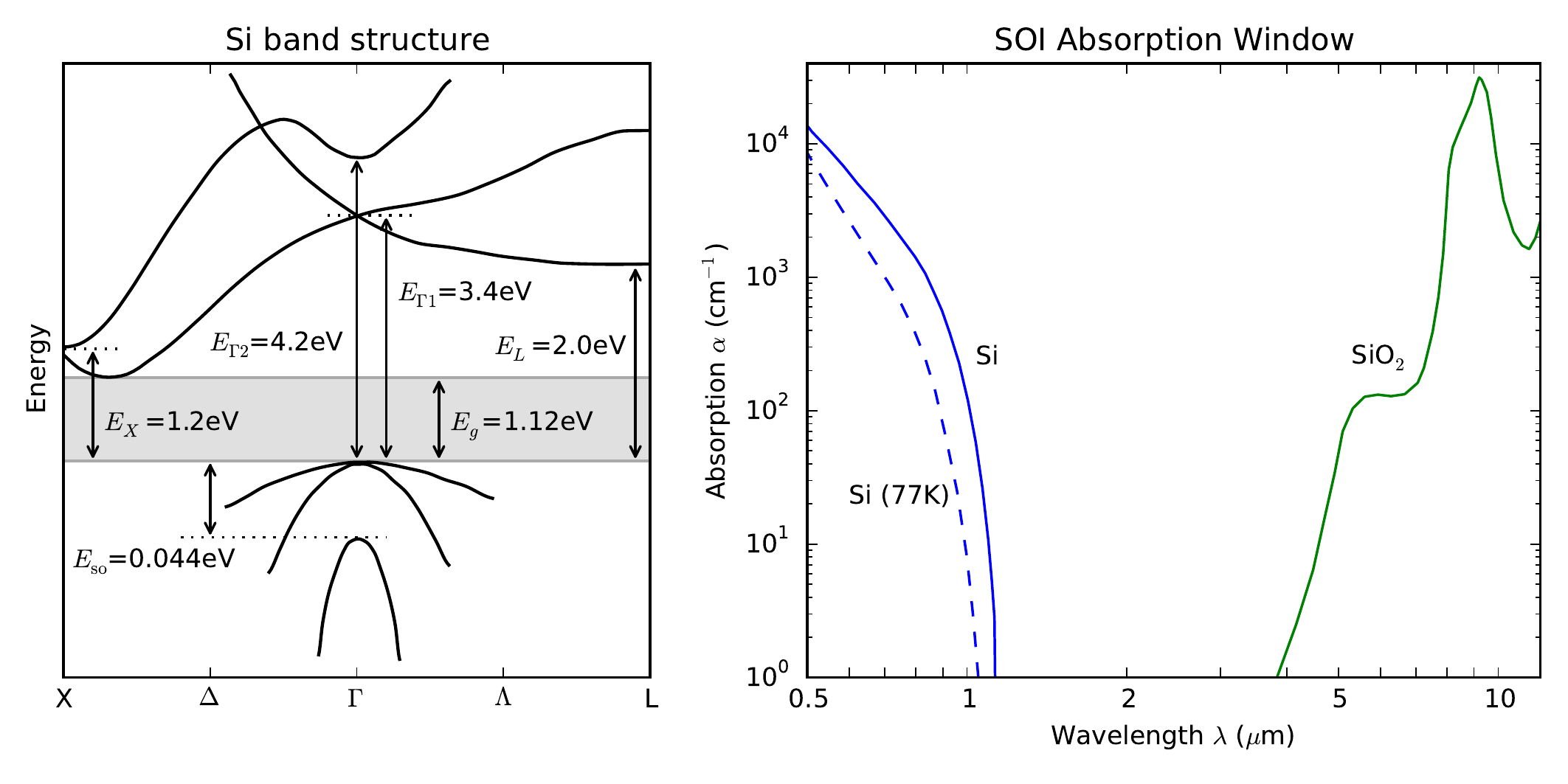}
\caption{Left: electronic band structure of silicon \cite{LundstromBook}.  Right: transmission window of SOI devices, limited by optical absorption in silicon core \cite{SzeBook} and SiO$_2$ substrate \cite{Kitamura2007}.}
\label{fig:12-f9}
\end{center}
\end{figure}

Loss in optical waveguides is limited by absorption in the silicon core and SiO$_2$ substrate.  Silicon has an indirect band gap $E_g = 1.12 eV$ associated with phonon-mediated transitions from the $\Gamma$- and $X$-points (Fig.~\ref{fig:12-f9}).  This causes absorption for $\lambda < 1.10\mu$m.  Note, however, that because of the indirect band gap, the absorption coefficient is not particularly large: the extinction length at 780 nm is around 5$\mu$m, increasing to 100$\mu$m at $\lambda = 1\mu$m.  Absorption is significantly reduced at cryogenic temperatures.

In the mid- and far-IR, there is significant absorption in silica due to Si--O vibrational modes.  This limits the wavelength to under 4--5$\mu$m \cite{Kitamura2007}.  

Initially, researchers were skeptical of mid-IR photonics in SOI because of the strong absorption features seen in silica fibers, attributed to Si--O harmonics and O--H stretching in impurities \cite{Izawa1977, Osanai1976, Mashanovich2011}.  However, fibers generally travel kilometers whereas a typical SOI device is at most centimeters long; moreover, the silica used in SOI is sufficiently high purity that devices can be transparent all the way out to 4--5$\mu$m.  In recent years, devices at 2$\mu$m \cite{Zlatanovic2010}, 3.3$\mu$m \cite{Mashanovich2011, Milovsevic2009}, and 4.4$\mu$m \cite{Shankar2011}, evidence that the full transparency window in Fig.~\ref{fig:12-f9} can be accessed.

If we operate in the regime $1.2\mu{\rm m} < \lambda < 4.5\mu{\rm m}$, then, material absorption will be negligible.  On the other hand, certain devices we might want to make (for example: up-conversion detectors, OPOs with far-IR idler fields) involve light in the absorbing regions.  To calculate the guided-wave absorption exactly, we use Eq.~(\ref{eq:12-wg-abs}) to decompose it into a weighted sum of the cladding, core and substrate absorption coefficients:
\beq
	\alpha = \frac{n_g}{n_{g,0}} \Gamma_0 \alpha_0 
		+ \frac{n_g}{n_{g,1}} \Gamma_1 \alpha_1 
		+ \frac{n_g}{n_{g,s}} \Gamma_s \alpha_s \label{eq:12-alphamat}
\eeq
where $n_g$ is the guided-mode group velocity, $n_{g,0}, n_{g,1}, n_{g,s}$ are the cladding, core and substrate (bulk) group velocities, $\alpha_0, \alpha_1, \alpha_s$ are the absorption coefficients, and $\Gamma_0, \Gamma_1, \Gamma_s$ are filling factors that sum to one (Eq.~(\ref{eq:12-dni})):
\beq
	\Gamma_{\textbf{r}} = \frac{\int_{R_{\textbf{r}}}{n n_g |E|^2\d A}}{\int{n n_g |E|^2\d A}}
\eeq
For tightly confined modes, typically $\Gamma_1 \approx 1$ and $\Gamma_0, \Gamma_s \ll 1$, reducing the effect of cladding and substrate absorption.  On the other hand, for $\lambda \gtrsim 4.5$ when substrate absorption sets in, the mode will be less tightly confined and the value of $\Gamma_s$ matters.  Figure \ref{fig:12-f14} plots the $\Gamma_{\textbf{r}}$ as a function of waveguide width and wavelength for a rib waveguide, $H = 220$, $h = 70$.  For most waveguide widths at $\lambda < 3\mu$m, a majority of the power is confined to the silicon.  However, for $\lambda \gtrsim 3\mu$m, a significant fraction ($\gtrsim 10\%$) leaks into the substrate.

\begin{figure}[tbp]
\begin{center}
\includegraphics[width=1.00\textwidth]{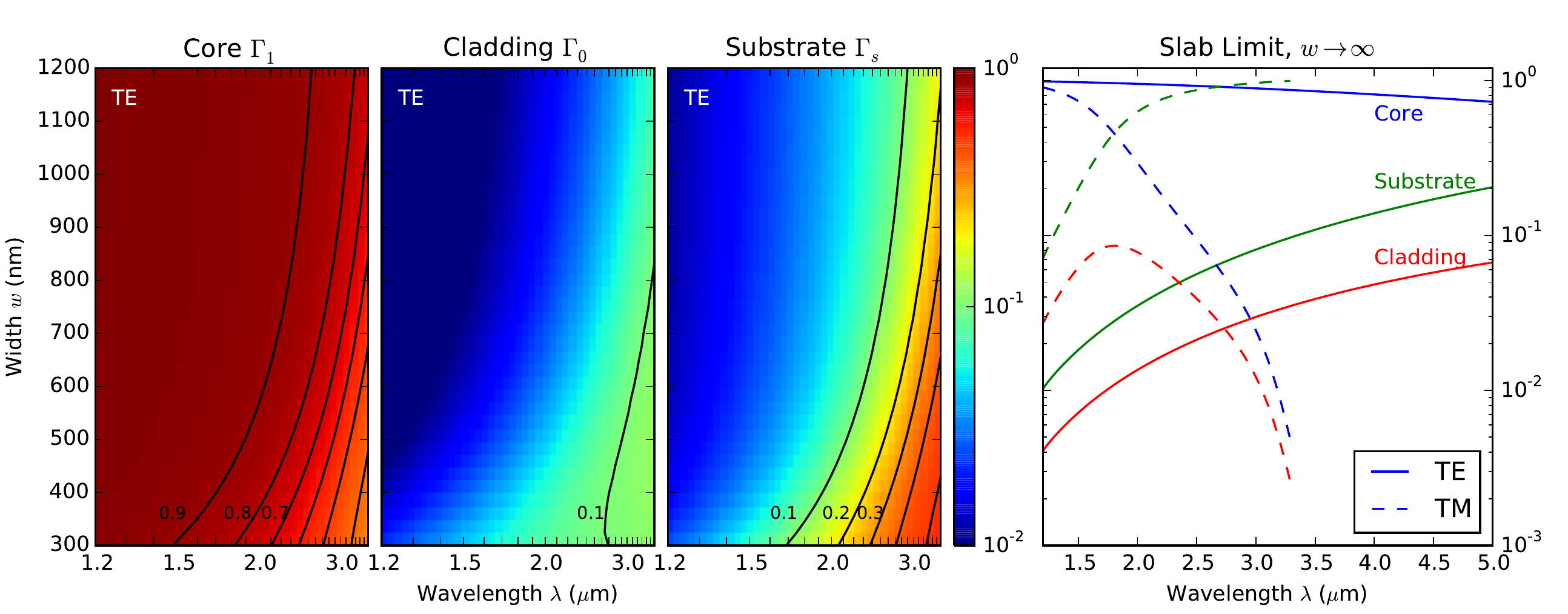}
\caption{Left: filling factors $\Gamma_1$, $\Gamma_0$, $\Gamma_s$ for core, cladding and substrate of the rib waveguide ($H = 220, h = 70$), TE-like mode.  Right: values in the slab-waveguide limit $w \rightarrow \infty$, for both TE and TM modes.}
\label{fig:12-f14}
\end{center}
\end{figure}

The TM mode is much less confined than the TE.  The right plot of Fig.~\ref{fig:12-f14} compares TE and TM filling factors for the slab waveguide ($w \rightarrow \infty$).  While the TE mode is strongly confined for all $\lambda \lesssim 5\mu$m, for the TM mode, a majority of the power leaks into the substrate at 1.8$\mu$m, and at 3.2$\mu$m the mode becomes unbound.  To better confine the TM mode, we would need a larger slab; since the eigenvalue equations are approximately scale-invariant (if the index is slowly varying in $\lambda$), doubling the slab height correspondingly doubles the wavelengths it can confine.  Thick slabs, unavailable at facilities such as IMEC, must be used to confine TM modes in the mid-IR.

\subsection{Free-carrier absorption}

It has long been known that doped silicon absorbs light below its band gap, and this absorption increases with doping \cite{Spitzer1957, Hara1966}.  This {\it free-carrier absorption} is typically explained using the Drude model, in which the dielectric constant has the plasma oscillation peak \cite{Schroder1978, YuBook}:
\beq
	\epsilon(\omega) = \epsilon_\infty - \frac{N_e e^2}{\epsilon_0 m_e (\omega^2 + i\omega\gamma_e)} - \frac{N_h e^2}{\epsilon_0 m_h (\omega^2 + i\omega\gamma_h)}
\eeq
where $m_{e,h}, \gamma_{e,h}$ are the carrier mass and relaxation time constant (electrons and holes have different values).  Because of the finite relaxation time, plasma oscillations are dissipative, leading (in the limit $\omega \gg \omega_p, \gamma$ generally applicable to semiconductors) to free-carrier absorption:
\beq
	\alpha(\omega) = \frac{e^2}{n(\omega) c\epsilon_0} \frac{1}{\omega^2} \left[\frac{\gamma_e}{m_e} N_e + \frac{\gamma_h}{m_h} N_h\right] \label{eq:12-fca}
\eeq
From Eq.~(\ref{eq:12-fca}), we expect the absorption to vary linearly with carrier density, and scale as $\lambda^2$.  Thus, free-carrier absorption should be more significant for longer wavelengths.  Because carriers in silicon are not an ideal plasma, Eq.~(\ref{eq:12-fca}) is only qualitatively correct.  Fig.~\ref{fig:12-f11} plots the absorption coefficient for a range of wavelengths and doping densities, comparing to the Drude-model result $\alpha \sim N \lambda^2$.  While $p$-type silicon has the Drude-model wavelength dependence, $n$-type silicon does not, exhibiting a ``plateau'' at 2--5$\mu$m.

\begin{figure}[t!]
\begin{center}
\includegraphics[width=1.00\textwidth]{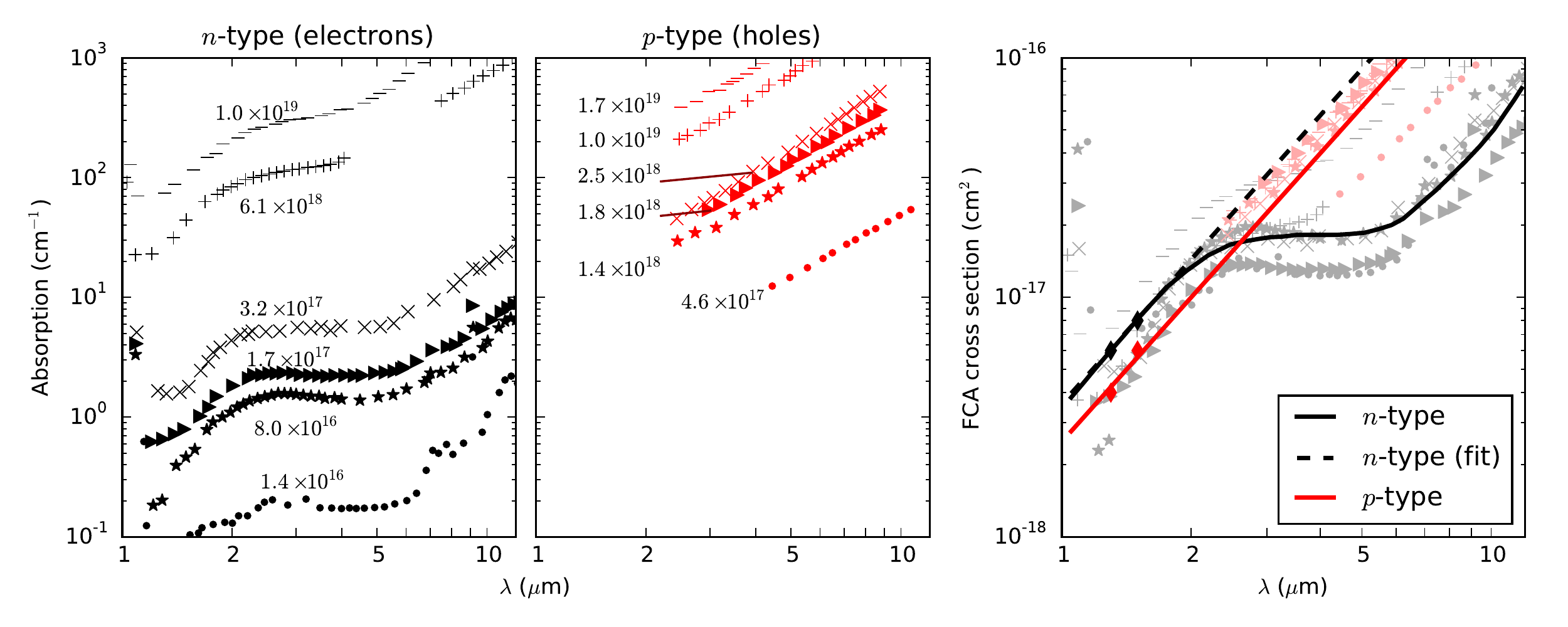}
\caption{Left: measure free-carrier absorption as a function of wavelength and doping \cite{Spitzer1957, Hara1966, Soref1987}.  Right: calculated absorption cross section and Drude-model extrapolation \cite{YuBook}.}
\label{fig:12-f10}
\end{center}
\end{figure}

\begin{figure}[t!]
\begin{center}
\includegraphics[width=0.80\textwidth]{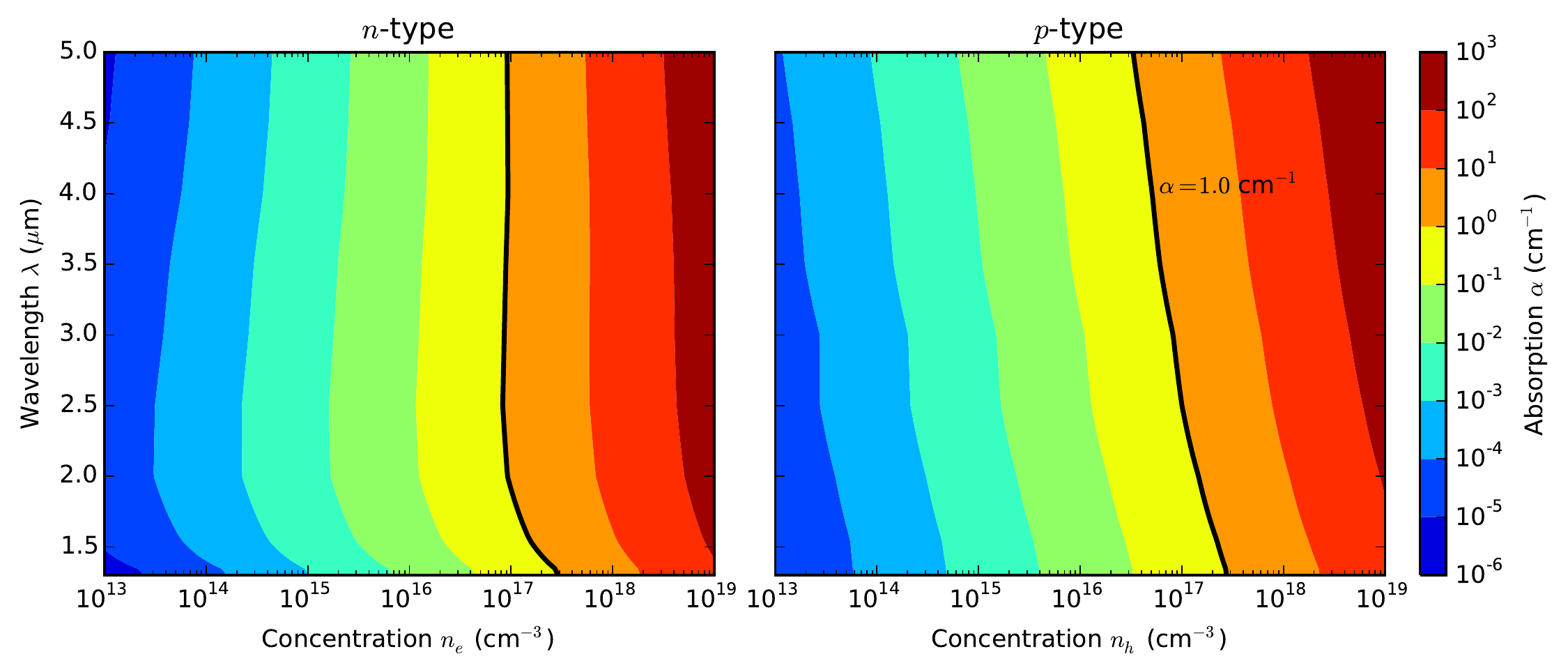}
\caption{Absorption coefficient of doped silicon as function of doping and wavelength.  Model parameters from Ref.~\cite{Nedeljkovic2011}}
\label{fig:12-f11}
\end{center}
\end{figure}

A more accurate power-law formula, proposed by Soref and Bennett for 1.3- and 1.55-$\mu$m \cite{Soref1987}, has become commonplace in the literature.  Recently it has been extended to the full transparency window of silicon, $1.2\mu{\rm m} < \lambda < 14\mu{\rm m}$ \cite{Nedeljkovic2011}:
\beq
	\alpha = a N_e^b + c N_h^d
\eeq
where the constants $a, b, c, d$ are wavelength-dependent \cite[Table 1]{Nedeljkovic2011}.

Figure \ref{fig:12-f11} plots the absorption as a function of both wavelength and carrier concentration, with the value $\alpha = 1.0\mbox{cm}^{-1}$ shown in black.  Typical SOI waveguides have 2--4 dB/cm loss ($\alpha = 0.5$--$1.0$cm$^{-1}$), \cite{Vlasov2004, Lee2000}, although smaller values around 0.5 dB/cm ($\alpha = 0.14$cm$^{-1}$) have been reported \cite{Mashanovich2011}.  To the left of the solid line in Fig.~\ref{fig:12-f11}, free-carrier absorption is weak compared to other waveguide loss terms.  To the right of the line, it dominates.

Note that, as with material absorption, the actual waveguide free-carrier absorption is multiplied by a filling factor $\Gamma$ as in Eq.~(\ref{eq:12-alphamat}).

\subsection{Substrate Loss}
\label{sec:12-leak}

The oxide layer in SOI has a finite thickness, typically around 1--3$\mu$m.  Because the wafer lying underneath the oxide is also silicon, light can leak out from the waveguide through the oxide layer (Fig.~\ref{fig:12-f15}).  In practice, the leakage rate is proportional to the waveguide mode's evanescent tail at the bottom of the oxide layer, which is exponentially small for near-IR light if the oxide thickness is $S \gtrsim 1\mu$m.  For longer wavelengths, the evanescent tail penetrates much deeper into the oxide layer, so substrate leakage is more problematic.

\begin{figure}[b!]
\begin{center}
\includegraphics[width=0.9\textwidth]{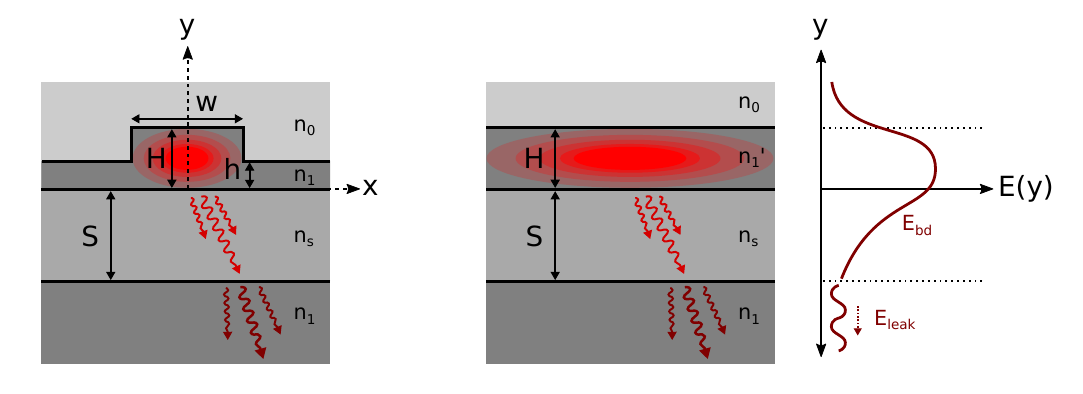}
\caption{Leakage from a rib waveguide through the oxide layer (left), and the equivalent slab waveguide (center).  Transverse field profile (right).}
\label{fig:12-f15}
\end{center}
\end{figure}

Mode solvers find leaky modes using perfectly-matched layers (PMLs) \cite{Berenger1994}.  Analytically, one can obtain good estimates for the leakage rate using effective index theory to convert the rib waveguide to the equivalent slab, and finding the slab loss rate by the transfer matrix method \cite{Bienstman2006, Ghatak1987}.

The transfer-matrix method gives exact solutions for 1D structures with a piecewise constant index of refraction.  The structure is divided into regions of constant index, and in each region, the transverse field ($\Psi = E$ for TE, $H$ for TM) can be expressed as:
\beq
    a^+ e^{iks} + a^- e^{-iks}
\eeq
(For evanescent fields, $k = -i\kappa$ is imaginary, so the field goes as $a^+ e^{\kappa s} + a^- e^{-\kappa s}$).  At boundaries, $\Psi$ is continuous.  The derivative boundary condition depends on polarization: $d\Psi/ds$ is continuous at the boundary for TE modes, $n^{-2} d\Psi/ds$ for TM.  Using these boundary conditions, we can relate the $a^\pm$ on the right side to the left:
\bea
    \begin{bmatrix} a^+_2 \\ a^-_2 \end{bmatrix}_{\rm TE} & = & \underbrace{\frac{1}{2} \begin{bmatrix} 1 + (k_1/k_2) & 1 - (k_1/k_2) \\ 1 - (k_1/k_2) & 1 + (k_1/k_2) \end{bmatrix}}_{T_{21}} \begin{bmatrix} a^+_1 \\ a^-_1 \end{bmatrix}_{\rm TE} \nonumber\\
    \begin{bmatrix} a^+_2 \\ a^-_2 \end{bmatrix}_{\rm TM} & = & \underbrace{\frac{1}{2} \begin{bmatrix} 1 + (n_2^2 k_1/n_1^2 k_2) & 1 - (n_2^2 k_1/n_1^2 k_2) \\ 1 - (n_2^2 k_1/n_1^2 k_2) & 1 + (n_2^2 k_1/n_1^2 k_2) \end{bmatrix}}_{T_{21}} \begin{bmatrix} a^+_1 \\ a^-_1 \end{bmatrix}_{\rm TM}
\eea
Passing from the left side of a region to the right simply changes the phase of the waves:
\beq
    \begin{bmatrix} a^+_i \\ a^-_i \end{bmatrix}_{\rm TE/TM} \rightarrow \underbrace{\begin{bmatrix} e^{ik_i L_i} & 0 \\ 0 & e^{-ik_i L_i} \end{bmatrix}}_{T_{i}} \begin{bmatrix} a^+_i \\ a^-_i\end{bmatrix}_{\rm TE/TM}
\eeq
Start at the top of the waveguide and work down (thus $s = -y$ in our usual coordinates).  Regions 1, 2, 3 and 4 are the air cladding, silicon core, SiO$_2$ substrate, and silicon wafer.  In the absence of the wafer layer (Region 4), there is no leakage from the slab.  Modes are found by solving for the boundary condition $\Psi(s) \rightarrow 0$ as $|s| \rightarrow \infty$.  Thus for region 1, $a_1^- = 0$ (set $a_1^+ = 1$ for convention).  In region 3, $a_3^+ = 0$.  However, we can relate $a_3^\pm$ to $a_1^\pm$ using the transfer matrices:
\beq
    \begin{bmatrix} a_3^+ \\ a_3^- \end{bmatrix} = T_{32} T_2 T_{21} \begin{bmatrix} 1 \\ 0 \end{bmatrix} 
\eeq
Setting $a_3^+ = 0$ constrains $\beta$, since it depends on $\beta$ through the transfer matrices.  This gives the same solutions found in Sec.~\ref{sec:12-slab}.

Once the modes is computed in the absence of a wafer layer, we can use the transfer matrix approach to compute the substrate leakage loss.  First, fields $a_4^\pm$ in the silicon wafer are computed:
\beq
	\begin{bmatrix} a_4^+ \\ a_4^- \end{bmatrix} = T_{43} T_3 T_{32} T_2 T_{21} \begin{bmatrix} 1 \\ 0 \end{bmatrix} = T_{43} \begin{bmatrix} e^{\kappa S} a_3^+ \\ e^{-\kappa S} a_3^- \end{bmatrix}
\eeq
Since the field goes as $e^{i\beta z}$, $a_4^+$ is the incoming wave, while $a_4^-$ is outgoing.  The correct boundary condition is $a_4^+ = 0$, which gives (for TE modes):
\beq
	(1 - i P\xi/\kappa) a^{\kappa S} a_3^+(\beta) + (1 + i P\xi/\kappa) a^{\kappa S} a_3^-(\beta) = 0 \label{eq:12-bdry-leak}
\eeq
where $P = 1$ for TE modes and $P = n_1^2/n_s^2$ for TM modes.

\begin{figure}[b!]
\begin{center}
\includegraphics[width=0.6\textwidth]{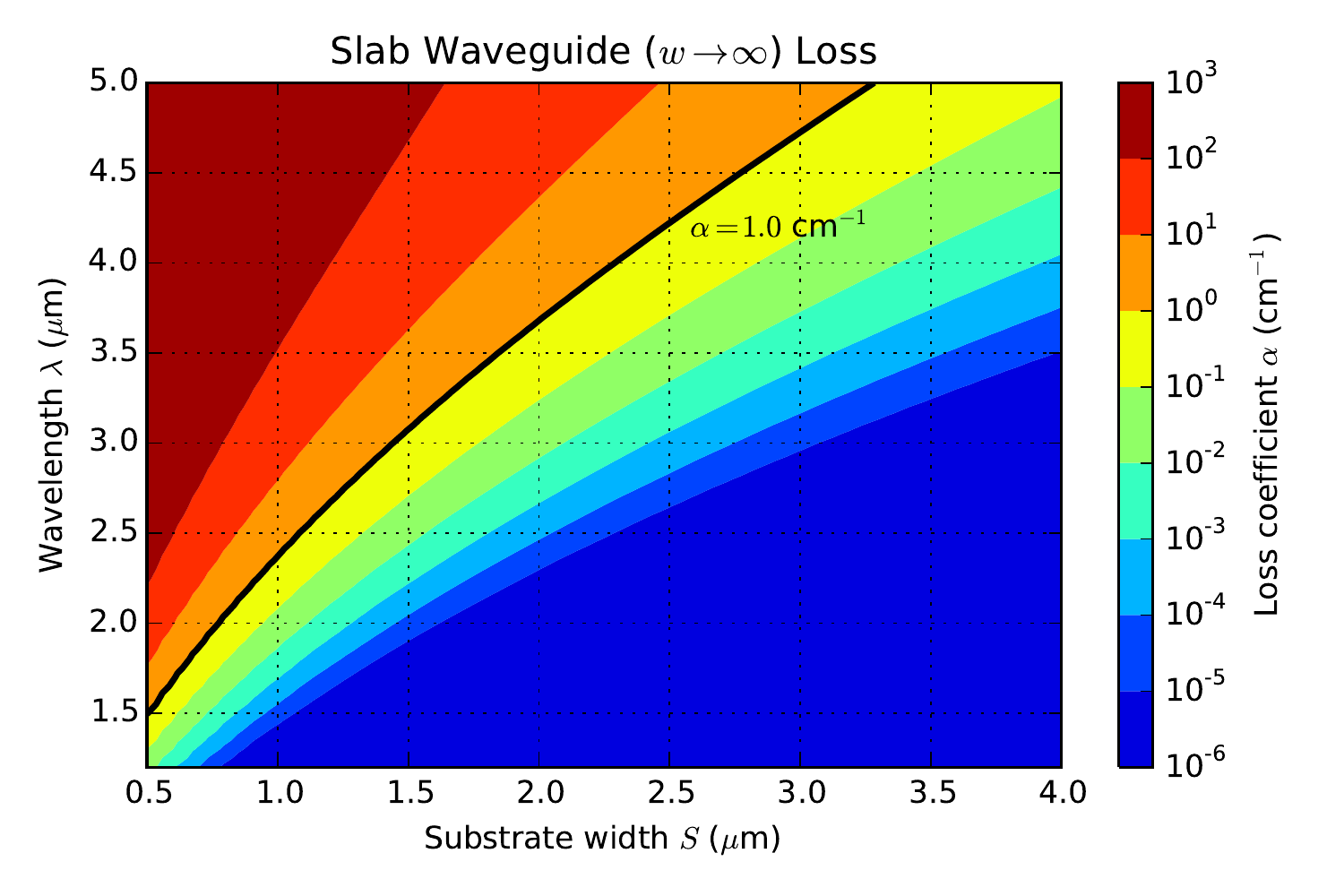}
\caption{Absorption coefficient due to substrate leakage from a slab waveguide with $H = 220$ nm.}
\label{fig:12-f12}
\end{center}
\end{figure}

Condition (\ref{eq:12-bdry-leak}) is not satisfied for the lossless $\beta$ computed in Sec.~\ref{sec:12-slab}, since $a_3^+$ vanishes in that case.  Adding a perturbation $\delta\beta$ and propagating the result through to first order, we find that:
\beq
    \delta\beta = e^{-2\kappa\xi} \frac{1+iP\xi/\kappa}{1-iP\xi/\kappa} \frac{a_3^-}{\d a_3^+/\d\beta} \label{eq:12-loss-dbeta}
\eeq
This expression is complex, indicating a lossy mode.  The loss can be computed from the imaginary part of Eq.~(\ref{eq:12-loss-dbeta}):
\beq
	\alpha = 2\,\mbox{Im}(\delta\beta) = e^{-2\xi S} \frac{4(P\xi/\kappa)}{1 + (P\xi/\kappa)^2} \frac{a_3^-}{\d a_3^+/\d\beta},\ \ \ 
	P \equiv \left\{\begin{array}{cc} 1 & \mbox{(TE)} \\ n_1^2/n_s^2 & \mbox{(TM)} \end{array}\right.
	\label{eq:12-leak-alpha}
\eeq
The absorption coefficient is proportional to the complex refractive index $n_{\rm i} = \mbox{Im}(\delta\beta)c/\omega = (\mbox{Im}(\delta\beta)/\beta) n_{\rm r}$.  Thus, (\ref{eq:12-leak-alpha}) can be recast into an equation for $n_{\rm i}$:
\beq
	n_{\rm i} = n_{\rm r} e^{-2\xi S} \frac{4(P\xi/\kappa)}{1 + (P\xi/\kappa)^2} \frac{a_3^-/\beta}{\d a_3^+/\d\beta}
	\label{eq:12-leak-ni}
\eeq
The two fractions in (\ref{eq:12-leak-ni}) are of order unity, so the dominant contribution to $n_{\rm i}$ is the exponential $e^{-2\xi S}$, the exponential attenuation of the evanescent field.  As a result, the leakage loss should decrease exponentially with increasing substrate width.  On the other hand, increasing wavelength decreases $\xi$, which will increase the loss.  Since TM and higher-order modes have smaller $\beta$ and also smaller $\xi$, they should be much lossier than the fundamental TE mode.

\begin{figure}[b!]
\begin{center}
\includegraphics[width=1.0\textwidth]{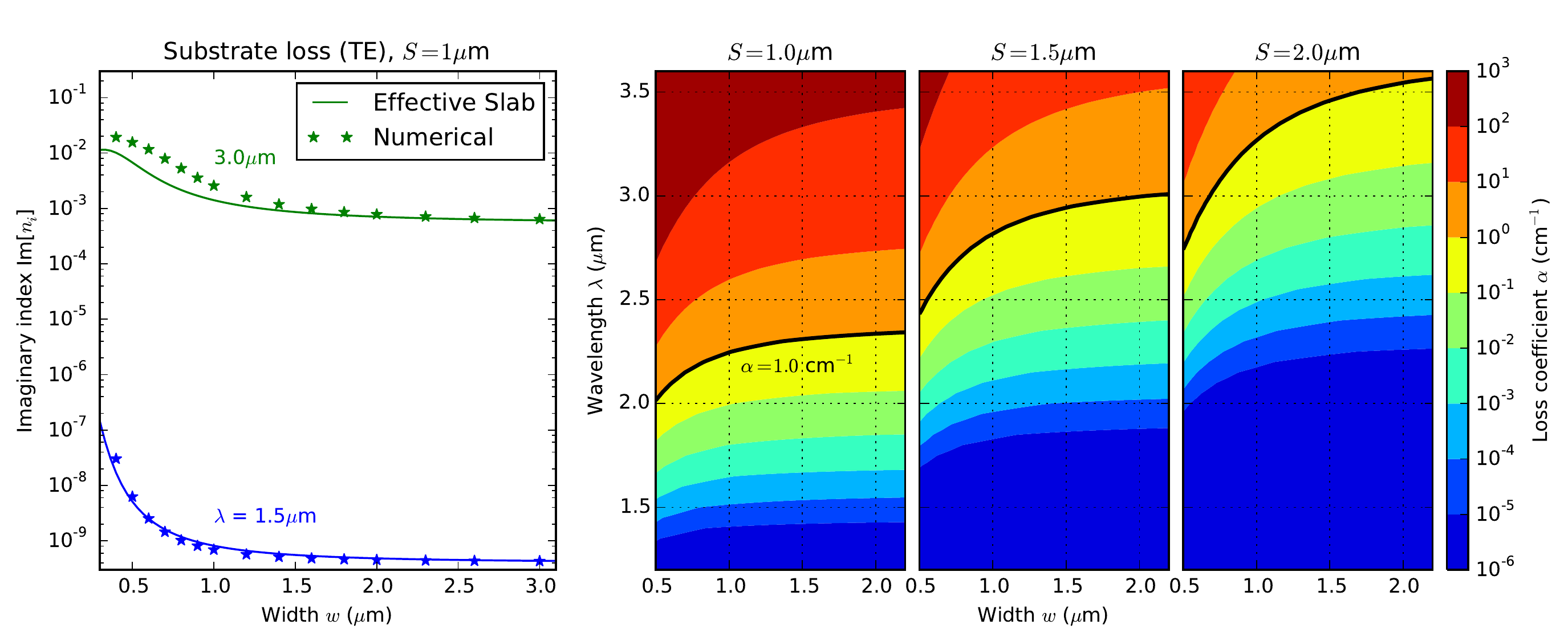}
\caption{Left: Loss for a rib waveguide with $h = 70$, $H = 220$ nm, as a function of width $w$.  Right: loss computed from effective index theory, in terms of width and wavelength.  The $w \rightarrow \infty$ limit was plotted in Fig.~\ref{fig:12-f12}.}
\label{fig:12-f13}
\end{center}
\end{figure}

Fig.~\ref{fig:12-f12} gives the substrate leakage loss for a slab waveguide as a function of substrate width and wavelength.  There is a strong exponential dependence on both parameters, due to the $e^{-2\xi S}$ term in Eq.~(\ref{eq:12-leak-alpha}).  A good analytic fit to Fig.~\ref{fig:12-f12} is:
\beq
	\frac{\alpha}{\mbox{cm}^{-1}} = \exp\left[10.55 - 37.87 \frac{S/\mu\mbox{m}}{(\lambda/\mu\mbox{m})^{1.5}}\right] \label{eq:12-leak-analytic}
\eeq
This fits the data to within a factor of 3.  Substrate leakage can be ignored if $\alpha < 1.0\,\mbox{cm}^{-1}$, since roughness loss is of this order.  Using (\ref{eq:12-leak-analytic}), this condition for negligible leakage is:
\beq
	(S/\mu\mbox{m}) > 0.28 (\lambda/\mu\mbox{m})^{1.5}
\eeq
Ridge and rib waveguides can be treated using the effective index method \cite{Kogelnik1988}.  First, we solve for the mode of a symmetric slab with width $w$, core index $n_1$, and cladding index $n'_0$, where $n'_0 = n_0$ for the ridge geometry ($h = 0$), and for $h \neq 0$ we can take a weighted sum: $(n'_0)^2 = (h/H) n_1^2 + (1-h/H) n_0^2$.  The mode index $n'_1 = \beta/(\omega/c)$ becomes the core index of our effective slab; see Fig. \ref{fig:12-f15}.  Then the loss is computed using transfer matrix theory, as discussed above.

Fig.~\ref{fig:12-f13} (left) gives the waveguide leakage loss, computed with both the numerical mode solver and effective-index theory.  The two agree to a factor of 2--3 over a wide range of widths and wavelengths, a fact that has been noted in the literature \cite{Bienstman2006}.  This is fortunate, since each mode-solver computation takes around 100 seconds on a 10-core machine, and in practice multiple runs are needed to verify the accuracy of the mesh and PML; on the other hand, the effective-index calculation takes milliseconds.  The right-side plots were computed using effective index theory, as a full numerical simulation would have taken days.

\begin{figure}[t!]
\begin{center}
\includegraphics[width=1.0\textwidth]{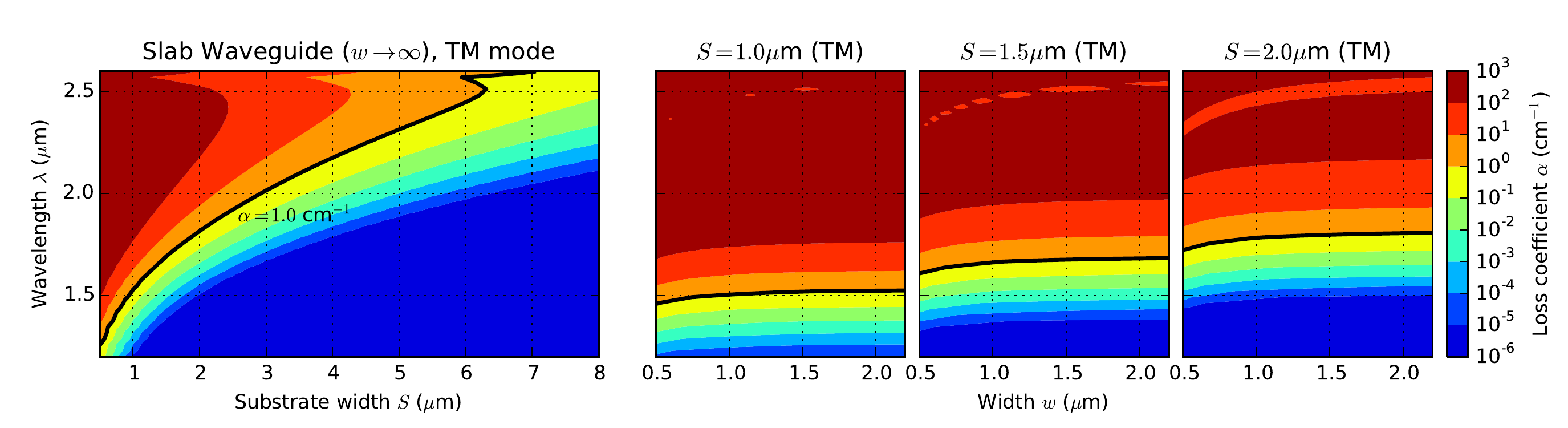}
\caption{Left: Loss for a slab waveguide with $H = 220$ (Fig.~\ref{fig:12-f12}), TM mode.  Right: rib waveguide, $H = 220, h = 70$ (Fig.~\ref{fig:12-f13}), TM mode}
\label{fig:12-f20}
\end{center}
\end{figure}

Compared to the TE mode, the TM mode is extremely lossy (Fig.~\ref{fig:12-f20}).  This is a consequence of the smaller effective index for TM and the TM boundary conditions, both which cause the field to extend deep into the substrate.  While leakage is insignificant for telecom frequencies, in the mid-IR losses are well over 10--100 cm$^{-1}$.  To reduce the leakage losses for TM modes, very thick oxide layers ($\gtrsim 6\mu$m) are needed to confine the light in the mid-IR.  Another way to reduce TM losses would be to increase the silicon thickness $H$.

\subsection{Scattering Loss}

In many waveguides, material and substrate absorption are negligible, and the dominant source of waveguide loss is scattering.  Modern SOI has advanced to the point that the top and bottom surfaces are atomically smooth, but roughness in the sidewalls is still significant \cite{Lee2000}.

The first treatment of the scattering problem was by Marcuse, who modeled it using coupled-mode perturbation theory \cite{Marcuse1969}; however, this approach requires detailed knowledge of both the bound and radiation modes, so is difficult to use in practice.  I will follow the approach of Payne and Lacey \cite{Lacey1990, Payne1994}, who treat scattering using radiation theory, where the rough waveguide surface acts as an antenna, and the far-field radiation is computed.  Their treatment was only concerned with TE modes in a slab waveguide, but is straightforward to extended to TM modes, and can be applied to ridge / rib waveguides using effective index theory \cite{Lee2000}.

\begin{figure}[tbp]
\begin{center}
\includegraphics[width=0.7\textwidth]{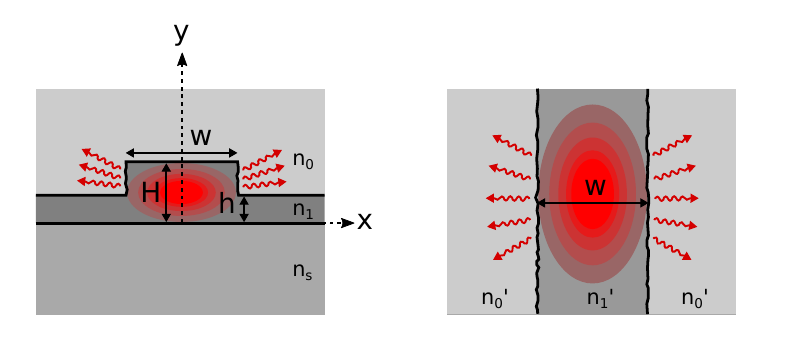}
\caption{Left: sidewall roughness scattering from a rib waveguide.  Right: equivalent effective slab.}
\label{fig:12-f23}
\end{center}
\end{figure}

Starting with a rib waveguide, we apply effective index theory to model the field.  Depending on the mode, the transverse field takes the form:
\begin{align}
	& \mbox{(TE-like)} & \vec{E} \sim \hat{x}, \vec{H} \sim \hat{y}, \hat{z}\ \ \ & E_x(x,y,z) \approx E_x(x,z)Y(y) \\
	& \mbox{(TM-like)} & \vec{H} \sim \hat{x}, \vec{E} \sim \hat{y}, \hat{z}\ \ \ & H_x(x,y,z) \approx H_x(x,z)Y(y)
\end{align}
where $Y(y)$ is the slab mode.  First one obtains the field in the absence of scattering: $E_{x0} = \Phi(x) e^{i(\omega t - \beta z)}$ (or likewise for $H_x$ in TM-like); $\Phi(x)$ is found by solving the mode equations for the effective slab; see Fig.~\ref{fig:12-f23}.  Recall that $E$ and $H$ in a slab are related by:
\begin{align}
	& \mbox{(TE-like)} & H_y & = \frac{n'_1}{Z_0} E_x \label{eq:12-exy-te} \\
	& \mbox{(TM-like)} & E_y & = \frac{n'_1 Z_0}{n^2} H_x \label{eq:12-exy}
\end{align}
where $Z_0 = \sqrt{\mu_0/\epsilon_0}$ is the impedance of free space.

Now in the effective slab (Fig.~\ref{fig:12-f23}, right) add roughness to the sidewalls, so that the sidewall width is a function of $z$.  The index profile then becomes $n = n'_1 + (n'_0 - n'_1) U\bigl(|x| - a - f(z)\bigr)$, where $U(x)$ is the Heaviside step function.  This breaks the $z$-translation symmetry, so solutions of the form $E(x, y) e^{i\beta z}$ are no longer exact.  Payne and Lacey \cite{Lacey1990} worked out the solution for TE polarization (relative to the effective slab, $\vec{E} \sim \hat{y}$) -- first one must solve the equation:
\beq
    \left(\frac{\d^2}{\d x^2} + \frac{\d^2}{\d z^2} + k_0^2 \right) E_y(x, z) = k_0^2 ({n'_0}^2 - {n'_1}^2) U\bigl(|x| - a - f(z)\bigr) E_y(x, z)
\eeq
where $k_0 = \omega/c$ is the free-space wavenumber.

Recall that in the absence of scattering, the mode looks like $E_{x0} = \Phi(x) e^{i(\omega t - \beta z)}$.  One then perturbs about that solution to find the scattering field.  The result depends on the surface-roughness autocorrelation function
\beq
	R(u) = \bigl\langle f(z)f(z+u) \bigr\rangle
\eeq
and its Fourier transform $\tilde{R}(k)$, and is given as an ``ensemble-average magnitude-squared radiated field per unit length'' \cite{Lacey1990}; the result was originally derived for TE polarization; here we assume that the TM formula is analogous:
\beq
	\frac{1}{2L} \bigl\langle |E_y(r, \theta)|^2 \bigr\rangle = \Phi(a)^2 ({n'_1}^2 - {n'_0}^2)^2 \frac{k_0^3}{4\pi n'_0 r} \tilde{R}(\beta - n'_0k_0\cos\theta) \label{eq:12-prad}
\eeq
The radiated power per unit length is
\beq
    \frac{1}{2L} P_{\rm rad} = \frac{n'_0}{2Z_0} \int_0^\pi \frac{\langle |E_y(r,\theta)|^2\rangle}{2L} r\,\d\theta
\eeq
and the guided power is given by the integral of the Poynting vector.  Since the vertical confinement is dominant, we use Eq.~(\ref{eq:12-exy-te}) to relate $H_y$ to $E_x$; the guided power becomes \cite[below Eq.~(15)]{Lacey1990}:
\beq
	P_g = \int{(\vec{E}\times \vec{H})\cdot \hat{z}\,\d A} = 
		\int{E_x H_y \d x} = \frac{n'_1}{Z_0} \int{\Phi(x)^2\d x}
	\label{eq:12-pguided}
\eeq
For convenience, we choose the following normalization for $\Phi$:
\beq
	\int \Phi(y)^2\d y = 1
\eeq
The loss coefficient $\alpha = (P_{\rm rad}/2L)/P_g$ then takes the following form:
\beq
	\alpha = \underbrace{\vphantom{\int_0^\pi}\Phi(a)^2}_{\rm (field)} \underbrace{\vphantom{\vphantom{\int_0^\pi}} ({n'_1}^2 - {n'_0}^2)^2 \frac{k_0^3}{4\pi n'_1}}_{\rm (constants)} \underbrace{\int_0^\pi {\tilde{R}(\beta - n'_0k_0\cos\theta)\d\theta}}_{S\ \rm (roughness)} \label{eq:12-alpha-sc}
\eeq
Eq.~(\ref{eq:12-alpha-sc}) shows that the loss depends on the surface field intensity through $\Phi(a)^2$, and the surface-roughness through the integral $S$.  To obtain $\Phi(a)^2$, one solves for the modes of the slab waveguide (Sec.~\ref{sec:12-slab}).  Defining the dimensionless slab parameters
\beq
	u \equiv \kappa a = a\sqrt{{n'_1}^2 k_0^2 - \beta^2},\ \ \ w \equiv \xi a = \sigma a = a\sqrt{\beta^2 - {n'_0}^2 k_0^2},\ \ \ v \equiv \sqrt{u^2 + w^2} = k_0 a \sqrt{{n'_1}^2 - {n'_0}^2} \label{eq:12-uvw}
\eeq
one can show that the surface-field intensity is \cite{AdamsBook}
\beq
	\Phi(a)^2 = \frac{1}{a}\frac{u^2}{v^2} \frac{w}{w+1} \label{eq:12-phisq}
\eeq
The roughness term $S$ depends on the autocorrelation function $R(u)$.  Different forms for the autocorrelation exist, but exponentials and Gaussians are most commonly cited in the literature \cite{Lee2000, Payne1994}.  The exponential agrees with AFM measurements of roughness \cite{Ladouceur1992} and gives a simpler expression for $S$, so I use it in what follows:
\begin{figure}[b!]
\begin{center}
\includegraphics[width=0.55\textwidth]{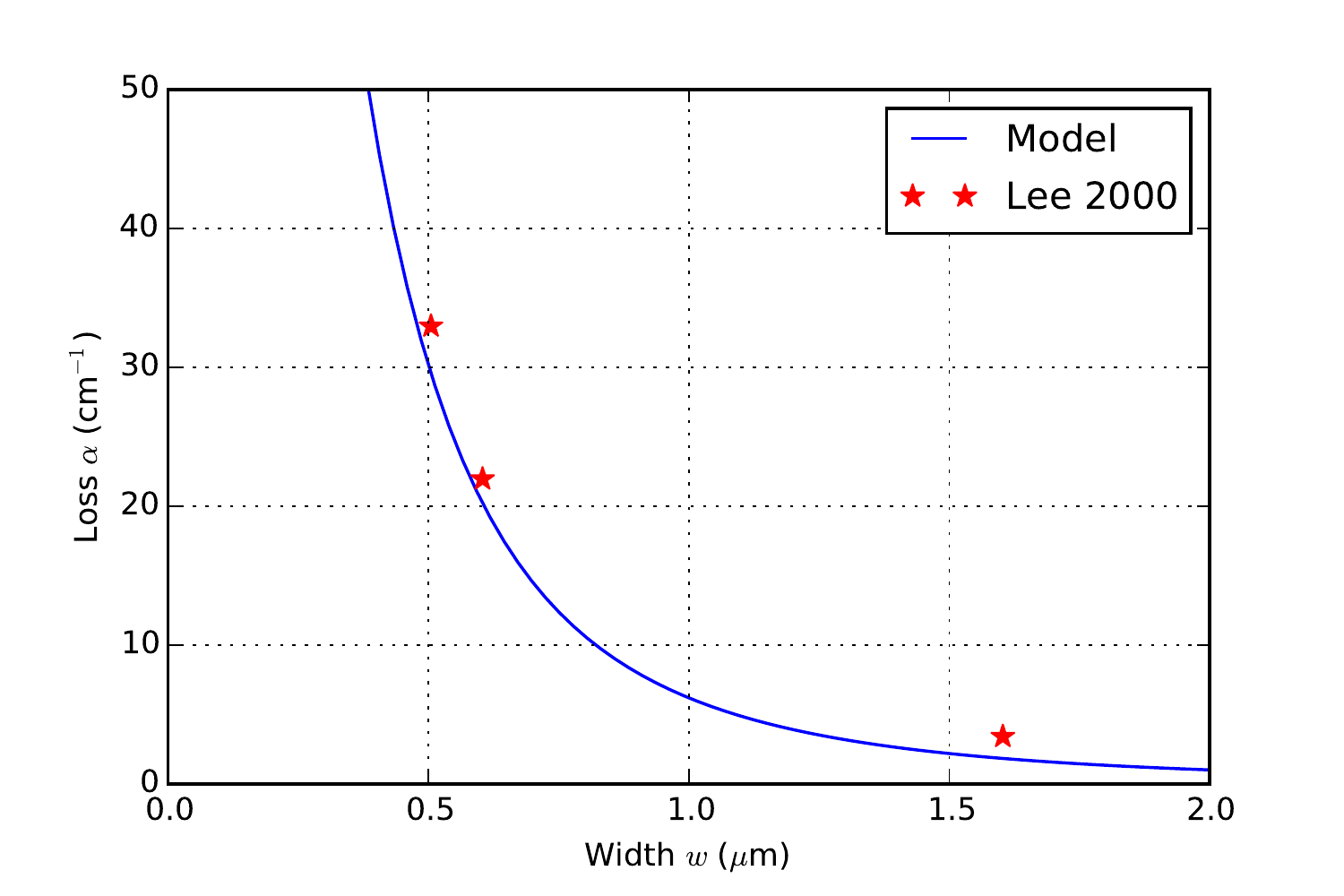}
\caption{Comparison between model (\ref{eq:alpha-sc-final}) and Ref.~\cite{Lee2000}}
\label{fig:12-f25}
\end{center}
\end{figure}
\beq
	R(u) = \sigma^2 e^{-|u|/L_c}
\eeq
Following Ref.~\cite{Payne1994}, $S$ is found to be:
\beq
	S = \frac{\sqrt{2}\pi \sigma^2 a}{w} f_e,\ \ \ f_e \equiv \frac{x \left[\bigl((1+x^2)^2 + 2x^2\gamma^2\bigr)^{1/2} + (1-x^2)\right]^{1/2}}{\bigl((1+x^2)^2 + 2x^2\gamma^2\bigr)^{1/2}}
\eeq
where $x = L_c w/a$ and $\gamma = (v/w) \sqrt{2n_0^2/(n_1^2 - n_0^2)}$.  Applying Eq.~(\ref{eq:12-alpha-sc}), the loss coefficient is found to be\footnote{Note that \cite[Eq.~(9)]{Payne1994} (which cites \cite{AdamsBook}) is off by a factor of two, probably mistaking waveguide width for half-width (corrected in Eqs.~(\ref{eq:12-phisq}, \ref{eq:alpha-sc-final}))} \cite[Eq.~(10)]{Payne1994}:
\beq
	\alpha = \frac{\sigma^2}{2\sqrt{2} k_0 a^4 n'_1} f_e g,\ \ \ \ g = \frac{u^2 v^2}{w+1} \label{eq:alpha-sc-final}
\eeq
Scattering loss from TM modes is likely very close to this, since TE and TM modes tend to have similar profiles for weak-guiding waveguides, when either the ratio $n'_1/n'_0 \approx 1$ or the width $w \gg \lambda/n'_1$.  In the present case, the waveguide is strongly confining in $y$ but only weakly in $x$, so we expect TE and TM losses to be similar.

To summarize, the procedure for finding surface-roughness loss is the following:

\begin{enumerate}
	\item Using effective index theory, compute $n'_0$ and $n'_1$ for the effective slab of width $w$ (Fig.~\ref{fig:12-f23}).
	\item Solve for the effective slab field, TE polarization, and obtain the dimensionless constants $u, v, w$ from Eq.~(\ref{eq:12-uvw}).
	\item Calculate $\alpha$ through Eq.~(\ref{eq:alpha-sc-final})
\end{enumerate}

A good reference for this procedure is Lee et al.\ \cite{Lee2000}.  In this paper, waveguide loss was measured in a number of devices of differing lengths, and compared to the theory.  As Figure \ref{fig:12-f25} shows, the theory match experimental data well for $\sigma = 9$nm, $L_c = 50$nm; by reducing both $\sigma$ and $L_c$, it was predicted that losses could be reduced by over an order of magnitude.

\begin{figure}[tbp]
\begin{center}
\includegraphics[width=1.00\textwidth]{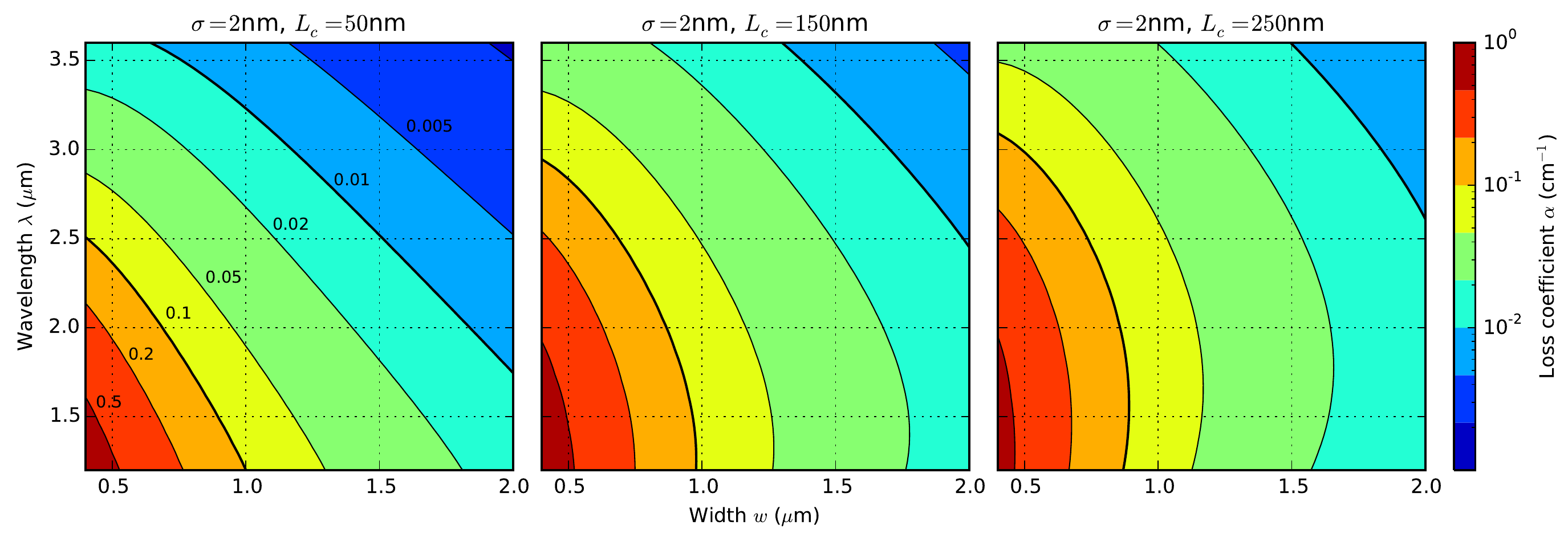}
\caption{Calculated surface-roughness scattering loss for $\sigma = 5$nm, $L_c = 50, 150, 250$nm, for a ridge waveguide ($H = 220$nm, $h = 0$), $n_0 = 1.0$, $n_1 = 3.5$, $n_s = 1.5$.}
\label{fig:12-f19}
\end{center}
\end{figure}

Figure \ref{fig:12-f19} gives the scattering loss for an SOI ridge waveguide as a function of width and wavelength.  The values for $\sigma, L_c$ vary from one process to the next; reasonable values with present technology are $\sigma = 2$nm, $L_c = 50$nm are shown here.  The absorption is strongly sensitive to both $\sigma$ and $L_c$ (compare Refs.~\cite{Lee2000, Vlasov2004, Yap2009}), and calculated values of scattering loss are not always quantitatively accurate, and fully three-dimensional calculations may be necessary if quantitative results are needed \cite{Barwicz2005}.  However, the qualitative trends seem to be reliable, and Fig.~\ref{fig:12-f19} suggests that scattering losses for long-wavelength light are not much greater than for short-wavelength light.

On the other hand, long-wavelength calculations for thin waveguides should not be trusted, because the field profile no longer resembles the field predicted by effective index theory (see Fig.~\ref{fig:12-f18}), the field may be much more weakly bound and losses may be much higher \cite{Vlasov2004}.

\begin{figure}[tbp]
\begin{center}
\includegraphics[width=1.00\textwidth]{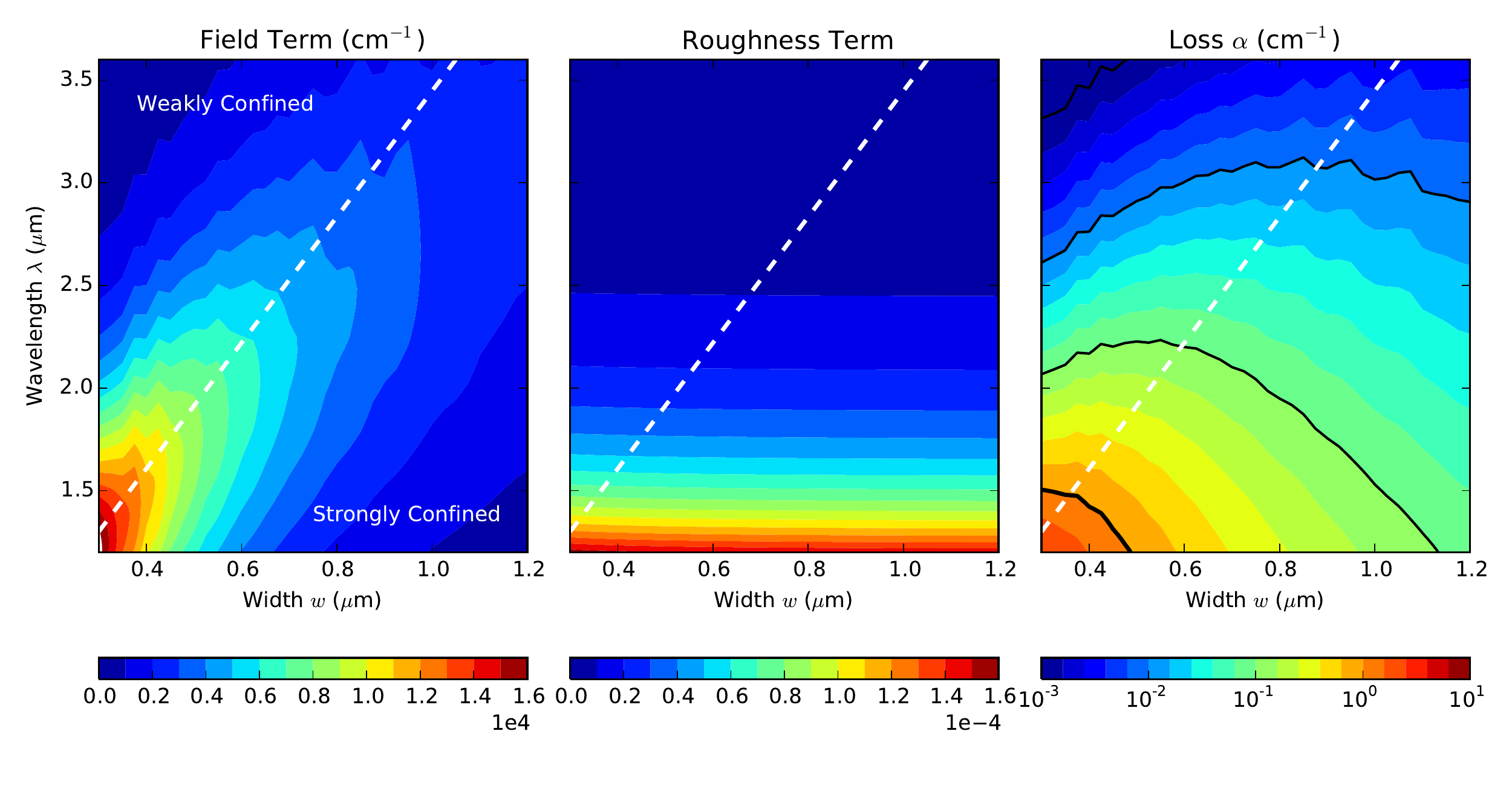}
\caption{Left: field term and roughness term from Eq.~(\ref{eq:12-alpha-sc2d}), using numerically computed field profiles.  Right: loss coefficient.}
\label{fig:12-f24}
\end{center}
\end{figure}

The most straightforward way to refine the results of Fig.~\ref{fig:12-f19} would be to use full numerical simulations to compute the waveguide mode, rather than relying on the effective index model.  Note from \ref{eq:12-alpha-sc} that the loss coefficient is a product of a surface field and some roughness factors.  This formula suggests an extension to 2D waveguides with numerically-computed modes:
\beq
	\alpha \stackrel{?}{=} \underbrace{\vphantom{\int_0^\pi}\oint\Phi(x, y)^2 ds}_{\rm (surface\,field)}\  \underbrace{\vphantom{\vphantom{\int_0^\pi}} ({n'_1}^2 - {n'_0}^2)^2 \frac{k_0^3}{4\pi n'_1} \int_0^\pi {\tilde{R}(\beta - n'_0k_0\cos\theta)\d\theta}}_{\rm (roughness\,constants)} \label{eq:12-alpha-sc2d}
\eeq
Here, the integral is taken over all rough surfaces of the waveguide (in a $z$-cross section, so the integral is one-dimensional).  Since the field is normalized to unity, this surface term has units of cm$^{-1}$.  The roughness term, by contrast, is unitless.

Figure \ref{fig:12-f24} shows the field and roughness terms as a function of waveguide width and wavelength.  The roughness term is largest for small wavelengths, consistent with the observation from Rayleigh theory that subwavelength objects preferentially scatter blue light.  The field term is small whenever the field is ``strongly confined'' inside the waveguide, or ``weakly confined,'' living primarily in the cladding and substrate.  Strongly-confined modes genuinely have low loss; the low loss for weakly-confined modes is illusory, since their substrate leakage loss very large (Sec.~\ref{sec:12-leak})

\section{Kerr Nonlinearity}

The Kerr nonlinearity is responsible for self-phase modulation, cross-phase modulation and four-wave mixing.  Due to silicon's $m3m$ point-group symmetry and the permutation relations of the $\chi^{(3)}$ tensor, one can show that the $\chi^{(3)}$ tensor takes the form \cite{Lin2007}:
\beq
	\chi_{ijkl} = \chi_{1111} \left[\frac{\rho}{3}(\delta_{ij}\delta_{kl} + \delta_{ik}\delta_{jl} + \delta_{il}\delta_{jk}) + (1-\rho)\delta_{ijkl}\right]
\eeq
where $\rho \approx 1.27$ is a parameter characterizing the anisotropy of the crystal ($\rho = 1$ is fully isotropic).  The effective $\chi^{(3)}$ depends on the field orientation, and is given by $\chi_{ijkl} \hat{e}_i \hat{e}_j \hat{e}_k \hat{e}_l$.  Typical values are given below; as one can see, the directional variation is only around 20\%, so the Kerr nonlinearity can typically be approximated as isotropic.
\begin{center}
\begin{tabular}{c|l}
\hline\hline
Field & $\chi^{(3)}_{\rm eff}$ \\ \hline
$\vec{E} \sim \avg{100}$ & $\chi_{1111}$ \\
$\vec{E} \sim \avg{110}$ & $\frac{1}{2}(1+\rho)\chi_{1111} = 1.14\chi_{1111}$ \\
$\vec{E} \sim \avg{111}$ & $\frac{1}{3}(1+2\rho)\chi_{1111} = 1.18\chi_{1111}$ \\ \hline\hline
\end{tabular}
\end{center}
  The Kerr and TPA parameters $n_2, \beta$ are related to $\chi_{1111}$ by:
\beq
	n_2 + i\frac{c}{2\omega} \beta = \frac{3}{4\epsilon_0 c n_0(\omega)^2} \chi_{1111}(-\omega;\omega,-\omega,\omega) \label{eq:12-chi1111}
\eeq
Typically, $n_2$ is quoted in units of cm$^2$/W, and $\beta$ has units of cm/W.

Measured values of $n_2$ and $\beta$ are reported in Table \ref{tab:12-t1}.  Clearly not all wavelengths have been studied equally.  There have been a large number of measurements at telecom wavelengths, which is understandable given that many silicon devices are built for those wavelengths.  However, later studies extended our knowledge to the near- and mid-IR, and presently the whole SOI transparency range has been measured.

\begin{figure}[p]
\begin{center}
\includegraphics[width=1.00\textwidth]{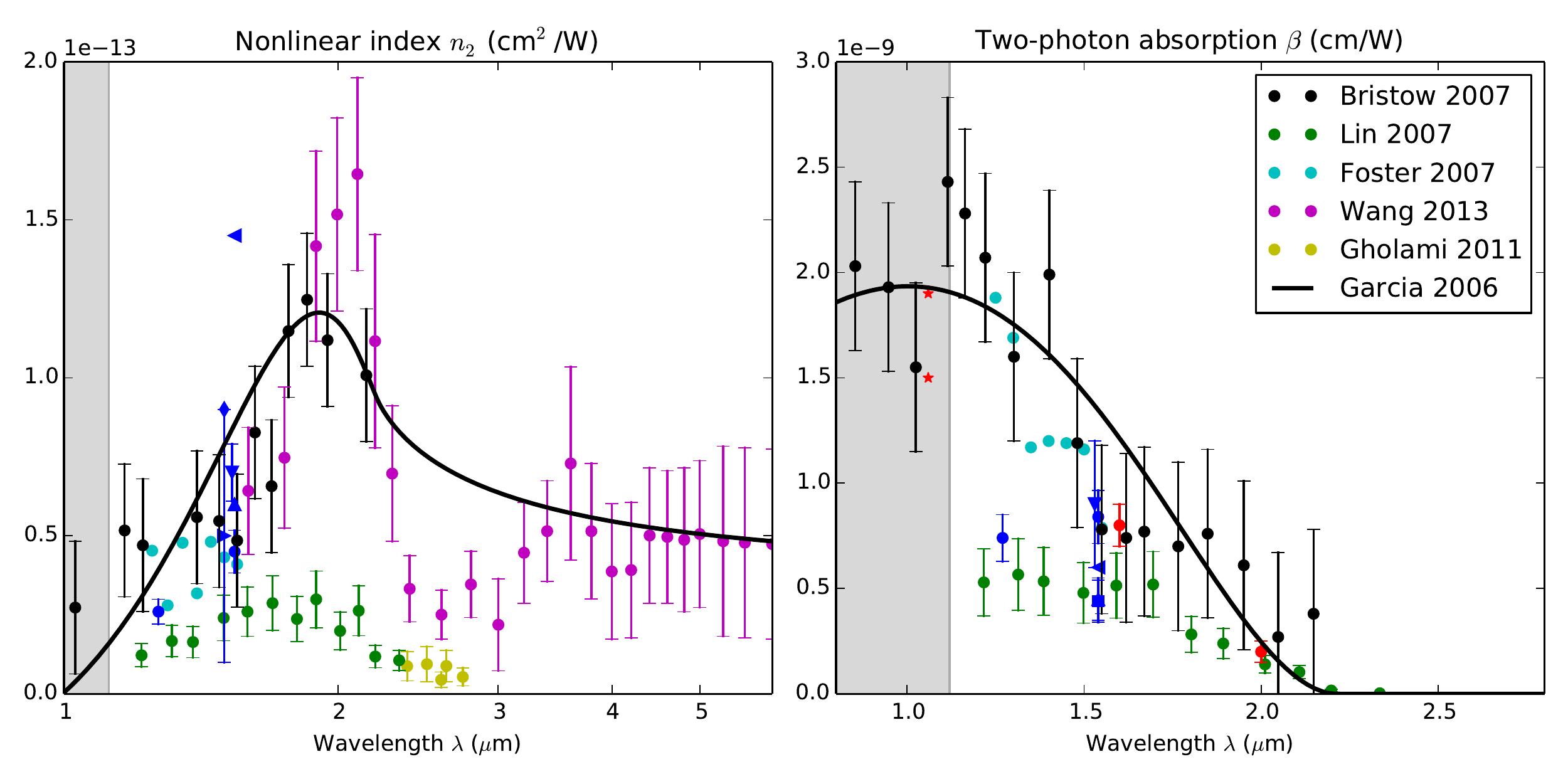}
\caption{Comparison of silicon $n_2$ and $\beta$ values reported in the literature.  See Table \ref{tab:12-t1}.}
\label{fig:12-f21}
\end{center}
\end{figure}

\begin{table}[p]
\begin{center}
\begin{tabular}{cc|ccc}
\hline\hline
\multicolumn{2}{c}{Reference} & $\lambda$ ($\mu$m) & $n_2$ ($10^{-14}\,$cm$^2$/W) & $\beta$ ($10^{-9}\,$cm/W) \\ \hline
\textcolor{blue}{$\bullet$} & \cite{Dinu2003}   & 1.54 & $4.5 \pm 0.07$ & $0.84 \pm 0.13$ \\
&                   & 1.27 & $2.6 \pm 0.04$ & $0.74 \pm 0.11$ \\ \hline
\textcolor{blue}{*} & \cite{Claps2003}  & 1.54 & -- & $0.44 \pm 0.10$ \\ \hline
\textcolor{blue}{$\blacktriangle$} & \cite{Tsang2002}  & 1.54 & $6$            & $0.45 \pm 0.10$ \\ \hline
\textcolor{blue}{$\blacktriangledown$} & \cite{Rieger2004} & 1.53 & $7.0 \pm 1.4$  & $0.9 \pm 0.3$ \\ \hline
\textcolor{blue}{$\blacktriangleleft$} & \cite{Yamada2005} & 1.54 & $14.5$ & $0.6$ \\ \hline
\textcolor{blue}{$\blacktriangleright$} & \cite{Dulkeith2006} & 1.50 & $5 \pm 4$ & -- \\ \hline
\textcolor{blue}{$\blacklozenge$} & \cite{Fukuda2005} & 1.54 & $9$ & -- \\ \hline
\textcolor{red}{$\bullet$} & \cite{Euser2005}  & 2.00 & -- & $0.20 \pm 0.05$ \\
&				  & 1.60 & -- & $0.80 \pm 0.10$ \\ \hline
\textcolor{red}{*} & \cite{Reintjes1973}  & 1.06 & -- & $1.9$ (20 K) \\
&				  & 1.06 & -- & $1.5$ (100 K) \\ \hline
& \cite{Wynne1969} & 10.6 & 8.2 & -- \\ \hline
$\bullet$ & \cite{Bristow2007} & 0.85--2.15 & \multicolumn{2}{c}{Fig.~\ref{fig:12-f21}} \\ \hline
\textcolor{green}{$\bullet$} & \cite{Lin2007b}    & 1.20--2.35 & \multicolumn{2}{c}{Fig.~\ref{fig:12-f21}} \\ \hline
\textcolor{cyan}{$\bullet$} & \cite{Foster2007}   & 1.25--1.55 & \multicolumn{2}{c}{Fig.~\ref{fig:12-f21}} \\ \hline
\textcolor{magenta}{$\bullet$} & \cite{Wang2013b}    & 1.60--6.00 & \multicolumn{2}{c}{Fig.~\ref{fig:12-f21}} \\ \hline
\textcolor{yellow}{$\bullet$} & \cite{Gholami2011}    & 2.38--2.74 & \multicolumn{2}{c}{Fig.~\ref{fig:12-f21}} \\ \hline\hline
\end{tabular}
\caption{Tabulated values of $n_2$ and $\beta$ for silicon.  See Fig.~\ref{fig:12-f21}.}
\end{center}
\label{tab:12-t1}
\end{table}

Unfortunately, there is considerable disagreement over the magnitude of $\beta$ and $n_2$.  While all studies confirm the same qualitative behavior ($\beta_2$ drops to zero as one reaches the half-bandgap frequency $\hbar\omega = E_g/2$ (near 2.3$\mu$m); at the same wavelength $n_2$ reaches its maximum), values of $n_2$ differ between studies by a factor of $\sim 6$.  Fig.~\ref{fig:12-f21} shows that the data cluster into two groups: ``large-$\chi^{(3)}$'' studies \cite{Bristow2007, Foster2007, Wang2013b} that predict values of $n_2$ up to $1.7\times 10^{-13}\mbox{cm}^2/\mbox{W}$, and ``small-$\chi^{(3)}$'' studies \cite{Lin2007b, Gholami2011} that find a maximum of $0.3\times 10^{-13}\mbox{cm}^2/\mbox{W}$.  Since $n_2$ values for the large-$\chi^{(3)}$ data appear to agree better with the results at telecom wavelengths, which are more numerous and consistent (except for a few outliers), I think it is best to treat the large-$\chi^{(3)}$ data as more reliable, and use them as a basis for future calculations.

Bristow et al.\ showed that the two-photon absorption data fit well to the model by Garcia and Kalyanaraman \cite{Bristow2007, Garcia2006} for indirect band-gap semiconductors with parabolic bands:
\beq
	\beta = 2C \sum_{n=0}^2 \frac{\pi(2n+1)!!}{2^{n+2}(n+2)!} \left.\frac{(2x-1)^{n+2}}{(2x)^5}\right|_{x=\hbar\omega/E_g}
\eeq
Setting $C = 4.3\times 10^{-8}\mbox{cm}/\mbox{W}$ fits the TPA data reasonably well (Fig.~\ref{fig:12-f21}, right).

\begin{figure}[tbp]
\begin{center}
\includegraphics[width=0.55\textwidth]{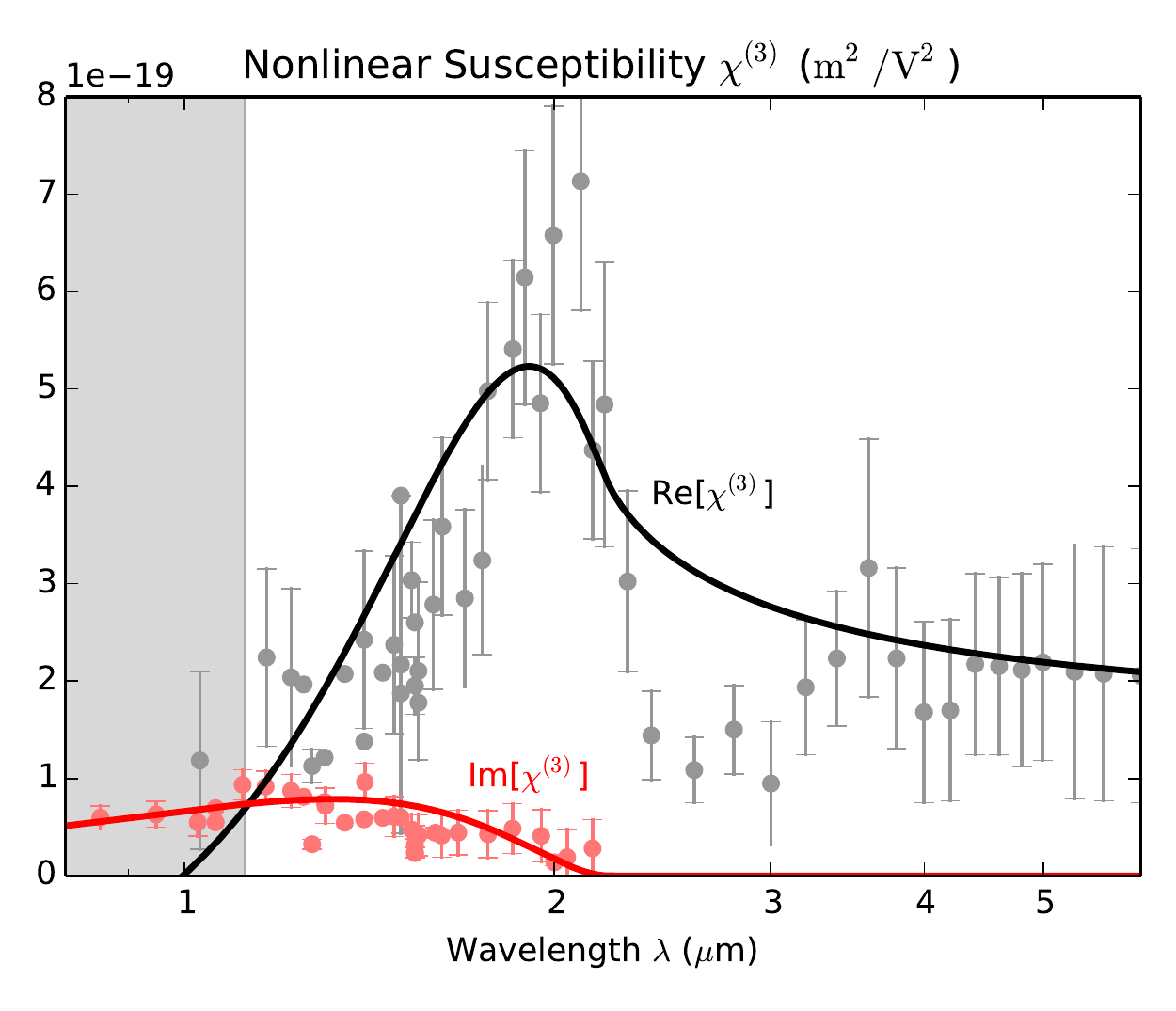}
\caption{Real and imaginary parts of silicon $\chi^{(3)}$.}
\label{fig:12-f22}
\end{center}
\end{figure}

The nonlinear index is related to the $\beta$ through the Kramers-Kr\"{o}nig transformation \cite{SheikBahae1991, Soh2016}.  A full calculation of $n_2$ must also include Raman, linear and quadratic-Start terms, but the TPA term is usually dominant, so we ignore the other terms here.  The formula for $n_2$ is \cite{Hon2011}:
\beq
	n_2 = \frac{c}{\pi}\,\mbox{P}\!\int_0^\infty {\frac{\beta(\omega, \omega')}{(\omega')^2 - \omega^2} \d\omega'} \label{eq:12-n2-kk}
\eeq
It is standard to approximate the nondegenerate TPA term in (\ref{eq:12-n2-kk}) by $\beta(\omega, \omega') = \beta\bigl((\omega+\omega')/2\bigr)$ \cite{SheikBahae1991, Hon2011}.  In this case, the integral in (\ref{eq:12-n2-kk}) can be evaluated analytically, arriving at the result:
\beq
	n_2 = C' \left[-\frac{464x^4-850x^3+740x^2-645x+195}{30720 x^5} - \frac{(1-2x)^2 \left(20x^2-4x+13\right) \log|1-2x|}{4096x^6}\right]_{x=\hbar\omega/E_g}
\eeq
where $C' = C(\hbar c/E_g) = 7.7\times 10^{-13}\mbox{cm}^2/\mbox{W}$ if one follows the Kramers-Kr\"{o}nig relation exactly, although it is often treated as a fitting parameter instead.  The value plotted in Fig.~\ref{fig:12-f21} is $C' = 4.5\times 10^{-12}\mbox{cm}^2/\mbox{W}$.

Using Eq.~(\ref{eq:12-chi1111}), the susceptibility $\chi^{(3)}$ (technically $\chi_{1111}$, but the difference between $\chi_{1111}$ and $\chi_{\rm eff}$ is smaller than the error bars) is plotted in Fig.~\ref{fig:12-f22}.

\ifstandalone{\bibliography{NoteRefs}{}
\bibliographystyle{alpha}
}

\appendix

\ifdefined\multidoc\else\input{Header}\fi

\ifstandalone{\setcounter{chapter}{9}}
\chapter{Material Nonlinearities}
\label{ch:01a}

Logic is inherently nonlinear.  Even the simplest digital algorithms -- binary gates -- are nonlinear by design.  As such, photonic logic will only be useful if it can harness strong, reliable optical nonlinearities.

Optical nonlinearities can be divided into two classes:

\begin{itemize}
	\item ``Defect'' nonlinearities -- trapped atoms, quantum dots, plasmons, etc.  In the simplest treatment, one has a two-level system coupled to a high-$Q$ cavity.  The two-level system generates the nonlinearity.
	\item ``Bulk'' nonlinearities -- $\chi^{(2)}$, $\chi^{(3)}$, free carriers, excitons, thermal effects, optomechanics, etc.  The system consists of a high-$Q$ cavity and the nonlinearity adds anharmonic terms to the Hamiltonian.
\end{itemize}

The first type are stronger and more ``quantum'', but they are not scalable with current technology.  The second type are weaker but more scalable.  But they are not too weak.  With sufficiently small high-$Q$ optical cavities such as rings or photonic crystals, they can be used to design photonic switches, amplifiers and other nonlinear devices that operate at high speeds (GHz--THz) and low powers (0.01--0.1 fJ), and may be a viable alternative to electronic computing \cite{Notomi2011}.

In this chapter, I discuss two nonlinearities two important nonlinearities in semiconductors: the Kerr effect and free-carrier effect.  Simple scaling laws can be described for direct-bandgap semiconductors that relate the size of the nonlinearity to basic properties like the electron and hole masses, band gap, and index of refraction.  This allows different materials to be compared in a consistent and intuitive way.  Similar results can also be obtained for indirect-gap semiconductors like silicon, but the strongest nonlinearities exist for direct-gap materials, which are discussed here.

All the results in this section are classical.  But see Ch.~\ref{ch:05b} and Sec.~\ref{sec:02-kerr} for a quantum-mechanical treatment.

\section{Linear Absorption}

It is worth starting our discussion with linear absorption.  Linear absorption isn't a nonlinear effect, but it is important for other effects like bandfilling, so it is worth discussing here.  Additionally, discussing linear absorption gives us an opportunity to discuss the semiconductor band model we use in the nonlinear sections below.

\subsection{Above-Bandgap Absorption}

We are interested in direct band-gap semiconductors here.  Many direct-gap semiconductors used in photonics, like GaAs, InP, and InGaAsP, crystallize into a zinc-blende structure.  The $s$ and $p$ orbitals hybridize into four bands, but spin-orbit coupling removes one of the bands, leaving two valence bands (``heavy'' and ``light'' holes) and one conduction band.

\begin{figure}[htbp]
\begin{center}
\subfigure{\includegraphics[width=0.60\textwidth]{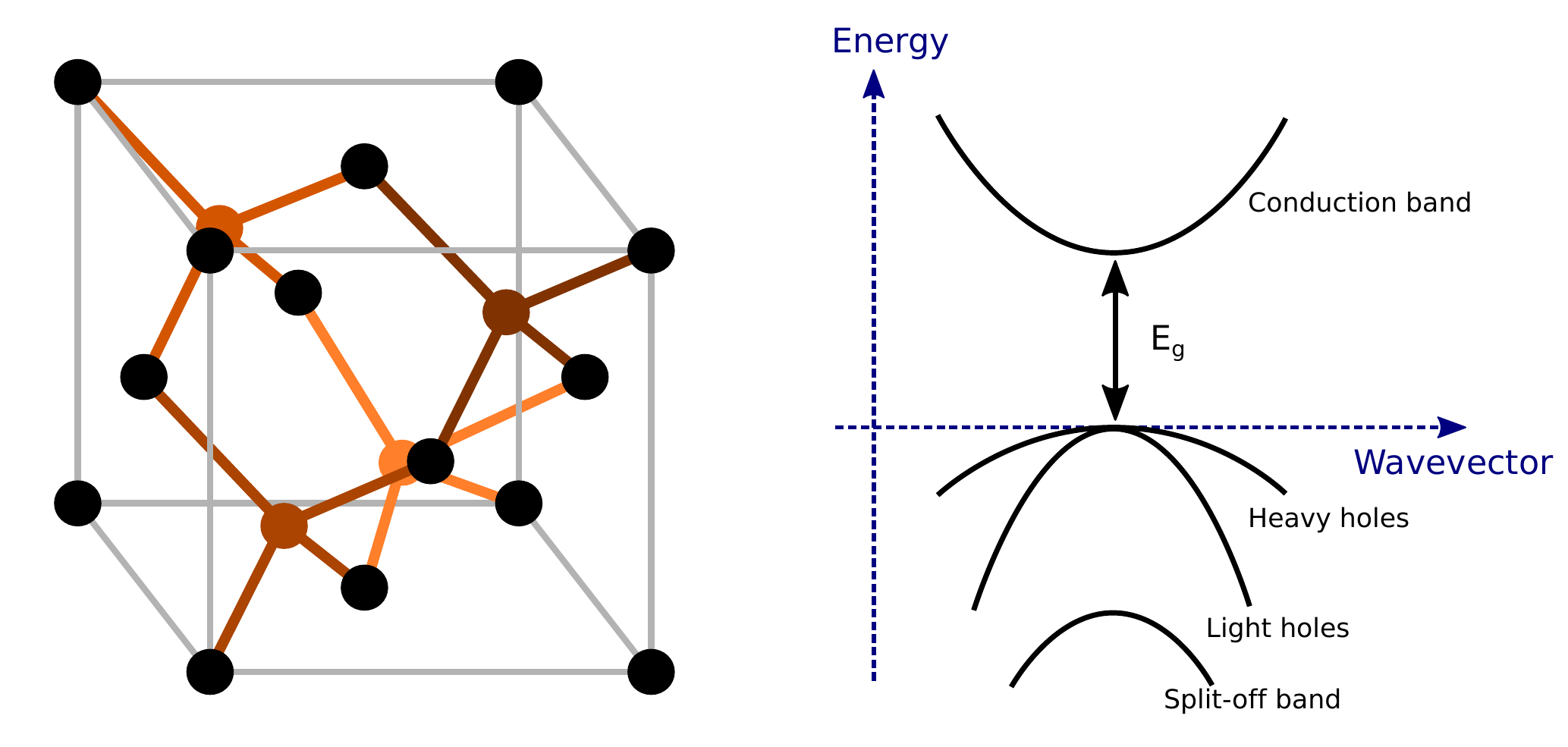}} \qquad
\subfigure{\includegraphics[width=0.30\textwidth]{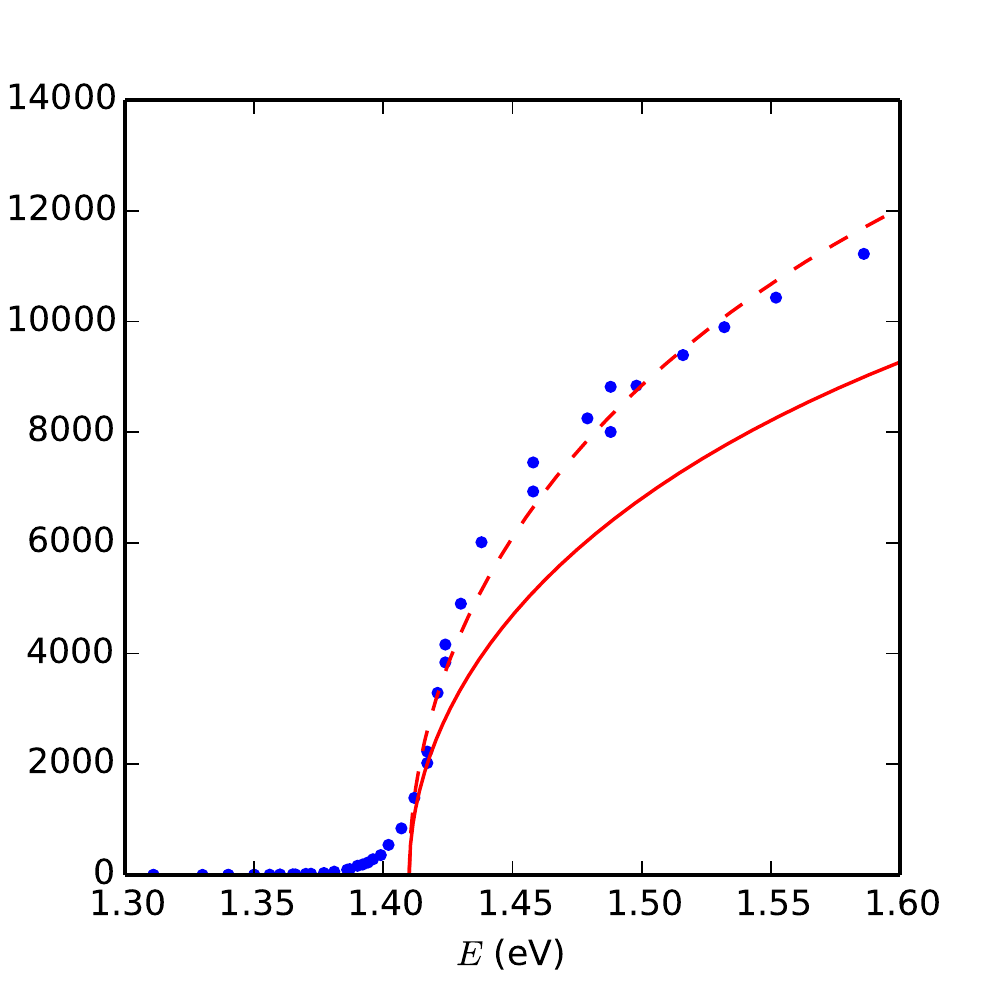}}
\caption{Left: Zinc-blende lattice structure.  Center: band diagram for GaAs, which crystallizes into a zinc-blende structure.  Right: Measured absorption of GaAs plotted against band theory prediction (solid) and with a fudge-factor (dashed).}
\label{fig:01a-f8}
\end{center}
\end{figure}


Using band theory, we can calculate the linear absorption spectrum.  This calculation assumes a perfect material, ignores imperfections and interactions between electrons, and does not capture excitons or the Urbach tail.  It also assumes parabolic bands so it will not work far above the band edge.  The result is \cite{Moss2013}:
\beq
	\boxed{\alpha(\omega) = K\frac{(\hbar\omega/E_g-1)^{1/2}}{\hbar\omega/E_g}}
\eeq
with the constant $K$ given by
\beq
	K = \frac{2\pi e^2 (2m_r)^{3/2}|p_{cv}|^2}{3m_0^2 n \epsilon_0 c h^2 \sqrt{E_g}}
\eeq
Here $|p_{cv}|$ is a matrix element from the valence to conduction-band states.  Since there are two hole bands, there are two linear absorption processes to keep track of, and the linear absorption is the sum of these two contributions.  For each process, using $k \cdot p$ theory and some assumptions, we can replace $|p_{cv}|^2 \rightarrow E_g m_0^2/2m_e$, giving a constant of:
\beq
	K = \frac{2\pi e^2 (2m_r)^{3/2} \sqrt{E_g}}{6m_e n \epsilon_0 h^2 c}
\eeq
For light-hole excitations, we have $m_e \approx m_h$ so $m_r = m_e/2$.  For heavy-hole excitations, $m_e \ll m_h$ so $m_r = m_e$.  Adding these together and multiplying by a factor of 2 for spin, we get:
\beq
	\boxed{K = \frac{2\pi e^2 (2^{3/2} + 1) m_e^{1/2} \sqrt{E_g}}{6 n \epsilon_0 h^2 c} = 3.38 \times 10^7 \frac{(m_e/m_0)^{1/2}(E_g/eV)^{1/2}}{n} \mbox{m}^{-1}}
\eeq
In the right pane of Fig. \ref{fig:01a-f8}, experimental data for GaAs are plotted against the model.  Up to a small constant factor likely due to the approximation we made to estimate the matrix element, the data and model agree.

\subsection{Below-Bandgap Absorption}
\label{sec:01a-urbach}

The spectrum calculated previously is reasonably accurate for $\hbar\omega > E_g$.  Below the band gap, however, it predicts zero linear absorption, consistent with the lack of transitions below the gap.  Experimentally, this is not true -- the absorption tends to fall off exponentially for $\hbar\omega < E_g$, according to ``Urbach's Rule'':
\beq
	\alpha(\omega) \sim e^{(\hbar\omega - E_g)/E_u}
\eeq
This exponential tail is a universal phenomenon, existing in III-V semiconductors, glasses, salts, and many other materials.  It does not appear to have a universal explanation, excitons and electric field fluctuations are thought to play a role \cite{Dow1972}.

Experimental absorption data for GaAs, InP, and InGaAsP near the band gap are plotted in Figure~\ref{fig:01a-f2}.  All three materials show an exponential tail -- and the constant $E_u$ is similar for all three of them -- it is around 0.01 eV.

\begin{figure}[t]
\begin{center}
\includegraphics[width=0.70\textwidth]{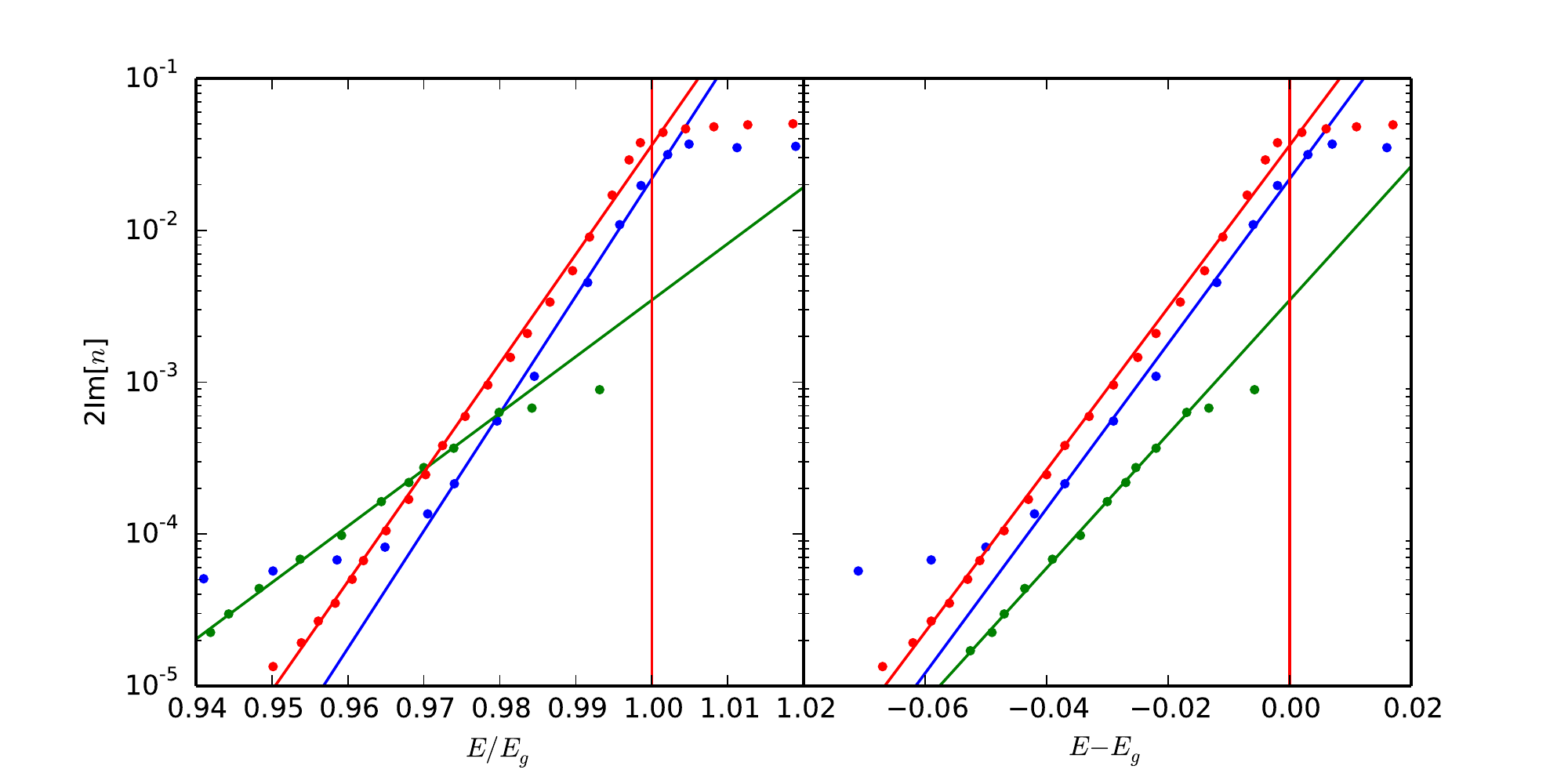}
\caption{Urbach tails compared for GaAs (blue) \cite{Casey1975}, InP (red) \cite{Turner1964}, and InGaAsP (green) \cite{Nozaki2010}.}
\label{fig:01a-f2}
\end{center}
\end{figure}

We will show in the next sections that the Kerr and free-carrier effects become very strong near the band gap.  To maximize the nonlinear effect, we should operate as close to the bandgap as feasible, where feasibility is limited by linear absorption in the Urbach tail.  Since the absorption depends on $E - E_g$, this favors wide-bandgap semiconductors -- having a large band gap lets you come closer to $E_g$ (as measured by the dimensionless quantity $x = E/E_g$) without strong linear absorption.

\section{Two-Photon Absorption}

Very general results for two-photon absorption are derived in M. Sheik-Bahae et al.\ \cite{SheikBahae1991}.  In that paper, the amplitudes for two-photon processes are computed from first principles -- and $\beta$ and $n_2$ are given by a scaling factors that depends on the effective electron mass $m_e$ and band gap $E_g$, times a universal function of the dimensionless quantity $x = E/E_g$.  Most materials agree with the Sheik-Bahae  results to within a factor of two.

Quoting from the paper, the two-photon absorption is given by:
\beq
	\beta(E) = K' \left(\frac{E_p}{\rm eV}\right)^{1/2}\left(\frac{E_g}{\rm eV}\right)^{-3} n_0^{-2} \left.\frac{(2x-1)^{3/2}}{(2x)^5}\right|_{x=E/E_g}
\eeq
Here, $E_p \equiv 2|p_{cv}|^2/m_0$, where $p_{cv}$ is the momentum matrix element.  Using the $k\cdot p$ approximation $|p_{cv}|^2 = E_g m_0^2/2m_e$ (discussed above), we find $E_p = E_g m_0/m_e$.  This gives:
\beq
	\beta(\omega) = K' \left(\frac{m_e}{m_0}\right)^{-1/2} \left(\frac{E_g}{\rm eV}\right)^{-5/2} n_0^{-2} \left.\frac{(2x-1)^{3/2}}{(2x)^5}\right|_{x=E/E_g}
\eeq
The value of $K'$ depends on the theory.  A simple two-band model gives $K' = 2^9\pi e^4/5\sqrt{m_0} c^2$, which according to the paper works out to $K' = 1940$ when $\beta$ is in units of cm/GW and $E_g$ and $E_p$ are in eV.  But a four-band model gives $K' = 5200$, and a fit to the data suggests a value of $3100$.  Following the Goldilocks principle, I stick with the 3100 figure.  Thus the two-photon absorption for direct band-gap materials is given by:
\beq
	\boxed{\beta = 3100 \left(\frac{m_e}{m_0}\right)^{-1/2} \left(\frac{E_g}{\rm eV}\right)^{-5/2} n_0^{-2} \left.\frac{(2x-1)^{3/2}}{(2x)^5}\right|_{x=E/E_g} \mbox{cm}/\mbox{GW}}
\eeq

\begin{figure}[t!]
\begin{center}
\includegraphics[width=0.80\textwidth]{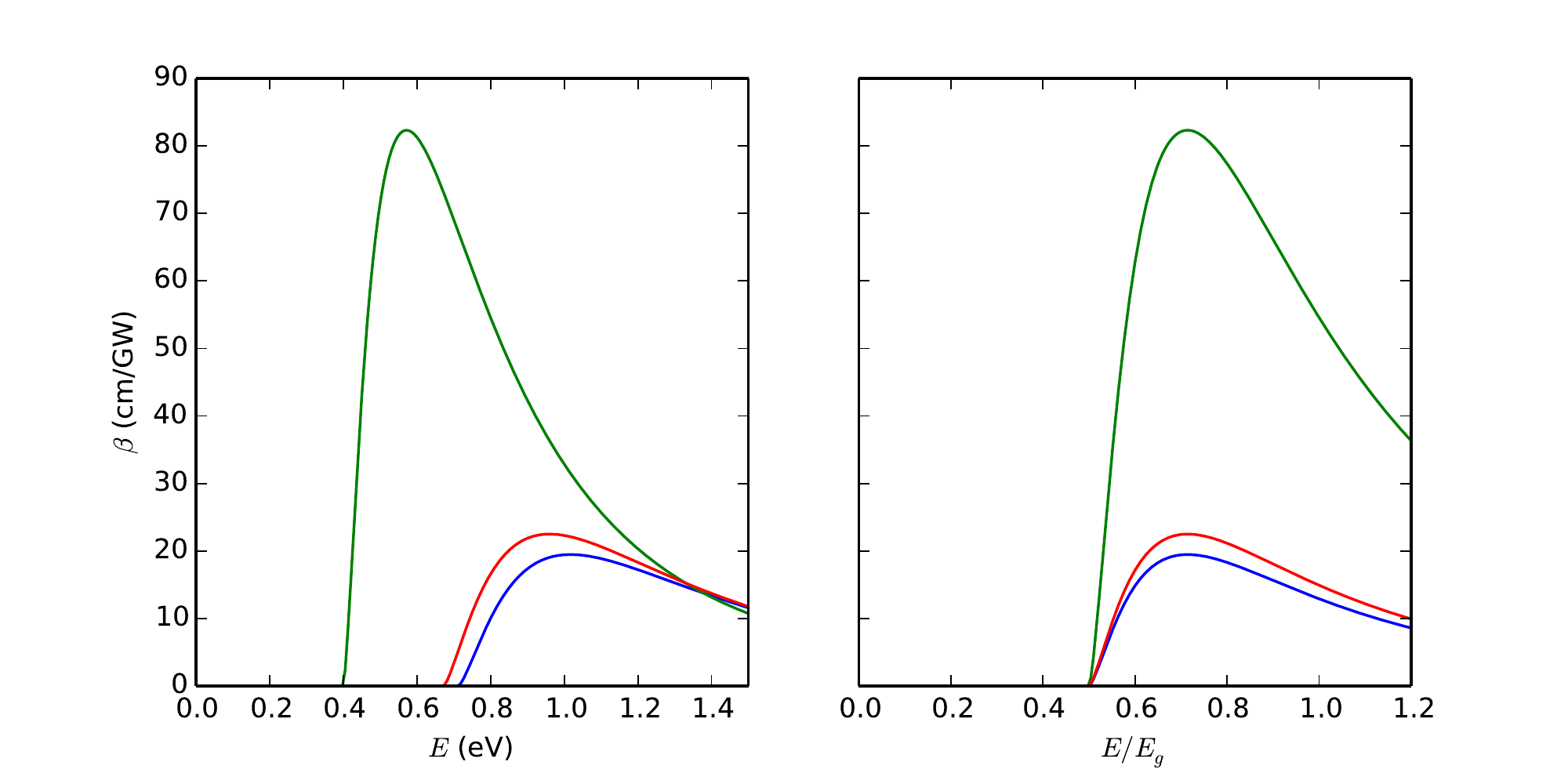}
\caption{Two-photon absorption $\beta$ for GaAs (blue), InP (red), and InGaAsP (green).}
\label{fig:01a-f3}
\end{center}
\end{figure}

In Fig. \ref{fig:01a-f3} for three materials: GaAs (blue), InP (red), and InGaAsP (1.47Q, green).  In \cite[Table 1]{Nozaki2010}, values for $\beta$ at $E = 0.8$ eV are quoted: 0.4-0.9 cm/GW for Si, 10 cm/GW for GaAs, and 40-80 cm/GW for InGaAsP.  The values for GaAs and InGaAsP agree with those in the plot above.  Since the \cite{SheikBahae1991} does not apply to silicon because of its indirect band gap, its $\beta$ cannot be calculated here.

\section{Kerr Effect}

The nonlinear refraction is obtained by a Kramers-Kr\"{o}nig transformation of the two-photon absorption, Raman effects, and various Stark shift effects \cite{SheikBahae1991, Soh2016}.  The result is:
\beq
	n_2 = K' \frac{\hbar c\sqrt{E_p}}{n_0^2 E_g^4} G(E/E_g) = 0.0612 \left(\frac{m_e}{m_0}\right)^{-1/2} \left(\frac{E_g}{\rm eV}\right)^{-7/2} n_0^{-2} G(E/E_g) \mbox{cm}^2/\mbox{GW}
\eeq
where $G(x)$ is the sum of four terms \cite{SheikBahae1991}:
\begin{eqnarray}
    G_{2PA} & = & \frac{1}{(2x)^6} \left[-\frac{3}{8}\frac{x^2}{\sqrt{1-x}} + 3x\sqrt{1-x} - 2(1-x)^{3/2} + 2\Theta(1-2x)(1-2x)^{3/2}\right] \label{eq:01a-g2pa}\\
    G_{RAM} & = & \frac{1}{(2x)^6} \left[-\frac{3}{8}\frac{x^2}{\sqrt{1+x}} - 3x\sqrt{1+x} - 2(1+x)^{3/2} + 2(1+2x)^{3/2}\right] \\
    G_{LSE} & = & \frac{1}{(2x)^6} \left[2 - (1-x)^{3/2} - (1+x)^{3/2}\right] \\
    G_{QSE} & = & \frac{1}{2^{10}x^5} \left[\frac{1}{\sqrt{1-x}} - \frac{1}{\sqrt{1+x}} - \frac{x}{2(1-x)^{3/2}} - \frac{x}{2(1+x)^{3/2}}\right]
\end{eqnarray}

For reasons explained in the paper, we have to subtract off a ``divergent term'', which goes like this:
\beq
    G_{div} = \frac{1}{(2x)^6} \left[-2 - \frac{35x^2}{8} + \frac{x}{8} \frac{3x-1}{\sqrt{1-x}} - 3x\sqrt{1-x} + (1-x)^{3/2} + \frac{x}{8} \frac{3x+1}{\sqrt{1+x}} + 3x\sqrt{1+x} + (1+x)^{3/2}\right] \label{eq:01a-gdiv}
\eeq

\begin{figure}[t]
\begin{center}
\includegraphics[width=0.90\textwidth]{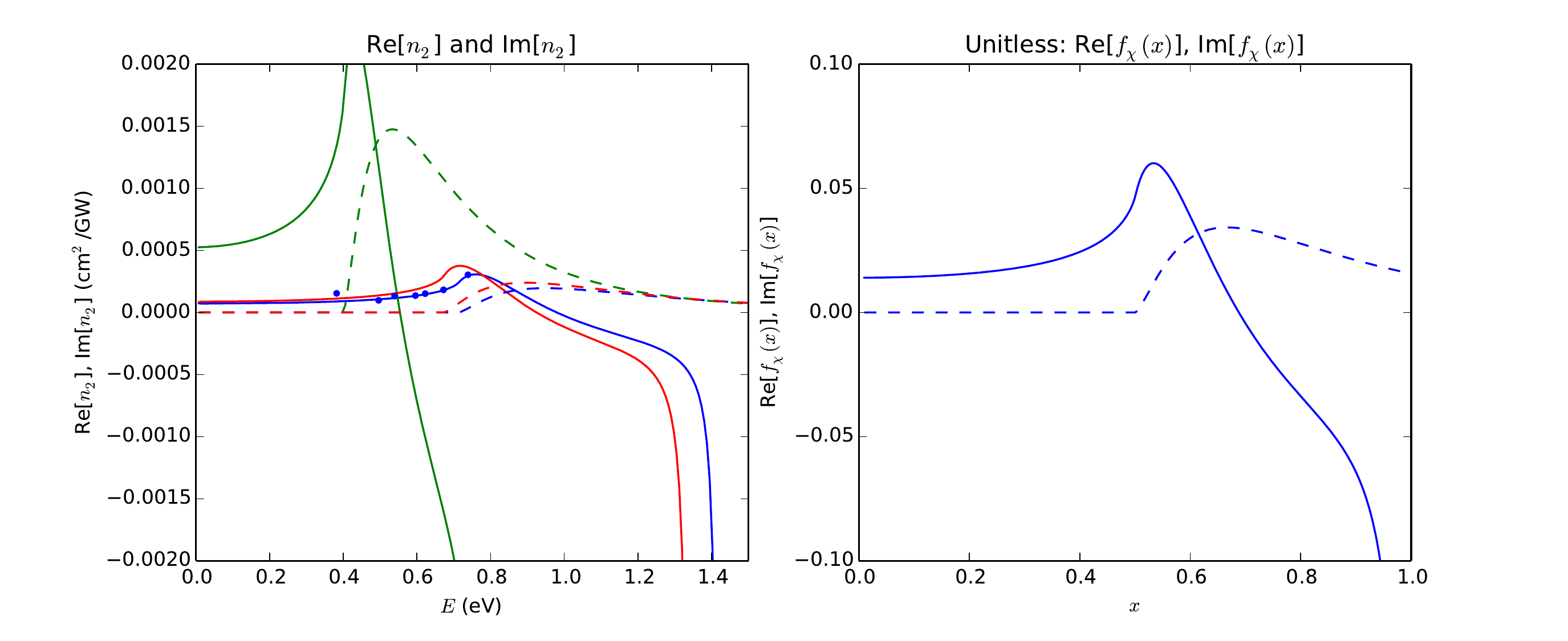}
\caption{Nonlinear dispersion and absorption for GaAs (blue), InP (red), and InGaAsP (green), with the imaginary part of $n_2$ dashed.  Dots are experimental data from \cite{Hurlbut2007}.}
\label{fig:01a-f5}
\end{center}
\end{figure}

The Kerr coefficient is shown in Figure \ref{fig:01a-f5} for GaAs (blue), InP (red) and InGaAsP (green).  The experimental values (blue dots, from \cite{Hurlbut2007}) fit quite well.  You would think that InGaAsP would be the ideal material by a long shot, but once we factor in the device size (limited by the photon wavelength, which is larger for InGaAsP) the picture is more nuanced.  All three materials will end up looking more-or-less equally good.

The nonlinear absorption coefficient $\beta$ can be related to the imaginary part of $n_2$ as follows:  Consider a beam of intensity $I_0$ which propagates through the medium with a beam profile $E_0 e^{i (n\omega/c)x-\omega t}$.  The power falls off as $e^{-2{\rm Im}[n](\omega/c)}x$, which implies that the two-photon absorption is related to the imaginary part of $n_2$ is:
\beq
	\mbox{Im}[n_2] = \frac{\beta c}{2\omega} = \frac{\beta\lambda}{4\pi} = \frac{\hbar c}{2(E/E_g)E_g} \beta
\eeq
This allows us to write $n_2$ as a complex number, i.e.
\begin{empheq}[box=\fbox]{align}
    n_2 & = K' \frac{\hbar c\sqrt{E_p}}{n_0^2 E_g^4} G(E/E_g) = \frac{0.0612\ \mbox{cm}^2/\mbox{GW}}{n_0^2 (m_e/m_0)^{1/2} (E_g/\mbox{eV})^{7/2}} f_\chi(E/E_g)
\end{empheq}
where $f_\chi(x)$ is given by:
\beq
	f_\chi = G(t) + \frac{(2x-1)^{3/2}}{(2x)^6}i
\eeq
and $G(x)$ is given by Eqs. (\ref{eq:01a-g2pa}-\ref{eq:01a-gdiv}), above.  The real part of $n_2$ gives the dispersive Kerr effect, while the imaginary part gives the two-photon absorption.  We see that $n_2$ consists of a scaling term multiplied by a function of $x$.  This dimensionless function is plotted in the right pane of Figure~\ref{fig:01a-f5}.

\section{Band-Filling}

The Kerr nonlinearity is due to virtual two-photon transitions.  Effects due to real transitions, the so-called free-carrier effects, are often much stronger.  There are two separate free-carrier effects: band-filling, which results from saturation of the absorption near the band edge, and free-carrier dispersion, which results from the collective motion of the carriers.  The latter effect is most important for photons far from the band edge, but near the band edge, the former matters most.

My results here are based off of Bennett et al.\ \cite{Bennett1990}.  This paper considers both bandfilling and free-carrier dispersion.  It also considers bandgap shrinkage, but that effect only matters at very high carrier densities.

When carriers are present in the valence or conduction band, the absorption is altered.  The usual absorption is given by the square-root law above:
\beq
	\alpha_0(E) = (K_{hh} + K_{lh}) \frac{\sqrt{E/E_g-1}}{E/E_g}
\eeq
Here we have split $K = K_{hh} + K_{lh}$ into its heavy-hole and light-hole components.  Each component is proportional to the 3/2 power of the reduced mass $\mu = (m_e^{-1}+m_h^{-1})^{-1}$, so that 
\beq
	K_{hh} = K\mu_{hh}^{3/2}/(\mu_{hh}^{3/2}+\mu_{lh}^{3/2}),\ \ \ 
	K_{lh} = K\mu_{lh}^{3/2}/(\mu_{hh}^{3/2}+\mu_{lh}^{3/2})
\eeq
Band-filling works by filling the valence and conduction bands with carriers, blocking additional abosrption which would create more carriers.  As before, the absorption is a sum of heavy-hole and light-hole components:
\beq
	\alpha(n,p,E) = \sum_{x = hh,lh} \frac{K_x \sqrt{E/E_g-1}}{E/E_g}\left[1 - f_{FD}(E_{e|x},E_{F,e}(n)) - f_{FD}(E_{h|x},E_{F,h}(p))\right]
\eeq
\begin{figure}[tbp]
\begin{center}
\includegraphics[width=0.50\textwidth]{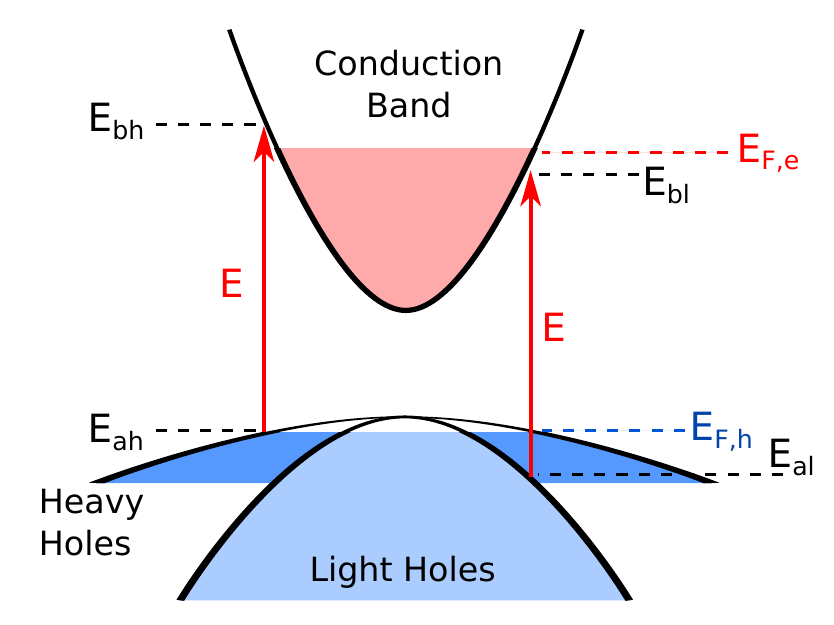}
\caption{Band diagram defining $E_{ah}, E_{al}, E_{bh}, E_{bl}$.}
\label{fig:01a-f9}
\end{center}
\end{figure}

Here $E_{ah}, E_{bh}, E_{al}$ and $E_{bl}$ are shown in Figure \ref{fig:01a-f9}.  For excitation from the heavy-hole band, an electron is promoted from energy $-E_{ah}$ to $E_g + E_{bh}$; likewise for the light hole band it is promoted $-E_{al} \rightarrow E_g + E_{bl}$.  From energy conservation, we can see that $E_{a,x} + E_{b,x} = \hbar\omega - E_g$, where $x$ is either $h$ or $l$.  From the band diagram we find that:
\beq
	E_{a} = (E - E_g) \frac{m_e}{m_e + m_{h}},\ \ \ 
	E_{b} = (E - E_g) \frac{m_{h}}{m_e + m_{h}}
\eeq
Excited electrons or holes will suppress absorption -- thus $\alpha$ contains terms proportional to the Fermi function $f_{FD}$.  The Fermi energies $E_{F,e}, E_{F,n}$ are estimated by the Nilsson approximation \cite{Nilsson1978}:
\beq
	E_F = \left[\log(N/N_0) + \frac{N}{N_0} \left[64 + 0.05524\frac{N}{N_0}\left(64 + \sqrt{N/N_0}\right)\right]^{-1/4}\right]kT
\eeq
where $N$ is the carrier number ($n$ for electrons, $p$ for holes) and $N_0$ is the effective density of states:
\beq
	N_0 = 2\left(\frac{m k T}{2\pi\hbar^2}\right)^{3/2} = 2.48 \times 10^{25} \left(\frac{m}{m_0}\right)^{3/2} \left(\frac{T}{298\,\mbox{K}}\right)^{3/2} \mbox{m}^{-3}
\eeq
and $m$ is either the electron mass or the effective hole mass $m_{dh} = (m_{hh}^{3/2}+m_{lh}^{3/2})^{2/3}$.

Any absorption effect will also generate dispersive effects through the Kramers-Kr\"{o}nig relations.  The dispersion and absorption changes are related by:
\beq
	\Delta n(E) = \frac{\hbar c}{\pi} \int_{E_g}^\infty{\frac{\Delta\alpha(E')}{(E')^2 - E^2}dE'}
\eeq
By calculating the absorption saturation and applying the Kramers-Kr\"{o}neg relation, above, we can numerically integrate to obtain the band-filling dispersion for all frequencies.

\subsection{Nondegenerate Case}

We are most interested in the nondegenerate case, where the carrier density is rather small.  The index change will also be small, but with a reasonably good cavity ($Q \sim 10^4$), even small index changes can be substantial.  In the nondegenerate case, we can replace the Fermi-Dirac filling factors with Boltzmann distributions, and derive an analytic form for the band-filling effect.

Recall that, near the band edge, the absorption goes as
\begin{eqnarray}
    \alpha & = & K\sqrt{E/E_g - 1} \\
    K & = & \frac{2\pi e^2 (2\mu)^{3/2} \sqrt{E_g}}{6m_e n \epsilon_0 h^2 c} = 2.49 \times 10^7 \frac{(\mu/m_0)^{3/2} (E_g/eV)^{1/2}}{n} \mbox{m}
\end{eqnarray}
There is a contribution due to excitation from the heavy-hole band (with $\mu^{-1} = m_{hh}^{-1} + m_e^{-1}$) as well as a contribution from the light-hole band ($\mu^{-1} = m_{lh}^{-1} + m_e^{-1}$).  This does not include absorption in the Urbach tail; this is only a small correction and does not affect the free-carrier dispersion.

Free carriers modify the absorption.  With free carriers present, we must adjust $\alpha$ by the filling factors:
\beq
    \alpha(E) \rightarrow \alpha(E) \left[1 - f_n\left(\frac{m_h(E-E_g)}{m_e+m_h}\right) - f_p\left(\frac{m_e(E-E_g)}{m_e+m_h}\right)\right]
\eeq
Let $n$ and $p_{hh}, p_{lh}$ be the electron and (heavy, light) hole concentrations.  In thermal equilibrium, the two hole concentrations are weighted by the density of states, which goes as $m_h^{3/2}$, so $p_{hh} = p\,m_{hh}^{3/2}/(m_{hh}^{3/2}+m_{lh}^{3/2})$ and likewise for $p_{lh}$.  The filling factors are given by:
\begin{eqnarray}
    f_{n,p}(E) & = & F_{n,p} e^{-E/kT} \\
    F_n & = & \frac{n}{2} \left(\frac{2\pi\hbar^2}{m_e kT}\right)^{3/2} = \frac{n}{2.48 \times 10^{25}(m_e/m_0)^{3/2}\mbox{m}^{-3}} \\
    F_{p(hh,lh)} & = & \frac{p_{hh,lh}}{2} \left(\frac{2\pi\hbar^2}{m_{hh,lh} kT}\right)^{3/2} = \frac{n}{2.48 \times 10^{25}(m_{hh,lh}/m_0)^{3/2}\mbox{m}^{-3}}
\end{eqnarray}
The absorption changes as follows:
\beq
	\Delta\alpha(E) = -\sum_{h=hh,lh} K_h\left[F_n e^{-\frac{m_h}{(m_e+m_h)kT}(E-E_g)} + F_p e^{-\frac{m_e}{(m_e+m_h)kT}(E-E_g)}\right]\sqrt{E/E_g - 1}
\eeq
This has a contribution from both heavy holes and light holes, as before.  To get the dispersion we have to take the Kramers-Kroneig transformation of this.  There will be four terms: $\Delta n_{n,hh}, \Delta n_{p,hh}, \Delta n_{n,lh}, \Delta n_{n,lh}$ -- the first two are due to the $FNn$ term above; the second two are due to the $F_p$ term.  Each is given by an integral of the following form:
\begin{eqnarray}
    \Delta n_{n|p,h}(E) & = & \frac{\hbar c}{\pi} \int_{E_g}^\infty{\frac{\Delta\alpha(E')}{(E')^2 - E^2}dE'} \nonumber \\
    & = & -\frac{\hbar c K_h F_{n|p}}{\pi} \int_{E_g}^\infty{\frac{\sqrt{E'/E_g-1}}{(E')^2 - E^2} e^{-\frac{m_{h|e}}{(m_e+m_h)kT}(E'-E_g)}dE'} \nonumber \\
    & \stackrel{x=E/E_g}{=} & -\frac{\hbar c K_h F_{n|p}}{\pi E_g} \int_{1}^\infty{\frac{\sqrt{x'-1}}{(x'-x)(x'+x)} e^{-\frac{m_{h|e}E_g}{(m_e+m_h)kT}(x'-1)}dx'} \nonumber \\
    & \stackrel{y=x-1}{=} & -\frac{\hbar c K_h F_{n|p}}{\pi E_g} \int_{0}^\infty{\frac{\sqrt{y'}}{(y'-(x-1))(y'+(x+1))} e^{-\frac{m_{h|e}E_g}{(m_e+m_h)kT}y'}dy'} \nonumber \\
    & \stackrel{r=\frac{m_{h|e}E_g}{(m_e+m_h)kT}}{=} & -\frac{\hbar c K_h F_{n|p}}{\pi E_g x} \int_{0}^\infty{\left[\frac{\sqrt{y'}}{y'-(x-1)} - \frac{\sqrt{y'}}{y'+(x+1)}\right] e^{-ry'}dy'} \nonumber \\
    & = & -\frac{\hbar c K_h F_{n|p}}{2\pi E_g x \sqrt{r}} \int_{0}^\infty{\left[\frac{\sqrt{z}}{z+r(1-x)} - \frac{\sqrt{z}}{z+r(1+x)}\right] e^{-z}dz}  \nonumber \\
    & = & -\frac{\hbar c K_h F_{n|p}}{2E_g x \sqrt{r}} \left[e^{r(1+x)}\sqrt{r(1+x)}\,\mbox{erfc}\sqrt{r(1+x)} - e^{r(1-x)}\sqrt{r(1-x)}\, \mbox{erfc}\sqrt{r(1-x)}\right] \nonumber \\
\end{eqnarray}
The rest follow by replacing $F_n$ with $F_p$ and $r$ with the appropriate quantity.  The result is an analytic (albeit messy) form for the nondegenerate band-filling effect.

\subsection{Asymptotic Solution}

A much simpler solution can be obtained by approximating $r \gg 1$ such that both $r(1+x)$ and $r(1-x)$ are large.  In practice, this can be a very good assumption.  The smallest value of $r$ is found with heavy holes for $F_p$ -- it is about $r = 8$.  For values of $x$ not too close to 1, both $r(1+x)$ and $r(1-x)$ are large enough that one can make the approximation:
\beq
	e^z \sqrt{z}\,\mbox{erfc}\sqrt{z} \sim \frac{1}{\sqrt{\pi}} - \frac{1}{2\sqrt{\pi}x}
\eeq
This gives the following $\Delta n_n$:
\bea
	\Delta n_{n,h}(E) & = & \frac{\hbar c K_h F_n}{2E_g x \sqrt{r}} \frac{1}{2\sqrt{\pi}} \left[\frac{1}{r(1+x)} - \frac{1}{r(1-x)}\right] \nonumber \\
	& = & -\frac{\hbar c K_h F_n}{2E_g x r^{3/2}} \frac{1}{2\sqrt{\pi}} \left[\frac{1}{1-x} - \frac{1}{1+x}\right] \nonumber \\
	& = & -\frac{\hbar c K_h F_n}{2\sqrt{\pi}E_g r^{3/2}} \frac{1}{1-x^2}
\eea

\begin{figure}[tb]
\begin{center}
\includegraphics[width=1.00\textwidth]{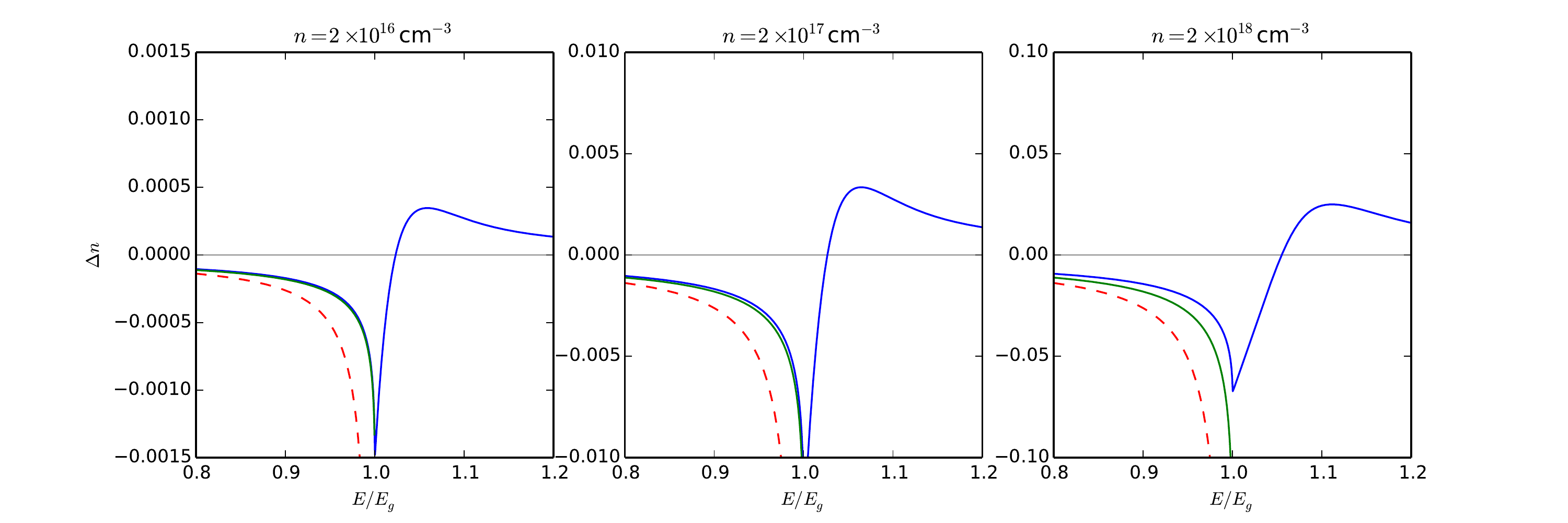}
\caption{Index change in GaAs for $n = 2 \times 10^{16},\ 2 \times 10^{17},\ 2 \times 10^{18}$cm${}^{-3}$.  Blue is full model, green is analytic model, red is asymptotic approximation.}
\label{fig:01b-f6}
\end{center}
\end{figure}

Now we plug in the values of the constants.  The free-carrier index change is:
\begin{eqnarray}
\Delta n_{n,h}(E) & = & -\frac{\hbar c K F_n}{2\sqrt{\pi}E_g r^{3/2}} \frac{1}{1-x^2} \nonumber \\
& = & -\frac{\hbar c}{2\sqrt{\pi}E_g} 
    \frac{2\pi e^2 (2\mu)^{3/2} \sqrt{E_g}}{6m_e n \epsilon_0 h^2 c} 
    \frac{n}{2} \left(\frac{2\pi\hbar^2}{m_e kT}\right)^{3/2}
    \left(\frac{(m_e+m_h)kT}{m_h E_g}\right)^{3/2}
    \frac{1}{1-x^2} \nonumber \\
& = & -\frac{\hbar^2 e^2 n}{6 m_e n_0 \epsilon_0E_g^2} 
    \frac{1}{1-x^2}
\end{eqnarray}
To calculate $\Delta n_p$, we multiply by $F_p$ rather than $F_n$.  This changes the result by a factor of $(m_e/m_h)^{3/2}$.  But we also use $r = m_e E_g/(m_e+m_h)kT$, and since $\Delta n \sim r^{-3/2}$, this changes the result by an opposite factor of $(m_h/m_e)^{3/2}$.  The result is that the final quantity is unchanged, except that $n \rightarrow p$, as follows:
\beq
	\Delta n_{p,h} = -\frac{\hbar^2 e^2 p_h}{6 m_e n_0 \epsilon_0E_g^2} \frac{1}{1-x^2}
\eeq
Thus the total contribution is:
\beq
	\Delta n = \Delta n_{n,hh} + \Delta n_{n,lh} + \Delta n_{p,hh} + \Delta n_{n,lh}
	= -\frac{\hbar^2 e^2}{6 m_e n_0 \epsilon_0E_g^2} \frac{n + n + p_{lh} + p_{hh}}{1-x^2}
\eeq
and since $p_{lh} + p_{hh} = p$, this sums to:
\beq
	\boxed{\Delta n = -\frac{\hbar^2 e^2}{2m_e n_0 \epsilon_0 E_g^2} \frac{1}{1-x^2} \frac{2n+p}{3}} \label{eq:01a-dnbf}
\eeq
This is an interesting result.  The electrons contribute twice as much to the bandfilling effect because each electron contributes to the absorption via two excitation channels, but each hole only contributes to one.

In Figure \ref{fig:01b-f6} we plot the full numerical result (blue) is plotted against the nondegenerate analytical result (green, only defined for $E < E_g$) and the asymptotic solution (red).  The asymptotic solution always does poorly near the band gap, but is a good approximation far from it, i.e.\ for $x < 0.9$.  The nondegenerate result is almost always a good approximation, unless the carrier concentration is so high that the valence and conduction bands are degenerately filled.

\section{Discrete-Carrier Derivation}

The same effect can be studied by treating the individual carriers discretely rather than as a distribution.  The analysis is simpler this way.  To start, the absorption without carriers is given by Fermi's golden rule, as follows:
\beq
	\alpha = \frac{\pi\hbar e^2}{m_0^2 n_0 \epsilon_0 e E} |p_{cv} \cdot \hat{E}|^2 \rho_j(E)
\eeq
where $\rho_j(E)$ is the joint density of states.  Here, let us look at the carriers as individual particles rather than a distribution.  As shown in Figure~\ref{fig:01a-f11}, carriers can ``block'' optical transitions, reducing the absorption at certain wavelengths.  Each electron blocks two transitions, while each hole can only block one.  The absorption is modified as follows:
\beq
	\Delta\alpha = -\sum_{E_i \in E_{\rm blocked}} \frac{\pi\hbar e^2 |p_{cv}\cdot \hat{E}|^2}{m_0^2 n_0 \epsilon_0 c E} \delta(E - E_i)
\eeq
Let $E_{eh,i}$ and $E_{el,i}$ be the blocked transitions for an electron or energy $\epsilon_i$, and let $E_{h,j}$ and $E_{l,k}$ be the blocked transitions for heavy holes of energy $\epsilon_j$ and light holes or energy $\epsilon_k$.  The change in absorption is:
\bea
	\Delta\alpha & = & -\frac{\pi\hbar e^2 |p_{cv}|^2}{m_0^2 n_0 \epsilon_0 c E} \biggl[\sum_{i,\rm el}  |\hat{p}_{cv}\cdot \hat{E}|^2\Bigl[\delta(E - E_{eh,i}) + \delta(E - E_{el,i})\Bigr] \nonumber \\
	& & \qquad\qquad\qquad +\sum_{j,\rm hh} |\hat{p}_{cv}\cdot \hat{E}|^2 \delta(E - E_{h,j})
	+\sum_{k,\rm lh} |\hat{p}_{cv}\cdot \hat{E}|^2 \delta(E - E_{l,k})\biggr]	
\eea
Applying the Kramers-Kr\"{o}nig theorem, the delta functions are integrated out, yielding the following nonlinear dispersion:
\bea
	\Delta n & = & -\frac{\hbar^2 e^2 |p_{cv}|^2}{m_0^2 n_0 \epsilon_0} \biggl[\sum_{i,\rm el}  |\hat{p}_{cv}\cdot \hat{E}|^2\Bigl[\frac{1}{E_{eh,i}(E_{eh,i}^2 - E^2)} + \frac{1}{E_{el,i}(E_{el,i}^2 - E^2)}\Bigr] \nonumber \\
	& & \qquad\qquad\qquad+\sum_{j,\rm hh} \frac{|\hat{p}_{cv}\cdot \hat{E}|^2}{E_{h,j}(E_{h,j}^2 - E^2)}
	+\sum_{k,\rm lh} \frac{|\hat{p}_{cv}\cdot \hat{E}|^2}{E_{h,k}(E_{h,k}^2 - E^2)}\biggr]	
\eea
\begin{figure}[tbp]
\begin{center}
\includegraphics[width=0.50\textwidth]{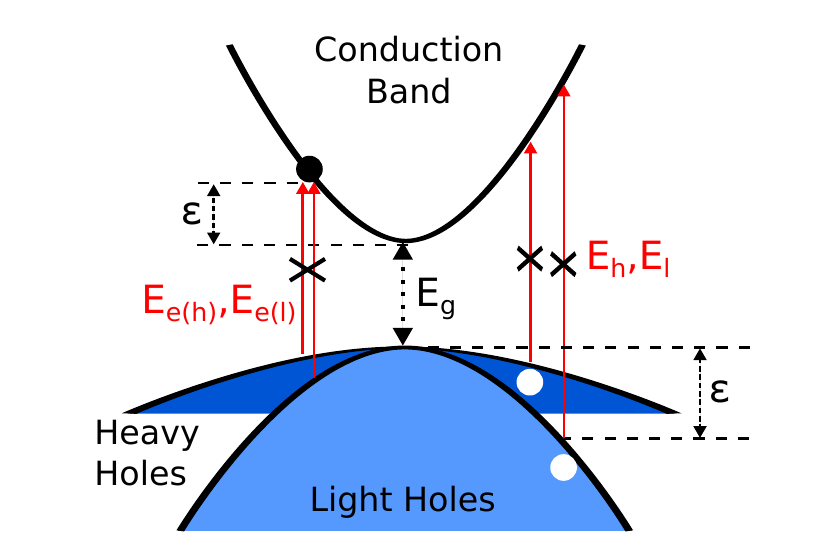}
\caption{Each conduction-band electron blocks two transitions, $E_{eh}$ and $E_{el}$, while each valence-band hole only blocks one -- $E_h$ or $E_l$, depending on the band.}
\label{fig:01a-f11}
\end{center}
\end{figure}

If there are many carriers, we can average over the geometric factor $|\hat{p}_{cv}\cdot \hat{E}|^2$ to get $1/3$.  If not too close to the band gap, i.e. $E_g - E \gg kT$, then we can replace $E_{eh}, E_{el}, E_h, E_l \rightarrow E_g$ without too much trouble.  All of the terms inside the sums become the same and independent of the particular carrier's energy, giving the following dispersion:
\bea
	\Delta n & = & -\frac{\hbar^2 e^2 |p_{cv}|^2}{m_0^2 n_0 \epsilon_0} \frac{1}{3} \frac{1}{E_g(E_g^2 - E^2)} \left[n + n + p_{hh} + p_{lh}\right] \nonumber \\
	& = & \frac{\hbar^2 e^2 |p_{cv}|^2}{m_0^2 n_0 \epsilon_0 E_g^3} \frac{1}{1-x^2} \frac{2n+p}{3} \nonumber \\
	& = & \frac{\hbar^2 e^2}{2m_e n_0 \epsilon_0 E_g^2} \frac{1}{1-x^2} \frac{2n+p}{3} 
\eea
This matches the asymptotic result obtained in Eq.~(\ref{eq:01a-dnbf}).  The results agree because they are both based on the same theory.  The previous one was derived for a continuous distribution of carriers, which requires more work but gives useful and more accurate results when the photon energy is close to $E_g$ or when the bands become degenerately filled.  The current derivation is simpler, but it requires a large sum and only simplifies when we are reasonably far from the band gap.

\section{Free-Carrier Dispersion}

Free carrier dispersion is treated using the Drude model.  Given a carrier density $N$, the index of refraction is modified as follows:
\beq
	n^2 \rightarrow n_0^2\left(1 - \frac{N e^2/m_e n_0^2 \epsilon_0}{\omega^2 + i\omega/\tau}\right)
\eeq
In the high-frequency limit ($\omega \gg \omega_p$) this becomes:
\beq
	n^2 = n_0^2 - \frac{N e^2}{m_e \epsilon_0} \left(\frac{1}{\omega^2} + \frac{i}{\omega^3\tau}\right) \Rightarrow \Delta n = -\frac{N e^2}{2m_e n_0 \epsilon_0} \left(\frac{1}{\omega^2} + \frac{i}{\omega^3\tau}\right)
\eeq
There will be two contributions -- one to electrons and one to holes.  The effective hole mass (due to two different valence bands with two different populations), is $\bar{m}_h = (m_hh^{3/2}+m_lh^{3/2})/(m_hh^{1/2}+m_lh^{1/2})$.  

\begin{figure}[t]
\begin{center}
\includegraphics[width=1.00\textwidth]{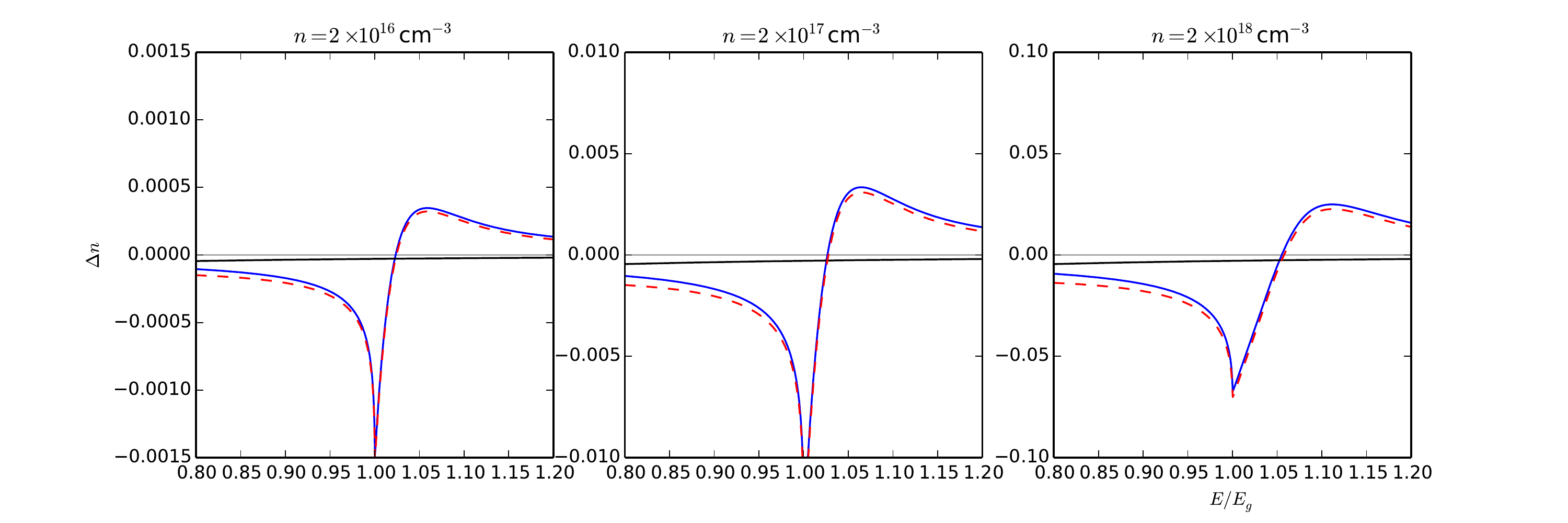}
\caption{Index change in GaAs for $n = 2 \times 10^{16},\ 2 \times 10^{17},\ 2 \times 10^{18}$cm${}^{-3}$.  Blue is band-filling effect, black is free-carrier dispersion, red is combined effect.}
\label{fig:01b-f7}
\end{center}
\end{figure}

The carrier index change can be expressed in terms of $x = E/E_g$:
\beq
	\Delta n = -\frac{\hbar^2 e^2}{2n_0 \epsilon_0 E_g^2} \left[\frac{n}{m_e} + \frac{p}{\bar{m}_h}\right] \left(\frac{1}{x^2} + \frac{1}{x^3}\frac{i\hbar}{E_g\tau}\right)
\eeq
The effective hole mass $\bar{m}_h$ is dominated by heavy holes, so if $n = p$ we can just ignore this part.  It is also often justifiable to ignore the absorption term.  The resulting equation is:
\beq
	\boxed{\Delta n = -\frac{\hbar^2 e^2}{2 m_e n_0 \epsilon_0 E_g^2} \frac{1}{x^2} n = \frac{6.89 \times 10^{-28}\mbox{m}^3}{(E_g/\mbox{eV})^2 (m_e/m_0) n_0} \frac{1}{x^2}} \label{eq:01a-fcd-asym}
\eeq
The scaling with material parameters -- band gap, index, electron mass -- is the same as for BFD.  Unlike BFD, FCD is most prominent when the frequency is very small, $x \ll 1$.  Unfortunately, light with $x \ll 1$ will not show strong free-carrier effects because it will never excite free carriers!

If $n = p \equiv N_c$, we can combine the band-filling and free-carrier effects.  In the asymptotic limit, the prefactors are the same, so all we need to do is combine $1/x^2$ and $1/(1-x^2)$.  This gives the equation:
\beq
	\Delta n = -\frac{\hbar^2 e^2}{2 m_e n_0 \epsilon_0 E_g^2} \frac{1}{x^2(1-x^2)} N_c
\eeq
This equation, which includes {\it both} free-carrier and band-filling effects, agrees with results in the literature \cite{Said1992}.

\section{Thermal Dispersion: Band-Gap Shrinkage}
\label{sec:01a-thermooptic}

The \textit{thermo-optic effect} causes a material's index of refraction to change with temperature.  Like any other dispersive effect, this can be related to a change in the absorption through Kramers-Kroneig.  In this case, the absorption is due to band-gap shrinkage.

The band gap of a material is a function of the temperature.  For most materials, the relationship is:
\beq
	E_g(T) = E_g(0) - \frac{a T^2}{T + \theta}
\eeq
where $a$ is an empirical constant and $\theta$ is the Debye temperature  \cite{Moss2013}.  This is roughly linear for most materials around 300 K, and usually has a negative slope.  

\begin{table}[htbp]
\centering
\begin{tabular}{c|ccc|cc}
	Material & $E_g(0)$ (eV) & $a$ (eV/K) & $\theta$ (K) & $E_g|_{300K}$ (eV) & $dE_g/dT|_{300K}$ (eV/K) \\ \hline
	Si & $1.17$ & $4.7 \times 10^{-4}$ & $636$ & $1.12$ & $-2.5 \times 10^{-4}$ \\
	GaAs & $1.52$ & $5.4\times10^{-4}$ & $204$ & $1.42$ & $-4.5 \times 10^{-4}$ \\
	InP & $1.42$ & $4.9 \times 10^{-4}$ & $327$ & $1.34$ & $-3.6 \times 10^{-4}$
\end{tabular}
\caption{Temperature-dependence parameters for Si, GaAs, and InP \cite{Moss2013}.}
\label{tab:01a-t1}
\end{table}

\begin{figure}[tbp]
\begin{center}
\includegraphics[width=0.6\textwidth]{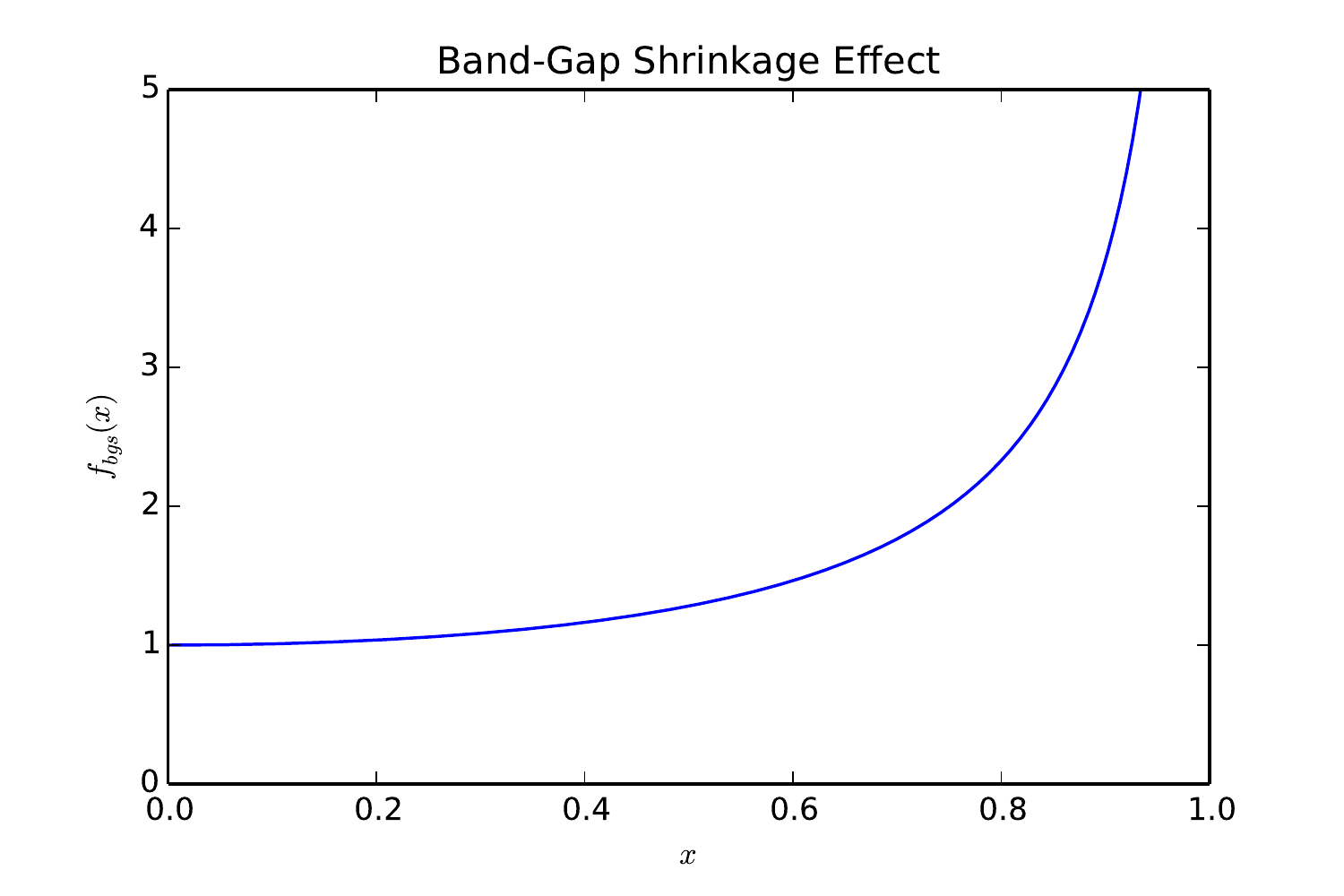}
\caption{Dimensionless band-gap shrinkage dispersion function $f(x)$.  See Eq.~(\ref{eq:01a-bgs})}
\label{fig:01a-f10}
\end{center}
\end{figure}

As we can see, all materials have similar slopes -- around $-3 \times 10^{-4}$ eV/K.  The change in band gap can, of course, be approximated as $\Delta E_g = (dE_g/dT)\Delta T$.  By shifting the whole absorption curve to the left, this induces a change in the absorption spectrum, as follows:
\beq
	\Delta \alpha(E) 
	= -\Delta E_g \frac{\partial\alpha}{\partial E} 
	= -\Delta E_g \frac{\partial}{\partial E}\left[K\frac{(E/E_g-1)^{1/2}}{E/E_g}\right]
	= \left.\frac{K\,\Delta E_g}{2E_g} \frac{x-2}{x^2\sqrt{x-1}}\right|_{x=E/E_g}
\eeq
which causes:
\begin{eqnarray}
    \Delta n(E) & = & \frac{\hbar c}{\pi} \int_{E_g}^\infty{\frac{\Delta\alpha(E')}{(E')^2 - E^2}dE'} \nonumber \\
    & = & \frac{\hbar c}{\pi} \frac{K\,\Delta E_g}{2E_g^2} \int_1^{\infty}{\frac{x'-2}{((x')^2-x^2)(x')^2\sqrt{x'-1}}} \nonumber \\
    & = & -\frac{\hbar c}{\pi} \frac{K\,\Delta E_g}{2E_g^2} \frac{\pi}{4x^3} \left[\frac{2-x}{\sqrt{1-x}} - \frac{2+x}{\sqrt{1+x}}\right] \nonumber \\
    & = & -\frac{\hbar c K\,\Delta E_g}{16E_g^2} \frac{2}{x^3} \left[\frac{2-x}{\sqrt{1-x}} - \frac{2+x}{\sqrt{1+x}}\right]_{x=E/E_g}
\end{eqnarray}
which may be written as
\beq
	\boxed{\Delta n(E) = -0.411 \frac{\Delta E_g/\mbox{eV}}{(E_g/\mbox{eV})^{3/2} n_0} f_{bgs}(E/E_g),\ \ \ f_{bgs}(x) = \frac{2}{x^3} \left[\frac{2-x}{\sqrt{1-x}} - \frac{2+x}{\sqrt{1+x}}\right]} \label{eq:01a-bgs}
\eeq
where $f(x)$ is plotted in Figure \ref{fig:01a-f10}, above.

Plugging in GaAs parameters, we arrive at $\Delta n = (3.0\times 10^{-5} \Delta T/\mbox{K}) f(x)$.  This is within a factor of 1.5 of the reported value.  I believe that the discrepancy arises from an inaccurate value of $K$ (the formula for $K$ is based on an approximation, and is off by a factor of 1.73) and non-parabolicity of the bands.  But the general shape of $f_{bgs}(x)$ will not change.

It is important to notice that the thermo-optic coefficient $dn/dT$ diverges as we approach the band edge.  So near the band edge, \textit{both} band-filling and thermal effects become very large.  In fact, comparing (\ref{eq:01a-fcd-asym}), we see that both diverge as $(1-x)^{-1}$ in this limit.

\ifstandalone{\bibliography{NoteRefs}{}
\bibliographystyle{alpha}
}
\ifdefined\multidoc\else\input{Header}\fi

\ifstandalone{\setcounter{chapter}{1}}
\chapter{Classical Coupled Mode Theory}

Many photonic devices are constructed from solid-state materials and owe their behavior to bulk optical nonlinearities.  In the classical, many-photon limit, the optical field can be modeled with Maxwell's equations using coupled mode theory.  Coupled mode theory is an approximation -- assuming that the system and its inputs vary on timescales very slow compared to the optical timescale, we can model the electromagnetic field, which has an infinite number of degrees of freedom, in terms of a small number of eigenmodes.  In an optical cavity, these modes can couple to each other if there are imperfections or nonlinearities in the cavity.  In addition, they couple to input-output fields.

Classical coupled-mode theory is a very well-studied subject.  None of the information in this chapter is new; rather, it serves to complement the quantum coupled-mode equations derived in Chapters \ref{ch:01} and \ref{ch:02}.  We expect a quantum-classical ``correspondence'' between these theories at high photon numbers.  Many of the quantum models were highly abstract and simplified, using dimensionless units and parameters that have no obvious connection to reality.  Making a correspondence with the classical models lets us tie those abstract models to real device properties -- materials, geometry, etc. -- and obtain accurate quantum models for real devices that can be built in the lab.

\section{Maxwell's Equations}

In any device, the electromagnetic field evolves according to Maxwell's Equations, which take the following form \cite{Landau1971, Griffiths1999}:
\begin{align}
	\nabla \cdot (\epsilon E) &= 0 & \nabla \times E &= -\frac{\partial B}{\partial t} \\
	\nabla \cdot B &= 0 & \nabla \times B &= \mu\epsilon\frac{\partial E}{\partial t}
\end{align}
Combining these, we see that the electric field satisfies the Helmholtz equation:
\beq
	\nabla\times\nabla\times E = -\mu\epsilon \frac{\partial^2 E}{\partial t^2}
\eeq
Usually the magnetic field $B$ does not play a major role in the dynamics, so modes are usually defined in terms of the electric field.  Say that:
\beq
	E = \mbox{Re}\left[\sum_\omega{\sqrt{2\hbar\omega/\epsilon_0} A_\omega E_\omega(x)e^{-i\omega t}}\right],\ \ \ 
	B = \mbox{Re}\left[\sum_\omega{\sqrt{2\hbar\omega/\epsilon_0} A_\omega B_\omega(x)e^{-i\omega t}}\right] \label{eq:01b-ebmodes}
\eeq
Maxwell's equations become:
\begin{align}
	\nabla \cdot (\epsilon E_\omega) &= 0 & \nabla \times E_\omega &= i\omega B_\omega \\
	\nabla \cdot B_\omega &= 0 & \nabla \times B_\omega &= -i\omega\,\mu\epsilon E_\omega
\end{align}
with the Helmholtz equation:
\beq
	\nabla\times\nabla\times E_\omega = \frac{\omega^2}{c^2}\epsilon_r E_\omega \label{eq:01b-hh-ft}
\eeq
Define an inner product for modes:
\beq
	\ip{E_\omega}{E_{\omega'}} = \int{\epsilon_r\,E_\omega^* E_{\omega'}}
\eeq
One can show using the Helmholtz equation that fields of different $\omega$ are orthogonal.  With a little extra work, we can compute the total electromagnetic energy.  Adopting the normalization $\ip{E_\omega}{E_\omega'} = \delta_{\omega,\omega'}$, this sharply resembles the harmonic-oscillator potential:
\beq
	H = \int{\frac{B^2}{2\mu} + \frac{\epsilon}{2} E^2} = \frac{1}{2}\sum_\omega 2\hbar\omega A_\omega^* A_{\omega} \ip{E_\omega}{E_\omega} = \sum_\omega \hbar\omega A_\omega^*A_\omega
\eeq
Going from classical to quantum mechanics, the c-number $A_\omega$ is replaced with the annihilation operator $a_\omega$, and the electromagnetic energy becomes a quadratic Hamiltonian $H = \sum_\omega \hbar\omega a_\omega^\dagger a_\omega$.  Everything else is just quantum mechanics.  Maxwell's equations and coupled mode theory are relevant because they dictate the spectrum of allowed modes $\omega$.

\section{Time-Dependent Perturbation Theory}

Perturbation theory will be used to treat the behavior of defective optical cavities, and cavities with nonlinearities, under the assumption that manufacturing defects and optical nonlinearities are small effects that can be treated as perturbations.  This is almost always a good approximation.

To apply perturbation theory, the amplitudes $A_\omega$ must be promoted to slowly-varying, time-dependent quantities: $A_\omega \rightarrow A_\omega(t)$.  The electric and magnetic fields become:
\beq
	E = \mbox{Re}\left[\sum_\omega{\sqrt{2\hbar\omega/\epsilon_0} A_\omega(t) E_\omega(x)e^{-i\omega t}}\right],\ \ \ 
	B = \mbox{Re}\left[\sum_\omega{\sqrt{2\hbar\omega/\epsilon_0} A_\omega(t) B_\omega(x)e^{-i\omega t}}\right]
\eeq
The Helmholtz equation here becomes:
\begin{align}
	& \mbox{Re}\left[\sum_\omega \sqrt{2\hbar\omega/\epsilon_0} A_\omega(t) E_\omega(x) e^{-i\omega t}\right] \nonumber \\
	& \quad = \mbox{Re}\left[-\mu\epsilon \sum_\omega{\sqrt{2\hbar\omega/\epsilon_0} \left[\ddot{A}_\omega(t) - 2i\omega \dot{A}_\omega(t) - \omega^2 A_\omega(t)\right] E_\omega(x) e^{-i\omega t}}\right] + \mu \frac{\partial^2 P(x,t)}{\partial t^2}
\end{align}
We can drop the $\ddot{A}_\omega(t)$ term because $A_\omega(t)$ varies slowly enough that $\ddot{A}_\omega(t) \ll \omega \dot{A}_\omega(t)$.  Likewise, the frequency-domain Helmholtz equation (\ref{eq:01b-hh-ft}) can be used to eliminate the left-hand term and the $\omega^2$ term.  This gives rise to the following equation:
\beq
	\mbox{Re}\left[-2i\omega \sum_\omega{\sqrt{2\hbar\omega/\epsilon_0} \dot{A}_\omega(t) E_\omega(x) e^{-i\omega t}}\right] = \frac{1}{\epsilon}\frac{\partial^2 P(x,t)}{\partial t^2} \label{eq:01b-hhtz}
\eeq
$P(x,t)$ is the perturbation polarization, which takes the form
\beq
	P_i = \epsilon_0\Bigl[ \underbrace{\delta\chi^{(1)}_{ij} E_j}_{P^{(1)}} + \underbrace{\chi^{(2)}_{ijk} E_j E_k}_{P^{(2)}} + \underbrace{\chi^{(3)}_{ijkl} E_j E_k E_l}_{P^{(3)}}\Bigr] \equiv \epsilon_0 \left[\delta\chi^{(1)}:E + \chi^{(2)}:E\,E + \chi^{(3)}:E\,E\,E\right]
\eeq
The first term is due to imperfections in the device; the next two terms are the $\chi^{(2)}$ and $\chi^{(3)}$ nonlinearities.  $P$ itself can be written as a sum of slowly-varying wave trains $P_\omega(x, t) e^{-i\omega t}$, in a manner analogous to $E$:
\beq
	P(x, t) = \mbox{Re}\left[\sum_\omega{\sqrt{2\hbar\omega/\epsilon_0} P_\omega(x, t)e^{-i\omega t}}\right] \label{eq:01b-pmodes}
\eeq
Here, $P_\omega(x, t)$ is a slowly-varying amplitude function in $t$.  Equation (\ref{eq:01b-hhtz}) can be re-expressed as follows:
\beq
	\mbox{Re}\left[-2i\sum_\omega{\omega\sqrt{2\hbar\omega/\epsilon_0} \dot{A}_\omega(t) E_\omega(x) e^{-i\omega t}} + \sum_\omega{\frac{\sqrt{2\hbar\omega/\epsilon_0}}{\epsilon} (-\omega^2 P_\omega(x, t) - 2i\omega\dot{P}_\omega(x,t) + \ddot{P}_\omega(x,t))e^{-i\omega t}}\right] = 0
\eeq
Since $A_\omega$ and $P_\omega$ are slowly varying and holds for all $t$, we can ignore the $\dot{P}_\omega$ and $\ddot{P}_\omega$ terms, and can assume that the equality holds for both real and imaginary parts, giving:
\beq
	-2i\sum_\omega{\omega\sqrt{2\hbar\omega/\epsilon_0} \dot{A}_\omega(t) E_\omega(x) e^{-i\omega t}} - \sum_\omega{\frac{\omega^2\sqrt{2\hbar\omega/\epsilon_0}}{\epsilon} P_\omega(x, t)e^{-i\omega t}} = 0
\eeq
Pre-multiplying by $E_\omega^*$ and integrating, one finds:
\beq
	-2i\dot{A}_\omega(t) = -\sum_{\omega'} \frac{\omega'(\omega'/\omega)^{3/2}}{\epsilon} \ip{E_\omega(x)}{P_\omega(x,t)} e^{i(\omega-\omega')t}
\eeq
Now we invoke the rotating wave approximation and ignore all terms with $\omega \neq \omega'$.  This is valid whenever the envelope functions $A_\omega, P_\omega$ vary on timescales much slower than $\omega$.  The result is:
\beq
	\boxed{\frac{dA_\omega}{dt} = \frac{i\omega}{2\epsilon_0} \int{E_\omega(x)^*P_\omega(x,t)}} \label{eq:01b-env}
\eeq
It remains to compute $P_\omega(x, t)$.  This depends on the kind of perturbation we are looking at.

\section{Linear Perturbations}
\label{sec:01b-lin}

\subsection{Nondegenerate Modes}

First, nondegenerate linear perturbations.  It should be pretty obvious from
\beq
	P^{(1)}_i = \mbox{Re}\left[\sum_\omega{\epsilon_0\sqrt{2\hbar\omega/\epsilon_0} A_\omega(t) \delta\chi^{(1)}:E_\omega(x)}\right]
\eeq
that $P_\omega = \epsilon_0 (\delta\chi^{(1)}:E_\omega)$.  From this we find:
\beq
	\left.\frac{dA_\omega}{dt}\right|_{\chi^{(1)}} = \left[\frac{i\omega}{2} \int{E_\omega^* (\delta\chi^{(1)}:E_\omega)}\right]A_\omega
\eeq
If $\chi^{(1)}$ is isotropic, that is, $\chi^{(1)}_{ij} = \chi^{(1)} \delta_{ij}$, the material has an isotropic dielectric constant related to the index of refraction by $\epsilon \equiv 1 + \chi^{(1)} = n^2$.  This means that $\delta\chi = 2n\,\delta n$.  The equation for $A_\omega$ is:
\beq
	\boxed{\left.\frac{dA_\omega}{dt}\right|_{\chi^{(1)}} = \left[i\omega \int{\delta(\log n)\ \epsilon_r E_\omega^* E_\omega}\right]A_\omega \equiv i\omega \bbracket{E_\omega}{\delta(\log n)}{E_\omega} A_\omega} \label{eq:01b-lin}
\eeq
where $\bracket{E_i}{f}{E_j} \equiv \int{\epsilon_r E_i^* E_j f(x) d^3x}$ is the matrix element of $f(x)$ with respect to modes $E_i$ and $E_j$.

Here's a good sanity check -- given the scaling of the Helmholtz equation, we know that increasing $n$ uniformly by a small amount, $n \rightarrow (1 + \epsilon)n$, will \textit{decrease} the frequency by the same factor: $\omega \rightarrow (1 - \epsilon)\omega$.  Compare this to Eq.~(\ref{eq:01b-lin}), where $\delta(\log n) = \delta n/n = \epsilon$.  One finds:
\beq
	\left.\frac{dA_\omega}{dt}\right|_{(1+\epsilon)n} = i\omega \bracket{E_\omega}{\epsilon}{E_\omega} A_\omega = i\epsilon\omega A_\omega \Rightarrow A_\omega(t) = e^{i\epsilon\omega}
\eeq
So instead of having a time dependence $e^{-i\omega t}$, it goes as $A_\omega(t)e^{-i\omega t} = e^{-i(1-\epsilon)\omega t}$, so the frequency has changed as expected: $\omega \rightarrow (1-\epsilon)\omega$, as predicted by the scaling argument.

\subsection{Degenerate Modes}

In the degenerate case, there are $N$ modes $E_1, \ldots, E_N$ with the same frequency $\omega$.  One finds:
\beq
	\frac{dA_i}{dt} = \left[\frac{i\omega}{2} \int{E_i^* (\delta\chi^{(1)}:E_j)}\right]A_j
\eeq
Or for isotropic materials:
\beq
	\frac{dA_i}{dt} = \sum_j i\omega \bbracket{E_i}{\delta(\log n)}{E_j} A_j \label{eq:01b-lin1}
\eeq

\subsection{Linear Absorption}

Linear absorption in a material is gives rise to a small imaginary contribution to the index of refraction.  This can be deduced by considering a traveling wave, for which the electric field takes the form:
\beq
	E(t) \sim e^{i(kx-\omega t) - \alpha x/2} = e^{i((k+i\alpha/2)x-\omega t)}
\eeq
where $\alpha \ll k$ is the absorption coefficient.  Most people in the literature quote absorption in terms of $\alpha$ rather than Im[$n$].  The complex index of refraction is given by:
\beq
	n = \frac{c(k+i\alpha/2)}{\omega} = \frac{ck}{\omega} + i\frac{c\alpha}{2\omega} 
	\Rightarrow \delta n = i\frac{c\alpha}{2\omega}
\eeq
from which we can derive the absorption law:
\beq
	\frac{dA_i}{dt} = -\frac{c\alpha}{2n} A_i \label{eq:01b-lin2}
\eeq

\section{$\chi^{(2)}$ Effects}

\subsection{Degenerate (SHG)}

In the degenerate case, there are two modes: $E_\omega$ and $E_{2\omega}$, and the $\chi^{(2)}$ effect connects them.  None of the modes are degenerate -- the term ``degenerate'' comes about because the signal and idler mode are the same (unlike sum and difference-frequency generation).  The polarization takes the following form:
\bea
	P^{(2)} & \!\!\!=\!\!\! & \epsilon_0 \chi^{(2)}:\mbox{Re}\left[\sum_\omega{\epsilon_0\sqrt{2\hbar\omega/\epsilon_0} A_\omega(t) E_\omega(x)e^{-i\omega t}}\right]\mbox{Re}\left[\sum_{\omega'}{\epsilon_0\sqrt{2\hbar\omega'/\epsilon_0} A_{\omega'}(t) E_{\omega'}(x)}e^{-i\omega't}\right] \nonumber \\
	& \!\!\!=\!\!\! & \mbox{Re}\left[2\sqrt{2}\hbar\omega\, A_{2\omega}A_\omega^* (\chi^{(2)}:E_\omega^*E_{2\omega})e^{-i\omega t}\right] + \mbox{Re}\left[\hbar\omega\,A_\omega^2 (\chi^{(2)}:E_\omega E_\omega)e^{-2i\omega t}\right] \nonumber \\
	& \!\!\!=\!\!\! & \mbox{Re}\left[\epsilon_0\sqrt{\frac{2\hbar\omega}{\epsilon_0}} 2\sqrt{\frac{\hbar\omega}{\epsilon_0}} A_{2\omega}A_\omega^* (\chi^{(2)}:E_\omega^*E_{2\omega})e^{-i\omega t}\right] \nonumber \\
	& & + \mbox{Re}\left[\epsilon_0\sqrt{\frac{2\hbar(2\omega)}{\epsilon_0}} \frac{\sqrt{\hbar\omega/\epsilon_0}}{2}\,A_\omega^2 (\chi^{(2)}:E_\omega E_\omega)e^{-2i\omega t}\right]	
\eea
from which we conclude tha
\bea
	P_\omega & = & 2\epsilon_0\sqrt{\frac{\hbar\omega}{\epsilon_0}} A_{2\omega}A_\omega^* (\chi^{(2)}:E_\omega^*E_{2\omega}) \\
	P_{2\omega} & = & \frac{1}{2} \sqrt{\hbar\omega/\epsilon_0}\,A_\omega^2 (\chi^{(2)}:E_\omega E_\omega)
\eea
Applying the envelope equation (\ref{eq:01b-env}), we find:
\bea
	\left.\frac{dA_{2\omega}}{dt}\right|_{\chi^{(2)}} & = & \frac{i\omega}{2} \sqrt{\frac{\hbar\omega}{\epsilon_0}} \left(\int{\chi^{(2)}:E_{2\omega}^* E_\omega E_\omega}\right)A_\omega A_\omega \\
	\left.\frac{dA_\omega}{dt}\right|_{\chi^{(2)}} & = & i\omega \sqrt{\frac{\hbar\omega}{\epsilon_0}} \left(\int{\chi^{(2)}:E_\omega^* E_\omega^*E_{2\omega}}\right) A_{2\omega}A_\omega^*
\eea
Define the a dimensionless SHG mode coupling $\epsilon$ as follows.
\beq
	\boxed{\epsilon \equiv i\omega \sqrt{\frac{\hbar\omega}{\epsilon_0}} \left(\int{\chi^{(2)}: E_\omega^* E_\omega^* E_{2\omega}}\right)}
\eeq
In terms of this $\epsilon$, the field equations become:
\begin{empheq}[box=\fbox]{align}
	\left.\frac{dA_{2\omega}}{dt}\right|_{\chi^{(2)}} & = -\frac{1}{2}\epsilon^* A_\omega^2 \\
	\left.\frac{dA_\omega}{dt}\right|_{\chi^{(2)}} & = \epsilon A_{2\omega} A_\omega^*
\end{empheq}
These are the classical field equations for a degenerate OPO.  Notice how they conserve energy -- for every $2\omega$ photon created, two $\omega$ photons must be annihilated:
\beq
	\frac{dN_{\omega}}{dt} = 2\mbox{Re}\left[A_\omega^* \frac{dA_\omega}{dt}\right] = 2\mbox{Re}\left[\epsilon A_{2\omega}(A_\omega^*)^2\right] = 4 \mbox{Re}\left[\frac{1}{2}\epsilon^* A_{2\omega}^*A_\omega^2\right] = -4\mbox{Re}\left[A_{2\omega}^* \frac{dA_{2\omega}}{dt}\right] = -2\frac{dN_{2\omega}}{dt}
\eeq
Depending on which field acts as the pump, the device can work as either a frequency doubler or a frequency-halver (degenerate OPO).  In the limit of many photons, the amplitude equations above are consistent with the quantum OPO equations derived in Section \ref{sec:02-opo}.
	
\subsection{Nondegenerate (SFG, DFG)}

In a nondegenerate $\chi^{(2)}$ device, there are three resonant fields -- $\omega_1, \omega_2$, and $\omega_3$, that satisfy the frequency-sum relation: $\omega_1 + \omega_2 = \omega_3$.  It is typically very difficult to design a device where all three modes resonate strongly, so in realistic devices the decay constant for one of the modes will be much larger than the other two.  But this does not affect the theory, which works for both good resonators and poor ones, as long as the mode lifetime is long compared to $1/\omega$.
\bea
	P^{(2)} & = & \epsilon_0 \chi^{(2)}:\mbox{Re}\left[\sum_\omega{\epsilon_0\sqrt{2\hbar\omega/\epsilon_0} A_\omega(t) E_\omega(x)e^{-i\omega t}}\right]\mbox{Re}\left[\sum_{\omega'}{\epsilon_0\sqrt{2\hbar\omega'/\epsilon_0} A_{\omega'}(t) E_{\omega'}(x)}e^{-i\omega't}\right] \nonumber \\
	& = & \mbox{Re}\left[2\hbar\sqrt{\omega_2\omega_3}\, A_{\omega_3} A_{\omega_2}^* \left(\chi^{(2)}:E_{\omega_2}^* E_{\omega_3} e^{-i\omega_1 t}\right)\right] 
		+ \mbox{Re}\left[2\hbar\sqrt{\omega_1\omega_3}\, A_{\omega_3} A_{\omega_1}^* \left(\chi^{(2)}:E_{\omega_1}^* E_{\omega_3} e^{-i\omega_2 t}\right)\right] \nonumber \\
	& & + \mbox{Re}\left[2\hbar\sqrt{\omega_1\omega_2}\, A_{\omega_1} A_{\omega_2} \left(\chi^{(2)}:E_{\omega_1} E_{\omega_2} e^{-i\omega_2 t}\right)\right] \nonumber \\
	& = & \mbox{Re}\left[\epsilon_0 \sqrt{\frac{2\hbar\omega_1}{\epsilon_0}}\sqrt{\frac{2\hbar\omega_2\omega_3}{\omega_1\epsilon_0}}\, A_{\omega_3} A_{\omega_2}^* \left(\chi^{(2)}:E_{\omega_2}^* E_{\omega_3} e^{-i\omega_1 t}\right)\right] \nonumber \\ 
	& & + \mbox{Re}\left[\epsilon_0 \sqrt{\frac{2\hbar\omega_2}{\epsilon_0}}\sqrt{\frac{2\hbar\omega_1\omega_3}{\omega_2\epsilon_0}}\, A_{\omega_3} A_{\omega_1}^* \left(\chi^{(2)}:E_{\omega_1}^* E_{\omega_3} e^{-i\omega_2 t}\right)\right] \nonumber \\
	& & + \mbox{Re}\left[\epsilon_0 \sqrt{\frac{2\hbar\omega_3}{\epsilon_0}}\sqrt{\frac{2\hbar\omega_1\omega_2}{\omega_3\epsilon_0}}\, A_{\omega_1} A_{\omega_2} \left(\chi^{(2)}:E_{\omega_1} E_{\omega_2} e^{-i\omega_2 t}\right)\right]
\eea
from which we may read off
\bea
	P_{\omega_1} & = & \epsilon_0 \sqrt{\frac{2\hbar\omega_2\omega_3}{\omega_1\epsilon_0}}\, A_{\omega_3} A_{\omega_2}^* \left(\chi^{(2)}:E_{\omega_2}^* E_{\omega_3}\right) \\
	P_{\omega_2} & = & \epsilon_0 \sqrt{\frac{2\hbar\omega_1\omega_3}{\omega_2\epsilon_0}}\, A_{\omega_3} A_{\omega_1}^* \left(\chi^{(2)}:E_{\omega_1}^* E_{\omega_3}\right) \\
	P_{\omega_3} & = & \epsilon_0 \sqrt{\frac{2\hbar\omega_1\omega_2}{\omega_3\epsilon_0}}\, A_{\omega_1} A_{\omega_2} \left(\chi^{(2)}:E_{\omega_1} E_{\omega_2}\right)
\eea
Now we may define a coupling constant
\beq
	\boxed{\epsilon = i\sqrt{\frac{2\hbar\omega_1\omega_2\omega_3}{\epsilon_0}} \int{\chi^{(2)}:E_{\omega_1}^*E_{\omega_2}^*E_{\omega_3}}}
\eeq
and use (\ref{eq:01b-env}) to get the equations of motion
\begin{empheq}[box=\fbox]{align}
	\left.\frac{dA_{\omega_1}}{dt}\right|_{\chi^{(2)}} &= \frac{1}{2}\epsilon A_{\omega_3}A_{\omega_2}^* \\
	\left.\frac{dA_{\omega_2}}{dt}\right|_{\chi^{(2)}} &= \frac{1}{2}\epsilon A_{\omega_3}A_{\omega_1}^* \\
	\left.\frac{dA_{\omega_3}}{dt}\right|_{\chi^{(2)}} &= -\frac{1}{2}\epsilon^* A_{\omega_1}A_{\omega_2}
\end{empheq}
As in the degenerate case, this system of equations conserves energy -- for every $\omega_3$ photon created, one $\omega_1$ photon and one $\omega_2$ photon must be annihilated:
\beq
	\frac{dN_{\omega_1}}{dt} = \frac{dN_{\omega_2}}{dt} = -\frac{dN_{\omega_3}}{dt}
\eeq

\section{$\chi^{(3)}$ Effects}

\subsection{Frequency-matched: THG, SPM, XPM}

In the frequency-matched case, two modes are on resonance -- one at $\omega$ and another at $3\omega$.  The $\chi^{(3)}$ effect can cause frequency up-conversion $\omega + \omega + \omega \rightarrow 3\omega$ and down-conversion: $3\omega \rightarrow \omega + \omega + \omega$.  Slightly more subtle, but probably more useful for devices, are self-phase modulation and cross-phase modulation, driven by the processes $\omega + \omega \rightarrow \omega + \omega$, $\omega + 3\omega \rightarrow \omega + 3\omega$, and $3\omega + 3\omega \rightarrow 3\omega + 3\omega$.  These effects create a power-dependent phase shift useful for, among other things, Kerr switching.

The $\chi^{(3)}$ polarization takes the following form:
\beq
	P^{(3)} = \epsilon_0 \chi^{(3)}:E\,E\,E \label{eq:01b-p3}
\eeq
where $E$ is has the mode decomposition (\ref{eq:01b-ebmodes}).  The full form of $P^{(3)}$ is rather cumbersome to write out, so I omit it here.  Suffice to say, I use Eqs.~(\ref{eq:01b-pmodes}) and (\ref{eq:01b-p3}) to compute $P_\omega$ and $P_{3\omega}$ in Mathematica.  The results are:
\bea
	 P_\omega & = & \frac{3\hbar\omega}{2\epsilon_0} (A_\omega^*A_\omega)A_\omega \left(\chi^{(SPM),\omega}:E_\omega^* E_\omega E_\omega\right) \nonumber \\
	 & & + \frac{9\hbar\omega}{\epsilon_0} (A_{3\omega}^*A_{3\omega})A_\omega \left(\chi^{(XPM)}:E_{3\omega}^* E_{3\omega} E_\omega\right) \nonumber \\
	 & & + \frac{3\sqrt{3}\hbar\omega}{2\epsilon_0} A_{3\omega}(A_\omega^*)^2 \left(\chi^{(THG)}:E_{\omega}^* E_{\omega}^* E_{3\omega}\right) \\
	 P_{3\omega} & = & \frac{9\hbar\omega}{2\epsilon_0} (A_{3\omega}^*A_{3\omega})A_{3\omega} \left(\chi^{(SPM),3\omega}:E_{3\omega}^* E_{3\omega} E_{3\omega}\right) \nonumber \\
	 & & + \frac{3\hbar\omega}{\epsilon_0} (A_{\omega}^*A_{\omega})A_{3\omega} \left(\chi^{(XPM)}:E_{\omega}^* E_{\omega} E_{3\omega}\right) \nonumber \\
	 & & + \frac{\hbar\omega}{2\sqrt{3}\epsilon_0} A_{\omega}^3 \left(\chi^{(THG)}:E_{\omega} E_{\omega} E_{\omega}\right)
\eea
Define the following dimensionless quantities:
\begin{empheq}[box=\fbox]{align}
	\chi_{\omega} &= -\frac{3\hbar\omega^2}{4\epsilon_0} \int{\chi^{(SPM),\omega}:E_\omega^*E_\omega^*E_\omega E_\omega} \\
	\chi_{3\omega} &= -\frac{3\hbar(3\omega)^2}{4\epsilon_0} \int{\chi^{(SPM),3\omega}:E_{3\omega}^*E_{3\omega}^*E_{3\omega} E_{3\omega}} \\
	\chi_{x} &= -\frac{3\hbar\omega(3\omega)}{4\epsilon_0} \int{\chi^{(XPM)}:E_{3\omega}^*E_{\omega}^*E_{3\omega} E_{\omega}} \\
	\chi_{h} &= i\frac{\sqrt{3}\hbar\omega}{2\epsilon_0} \int{\chi^{(THG)}:E_{3\omega}^*E_\omega E_\omega E_\omega}
\end{empheq}

Here, $\chi_\omega$ and $\chi_{3\omega}$ are the dimensionless self-phase modulation strengths for the $\omega$ and $3\omega$ fields and $\chi_x$ is the cross-phase modulation term (which is the symmetric), and $\chi_h$ is the third-harmonic generation term.  Using the envelope equations, (\ref{eq:01b-env}), the equations of motion for this system are:

\begin{empheq}[box=\fbox]{align}
	\frac{dA_\omega}{dt} &= -i\left[\chi_\omega (A_\omega^*A_\omega) + 2\chi_{x} (A_{3\omega}^*A_{3\omega})\right]A_\omega - 3\chi_{h}^* A_{3\omega}(A_\omega^*)^2 \\
	\frac{dA_{3\omega}}{dt} &= -i\left[\chi_{3\omega} (A_{3\omega}^*A_{3\omega}) + 2\chi_{x} (A_{\omega}^*A_{\omega})\right]A_{3\omega} + \chi_{h} A_\omega^3
\end{empheq}

When the $\chi_\omega$, $\chi_{3\omega}$, and $\chi_x$ are real, this conserves energy.  The third-harmonic term can trade a single $3\omega$ photon for three $\omega$ photons, and the rest of the terms do not create or annihilate photons at all -- they just dephase them.

However, in real systems, $\chi_\omega$, $\chi_{3\omega}$, and $\chi_x$ are complex and energy conservation is violated.  They will always have a negative complex part, giving rise to absorption.  Because this is a $\chi^{(3)}$ effect, it will be two-photon absorption.

\subsection{Unmatched: SPM, XPM}

Now consider the unmatched case -- there are two resonant modes $\omega_1$ and $\omega_2$, but $\omega_1 \neq 3\omega_2$ (or the other way around).  Third harmonic generation will not occur, but $\chi^{(3)}$ effects are still relevant because the self- and cross-phase terms do not require frequency matching.

As before, the polarization is given by $P^{(3)} = \epsilon_0 \chi^{(3)}:E\,E\,E$.  In this case, $P_{\omega_1}$ and $P_{\omega_2}$ are:
\begin{align}
	P_{\omega_1} & = \frac{3\hbar\omega_1}{2\epsilon_0} (A_{\omega_1}^*A_{\omega_1})A_{\omega_1} \left(\chi^{(SPM),\omega_1}:E_{\omega_1}^* E_{\omega_1} E_{\omega_1}\right) + \frac{3\hbar\omega_2}{\epsilon_0} (A_{\omega_2}^*A_{\omega_2})A_{\omega_1} \left(\chi^{(XPM)}:E_{\omega_2}^* E_{\omega_2} E_{\omega_1}\right) \\
	P_{\omega_2} & = \frac{3\hbar\omega_2}{2\epsilon_0} (A_{\omega_2}^*A_{\omega_2})A_{\omega_2} \left(\chi^{(SPM),\omega_2}:E_{\omega_2}^* E_{\omega_2} E_{\omega_2}\right) + \frac{3\hbar\omega_1}{\epsilon_0} (A_{\omega_1}^*A_{\omega_1})A_{\omega_2} \left(\chi^{(XPM)}:E_{\omega_1}^* E_{\omega_1} E_{\omega_2}\right)
\end{align}
Like before, we can define dimensionless self- and cross-phase modulation constants
\begin{empheq}[box=\fbox]{align}
	\chi_{11} &= -\frac{3\hbar\omega_1^2}{4\epsilon_0} \int{\chi^{(SPM),\omega_1}:E_{\omega_1}^*E_{\omega_1}^*E_{\omega_1} E_{\omega_1}} \\
	\chi_{22} &= -\frac{3\hbar\omega_2^2}{4\epsilon_0} \int{\chi^{(SPM),\omega_2}:E_{\omega_2}^*E_{\omega_2}^*E_{\omega_2} E_{\omega_2}} \\
	\chi_{12} &= -\frac{3\hbar\omega_1\omega_2}{4\epsilon_0} \int{\chi^{(XPM)}:E_{\omega_1}^*E_{\omega_2}^*E_{\omega_1} E_{\omega_2}}
\end{empheq}
and from these derive the field equations:
\begin{empheq}[box=\fbox]{align}
	\frac{dA_{\omega_1}}{dt} &= -i\left[\chi_{11} (A_{\omega_1}^*A_{\omega_1}) + 2\chi_{12}(A_{\omega_2}^*A_{\omega_2})\right] A_{\omega_1} \\
	\frac{dA_{\omega_2}}{dt} &= -i\left[\chi_{22} (A_{\omega_2}^*A_{\omega_2}) + 2\chi_{12}(A_{\omega_1}^*A_{\omega_1})\right] A_{\omega_2}
\end{empheq}

\subsection{Degenerate: Kerr Effect}

In the degenerate case, there are many modes, but they all have the same frequency.  This allows for additional processes not found in the SPM / XPM case.  For example, two photons can jump from mode 1 into mode 2 through the $\chi^{(3)}$ nonlinearity.  Such a process would be disallowed if the modes were not degenerate, since it violates conservation of energy.

Let $E_1, \ldots E_N$ be the set of degenerate modes.  The expression for $P_\omega$ has many terms (scales as $O(N^4)$) and is not shown here.  The equations of motion can be expressed in terms oa a dimensionless Kerr coupling tensor $\chi_{ijkl}$.  This is defined as follows:
\beq
	\boxed{\chi_{ijkl} = -\frac{3\hbar\omega^2}{4\epsilon_0} \int{\chi^{(SPM),\omega}:E_i^* E_j^* E_k E_l}}
\eeq	
Kerr nonlinearities for isotropic materials are usually quoted in terms of the nonlinear index $n_2$.  The full tensor form of $\chi^{(3)}$ can be complicated, since even in the isotropic case there are multiple tensor components.  But if all of the fields have the same polarization, this complication is avoided, and a scalar $\chi^{(3)}$ can be related to $n_2$ as follows:
\beq
	n_2 = \frac{3\chi^{(3)}}{4n^2c\epsilon_0}
\eeq
and the $\chi_{ijkl}$ coefficient becomes:
\beq
	\chi_{ijkl} = -\hbar\omega^2c \int{n_2 \epsilon_r E_i^* E_j^* E_k E_l} \label{eq:01b-chi-int}
\eeq
The equations of motion take a very elegant form:
\beq
	\boxed{\frac{dA_i}{dt} = -i \sum_{jkl}\chi_{ijkl} A_j^* A_k A_l}
\eeq
In the single-mode case, this reverts to the self-phase modulation effect discussed above.  In Section \ref{sec:02-kerr}, a very similar result is derived for Kerr resonators in the quantum regime.

\subsection{Size of Kerr Nonlinearity}
\label{sec:01b-kerrsize}

In Appendix \ref{ch:01a} we obtained a universal formula for $n_2$ in direct-gap semiconductors.  This formula relates $n_2$ to powers of the band gap and electron mass, times a universal function of $x \equiv E/E_g$:
\beq
	n_2 = K' \frac{\hbar c\sqrt{E_p}}{n_0^2 E_g^4} f_\chi(E/E_g) = \frac{0.0612\ \mbox{cm}^2/\mbox{GW}}{n_0^2 (m_e/m_0)^{1/2} (E_g/\mbox{eV})^{7/2}} f_\chi(E/E_g)
\eeq
where $f_\chi(x)$ is plotted in Figure \ref{fig:01b-f1}.  Consider a cavity with a single optical mode of volume $V = \tilde{V}(\lambda/n)^3$.  The Kerr constant $\chi$ is $n_2$ times the integral and the other factors in (\ref{eq:01b-chi-int}).  Suppose that $n_2$, $\epsilon_r$, and $E$ were constant-valued over this mode-volume.  Then $|E|^2 = 1/(\epsilon_r V)$ and therefore:
\beq
	\int{n_2\epsilon_r E^* E^* E E\,d^3x} = \frac{n_2}{n_0^2 V}
\eeq
Indeed, for a non-constant mode volume, this is a good way to \textit{define} $V$.  Using this definition, $\chi$ works out to:
\bea
	\chi & = & -\frac{\hbar\omega^2 c}{n_0^2 V} n_2 = \left(\frac{3.83\times 10^{9}\mbox{GW}/\mbox{cm}^2\mbox{s} \,n_0(x\,E_g/\mbox{eV})^5}{\tilde{V}}\right)\left(\frac{0.0612\ \mbox{cm}^2/\mbox{GW}}{n_0^2 (m_e/m_0)^{1/2} (E_g/\mbox{eV})^{7/2}} f_\chi(E/E_g)\right) \nonumber \\
	& = & \frac{2.34 \times 10^8 \mbox{s}^{-1} (E_g/\mbox{eV})^{3/2} x^5}{n_0 \tilde{V} \sqrt{m_e/m_0}}
\eea

\begin{figure}[t]
\begin{center}
\includegraphics[width=0.50\textwidth]{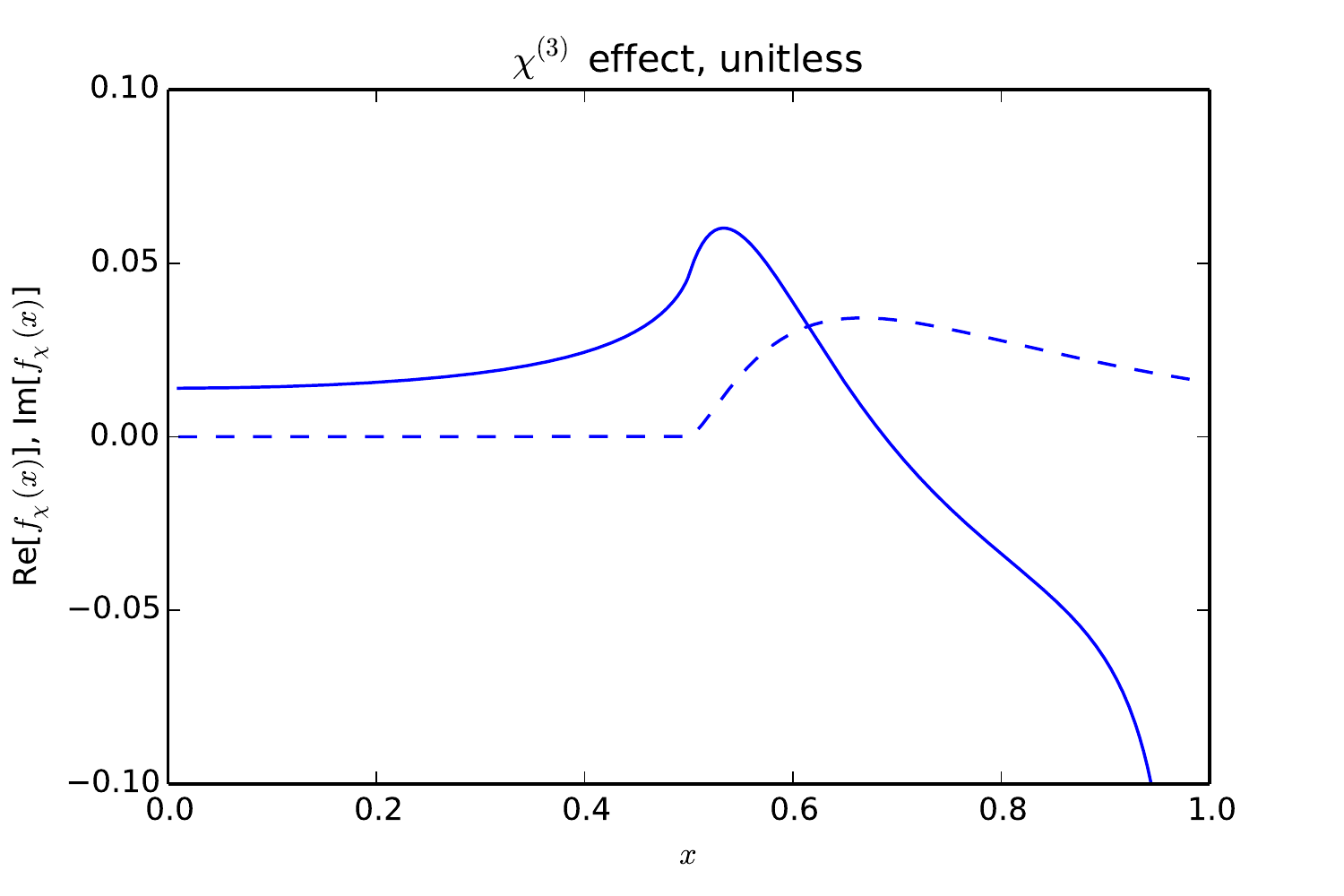}
\caption{Real (solid) and imaginary (dashed) parts of the universal Kerr function $f_\chi(x)$.}
\label{fig:01b-f1}
\end{center}
\end{figure}

A more relevant figure is $\chi/\kappa$, where $\kappa$ is the total cavity linewidth (radiative loss plus absorptive loss).  This is a dimensionless quantity, and as such, gives a good scale-free measure of how ``strong'' the nonlinearity is.  Since the quality factor is defined as $Q = \omega/\kappa$, we can write $\kappa = \omega/Q$, from which we obtain:
\bea
	\frac{\chi}{\kappa} = 1.54 \times 10^{-7} \frac{Q x^4 \sqrt{E_g/\mbox{eV}}}{n_0 \tilde{V} \sqrt{m_e/m_0}} f_\chi \label{eq:01b-chidim}
\eea
The field-dependent detuning shift is $\chi A_\omega^*A_\omega$, so the number of photons needed to shift the cavity by one linewidth is $N_{sw} = \kappa/\chi$.  When $\chi/\kappa \ll 1$, the nonlinearity is weak and only classical, many-photon states will experience nonlinear behavior.  If $\chi/\kappa \gtrsim 1$, the nonlinearity is strong and is important even for few-photon states.  Since the prefactor is very small and all of the dimensional terms (except $Q$) are of order unity, very high-$Q$ resonators will be needed to demonstrate low-photon Kerr switching.

Consider a hypothetical GaAs cavity with $E_g = 1.41$ eV, $\tilde{V} = 0.25$, $m_2 = 0.066m_0$, and $n_0 = 3.6$.  To maximize $n_2$ we could either work at $x \approx 0.5$ or $x \approx 1$.  The former case will give a weaker but ``cleaner'' Kerr nonlinearity, with less two-photon absorption.  The latter will give a stronger nonlinearity, but absorption and free-carrier effects (see below) will likely dominate.  I consider both cases below -- computed from Eq.~(\ref{eq:01b-chidim}) using the GaAs parameters:
\begin{align}
	x & = 0.5: & \frac{\chi}{\kappa} &= (1.20 \times 10^{-10})Q,\ \ \ & N_{sw} = \frac{1.20 \times 10^{9}}{Q} & \nonumber \\
	x & = 0.95: & \frac{\chi}{\kappa} &= ((-4.82 + 0.76i) \times 10^{-8}){Q},\ \ \ & N_{sw} = \frac{2.07 \times 10^7}{Q} & \label{eq:01b-kerrsize}
\end{align}
The $x = 0.95$ values might be promising -- $Q$'s of only 200,000 or so are required for Kerr switching at $N_{sw} = 100$.  But as I discuss below, the Kerr effect is probably masked by free-carrier effects which become very strong for driving fields near the band gap.  The cleaner $x = 0.5$ effect is much weaker, requiring very high $Q$ factors (tens of millions) to achieve switching at $N_{sw} = 100$.

From this we can conclude that the Kerr effect in bulk semiconductors is very weak in even very high-$Q$ cavities, and that ``strong'' / ``quantum'' Kerr effects will not be seen unless $Q$ is made extremely large, or new materials with larger $n_2$'s are used.

\section{Free-Carrier Effects}

Free carrier effects are another promising source of optical nonlinearity.  Because they involve real rather than virtual excitations, free carriers often give rise to much stronger optical nonlinearities, pushing useful phenomena like switching to lower powers.  In addition, the additional dynamical degree of freedom -- the carrier number -- allows for some new dynamics that are not possible with $\chi^{(2)}$ or $\chi^{(3)}$ systems.

Four effects are present in systems with free carriers:

\begin{enumerate}
	\item Excitation
	\item Dispersion / Absorption
	\item Decay
	\item Diffusion
\end{enumerate}

The strength of the free-carrier interaction is determined by the cavity geometry and a number of material parameters.  The most relevant of these is the carrier index change $\sigma_c = dn/dn_c$.  This is usually negative, and has units of $m^3$ since the carrier density has units of $m^{-3}$.  The material absorption coefficients $\alpha$ (single-photon) and $\beta$ (two-photon, equivalently Im[$n_2$]) are also very important, as they set the carrier excitation rate.

In addition to the optical fields, $n_c(x,t)$, the carrier density, will become a relevant dynamical field.  But because carriers diffuse quickly on relevant cavity timescales (even for very poor cavities).

\subsection{Excitation}

For linear absorption, the rate of carrier excitation is equal to the rate of photon absorption.  For two-photon absorption, it is half the rate of photon absorption (two photons needed to excite one carrier).  This gives:
\beq
	\left.\frac{dN_c}{dt}\right|_{\rm exc} = \frac{c\alpha}{n} \sum_{i} A_i^* A_i + \frac{1}{2} \mbox{Im}\left[\sum_{ijkl} \chi_{ijkl} A_i^* A_j^* A_k A_l\right]  \label{eq:01b-excite}
\eeq
Usually, the carriers will diffuse around the cavity on timescales fast compared to the cavity dynamics.  This means that they quickly equilibrate to a fixed distribution $n_c(x,t) = N_c(t) n_c(x)$, so all we need to keep track of is the carrier number, not their distribution.  When this is the case Eq.~(\ref{eq:01b-excite}) is sufficient to describe the dynamics.  When not, we'll need a more detailed model.

\subsection{Dispersion / Absorption}

Let $n_c(x)$ be the carrier density (not distinguishing between particles and holes here), and let $\sigma_c = dn/dn_c$ be the index change as a function of carrier density.  Applying the perturbation equations, one finds that the optical fields evolve as follows:
\beq
	\frac{dA_i}{dt} = \left(i\omega \sigma_c \int{n\,n_c E_i^* E_j}\right)A_j
\eeq
If the field is mainly confined to within the free-carrier material, and the material's properties are homogeneous, then this becomes:
\beq
	\frac{dA_i}{dt} = \left(\frac{i\omega\sigma_c}{n} \int{n_c \epsilon E_i^* E_j}\right)A_j
\eeq
The question is how $n_c$ plays into this equation.  In general, $n_c$ is not a constant, but depends on both space and time.  It will increase when carriers are excited, spread out due to diffusion, and decrease due to carrier decay.  But if we treat the distribution as constant, $n_c(x, t) = N_c(t) n_c(x)$, then we can write $dA_i/dt$ in terms of the total carrier number:
\beq
	\frac{dA_i}{dt} = \left(\frac{i\omega\sigma_c}{n} \int{n_c(x) \epsilon E_i^* E_j}\right)N_c A_j \equiv -i\Delta_{ij} N_c A_j \label{eq:01b-delta}
\eeq
Since free carriers are excited from the optical field, the free-carrier density $n_c(x)$ is always at least as spread-out as the optical field intensity $|E(x)|^2$.  Thus, $n_c(x)$ has is spread out over some volume $V_c \geq V_{ph}$, where $V_{ph}$ is volume occupied by the optical field.  From dimensional reasoning, one finds that the integral in (\ref{eq:01b-delta}) takes the form:
\beq
	\int{n_c(x) \epsilon E_i^* E_j} \sim \frac{1}{V_c} \label{eq:01b-vc}
\eeq
We can, in fact, \textit{define} $V_c$ so that Equation (\ref{eq:01b-vc}) is exact.  If all cavity modes are influenced equally by the carrier distribution, then $\Delta_{ij} = \delta \times \delta_{ij}$ and the carrier-dependent detuning $\delta$ becomes:
\beq
	\delta = -\frac{\omega \sigma_c}{n V_c}
\eeq
Not surprisingly, it is stronger for materials with stronger carrier effects, and for cavities with smaller modes.

\subsection{Decay / Diffusion}

Free carriers can decay through bulk recombination, stimulated emission, surface recombination, and diffusion out of the cavity.  The first two mechanisms are usually negligible, and carrier decay is usually dictated by the second two.

Treating carrier diffusion and surface recombination rigorously is a challenging task; see, e.g.\ \cite{Nozaki2010, JohnsonThesis}.  In summary, what happens is that the carrier distribution follows the diffusion equation:
\beq
	\frac{\partial n_c}{\partial t} = D\nabla^2 n_c + G
\eeq
where $G$ is the generation rate, and the surface-recombination boundary condition is satisfied:
\beq
	D\nabla n_c = v_s n_c \hat{n}
\eeq
Here, $D$ is the carrier diffusion constant and $v_s$ is the surface recombination velocity.  If a given profile for $G$ is assumed (the exact form is not too important), one can calculate the equilibrium $n_c(x, t)$.  From this one can derive an approximate exponential decay law for the carrier number $N_c$.  It will not be exact, but an exact treatment needs to take into account the entire carrier distribution, which is extremely cumbersome.  What we derive from the approximate method, which can be simulated with either finite-difference or Monte Carlo, is a carrier decay time constant $\gamma$, 
\beq
	\left.\frac{\d N_c}{\d t}\right|_{\rm decay} = -\gamma N_c
\eeq
While there is no analytic formula for $\gamma$, there are ways to estimate it.  If decay is limited by surface recombination, the decay constant will be roughly $\gamma \sim v_s/L$, where $L$ is the typical distance between adjacent surfaces.  In a silicon microring, $L \sim 0.35 \mu$m and $v_s = 0.24 \mu$m/ns, so $\gamma \sim 1.45\,\mbox{ns}^{-1}$.

If decay is limited by diffusion out of the cavity, the decay constant goes as $\gamma \sim D/L^2$, where $L$ is the cavity (field) dimension.  In an H0 cavity in InGaAsP, $L \sim 0.1 \mu$m and $D \sim 2 \mu\mbox{m}^2$/ns, giving a decay time of $\gamma \sim 200\,\mbox{ns}^{-1}$.

\subsection{Size of Free-Carrier Nonlinearity}
\label{sec:01b-fcsize}

The important figure of merit for the free-carrier cavity is the ratio of the carrier-dependent detuning to the linewidth, $\delta/\kappa$.  The number of carriers needed to switch the cavity by one linewidth is given by:
\beq
	N_{sw,c} = \frac{\kappa}{\delta} = \frac{1}{\delta/\kappa}
\eeq
This is an absolute lower bound to the energy required for switching.  Other factors may make the switching energy higher, but not lower.  One finds that $\delta/\kappa$ is given by:
\beq
	\frac{\delta}{\kappa} = \frac{Q \sigma_c}{n_0 V_c}
\eeq
An approximate formula for $\sigma_c$ was obtained in the previous section:
\beq
	\sigma_c = -\frac{\hbar^2 e^2}{2m_e n_0 \epsilon_0 E_g^2} \frac{1}{x^2(1-x^2)}
\eeq
$V_c$ can be approximated as $\tilde{V} (\lambda/n)^3$, where $\tilde{V} \sim 1$.  Plugging this in, one finds:
\beq
	\boxed{\frac{\delta}{\kappa} = \frac{e^2 n_0 E_g}{16 \pi^3 \hbar c^3 \tilde{V} m_e \epsilon_0}\frac{x}{1-x^2}Q = \bigl(3.62 \times 10^{-10}\bigr) \frac{n_0(E_g/\mbox{eV})}{\tilde{V}(m_e/m_0)} \frac{x}{1-x^2} Q} \label{eq:01b-dk}
\eeq
Inserting GaAs parameters $E_g = 1.41$ eV, $n_0 = 3.6$, $m_e = 0.066$ \cite{SheikBahae1991}, working very close to the band gap ($x = 0.95$) with a small cavity ($\tilde{V} = 0.25$ \cite{Nozaki2010}), we obtain:
\beq
	\frac{\delta}{\kappa} = (1.0 \times 10^{-6}) Q,\ \ \ N_{sw,c} = \frac{10^6}{Q}
\eeq
If the cavity photon number is limited by linear absorption, the steady-state carrier number and photon number will be related to each other, roughly:
\beq
	\frac{N_c}{\tau_c} \sim \frac{N_{ph}}{\tau_{ph}}
\eeq
The cavity photon number needed to achieve switching is therefore:
\beq
	N_{sw,ph} = \frac{\tau_{ph}}{\tau_c} N_c = \frac{16 \pi^3 \hbar c^3 \tilde{V} m_e \epsilon_0}{e^2 n_0 E_g} \frac{1-x^2}{x Q} \frac{\hbar Q}{x E_g\tau_c} = \frac{16 \pi^3 \hbar^2 c^3 \tilde{V} m_e \epsilon_0}{e^2 n_0 E_g^2 \tau_c} \frac{1-x^2}{x^2} = 1.82 \times 10^6 \frac{1-x^2}{x^2} \frac{\tilde{V}(m_e/m_0)}{n_0(E_g/\mbox{eV})^2 (\tau_c/\mbox{ps})}
\eeq
For a cavity of this size, $\tau_c \sim 2$ ps is a reasonable approximation.  Note that, if the carrier lifetime is limited by diffusion, then $\tau_c \sim \lambda^2 \sim E_g^{-2}$, so the switching photon number will ultimately be independent of $E_g$.  Plugging in GaAs parameters, with $x = 0.95$, we obtain
\beq
	N_{sw,ph} = 240
\eeq
This is independent of $Q$ or anything else.  Compare it to the Kerr result derived in (\ref{eq:01b-kerrsize}), $N_{sw} = 2 \times 10^7/Q$.  The Kerr effect will be dominant when $Q > 10^5$; free carriers will be dominant for $Q < 10^5$.  For the free-carrier effect, it is ideal to operate around $Q \sim 10^4$, since this is where $\tau_c$ and $\tau_{ph}$ will be comparable, so the carrier effect is probably dominant.  But the Kerr effect could be dominant if it were possible to construct a very high-$Q$ cavity.

As we can see from these figures, both the cavity photon number and the free-carrier number are quite large, of order 200 or so.  Since we are using state-of-the-art cavity parameters, this may cast doubt on our ability to reduce the power consumption of to truly quantum levels, where quantum noise effects become relevant.

However, if it were possible to confine the carriers and increase the carrier lifetime beyond the diffusion time, by a factor of say 10 or 100, say, then the switching photon number could be driven down to 100 or 10 provided we can build cavities with high enough $Q$.  The extreme limit of this is a quantum dot.  Which brings us back full circle to cavity QED.


\ifstandalone{\bibliography{NoteRefs}{}
\bibliographystyle{alpha}
}

\bibliography{NoteRefs}{}
\bibliographystyle{alpha}
\end{document}